\documentclass[12pt]{report}
\usepackage{xspace}
\usepackage{amsmath,mathtools}
\usepackage{amssymb}
\usepackage{graphicx}
\usepackage{url}
\usepackage[breaklinks=true,colorlinks=true,linkcolor=blue,urlcolor=blue,citecolor=blue]{hyperref}
\usepackage{subfig}
\usepackage{slashed}
\usepackage{enumitem}
\usepackage{authblk}
\usepackage{wrapfig}
\usepackage{hhline}
\usepackage{multirow}

\newcommand{\GeV}{\,\mathrm{GeV}}
\newcommand{\TeV}{\,\mathrm{TeV}}
\newcommand{\ie}{i.e.\ }
\newcommand{\eg}{e.g.\ }
\newcommand{\order}[1]{{\cal O}\left(#1\right)}
\newcommand{\avg}[1]{\left\langle\smash{#1}\right\rangle}
\newcommand{\as}{\alpha_s}

\newcommand{\zcut}{z_{\text{cut}}}

\newcommand{\ftrim}{f_{\text{trim}}}
\newcommand{\Rtrim}{R_{\text{trim}}}
\newcommand{\rtrim}{r_{\text{trim}}}
\newcommand{\zprune}{z_{\text{prune}}}
\newcommand{\Rprune}{R_{\text{prune}}}
\newcommand{\rprune}{r_{\text{prune}}}
\newcommand{\e}{\varepsilon}
\newcommand{\cf}{C_{F}}
\newcommand{\ca}{C_{A}}

\newcommand{\nc}{N_C}
\newcommand{\kt}{k_t}
\newcommand{\MSb}{\overline{\rm MS}}

\newcommand{\ord}{\mathcal{O}}
\newcommand{\gstrong}{g_s}
\newcommand{\muNP}{\mu_\text{NP}}

\newcommand{\tlambda}{\tilde{\lambda}}

\newcommand{\SD}{SoftDrop\xspace}
\newcommand{\ttt}[1]{{\small\texttt{#1}}}
\newcommand{\fastjet}{\texttt{FastJet}\xspace}

\usepackage{color}
\definecolor{darkgreen}{rgb}{0,0.5,0}
\definecolor{darkblue}{rgb}{0,0,0.7}
\definecolor{darkred}{rgb}{0.5,0,0.0}

\def\caesar{\textsc{Caesar} }

\def\jet{{\text{jet}}}
\DeclareMathOperator*{\argmin}{arg\,min}

\makeatletter
\g@addto@macro\bfseries{\boldmath}
\makeatother

\title{{\bf Looking inside jets}: an introduction to jet substructure and boosted-object phenomenology}
\author[1]{Simone Marzani}
\author[2]{Gregory Soyez}
\author[3]{Michael Spannowsky}
\date{}
\affil[1]{\emph{Dipartimento di Fisica, Universit\`a di Genova and INFN, Sezione di Genova, Via Dodecaneso 33, 16146, Italy}}
\affil[2]{\emph{IPhT, CNRS, CEA Saclay, Universit\'e Paris-Saclay, F-91191 Gif-sur-Yvette, France}}
\affil[3]{\emph{Institute of Particle Physics Phenomenology, Physics Department, Durham University, Durham DH1 3LE, UK}}
\begin{document}

\maketitle

%\begin{abstract}
%to be filled
%\end{abstract}

\chapter*{Preface}
The study of the internal structure of hadronic jets has become in recent years a very active area of research in particle physics. Jet substructure techniques are increasingly used in experimental analyses by the Large Hadron Collider collaborations, both in the context of searching for new physics and for Standard Model measurements. 
On the theory side, the quest for a deeper understanding of jet
substructure algorithms has contributed to a renewed interest in
all-order calculations in Quantum Chromodynamics (QCD).
%
% , attracting new people to the field, which
This has resulted in new ideas about how to design better observables
and how to provide a solid theoretical description for them.
In the last years, jet substructure has seen its scope extended, for
example, with an increasing impact in the study of heavy-ion
collisions, or with the exploration of deep-learning techniques.
Furthermore, jet physics is an area
% of research in particle physics
in which experimental and theoretical approaches meet together, where cross-pollination and collaboration between the two communities often bear the fruits of innovative techniques.
The vivacity of the field is testified, for instance, by the very successful series of BOOST conferences together with their workshop reports, which constitute a valuable picture of the status of the field at any given time. 

However, despite the wealth of literature on this topic, we feel that a comprehensive and, at the same time, pedagogical introduction to jet substructure is still missing. This makes the endeavour of approaching the field particularly hard, as newcomers have to digest an increasing number of substructure algorithms and techniques, too often characterised by opaque terminology and jargon. Furthermore, while first-principle calculations in QCD have successfully been applied in order to understand and characterise the substructure of jets, they often make use of calculational techniques, such as resummation, which are not the usual textbook material.
This seeded the idea of combining our experience in different aspects of jet substructure phenomenology to put together this set of lecture notes, which we hope could help and guide someone who moves their first steps in the physics of jet substructure.  

\pagebreak

\centerline{\bf Acknowledgements}
\vspace*{0.5cm}

Most of (if not all) the material collected in this book comes from years of collaboration and discussions with excellent colleagues that helped us and influenced us tremendously. In strict alphabetical order, we wish to thank %\sm{please add!}
Jon Butterworth,
Matteo Cacciari,
Simone Caletti,
Mrinal Dasgupta,
Frederic Dreyer,
Danilo Ferreira de Lima,
Steve Ellis,
Deepak Kar,
Roman Kogler,
Andrea Ghira,
Phil Harris,
Andrew Larkoski,
Matt LeBlanc,
Peter Loch,
David Miller,
Ian Moult,
Ben Nachman,
Tilman Plehn,
Sal Rappoccio,
Daniel Reichelt,
Jennifer Roloff,
Gavin Salam,
Steffen Schumann,
Lais Schunk,
Dave Soper,
Michihisa Takeuchi,
Jesse Thaler, and
Nhan Tran.

We would also like to thank Frederic Dreyer, Andrew Lifson, Ben
Nachman, Davide Napoletano, Gavin Salam and Jesse Thaler for helpful suggestions and
comments on the manuscript.

\vspace{0.5cm}

\begin{center}
Finally, we wish to remember Deepak, who left us too soon.
\end{center}

%  LocalWords:  calculational Butterworth Mrinal Dasgupta Larkoski
%  LocalWords:  Loch Nachmann Tilman Plehn Rappoccio Soper Thaler
%  LocalWords:  Nhan

\tableofcontents
% $Id: introduction.tex 610 2026-08-04 15:41:32Z smarzani $
%
% Introduction to jet physics, substructure and what is (and is not)
% covered in these lecture notes
% ------------------------------------------------------------------------
\chapter{Introduction and motivation}\label{chap:introduction}

The Large Hadron Collider (LHC) at CERN is the largest and most
sophisticated machine to study the elementary building blocks of
nature ever built.
At the LHC protons are brought into collision with a large
centre-of-mass energy --- 7~and 8~TeV for Run I (2010-13), 13~TeV for
Run~II (2015-18) and 13.6~TeV for Run III (2022-26)
--- to resolve the smallest structures in a controlled and
reproducible environment. As protons are not elementary particles
themselves, but rather consist of quarks and gluons, their
interactions result in highly complex scattering processes, often with
final state populated with hundreds of particles, which are measured
via their interactions with particle detectors.

Jets are collimated sprays of hadrons, ubiquitous in collider
experiments, usually associated with the production of an elementary
particle that carries colour charge, e.g.\ quarks and gluons. Their
evolution is governed by the strong force, which within the Standard
Model of particle physics is described by Quantum Chromodynamics
(QCD).  The parton (\ie quark or gluon) that initiates a jet may
radiate further partons and produce a (collimated) shower of quarks
and gluons, a so-called parton shower, that eventually turn into the
hadrons ($\pi$, $K$, $p$, $n$,...) observed in the detector.  The vast
majority of LHC events (that one is interested in) contain jets. They
are the most frequently produced and most complex objects measured at
the LHC multipurpose experiments, ATLAS and CMS.

When protons collide inelastically with a large energy transfer
between them, one can formally isolate a hard process at the core of
the collision, which involves one highly-energetic parton from each of
the two protons. These two partons interact and produce a few
elementary particles, like two partons, a Higgs boson associated with
a gluon, a top--anti-top pair, new particles, ... Since the energy of
this hard process is large, typically between 100~GeV and several TeV,
there is a large gap between the incoming proton scale and the hard
process on one hand, and between the hard process and the hadron scale
on the other. This leaves a large phase-space for parton showers to
develop both in the initial and final state of the collision.
This picture is clearly a simplification because we can imagine that
secondary parton-parton interactions might take place. These
multi-parton interactions constitute what is usually referred to as
the \emph{Underlying Event}.
To complicate things further, the LHC does not collide individual
protons, but bunches of $\mathcal{O}(10^{11})$ protons. During one
bunch crossing it is very likely that several of the protons scatter
off each other. While only one proton pair might result in an event
interesting enough to trigger the storage of the event on tape, other
proton pairs typically interact to give rise to hadronic activity in
the detectors. This additional hadronic activity from multiple proton
interactions is called \emph{pileup}. On average, radiation from
pileup is much softer than the jets produced from the hard
interaction, but for jet (and jet substructure) studies it can have a
significant impact by distorting the kinematic relation of the jet
with the hard process.

In recent years the detailed study of the internal structure of jets
has gained a lot of attention.  At LHC collision energy electroweak
(EW) scale resonances, such as the top quark, W/Z bosons and the Higgs
boson, are frequently produced beyond threshold, i.e.\ their energy
(transverse momentum) can significantly exceed their mass.
Therefore, analyses and searching strategies developed for earlier
colliders, in which EW-scale particles were produced with small
velocity, have to be fundamentally reconsidered.
Because EW resonances decay dominantly into quarks, when they are
boosted, their
decay products can become collimated in the lab-frame and result in one large and massive jet, often referred to
as a fat jet. Initially such a configuration was considered
disadvantageous in separating processes of interest (\ie processes
which included EW resonances) from the large QCD backgrounds (where
jets are abundantly produced from high-energy quarks and
gluons).
However, with the popularisation of sequential jet clustering
algorithms retaining the full information of the jet's recombination
history, it transpired that one can use the internal structure of jets
to tell apart jets that were induced by a decaying boosted EW
resonance or by a QCD parton.
{\em This investigation of the internal structure of jets is what one
  refers to as jet substructure}.

While the first jet substructure methods have been put forward in the
1990s and early
2000s~\cite{Seymour:1991cb,Seymour:1993mx,Seymour:1994by,Butterworth:2002tt},
it was only in 2008, with the proposal to reconstruct the Higgs boson
in vector-boson associated production~\cite{Butterworth:2008iy}, that
the interest in understanding and utilising jet substructure surged
tremendously~\cite{Abdesselam:2010pt,Altheimer:2012mn,Altheimer:2013yza,Adams:2015hiv,Larkoski:2017jix,Asquith:2018igt}. If
the Higgs boson, being spin and colour-less, the perfect prototype of
a featureless resonance could be reconstructed, surely other
EW-scale resonances proposed in many extensions of the Standard Model could be
discovered as well. 
Furthermore, jet substructure can be exploited in searches of physics beyond the Standard Model (BSM) not necessarily restricted to the EW scale. 
For instance, in many such extensions TeV-scale resonances are
predicted which decay subsequently into EW particles, which could either be Standard Model or BSM resonances.
Because of the mass differences, these EW-particles are typically
boosted and their hadronic decay might be reconstructed as a fat jet.
Thus, scenarios where jet substructure methods can benefit searches
for BSM physics are rather frequent.

\begin{figure}
\begin{center}
\includegraphics[width=0.9\textwidth]{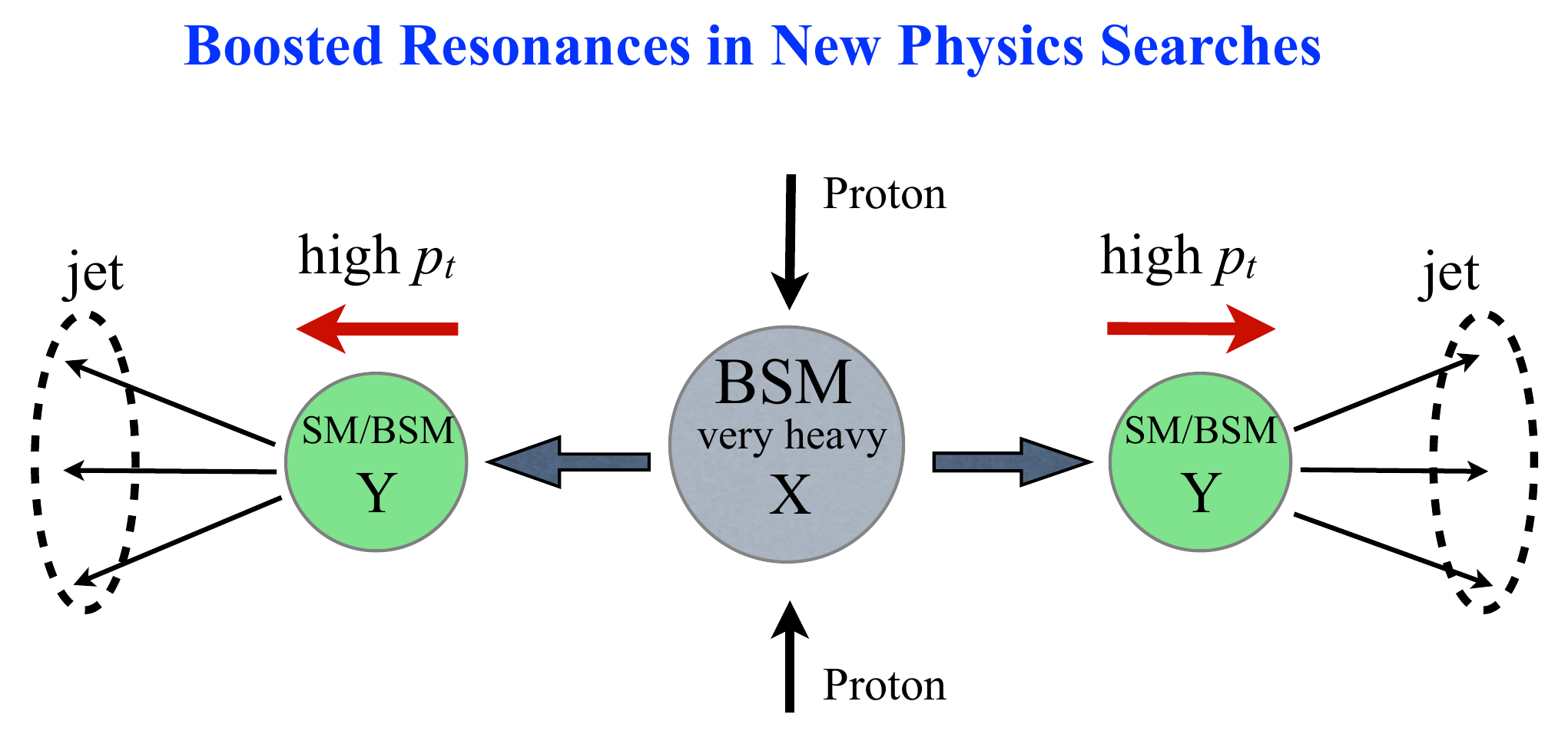}
\caption{Generic interaction sequence for the search of a very heavy BSM
  resonance that decays into electroweak-scale particles that
  subsequently decay hadronically.}
\label{fig:resonance} 
\end{center}
\end{figure}
A typical situation of interest for BSM searches using jet
substructure is illustrated in Fig.~\ref{fig:resonance}. A heavy new
resonance X with a mass of $\mathcal{O}{(1)}$ TeV is produced in a
proton-proton collision. This heavy BSM resonance quickly decays into
lighter states Y --- \eg W/Z/H bosons or lighter BSM particles ---,
with a mass around the EW scale. Particles Y are
typically produced with large transverse momentum ($p_t$) because
their mass is much smaller than the mass of the decaying particle
X. Finally, if a particle Y decays hadronically, because of its
large boost, its decay products in the lab frame are collimated and
reconstructed into a jet. The aim of jet substructure is therefore to
distinguish a \emph{signal jet}, originated from a boosted massive
particle, such as Y, from \emph{background jets}, which typically are
QCD jets originated from quarks and gluons.

Consequently various ways of discriminating the sources of jets have
been devised, with the aim of classifying a jet as of interest for a
specific search or measurement or not. Most methods to achieve this
classification task follow a two-step approach: firstly, the jet is
cleaned up ({\it groomed}), i.e.\ soft radiation which is unlikely to
come from the decaying resonance is removed, and, secondly, one
computes observables specifically designed to separate signal and
background jets based on the energy distribution amongst the remaining
jet constituents. Step two could be subdivided further into two
classes of classifiers: {\it jet-shape observables} and {\it
  prong-finders}. Jet-shape observables only consider the way the
energy is spatially distributed inside a jet, e.g.\ they do not take
into account the recombination history of the fat jet
itself. Prong-finders instead aim to construct hard subjets inside a
fat jet, i.e.\ isolated islands of energy inside the jet, and compare
properties of subjets, potentially including information on their
formation in the fat jet's recombination history.

Both jet shapes and prong finders aim to disentangle the different
topologies that characterise signal and background jets. For instance,
QCD jets are characterised by a hard core surrounded by soft/collinear
radiation, leading predominantly to jets with a one-prong
structure. EW bosons instead, such as W/Z and the Higgs, decays in a
quark-antiquark pair, which roughly share equal fractions of the heavy
particle momentum, leading to a two-prong structure. Finally, the top
quark preferentially decays into a bottom quark and a W boson, which
then decays into a pair of light quarks. Hence, top-initiated jets
feature a three-pronged structure.
It has been shown that grooming techniques and jet substructure
observables are sensitive to different effects during the complex
evolution of a jet, hence the classification of jets benefits
from combining several of these techniques~\cite{Soper:2010xk,
  Adams:2015hiv}.\footnote{Finding hard subjets (the task of
  prong-finders) and removing soft contamination (the task of
  groomers) are similar in practice. This means that tools which do
  one, very often do the other as well.} Thus, by combining groomers
and different subjet observables, high-level tagging methods can be
constructed for the reconstruction of top quarks, W/Z and Higgs bosons
and new-physics resonances.

Nowadays, the application of jet substructure techniques has
considerably widened and goes well beyond the identification of massive
boosted particles.  A specific example particularly relevant for this
book is that because grooming techniques reduce an observable's
sensitivity to soft physics, comparisons between experimental data and
first-principle calculations are less affected by non-perturbative
contamination. Consequently, the catalogue of Standard Model
measurements with jet substructure techniques keeps
growing. Furthermore, jet substructure techniques have found
applications also in initially unexpected ways. For instance, it has
been realised that substructure variables can be used to probe the jet
interaction with the quark-gluon plasma in heavy-ion collisions,
providing new observables helping to improve our understanding of this
difficult question. Finally, particle physics in general, and jet
physics in particular, is enjoying a period of rapid development as
innovative ideas and techniques exploiting machine-learning are poured
into the field. Unfortunately, this topic goes beyond the scope of
this book and we refer the interested reader to the recent
review~\cite{Larkoski:2017jix}.

Although this book focuses on LHC physics, it is worth pointing out
that jet substructure techniques have also been used at other
colliders, such as the Tevatron or RHIC. Due to the lower collision
energy, the scope of substructure studies is more limited. We can
however point the readers to
Refs.~\cite{Almeida:2011ud,Altheimer:2012mn} for reviews of
substructure studies at the Tevatron and to Ref.~\cite{Kauder:2017mhg}
for an explicit measurement by the STAR collaboration at RHIC.
 
%\vspace{0.2 cm}

These lecture notes aim to provide an accessible entry --- at the
level of graduate students with some expertise in collider
phenomenology --- to the quickly growing field of jet substructure
physics. Due to the complexity of the internal structure of jets, this
topic connects to subtle experimental and quantum-field theoretical
questions. In order to make these notes as self-contained as possible,
the first four chapters will provide a broad introduction to jet
physics and related QCD ingredients. First, we will give a brief
introduction into QCD and its application to collider phenomenology in
Chapter~\ref{chap:qcd-colliders}, focusing on those aspects that are
needed the most in jet physics. Chapter~\ref{chap:jets-and-algs} will
introduce the basics of jet definition and jet algorithms, including
some of the experimental issues related to defining and measuring
jets.  In Chapter~\ref{chap:calculations-jets}, we will discuss in
some detail a key observable in jet physics, namely the jet invariant
mass. We will show how its theoretical description requires an
all-order perturbative approach, and we will discuss various aspects of
this resummation.
We will dive into the topic of modern jet substructure in
Chapter~\ref{tools} where we will  first describe the main concepts and
ideas behind substructure tools and then try to give a comprehensive
list of the different approaches and tools which are currently
employed by the substructure community (theoretical and experimental).
Chapters~\ref{calculations-substructure-mass}-\ref{sec:curiosities}
explore our current first-principle understanding of jet
substructure with each chapter addressing a different application.
First, in Chapter~\ref{calculations-substructure-mass} we discuss
groomers, which have been the first tools for which an analytic
understanding became available. In particular, we will go back to the
jet mass and we will study in detail how its distribution is modified
if grooming techniques are applied.  In the remaining chapters, we
will discuss more advanced topics such as quark/gluon discrimination
in Chapter~\ref{sec:calc-shapes-qg}, two-prong taggers in
Chapter~\ref{chap:calc-two-prongs}, Sudakov safety in
Chapter~\ref{sec:curiosities}, and the Lund jet plane in Chapter~\ref{lundplane}.
Finally, in the last part of this book, we will discuss the current
status of searches and measurements using jet substructure in
Chapter~\ref{searches-measurements}.

%\vspace{0.2 cm}

A large part of these lecture notes will focus on our current
first-principle understanding of jet substructure in QCD.
The key observation to keep in mind in this context is the fact that
substructure techniques are primarily dealing with boosted jets, for
which the transverse momentum, $p_t$, is much larger than the mass,
$m$.
From a perturbative QCD viewpoint, this means that powers of the
strong coupling will be accompanied with large logarithms of $p_t/m$,
a common feature of QCD whenever we have two largely disparate scales.
For these situations, a fixed-order perturbative approach is not
suited and one should instead use all-order, {\em resummed},
calculations which focus on including the dominant
logarithmically-enhanced contributions at all orders in the strong
coupling.
Chapter~\ref{chap:calculations-jets} will present a basic
introduction to resummation taking the calculation of the jet mass as
a practical example.

There exist different approaches on how to tackle this type of
calculations. On the one hand, one could analyse the structure of
matrix elements for an arbitrary number of quark and gluon emissions
in the soft/collinear limit and from that derive the all-order
behaviour of the distribution of interest. In this context, the
coherent branching algorithm~\cite{Catani:1990rr,Catani:1992ua}
deserves a special mention because not only it is the basis of
angular-ordered parton showers, but it also constitutes the foundation
of many resummed calculations (for a review see
e.g.~\cite{Luisoni:2015xha}). Other approaches to all-order
resummation instead take a more formal viewpoint and try to establish
a factorisation theorem for the observable at hand, therefore
separating out the contribution from hard, soft and collinear
modes. This point of view is, for instance, the one taken when
calculations are performed in Soft-Collinear Effective Theory
(SCET). For a pedagogical introduction to SCET, we recommend
Ref.~\cite{Becher:2014oda}.

In this book, we will use the former approach, but we will try to
point out the relevant literature for SCET-based calculations too.
That said, our aim is not to present a rigorous and formal proof of
resummed calculations, but rather to lay out the essential ingredients
that go into these theoretical predictions, while keeping the
discussion at a level which we think it is understandable for readers
with both theoretical and experimental backgrounds.
In particular, even though
Chapters~\ref{calculations-substructure-mass}-\ref{lundplane}
start with (sometimes heavy) analytic QCD calculations, we will always
come back to comparisons between these analytic calculations and Monte
Carlo simulations in the end.
This will allow us to discuss the main physical features of the
observed distributions and how they emerge from the analytic
understanding. It will also allow us to discuss how the analytic
results obtained in perturbative QCD are affected by non-perturbative
corrections.

%% GS helper for auctex
%%% Local Variables:
%%% mode: latex
%%% TeX-master: "notes"
%%% End:

%  LocalWords:  EW BSM SCET

% $Id: qcd-colliders.tex 610 2026-08-04 15:41:32Z smarzani $
%
% Introduction to QCD at colliders (not exactly clear what this means)
% ------------------------------------------------------------------------
\chapter{Introduction to QCD at Colliders}\label{chap:qcd-colliders}
Jet physics is QCD physics. Therefore, a solid and insightful
description of jets and their substructure relies on a deep understanding of the dynamics of strong interactions in collider experiments. 
QCD is an incredibly rich but, at the same time, rather complicated theory and building up a profound knowledge of its workings goes beyond the scope of this book. At the same time, some familiarity with perturbative calculations in quantum field theory is necessary in order to proceed with our discussion.
Therefore, in this chapter we recall the essential features of the theory of strong interactions that are needed in jet physics.
Because we aim to make this book accessible to both theorists and experimenters who want to take their first steps in jet substructure, we are going to take a rather phenomenological approach and we will try to supplement the lack of theoretical rigour with physical intuition. QCD itself helps us in this endeavour because the dynamics that characterises jet physics is often dominated by soft and collinear radiation, \ie emissions of partons that only carry a small fraction of the hard process energy or that are emitted at small angular distances. The structure of the theory greatly simplifies in this limit and many results can be interpreted using semi-classical arguments. The price we have to pay is that, if we want to achieve a reliable description of observables in the soft and collinear regions of phase space, we have to go beyond standard perturbation theory and consider the summation of some contributions to all orders in the perturbative expansion.

\section{The theory of strong interactions}
Let us begin our discussion with a historical detour. 
The quest for a coherent description of strong interactions started in the 1960s and had the principal aim of understanding and classifying the plethora of new
particles produced at the first particle colliders. 
Indeed, as machines to accelerate and collide particles were becoming more powerful, many new strongly-interacting particles, collectively referred to as hadrons, were produced, leading to what was defined as a particle zoo. Some of these particles shared many similarities to the well-known protons, neutrons and pions and could therefore be interpreted as excited states of the former. Other particles instead presented new and intriguing properties. 
A major breakthrough was realised with the \emph{quark model}.
This model successfully applied the formalism of group theory to describe the quantum numbers of the hadrons known at that time. It introduced fundamental constituents with fractional electric charge called quarks and described mesons and baryons in terms of the different combinations of these constituents. However, the model made no attempt to describe the dynamics of these constituents. 
The quark model led to another important discovery: the introduction of a new degree of freedom, which was termed colour. Its introduction was made necessary in order to recover the symmetry properties of the wave-function of some baryonic states such as the $\Delta^{++}$ or the $\Omega^-$.

Alongside hadron spectroscopy, scattering processes were used to study the structure of the hadrons. In this context, experiments where beams of electrons were scattered off protons played a particularly important role, as they were used to probe the structure of the protons at increasingly short distances. 
The experiments in the deep-inelastic regime, where the target protons were destroyed by the high-momentum-transfer interaction with the electron, pointed to peculiar results. The interaction was not between the electron and the proton as a whole, but rather with pointlike constituents of the proton, which behaved as almost-free particles. 
In order to explain these experimental data, the \emph{parton model} was introduced in the late Sixties.
 The basic assumption of this model is that in high-energy interactions, hadrons behave as made up of almost free constituents, the partons, which carry a fraction of the hadron momentum.
 Thus, the description of the hadron is given in terms of partonic
distributions that represent the probability of having a particular
parton which carries a fraction of the total hadron's momentum.

The quark model and the parton model aim to describe rather different physics: the former classifies the possible states of hadronic matter, while the latter applies if we want to describe how a hadron interacts at high energy. However, it is very suggestive that they both describe hadronic matter as made up of more elementary constituents. 
A successful theory of the strong force should be able to accommodate both models. Nowadays Quantum Chromo-Dynamics (QCD) is accepted as the theory of strong interactions.
It is a non-Abelian gauge theory and the symmetry group is the local version of the colour symmetry group SU(3). The theory describes the interaction between fermionic and bosonic fields
associated to quarks and gluons respectively (see for instance~\cite{Peskin:257493,Schwartz:2013pla,Collins:1350496,Ellis:1991qj,Campbell:2286381} and references therein).

The QCD Lagrangian
\begin{equation}\label{eq:qcd-lagrangian}
\mathcal{L}= -\frac{1}{4}F^A_{\mu\nu}F_A^{\mu\nu}+
\sum_\text{flavours}\bar{\psi}_a (i \gamma_\mu D^\mu-m)_{ab}\psi_b \,,
\end{equation}
where  $F^A_{\mu \nu}$ is
the gluon field strength, defined by:
\begin{equation}\label{eq:qcd-fs}
F_{\mu\nu}^A= \partial_\mu A_\nu^A - \partial_\nu A_\mu^A +
g_sf^{ABC}A_\mu^BA_\nu^C \,.
\end{equation}
and $D_\mu$ is the covariant derivative
\begin{equation}
\left( D_\mu\right)_{ab}=\partial_\mu \delta_{ab}- i g_s A_\mu^A t^A_{ab},
\end{equation}
where $t^A$ are the algebra generators.
In the above equations both lower-case and upper-case indices refer to SU(3), the formers denote indices in the (anti)-fundamental representation, while the latter in the adjoint one. We note that a sum over quark flavours, namely up, down, charm, strange, top, and bottom is indicated. Strong interactions are completely blind to this quantum number and therefore the only distinction between different quark flavours in this context comes about only because of the mass. Note that the quark masses span several orders of magnitude and therefore the related phenomenology is extremely different!

\begin{figure}
\begin{center}
\includegraphics[scale=.4]{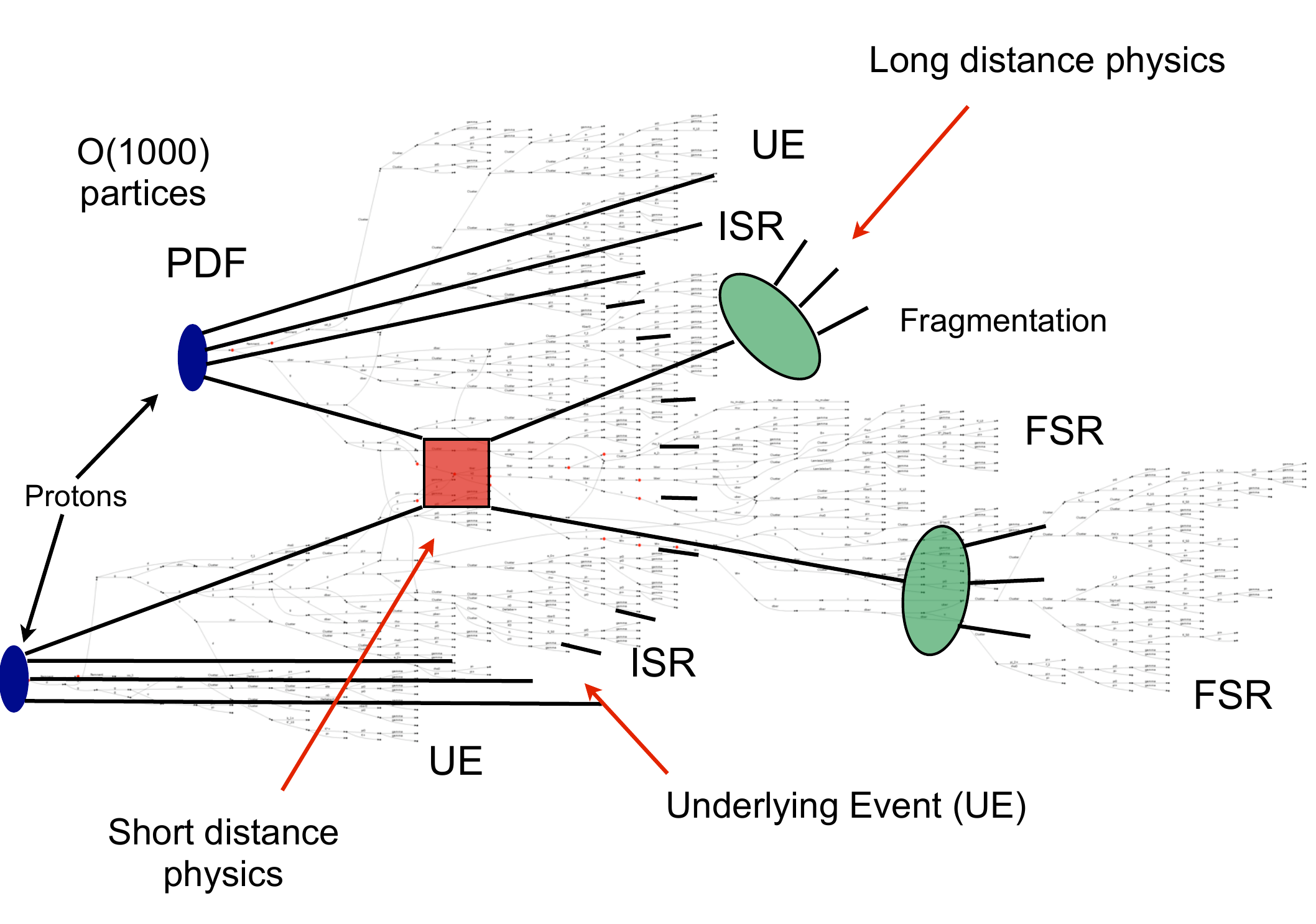}
\caption{A schematic representation of a typical high-energy proton-proton collision.}
\label{fig:tth_event}
\end{center}
\end{figure}

A remarkable feature of QCD is the fact that the strong coupling $\as= g_s^2/4 \pi$ is a decreasing function
of the energy involved in the process.
For this reason QCD has a
low energy regime, in which the theory is strongly-interacting and
a high-energy one, in which it is asymptotically free. This implies
that strong processes are computable in perturbation theory
if a sufficiently high-energy scale is involved. 
Thus, asymptotic freedom provides the theoretical justification of the parton model, which can be understood as the lowest
order approximation of a perturbative QCD calculation.

The theoretical description of high energy collisions of protons is fairly complex. In a typical event, hundreds of particles are produced, as depicted in Fig.~\ref{fig:tth_event}.
The short-distance, \ie high-energy, part of the process can be computed using perturbation theory; however, long-distance physics is driven by the non-perturbative nature of QCD at low energy scales. 
Fortunately, there exists a theorem in QCD that enables us to separate
the perturbative, \ie calculable, part of a process from the
non-perturbative one, which can be described in terms of parton
distribution (or fragmentation) functions. These objects essentially
generalise the probability distributions introduced by the parton
model.  Parton distributions are universal, \ie they do not depend on
the particular process, and they can be determined by fitting data
from previous experiments. This is the collinear factorisation theorem
and although it has been explicitly proven only for a few processes
(deep inelastic scattering of an electron off a proton and the
Drell-Yan process), it is usually considered valid and is used
ubiquitously in perturbative QCD calculations.\footnote{However,
  examples of short-distance processes that exhibit collinear
  factorisation breaking have been identified and
  studied~\cite{Forshaw:2006fk,Forshaw:2008cq,Catani:2011st,Forshaw:2012bi}.}
In collinear factorisation, the total cross section of inelastic proton-proton scattering to produce a final state $n$ can be calculated with the formula 
\begin{equation}
\sigma = \sum_{a,b} \int_0^1 dx_a dx_b \int d\Phi_n f_a^{h_1}(x_a,\mu_F) f_b^{h_2}(x_b, \mu_F) \frac{1}{2 \hat{s}} |\mathcal{M}_{ab \to n}|^2(\Phi_n;\mu_F,\mu_R)\;,
\label{eq:master}
\end{equation}
where $f_a^h(x,\mu)$ denotes the parton distribution functions, which depend on the longitudinal momentum fraction $x$ of parton $a$ with respect to its parent hadron $h$, and on an arbitrary energy scale called factorisation scale $\mu_F$. In the above equation, $d\Phi_n$ denotes the differential phase space element over $n$ final-state particles,
\begin{equation}
d\Phi_n = \prod_{i=1}^{n} \frac{d^3p_i}{(2\pi)^3 2 E_i} (2 \pi)^4 \delta^{(4)}(p_a + p_b - \sum_{i=1}^{n} p_i)\;,
\label{eq:dLIPS}
\end{equation}
where $p_a$ and $p_b$ are the initial-state momenta.
The convolution of the squared matrix element $|\mathcal{M}_{ab\to
  n}|^2$, averaged over initial-state spin and colour degrees of
freedom, with the Lorentz-invariant phase space $\Phi_n$ and
multiplied by the flux factor $1/(2 \hat{s}) = 1/(2 x_a x_b s)$
results in the calculation of the parton-level cross section
$\hat{\sigma}_{ab \to n}$. The {\it cross section master formula} of
Eq.~(\ref{eq:master}) holds to all orders in perturbation theory, up
to terms which are suppressed by
$\left(\frac{\Lambda_\text{QCD}^2}{Q^2_\text{min}}\right)^p$, where $\Lambda_\text{QCD}$ is the non-perturbative QCD scale, $Q_\text{min}$ is the minimum hard energy scale
probed by the process, and typically $p=1$.  
For instance, in the case of the inclusive jet cross-section, we typically have $Q_\text{min}=p_t$, the jet transverse momentum. 
In what follows we will spend plenty of time discussing the invariant mass $m$ of a jet with
large transverse momentum $p_t$. In that case, we will be able to
identify $Q_\text{min}=m$.

Protons consist of many partons, each carrying a fraction of the
total proton energy. The partons of the two protons that interact with
each other via a large momentum transfer and the wide gap between this hard scale the proton mass scale is typically filled by the emission of extra partons, which is usually referred to as {\it initial state radiation}. 
Furthermore, because the hard momentum transfer can be much smaller than the proton collision energy (13 TeV), initial-state radiation is not necessarily soft.
In the hard process, large interaction scales and momentum transfers are probed. New heavy particles can be produced and novel interactions can be tested. Thus, the nature of the hard interaction process leaves a strong imprint in the topological structure and the composition of the whole final state.
However, if colour-charged particles\footnote{We are focusing here on QCD-induced parton showers. EW interactions can also give rise to parton showers, however, due to $\alpha \ll \alpha_s$ their contributions are suppressed. 
However, it should be noted the impact of EW corrections increases with the energy and so it becomes imperative to consistently include them in order to perform accurate phenomenology  at future higher-energy colliders.} are produced during the hard interaction process, they are likely to emit further partons, \ie {\it final state radiation}, to evolve from the hard interaction scale down to the hadronisation scale $\mathcal{O}(\Lambda_\mathrm{QCD})$, where non-perturbative processes rearrange the partons into colour-neutral hadrons. 

The proton's energy carried by the
spectator partons, \ie partons of the proton that are not considered
initial states of the hard interaction process, is mostly directed
into the forward direction of the detector, but a non-negligible
amount of radiation off these spectator partons can still end up in
the central region of the detector. This so-called Underlying
  Event (UE), contributing to the measured radiation in a detector,
is, on average, softer, \ie has lower transverse momentum, than for
example the decay products of the hard process or initial state
radiation. For jet substructure observables, however, it plays an
important role as it can complicate the extraction of information from
observables that rely on the details of the energy distribution inside
a jet.

Furthermore, protons are accelerated and collided in bunches.
When two bunches of protons cross at an interaction point, multiple
proton-proton collisions can occur simultaneously. What is observed in
the detectors is therefore a superposition of these many events.
When one of these collisions is hard and deemed interesting enough by
the experiments' triggers to be stored on tape, it therefore
overlays in the detector with all the other simultaneous, mostly
soft, collisions.
This effect is known as pileup and presents a challenge to the
reconstruction of the objects seen in the detectors in general and of
the hadronic part of the event, in particular.
To give a quantitative estimate, at the end of Run II of the LHC (late
2018), the machine delivers a luminosity
${\cal L}\sim 2\times 10^{34}\:\text{cm}^{-2}\text{s}^{-1}$ which, for a
bunch spacing of $25$ nanoseconds and a typical total proton-proton
cross-section of $100$~mb, corresponds to an average of 50 interactions
per bunch-crossing (assuming that they are Poisson-distributed).
We refer the interested reader to a recent review on this subject in
the context of jet physics, written by one of us~\cite{Soyez:2018opl}.

\section{Generalities on perturbative calculations}\label{sec:generalities}

The calculation of the matrix element in Eq.~(\ref{eq:master}) is
usually approximated by a perturbative series in powers of the strong
coupling, henceforth the \emph{fixed-order} expansion. The evaluation
of such perturbative expansion, and more generally the development of improved
techniques to compute {\em amplitudes}, is one of the core activities of QCD phenomenology. 
In this framework, theoretical precision is achieved by computing cross-sections $\sigma$ including increasingly higher-order corrections in the strong coupling $\as$
\begin{equation} \label{fixedorder_sigma}
\sigma\left(v\right) = \sigma_0 + \as \, \sigma_1 +\as^2 \, \sigma_2 +\as^3 \, \sigma_3 + \ord(\as^4),
\end{equation}
where $v$ is a generic observable, which for definiteness we take dimensionless. In the above expression leading order (LO) contribution $\sigma_0$ is the Born-level cross section for the scattering process of interest. Subsequent contributions in the perturbative expansion $\sigma_i$ constitute the next-to$^i$-leading order (N$^i$LO) corrections. In the language of Feynman diagrams, each power of $\as$ corresponds to the emission of a QCD parton, either a quark or a gluon, in the final state or to a virtual correction. 
The theoretical community has put a huge effort in computing higher-order corrections. LO cross-sections can be computed for an essentially arbitrary number of external particles. Automation has been achieved in recent years also for NLO calculations and an increasing number of NNLO calculations is now available in computer programs. 
Moreover, for hadron-collider processes with simple topologies, recent
milestone calculations have achieved N$^3$LO
accuracy~\cite{Anastasiou:2015ema,Dreyer:2016oyx}. A particularly
important example which  falls under this category is the main
production channel of the Higgs boson (through gluon-gluon fusion).
One of the main challenges in this enterprise is the treatment of the infra-red region. As it is going to be discussed in the following, the emissions of soft and/or collinear partons is also problematic because it can generate large logarithmic terms in the perturbative coefficients, thus invalidating the fixed-order approach.

It is well known that calculations of Feynman diagrams are plagued by the appearance of divergences of different nature. 
Loop diagrams can exhibit ultraviolet singularities. Because QCD is a renormalisable theory, such infinities can be absorbed into a redefinition of the parameters that enter the Lagrangian, \eg the strong coupling $\as$.
Moreover, real-emission diagrams exhibit singularities in particular
corners of the phase-space. More specifically, the singular
contributions have to do with collinear, \ie small angle, splittings
of massless partons and emissions of soft gluons, off both massless and
massive particles. Virtual diagrams also exhibit analogous infra-red
and collinear (IRC) singularities and rather general
theorems~\cite{Blochnordsieck,Kinoshita,Lee} state that such
infinities cancel at each order of the perturbative series Eq.~(\ref{fixedorder_sigma}), when real and virtual corrections are added together, thus leading to observable transition probabilities that are free of IRC singularities.
We will explicitly discuss infra-red singularities in an NLO calculation in the next section.
Moreover, in order to be able to use the perturbative expansion of
Eq.~(\ref{fixedorder_sigma}), one has to consider observables
$v$ that are infra-red and collinear (IRC) safe, \ie measurable quantities that do not spoil the above theorems. We will come back to a more precise definition of IRC safety in Sec.~\ref{sec:IRC-safety}.

It is worth pointing out that, in
  practice, non-perturbative effects like hadronisation regulate
  soft and collinear divergences, so that cross-sections are
  finite. The requirement of IRC safety means that an observable can
  be computed reliably in perturbative QCD, up to non-perturbative
  power corrections, which decrease as the hard scale of the process increases.
  Moreover, from an
  experimental viewpoint, the finite resolution of the detectors also acts
as a regulator, thus preventing the occurrence of actual
singularities. However, this in turn would be reflected on a possibly
strong dependence of theoretical predictions on the detector
resolution parameters, which one wishes to avoid.

The fixed-order expansion of Eq.~(\ref{fixedorder_sigma}) works well
if the measured value of the observable is $v\simeq 1$, a situation in which there is no significant hierarchy of scales.
However, it loses its predictive power if the measurement of
$v\ll 1$ confines the real radiation into a small corner of
phase-space, while clearly leaving virtual corrections
unrestricted. 
For IRC safe observables the singular terms still
cancel, but logarithmic corrections in $v$ are
left behind, causing the coefficients $\sigma_i$ to become large, so
that $\as^i \sigma_i \sim 1$. 
Because these logarithmic corrections
are related to soft and/or collinear emissions, one can expect at most
two powers of $L=\log\big(\frac{1}{v}\big)$~\footnote{Throughout this book we denote with $\log(x)$ the natural logarithm of $x$.} for each power of the
strong coupling.
For example, when $v$ is sensitive
only to angles up to $\theta_\text{cut}\ll 1$, one should expect large
(collinear) logarithms of $1/\theta_\text{cut}$, and when
$v$ is sensitive only to $|k_{3\perp}|$ up to
$|k_{3\perp}^\text{cut}|\ll 1$, one should expect large (soft) logarithms of $Q/|k_{3\perp}^\text{cut}|$.

Let us consider the {\it cumulative} cross-section for measuring a
value of the observable of interest which is less than a given value
$v$, normalised to the inclusive Born-level cross-section
$\sigma_0$.\footnote{Note that in the literature, $\Sigma$
  sometimes refers to the un-normalised cumulative cross-section.}
% i.e.\ the so-called cumulative, then
We have
\begin{align} 
  \Sigma\left(v\right)
  & = \int_0^v dv' \frac{1}{\sigma_0}\frac{d\sigma}{dv'} \\
  & = 1 + \as \, \left( \sigma_{12} L^2+\sigma_{11} L+ \dots \right)  
+\as^2 \,\left( \sigma_{24} L^4+\sigma_{23} L^3+ \dots \right)   + \ord(\as^n L^{2n}).\label{expanded_sigma}
\end{align}
All-order resummation is then a re-organisation of the above perturbative series. For many observables of interest, the resummed expression exponentiates, leading to
\begin{equation} \label{resummed_sigma}
\sigma\left(v \right) = \sigma_0 \, g_0 \exp \left[ L g_1(\as L)+g_2(\as L)+\as g_3(\as L)+ \dots \right],
\end{equation}
where $g_0$ is a constant contribution which admits an expansion in $\as$. In analogy to the fixed-order terminology, the inclusion of the contribution $g_{i+1}$, $i\ge0$, leads to next-to$^i$-leading logarithmic (N$^i$LL) accuracy.

Fixed-order Eq.~(\ref{fixedorder_sigma}) and resummed Eq.~(\ref{resummed_sigma}) expansions are complementary.
On the one hand, fixed-order calculations fail in particular limits of
phase-space, indicating the need for an all-order approach. On the
other hand, all-order calculations are only possible if particular
assumptions on the emission kinematics are made. Thus, the most
accurate theoretical description for the observable $v$ is
achieved by matching the two approaches \eg using (other so-called
{\em matching schemes} exist)
\begin{equation}\label{eq:matching-add}
\sigma^\text{matched}(v)=\sigma^\text{fixed-order}(v)+\sigma^\text{resummed}(v)-\sigma^\text{double counting}(v).
\end{equation}

\section{Factorisation in the soft and collinear limits}\label{sec:qcd_soft_fact}
\begin{figure}
\begin{center}
  \subfloat[]{\includegraphics[scale=0.6,page=1]{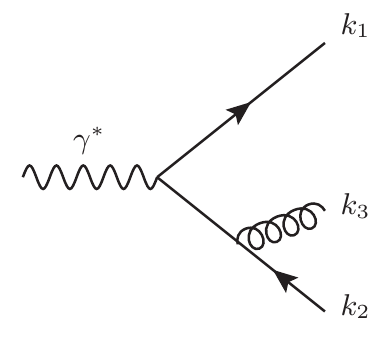} \label{eenloR1}} \hfill
  \subfloat[]{\includegraphics[scale=0.6,page=2]{figures/eenlo} \label{eenloR2}} \hfill
  \subfloat[]{\includegraphics[scale=0.6,page=3]{figures/eenlo} \label{eenloV}} 
\caption{Feynman diagrams contributing to the cross-section of
  $e^{+}e^{-}\to q\bar{q}$ at $\mathcal{O}(\alpha_s)$.}\label{fig:sigma_nlo}
\end{center}
\end{figure}

In order to highlight the structure of IRC singularities in matrix elements, we consider the calculation of the NLO QCD corrections in the soft limit. For this presentation we closely follow the review~\cite{Luisoni:2015xha}.
In order to simplify our discussion, rather than presenting a
calculation for proton-proton collisions, for which we would have to
include parton distribution functions and  discuss how to treat
initial-state radiation, we focus our discussion on a process in
electron-positron collisions, for which we can concentrate on QCD
radiation off the final-state quarks. We will show that the requirement of
  IRC safety implies with some constraints on observables to guarantee
  the cancellation of divergences when combining real and virtual
  diagrams. Furthermore, we will also see that, if we consider an
inclusive observable, we obtain an NLO correction which is free of
large logarithms.

Let us therefore consider the $\mathcal{O}(\as)$ correction to the process
\begin{equation}
e^+ e^- \to \gamma^* \to  q \bar q.
\end{equation}
The relevant Feynman diagrams are shown in Fig.~\ref{fig:sigma_nlo}, where for convenience we have dropped the initial-state lepton line. 
We label the momentum of the quark and anti-quark $k_1$ and $k_2$,
respectively, and we start by considering the real emission of a soft
gluon with momentum $k_3$, \ie diagrams in Fig.~$\ref{eenloR1}$ and Fig.~$\ref{eenloR2}$.
The matrix element for diagram $(b)$ can be written as
\begin{eqnarray}
M_{3}^{(b)}&=&\gstrong\, \bar{u}\left(k_1\right)\gamma^{\mu}\e^*_{\mu}\left(k_3\right)\frac{\slashed{k}_1+\slashed{k}_{3}}{\left(k_1+k_3\right)^{2}+i\epsilon}t_{1}^{a}\, \tilde{M}_{2}\nonumber\\
&\stackrel{k_3\to 0}{\longrightarrow}&\gstrong\, \bar{u}\left(k_1\right)\gamma^{\mu}\e^*_{\mu}\left(k_3\right)\frac{\slashed{k}_{1}}{2 k_1 \cdot k_3+i\epsilon}t_{1}^{a}\, \tilde{M_{2}}\nonumber\\%\angIn{M_{k_3\bar{k_3}}'}\nonumber\\
&=&\gstrong\, \frac{k_1^{\mu}}{k_1\cdot k_3}\e^*_{\mu}\left(k_3\right)  \bar{u}\left(k_1\right) t_{1}^{a}\, \tilde{M_{2}},
%=\gstrong\, \frac{k_1^{\mu}}{k_1\cdot k_3}\e^*_{\mu}\left(k_3\right)  {M_{2}},
\end{eqnarray}
where we have used anti-commutation relations of the Dirac matrices and $\slashed{k_1}u(k_1)=0$ to get the last line.
The factor $k_1^{\mu}/(k_1\cdot k_3)$ is called eikonal factor and $t_{1}^{a}$ is the colour charge associated to the emission of a gluon off a quark line, \ie it is a generator of SU(3) in the fundamental representation. We have also used fairly standard notation for the Dirac spinor $\bar u(k)$ and for the gluon polarisation vector $\e_\mu(k_3)$. %In the last step the Dirac spinor was absorbed in the 2-parton matrix element $M_2$ and therefore we dropped the tilde on it. 
%
%For the full real-emission amplitude we find
%%
%\begin{equation}\label{eq:softfactorisation}
%M_{3}=M_{3}^{(a)}+M_{3}^{(b)}\stackrel{k_3\to 0}{\longrightarrow} g_s  J^{\mu}\left(k_3\right)\e_{\mu}\left(k_3\right)M_{2},
%\end{equation}
%where we have introduced the eikonal current
%\begin{equation}\label{eq:eikonalcurrent}
%J^{\mu}\left(k\right)=\sum_{i=1}^{2}t_i^{a}\frac{k_i^{\mu}}{k\cdot k_i}.
%\end{equation}
It is important to note that the factorisation does not depend on the internal structure of the amplitude. From the physical point of view, this reflects the fact that the large wavelength of the soft radiation cannot resolve the details of the short distance interactions. However, the proof of this statement to any perturbative orders is highly non-trivial and it heavily relies on gauge invariance. Furthermore, we note that, unlike in the case of Abelian gauge theories, such as QED, in QCD we do not have a complete factorisation of the amplitude because of the presence of the colour matrix $t^a_1$. Indeed, the treatment of colour correlations induced by soft gluons is highly nontrivial.

We now square the amplitude and we arrive at the following factorised
expression for the emission of a soft real gluon
\begin{eqnarray}\label{eq:eikonalfactorisation}
\left|M_3\right|^{2}&=&\left |M_{3}^{(a)}+M_{3}^{(b)}\right|^2 \\ \nonumber
&\stackrel{k_3\to 0}{\longrightarrow} &g_s^{2} \frac{k_1\cdot k_2}{(k_1\cdot k_3)(k_2\cdot k_3)} \;{\rm tr}\left[ C_{12}\,
 \bar{u}\left(k_1\right) \, \tilde{M_{2}} \, \tilde{M^*_{2}} \, v\left(k_2\right)\, \right],%\nonumber\\
%&=&\left|M_2\right|^{2}g_s^{2}  C_{12}\frac{k_1\cdot k_2}{(k_1\cdot k_3)(k_2\cdot k_3)},
\end{eqnarray}
where we have introduced the effective colour charge
\begin{equation}\label{eq:eff-col-def}
C_{ij}= - 2 \, t_{i}^{a}\, t_{j}^{a},
\end{equation}
and the trace is taken over the colour indices. 
We note that the effective colour charge is a matrix which has in principle non-zero entries also away from the diagonal. It is easy to show using colour conservation that its structure noticeably simplifies in the case under consideration, because we only have two hard legs which carry colour:
\begin{equation}
t^a_1+t^a_2=0 \Longrightarrow \left({t_1^a}\right)^2+\left({t_2^a}\right)^2= - 2 t_1^a t_2^a  \Longrightarrow C_{12}= 2\cf,
\end{equation}
where all the above equalities are meant to hold when the matrices act on physical states.
The effective colour charge turns out to be diagonal also in the case of three hard coloured legs, as we shall see in Sec.~\ref{sec:Zjet}, while with four or more hard partons a non-trivial matrix structure emerges.
For the interest reader, we point out  that a general and rather powerful colour-operator formalism to deal with this issue exists~\cite{Catani:1985xt,Catani:1996vz,Catani:1996jh,Dokshitzer:2005ig,Dokshitzer:2005ek}. Within this framework, we introduce colour-operators $T_i$ that enable us to write the QCD eikonal current as
\begin{equation}\label{eq:eikonalcurrent}
J^{\mu}\left(k\right)=\sum_{i=1}^{2} T_i  \frac{k_i^{\mu}}{k\cdot k_i}.
\end{equation}
which closely resembles the QED current.

%%%%%%%%%%%%HERE%%%%%%%%%%%%%

The soft approximation can also be applied to the virtual
corrections. i.e.\ to the diagram in Fig.~\ref{eenloV}. In this limit we can in general neglect powers of the
loop momentum $k_3$ in the numerator.  Moreover, in the denominator we can use the fact that $k_3^{2}\ll k_i\cdot k_3$. The loop correction to quark-antiquark pair production is therefore proportional to
\begin{eqnarray}
I&=&g_s^{2} \cf(-i)\int\frac{d^{4}k_3}{(2\pi)^{4}}\frac{\bar{u}\left(k_1\right)\gamma^{\mu}\left(\slashed{k}_1+\slashed{k}_3\right)\gamma^{\rho}\left(\slashed{k}_3-\slashed{k}_2\right)\gamma_{\mu}v\left(k_2\right)}{\left[\left(k_3+k_1\right)^{2}+i\epsilon\right]\left[\left(k_3-k_2\right)^{2}+i\epsilon\right]\left[k_3^{2}+i\epsilon\right]}\nonumber\\
&\to&i g_s^{2} \cf\int\frac{d^{4}k_3}{(2\pi)^{4}}\frac{\left(k_1\cdot k_2\right)\left[\bar{u}\left(k_1\right)\gamma^{\rho}v\left(k_2\right)\right]} {\left[k_3\cdot k_1+i\epsilon\right]\left[-k_3\cdot k_2+i\epsilon\right]\left[k_3^{2}+i\epsilon\right]},
\end{eqnarray}
where we have written the result in $d=4$ space-time dimensions
because we are going to combine it with the real-emission
part before calculating the divergent integrals.

It is helpful to use the following parametrisation of the four-momenta:
\begin{equation}
k_1^{\mu}=E_1\left(1,0,0,1\right)\,,\quad k_2^{\mu}=E_2\left(1,0,0,-1\right)\,,\quad k_3^{\mu}=\left(k_{3}^0,\vec{k}_3\right)\,\textrm{ with }\, \vec{k}_3=\left(\vec{k}_{3\perp},k_{3}^z\right)\,,
\end{equation}
where $\vec{k}_{3 \perp}$ is the vectorial transverse loop momentum and $k_{3\perp} \equiv | \vec{k}_{3\perp} |$.
We note that
\begin{equation}\label{eq:dipole-kperp}
  k_{3\perp}^2=\frac{2(k_1.k_3)(k_3.k_2)}{(k_1.k_2)}.
\end{equation}
We thus obtain
\begin{equation}\label{eq:loopintsoft}
I=i g_s^{2} \cf\int\frac{d^{3}k_3}{(2\pi)^{4}}\frac{2\:d k_3^0 \left[\bar{u}\left(k_1\right)\gamma^{\rho}v\left(k_2\right)\right]}{\left(k_3^0-k_3^z+i\epsilon\right)\left(-k_3^0-k_3^z+i\epsilon\right)\left({k_3^0}^{2}-{k_3^z}^{2}-k_{3 \perp}^{2}+i\epsilon\right)}
\end{equation}
When performing loop calculations, one usually introduces a regulator,
such as, for instance, dimensional regularisation, and then evaluates
the integrals in Eq.~(\ref{eq:loopintsoft}) directly. Here we take
another approach which allows us to highlight the similarities between loop integrals for the
  virtual terms and phase-space integrals for the real
  contributions. We want first to evaluate the integral in
$k_3^0$. We note that the integrand has four poles in the complex
$k_3^0$ plane, which are located at 
\begin{eqnarray}
k_3^0=k_3^z-i\epsilon,\qquad k_3^0=-k_3^z+i\epsilon, \qquad k_3^0=\pm\left(|\vec{k_3\,}|-i\epsilon\right)\,.
\end{eqnarray}
Closing the contour from below, we find
\begin{equation}\label{eq:virtual-antenna+glauber}
I=- g_s^{2} \cf\left[\bar{u}\left(k_1\right)\gamma^{\rho}v\left(k_2\right)\right]\int\frac{d^{3}k_3}{(2\pi)^{3}}\left[\frac{k_1\cdot k_2 }{2|\vec{k_3\,}|\left(k_1\cdot k_3\right)\left(k_2\cdot k_3\right)}+\frac{1}{\left(k_3^z-i\epsilon\right)\left(k_{3\perp}^{2}\right)}\right]\,,
\end{equation}
where the second integral is a pure phase
\begin{equation}
\int\frac{d k_3^z\, d^2 k_{3\perp}}{\left(2\pi\right)^{3}}\frac{1}{\left(k_3^z-i\epsilon\right)\left(k_{3\perp}^2 \right)}=\int d k_3^z\frac{k_3^z+i\epsilon}{{k_3^z}^{2}+\epsilon^{2}}\int\frac{d k_{3\perp}}{\left(2\pi\right)^{2}}\frac{1}{k_{3\perp}}=-\frac{i\pi }{\left(2\pi\right)^{2}}\int \frac{d k_{3\perp}}{k_{3\perp}}\,.
\end{equation}
This contribution is usually referred to as the Coulomb, or Glauber,
phase. We note that the above phase always cancels when considering
physical cross-sections in Abelian theories like QED. However, it can
have a measurable effect in QCD cross-sections, in the presence of a
high enough number of harder coloured legs, which lead to a
non-trivial matrix structure for the effective colour charges
Eq.~(\ref{eq:eff-col-def}).

Collecting real and virtual contributions together, we can compute the
NLO distribution of an observable $v$ by introducing an
appropriate measurement function
$V_{n}\left(\left\{k_i\right\}\right)$, which describes
the value of the observable for a set of $n$ final-state particles
$k_1,\dots,k_n$.
The measurement function can contain Dirac delta corresponding to
constraints imposed in differential distributions, and/or Heaviside
$\Theta$ functions, for example when one imposes cuts on the
final-state or if one works with cumulative distributions.
Furthermore, if we are dealing with jet observables, the measurement
functions must also tell us how to combine particles in a jet, \ie it
must specify the jet algorithm
(cf.~Chapter~\ref{chap:jets-and-algs}).\footnote{Technically, the jet
  clustering can usually be written as a series of $\Theta$
  functions.}  With this in mind, we can write the cross-section for
an observable $v$ to NLO accuracy as the sum of three contribution:
Born, real emission and virtual corrections:
\begin{align}\label{eq:real-virtual-together}
\sigma\left(v\right)&=\frac{1}{2s}\int d\Phi_2\left|M_2\right|^{2}V_2\left(k_1,k_2\right)\\
&+\frac{1}{2s}\int d\Phi_2\left|M_2\right|^{2}\int\frac{d^{3}k_3}{(2\pi)^{3}2 | \vec{k}_3|}2 g_s^{2}\cf\frac{\left(k_1\cdot k_2\right)}{\left(k_1\cdot k_3\right)\left(k_2\cdot k_3\right)}
  \left[V_{3}\left(k_1,k_2,k_3\right)-V_2\left(k_1,k_2\right)\right].\nonumber
\end{align}
We note that the Born contribution and the one-loop corrections live in the two-particle phase-space and are characterised by the same measurement function. Instead, the real emission contribution live in a three-body phase-space and, consequently, the measurement function is the three-particle one.

The main result of our discussion so far is Eq.~(\ref{eq:real-virtual-together}), which describes the behaviour of a typical NLO cross-section in the limit where the radiated parton (gluon) is soft. However, if we take a closer look we note that 
\begin{equation}
k_i \cdot k_3 = E_i E_3 (1-\cos \theta_{i3}), \quad i=1,2.
\end{equation}
Thus, the eikonal factor exhibits a singularity not only in the soft limit but also when the parton with momentum $k_3$ becomes collinear with either $k_1$ or $k_2$.
It is clear that, while the eikonal approximation is sufficient to
correctly capture both the soft-collinear and soft wide-angle, we have to extend our formalism in order to include also the relevant hard-collinear terms. It must be noted that the collinear limit is in many respects easier than the soft limit discussed so far, essentially because the collinear factorisation emerges from a semi-classical picture whereby a parent parton splits into two daughters. An important consequence of this fact is that collinear singularities are always accompanied by diagonal colour charge $C_{ii}$, which is the Casimir of the relevant splitting, \ie  $C_F$ for quark splittings and $C_A$ for gluon splittings.\footnote{We warn the reader that although physically motivated, this statement is all but trivial to show!
 After the first splitting the total colour charge will be shared among the two partons and further radiation can be emitted from either of them.
 This leads to a colour radiation pattern which is in principle rather complicated.
 However, soft radiation cannot resolve the details of the interaction
 which happens at shorter distance and higher momentum scale, a
 phenomenon called \emph{coherence}. Therefore a soft gluon emitted at
 an angle $\theta$ will only see the total colour charge of the
 radiation emitted at smaller angles~\cite{Ermolaev:1981cm,Mueller:1981ex,Bassetto:1982ma}. 
 The iteration of this argument essentially leads angular-ordered
 parton showers and to the resummation of large logarithms in the
 framework of the coherent branching
 algorithm~\cite{Catani:1990rr,Catani:1992ua}.} 
 The splitting of a quark into a gluon with momentum fraction $z$ and
 a quark with momentum fraction $1-z$ $q \to q g$ is described at LO by the
 splitting function
 \begin{equation}\label{eq:quarksplitting}
 P_q(z)=C_F\frac{1+(1-z)^2}{z},
 \end{equation}
 while the gluon splitting into a pair of gluons or a quark-antiquark pair reads
 \begin{equation}\label{eq:gluonsplitting}
% P_g (z)=C_A \left[ \frac{1-z}{z}+\frac{z(1-z)}{2}+\frac{n_f T_R}{2C_A}\left(z^2+(1-z)^2\right),
  P_g (z)=C_A \left[ 2\frac{1-z}{z}+z(1-z)+\frac{n_f T_R}{C_A}\left(z^2+(1-z)^2\right)
 \right].
 \end{equation}
 where the first contribution describes the splitting $g \to gg$,
 while the second one, proportional to $n_fT_R$ with $n_f$ the number
 of massless flavours, corresponds to the splitting $g \to q \bar q$.
 We note that both Eqs.~(\ref{eq:quarksplitting})
 and~(\ref{eq:gluonsplitting}) exhibit a $z \to 0$ singularity which
 is the soft singularity of Eq.~(\ref{eq:real-virtual-together}),
 while the finite-$z$ part of the splitting functions describe the
 hard-collinear contribution.
 
 %%%%%%%%%%%%%%%%%%%%%%%%

 \section{Infra-red and collinear safety.}\label{sec:IRC-safety}

We are now ready to discuss infra-red and collinear safety in a more detailed way. Let us go back to Eq.~(\ref{eq:real-virtual-together}), or alternatively we could consider its extension in the collinear limit.  In order to achieve a complete cancellation of the IRC singularities, we must consider observables $V$ that satisfy the following properties, which we take as the definition of IRC safety~\cite{Sterman:1977wj}:
\begin{align}
\text{collinear safety:}&&
V_{m+1}\left(\ldots,k_i,k_j,\ldots\right)&\longrightarrow V_{m}\left(\ldots,k_i+k_j,\ldots\right) &&\mathrm{if}\;k_i\parallel k_j ,\\
\text{infrared safety:}&&
V_{m+1}\left(\ldots,k_i,\ldots\right)&\longrightarrow V_{m}\left(\ldots,k_{i-1},k_{i+1},\ldots\right) &&\mathrm{if}\;k_i\to 0.
\end{align}
In words, whenever a parton is split into two collinear partons, or
whenever an infinitesimally soft parton is added --- \ie in situations
where an extra emission makes the real amplitude divergent --- the value
of the observable must remain unchanged, in order to guarantee a
proper cancellation of the divergence against virtual corrections.
The above limits have to hold not only for a single particle, but for an ensemble of partons becoming soft and/or collinear.
IRC safe properties of jet cross-sections and related variables, such as event shapes and energy correlation functions were first studied in Refs.~\cite{Sterman:1978bi,Sterman:1978bj,Sterman:1979uw}.

Let us consider first the case of inclusive observables, \ie observables
that do not constrain additional radiation. We then have
$V_m\left(k_1,\ldots,k_m\right)=1$ for all $m$ and the cancellation is
complete. Consequently, the total cross-section remains unchanged by
the emission of soft particles, as it should. 
Note that
  Eq.~(\ref{eq:real-virtual-together}) is computed in the soft
  limit. An exact calculation involves additional corrections,
  non-divergent in the soft limit, so that the NLO contribution is a
  finite ${\cal{O}}(\alpha_s)$ correction.
Finally, and more interestingly for the topic of this book, let us consider the case of an exclusive (but IRC safe) measurement. Although the singularities cancel, the kinematic dependence of the observable can cause an imbalance between real and virtual contributions, which manifests itself with the appearance of potentially large logarithmic corrections to any orders in perturbation theory.
As we have previously mentioned, these logarithmic become large if
$v\ll 1$, \ie if the measurement function constrains real
radiation in a small corner of phase-space.
These contributions spoil
the perturbative expansion in the strong coupling and must be resummed
to all orders in order to obtain reliable theoretical predictions for
exclusive measurements. A typical observable in jet physics
is the jet invariant mass $m$ indeed suffers from these large
logarithmic corrections, if we are to consider the boosted regime
$p_t\gg m$, where $p_t$ is the jet transverse momentum. We will study
the jet mass distribution in great detail in
Chapter~\ref{chap:calculations-jets} and discuss how its behaviour is
modified by jet substructure algorithms called \emph{groomers}, in
Chapter~\ref{calculations-substructure-mass}.

We note here that there exists a wealth of observables that are of great interest despite them being IRC unsafe. Generally speaking, these observables require the introduction of non-perturbative functions to describe their soft and/or collinear behaviour. For example, lepton-hadron and hadron-hadron cross-sections are written as a momentum-fraction convolution of partonic cross-sections and parton distribution functions. Arbitrary collinear emissions change the value of the momentum fraction that enters the hard scattering, resulting in un-cancelled collinear singularities. Finite cross-sections are then obtained by a renormalisation procedure of the parton densities.  Similar situations are also encountered in final-state evolution, if one is interested in measuring a particular type of hadron (see \eg~\cite{Collins:1987pm}) or if the measurement only involves charged particles~\cite{Chang:2013rca,Chang:2013iba}.
Furthermore, we mention that recent work~\cite{Larkoski:2013paa, Larkoski:2014wba,Larkoski:2015lea} has introduced the concept of Sudakov safety, which enables to extend the reach of (resummed) perturbation theory beyond the IRC domain. We will come back to this in Chapter~\ref{sec:curiosities}.

\section{Hadron collider kinematics}\label{sec:hadron-collider-kinematics}

Although we have so far considered $e^+e^-$ collisions, which provide
an easy framework for QCD studies, the majority of this book will
focus on hadron-hadron colliders, with the LHC and possible future
hadronic colliders in mind.
All the concepts and arguments discussed above remain valid either
straightforwardly, or with little adjustments.
One of these adjustments is the choice of kinematic variables.
This is what we discuss in this section, so as to make our notations
clear for the rest of this book.

In the factorised picture described earlier,
cf.~Eq.~(\ref{eq:master}), the hard interaction of a hadron-hadron
collisions is really an interaction between two high-energy partons,
one from each beam.
These two partons carry respectively a fraction $x_1$ and $x_2$ of the
proton's momentum. Since in general, $x_1$ and $x_2$ are different,
the centre-of-mass of the hard interaction is longitudinally boosted
(along the beam axis) compared to the lab frame.  
We therefore need to use a set of kinematic variables which is
well-behaved with respect to longitudinal boosts.
Instead of using energy and polar angles, one usually prefers to use
{\em transverse momentum} $p_t$, {\em rapidity} $y$ and {\em azimuthal
  angle} $\phi$.
For a four-vector $(E,p_x,p_y,p_z)$, $p_t$ and $\phi$ are defined as the
modulus and azimuthal angle in the transverse plane $(p_x,p_y)$, \ie
we have
\begin{equation}
  p_t = \sqrt{p_x^2+p_y^2},
\end{equation}
and rapidity is defined as
\begin{equation}\label{eq:def-rapidity}
y = \frac{1}{2}\log\bigg(\frac{E+p_z}{E-p_z}\bigg).
\end{equation}
In other words, a four-vector of mass $m$ can be represented as
\begin{equation}
p^\mu\equiv (m_t\cosh y,p_t\cos\phi,p_t\sin\phi,m_t\sinh y),
\end{equation}
with $m_t=\sqrt{p_t^2+m^2}$ often referred to as the transverse
mass.
As for the $e^+e^-$ case, a particle of mass $m$ is described with
one dimensionful (energy-like) variable, $p_t$, and two dimensionless
variables with a cylindrical geometry: $y$ and $\phi$.
One can then define a distance (extensively used in this book) between two
particles in the $(y,\phi)$ plane:
\begin{equation}\label{eq:DeltaR-def}
\Delta R_{12} = \sqrt{\Delta y_{12}^2+\Delta\phi_{12}^2}.
\end{equation}

Since we shall integrate over particles produced in the final-state,
it is helpful to mention that with the above parametrisation, we have
\begin{equation}\label{eq:4vect-phase-space}
  \int \frac{d^4k}{(2\pi)^4}(2\pi)\delta(k^2)
  = \frac{1}{16\pi^2}\int dk_t^2\,dy\int_0^{2\pi}\frac{d\phi}{2\pi} 
\end{equation}

It is a straightforward exercise in relativistic kinematics to show that for two four-vectors of rapidities
$y_1$ and $y_2$, the difference $y_1-y_2$ remains invariant upon a
longitudinal boost of the whole system.
Additionally, if we come back to the two incoming partons carrying
respective fractions $x_1$ and $x_2$ of the beam energies, it is easy
to show that the centre-of-mass of the collisions has a rapidity
$y_\text{collision} =  \frac{1}{2}\log\big(\frac{x_1}{x_2}\big)$ with respect to the lab frame.

Finally, in an experimental context, one often makes use of the {\em
  pseudo-rapidity} $\eta$ instead of rapidity. The former is directly
defined either in terms of the modulus $|\vec{p}|$ of the 3-momentum,
or in terms of the polar angle $\theta$ between the direction of the
particle and the beam:
\begin{equation}\label{eq:def-pseudo-rapidity}
\eta = \frac{1}{2}\log\bigg(\frac{|\vec{p}|+p_z}{|\vec{p}|-p_z}\bigg) = -\log\bigg(\tan\frac{\theta}{2}\bigg).
\end{equation}
Contrary to rapidity differences, pseudo-rapidity differences are
generally \emph{not} invariant under longitudinal boosts, meaning that one
should use rapidity whenever possible.
For massless particles $y=\eta$ but this does not hold for massive
particles. Hence, for a final-state of massless particles
pseudo-rapidity and rapidity can be swapped, but they differ for more
complex objects like jets (see next chapter) which have acquired a
mass.
For these objects, it is recommended to use rapidity whenever possible.
% (see also~\cite{Schegelsky:2010xi,Gallicchio:2018elx} for related
% discussions).

%% GS helper for auctex
%%% Local Variables:
%%% mode: latex
%%% TeX-master: "notes"
%%% End:

%  LocalWords:  pions formers baryonic fermionic bosonic covariant Eq
%  LocalWords:  adjoint computable UE NNLO eikonal spinor Glauber
%  LocalWords:  Casimir dimensionful

% $Id: jets-and-algs.tex 531 2022-01-31 11:32:19Z smarzani $ 
%
% Descriptino of what jets are and what algoriotyhms are used to
% reconstruct them
%------------------------------------------------------------------------
\chapter{Jets and jet algorithms}\label{chap:jets-and-algs}

\section{The concept of jets}\label{sec:jet-concept}

When studying high-energy collisions one often has to consider
processes where quarks and gluons are produced in the final-state.
For $e^+e^-$ collisions, the study of hadronic final-states has been a
major source of information, helping to establish QCD as the
fundamental theory of strong interactions, but also providing a clean
playground for the study of perturbative QCD and the tuning of
Monte-Carlo event generators.
At the LHC, the list of processes involving high-energy quarks and/or
gluons in their final state is even longer. First, since we collide
protons, a hard QCD parton can be radiate from the incoming
partons. Then, other particles like W, Z and Higgs bosons can
themselves decay to quarks. And, finally, when searching for new
particles, one often has to consider decay chains involving quarks and
gluons.

However, these high-energy quarks and gluons are not directly observed
in the final state of the collision. First of all, as mentioned in the
previous chapters, they tend to undergo successive branchings at small
angles, producing a series of collimated quarks and gluons. The fact that this parton shower is
collimated traces back to the collinear divergence of QCD.
Starting from a parton with high virtuality (of the order of the hard
scale of the process), the parton shower will produce branchings into
further partons of decreasing virtuality, until one reaches a
non-perturbative (hadronisation) scale, typically of order
$\Lambda_\text{QCD}$ or $1$~GeV.
At this stage, due to confinement, these quarks and gluons will form
hadrons. Although some analytic approaches to hadronisation exist,
this non-perturbative step often relies on models implemented in Monte
Carlo Event generators.

Overall, the high-energy partons produced by the collision appear in
the final state as a collimated bunch of hadrons that we call {\em
  jets}.
Conceptually, {\it jets are collimated flows of hadrons and they can
  be seen as proxies to the high-energy quarks and gluons produced in
  a collision}.
This behaviour is observed directly in experiments where the hadronic
final state appears to be collimated around a few directions in the
detector.

\subsection{Jet definitions and algorithms}\label{sec:jet-algs}

The above picture is over-simplified in a few respects.
First of all, partons are ill-defined objects, \eg due to
higher-order QCD corrections where additional partons, real or
virtual, have to be included.
Then, whether two particles are part of the same jet or belong to two
separate jets also has some degree of arbitrariness, related to what
we practically mean by ``collimated''. 

The simple concept of what a jet is meant to represent is therefore
not sufficient to practically identify the jets in an event.
To do that, one relies on a {\em jet definition}, \ie a well-defined
procedure that tells how to reconstruct the jets from the set of
hadrons in the final state of the collision.

A jet definition can be seen as made of a few essential building
blocks: the {\em jet algorithm}, which is the recipe itself and a set
of parameters associated with free knobs in the algorithm. A typical
parameter, present in almost all jet definitions used in hadron
colliders is the {\em jet radius} which essentially provides a
distance in the rapidity-azimuth ($y-\phi$) plane above which two
particles are considered as no longer part of the same jet, \ie no
longer considered as collinear.

In addition, a jet definition uses a {\em recombination scheme} which
specifies how the kinematic properties of the jet are obtained from
its constituents.
Most applications today use the {\em ``$E$-scheme''} recombination
scheme which simply sums the components of the four-vectors. Other
recombination schemes, like the massless $p_t$ or $E_t$ schemes, have
been used in the past but are not discussed here.
Several jet-substructure applications make use of the {\em
  winner-take-all} (WTA) recombination scheme \cite{Larkoski:2014uqa}
where the result of the recombination of two particles has the
rapidity, azimuth and mass of the particle with the larger $p_t$, and
a $p_t$ equal to the sum of the two $p_t$'s. As we will further
discuss later in this book, this approach has the advantage that
it reduces effects related to the recoil of the jet axis when
computing jet observables that share similarities with the event-shape broadening~\cite{Rakow:1981qn}.

Over the past few decades, a number of jet algorithms have been
proposed. They typically fall under two big categories: {\em cone
  algorithms} and {\em sequential-recombination algorithms}. We
discuss them both separately below, focusing on the algorithms that
have been most commonly used recently at hadronic colliders.
For an extensive review on jet definitions, we highly recommend the
reading of Ref.~\cite{Salam:2009jx}.

\subsection{Basic requirements}\label{sec:jetalgs-snowmass}

Before giving explicit descriptions of how the most commonly-used jet
algorithms are defined, we briefly discuss what basic properties we do
expect them to satisfy.
In the 1990s a group of theorists and Tevatron experimentalists
formulated what is known as the Snowmass accord~\cite{Huth:1990mi}. This document
listed the fundamental criteria that any jet algorithm should
satisfy.

\vspace*{0.3cm}\noindent\centerline{\fbox{
\begin{minipage}{0.9\textwidth}
%\sf 
Several important properties that should be met by a jet definition are:
\vspace*{-0.2cm}
\begin{enumerate}
  \itemsep-0.1cm
\item Simple to implement in an experimental analysis;
\item Simple to implement in the theoretical calculation;
\item Defined at any order of perturbation theory;
\item Yields finite cross sections at any order of perturbation theory;
\item Yields a cross section that is relatively insensitive to hadronisation.
\end{enumerate}
\end{minipage}
}}\vspace*{0.3cm}

The first two criteria are mostly practical aspects. For example, if
an algorithm is too slow at reconstructing jets in an experimental
context, it would be deemed impractical. These two conditions also
mean that the algorithm should be applicable to an input made either
of partons (in a theoretical calculation), or of tracks and
calorimeter towers (in an experiment analysis).
The third and fourth conditions are mainly those of
IRC safety, a requirement that, as we have already seen, is at the core of
perturbative QCD calculations.
The fifth condition is a little bit more subjective.  We have already
seen that the description of a particle-collision event relies upon
several building blocks: the short-distance interaction computed in
fixed-order perturbation theory, the parton shower, the hadronisation
process and multi-parton interactions.
Since jets are supposed to capture the ``hard partons in an event'',
one should hope that the jets which come out of each of these
different steps of an event simulation are in good agreement. In
particular, this means that observables built from jet quantities
should be as little sensitive as possible to non-perturbative effects
like hadronisation and the Underlying Event.
Furthermore, to be simple to implement in an experimental analysis,
the jets should also be as little sensitive as possible to detector
effects and pileup.

The question of the sensitivity of different jet definitions to
non-perturbative effects, pileup and detector effects has been an
active topic of discussion when deciding which algorithm to use at
Tevatron and the LHC. A complete assessment of this question is
clearly beyond the scope of the present lecture notes. We will however
come back to a few crucial points when introducing the different
relevant jet definitions below.

\section{Sequential recombination algorithms}\label{sec:jetalgs-recombination}

Sequential recombination algorithms are based on the concept that,
from a perturbative QCD viewpoint, jets are the product of successive
parton branchings. These algorithms therefore try to invert this
process by successively recombining two particles into one. This
recombination is based on a distance measure that is small when the
QCD branching process is kinematically enhanced. Thus, one
successively recombine particles which minimise the distance in order
to mimic the QCD dynamics of the parton shower.
It is easy to check that all the recombination algorithms described
below are infrared-and-collinear safe.

\paragraph{Generalised-$k_t$ algorithm.}
Most of the recombination algorithms used in the context of hadronic
collisions belong to the family of the {\em generalised-$k_t$
  algorithm}~\cite{Cacciari:2011ma} which clusters jets as follows.
\begin{enumerate}
\item Take the particles in the event as our initial list of objects.
\item From the list of objects, build two sets of distances: an
  {\em inter-particle distance}
  \begin{equation}
    d_{ij} = \text{min}(p_{t,i}^{2p},p_{t,j}^{2p})\Delta R_{ij}^2,
  \end{equation}
  where $p$ is a free parameter and $\Delta R_{ij}$ is the geometric
  distance in the rapidity-azimuthal angle plane
  (Eq.~(\ref{eq:DeltaR-def}), and a {\em beam distance}
  \begin{equation}
    d_{iB} = p_{t,i}^{2p}R^2,
  \end{equation}
  with $R$ a free parameter usually called the {\em jet radius}.
\item Iteratively find the smallest distance among all the
  $d_{ij}$ and $d_{iB}$
  \begin{itemize}
  \item If the smallest distance is a $d_{ij}$ then objects $i$ and
    $j$ are removed from the list and recombined into a new object $k$
    (using the recombination scheme) which is itself added to the
    list.
  \item If the smallest is a $d_{iB}$, object $i$ is called a {\em
      jet} and removed from the list.
  \end{itemize}
  Go back to step 2 until all the objects in the list have
  been exhausted.
\end{enumerate}

In all cases, we see that if two objects are close in the
rapidity-azimuth plane, as would be the case after a collinear parton
splitting, the distance $d_{ij}$ becomes small and the two objects are
more likely to recombine. Similarly, when the inter-particle distances
are such that $\Delta R_{ij}>R$, the beam distance becomes smaller
than the inter-particle distance and objects are no longer recombined,
making $R$ a typical measure of the size of the jet.

\paragraph{$k_t$ algorithm.}
Historically, the best-known algorithm in the generalised-$k_t$ family
is the {\em $k_t$ algorithm}~\cite{Catani:1993hr,Ellis:1993tq},
corresponding to $p=1$ above. In that case, a soft emission, \ie one
with small $p_t$, would also be associated a small distance and
therefore recombine early in the clustering process. This is
motivated by the fact that soft emissions are also enhanced in
perturbative QCD.\footnote{Note that the presence of the ``min'' in
  the distance measure, instead of a product, guarantees that two soft
  objects far apart are not recombined. This would lead to undesired
  behaviours and complex analytic structures, as it is the case with
  the JADE algorithm~\cite{Bartel:1986ua,Bethke:1988zc}.}
Its sensitivity to soft emissions, while desirable from a perturbative
QCD standpoint, has the disadvantage that jets become more sensitive
to extra soft radiation in the event, typically like the Underlying
Event or pileup.
Although the Tevatron experiments have sometimes resorted to the $k_t$
algorithm, they have predominantly used cone algorithms (see below)
for that reason.

\paragraph{Cambridge/Aachen algorithm.}
Another specific cases of the generalised-$k_t$ algorithm is the {\em
  Cambridge/Aachen algorithm}~\cite{Dokshitzer:1997in,Wobisch:1998wt},
obtained by setting $p=0$ above.
In this case, the distance becomes purely geometrical and suffers less
from the contamination due to soft backgrounds than the $k_t$
algorithm does.

\paragraph{Anti-$k_t$ algorithm.}
In the context of LHC physics, jets are almost always reconstructed
with the {\em anti-$k_t$ algorithm}~\cite{Cacciari:2008gp}, which
corresponds to the generalised-$k_t$ algorithm with $p=-1$. The
primary advantage of this choice is that it favours hard particles
which will cluster first.
A hard jet will grow by successively aggregating soft particles around
it until it has reached a (geometrical) distance $R$ away from the jet
axis. This means that hard jets will be insensitive to soft radiation
and have a circular shape in the $y-\phi$ plane.
This soft-resilience of the anti-$k_t$ algorithm largely
facilitates its calibration in an experimental context and is the main
reason why it was adopted as the default jet clustering algorithm by
all the LHC experiments.

\begin{figure}
  \centering
  \raisebox{1.8cm}{\phantom{$\!\to\!$}}{\includegraphics[width=0.24\textwidth,page=1]{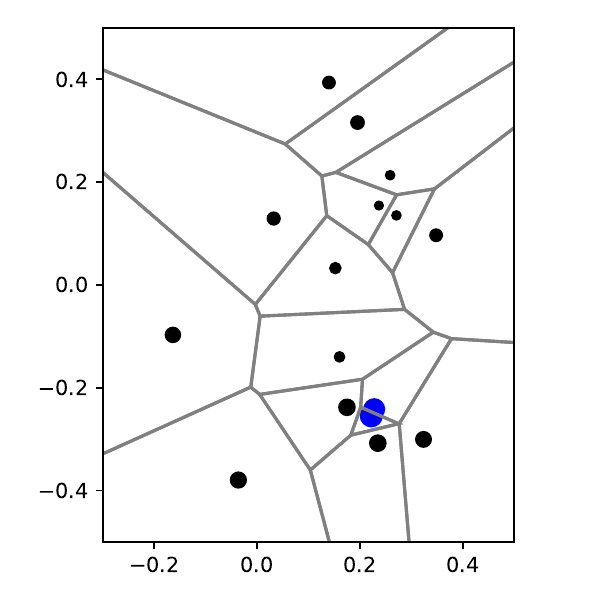}}%
  \raisebox{1.8cm}{$\!\!\to$}{\includegraphics[width=0.24\textwidth,page=2]{figures/jet-clustering-cartoon.pdf}}%
  \raisebox{1.8cm}{$\!\!\to$}{\includegraphics[width=0.24\textwidth,page=3]{figures/jet-clustering-cartoon.pdf}}%
  \raisebox{1.8cm}{$\!\!\to$}{\includegraphics[width=0.24\textwidth,page=4]{figures/jet-clustering-cartoon.pdf}}\\
  \raisebox{1.8cm}{$\!\!\to$}{\includegraphics[width=0.24\textwidth,page=5]{figures/jet-clustering-cartoon.pdf}}%
  \raisebox{1.8cm}{$\!\!\to$}{\includegraphics[width=0.24\textwidth,page=6]{figures/jet-clustering-cartoon.pdf}}%
  \raisebox{1.8cm}{$\!\!\to$}{\includegraphics[width=0.24\textwidth,page=7]{figures/jet-clustering-cartoon.pdf}}%
  \raisebox{1.8cm}{$\!\!\to$}{\includegraphics[width=0.24\textwidth,page=8]{figures/jet-clustering-cartoon.pdf}}\\
  \raisebox{1.8cm}{$\!\!\to$}{\includegraphics[width=0.24\textwidth,page=9]{figures/jet-clustering-cartoon.pdf}}%
  \raisebox{1.8cm}{$\!\!\to$}{\includegraphics[width=0.24\textwidth,page=10]{figures/jet-clustering-cartoon.pdf}}%
  \raisebox{1.8cm}{$\!\!\to$}{\includegraphics[width=0.24\textwidth,page=11]{figures/jet-clustering-cartoon.pdf}}%
  \raisebox{1.8cm}{$\!\!\to$}{\includegraphics[width=0.24\textwidth,page=12]{figures/jet-clustering-cartoon.pdf}}\\
  \raisebox{1.8cm}{$\!\!\to$}{\includegraphics[width=0.24\textwidth,page=13]{figures/jet-clustering-cartoon.pdf}}%
  \raisebox{1.8cm}{$\!\!\to$}{\includegraphics[width=0.24\textwidth,page=14]{figures/jet-clustering-cartoon.pdf}}%
  \raisebox{1.8cm}{$\!\!\to$}{\includegraphics[width=0.24\textwidth,page=15]{figures/jet-clustering-cartoon.pdf}}%
  \raisebox{1.8cm}{$\!\!\to$}{\includegraphics[width=0.24\textwidth,page=16]{figures/jet-clustering-cartoon.pdf}}
  \caption{
    Illustration of a step-by-step clustering using the anti-$k_t$
    algorithm with $R=0.4$.
    The axes of each plot are rapidity and azimuthal angle. 
    Each particle is represented by a cross with a size increasing
    with the $p_t$ of the particle.
    To help viewing the event, we also draw in grey lines the Voronoi
    cells obtained for the set of particles in the event (\ie cells
    obtained from the bisectors of any pair of points).
    Each panel corresponds to one step of the clustering.
    At each step, the dots represent the objects which are left 
    for clustering (again, with size increasing with $p_t$).
    Pairwise clusterings are indicated by a blue pair of dots, while
    red dots correspond to final jets (\ie beam clusterings).
    The shaded areas show the cells included in each of the three jets
    which are found ultimately. 
  }\label{fig:jet-clustering-steps}
\end{figure}

\begin{figure}
  \centering
  \includegraphics[width=0.48\textwidth]{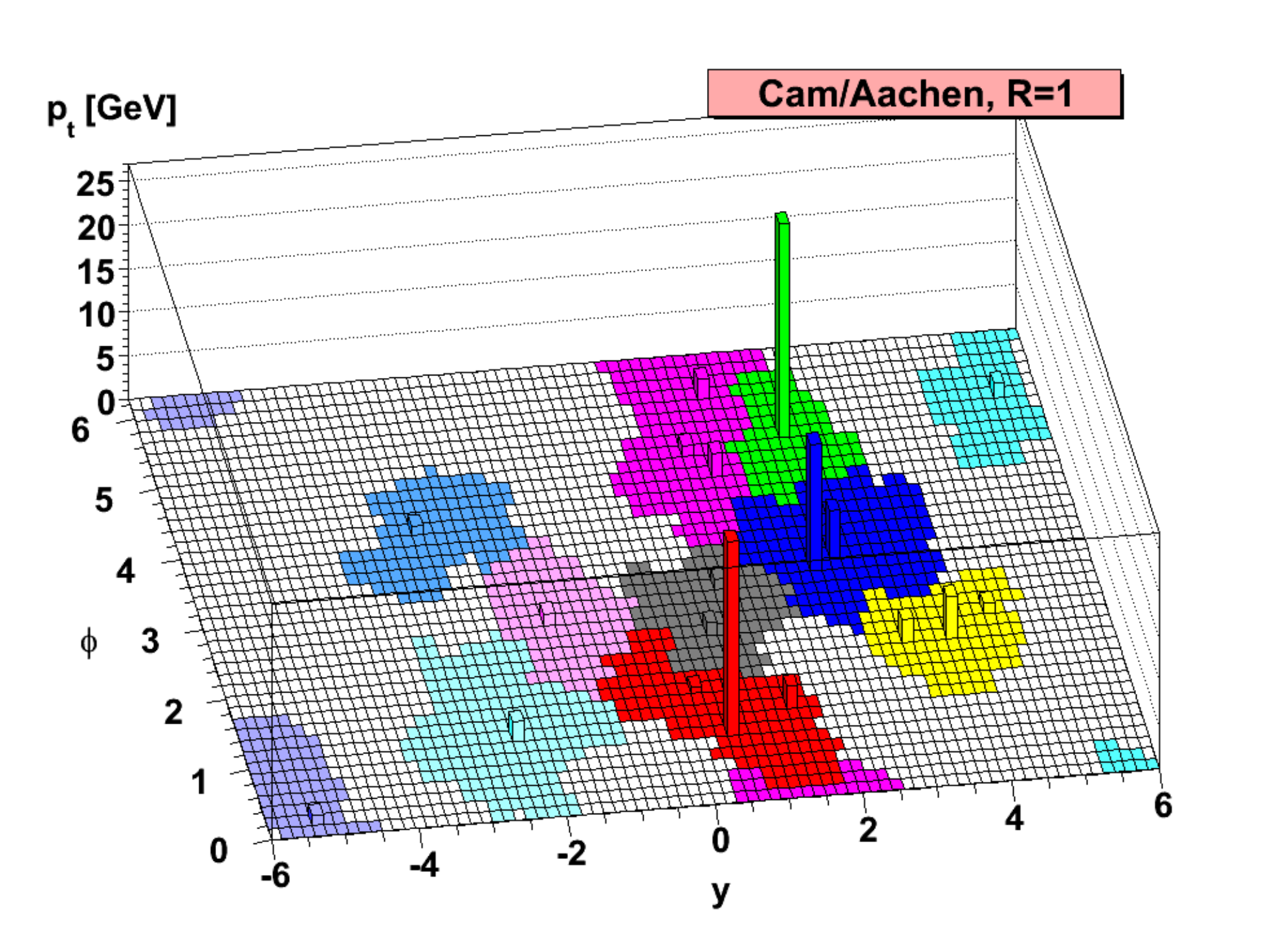}%
  \hfill%
  \includegraphics[width=0.48\textwidth]{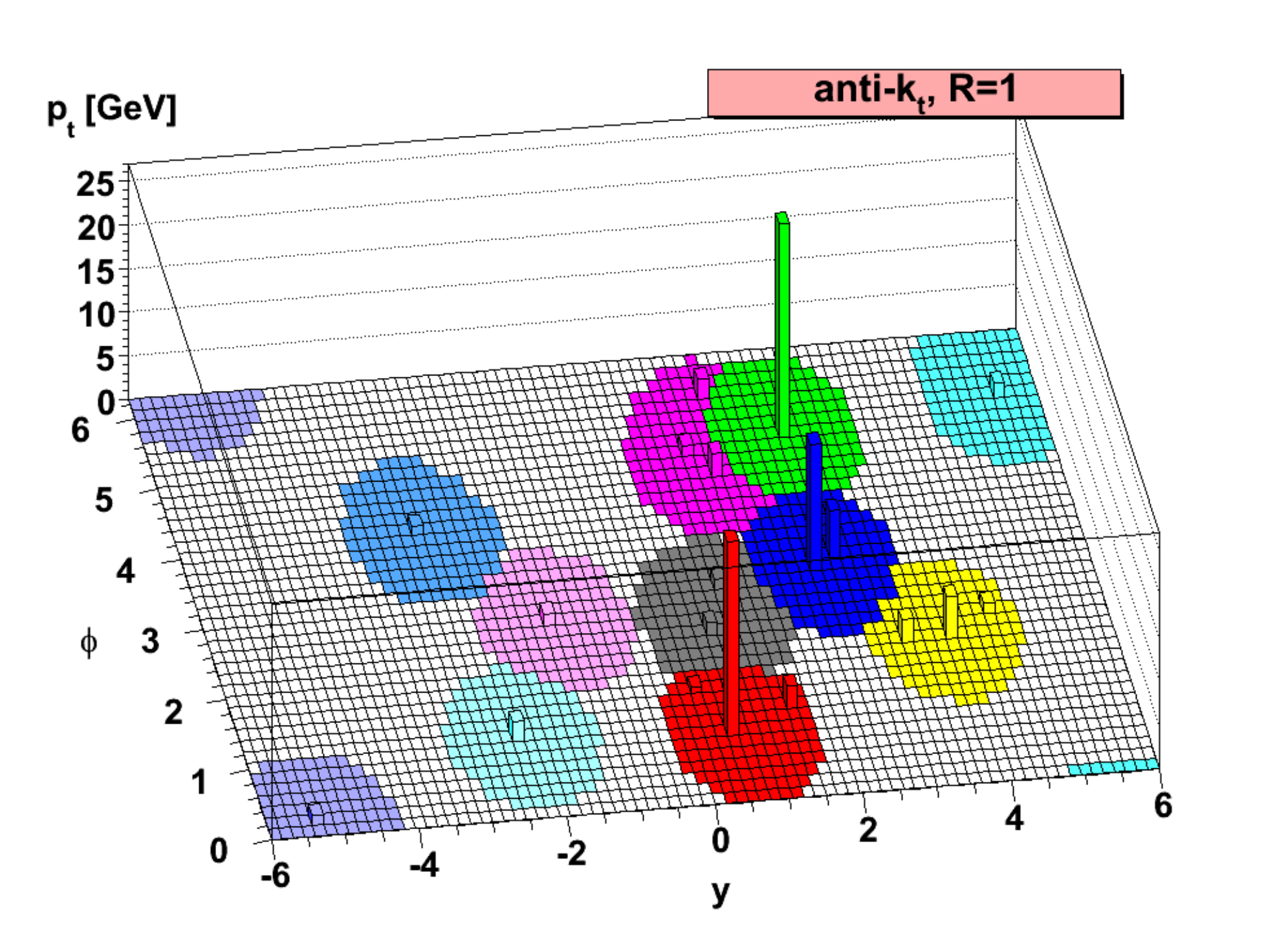}  
  \caption{Jets obtained with the Cambridge/Aachen (left) and
    anti-$k_t$ (right) algorithms with $R=1$. The shaded regions
    correspond to the (active) catchment
    area~(see~\cite{Cacciari:2008gn}) of each jet. While the jets
    obtained with the Cambridge/Aachen algorithm have complex
    boundaries (a similar property would be seen on $k_t$ jets), the
    hard jets obtained with anti-$k_t$ clustering are almost
    perfectly circular. This figure has been taken from~\cite{Cacciari:2008gp}.}\label{fig:jet-areas}
\end{figure}

To make things more concrete, we show in
Fig.~\ref{fig:jet-clustering-steps} a step-by-step example of a
clustering sequence with the anti-$k_t$ jet algorithm on a small set
of particles.
The successive pairwise recombinations, and beam recombination
giving the final jets, is clearly visible on this figure.
Finally, the resilience of anti-$k_t$ jets with respect to soft radiation is shown
in Fig.~\ref{fig:jet-areas}, where we see that anti-$k_t$ jets have a
circular shape while Cambridge/Aachen jets have complex
boundaries.\footnote{In practice, the jet areas are obtained by adding
  a infinitely soft particles, {\em aka} ghosts, to each calorimeter
  tower, These are clustered with the hard jets, indicating the
  boundaries of the jets.}

\paragraph{Relevance for jet substructure.}
In the context of jet substructure studies, several recombination
algorithms are used. Initially, jets are usually reconstructed using
the anti-$k_t$ algorithm with a large radius (typically $R$ in the
0.8--1.2 range). Many substructure tools then rely on reclustering the
constituents of that jet with another sequential-recombination jet
algorithm (or jet definition), allowing one to have a convenient view
of the jet clustering as a tree structure.
The most commonly used algorithm is probably Cambridge/Aachen since it
gives a natural handle on the structure of the jet at different
angular scales, in a way that respects the angular ordering of parton
showers (see.\ also~\cite{Dreyer:2018nbf}).
One also relies on the $k_t$ algorithm used \eg to split the jet into
subjets, or the generalised-$k_t$ algorithm with $p=1/2$, used because
it mimics an mass/virtuality ordering of the subjets.
More details will be given later when we review the main substructure
tools.

\section{Cone algorithms}\label{sec:jetalgs-cone}

Cone algorithms were first introduced in
1979~\cite{Sterman:1977wj}. They are based on the
idea that jets represent dominant flows of energy in an event.
Modern cone algorithms rely on the concept of a {\em
  stable cone}: for a given cone centre $y_c,\phi_c$ in the
rapidity-azimuth plane, one sums the 4-momenta of all the particles
with rapidity and $\phi$ within a (fixed) radius $R$ around the cone
centre; if the 4-momentum of the sum has rapidity $y_c$ and
azimuth $\phi_c$ --- \ie the sum of all the momenta in the cone points
in the direction of the centre of the cone --- the cone is called {\em
  stable}.
This can be viewed as a self-consistency criterion.

In order to find stable cones, the JetClu~\cite{Abe:1991ui} and
(various) midpoint-type~\cite{Blazey:2000qt,Abazov:2011vi} cone
algorithms use a procedure that starts with a given set of
seeds. Taking each of them as a candidate cone centre, one calculates
the cone contents, find a new centre based on the 4-vector sum of the
cone contents and iterate until a stable cone is found.
The JetClu algorithm, used during Run I at the Tevatron, takes the set
of particles as seeds, optionally above a given $p_t$ cut. This can be
shown to lead to an infrared unsafety when two hard particles are
within a distance $2R$, rendering JetClu unsatisfactory for theoretical
calculations.

Midpoint-type algorithms, used for Run II of the Tevatron, added to
the list of seeds the midpoints between any pair of stable cones found
by JetClu. This is still infrared unsafe, this time when 3 hard
particles are in the same vicinity, \ie one order later in the
perturbative expansion than the JetClu algorithm.
This infrared-unsafety issue was solved by the introduction of the
SISCone~\cite{Salam:2007xv} algorithm. It provably finds all possible
stable cones in an event, making the stable cone search
infrared-and-collinear safe.

Finally,  note that finding the stable cones is
not equivalent to finding the jets since stable cones can overlap.
The most common approach is to run a split--merge procedure once the
stable cones have been found. This iteratively takes the most
overlapping stable cones and either merges them or splits them depending
on their overlapping fraction.

% $Id: experimental.tex 551 2025-02-11 17:22:20Z mspannow $
%
% Basic experimental aspects of jets
%------------------------------------------------------------------------
\section{Experimental aspects}\label{chap:experimental}

The experimental input to the jet algorithms previously discussed is
reconstructed from energy deposits of elementary particles within the
different detector components. The details of the reconstruction
differ between the four LHC experiments For instance, for the first runs of the LHC,  the ATLAS collaboration mostly used  topoclusters, while CMS developed particle-flow objects as inputs to their jet
recombination algorithms.
All majors experimental collaborations have dedicated groups actively working on the performance of jet definitions.  
While details of how jet constituents are reconstructed can affect the
properties of the jets, we will
constrain our discussion here to a generic description of qualitative
features in the process of measuring them.

\begin{figure}
  \centerline{\includegraphics[width=0.90\textwidth]{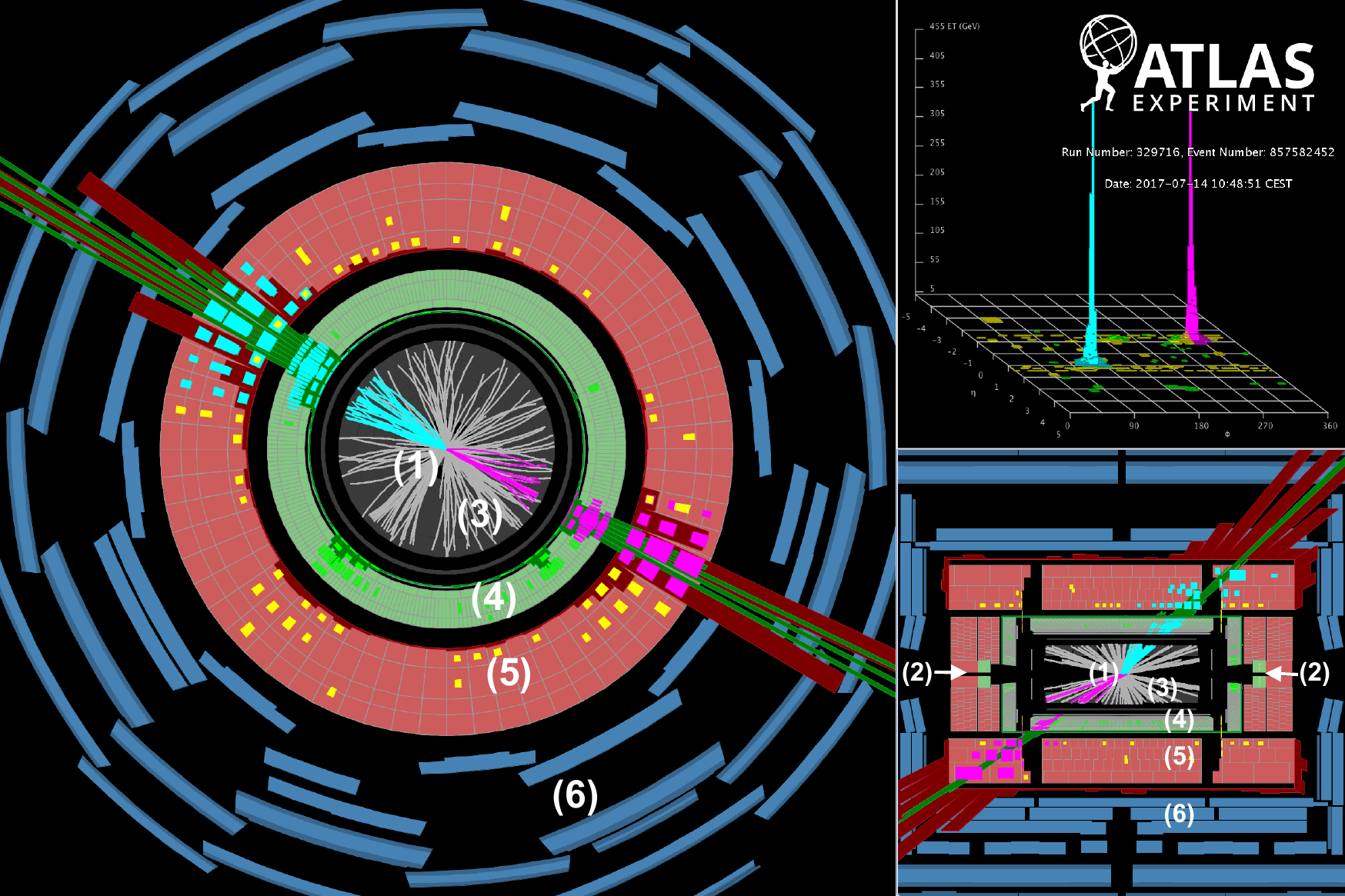}}
  \caption{
 Display of a dijet event recorded by ATLAS in proton-proton collisions at centre-of-mass energy 13~TeV. 
The two high-$p_t$ jets have both transverse momentum of 2.9~TeV and the dijet system exhibit an invariant mass of 9.3~TeV. 
The different panels correspond to the view of the event in the plane transverse to the beam direction (large figure on the left-hand side). The two smaller figures on the right-hand side show the calorimeter clusters transverse energies in the $(\eta,\phi)$ plane on the top and the longitudinal view of the event on the bottom. 
The numbers corresponds to different detectors components, as discussed in the text. 
%The figure has been taken and modified from~\cite{dijet-event}.
ATLAS Experiment~\copyright~2018 CERN.}
  \label{fig:detector}
\end{figure}

Multi-purpose detectors at the LHC are cylinder-shaped highly-complex
objects consisting of layers of different components, as depicted in
Fig.~\ref{fig:detector}, each component measuring a certain way a
particle can interact with the detector. Fig.~\ref{fig:detector} shows
a dijet event with an invariant mass of the two jets of $m_{jj} = 9.3$
TeV, measured by ATLAS and consists of three different images. In the
large image on the left the detector plane transverse to the beam axis
is shown. In the lower image on the right we see a lengthwise slice of
the ATLAS detector. The upper image on the right shows the energy
deposits of particles transverse to the beam axis in the so-called
{\it lego-plot} plane. In the lego plot the cylinder shape of the
detector is projected onto a 2-dimensional plane, consisting of the
variables $\eta \in (-\infty, \infty)$, the pseudo-rapidity, cf.\
Eq.~\eqref{eq:def-pseudo-rapidity}, and the azimuthal angle $\phi \in
[0,2 \pi]$.
$\eta$ measures how forward a particle is emitted during the proton-proton interaction. 
Note the similarities between the pseudo-rapidity and the rapidity defined in Eq.~(\ref{eq:def-rapidity}): the two coincide for massless particles. 
Distances between two cells or particles $i$ and $j$ on the lego plane are measured via 
\begin{equation}
\Delta R_{ij}^{\text{(detector)}} = \sqrt{(\phi_i - \phi_j)^2 + (\eta_i -\eta_j)^2 }.
\label{eq:dr_eta}
\end{equation}
Note that the topoclusters are assumed massless, i.e. their rapidity equates their pseudo-rapidity. Thus, for detector cells the definitions of Eqs.~(\ref{eq:dr_eta}) and~(\ref{eq:DeltaR-def}) agree.
The different detector components are labelled in Fig.~\ref{fig:detector} in the following way:
\begin{itemize}
\item[(1)] Interaction point of the proton beams.
\item[(2)] The arrows indicate the direction of the particle beams. The proton beams are entering from either side of the detector and exit on the opposite side after crossing at the collision point.
\item[(3)] The innermost part of the ATLAS and CMS detectors consists of the tracking detectors which measure the momentum of charged particles. Strong magnetic fields bend the particles when traversing through the detectors. The way the tracks are bent is indicative of the particle's charge, mass and velocity.
\item[(4)] The electromagnetic calorimeter measures predominantly the energies of electrons and photons. Such particles are stopped and induce a cascade of particles, \emph{a shower}, in the calorimeter. Charged particles can be discriminated from photons by the presence or absence of tracks in the tracking detectors. Cell sizes for this calorimeter vary between the central and forward direction of the detector. In the central part they are roughly $(0.025 \times 0.025)$ in the $\phi-\eta$ plane.
\item[(5)] The hadronic calorimeter measures the energies of hadronic particles, e.g. protons and neutrons. As in the case of the electromagnetic calorimeter, charged hadrons can be discriminated from neutral ones due to their energy loss in the tracking detectors. The cells that make the hadronic calorimeter have in the central region of the detector a size of roughly $(0.1 \times 0.1)$ in the $\phi-\eta$ plane.
\item[(6)] The most outer layer of the detector is the muon spectrometer. Muons, produced with characteristic LHC energies, are weakly interacting with the detector material and are consequently not stopped. However, they may leave tracks in the tracking system, undergo energy loss in the electromagnetic and hadronic calorimeter and may eventually interact with the muon spectrometer. 
\end{itemize}

\begin{figure}
  \centerline{\includegraphics[width=0.95\textwidth]{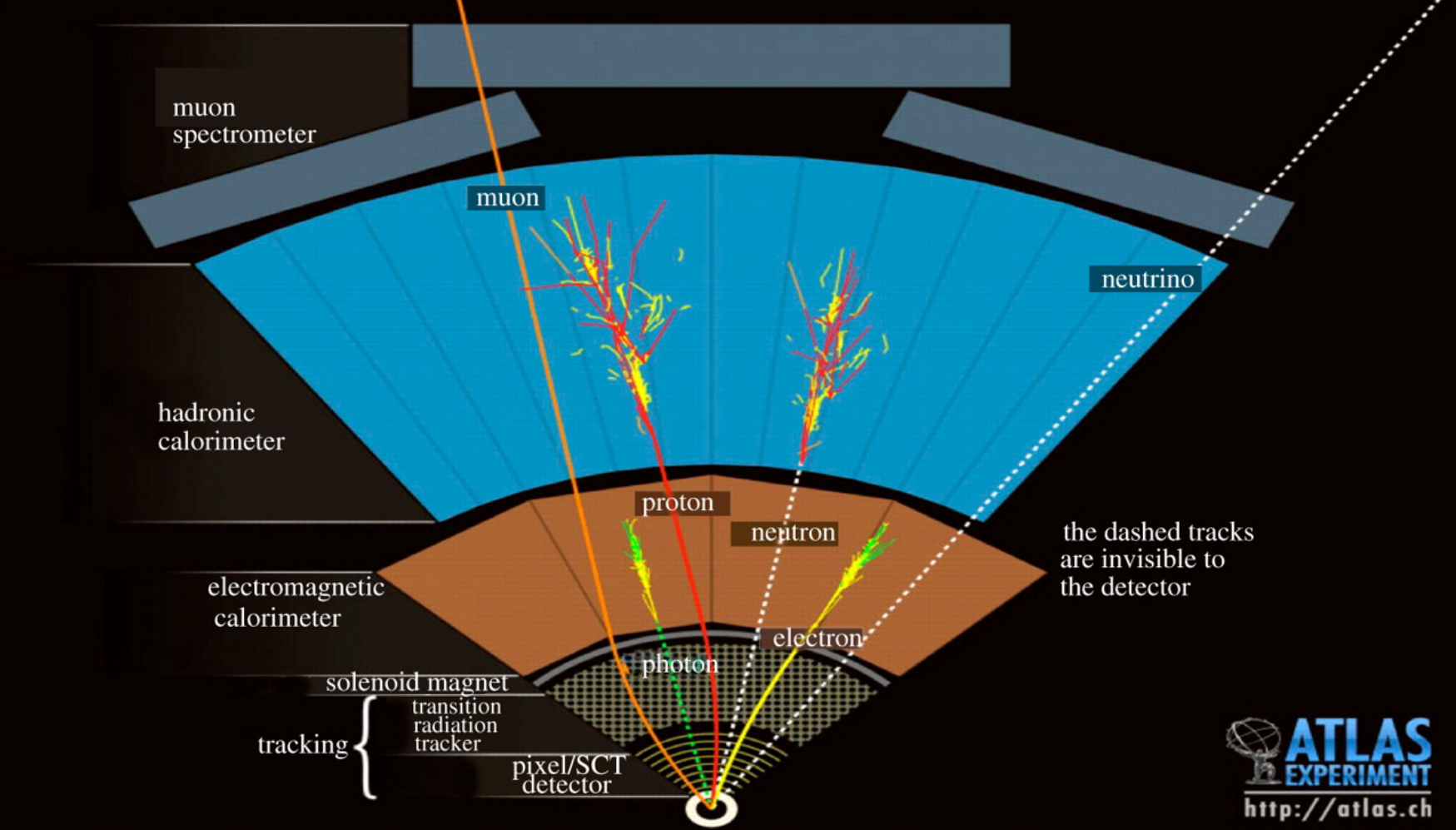}}
  \caption{Schematic depiction of a multi-purpose detector, here ATLAS. The picture illustrates how different particles interact with the various layers of the detector.  
  ATLAS Experiment~\copyright~2018 CERN.
%  The figure has been taken from Ref.~\cite{Kourkoumelis:2013vpa}).
}
  \label{fig:interaction}
\end{figure}

In Fig.~\ref{fig:interaction} we show a segment of a slice of the transverse plane and how classes of particles interact with the individual detector components. For each high-energy Standard Model event we expect of $\mathcal{O}(500)$ resulting particles, which we can classify into photons, charged leptons, neutral and charged hadrons and non-interacting particles, i.e.\ neutrinos. In a typical proton-proton collision, about 65\% of the jet energy is carried by charged particles, 25\% by photons, produced mainly from $\pi_0$ decays, and only 10\% by neutral hadrons (mostly neutrons and $K_{L} $) \cite{CMS:2010byl, CMS:2010eua}. However, these fractions can vary significantly from event to event.

Charged particles loose energy when traversing the detector material in various ways. One mechanism is ionisation and excitation interactions with the detector material, e.g. $\mu^- + \mathrm{atom} \to \mathrm{atom}^* + \mu^- \to \mathrm{atom} + \gamma + \mu^-$, where their energy loss per distance is governed by the Bethe equation \cite{Tanabashi:2018oca}. Further mechanisms for charged particles to interact with the detector material are {\it bremsstrahlung}, {\it direct electron-pair production} and {\it photonuclear interactions
}. 
Photons interact with the detector material through {\it photoelectric effect}, {\it Compton scattering} and {\it electron-pair production}. The latter being dominant for $E_\gamma \gg 1$ MeV.
In the case of hadron-detector interactions, we are dealing mostly with inelastic processes, where secondary strongly interacting particles are produced in the collision.

\vspace{0.3cm}

\begin{wrapfigure}{R}{0.5\textwidth}
 \centering
  \includegraphics[width=0.5\textwidth]{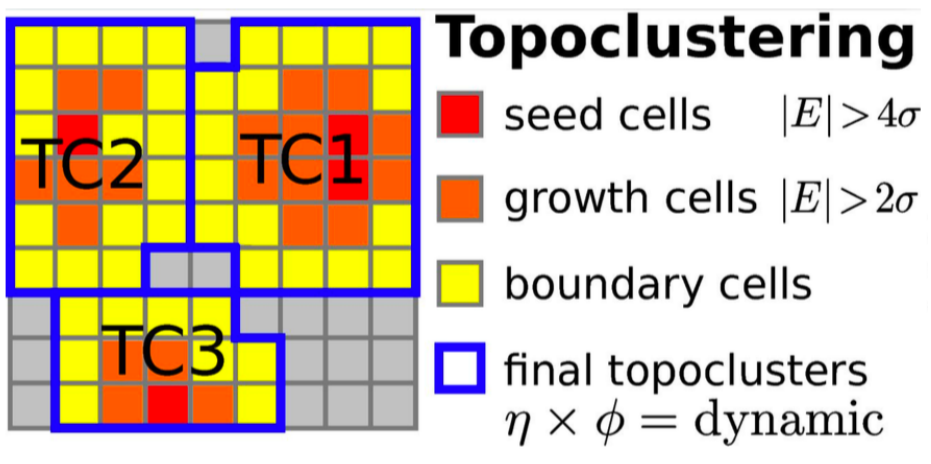}
  \caption{The figure shows how calorimetric information is used by ATLAS to construct jet constituents (taken from~\cite{boosttalk}).}
  \label{fig:topoclusters}
\end{wrapfigure}

Information gathered from the detector components (3)-(6) allow to
obtain a global picture of the particles produced in the
event. However, particles are not directly used as input to construct
jets using the algorithms previously discussed. As previously mentioned, ATLAS and CMS used traditionally
different approaches to construct jet constituents. The former is
using topological clusters, or, in short, topoclusters, which are
mainly based on calorimeter objects, while the latter use so-called
particle flow objects, which combine information from the tracker and
the calorimeter to build a coherent single object. 
The
benefit of using calorimeter objects is a good calibration of the
energy component of the topoclusters. On the other hand, the cell size
of the hadronic calorimeter is $0.1 \times 0.1$ in $(\eta, \phi)$ and
topological cell clusters are formed around seed cells with an energy
$|E_\mathrm{cell}|$ at least $4 \sigma$ above the noise by adding the
neighbouring cells with $|E_\mathrm{cell}|$ at least $2\sigma$ above
the noise, and then all surrounding cells \cite{Aad:2012vm}, see
Fig.~\ref{fig:topoclusters}. The minimal transverse size for a cluster
of hadronic calorimeter cells is therefore $0.3 \times 0.3$ and is
reached if all significant activity is concentrated in one cell. Two
energy depositions leave distinguishable clusters if each one hits
only a single cell and their individual axes are separated by at least
$\Delta R = 0.2$, so that there is one empty cell between the two seed
cells. In the context of this book, it means that if important
characteristics of the substructure in a jet are so close that it does
not leave separate clusters in the jet, it is impossible to resolve
it. This leaves a residual lower granularity scale when using
topocluster as fundamental objects to form jets. Thus, in particular
when a fine-grained substructure in the jet is of importance, e.g. in
the reconstruction of highly boosted resonances, the benefit of
particle flow objects is widely appreciated across both multi-purpose
experiments.

Indeed,  the ATLAS collaboration has introduced for LHC Run 3 new Unified Flow Objects (UFOs) that aim to maximise performance across many orders of magnitude in the jet transverse momentum by combing the virtues of calorimetric and particle-flow approaches~\cite{ATLAS:2020gwe}. 
UFOs are a hybrid data structure that enhances jet and event reconstruction by integrating information from multiple detector components. Unlike purely calorimeter-based topoclusters~\cite{Aad:2012vm} or particle-flow objects that rely heavily on tracker input~\cite{CMS:2010byl}, UFOs are constructed by associating charged tracks from the inner detector with their corresponding calorimeter deposits. The process begins with identifying charged-particle tracks in the inner detector, which are then extrapolated to the calorimeters. The energy deposits attributed to these charged tracks are subtracted from the total calorimetric energy, leaving residual deposits corresponding to neutral particles such as photons and neutral hadrons. 
The primary goal of this approach is to leverage the superior energy resolution of calorimetric measurements while maintaining the fine spatial resolution of tracker-based methods. In topological cluster formation, granularity limitations arise from the hadronic calorimeter cell size of $0.1 \times 0.1$ in $(\eta, \phi)$~\cite{ATLAS:2020gwe}. Consequently, closely spaced particles within a jet may produce overlapping clusters, complicating jet substructure analysis. By incorporating tracker information, UFOs provide finer spatial discrimination, allowing for the resolution of complex substructure, which is critical for identifying multi-prong decays of highly boosted particles.
Additionally, UFOs improve pile-up mitigation by exploiting the vertex information from charged-particle tracks~\cite{CMS:2010eua}. This vertex association ensures that only particles originating from the primary interaction contribute to the jet constituents, thereby reducing contamination from secondary interactions. This feature is particularly advantageous for analyses sensitive to missing transverse momentum ($E_T^{\text{miss}}$), where pile-up can significantly degrade the resolution and accuracy of the measurement~\cite{Aad:2010ac}.
In practical applications, UFOs enhance the precision and robustness of key observables in ATLAS physics analyses. For example, in searches for new resonances involving boosted hadronic final states, UFOs allow for a more detailed characterisation of jet shapes and substructure variables such as N-subjettiness and energy correlation functions~\cite{Thaler:2011gf}. These improvements increase the sensitivity to signals, which is essential when distinguishing between decay topologies. Similarly, flavor-tagging algorithms benefit from the improved reconstruction of charged and neutral particles within jets, leading to more accurate identification of $b$-jets and other jet flavors.
Preliminary studies indicate that UFOs achieve stable performance across a wide range of transverse momentum scales, from a few GeV to multiple TeV. This versatility makes them well-suited to handle low-energy soft radiation and high-energy jets produced in extreme kinematic conditions. By combining the best aspects of calorimetric and particle-flow reconstruction, UFOs provide a unified framework that addresses the challenges posed by the high-luminosity environment of the LHC, ensuring that ATLAS remains at the forefront of precision measurements and searches for new physics in Run 3 and beyond.

Focusing exclusively on the tracking detectors when reconstructing jets is an even more radical approach to optimising the spatial resolution of a final state. Tracking detectors can reconstruct the trajectories of a charged particles, which carry $\sim 65\%$ of the final state's energy, and can specify the direction of the particle at any point of the trajectory with a precision much better than the granularity of the calorimeter. For example, the angular resolution of the ATLAS inner tracking detector for charged particles with $p_T = 10$ GeV and $\eta  = 0.25$ is $\sim 10^{-3}$ in $\eta$ and $\sim 0.3$ mrad in $\phi$ \cite{Aad:2008zzm} with a reconstruction efficiency of $> 78\%$ for tracks of charged particles with $p_T > 500$ MeV \cite{Aad:2010ac}. Further, the momentum resolution for charged pions is 4\% for momenta $|p| < 10$ GeV, rising to 18\% at $|p| = 100$ GeV \cite{Aad:2008zzm}.
Note that, generally speaking, the energy resolution tends to degrade
with energy in for calorimeters, but improves with energy for trackers.

%  LocalWords:  topoclusters Eq eq lego Eqs

\section{Implementation}\label{sec:jetdefs-implementation}
\begin{figure}
  \includegraphics[width=0.53\textwidth]{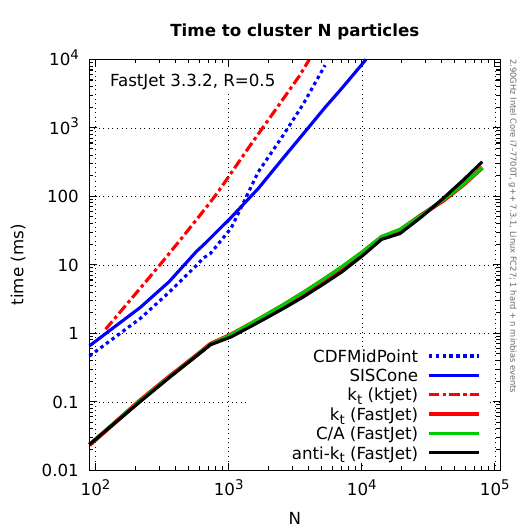}
  \hspace*{0.4cm}
  \begin{minipage}[b]{0.38\textwidth}
    \caption{Average clustering time as a function of the event
      multiplicity $N$, obtained with the FastJet implementation of
      several representative algorithms.}\label{fig:fastjet-timings}
    \vspace*{1.4cm}
  \end{minipage}  
\end{figure}

Most of the practical applications of jets use numerical inputs,
either from (fixed-order or parton-shower) Monte Carlo simulations, or
directly from experimental data. It is therefore important to have a
numerical implementation of the jet algorithms.
Furthermore, this implementation needs to be fast enough for practical
usability in an experimental (and, to a lesser extent, theoretical)
context.
Currently, the standard package for jet clustering is
{\tt{FastJet}}~\cite{Cacciari:2005hq,Cacciari:2011ma},\footnote{See
  also \url{http://fastjet.fr}.} used by both the experimental and
theoretical communities at the LHC. It provides a native
implementation of all the recombination algorithms introduced in
Sec.~\ref{sec:jetalgs-recombination} and plugins for a series of
other jet algorithms, including the cone algorithms discussed in
Sec.~\ref{sec:jetalgs-cone}.
As an illustration, we show in Fig.~\ref{fig:fastjet-timings} the
average time it takes to cluster an event with $N$ particles for a few
representative algorithms.
For the specific case of the $k_t$ algorithm, we show the timings for
two different implementations: the initial {\tt{ktjet}}
implementation~\cite{Butterworth:2002xg} available at the time of the
Tevatron and deemed too slow, and the {\tt{FastJet}} implementation
which is faster by 2-3 orders of magnitude in the region relevant for
phenomenology (around a few thousands particles).
Regarding cone algorithms, this plot shows that
infrared-and-collinear SISCone has clustering times similar to the
unsafe MidPoint.\footnote{MidPoint has here been used with a seed
  threshold of 1~GeV. Without a seed threshold, it would be slower by
  about an order of magnitude.}
Finally, if one keeps in mind that in practical (trigger-level) jet
reconstruction at the LHC, one has a few tens of milliseconds for
clustering, Fig.~\ref{fig:fastjet-timings} shows that the
recombination algorithms (and their {\tt{FastJet}} implementation) are
currently clearly preferred.

%% GS helper for auctex
%%% Local Variables:
%%% mode: latex
%%% TeX-master: "notes"
%%% End:
%  LocalWords:  WTA Snowmass kinematically iB ij JetClu ktjet
%  LocalWords:  MidPoint

%\include{experimental}
% $Id: calculations-jets.tex 610 2026-08-04 15:41:32Z smarzani $
%
% This contains basic calculations for jet physics, in particulwar the
% jet mass at fixed order and resummed
%------------------------------------------------------------------------
\chapter{Calculations for jets: the jet mass distribution}\label{chap:calculations-jets}
In this chapter we begin our discussion about the calculation of jet properties in perturbative QCD.
We start by considering  an important observable in jet physics, namely the jet invariant mass
\begin{equation}\label{eq:jet-mass-def}
m^2=\bigg( \sum_{i \in \text{jet}}k_i\bigg)^2,
\end{equation}
where the sum runs over all the particles $i$ which are clustered in the jet. 
In these lecture notes, because of its simple definition, we are going to take the jet mass as the prototype of a jet substructure observable.
This observable will be discussed in detail in this
  chapter and we will again come back to it in
  Chapter~\ref{calculations-substructure-mass} where we are going to compute the jet
  mass for jets modified by substructure techniques, a case
  particularly relevant for phenomenological applications at the
  LHC.

In our discussion, we shall focus on QCD jets, i.e.\ jets which are initiated by a hard parton and subsequently evolve through parton shower. 
Our perturbative analysis will mostly be performed at parton level, i.e.\
we will consider quarks and gluons to be the jet's
constituents. Perturbation theory is not able to describe the
transition to particle level and hadronisation models are usually
employed in event generators to describe the parton-to-hadron
transition.  In this chapter, we will only briefly comment on these
non-perturbative issues, postponing a numerical analysis of their impact to Chapter~\ref{calculations-substructure-mass}.
Even if we remain within the regime of perturbative QCD, we will see
that the fixed-order methods are not adequate in order to capture the
relevant dynamics of the jet mass, especially in the boosted regime
where emissions are accompanied by large logarithms. Thus, we will
exploit all-order resummation techniques to better handle the theoretical description of this observable.
In order to maintain our presentation as simple as possible, while discussing most of the relevant features, we are going to still focus our discussion on jets produced in $e^+e^-$. 
We shall comment on the complications that arise when considering hadron-hadron collisions in Sec.~\ref{sec:pp-collisions}.
In order to make the connection between the $e^+e^-$ and the $pp$ discussion as close as possible, we consider in both cases jets clustered with a generalised $k_t$ algorithm with radius $R$, in its $e^+e^-$ and $pp$ adaptations, respectively~\cite{Cacciari:2011ma}.

\section{The one-loop calculation}
We start by considering the so-called cumulative distribution, which is defined as the normalised cross-section for measuring a value of the jet mass below a certain $m^2$:
\begin{equation}\label{eq:cumu-def}
\Sigma(m^2)= \frac{1}{\sigma_0}\int_0^{m^2} d {m'}^2 \frac{d \sigma}{d {m'}^2}=1+\as \Sigma^{(1)}+\ord \left( \as^2\right),
\end{equation}
where following common practice in the literature, we have chosen to use the Born cross-section as a normalisation factor. The cumulative distribution is a dimensionless quantity and so we can anticipate that its dependence on the jet mass must come as a ratio to another energy scale, which is typically the jet energy (or in proton-proton collision the jet transverse momentum).

We first tackle the calculation of Eq.~(\ref{eq:cumu-def}) to $\ord
\left(\as \right)$, in the soft limit. Thus, we consider the eikonal
factor for the quark-antiquark dipole (cf.~Eq.~(\ref{eq:real-virtual-together}))
\begin{align}\label{eq:dipole-quark-antiquark}
W_{12}&= \frac{\as}{2 \pi} (2 C_F) \frac{k_1 \cdot k_2}{(k_1 \cdot k_3)(k_2 \cdot k_3)},
\end{align}
where $k_1$ and $k_2$ are the momenta of the quark and antiquark, respectively and $k_3$ is the momentum of the soft gluon. For instance, we can choose to parametrise them as
\begin{align}
k_1&=\frac{Q}{2} \left(1,0,0,1 \right), \quad k_2=\frac{Q}{2} \left(1,0,0,-1 \right), \nonumber\\
k_3&= \omega \left(1, \sin \theta \cos \phi, \sin \theta \sin \phi, \cos \theta \right).
\end{align}
In terms of the above parametrisation of the kinematics, the Lorentz-invariant phase-space becomes
\begin{equation}\label{eq:phase-space-integration}
\int d\Phi\equiv \int_0^\infty \omega\, d\omega \int_{-1}^1 d\cos \theta \int_0^{2\pi}\frac{d\phi}{2\pi}.
\end{equation}
This is equivalent to $\tfrac{1}{\pi}\int d^4k_3\delta(k_3^2)$, which
for simplicity has a slightly different normalisation convention
than Eq.~(\ref{eq:4vect-phase-space}).\footnote{Watch out that
  different conventions are present in the literature. For example
  (see \eg~\cite{Dasgupta:2007wa}), one sometimes uses
  $\int d^4k_3\delta(k_3^2)$ as a phase-space integration, in which
  case Eq.~\eqref{eq:dipole-quark-antiquark} has a
  $\tfrac{\alpha_s}{2\pi^2}$ factor instead of
  $\tfrac{\alpha_s}{2\pi}$.}  Note that in the above expression we are
allowed to ignore any recoil of the quarks against the gluon because
we work in the soft limit. Furthermore, in this limit, the energy of
the jet is simply $Q/2$.
The colour factor $2 C_F$ in Eq.~(\ref{eq:dipole-quark-antiquark})
emerges because we are in the presence of only one dipole. For a
process with more hard partonic lines, we should sum over all possible
dipoles each of which is accompanied by the effective colour factors
introduced in Eq.~(\ref{eq:eff-col-def}).

The one-loop evaluation of the cumulative distribution is then obtained by adding together real and virtual corrections. At one loop these contributions are both given by the eikonal factor $W_{12}$, but with opposite sign. Another crucial difference is that when the emitted gluon is real, then we have to impose the appropriate phase-space constraints. 
In particular, if the gluon is clustered in the jet seeded by the hard parton $k_1$, then its contribution to the jet mass is constrained to be less than $m^2$. 
If instead it falls outside the jet, then it only contributes to the zero-mass bin. 
In formulae, we have~\footnote{For simplicity, we introduce the
  following notation for the Heaviside step function:
  $\Theta\left( a>b\right)\equiv \Theta\left( a-b\right)$,
  $\Theta\left( a<b\right)\equiv \Theta\left( b-a\right)$, and
  $\Theta\left( a<b<c\right)\equiv \Theta\left(b-a \right)
  \Theta\left(c-b \right)$.  }
\begin{align}\label{eq:sigma-1-loop-begin}
\as \Sigma^{(1)}(m^2)&=\int_{-1}^1 d \cos \theta \int_0^{2 \pi} \frac{d \phi}{2 \pi} \int_0^{Q/2} \omega d \omega
\frac{2 C_F \as}{\pi} \frac{1}{\omega^2 (1-\cos \theta)(1+\cos \theta)} 
\nonumber \\ & \times
\left[ \Theta_\text{in jet} \Theta \left(2\frac{Q \omega}{2}(1-\cos \theta) < m^2 \right)
+\Theta_\text{out jet} -1\right]\nonumber\\
&= - \frac{2 \as C_F}{\pi} \int_{\cos R}^{1- 2 \frac{m^2}{Q^2}} 
\frac{d \cos \theta }{(1-\cos \theta)(1+\cos \theta)} \log\left( \frac{Q^2 (1- \cos \theta)}{2 m^2}\right),
\end{align}
In the above equation, we have used $\Theta_\text{in jet}=\Theta\left(1-\cos \theta< 1-\cos R \right)$ and $\Theta_\text{out jet}=1-\Theta_\text{in jet}$ , which, for a jet made up of two particles, is the condition to be satisfied for any clustering algorithm of the generalised $k_t$ family. We will see in Sec.~\ref{sec:clustering} that beyond one loop the details of the clustering algorithm affect the single-logarithmic structure of the jet mass distribution.

The integral over the gluon angle is fairly straightforward. Since we
are interested in the logarithmic region, we neglect powers of the jet
mass divided by the hard scale $Q$:
\begin{align}\label{eq:sigma-1-loop-cntd}
\as \Sigma^{(1)}&=- \frac{\as C_F}{2 \pi} 
\left[\log^2 \left( \frac{Q^2}{m^2}\tan^2 \frac{R}{2} \right) -\log^2 \left(\cos^2 \frac{R}{2}\right)- 2 \text{Li}_2\left(\sin^2\frac{R}{2} \right) \right] +\ord \left(\frac{m^2}{Q^2} \right),
\end{align}
which is valid for $\frac{m^2}{Q^2}< \sin^2\frac{R}{2}$.
Thus, we see that the jet mass distribution exhibits a double
logarithmic behaviour in the ratio of the jet mass to the hard
scale. 
We note that these logarithmic contributions are large if the characteristic energy scale of the jet is much bigger than the jet invariant mass.
This situation is precisely what defines boosted topologies and therefore reaching a quantitative understanding boosted-object phenomenology requires dealing with these potentially large logarithmic corrections.
As we discussed before, these double logarithms arise from the
emission gluons which are both soft and collinear and we therefore
expect their presence to any order in perturbation theory. This $\as^n
L^{2n}$ behaviour jeopardises our faith in the perturbative expansion
because the suppression in the strong coupling is compensated by the
presence of the potentially large logarithm $L$. In the next section,
we will discuss how to resum this contributions, i.e. how to
reorganise the perturbative expansions in such a way that logarithmic
contributions are accounted for to all orders. We also note that this
is necessary only if we are interested in the region $m^2/Q^2\ll1$,
where the logarithms are large. In the large-mass tail of the distribution instead $m^2/Q^2\sim 1$ and fixed-order perturbation theory is the appropriate way to capture the relevant physics. Ideally, we would then match resummation to fixed-order to obtain a reliable prediction across the whole range, as shown for instance in Eq.~(\ref{eq:matching-add}).

Before moving to the resummed calculation, we want point out two more considerations. 
First,  we can consider a further simplification to Eq.~(\ref{eq:sigma-1-loop-cntd}), namely we can expand it in powers of the jet radius $R$, which is appropriate for narrow jets
\begin{align}\label{eq:sigma-1-loop-smallR}
\as \Sigma^{(1)}(m^2)&=- \frac{\as C_F}{2 \pi}  \log^2 \left( \frac{Q^2R^2}{4 m^2}  \right) +\ord \left( R^2\right)
=- \frac{\as C_F}{2 \pi}  \log^2 \left( \frac{1}{\rho}  \right),
\end{align}
where we have introduced $\rho=\frac{4 m^2}{Q^2R^2}$.
Second, we want to discuss further the collinear limit. 
The starting point of our discussion so far has been the eikonal factor
$W_{12}$ in Eq.~(\ref{eq:dipole-quark-antiquark}), which means that we
have only considered the emission of a soft gluon. However, as we
discussed in Sec.~\ref{sec:qcd_soft_fact}, there is another region
of the emission phase-space which can produce logarithmic
contributions, namely collinear emissions with finite energy
$\omega$. We expect this region to be single-logarithmic with the
logarithms originating because of the $\cos \theta \to 1$ singularity
of the matrix element. The residue of this singularity is given by the
appropriate splitting function $P_i(z)$, with $z= \frac{2 \omega}{Q}$,
which were given in Eq.~(\ref{eq:quarksplitting})
and~(\ref{eq:gluonsplitting}). Our one-loop result is modified
accordingly and we get
\begin{align}\label{eq:sigma-1-loop-smallR-coll}
\as \Sigma^{(1)}(\rho)&
=- \frac{\as C_F}{ \pi}  \left [\frac{1}{2} \log^2\left(\frac{1}{\rho}\right)  +B_q \log \left(\frac{1}{\rho}\right) \right],
\end{align}
with 
\begin{align}\label{eq:B1}
B_q=\int_0^1 d z \left[ \frac{P_{q}(z)}{2C_F}-\frac{1}{z}  \right]&= -\frac{3}{4}.
\end{align}
The collinear limit is of particular relevance when discussing
boosted-objects, as radiation is typically collimated along the jet
axis.  Furthermore, it is often easier from a computational viewpoint
to work in such limit because collinear emissions essentially
factorise at the cross-section level, while we need to take into
account colour correlation at the amplitude level to correctly
describe soft emissions at wide angle. Therefore, unless explicitly
stated, from now on, we are going to present first calculations in the
collinear (and optionally soft) limit and then comment to their
extension to include wide-angle soft emission. However, we stress that
in general both contributions are necessary to achieve a given
(logarithmic) accuracy in the theoretical description of a process.
%we are interested in.

\section{Going to all orders}\label{sec:jet-mass-res}
In order to obtain theoretical predictions that can be applied in the regime $\rho \ll1$, we have to move away from fixed-order predictions and resum parton emission to all orders in perturbation theory. 
Inevitably, we are only going to scratch the surface of the all-order formalism behind resummed calculations and we encourage the interested readers to study more specialised reviews and the original literature on the topic.
In particular, the resummation framework that we adopt heavily relies on the so-called \caesar approach~\cite{Banfi:2004yd}, originally developed for the NLL resummation of event shapes both in $e^+e^-$ and hadron collisions~\cite{Banfi:2001bz,Banfi:2003je,Banfi:2004nk,Banfi:2010xy} and subsequently extended to jet observables~\cite{Dasgupta:2012hg,Dasgupta:2013ihk}. 
A numerical implementation of this resummation formalism exists~\cite{Reichelt:2021eru} and it has been used for several phenomenological studies, see for instance~\cite{Hoche:2017kst,Marzani:2019evv,Baberuxki:2019ifp,Baron:2020xoi,Caletti:2021oor,Caletti:2021ysv,Reichelt:2021svh}.
We mention that this formalism can also be extended beyond NLL~\cite{Banfi:2014sua}, but this goes beyond what is discussed in this book.

For our discussion, we are going to consider a quark-initiated jet in the presence of many collinear (hard or soft) partons. As discussed above, the complete resummed calculation must also consider soft gluons at large angle, while the soft quarks at large angle do not give rise to logarithmic contributions.
Let us begin with some consideration on the observable. We want to recast the definition Eq.~(\ref{eq:jet-mass-def}) in a form which is suitable for the all-order treatment. 
In the collinear limit, the angular separation between any two jet constituents is small, so we have
\begin{align} \label{eq:jetmass-many-emissions}
m^2&= 2\sum_{(i<j)\in \text{jet}} k_i \cdot k_j
=\sum_{(i<j)\in \text{jet}}  \omega_i \omega_j \theta_{ij}^2+ \mathcal{O}\left(\theta_{ij}^4\right).
\end{align}
Any pair-wise distance can be written in terms of each particle's
distance from the jet axis and the azimuth in the plane transverse to
the jet axis: $\theta_{ij}^2= \theta_i^2+\theta_j^2-2 \theta_i \theta_j \cos \phi_{ij}$.
% \begin{align}\label{eq:thetaij}
% \theta_{ij}^2= \theta_i^2+\theta_j^2-2 \theta_i \theta_j \cos \phi_{ij}.
% \end{align}
Substituting the above expression in Eq.~(\ref{eq:jetmass-many-emissions}), we obtain
\begin{equation}\label{eq:jetmass-many-emissions-2}
m^2=
\frac{1}{2} \sum_{(i,j) \in \text{jet}} \omega_i \omega_j \theta_{ij}^2
 = \frac{1}{2}
 \sum_{(i,j) \in \text{jet}} \omega_i \omega_j \left(\theta_i^2+\theta_j^2-2 \theta_i \theta_j \cos \phi_{ij} \right) =  \sum_{i \in \text{jet}} E_J \omega_i \theta_i^2,
\end{equation}
where $E_J= \sum_{i \in \text{jet}} \omega_i=\tfrac{Q}{2}$ is the jet energy and we have exploited that for each $i$,
\begin{equation}
\sum_{j \in \text{jet}} \omega_j \theta_j \cos \phi_{ij}=0,
\end{equation}
because of momentum conservation along $i$ in the plane transverse to the jet.

As before, we are going to consider the cumulative distribution, i.e.\ the probability  for a jet to have an invariant jet mass (squared) less than $m^2$.
We have to consider three cases. Real emissions that are clustered into the jet do contribute to the jet mass distribution, while real emissions outside the jet, as well as virtual corrections, do not change the jet mass. Thus, the cumulative distribution in this approximation reads:
\begin{align}\label{eq:res-mass-start}
\Sigma(\rho)&=\sum_{n=0}^\infty \frac{1}{n!} \prod_{i=1}^n \int \frac{d \theta_i^2}{\theta_i^2} \int d z_i P_q(z_i) \frac{\as(z_i \theta_i \frac{Q}{2})}{2\pi}
\Theta_{i \in \text{jet}} \Theta \left(\sum_{i=1}^n z_i \frac{\theta_i^2}{R^2} < \rho \right) \nonumber \\&
+\sum_{n=0}^\infty \frac{1}{n!} \prod_{i=1}^n \int \frac{d \theta_i^2}{\theta_i^2} \int d z_i P_q(z_i) \frac{\as(z_i \theta_i \frac{Q}{2})}{2\pi}
\Big [\Theta_{i \notin \text{jet}}-1\Big],
\end{align}
where the running coupling is evaluated at a scale which represents
the transverse momentum of emission $i$ with respect to the $q \bar q$
dipole, in the dipole rest frame, cf.~Eq.~(\ref{eq:dipole-kperp}).
The above expression deserves some comments.
In order to derive it, we have exploited the factorisation properties of QCD matrix elements squared in the collinear limit. 
We note that the $1/n!$ prefactor can be viewed as consequence of (angular) ordering.
Furthermore, we note that the argument of the each splitting function is energy fraction $z_i$. This is true if the fractional energy coming out of each splitting is computed with respect to the parent parton.
On the other hand, the energy fraction that enters the observable definition is calculated with respect to the jet energy, which in our approximation coincides with the energy of the initial hard quark $E_J=\frac{Q}{2}$.
In the collinear limit, these two fractions are related by a rescaling
factor $x_i$ that takes into account the energy carried away by
previous emissions $x_i=\prod_{k=1}^{i-1}(1-z_{k})$. However, this
rescaling only gives rise to subleading (NNLL) corrections and can
therefore be dropped in Eq.~(\ref{eq:res-mass-start}).
Furthermore, we have also written the jet clustering condition in a factorised form, essentially assuming $\Theta_{i \in \text{jet}} =\Theta(\theta_i<R)$. 
If the jet is made up of only two particles, this condition is exact for any member of the generalised $k_t$ clustering family. However, there is no guarantee that such condition can be written in a factorised form, in presence of an arbitrary number of particles.
Crucially, the widely used anti-$k_t$ algorithm does exhibit this
property in the soft limit. In other words, anti-$k_t$ behaves as a
perfectly rigid cone in the soft-limit, where all soft particles are
clustered first to the hard core, leading to a factorised
expression. This is not true with other jet algorithms, such as the
Cambridge/Aachen algorithm and the $k_t$ algorithm, for which
corrections to the factorised expression occur at NLL accuracy for
soft gluon emissions. We
will return to this point in Sec.~\ref{sec:clustering}.

With the above clarifications in mind, we can go back to Eq.~(\ref{eq:res-mass-start}).
While the second line of~(\ref{eq:res-mass-start}) is already in a
fully factorised form, the $\Theta$-function constraining the
observable in the first line spoils factorisation. The way around this
obstacle is to consider an appropriate integral representation of the
$\Theta$ function in order to obtain a factorised expression in a
conjugate space~\cite{Catani:1991kz,Catani:1992ua}. In other words, we
could compute Mellin moments of the cumulative distribution in order
to obtain a factorised expression.

At LL accuracy, where each emission comes with a maximal number of
logarithms, one can further assume {\em strong ordering}, \ie that the
$z_i\theta_i^2$ themselves are strongly ordered.
In this case, a single emission strongly dominates the sum and we
can write
\begin{equation}\label{eq:strong-rho-ordering-LL}
  \Theta \left(\sum_{i=1}^n \rho_i < \rho \right)
  \approx \Theta \left(\max_{i} \rho_i < \rho \right)
  =\prod_{i=1}^n \Theta \left( \rho_i < \rho \right), \qquad \rho_i =  z_i \frac{\theta_i^2}{R^2},
\end{equation}
The fact that, at LL accuracy, a single emission strongly dominates
the jet mass is an important result that we will use extensively
through this book.

\begin{figure}
    \centering
 \includegraphics[width=0.45\textwidth]{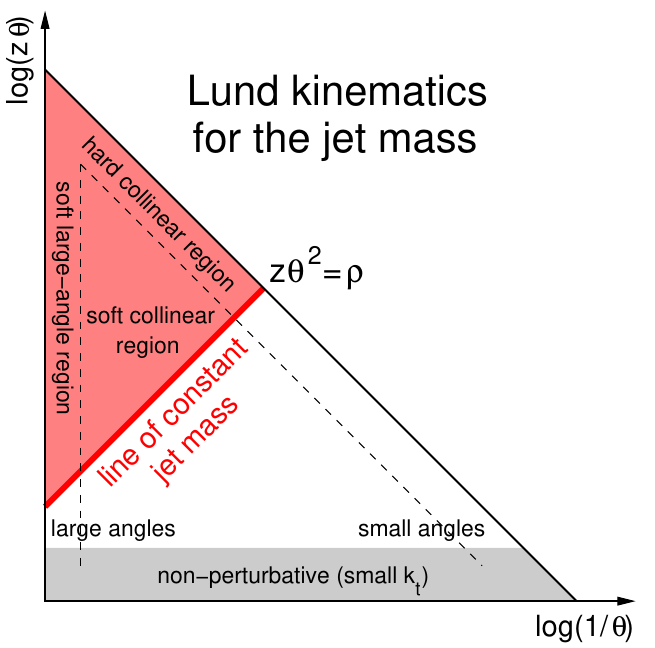}%
 \caption{Lund diagram for the jet mass distribution at LL. The solid
   red line corresponds to emissions yielding the requested jet mass,
   \ie with $z\theta^2=\rho$ (using angles rescaled by $R$). The
   shaded red area is the vetoed area associated with the Sudakov
   suppression. ``Soft, wide-angle'' emissions have a small $k_t$ and
   angles of order $R$, and ``hard collinear'' splittings have a small
   angle and a large $z$ fraction. The shaded grey region at the
   bottom of the plot corresponds to the non-perturbative,
   small-$k_t$, region.}\label{fig:lund-plain}
\end{figure}

With the above assumptions, it is now straightforward to perform the
sum over the number of emissions 
  \begin{align}\label{eq:res-mass-cont}
\Sigma^{(LL)}(\rho)&=-\sum_{n=0}^\infty \frac{1}{n!} \prod_{i=1}^n \int \frac{d \rho_i}{\rho_i} \int d z_i P_q(z_i) \frac{\as(\sqrt{z_i \rho_i} \frac{QR}{2})}{2\pi}
\Big[ \Theta(\theta<R) \Theta \left(\rho_i> \rho \right) \Big]\nonumber
 \\
&= \exp \bigg[ -  \int_\rho^1 \frac{d \rho'}{\rho'} \int d z_i P_q(z_i) \frac{\as(\sqrt{z \rho'}\frac{QR}{2})}{2\pi}
\Theta(\theta<R) \Theta \left(\rho_i> \rho \right)\bigg] \nonumber \\ &\equiv \exp \Big [ -R(\rho) \Big].
\end{align}
This is an interesting and important result: {\em the cumulative
  distribution can be written, at LL accuracy, in an exponential
  form}. At this accuracy, the exponent is determined by the one-gluon
contribution and, in particular, can be interpreted as the virtual
one-loop contribution, because of the negative sign, evaluated on the
region of phase-space where the real emission is vetoed. The function
$R(\rho)$ is usually referred to as the {\em Sudakov
  exponent}~\cite{Sudakov:1954sw} (or the radiator) and it represent
the no-emission probability. \footnote{Please note that throughout
  this book, $R$ can either denote the jet radius or the
  radiator/Sudakov exponent. In context, it should be trivial to tell
  one from the other.}
From the cumulative distribution, we can immediately obtain the resummed jet mass spectrum
\begin{align} \label{eq:spectrum-res}
\frac{\rho}{\sigma_0}\frac{d \sigma }{d \rho}&= \frac{d}{d \log(\rho)} \Sigma(\rho)= R'(\rho) e^{-R (\rho)},
\end{align}
where $R'= \frac{d}{d L}R$ and $L=\log\big( \frac{1}{\rho}\big)$.
It is useful to re-interpret this result in terms of Lund
diagrams~\cite{Andersson:1988gp}. These diagrams represent the emission kinematics
in terms of two variables: vertically, the logarithm of an emission's
transverse momentum $k_t$ with respect to the jet axis, and
horizontally, the logarithm of the inverse of the emission's angle
$\theta$ with respect to the jet axis, (alternatively, we could use
its rapidity with respect to the jet axis, if we want to work with
hadron colliders coordinates).\footnote{More generally, if one
  considers a gluon emitted from a dipole, as we did in
  Chapter~\ref{chap:qcd-colliders} and earlier in this chapter, one
  would consider the rapidity along the dipole direction,
  $-\log(\tan(\theta/2))$, and the transverse momentum $k_\perp$ with respect to the
  dipole, cf.~Eq.~(\ref{eq:dipole-kperp}).}
Note that, in Lund diagrams (and often in actual calculations) we make use of \emph{rescaled} variables, i.e.\ angles are given in units of the jet radius and the emission transverse momentum (or energy) in units of the jet transverse momentum (or energy).
The diagram in Fig.~\ref{fig:lund-plain} shows a line of constant jet
mass, together with a shaded (red) region corresponding to the part of
the kinematic plane where real emissions are vetoed because they would lead
to a value of the mass larger than $\rho$.
In this region, only virtual contributions are allowed, giving rise to
the Sudakov factor $\exp[-R(\rho)]$.
Outside the shaded (red) region, real and virtual contributions cancel.
Because QCD matrix elements are logarithmic in the soft/collinear
region, the no-emission probability is proportional to the area of the
shaded region (up to running-coupling corrections).
    
In order to obtain explicit resummed expressions, we have to evaluate
the integrals in Eq.~(\ref{eq:res-mass-cont}) to the required
accuracy. For instance, if we aim to NLL (in the small $R$ limit), we
have to consider the running of the strong coupling at two
loops. Furthermore, we have to include the complete one-loop splitting
function $P_q(z)$ as well its soft contribution at two loops, which
corresponds to the two-loop cusp anomalous dimension
$K=C_A \left (\frac{67}{18}- \frac{\pi^2}{6} \right ) - \frac{5}{9}
n_f$. We note that this contribution accounts for correlated gluon
emission which are unresolved at NLL accuracy. This correction can
therefore be absorbed into the running coupling, giving rise to the
so-called Catani-Marchesini-Webber (CMW) scheme~\cite{Catani:1990rr}:
\begin{equation}\label{eq:CMW} 
\frac{\as^\text{CMW}(\mu)}{2\pi }=\frac{\as(\mu)}{2 \pi}+ K \left(\frac{\as(\mu)}{2 \pi}\right)^2.
\end{equation}
We write the resummed exponent as
\begin{equation}\label{eq:radiator-nll-expansion}
R (\rho)= Lf_1(\lambda)+ f_2(\lambda),
\end{equation}
where $f_1$ and $f_2$ resum leading and next-to-leading logarithms,
respectively:
\begin{equation}
\label{eq:quark}
f_1(\lambda) =  \frac{C_F}{\pi \beta_0 \lambda} \left [ \left(1-\lambda \right ) 
\log \left(1-\lambda \right)-2 \left ( 1-\frac{\lambda}{2} \right ) \log \left
  (1-\frac{\lambda}{2} \right ) \right ],
\end{equation}
and
\begin{multline} \label{eq:radiator-nll-contribution}
f_2(\lambda) =  \frac{C_F K}{4 \pi^2 \beta_0^2} \left [2 \log \left 
(1-\frac{\lambda}{2} \right ) - \log \left (1-\lambda \right )\right ] -
\frac{ C_F B_q}{\pi \beta_0} \log \left ( 1-\frac{\lambda}{2} \right ) 
\\  +\frac{C_F \beta_1}{2 \pi \beta_0^3} \left [ \log \left (1-\lambda \right )-2 \log 
\left (1-\frac{\lambda}{2} \right ) + \frac{1}{2} \log^2 \left (1- \lambda \right ) 
- \log^2 \left (1-\frac{\lambda}{2} \right ) \right ],
\end{multline}
$\lambda = 2 \alpha_s\beta_0 L$, $B_q$ was defined in
Eq.~(\ref{eq:B1}), and $\alpha_s\equiv\alpha_s\left(QR/2\right)$ is
the $\overline{\text{MS}}$ strong coupling.
Since this kind of results will appear repeatedly throughout this
book, we give an explicit derivation of the above formul\ae\ in
Appendix~\ref{chap:app-analytic-details}. In the above results we have also
introduced the one-loop and two-loop coefficients of the QCD $\beta$-function, namely $\beta_0$ and $\beta_1$. 
Their explicit expressions are given in Appendix~\ref{chap:app-analytic-details}.

In order to achieve the complete NLL resummation formula for the
invariant mass distribution of narrow, i.e.\ small $R$, jets we need
to consider two additional contributions: multiple emissions and
non-global logarithms~\cite{Dasgupta:2001sh}. We have already
mentioned how to deal with the former: in the real-emission
contribution to Eq.~(\ref{eq:res-mass-start}), we can no longer apply
the strong-ordering simplification 
Eq.~\eqref{eq:strong-rho-ordering-LL} and the resummed calculation must
be done in a conjugate (Mellin) space in order to factorise the
observable definition. 
At the end of the calculation, the result must then brought back to
physical space. In case of jet masses this inversion can be done in
closed-form and, to NLL accuracy, it can be expressed as a correction
factor:
\begin{align}\label{eq:mult-emission-nll}
\mathcal{M}(\rho)&=  \frac{e^{-\gamma_E R'(\rho)}}{\Gamma(1 +R'(\rho))}.
\end{align}
Non-global logarithms are instead resummed into a factor $\mathcal{S}(\rho)$ which has a much richer (and complex) structure. We will discuss it in some detail in Sec.~\ref{sec:non-global}. Putting all things together the NLL result for the cumulative mass distribution reads
\begin{equation}
  \label{eq:Sigma-plain-jet-mass}
  \Sigma^{(\text{NLL})}(\rho) = \mathcal{M} \,\mathcal{S}\, e^{-R} .
\end{equation}

Thus far we have discussed the jet mass distribution in the context of perturbation theory. However, when dealing with soft and collinear emissions, we are probing the strong coupling deeper and deeper in the infra-red and we may become sensitive to non-perturbative contributions. This is clearly dangerous because as the coupling grows, perturbation theory becomes first unreliable and then meaningless. The presence of an infra-red singularity (Landau pole) for the coupling makes this breakdown manifest: at long distances we cannot use partons as degrees of freedom but we have to employ hadrons. 
From this point of view it is then crucial to work with IRC safe observables, for which we can identify regions in which the dependence on non-perturbative physics can be treated as a (small) correction.

\subsection{A sanity check: explicit calculation of the second order}\label{sec:sanity}
\begin{figure}[tb]
  \begin{center}
\includegraphics[width=0.8\textwidth]{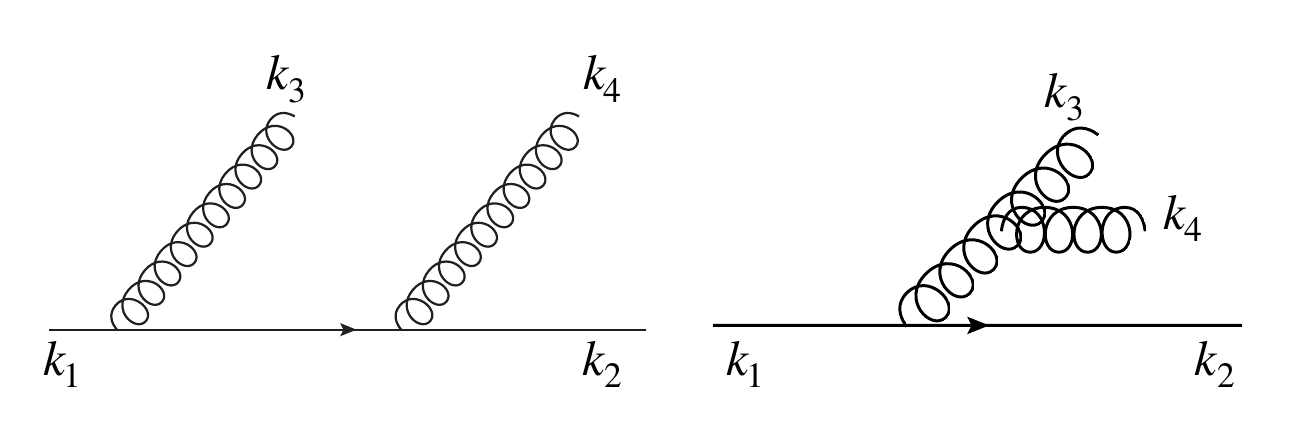}
    \caption{A schematic representation of the types of contributions to the strongly order emission of two soft gluons, in the double real-emission case: independent emission  of the left and correlated emission on the right.}
    \label{fig:ind-corr}
  \end{center}
\end{figure}

As a sanity check of the all-order calculation we have performed in the previous section, we explicitly calculate the double logarithmic contribution at two loops and compare it to the expansion of the resummation to second order.
Thus, we need to consider the squares matrix element for the emission
of two soft gluons with momenta $k_3$ and $k_4$, off a $q \bar q$
dipole, in the limit where both $k_3$ and $k_4$ are soft, with $k_4$ much softer than $k_3$~\cite{Catani:1983bz,Dokshitzer:1992ip}. This can be written as the sum of two pieces: independent and correlated emissions
\begin{equation}\label{eq:2gluons}
W= \cf^2 W^{(\rm ind)}+ \cf \ca W^{(\rm corr)},
\end{equation}
where
\begin{eqnarray}
W^{(\rm ind)}&=&   \frac{2\, k_1\cdot k_2}{k_1\cdot k_3 \, k_2 \cdot k_3}   \frac{2\, k_1\cdot k_2}{k_1\cdot k_4 \, k_2 \cdot k_4}, \label{eq:independent}\\
W^{(\rm corr)}&=&  \frac{2\, k_1\cdot k_2}{k_1\cdot k_3 \, k_2 \cdot k_3}
 \left( \frac{k_1\cdot k_3}{k_1\cdot k_4 \, k_3 \cdot k_4}+ \frac{k_2\cdot k_3}{k_2\cdot k_4 \, k_3 \cdot k_4}-
  \frac{k_1\cdot k_2}{k_1\cdot k_4 \, k_2 \cdot k_4}
 \right).\label{eq:correlated}
\end{eqnarray}
The two contributions are schematically shown in Fig.~\ref{fig:ind-corr}.
Because we are interested in the $\as^2 L^4$ contribution to the cumulative distribution, which is the most singular one, we expect it to originate from the independent emission  of two gluons in the soft and collinear limit. 
We have to consider three types of configuration: double real emission, double virtual and real emission at one loop. 
For each of the three types, the contribution to the squared matrix
element for ordered two-gluon emission is the same up to an overall sign. Focusing on the independent emission contribution, the result for the double real (RR) or double virtual (VV) is
\begin{equation}
W^{(\rm ind)} =   \frac{256}{Q^4} \frac{1}{z_3^2 z_4^2} \frac{1}{\theta_3^2 \theta_4^2 }\,.
\end{equation}
A similar result holds for the real emission at one loop, with a
relative minus sign. The latter has to be counted twice because the real
emission could be either $k_3$ (RV) or the softer gluon $k_4$ (VR).
 We are now in a position to compute the jet mass distribution at the two-gluon level for the independent emission $C_F^2$ term. 
To perform this calculation we note that it is actually more convenient to consider the differential jet mass distribution rather than the cumulative, as we usually do. In fact, if we demand $m^2>0$, then the double virtual configuration does not contribute because it lives at $m^2=0$.
Therefore we consider
\begin{equation}
\frac{d \tilde \sigma}{d \rho}=\frac{1}{\sigma_0}\frac{d \sigma}{d \rho}=
\as \frac{d \tilde \sigma^{(1)}}{d \rho}+\as^2 \frac{d \tilde \sigma^{(2)}}{d \rho}+ \ord \left(\as^3 \right).
\end{equation}

We start by noting that the phase space integration region for all configurations can be divided according to whether the real gluons $k_3$ and $k_4$ are inside or outside the jet of interest. We have four distinct regions: $k_3,k_4$ both outside the jet, $k_3,k_4$ both inside the jet or either of the gluons inside and the other outside the jet. The condition for a given gluon to end up inside or outside the jet depends on the jet definition. 
In the anti-$k_t$ algorithm with radius $R$ the condition is
particularly simple when considering only soft emissions: a soft
emission $k_i$ is inside the jet if it is within an angle $R$ of the
hard parton initiating the jet, otherwise it is outside. As we have already noted, the anti-$k_t$ algorithm in the soft limit works as a perfect cone.

Let us consider all four cases one by one.
The contribution where both $k_3$ and $k_4$ are outside the jet
trivially vanishes since it gives a massless jet.
We then consider the case where the harder emission $k_3$ is in the
jet and $k_4$ is out. Graphs RR and RV cancel since the real $k_4$ does not contribute to the jet mass exactly like the virtual $k_4$. This leaves diagram VR, which gives zero since the in-jet gluon $k_3$ is virtual and hence does not generate a jet mass.
Hence the region with $k_3$ in and $k_4$ out gives no contribution. 
The contribution where $k_4$ is in the jet and $k_3$ out vanishes for the same reason.
Hence we only need to treat the region with both gluons in the jet and
we shall show that this calculation correctly reproduces the result
based on exponentiation of the single gluon result. The sum of the RR,
RV and VR contributions can be represented as\footnote{Here with an
  abuse of notation we are indicating the LHS of the equation as
  $\as^2 \frac{d \tilde \sigma^{(2)}}{{d}\rho}$, while we really mean
  only its double leading contribution.} (with $d\Phi$ defined in Eq.~\eqref{eq:phase-space-integration})
\begin{equation}
\as^2 \frac{d \tilde \sigma^{(2)}}{{d}\rho} = \int d \Phi\, W \left[ \delta\left(\rho- z_3 \theta_3^2-z_4 \theta_4^2\right)-\delta\left(\rho-z_3 \theta_3^2\right)-\delta \left(\rho-z_4 \theta_4^2 \right) \right],
\end{equation}
where in order to keep our notation simple, we have switched to rescaled angular variables: $\theta_i  \to \frac{\theta_i}{R}$, so that now $\theta_i<1$.
To proceed, we note that in the leading-logarithmic approximation
emissions are also strongly ordered in $z\theta^2$, \ie we
have either $z_3\theta_3^3\gg z_4\theta_4^2$, or $z_4\theta_4^3\gg
z_3\theta_3^2$.
This means that only the largest of $z_3\theta_3^3$ and
$z_4\theta_4^2$ contributes to $\delta\left(\rho-z_3 \theta_3^2-z_4
  \theta_4^2 \right)$, with the other being much smaller. We can
therefore write
\begin{equation}\label{eq:delta-2em-simplifacation-LL}
  \delta\left(\rho-z_3 \theta_3^2-z_4 \theta_4^2\right) \to
  \delta\left(\rho-z_3 \theta_3^2\right)\Theta\left(\rho>z_4
    \theta_4^2 \right)+3\leftrightarrow 4\,.
\end{equation} 
Doing so and using the explicit forms of $W$ and the phase space $d \Phi$ in the small angle limit we get 
\begin{multline}
\as^2 \frac{d \tilde \sigma^{(2)}}{{d}\rho} = - \left ( \frac{\alpha_sC_F}{\pi} \right)^2 
\int \frac{d\theta_3^2}{\theta_3^2}\frac{d\theta_4^2}{\theta_4^2} 
\frac{d\phi}{2\pi} \frac{dz_3}{z_3} \frac{dz_4}{z_4} \left[ \delta\left(\rho-z_3 \theta_3^2\right)\Theta\left(z_4 \theta_4^2 >\rho \right)+3 \leftrightarrow 4 \right]\\ \Theta \left(z_3>z_4\right),
\end{multline}
where $\phi$ is the azimuthal angle between the two gluons (the other
azimuthal integration is trivial because the matrix element does not
depend on either $\phi_3$ or $\phi_4$).
We note that the overall factor
$-\Theta\left(z_4 \theta_4^2 >\rho \right)$ comes again from the
region where $k_4$ is virtual, while real and virtual emissions cancel
each other for $z_4 \theta_4^2 <\rho$.
Carrying out the integrals we obtain
\begin{equation}\label{eq:LL-result-alphas2}
\as^2 \frac{d \tilde \sigma^{(2)}}{{d}\rho}  = -\frac{1}{2} \left(\frac{\alpha_sC_F}{\pi} \right)^2 \frac{1}{\rho} \log^3 \left (\frac{1}{\rho} \right),
\end{equation}
which is precisely the result obtained by expanding the exponentiated double-logarithmic one-gluon result to order $\alpha_s^2$ and differentiating with respect to $\rho$.
Thus the standard double-logarithmic result for the jet-mass
distribution arises entirely from the region with both gluons in the
jet. Contributions from soft emission arising from the other regions
cancel in the sense that they produce no relevant logarithms.

We note that since we have used a soft-gluon approximation (with
gluons emitted from colour dipoles), the result above does not
include the contribution from hard-collinear splittings which, at this
order would give a contribution
$\frac{3}{2}\big(\frac{\alpha_sC_F}{\pi}\big)^2\frac{1}{\rho}\log^2(\rho)B_q$.
Finally, beyond the double-logarithmic approximation, the
approximation~\eqref{eq:delta-2em-simplifacation-LL} is no longer
valid. It does bring a correction to Eq.~\eqref{eq:LL-result-alphas2}
coming from the difference between the left-hand side and the right-hand side of
\eqref{eq:delta-2em-simplifacation-LL}.
In practice, we get
\begin{align}
& \frac{1}{2} \left ( \frac{\alpha_sC_F}{\pi} \right)^2 
\int \frac{d\theta_3^2}{\theta_3^2}\frac{d\theta_4^2}{\theta_4^2} 
  \frac{dz_3}{z_3} \frac{dz_4}{z_4} \left[
 \delta\left(\rho-z_3 \theta_3^2-z_4 \theta_4^2\right)
 -\delta\left(\rho-z_3 \theta_3^2\right)\Theta\left(z_4 \theta_4^2 >\rho \right)
 -3 \leftrightarrow 4 \right]\nonumber\\
  &= \left ( \frac{\alpha_sC_F}{\pi} \right)^2 
\int_0^\rho \frac{d\rho_3}{\rho_3}\frac{d\rho_4}{\rho_4} 
  \log\left(\frac{1}{\rho_3}\right)\log\left(\frac{1}{\rho_4}\right) \left[
 \delta\left(\rho-\rho_3-\rho_4\right) -\delta\left(\rho-\rho_3\right)\right]
    \Theta(\rho_3>\rho_4)\nonumber\\
  &= \left ( \frac{\alpha_sC_F}{\pi} \right)^2 \frac{1}{\rho}
\int_0^\rho \frac{d\rho_4}{\rho_4} 
  \log\left(\frac{1}{\rho_4}\right)
    \left[\log\left(\frac{1}{\rho-\rho_4}\right)-\log\left(\frac{1}{\rho}\right)\right]\nonumber\\
  &= \left ( \frac{\alpha_sC_F}{\pi} \right)^2 \frac{1}{\rho}
\frac{\pi^2}{6}  \log\left(\frac{1}{\rho}\right) + \text{(terms with no
    $\log(\rho)$ enhancements)}
\end{align}
where we have introduced $\rho_i=z_i\theta_i^2$ and used
$\int\frac{d\theta_i^2}{\theta_i^2}\frac{dz_i}{z_i}f(\rho_i) = \int\frac{d\rho_i}{\rho_i}\log(1/\rho_i)f(\rho_i)$.
It is easy to show that this contribution corresponds exactly to the
first non-trivial correction from $\mathcal{M}(\rho)$ in
Eq.~(\ref{eq:mult-emission-nll}), after differentiation with respect to $\rho$,
with $R'(\rho)=\frac{\alpha_sC_F}{\pi}\log\big(\tfrac{1}{\rho}\big)$.

\subsection{Non-global logarithms}\label{sec:non-global}
\begin{figure}[tb]
  \begin{center}
\includegraphics[width=0.7\textwidth]{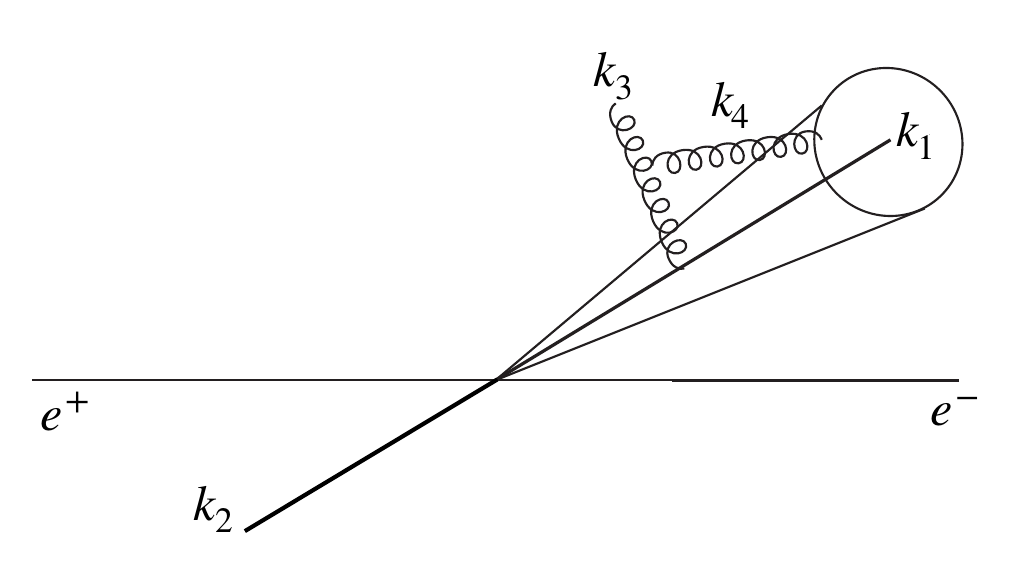}
\caption{Kinematic configuration that gives rise to non-global
  logarithms to lowest order in perturbation theory. The $k_3$ gluon
  is in the jet and does not contribute to the jet mass, while the
  $k_4$ gluon is in the jet and thus contributes to the jet mass.}
    \label{fig:non-globalcontribution}
  \end{center}
\end{figure}
In Sec.~\ref{sec:jet-mass-res} we have described an all-order calculation that aims to resum large logarithms of the ratio of the jet mass to the hard scale of the process to NLL. Furthermore, in Sec.~\ref{sec:sanity} we have verified the leading logarithmic behaviour predicted by the resummation by performing a two-loop calculation in the soft and collinear limit. 
In order to do that we have considered the independent emission contribution to the soft eikonal current Eq.~(\ref{eq:2gluons}).
For observables that are sensitive to emissions in the whole phase-space, such as for instance event shapes like thrust~\cite{Farhi:1977sg} a similar exercise can be also done for the correlated emission contribution to the soft current. Then we would find that these effects are fully accounted for by treating the running coupling in the CMW scheme, i.e.\ by considering the two-loop contribution to the cusp anomalous dimensions.
However, it turns out that for so called non-global observables, i.e.\ observables that are sensitive only to a restricted region of phase-space, the all-order calculation previously described is not enough to capture full NLL accuracy. Indeed, correlated gluon emissions generate a new tower of single-logarithmic corrections~\cite{Dasgupta:2001sh,Dasgupta:2002bw}  the resummation of which is far from trivial.

Let us focus our discussion on a fixed-order example, which illustrates how a single logarithmic contribution arises in non-global observables. 
Because we are dealing with an observable that is only sensitive to emissions in a patch of the phase-space, we can have a configuration where a gluon is emitted outside this patch, in this case outside the jet, and it re-emits a softer gluon inside the jet. 
Thus, we consider the correlated emission contribution to the matrix
element square for the emission of two soft gluons in the kinematic
region where the harder gluon $k_3$ is not recombined with the jet,
while the softer gluon $k_4$ is. 
In order to better illustrate the features of the calculation, in this section we are going to retain the full angular dependence, without taking the collinear limit. 
This makes sense because one of the gluons is emitted outside the jet, where the collinear approximation is less justified. 
Note that the integration over the gluon momentum $k_3$ is sensitive
to the rest of the event and it may depend, for instance, on the way
we select the jet, the mass of which we are measuring. For example, if
we only select the hardest jet in the event, then one would have to
prevent $k_3$ from clustering with $k_2$. For simplicity, in this example, we are going to
integrate $k_3$ over the whole phase-space outside the measured jet.
If we restrict ourselves to a jet algorithm, such as anti-$k_t$, which works as a perfect cone in the soft limit, this condition simply translates to $ 1-  \cos \theta_3> 1- \cos R$ and $1- \cos \theta_4<1- \cos R$.
This situation is depicted in Fig.~\ref{fig:non-globalcontribution}.
%
% Thus, we consider the first non-global contribution to the cumulative
% distribution. After adding together real and virtual corrections, we
% obtain
At order $\alpha_s^2$, the leading non-global contribution can be
written as
\begin{align}
\as^2 S^{(2)} = & -4 C_F C_A \left ( \frac{\alpha_s}{2\pi} \right)^2 \int \frac{d\omega_3}{\omega_3} \int \frac{d \omega_4}{\omega_4} 
\Theta\left(\omega_3>\omega_4 \right)
\int d\cos\theta_3 \int d\cos\theta_4  \, \Omega(\theta_3,\theta_4)\nonumber \\ 
& \,\Theta\left(\cos \theta_3<\cos R\right)
\, \Theta \left(\cos \theta_4> \cos R \right)
 \, \Theta \left(\omega_4 Q (1-\cos\theta_4)>m^2 \right),
\end{align}
In this expression, the last $\Theta$ constraint comes from adding the
real and virtual contributions for the gluon $k_4$.
The angular function $\Omega$ arises after integrating the correlated
matrix element square, Eq.~\eqref{eq:correlated}, over the azimuth
$\phi$. Its expression reads~\cite{Dasgupta:2001sh}
\begin{equation}
\Omega(\theta_3,\theta_4) = \frac{2}{\left(\cos\theta_4-\cos\theta_3 \right) \left(1-\cos\theta_3\right) \left(1+\cos \theta_4 \right)}.
\end{equation}
We first perform the integration over the energies of the two gluons, obtaining 
\begin{multline}
\as^2 S^{(2)} = - 2 C_F C_A \left ( \frac{\alpha_s}{2\pi} \right)^2 
\int d\cos\theta_3 \int d\cos\theta_4  
\, \Theta\left( \cos \theta_3< \cos R\right)
\, \Theta \left(\cos \theta_4>\cos R \right)
\\
\, \Omega(\theta_3,\theta_4)\log^2\left(\frac{2m^2}{Q^2(1-\cos\theta_4)}\right) \Theta\left( \frac{2m^2}{Q^2(1-\cos\theta_4)}>1\right).
\end{multline}
We can now perform the angular integrations and express the results in
terms of our rescaled variable $\rho$. The calculation can be
simplified by noting that, since we are interested only in the NLL
contribution, we can safely ignore the angular dependence in the
argument of the logarithm.
We obtain:
\begin{equation}
\label{eq:sigma_NG}
\as S^{(2)} = -2C_F C_A \left (\frac{\alpha_s}{2\pi}\right)^2  \frac{\pi^2}{6}\log^2\left(\frac{1}{\rho}\right) + \dots
\end{equation}
where the dots indicate subleading contributions.  It is interesting
to observe that the coefficient of the first non-global logarithm is
independent of $R$.\footnote{This result depends on the fact that we
  have integrated $k_3$ over the whole phase-space outside the
  jet. With additional constraints on the external region, the
  coefficient of $\log^2(\rho)$ would be more complex. However, in the
  small-$R$ limit, one would always obtain $\tfrac{\pi^2}{6}$ up to
  powers of $R$.}
This might seem counter-intuitive at first because we might naively think that the probability for $k_3$ to emit a softer gluon inside the jet must be proportional to the jet area. However, the calculation shows that there is a nontrivial and $R$-independent contribution arising from the region where both gluons are close to the jet boundary. This results in an integrable singularity which is the origin of the $\pi^2/6$ contribution.

The result in Eq.~(\ref{eq:sigma_NG}) represents only the leading term
at the first order at which non-global logarithms appear. In order to
achieve full NLL accuracy these contributions must be resummed to
all-orders. This is highly non-trivial, even if our aim is to resum
only the leading tower of non-global logarithms needed at NLL.
In order to perform an all-order analysis of non-global logarithms, we must consider configurations of many soft gluons. If we restrict ourselves to considering their leading contributions, which is single-logarithmic, we can assume energy-ordering; however, no collinear approximation can be made. Thus, we have to describe how an ensemble of an arbitrary number of soft gluons, all outside the jet, can emit an even softer gluon inside the jet. 

Colour correlations make the colour algebra very complex as every
emission increases the dimensionality of the relevant colour
space. Moreover, describing the geometry of such ensembles also
becomes difficult. The approach that was taken in the first analysis
of non-global logarithms~\cite{Dasgupta:2001sh} was to consider the
large-$N_C$ limit. Colour correlations becomes trivial in this limit
because the off-diagonal entries of the colour matrices vanish. Thus,
we are able to write the matrix element square for the $n$ gluon
ensemble in a factorised way~\cite{Bassetto:1984ik} and a simplified
physical picture emerges. An emission off an ensemble of $n-1$ gluons
(plus the two hard partons) reduces to the sum over the emission off
each of the $n$ dipoles.
When the dipole radiates a gluon, it splits into two dipoles, originating configurations which are determined by the history of the gluon branching.
This can be implemented as a Monte Carlo which enables one to deal
numerically with the second above-mentioned difficulty, namely the
complicated geometry of the multi-gluon final states. This solution
was first implemented in Ref.~\cite{Dasgupta:2001sh} and subsequently
used in a number of phenomenological applications,
e.g.~\cite{Dasgupta:2002dc,Banfi:2006gy,Banfi:2008qs,Banfi:2010pa,Dasgupta:2012hg}.

The numerical impact of non-global logarithm on jet mass spectra can
be large, see e.g.~\cite{Banfi:2010pa,Dasgupta:2012hg}, and because
their treatment at NLL is only approximate, they often represent the
bottleneck to reach perturbative precision in this kind of
calculations. Remarkably, as we will discuss in
Chapter~\ref{calculations-substructure-mass}, some grooming algorithms
greatly reduce or even get rid of non-global logarithms, thus paving
the way towards an improved perturbative accuracy of jet mass distributions.

Because of their complexity, a lot of effort has been invested in better
understanding and controlling non-global logarithms. In the rest of
this section, we highlight some of the main results for the reader
interested in a deeper exploration of non-global logarithms.
The resummation of non-global logarithms was formalised by Banfi, Marchesini and Smye. In Ref.~\cite{Banfi:2002hw}, they were able to derive an evolution equation, henceforth the BMS equation, which, equivalently to the Monte Carlo approach, resums the leading non-global logarithm, in the large-$N_C$ limit. 
More recently, this formalism has been extended to the next subleading order~\cite{Banfi:2021owj,Banfi:2021xzn}.

It has been noted~\cite{Marchesini:2003nh} that the BMS equation has the same form as the Balitsky-Kovchegov (BK) equation~\cite{Balitsky:1995ub,Kovchegov:1999yj} that describes non-linear small-$x$ evolution in the saturation regime. This correspondence has been studied in detail in Refs~\cite{Avsar:2009yb,Hatta:2009nd}, where BMS and BK were related via a stereographic projection. Because a generalisation of the BK equation to finite $N_C$ exists~\cite{JalilianMarian:1997gr,Iancu:2000hn}, the correspondence between non-global logarithms and small-$x$ physics was argued to hold at finite-$N_C$ and numerical solutions have been studied~\cite{Weigert:2003mm,Hatta:2013iba}. Very recently, this correspondence was indeed mathematically established~\cite{Caron-Huot:2015bja}. 
In this approach, a \emph{colour density matrix} is introduced, with the aim of describing soft radiation and an evolution equation is then derived for the colour density matrix, to all-loops, at finite $N_C$. The related anomalous dimension $K$ is  explicitly computed to one and two loops. The one-loop approximation to this evolution equation coincides with the BMS equation, once the large-$N_C$ limit is taken and it confirms on a firmer ground the results of Refs.~\cite{Weigert:2003mm,Hatta:2013iba} at finite $N_C$. More importantly, the explicit calculation of the two-loop contribution to $K$ paves the way for the resummation of non-global logarithms at higher-logarithmic accuracy, although computing solutions to the evolution equation remains a challenging task.

A different approach to the question of resumming non-global logarithms was developed in Refs.~\cite{Forshaw:2006fk,Forshaw:2008cq,Martinez:2018ffw} and applied to a phenomenological study of jet vetoes between hard jets in Refs.~\cite{Forshaw:2009fz,Delgado:2011tp}. In that context, because colour-correlations were of primary interest, the large-$N_C$ limit did not seem adequate.
We finish this discussion pointing out that other approaches similar in spirit were recently developed using techniques of SCET.~\cite{Larkoski:2015zka,Larkoski:2016zzc,Becher:2015hka,Becher:2016mmh,Balsiger:2019tne,Balsiger:2020ogy,Becher:2021zkk,Becher:2021urs,Becher:2023vrh,Becher:2023mtx}

\subsection{Dependence on the clustering algorithm}\label{sec:clustering}

In all the calculations performed thus far we have always treated the constraints originating from the jet algorithm in a rather simple way.
Essentially, we have always drawn a hard cone of radius $R$ centred on the hard parton and considered as clustered into the jet soft emissions laying within that cone. 
As already mentioned, this approach is justified if we are using the
anti-$k_t$ algorithm.
However, the situation changes for other members of the generalised $k_t$
family, such as the Cambridge/Aachen algorithm or the $k_t$
algorithms. Indeed, these clustering algorithms have a distance
measure which admits the possibility of two soft gluons being the
closest pair, thus combining them before they cluster with the hard
parton.

We now revisit the two-gluon calculation described in
Sec.~\ref{sec:sanity}, this time making use of the $k_t$ clustering
algorithm. We keep the same convention for the kinematics, i.e.\ the
soft gluon momenta are labelled $k_3$ and $k_4$, with $k_4$ much
softer than $k_3$.
As in the previous section, we should consider either the case where
both gluons are real, or the case where one of the gluons (either
$k_3$ or $k_4$) is real and the other is virtual.

We start by considering the RR contribution in different kinematic
configurations.  Clearly, when both $k_3$ and $k_4$ are beyond an angle
$R$ with respect to the hard parton there is no contribution from
either to the jet-mass.
When both $k_3$ and $k_4$ are within an angle $R$ of the hard parton,
both soft gluons get combined into the hard jet and this region
produces precisely the same result as the anti-$k_t$ algorithm,
corresponding to exponentiation of the one-gluon
result.\footnote{Remember that when a soft particle clusters with a
  much harder one, the resulting object has the $p_t$ and direction of
  the harder particle, up to negligible recoil.}
However, when $k_3$ is beyond an angle $R$ and $k_4$ is inside an angle $R$ the situation changes from the anti-$k_t$ case. 
This is because the $k_t$ distance between the two soft gluons can be
smaller than the $k_t$ distance between $k_4$ and the hard parton,
in which case $k_4$ clusters with $k_3$, resulting in a soft jet along
the direction of $k_3$.
Thus, when $k_3$ is beyond an angle $R$ it can pull $k_4$ out of the hard jet since the soft jet $k_3+k_4$ lays now at angle larger than $R$ with respect to the hard parton, i.e.\ outside the jet. 
Therefore, this kinematic configuration results in a massless jet.
In precisely the same angular region the VR configuration is obviously
unaffected by clustering and it does give a contribution to
$\tfrac{d\sigma}{d\rho}$. This contrasts with the anti-$k_t$ case
where the real and virtual contributions cancelled exactly at this
order.
Note also that the RV configuration gives no contribution (as in the
anti-$k_t$ case) because no real gluons are in the jet.
Finally, for the case were $k_3$ is inside the jet and $k_4$ is
outside the jet, a similar situation can happen where $k_3$ and $k_4$
are clustered first, pulling $k_4$ back in the jet. This case however
does not lead to an extra contribution because, since $k_4$ is much
softer than $k_3$, it does not affect the mass of the jet already
dominated by $k_3$.

Thus a new  contribution arises for the $k_t$ algorithm from the region where the two real
gluons $k_3$ and $k_4$ are clustered, where we only get a contribution
from the case where $k_3$ is virtual and $k_4$ is real.
We now carry out this calculation explicitly. 
We work in the small-$R$ limit and consider the angles $\theta_3$,
$\theta_4$ and $\theta_{34}$ as the angles between $k_3$ and the hard
parton, $k_4$ and the hard parton and $k_3$ and $k_4$ respectively. 
In order to apply the $k_t$-algorithm in $e^+ e^-$, we have to compare
the distances $\omega_3^2 \theta_3^2$, $\omega_4^2 \theta_4^2$ and
$\omega_4^2 \theta_{34}^2$.
Now since $\theta_3^2 > R^2$, $\theta_4^2 <R^2$ and
$\omega_4\ll\omega_3$, the only quantities that can be a candidate for
the smallest distance are $\omega_4^2 \theta_4^2$ and $\omega_4^2 \theta_{34}^2$. Thus
the gluons are clustered and $k_4$ is pulled out of the jet if
$\theta_{34}<\theta_4<R$.
Otherwise $k_4$ is in the jet and cancels against virtual corrections, precisely as it happened for the anti-$k_t$ algorithm.

Making use of the usual rescaling $\theta \to \theta/R$, we can then write the VR contribution in the clustering region as
\begin{multline} 
\frac{ d \tilde{\sigma}_2^{\mathrm{cluster}}}{d \rho}=-4 C_F^2 \left(\frac{\alpha_s}{2\pi}\right)^2 \int \frac{d\theta_3^2}{\theta_3^2}\frac{d\theta_4^2}{\theta_4^2} \frac{d\phi}{2\pi} \frac{d z_3}{z_3} \frac{dz_4}{z_4} \delta \left(\rho -z_4 \theta_4^2 \right)\Theta(z_3>z_4) \\  \Theta\left(\theta_3^2>1\right) \Theta \left(\theta_{34}^2<\theta_4^2\right)\Theta \left(\theta_{4}^2<1\right).
\end{multline}
Within our small-angle approximation, we can write
$\theta_{34}^2 =\theta_3^2 +\theta_4^2-2 \theta_3 \theta_4 \cos \phi$.
Integrating over $z_3$ and $z_4$ and using
$t= \frac{\theta_4^2}{\rho}$ one obtains
 \begin{multline} 
\frac{ d \tilde{\sigma}_2^{\mathrm{cluster}}}{d \rho}= -4 C_F^2 \left( \frac{\alpha_s}{2\pi} \right)^2 \frac{1}{\rho} \int \frac{d\theta_3^2}{\theta_3^2} \frac{dt}{t} \frac{d \phi}{2 \pi} \log(t) \\
 \Theta \left(t>1\right) \Theta \left(\theta_3^2>1\right) \Theta\left(4 \rho t \cos^2\phi>\theta_3^2\right)\Theta \left(t \rho<1 \right).
\end{multline}
Carrying out the integral over $\theta_3^2$ results in 
\begin{multline}\label{eq:logs-clustering-interm}
\frac{ d \tilde{\sigma}_2^{\mathrm{cluster}}}{d \rho}= -4 C_F^2 \left( \frac{\alpha_s}{2\pi} \right)^2 \frac{1}{\rho} \int \frac{dt}{t} \frac{d \phi}{2 \pi} \log \left(4 \rho t \cos^2 \phi \right) \log(t) 
\\ \Theta \left(t>1\right) \Theta\left(4 \rho t \cos^2\phi>1\right)\Theta \left(\rho t <1\right).
\end{multline}
Now we need to carry out the $t$ integral for which we note
$t > \mathrm{max}\left(1, \frac{1}{4\rho\cos^2 \phi}\right)$. In the
region of large logarithms which we resum one has however that
$\rho \ll 1$ and hence $4\rho\cos^2 \phi\ll 1$.
At NLL accuracy we can therefore take $t>\frac{1}{4\rho\cos^2 \phi}$
and replace the $\log(t)$ factor by $\log \big(\tfrac{1}{\rho}\big)$ in~\eqref{eq:logs-clustering-interm}.
It is then straightforward to carry out the $t$ integration to get
\begin{equation} 
 \frac{ d \tilde{\sigma}_2^{\mathrm{cluster}}}{d \rho}
 = -4 C_F^2 \left( \frac{\alpha_s}{2\pi} \right)^2 \frac{1}{\rho} \log\left(\frac{1}{\rho}\right) \int_{-\frac{\pi}{3}}^{\frac{\pi}{3}} \frac{d \phi}{\pi} \log^2(2 \cos \phi) = -\frac{2\pi^2}{27} C_F^2 \left( \frac{\alpha_s}{2\pi} \right)^2 \frac{1}{\rho} \log \left(\frac{1}{\rho}\right).
 \end{equation} 
This behaviour in the distribution corresponds to a single-logarithmic $\alpha_s^2 \log^2 \big(\frac{1}{\rho}\big)$ contribution to the cumulative, which is, as anticipated, necessary to claim NLL accuracy.
The all-order treatment of these clustering effects is far from trivial because of the complicated kinematic configurations, which results into many nested $\Theta$ function. Therefore, from this point of view, resummation of mass spectra for jet defined with the anti-$k_t$ algorithm appears simpler. Conversely, because of these clustering effects, the jet boundary becomes somewhat blurred, resulting in milder non-global contributions. 
For an analysis of clustering logarithms using SCET, we refer the reader to~\cite{Becher:2023znt}.

\subsection{Non-perturbative corrections:
  hadronisation}\label{sec:plain-mass-hadronisation}

Lund diagrams, such as the one in Fig.~\ref{fig:lund-plain}, turn out
to be particularly useful in order to determine the sensitivity of an
observable to non-perturbative dynamics. We can introduce a
non-perturbative scale $\mu_\text{NP}\sim 1$~GeV below which we enter
a non-perturbative regime. Because the running coupling in
Eq.~(\ref{eq:res-mass-start}) is evaluated at a scale that represent
the emission transverse momentum with respect to the jet, a horizontal
line $z \theta =\tilde{\mu}=\frac{\muNP}{E_J R}$  marks the boundary between
perturbative and non-perturbative dynamics (recall that
$\theta$ is measured in unit of the jet radius $R$).
It is then simple to calculate what is the corresponding value of the jet mass for which the integrals we have to perform have support on the non-perturbative region: we just have to work out where the line of constant $\rho$ first crosses into the non-perturbative region. This happens when $z \theta= \tilde{\mu} $ and $\theta=1$, which implies $\rho=\tilde{\mu}$.
Thus, this simple argument suggests that the mass distribution becomes sensitive to non-perturbative physics at 
\begin{equation}\label{eq:mass-NP-estimate}
m^2\simeq \frac{\muNP}{E_J R} E_J^2 R^2= \muNP E_J R.
\end{equation}
Note that this scale grows with the jet energy, so that even apparently
large masses, $m \gg \Lambda_\text{QCD}$, may in fact be driven by
non-perturbative physics.
For a $3\TeV$ jet with $R=1$, taking $\muNP = 1\GeV$, the
non-perturbative region corresponds to $m \lesssim 55\GeV$,
disturbingly close to the electroweak scale!

Experimentally jets can be thought of as a bunch of collimated hadrons
(mesons and baryons). However, we have so far considered jets from a
perturbative QCD perspective and used partons to describe their constituents. 
The parton-to-hadron transition, namely hadronisation, is a
non-perturbative phenomenon.
Non-perturbative corrections due to hadronisation can be treated,
within certain approximations, with analytic methods, see
e.g.~\cite{Lee:2006fn,Stewart:2014nna}. For the jet mass the leading correction turns
out to be a shift of the differential
distribution~\cite{Dokshitzer:1997ew,Salam:2001bd}. Furthermore, this type of analytic calculations can provide insights about the dependence of these corrections on the parameters of the jet algorithm, such as the jet radius~\cite{Dasgupta:2007wa}.
Alternatively, we can take a more phenomenological point of view and
use Monte Carlo parton showers to estimate non-perturbative
correction. For instance, we can either calculate a given observable
on a simulated event with hadrons in the final state, or stop the
event simulation before hadronisation takes place and compute the same
observable with partons. We can then take the bin-by-bin ratio of the
jet mass distribution computed with and without hadronisation as a
proxy for these corrections.
This is the path we are going to employ in this book to illustrate the
impact of non-perturbative corrections (both hadronisation and the
Underlying Event, which we must also include when considering hadron-hadron collisions).
 We will present such studies in
Chapter~\ref{calculations-substructure-mass}, where hadronisation
correction to the jet mass distribution discussed here will be
compared to the ones for jets with substructure (typically grooming)
algorithms.

\section{From $e^+e^-$ to hadron-hadron collisions}\label{sec:pp-collisions}

Thus far we have discussed the resummation of the invariant mass distribution of a jet produced in an electron-positron collision. In order to be able to perform jet studies in proton-proton collision we have to extend the formalism developed so far. A detailed derivation of the resummation formulae goes beyond the scope of this book and we refer the interested reader to the original literature, e.g.\ ~\cite{Catani:1996yz,Kidonakis:1998nf}. Here, instead we briefly sketch the issues that we have to tackle and how we can go about them.

\begin{enumerate}[label=\alph*)]
\item As discussed in Chapter~\ref{chap:qcd-colliders}, in proton-proton collision, we work
  in the collinear factorisation framework, Eq.~(\ref{eq:master}),
  where cross-sections are described as a convolution between a
  partonic interaction and universal parton distribution functions.
  Furthermore, we need to switch to the appropriate kinematic
  variables for proton-proton collisions, namely transverse momentum,
  rapidity and azimuthal angle
  (cf.~Sec.~\ref{sec:hadron-collider-kinematics}).
\item The complexity of resummed calculations increases in the case of
  hadronic process because we have to deal with many hard legs with
  colour, including the initial-state partons. As we have noted in Sec.~\ref{sec:qcd_soft_fact}, factorisation in the soft limit happens at the amplitude level and interference terms play a crucial role in the soft limit. 

As a consequence, resummed calculations that aim to correctly capture these effect must account for all non-trivial colour configuration. 
In particular, if we have a process that at Born level as more than two coloured hard legs, either in the initial or final state, then the one-gluon emission contribution in the soft limit Eq.~(\ref{eq:dipole-quark-antiquark}) can be generalised as follows
\begin{equation} \label{eq:sumdipoles}
W= \sum_{(ij)}\frac{\alpha_s(\kappa_{ij}) \, C_{ij}}{2 \pi} \frac{p_i \cdot p_j}{(p_i \cdot k) (p_j \cdot k)},
\end{equation}
where to avoid confusion we have labelled the momenta of the hard legs as $p_i$ (rather than $k_i)$ and $k$ is the soft gluon momentum. We note that the sum runs all over the dipoles $(ij)$, i.e.\ all pairs of hard legs $i$ and $j$ . To NLL accuracy, the running coupling in Eq.~(\ref{eq:sumdipoles}) must be evaluated at the scale $\kappa_{ij}^2= \frac{2(p_i \cdot k) (p_j \cdot k)}{(p_i \cdot p_j)}$, which is the transverse momentum of the emission with respect to the dipole axis, in the dipole rest frame. $C_{ij}$ is a generalisation of the effective colour charge, Eq.~(\ref{eq:eff-col-def}), which is not necessarily diagonal:
\begin{equation}\label{eq:effective_colour}
C_{ij}=-2 \, T_i \cdot T_j,
\end{equation}
where the colour matrices $T_i$ are not necessarily in the fundamental representation, as the gluon can be emitted off a gluon line as well.
We note that the expression above greatly simplifies in the collinear limit, where one recovers the usual colour factors $C_F$ and $C_A$. However, soft emissions at large angle do contribute beyond LL and therefore dealing with the sum over dipoles is mandatory in order to achieve NLL accuracy. 

It is possible to show that, even in the presence of many hard legs, the one-loop contribution above still exponentiates. However, one must keep track, for each dipole, of the different colour flow configurations. This results into a rather complex matrix structure in colour space~\cite{Catani:1996yz,Kidonakis:1998nf}.
As an example, in Sec.~\ref{sec:ISR}, we will evaluate the contribution to the jet mass distribution in $pp$ collision from a soft gluon emission emitted from the dipole made up of the incoming hard legs.

\item Finally, new sources of non-perturbative corrections arise in proton-proton collisions. Collinear factorisation assumes that only one parton from each proton undergoes a hard scattering. However, we can clearly have secondary, softer, scatterings between the protons' constituents. As we have mentioned at the beginning of this book, these multiple-parton interactions produce what is usually referred to as the Underlying Event. Furthermore, because protons are accelerated and collided in bunches, we also have multiple proton-proton interactions per bunch-crossing, leading to what we call pileup.
As a consequence hadronic collisions are polluted by radiation that
does not originated from the hard scattering. In the context of jet
physics this radiation has important consequences as it modifies the
jet properties, e.g.\ its transverse momentum or its mass, in a way
which is proportional to powers the jet radius $R$.
More specifically, corrections to the jet transverse momentum are proportional to $R^2$, while corrections to the jet mass exhibit a $R^4$ behaviour~\cite{Dasgupta:2007wa}.
Therefore large-$R$ jets are more significantly affected by these effects.

Some first-principle studies have been performed, mostly concentrating on double-parton scattering (see Ref.~\cite{Diehl:2017wew} for a recent review), however most phenomenological analyses rely on models of the underlying event which are usually incorporated in Monte Carlo simulations. These models are characterised by a number of free parameters which are determined by comparisons with experimental data with a process known as \emph{tuning}. We will come back to the numerical impact of the underlying event in Chapter~\ref{calculations-substructure-mass}, where we will discuss the ability of grooming techniques to reduce such contamination.

\end{enumerate} 

To illustrate the extra complications one has to deal with in
proton-proton collisions, we conclude this chapter by computing first
the effect of initial-state radiation and then the jet mass
distribution in Z+jet events.

\subsection{Initial-state radiation as an example}\label{sec:ISR}

In this section we sketch the calculation of the contribution to the jet mass distribution from the emission of a soft a gluon from the dipole formed by the two incoming hard legs. This can be taken as a good proxy to the effect of initial-state radiation.
As it is the first calculation we perform with hadron-collider
kinematic variables, let us explicitly specify the
kinematics:
\begin{align}
p_1 &= \frac{\sqrt{s}}{2} x_1 \left (1,0,0,1 \right),\quad
p_2 = \frac{\sqrt{s}}{2} x_2 \left (1,0,0,-1 \right), \nonumber \\
p_3 &= p_t \left( \cosh y, 1,0,\sinh y \right), \quad
 k=  k_t \left(\cosh \eta, \cos \phi, \sin \phi, \sinh \eta \right), 
\end{align}
where $p_1$ and $p_2$ denote the four-momenta of the incoming hard
partons, $p_3$ the momentum of the jet, and $k$ of the soft gluon. It
is understood that the jet must recoil against a system with momentum
$p_4$ (not specified above), over which we are inclusive. 
Note that we have used hadron-collider variables, \ie transverse
momenta $p_t$ and $k_t$, rapidities $y$ and $\eta$, and azimuthal
angle $\phi$, assuming without loss of generality that the jet is
produced at $\phi=0$. 
Provided the soft gluon is clustered with the jet, its contribution to the jet mass is
\begin{align}
m^2&=(p_3+k)^2=2 p_3 \cdot k = 2 p_t k_t \left(\cosh (\eta-y) -\cos \phi \right).
\end{align}
We can now write the contribution to the cumulative distribution from
the 12 dipole as
\begin{align}
\as \Sigma^{(1)}_{12} &= -C_{12} \int k_t dk_t d\eta \frac{d\phi}{2\pi} \frac{\alpha_s \left (\kappa_{12} \right) }{2 \pi} \frac{(p_1.p_2)}{(p_1.k)(p_2.k)} \Theta \left ((\eta-y)^2+\phi^2 <R^2 \right)
\nonumber \\ & \cdot \Theta \left 
( \frac{2 k_t}{p_t R^2} \left (\cosh (\eta-y) -\cos \phi \right) > \rho \right),
\label{eq:initial-state-rad-interm}
\end{align}
where the first $\Theta$ function is the jet clustering condition and
we have introduced $\rho= \frac{m^2}{p_t^2 R^2}$, analogously to the
$e^+ e^-$ case.
We next note that
\begin{equation}
\kappa^2_{12} =  2 \frac{(p_1.k)(p_2.k)}{(p_1.p_2 )} = k_t^2.
\end{equation}
Eq.~\eqref{eq:initial-state-rad-interm} therefore exhibits a
logarithmic enhancement at small $k_t$ as expected.
To isolate the leading (NLL) contribution, we can as usual just retain
the dependence of the jet mass on $k_t$ in the second line
of~\eqref{eq:initial-state-rad-interm}, and neglect the dependence on
$y$, $\eta$ and $\phi$ which produces terms beyond NLL accuracy.
We can then carry out the integration over $\eta$ and $\phi$ which
simply measures the jet area $\pi R^2$ and obtain
\begin{equation}
\as \Sigma^{(1)}_{12} = -C_{12} {R^2} \int_{\rho p_t}^{p_t} \frac{\alpha_s (k_t)}{2 \pi} \frac{dk_t}{k_t},
\end{equation}
where the lower limit of integration stems from the constraint on the
jet mass. 
The dipole consisting of the two incoming partons gives indeed rise to
a pure single-logarithmic behaviour. Since the emitted gluon is inside
the jet region, away from the hard legs constituting the dipole, there
are no collinear enhancements. Furthermore, the soft wide-angle single
logarithm we obtain is accompanied by an $R^2$ dependence on jet
radius, reflecting the integration over the jet area.

\subsection{The jet mass distribution in $pp \to$ Z+jet}\label{sec:Zjet}

\begin{figure}[tb]
  \begin{center}
\includegraphics[page=1,scale=0.85]{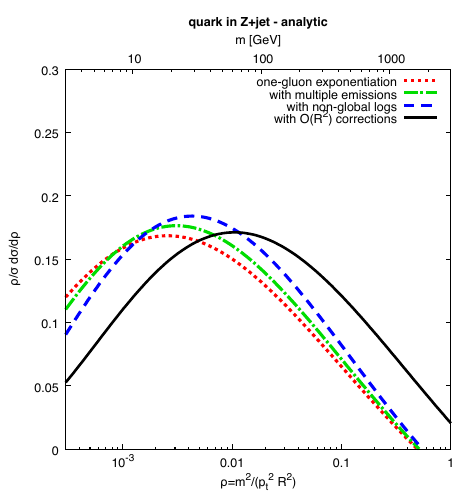}
\includegraphics[page=2,scale=0.85]{figures/plain-mass.pdf}
    \caption{The mass distribution of the quark-initiated  and gluon-initiated jets in Z+jet. The numerical impact of different contributions at NLL accuracy is shown.}
    \label{fig:z+jet}
  \end{center}
\end{figure}

We finish this chapter by showing how all the effects discussed so far affects the calculation of a jet mass distribution. 
We choose to study the jet mass distribution of the hardest jet produced in association with a Z boson. This process is of particular interest in the boosted regime $p_t\gg m$ or, equivalently, $\rho \ll1 $ because it is the main background for the production of a boosted Higgs boson, recoiling against the Z.
In practice, it also has a simpler structure than the jet mass in
dijet events since there are only three coloured hard legs.

At Born level we have to consider two partonic processes
$q g  \to Z q$ and $q \bar  q \to Z g$. 
We can think as the first process to describe the production of a quark-initiated jet, while the second one gives a gluon-initiated jet. We consider a very hard jet with $p_t=3$~TeV and jet radius $R=1$.
We plot in Fig.~\ref{fig:z+jet} the distribution of the variable
$\rho$ calculated to NLL in several approximations, on the left for a
quark-initiated jet, and on the right for a gluon-initiated one. We
start by considering  the exponentiation of the single gluon emission
Eq.~(\ref{eq:spectrum-res}), in the collinear, i.e. \ small $R$ limit
(dotted red curve). We then add the contribution to multiple
emission Eq.~(\ref{eq:mult-emission-nll}) (dash-dotted green
curve). We then add the correction due to non-global logarithms in the
large-$N_C$ limit~\cite{Dasgupta:2001sh} (dashed blue curve). Finally,
we include corrections which are suppressed by powers of the jet
radius (solid black curve).

To illustrate basic aspects of the colour algebra, we work out the
effective colour factors $C_{ij}$ associated to the colour dipoles
(cf.~Eq.~(\ref{eq:effective_colour})) of our Z+jet process.
Let us start with the $q g \to Z q$ process and label with 1 the
incoming quark, with 2 the incoming gluon and with 3 the outgoing
quark, we have that $T_1^2=T_3^2=C_F$ and $T_2^2=C_A$. Exploiting
colour conservation, \ie $T_1+T_2+T_3=0$ (with
all dipole legs considered outgoing), we find
\begin{equation}
%C_{12}=C_{23}=2C_A=N_C, \qquad C_{13}=2C_F-C_A=-\frac{1}{N_C}.
C_{12}=C_{23}=C_A=N_C, \qquad C_{13}=2C_F-C_A=-\frac{1}{N_C}.
\end{equation}
We then move to the gluon-initiated jet case, i.e.\ the Born process
$q \bar  q \to Z g$ and label with 1 the incoming quark, with 2 the
incoming antiquark and with 3 the outgoing gluon. We have that $T_1^2=T_2^2=C_F$ and $T_3^2=C_A$ and
\begin{equation}
%C_{12}=2C_F-C_A=-\frac{1}{N_C}, \qquad C_{13}=C_{23}=2C_A=N_C.
C_{12}=2C_F-C_A=-\frac{1}{N_C}, \qquad C_{13}=C_{23}=C_A=N_C.
\end{equation}

 We note that the $\mathcal{O}(R^2)$ corrections are rather sizeable because we are dealing with a jet with large radius. However, further corrections $\mathcal{O}(R^4)$ turn out to be very small and indistinguishable on the plot. The bulk of this large $\mathcal{O}(R^2)$ effect originates from the $12$ dipole studied above, i.e.\ it can be thought as the contribution of initial-state radiation to the jet mass. 
Finally, we remind the reader that the result in Fig.~\ref{fig:z+jet}
is not matched to fixed-order and therefore it is not reliable in the
$\rho\sim 1$ region. In particular, the resummation is not capable to
correctly capture the end-point of the distribution and matching to
(at least) NLO is mandatory to perform accurate phenomenology.

%% GS helper for auctex
%%% Local Variables:
%%% mode: latex
%%% TeX-master: "notes"
%%% End:

%  LocalWords:  Eq eikonal NNLL NLL Catani Marchesini Webber CMW Smye
%  LocalWords:  integrable Banfi BMS Baliksty Kochegov ij eq

% $Id: groomers-taggers.tex 608 2026-03-15 14:25:57Z smarzani $
%
% This contains an introduction to the jet substructure tools and a
% dictionary of commonly-used methods
%------------------------------------------------------------------------
\chapter{Jet substructure: concepts and tools}\label{tools}

The widest application of jet substructure tools is to disentangle
different kinds of jets. This typically includes separating quark and
gluon-initiated jets or isolating boosted W/Z/H or top jets (our
signal) from the much more abundant QCD background of ``standard''
quark and gluon-initiated jets.
In this chapter, we discuss these methods in some detail. We start by considering 
the guiding principles behind the different algorithms and how to assess their performance. 
Then, we will review some of
the most commonly-used jet substructure techniques over the past 10
years. 
Explicit examples on how these tools behave in Monte-Carlo
simulations and analytic calculations and how they are used in
experimental analyses will be given in the next Chapters.

\section{General guiding principles}\label{sec:tools-generic}

Jet substructure aims to study the internal kinematic 
properties of a high-$p_t$ jet in order to distinguish whether it is more likely to be a signal or  background jet.
Although a large variety of methods have been proposed over the last
ten years, they can be grouped into three wide categories, according to the physical observation that they mostly rely on.
\begin{description}
\item[{\bf Category I: prong finders.}]
  Tools in this category exploit the fact that when a boosted massive object decays
  into partons, all the partons typically carry a sizeable fraction of
  the initial jet transverse momentum, resulting in multiple hard
  cores in the jet. Conversely, quark and gluon jets are dominated by
  the radiation of soft gluons, and are therefore mainly single-core
  jets.
  Prong finders therefore look for multiple hard cores in a jet, hence
  reducing the contamination from ``standard'' QCD jets.
  This is often used to characterise the boosted jets in terms of their
  ``pronginess'', i.e.\ to their expected number of hard cores:
  QCD jets would be 1-prong objects, W/Z/H jets would be two-pronged,
  boosted top jets would be three-pronged, an elusive new resonance
  with a boosted decay into two Higgs bosons, both decaying to a
  $b\bar b$ pair would be a 4-prong object, ...
\item[{\bf Category II: radiation constraints.}]
  The second main difference between signal and background jets is
  their colour structure. This means that signal and background jets
  will exhibit different soft-gluon radiation patterns. For example, QCD radiation associated with 
  an EW-boson jet, which is colourless, it is expected to be less than what we typically find in a QCD jet. 
  Similarly, quark-initiated jets are expected to radiate less
  soft gluons than gluon-initiated jets. Many jet shapes have been
  introduced to quantify the radiation inside a jet and hence separate
  signal jets from background jets.  
\item[{\bf Category III: groomers.}]
  There is a third category of widely-used tools related to the fact
  that one often use large-radius jets for
  substructure studies. As we have already discussed, because of their large area, these jets are particularly sensitive to soft backgrounds, such as the UE and pileup. ``Grooming'' tools have therefore been
  introduced to mitigate the impact of these soft backgrounds on the
  fat jets.
  These tools usually work by removing the soft radiation far from the
  jet axis, where it is the most likely to come from a soft
  contamination rather than from QCD radiation inside the jet.
  In many respects, groomers share similarities with prong finders,
  essentially due to the fact that removing soft contamination and
  keeping the hard prongs are closely related.
  \end{description}
Additionally, we note that we might expect non-trivial interplay between groomers and radiation-constraint observables.
For instance, if we apply observables that exploit radiation constraints on soft
  radiation, to groomed jets, which precisely throws
  away soft radiation, we expect to obtain worse performance. 
  Therefore, we can anticipate that we will have to find a sweet spot between
  keeping the sensitivity to UE and pileup under control, while
  maintaining a large discriminating power.

\section{Assessing performance}\label{sec:performance-intro}

Even though they are based on only a handful of key concepts, a long
list of jet substructure tools have been introduced. Before we dive
into a description of these tools, it is helpful to briefly discuss
how one can compare their relative performance.
Note that, here, we are not referring to how the tools can be
validated, which is often do via Monte Carlo studies, direct
measurements in data or analytic studies. Instead, we would like to answer questions like 
 ``There are dozens of tools around, which one should I
use for my problem?'' or ``Which one has the largest performance?''.
It is of course impossible to give a definite answer to such
questions, but what we can at least provide is some key ideas of what
we mean by ``performant'' which can be properly tested and quantified.

The case of groomers is the probably the easiest to address, since
groomers have the specific purpose of suppressing the sensitivity to
the UE and pileup.
In the case of the UE, we can perform Monte Carlo studies, switching
multiple particle interactions on and off, to check how key distributions --- like the jet
mass distribution in QCD events, W$\to q\bar q$ or hadronic top decay
--- vary.
A similar approach applies for pileup, where we can perform Monte
Carlo studies, overlaying minimum bias events with the hard events.
In cases where we have access to both a reference event (\eg a hard
collision) and a modified event (\eg the same event overlaid with
pileup), quality measures can then involve average shifts and
dispersions of how jet quantities like the jet mass are affected event
by event.
More generally, we can study the position and width of peaks like the
reconstructed W or top mass, and study their stability with respect to the
UE or pileup multiplicity.
We refer to Section~4 of Ref.~\cite{Altheimer:2013yza} for an explicit
application of the above procedure.

\begin{figure}
  \centerline{\includegraphics[width=0.45\textwidth]{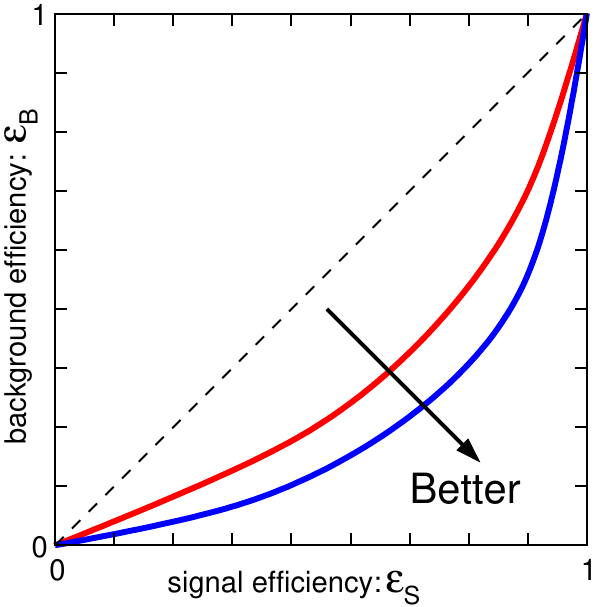}}
  \caption{A ROC curve represents the background efficiency
    $\epsilon_B$ as a function of the signal efficiency
    $\epsilon_S$. For a given signal efficiency, a lower background
    efficiency is better.}\label{fig:ROC-sketch}
\end{figure}

In the following, we are going to focus on
the case of boosted-object tagging.
In this case, there is again a very obvious meaning of what performant
means: the best tool is the one which keeps most of the signal and
rejects most of the background.
In practice, for a signal S and a background B, we define the signal
(respectively background) efficiency $\epsilon_S$ ($\epsilon_B$) as
the rate of signal (or background) jets that are accepted by the
tagger. For cases with limited statistics (which is often the case in
searches), the best tool is then the one that maximises the signal
significance, $\epsilon_S/\sqrt{\epsilon_B}$. More generally, for a
given signal efficiency, one would like to have the smallest possible
background rate, \ie for a given amount of signal kept by the tagger,
we want to minimise the rate of background events which wrongly pass
the tagger conditions.
This is usually represented by Receiver Operating Characteristic (ROC)
curves which show $\epsilon_B$ as a function of $\epsilon_S$, such as
represented in Fig.~\ref{fig:ROC-sketch}.
This can be used to directly compare the performance of different
substructure tools.

That said, signal significance is not the only criterion one may
desire from a jet substructure tagger.
Similarly to the properties of jet definitions discussed in
Chapter~\ref{chap:jets-and-algs}, we may want additional conditions
such as the following:
\vspace*{-0.2cm}
\begin{itemize}
\itemsep0.0cm
\item we would like to work with tools that are infrared and collinear
  safe, \ie which are finite at any order of the perturbation
  theory,\footnote{An interesting class of observables, known as {\it
      Sudakov safe}, fails to fully satisfy this condition but remain
    calculable once a proper all-order calculation is performed (see
    chapter~\ref{sec:curiosities}).}
\item we would like to work with tools that are as little sensitive as possible to model-dependent
  non-perturbative effects such as hadronisation and the Underlying
  Event,
\item we would like to work with tools that are as little sensitive as possible to
  detector effects and pileup.
\end{itemize}
In a way, the last two of the above criteria are related to the {\it
  robustness} of our tools, \ie we want to be able to assess how robust our conclusions are
against details of the more poorly-known (compared to
the perturbative part) aspects of high-energy collisions.
One should typically expect that a more robust tool would have a
smaller systematic uncertainty associated with theory modelling (\eg
the dependence on which Monte Carlo sample is used), pileup
sensitivity and detector sensitivity/unfolding.\footnote{Small
  systematic uncertainties is really the fundamental assessment of
  robustness. Asking, as we do here, for a small sensitivity to
  non-perturbative and detector effects, is a sufficient condition to
  achieve this, but it is not strictly necessary. One could for
  example imagime a situation where detector effects are large but
  perfectly well understood such that the resulting systematic
  uncertainty remains small.}

Robustness can be quantified in several ways, typically by measuring
how the signal and background efficiencies are affected by a given
effect (see
\eg~\cite{Dasgupta:2016ktv,Salam:2016yht,Bendavid:2018nar}).
Some concrete ideas about how to assess robustness were put forward in
Ref.~\cite{Bendavid:2018nar} (Section III.2). Let us say that we want to
test the sensitivity of a tagger with respect to the UE. From a
Monte Carlo simulation, we can compute the signal and background
efficiencies, first without UE, $\epsilon_{S,B}\equiv\epsilon_{S,B}^{(\text{no UE})}$, and then
with UE $\epsilon_{S,B}'\equiv\epsilon_{S,B}^{(\text{UE})}$. We define {\it resilience}, a measure
of robustness, as
\begin{equation}\label{eq:resilience}
  \zeta = \left(
      \frac{\Delta\epsilon_{S}^2}{\left\langle\epsilon\right\rangle_{S}^2}
    + \frac{\Delta\epsilon_{B}^2}{\left\langle\epsilon\right\rangle_{B}^2}
  \right)^{-1/2}
\end{equation}
where
\begin{equation}
  \Delta \epsilon_{S,B}  =\epsilon_{S,B}-\epsilon_{S,B}'
  \qquad\text{ and }\qquad
\left\langle\epsilon\right\rangle_{S,B} 
     =\frac{1}{2}\left(\epsilon_{S,B}+\epsilon_{S,B}'\right).
\end{equation}
With this definition, a large resilience means that the signal and
background efficiencies have not changed much when switching the UE on
and hence that the tool is robust.
Resilience can be defined for hadronisation, \ie when switching on
hadronisation and going from parton level to hadron level, for the UE,
as discussed above, for pileup sensitivity, \ie when overlaying the
event with pileup and applying a pileup mitigation technique, and for
detector sensitivity, \ie when running events through a detector
simulation.

To conclude, it is important to realise that the performance of a jet
substructure tagger is characterised by several aspects. Performance,
typically quantified by ROC curves of signal significance is certainly
the most regarded feature of a tagger. However, other requirements
like the robustness against non-perturbative effects, pileup and
detector effects are desirable as well. These can be quantified \eg
via resilience.

\section{Prong-finders and groomers}\label{sec:tools-prong-finders-groomers}

\paragraph{Mass-drop tagger.} The Mass-Drop
tagger was originally proposed~\cite{Butterworth:2008iy} as a tool to
isolate boosted Higgs bosons, decaying to $b\bar b$ pairs, from the
QCD background. 
In this procedure, one first reclusters the jet constituents of the fat
jet with the Cambridge/Aachen algorithm. One then iteratively undoes
the last step of the clustering $p_{i+j}\to p_i+p_j$ and check the
following criteria: (i) there is a ``mass drop'' \ie
${\rm max}(m_i,m_j)<\mu_{\rm cut}m_{i+j}$, (ii) the splitting is
sufficiently symmetric \ie
${\rm min}(p_{t,i}^2,p_{t,j}^2)\Delta R_{ij}^2>y_{\rm cut} m_{i+j}^2$.
When both criteria are met, we keep ``$i+j$'' as the result of the
mass-drop tagger, otherwise the least massive of $i$ and $j$ is
discarded and the procedure is repeated iteratively using the most
massive of $i$ and $j$.\footnote{If the procedure fails to find two
  subjets satisfying the conditions, \ie end up recursing until it
  reaches a single constituent which can not be further de-clustered,
  it is considered as having failed and returns an empty jet.}
The mass-drop tagger has two parameters: $\mu_{\rm cut}$, the
mass-drop parameter itself, and $y_{\rm cut}$, the symmetry cut.
The two conditions imposed by the mass-drop tagger exploit the
fundamental properties introduced above for tagging two-pronged
boosted objects: the symmetry cut requires that one indeed finds two
hard prongs and the mass-drop condition imposes that one goes from a
massive boson jet to two jets originated from massless QCD partons.
Although it was originally introduced as a tagger, the mass-drop
tagger also acts as a groomer since, following the declustering
procedure, it would iteratively remove soft radiation at the
outskirts of the jet, hence reducing the pileup/UE contamination.

\paragraph{modified Mass-Drop Tagger (mMDT).} When trying to understand the
analytic behaviour of the mass-drop tagger on QCD jets, it was
realised that following the most massive branch in the iterative
de-clustering procedure leads to pathological situations. It was
therefore suggested~\cite{Dasgupta:2013ihk} to adapt the procedure so
that it instead follows the hardest branch (in terms of $p_t$).
This modification makes the analytical calculation much easier and
more robust without affecting the performance of the method (even
improving it slightly).
The same study also added two more minor modifications. First, it was
realised that the symmetry condition could be replaced by
${\rm min}(p_{t,i},p_{t,j})>z_{\rm cut}(p_{t,i}+p_{t,j})$ which has
the same leading analytic behaviour as the $y_{\text{cut}}$ condition
and a slightly reduced sensitivity to non-perturbative
corrections. Second, the mass-drop condition would only enter as a
subleading correction in the strong coupling constant $\alpha_s$,
compared to the symmetry condition. It can therefore usually be
ignored.

\paragraph{\SD.} \SD~\cite{Larkoski:2014wba} can be seen a
generalisation of mMDT. It also proceeds by
iteratively declustering a jet reclustered with the Cambridge/Aachen
algorithm but replaces the symmetry condition for the
declustering of $p_{i+j}$ into $p_i$ and $p_j$, with
\begin{equation}\label{eq:soft-drop-condition}
  \frac{{\rm min}(p_{t,i},p_{t,j})}{p_{t,i}+p_{t,j}} 
    > \zcut \left(\frac{\Delta R_{ij}}{R}\right)^\beta,
\end{equation}
where $R$ is the jet radius.
\SD has two parameters. The $\zcut$ parameter plays the same role as
in the (m)MDT of keeping the hard structure and excluding soft
emissions, starting from large angles.
The $\beta$ parameter gives \SD some extra freedom in controlling how
aggressive the groomer is.
In the limit $\beta\to 0$, \SD reduces to the mMDT. Increasing $\beta$
leads to a less aggressive grooming procedure, with $\beta \to \infty$
corresponding to no grooming at all. Conversely, choosing a negative
value for $\beta$ would lead a more aggressive two-prong tagger than
mMDT.\footnote{The \SD procedure returns by default a single particle
  if it fails to find two subjets satisfying the \SD condition. This
  ``grooming mode'' is different from the default ``tagging mode'' of
  the mMDT which would fail, \ie return an empty jet, if no
  substructure are found.}
For practical applications, mMDT and \SD with negative $\beta$
(typically $\beta=-1$) would, alone, be perfectly adequate and
efficient taggers (see e.g. Section 7 of Ref.~\cite{Larkoski:2014wba})

\paragraph{Recursive \SD.} \SD typically finds two prongs in a jet. If we want to find more than two prongs, we can apply \SD
recursively. Recursive \SD~\cite{Dreyer:2018tjj} does this by
iteratively undoing the clustering with the largest $\Delta R$ in the
Cambridge/Aachen tree. Both branches are kept if the \SD
condition~(\ref{eq:soft-drop-condition}) is met and the softer branch
is dropped otherwise. The procedure stops when $N+1$ prongs have been
found, with $N$ an adjustable parameter that can be taken to infinity.

\paragraph{Filtering.} Filtering was first introduced in
Ref.~\cite{Butterworth:2008iy} as a grooming strategy to clean the jet
from UE after the mMDT has been applied. For a given jet, it
re-clusters its constituents with the Cambridge/Aachen algorithm with
a small radius $R_{\rm filt}$ and only keeps the $n_{\rm filt}$ larger
$p_t$ subjets. The subjets that have been kept constitute filtered jet.
This has two adjustable parameters: $R_{\rm filt}$ and $n_{\rm filt}$.
It is typically used to reduce soft contamination in situations where
we have a prior knowledge of the number of hard prongs in a
jet. For a jet with $n_{\rm prong}$ hard prongs --- $n_{\rm prong}=2$
for a $W/Z/H$ bosons and $n_{\rm prong}=3$ for a top --- we would
typically use $n_{\rm filt}=n_{\rm prong}+1$ which would also keep the
(perturbative) radiation of an extra gluon.

\paragraph{Trimming.} Trimming~\cite{Krohn:2009th} shares some
similarities with filtering. It also starts with re-clustering the jet
with a smaller radius, $R_{\rm trim}$, using either the $k_t$ or the
Cambridge/Aachen algorithm. It then keeps all the subjets with a
transverse momentum larger than a fraction $f_{\rm trim}$ of the
initial jet transverse momentum.
On top of the choice of algorithm, this also has two parameters:
$R_{\rm trim}$ and $f_{\rm trim}$.
It is often used both as a generic groomer and as a prong finder in boosted-jet
studies.

\paragraph{Pruning.} Pruning~\cite{Ellis:2009su} is similar in spirit
to trimming but it adopts a bottom-up approach (with trimming seen as
a top-down approach). Given a jet, pruning reclusters its constituents
using a user-specified jet definition (based on pairwise
recombinations) and imposes a constraint at each step of the
clustering: objects $i$ and $j$ are recombined if they satisfy at
least one of these two criteria: (i) the geometric distance
$\Delta R_{ij}$ is
smaller than
$R_{\rm prune}=2 f_{\rm prune} m_{\rm jet}/p_{t,\rm jet}$, with
$p_{t,\rm jet}$ and $m_{\rm jet}$ the original jet transverse momentum
and jet mass, (ii) the splitting between $i$ and $j$ is sufficiently
symmetric, \ie
${\rm min}(p_{t,i},p_{t,j})\ge z_{\rm prune}p_{t,(i+j)}$. If neither criteria are met, only the hardest of $i$ and $j$ (in terms of
their $p_t$) is kept for the rest of the clustering and the other is
rejected.
On top of the jet definition used for the re-clustering, which is
usually taken to be either $k_t$ or Cambridge/Aachen with a radius
much larger than the one of original jet, this has two parameters:
$f_{\rm prune}$ and $z_{\rm prune}$. 
$z_{\rm prune}$ plays the same role as $f_{\rm trim}$ for trimming and
$f_{\rm prune}$ plays a role similar to $R_{\rm trim}$. Note that, in
the case of pruning, $R_{\rm prune}$ is defined dynamically based on the
jet kinematics, while $R_{\rm trim}$ is kept fixed. This can have
important consequences both analytically and phenomenologically.
Pruning can be considered as a general-purpose groomer and tagger
and is often used in situations similar to trimming, although it tends
to be slightly more sensitive to pileup contamination.

\paragraph{I and Y-Pruning.} When pruning a jet, there might be
situations where a soft emission at large angle dominates the mass of
the jet, thus setting the pruning radius, but gets pruned away because it does not satisfy the pruning
conditions. The mass of the pruned jet is then determined by radiation
at smaller angle, typically within the pruning radius. This situation
where the jet mass and the pruning radius are determined by different
emissions in the jet would result in a jet with a single prong, and it usually referred to 
called ``I-pruning"~\cite{Dasgupta:2013ihk}. For I-pruning, the pruning radius does not have the
relation to the hard substructure of the jet it is intended to.

More precisely, I-Pruning is defined as the
subclass of pruned jets for which, during the sequential clustering,
there was never a recombination with $\Delta R_{ij}>R_{\rm prune}$ and
${\rm min}(p_{t,i},p_{t,j})> z_{\rm prune}p_{t,(i+j)}$.
The other situation, \ie a pruned jet for which there was at least one
recombination for which $\Delta R_{ij}>R_{\rm prune}$ and
${\rm min}(p_{t,i},p_{t,j})> z_{\rm prune}p_{t,(i+j)}$, corresponds to
a genuine two-prong structure and is called Y-Pruning.

This distinction between I- and Y-Pruning is mostly irrelevant for
boosted jet tagging. However, it has been shown to have an impact on
the analytical behaviour of Pruning, with Y-Pruning being under better
control and than I-Pruning, the latter adding an extra layer of
complexity to the calculation. If one's goal is to reach some level of
analytic control over groomed jets, Y-Pruning appears as a more
natural choice than Pruning which also includes the contribution from
I-Pruning.

\paragraph{Y-Splitter.} Y-Splitter is one of the very few tools
proposed for boosted W-boson tagging at the
LHC~\cite{Butterworth:2002tt}.
The idea is to recluster the constituents of the jet with the $k_t$
algorithm and to undo the last step of the clustering. This gives two
subjets $j_1$ and $j_2$. One then defines
\begin{equation}
y_{12} = \frac{k_{t,12}^2}{m_{12}^2} = \frac{{\rm
    min}(p_{t1}^2,p_{t2}^2)\Delta R_{12}^2}{m_{12}^2},
\end{equation}
similar to what has been used later in the MassDrop Tagger. One then
imposes the cut $y>y_\text{cut}$ to require to hard prongs in the
jet.\footnote{A cut on $y$ is roughly equivalent to a cut on the $p_t$
  fraction $z$. For example, for a jet made of two collimated partons
  carrying a momentum fraction $z$ and $1-z$ of the jet, one has
  $y=\tfrac{z}{1-z}$.}
Note that similar quantities have been introduced as event shapes in
$e^+e^-$ collisions.

\paragraph{Johns Hopkins top tagger.} As its name suggests, this is a
tagger meant to separate fat jets originating from the decay of
boosted top quarks from the background made of light-quark jets.
It was one of the first substructure techniques introduced in the
context of LHC physics.
The tagger aims at finding three hard prongs in the jet, corresponding
to the $q\bar q b$ hard quarks produced by the hadronic decay of the
top, adding constraints that two of the three prongs are compatible
with a hadronically-decaying W boson.
In practice, it proceeds as follows~\cite{Kaplan:2008ie}:
\begin{enumerate}
\item If the initial jet has not been obtained by the Cambridge/Aachen
  algorithm, re-cluster the jet constituents using this algorithm,
\item {\em Primary decomposition}: as for the mMDT, we iteratively
  undo the last step of the Cambridge/Aachen clustering. The softer of the two
  subjets is discarded if its transverse momentum divided by the {\em
    initial} jet $p_t$ is smaller than a parameter $\delta_p$. The
  de-clustering procedure then continues with the harder subjet.
  This is repeated until one of four things happens: (i) both subjets are
  above $\delta_p$, (ii) both subjets are below $\delta_p$, (iii) the
  two subjets satisfy $|\Delta y|+|\Delta \phi|<\delta_r$, with
  $\delta_r$ another parameter of the tagger, or (iv) the subjet can
  no longer be declustered. In case (i) the two hard subjets are kept
  and further examined, in the other three cases, the jet is not
  tagged as a top candidate.
\item {\em Secondary decomposition}: with the two prongs found by the
  primary decomposition, repeat the declustering procedure as for the
  primary decomposition, still defining the $\delta_p$ condition with respect to
  the original jet $p_t$.
  This can result in either both prongs from the primary decomposition
  being declustered into two sub-prongs, only one prong being
  declustered, or none.
  When no further substructure is found in a primary prong, the
  primary prong is kept intact in the final list of prongs. When two
  sub-prongs are found both are kept in the final list of prongs.
  Ultimately, this leads to two, three or four prongs emerging from
  the original jet. Only jets with three or four sub-prongs are then
  considered as top candidates, while the case with only two prongs is
  rejected.
\item {\em Kinematic cuts}: with the three or four prongs found from
  the secondary decomposition, impose additional kinematic
  conditions. First, the sum of the four-momenta of all the hard
  prongs should be close to the top mass. Then, there exists two
  prongs with invariant mass close to the W mass. Finally, we
  impose that the W helicity angle be consistent with a top
  decay.
  The W helicity angle, $\theta_h$, is defined as the angle between
  the top direction and one of the W decay products, in the rest
  frame of the W. We impose $\mathcal{O}s(\theta_h)<0.7$.\footnote{Top decays are almost isotropic and the helicity angle had
    an almost flat distribution, while for QCD jets, it diverges like
    $1/(1-\mathcal{O}s(\theta_h))$.}
\end{enumerate}
The original paper suggested that the parameters should be adjusted according to the
event's scalar $E_T$:
\begin{align}
  1~\text{TeV}<E_T<1.6~\text{TeV}:\;   & R=0.8, && \delta_p=0.10, && \delta_r=0.19, \\ 
  1.6~\text{TeV}<E_T<2.6~\text{TeV}:\; & R=0.6, && \delta_p=0.05, && \delta_r=0.19, \\
  2.6~\text{TeV}<E_T:\;                & R=0.4, && \delta_p=0.05, && \delta_r=0.19. 
\end{align}
The kinematic cuts are then adjusted based on the jet $p_t$:
\begin{align}
  p_t<1~\text{TeV: } & 145<m_\text{top}<205~\text{GeV}, && 65<m_{W}<95~\text{GeV}, \\ 
  p_t>1~\text{TeV: } & 145<m_\text{top}<p_t/20+155~\text{GeV}, && 65<m_{W}<70+p_t/40~\text{GeV},
\end{align}
where $m_\text{top}$ and $m_W$ are the reconstructed top and W mass respectively.

The prong decomposition of the Johns Hopkins top tagger shared
obvious similarities with the (modified) MassDrop Tagger introduced to
tag Higgs bosons, in the sense that it follows the hardest branch on a
Cambridge/Aachen clustering tree and imposes a hardness condition on
the subjets. Since we now want to require three hard prongs in the
jet, the de-clustering procedure is repeated twice. The main
noticeable differences between the (modified) MassDrop Tagger and the
Johns Hopkins top tagger is that the latter imposes a $\delta_p$
condition computed with respect to to the original jet $p_t$ while the mMDT
imposes its $\zcut$ condition computed as a fraction of the subjets
parent's $p_t$.
Note also the use of the Manhattan distance in the $\delta_r$
condition.

In practice, for a top efficiency between 20 and 40\%, the Johns
Hopkins top tagger achieves reductions of the background by a factor
$\sim 100$ (remember these numbers should be squared for the
efficiency to tag a $t\bar t$ pair).

\paragraph{CMS top tagger.}  The CMS top tagger is essentially an
adaptation of the Johns Hopkins top tagger proposed by the CMS
collaboration~\cite{CMS:2009lxa,CMS:2014fya}. Declustering proceeds analogously to the Johns Hopkins top tagger --- except for the
two-prongs distance condition which uses a $p_t$-dependent cut on the
standard $\Delta R_{ij}$ subjet distance ---, but the kinematic
conditions are different. The detailed procedure works as follows:
\begin{enumerate}
\item If needed, the initial jet is re-clustered using the Cambridge/Aachen
  algorithm.
\item {\em Primary decomposition}: the last step of the clustering is
  undone, giving two prongs. These two prongs are examined for the
  condition
  \begin{equation}\label{eq:cms-top-cdt}
    p_t^{\mathrm{prong}} > \delta_p \, p_t^{\text{jet}},
  \end{equation}
  where $p_t^{\text{jet}}$ refers to the hard jet transverse
  momentum. $\delta_p$ is a parameter which is usually taken as
  $0.05$.
  If both prongs pass the cut then the ``primary'' decomposition
  succeeds.
  If both prongs fail the cut then the jet is rejected \ie is not
  tagged as a top jet.
  If a single prong passes the cut the primary decomposition recurses
  into the passed prong, until the decomposition succeeds or the whole
  jet is rejected.
  Note that during the recurrence, $p_t^{\text{jet}}$ (used in
  (\ref{eq:cms-top-cdt})) is kept as the transverse momentum of the
  original jet.
\item {\em Secondary decomposition}: with the two prongs found by the
  primary decomposition, repeat the declustering procedure as for
  the primary decomposition, still defining the $\delta_p$
  condition~\eqref{eq:cms-top-cdt} with respect to the original jet $p_t$.
  This can result in either both prongs from the primary decomposition
  being declustered into two sub-prongs, only one prong being
  declustered, or none.
  When no further substructure is found in a primary prong, the
  primary prong is kept intact in the final list of prongs. When two
  sub-prongs are found both are kept in the final list of prongs.
  Ultimately, this leads to two, three or four prongs emerging from
  the original jet. Only jets with three or four sub-prongs are then
  considered as top candidates.
\item {\em Kinematic constraints}: taking the three highest $p_t$
  subjets (\ie prongs) obtained by the declustering, find the minimum
  pairwise mass and require this to be related to the W mass, $m_W$,
  by imposing the condition
  $\mathrm{min} \left(m_{12},m_{13},m_{23} \right) > m_{\mathrm{min}}
  $ with $m_{\mathrm{min}} \lesssim m_W$.
  For practical applications, $m_{\text{min}}$ is usually taken as
  $50$~GeV.
\item Note that in the second version of the
  tagger~\cite{CMS:2014fya}, the decomposition procedure also imposes
  an angular cut: when examining the decomposition of a subjet $S$
  into two prongs $i$ and $j$, the CMS tagger also requires
  $\Delta R_{ij} > 0.4 - A p_t^S$ where
  $\Delta R_{ij} = \sqrt{\Delta y_{ij}^2+ \Delta \phi_{ij}^2}$ and
  $p_t^S$ refers to the transverse momentum of the
  subjet.
  The default value for $A$ is $0.0004 \, \mathrm{GeV^{-1}}$.
  We note that without a $\Delta R$ condition in the decomposition of
  a cluster, the CMSTopTagger is collinear unsafe
  (see~\cite{Dasgupta:2018emf} for a discussion of this and proposed
  alternatives).
\end{enumerate}

\section{Radiation constraints}\label{sec:tools-radiation-constraints}

The standard approach to constraining radiation inside a jet is to
impose a cut on a {\em jet shape} which, similarly to event shapes in
electron-positron collisions, is sensitive to the distribution of the particles in the jet
(or in the event for the $e^+e^-$ case).
Over the past ten years, several jet shapes have been introduced. In what follows,  we
review the most common ones.

\subsection{Angularities and generalised angularities}\label{sec:def-angularities}

The simplest family of jet shapes is probably the {\em generalised
  angularities}~\cite{Larkoski:2014pca} defined as
\begin{equation}\label{eq:generalised-angularities}
  \lambda_\beta^\kappa = \sum_{i\in\jet} z_i^\kappa
    \left(\frac{\Delta R_{i,\jet}}{\\R}\right)^\beta,
\end{equation}
where $z_i$ is the jet transverse momentum fraction carried by the
constituent $i$ and $\Delta R_{i,\jet}$ its distance to the jet axis:
\begin{equation}
  z_i = \frac{p_{t,i}}{\sum_{j\in\jet}p_{t,j}}
  \qquad\text{ and }\quad
  \Delta R_{i,\jet}^2 = (y_i- y_{\jet})^2 + (\Delta \phi- \phi_{\jet})^2.
\end{equation}
Note that generalised angularities (and more in general, the other jet shapes
presented later) can also be used for jets in $e^+e^-$
collisions if we define $z_i=E_i/E_\jet$ and replace
$\Delta R_{i,\jet}$ either by $\theta_{i,\jet}$, the angle to the jet
axis, or by
$2\sin(\theta_{i,\jet}/2)=\sqrt{2(1-\mathcal{O}s \theta_{i,\jet})}$.

Generalised angularities are collinear unsafe, except for the special
case $\kappa=1$ which corresponds to the IRC safe 
{\em angularities}~\cite{Berger:2003iw,Almeida:2008yp}:
\begin{equation}\label{eq:angularities}
\lambda_\beta \equiv \lambda_\beta^{(\kappa=1)}.
\end{equation}
The specific case $\beta=1$ is sometimes referred to as {\em width} or
{\em girth} or {\em broadening}, while $\beta=2$ is closely related to
the jet mass.\footnote{It reduces to $\rho=m^2/(p_tR)^2$ in the limit
  of massless particles and small jet radius $R$.}

Obviously, the more radiation there is in a jet, the larger generalised
angularities are. 
Angularities and generalised angularities can therefore be seen as a
measure of QCD radiation around the jet axis, \ie as the radiation
in a one-pronged jet. They are often used as a quark-gluon
discriminator, where gluon-initiated jets would, on average, have
larger angularity values that quark-initiated
jets~\cite{Gallicchio:2011xq,Gallicchio:2012ez,Badger:2016bpw}.

For completeness, we note that the ``jet'' axis used to compute
angularities can differ from the axis obtained via the initial
jet clustering (usually the anti-$k_t$ algorithm with jet radius $R$
and $E$-scheme recombination). A typical example is the case of the jet
width where using an axis defined with the $E$-scheme recombination
introduces a sensitivity to recoil and complicates the analytic
calculations of width. The workaround is to use a recoil-free axis,
like the WTA recombination scheme. More generally, it is advisable to use the
WTA axis for angular exponents $\beta\le 1$. This is also valid for
the other shapes defined below and we will adopt this choice when presenting
analytic calculations.

There are at least two other examples of generalised angularities
that, despite being IRC unsafe, are widely used in applications. The
case $\beta=\kappa=0$ corresponds to the {\em jet multiplicity}, and
$\beta=0$, $\kappa=2$, which is related to
$p_t^D$~\cite{Pandolfi:1480598,Chatrchyan:2012sn}.
Finally, generalised angularities can be defined as track-based
observables by limiting the sum in
Eq.~(\ref{eq:generalised-angularities}) to the charged tracks (\ie
charged constituents) in the jet. Tracked-based angularities are
advantageous in the context of pileup mitigation, because compared to
neutral energy deposits in calorimeters, it is easier to separate
tracks that originate from pileup vertices from tracks from the
hard-interaction. The price we pay is that tracked-based observables
are not IRC safe and theoretical predictions involve non-perturbative
fragmentation
functions~\cite{Chang:2013rca,Chang:2013iba,Elder:2018mcr}.

\subsection{$N$-subjettiness}\label{sec:def-Nsubjettiness}
As the name suggests, $N$-subjettiness~\cite{Thaler:2010tr} is a jet shape that aims to discriminate jets according to the number $N$ of subjets they are made of. 
It takes inspiration from the event-shape $N$-jettiness~\cite{Stewart:2010tn}.
In order to achieve this, a set of axes $a_1,\dots,a_N$ is introduced (see below for a more precise
definition) and the following jet shape is introduced~\footnote{Eq.~\eqref{eq:Nsubjettiness} corresponds to the {\em
    un-normalised} definition of $N$-subjettiness. Alternatively, one
  can normalise $\tau_N$ by the jet scalar $p_t$,
  $\tilde p_t=\sum_{i\in\jet}p_{ti}$, or, more simply, the jet $p_t$.}
\begin{equation}\label{eq:Nsubjettiness}
  \tau_N^{(\beta)} = \sum_{i\in\jet} p_{ti}\,
  {\text{min}}(\Delta R_{ia_1}^\beta,\dots,\Delta R_{ia_N}^\beta),
\end{equation}
where $\beta$ is a free parameter.\footnote{Although it is strongly
advised to specify the value of $\beta$ one uses, $\beta=1$ is
often implicitly assumed in the literature.}
The axes $a_i$  can be defined in several ways, the most common
choices being the following:
\begin{itemize}
\item {\bf $k_t$ axes}: the jet is re-clustered with the $k_t$
  algorithm and the $a_i$ are taken as the  $N$ exclusive jets.
\item {\bf WTA $k_t$ axes}: the jet is re-clustered with the $k_t$
  algorithm, using the winner-take-all recombination scheme. The $a_i$
  are taken as the $N$ exclusive jets. As for angularities, the
  use of the WTA axes guarantees a recoil-free observable.
\item {\bf generalised-$k_t$ axes}: this is defined as above but now
  one uses the exclusive jets obtained with the generalised $k_t$
  algorithm.
  It is helpful to set the $p$ parameter of the generalised $k_t$
  algorithm to $1/\beta$, so as to match the distance measure used for
  the clustering with the one used to compute $\tau_N$.
  For $\beta<1$ one would again use the WTA generalised-$k_t$ axes.
\item {\bf minimal axes}: chose the axes $a_i$ which minimise the
  value of $\tau_N$. The minimum is found by iterating the
  minimisation procedure described in Ref.~\cite{Thaler:2011gf}
  starting with a set of seeds. It is often possible to find a less
  computer-expensive definition (amongst the other choices listed
  here) which would be as suitable to the minimal axes, both for
  phenomenological applications and for analytic calculations.
\item {\bf one-pass minimisation axes}: instead of running a full
  minimisation procedure as for the minimal axes, one can instead
  start from any other choice of axes listed above and run the
  minimisation procedure described in Ref.~\cite{Thaler:2011gf}.
\end{itemize}

As for the angularities discussed in the previous section, $\tau_N$ is
a measure of the radiation around the $N$ axes $a_1,\dots,a_N$. For a
jet with $N$ prongs, one expects $\tau_1,\dots,\tau_{N-1}$ to be
large and $\tau_{\ge N}$ to be small. The value of $\tau_N$ will also
be larger when the prongs are gluons. For these reasons, the
$N$-subjettiness ratio
\begin{equation}\label{eq:tau_ratios}
  \tau_{N,N-1}^{(\beta)} = \frac{\tau_N^{(\beta)}}{\tau_{N-1}^{(\beta)}}
\end{equation}
is a good discriminating variable for $N$-prong signal jets against
the QCD background.
More precisely, one would impose a cut
$\tau_{21}^{(\beta)}<\tau_{\text{cut}}$ to discriminate W/Z/H jets
against QCD jets and $\tau_{32}^{(\beta)}<\tau_{\text{cut}}$ to
discriminate top jets against QCD jets.
Although the most common use of $N$-subjettiness in the literature
takes $\beta=1$, there are also some motivations to use $\beta=2$, see
e.g.~\cite{Larkoski:2013eya,Salam:2016yht}.

\subsection{Energy-Correlation Functions}\label{sec:def-ecfs}

Energy-correlation functions (ECFs) achieve essentially the same objective
than $N$-subjettiness without requiring the selection of $N$
reference axes.
In their original formulation~\cite{Larkoski:2013eya}, they are defined as
\begin{align}
e_2^{(\beta)} &= \sum_{i<j\in\jet} z_i z_j \,\Delta R_{ij}^\beta,\label{eq:ecf-e2}\\
e_3^{(\beta)} &= \sum_{i<j<k\in\jet} z_i z_j z_k \,\Delta
                R_{ij}^\beta \Delta R_{jk}^\beta \Delta R_{ik}^\beta,\label{eq:ecf-e3}\\
 &\vdots\nonumber\\
e_N^{(\beta)} &= \sum_{i_1<...<i_N\in\jet} \bigg(\prod_{j=1}^Nz_{i_j}\bigg)
                \bigg(\prod_{k<\ell=1}^N\Delta R_{i_ki_\ell}^\beta\bigg),\label{eq:ecf-eN}
\end{align}
with $z_i=p_{t,i}/\sum_j p_{t,j}$.
Compared to $N$-subjettiness, energy-correlation functions have the
advantage of not requiring a potentially delicate choice of reference
axes. Furthermore, from an analytic viewpoint, they are insensitive to
recoil for all values of the angular exponent $\beta$, allowing for an easier analytic
treatment (although, as we have mentioned earlier, this issue can be alleviated in the
$N$-subjettiness case by using WTA axes).

Generalised versions of the
angularities have been introduced~\cite{Moult:2016cvt}. They still involve $p_t$ weighted
sums over pairs, triplets,... of particles but are built from other
angular combinations:
\begin{align}
  {}_1e_2^{(\beta)} &\equiv e_2,\\
  {}_3e_3^{(\beta)} &\equiv e_3,\\
  {}_2e_3^{(\beta)} &= \sum_{i<j<k\in\jet} z_i z_j z_k
            \:\text{min}\big(
            \Delta R_{ij}^\beta \Delta R_{ik}^\beta
            \Delta R_{ij}^\beta \Delta R_{jk}^\beta
            \Delta R_{ik}^\beta \Delta R_{jk}^\beta\big), \\
  {}_1e_3^{(\beta)} &= \sum_{i<j<k\in\jet} z_i z_j z_k
            \:\text{min}\big(
            \Delta R_{ij}^\beta,\Delta R_{ik}^\beta,\Delta R_{jk}^\beta\big), \\
 &\vdots\nonumber\\
  {}_ke_N^{(\beta)} &= \sum_{i_1<...<i_N\in\jet} \bigg(\prod_{j=1}^Nz_{i_j}\bigg)
                \bigg(\prod_{\ell=1}^k
            \underset{u<v\in\{i_1,...,i_N\}}{\overset{\ell}{\text{min}}} \Delta R_{uv}^\beta\bigg),
\end{align}
where $\overset{\ell}{\text{min}}$ denotes the $\ell$-th smallest number.

Similarly to $N$-subjettiness, in order to discriminate boosted massive particles from background QCD jets,
we again introduce ratios of (generalised-)ECFs. Over the past few years,
several combinations have been proposed. Examples of ratios of ECFs that are used as two-prong taggers include
\begin{align}
  C_2^{(\beta)} &= \frac{{}_3e_3^{(\beta)}}{\big({}_1e_2^{(\beta)}\big)^2}
                  \equiv
                  \frac{e_3^{(\beta)}}{\big(e_2^{(\beta)}\big)^2},
                  &
  D_2^{(\beta)} &= \frac{e_3^{(\beta)}}{\big(e_2^{(\beta)}\big)^3},\\
  N_2^{(\beta)} &=  \frac{{}_2e_3^{(\beta)}}{\big(e_2^{(\beta)}\big)^2},
                  &
  M_2^{(\beta)} &= \frac{{}_1e_3^{(\beta)}}{e_2^{(\beta)}},\nonumber
\end{align}
while for three-prong tagging, one introduces~\cite{Larkoski:2013eya,Larkoski:2014zma,Moult:2016cvt}
\begin{align}
  C_3^{(\beta)} &= \frac{e_4^{(\beta)}e_2^{(\beta)}}{\big(e_3^{(\beta)}\big)^2},\qquad\qquad\qquad
  N_3 = \frac{{}_2e_4^{(\beta)}}{\big({}_1e_3^{(\beta)}\big)^2},\qquad\qquad\qquad
  M_3 = \frac{{}_1e_4^{(\beta)}}{{}_1e_3^{(\beta)}},\\
  D_3^{(\alpha,\beta,\gamma)}
    &=\frac{e_4^{(\gamma)}\big(e_2^{(\alpha)}\big)^{\frac{3\gamma}{\alpha}}}
           {\big(e_3^{(\beta)}\big)^{\frac{3\gamma}{\beta}}}
     +\kappa_1\Big(\frac{p_t^2}{m^2}\Big)^{\frac{\alpha\gamma}{\beta}-\frac{\alpha}{2}}
      \frac{e_4^{(\gamma)}\big(e_2^{(\alpha)}\big)^{\frac{2\gamma}{\beta}-1}}
           {\big(e_3^{(\beta)}\big)^{\frac{2\gamma}{\beta}}}
     +\kappa_2\Big(\frac{p_t^2}{m^2}\Big)^{\frac{5\gamma}{2}-2\beta}
      \frac{e_4^{(\gamma)}\big(e_2^{(\alpha)}\big)^{\frac{2\beta}{\alpha}-\frac{\gamma}{\alpha}}}
           {\big(e_3^{(\beta)}\big)^2},\nonumber
\end{align}
where $\kappa_1$ and $\kappa_2$ are ${\cal{O}}(1)$ constants.

In this series, the $D$ family has typically a larger
discriminating power, at the expense of being more sensitive to
model-dependent soft contamination in the jet like the UE
or pileup.
Instead, the $N$ family is closer to $N$-subjettiness, and the $M$
family is less discriminating but more resilient against soft
contamination in the jet.

Finally, we note that Energy Correlation functions have recently been
extended into Energy Flow polynomials~\cite{Komiske:2017aww} which
provide a linear basis for all infrared-and-collinear-safe jet
substructure observables. These can then be used to design Energy Flow
Networks~\cite{Komiske:2018cqr} which are QCD-motivated
machine-learning substructure tools.

\subsection{Additional shapes}\label{sec:def-other-shapes}

Over the past decade, several other jet shapes have been introduced in
the literature and studied by the LHC experiments. Since they tend to
be less used than the ones introduced above, we just briefly list the
most common ones below, without entering into a more detailed
discussion.

\paragraph{Iterated \SD.} This is related to Recursive \SD
introduced earlier. The idea is still to apply \SD multiple
times except that this time we will only follow the hardest branch in
the recursion procedure~\cite{Frye:2017yrw}. This gives a list of
branchings which pass the \SD condition,
$(z_1,\theta_1), \dots, (z_n,\theta_n)$, from which we can build
observables. The most interesting observable is probably the {\em
  Iterated \SD multiplicity}, which is simply the number of branchings
which have passed the \SD condition and which is an efficient
quark-gluon discriminator as we will show in
chapter~\ref{sec:calc-shapes-qg}.
Alternatively, we can build Iterated \SD angularities from the set of
$(z_i,\theta_i)$.
We note that for the Iterated SoftDrop multiplicity to be infrared
and collinear safe, we need either to take a negative value of the \SD
parameter $\beta$ or impose an explicit cut (in $\theta$ or in $k_t$).

\paragraph{Planar flow.} Planar flow~\cite{Almeida:2008yp} (see also~\cite{Thaler:2008ju}) is defined as
\begin{equation}
  Pf = \frac{4\,\text{det}(I_\omega)}{\text{tr}^2(I_\omega)}
     = \frac{4\lambda_1\lambda_2}{(\lambda_1 + \lambda_2)^2}
  \qquad\text{ with }\quad
  I_\omega^{kl}  =  \sum_{i\in \text{jet}}
                  \omega_i \frac{p_{i,k}}{\omega_i} \frac{p_{i,l}}{\omega_i},
\end{equation}
where $m$ is the jet mass, $\omega_i$ is the energy of constituent $i$,
$p_{i,k}$ the $k^{\text{th}}$ component of its transverse momentum
with respect to the jet axis, and $\lambda_1$ and $\lambda_2$ are the eigenvalues
of $I_\omega$.

Planar flow is meant to tag object with 3-or-more-body decays. These
would appear as a planar configuration with large values of $Pf$,
while QCD jets tend to have a linear configuration and a small value
of $Pf$. This is similar to the $D$-parameter in $e^+e^-$ collisions.
A boost-invariant version of planar flow can be defined as
\begin{equation}
  Pf_{\text{BI}} = \frac{4\,\text{det}(I_{\text{BI}})}{\text{tr}^2(I_{\text{BI}})}
  \qquad\text{ with }\quad
  I_{\text{BI}} ^{\alpha\beta} = \sum_{i\in \text{jet}}
  p_{t,i} (\alpha_i-\alpha_{\text{jet}}) (\beta_i-\beta_{\text{jet}}),
\end{equation}
where, now, $\alpha$ and $\beta$ correspond either to the rapidity $y$
or azimuth $\phi$.
We note that $Pf$ and $Pf_{\text{BI}}$ are quite sensitive to
the UE and pileup activity in a jet (see
e.g.~\cite{Soyez:2012hv}) making them difficult to use in experimental
analyses. Since we will not come back to planar flow in our analytic
calculations in the following chapters, let us mention that some
fixed-order analytic results are available in the literature~\cite{Field:2012rw}.

\paragraph{Q-jet volatility.}
The main idea behind Q-jet~\cite{Ellis:2012sn,Ellis:2014eya} is to define jets as a set of multiple
clustering trees (weighted by an appropriate metric) instead of a
single one. A tree would be constructed using a modified
pairwise-recombination algorithm working as follows:
\begin{enumerate}
\item for a set of particles at a given stage of the clustering, we
  first compute the $k_t$ or Cambridge/Aachen set of distances
  $d_{ij}$. Let $d_{\text{min}}$ be their minimum.
\item We then compute a set of weights $w_{ij}$ for each pair and
  assign the probability $\Omega_{ij}=w_{ij}/\sum_{(ij)}w_{ij}$ to
  each pair. The weights are typically taken as
  \begin{equation}\label{eq:Qjet-rigidity}
    w_{ij} = \exp\bigg(-\alpha\frac{d_{ij}-d_{\text{min}}}{d_{\text{min}}}\bigg)
  \end{equation}
  where $\alpha$ is a parameter called rigidity.
\item we generate a random number used to select a pair $(ij)$ with
  probability $\Omega_{ij}$.
\item The pair is recombined and the procedure is iterated until no
  particles are left.
\end{enumerate}
The algorithm is then repeated $N_{\text{tree}}$ times. In the limit
$\alpha\to\infty$ one recovers the standard clustering. In practice
one usually takes $\alpha \simeq 0.01$ and $N_{\text{tree}}\gtrsim 50$
(typically 256).

Q-jets can then be used to compute jet physics observables, including
substructure variables, by taking the statistical average over the many
trees. New observables, related to the fact that we now have a
distribution of trees, can also be considered.
A powerful example is Q-jet volatility. It is defined by applying
pruning together with Q-jet, i.e.\ imposing the pruning condition (see
\ref{sec:tools-prong-finders-groomers} above) on each of the
clusterings trees, and then measuring the width of the resulting mass
distribution:
\begin{equation}\label{eq:Qjet-volatility}
  {\cal {V}} = \frac{\sqrt{\avg{m^2}-\avg{m}^2}}{\avg{m}}.
\end{equation}
When disentangling boosted W jets from background QCD jets, one
would expect ${\cal{V}}$ to be smaller in W jets than in QCD jets,
mostly because the former have a better-defined mass scale than the
latter.

\section{Combinations of tools}\label{sec:tools-combinations}

A few methods commonly used in recent substructure works can be seen as
combinations of ingredients borrowed from  the two categories above.
We list the most important ones in the next paragraphs.

Before doing so, we want to stress that substructure observables do not commute and therefore, when considering combinations tools, the order
in which we apply the different algorithms does matter. 
For instance, when imposing both a condition on the ``groomed'' jet mass and on a
jet shape, one would obtain different results if the jet shape is
computed on the plain jet or on the groomed jet.
A clear example of this is the combination of Y-splitter with
trimming or the mMDT, where imposing the Y-splitter cut on the
plain jet greatly improves performance.
It is therefore important that the description of the tagging strategy
clearly specify all the details of the combination including for
example what jet, groomed or ungroomed, is used to compute jet shapes.

That said, while several specific combinations are worth mentioning, we
limit ourselves to two-prong taggers:

\paragraph{ATLAS two-prong tagger.} The standard algorithm adopted by
ATLAS for Run-II of the LHC proceeds as follows.  Trimming is applied
to the jet, using the $k_t$ algorithm with a trimming radius
$R_{\text{trim}}=0.2$ and an energy cut $f_{\text{trim}}=0.05$. One
then requires the trimmed mass to be between 65 and 105~GeV. One then
computes $D_2^{(\beta=1)}$ on the trimmed jet and impose a cut on this
variable.
\paragraph{CMS two-prong taggers.}  At LHC Run-II, CMS has used
  two different two-prong taggers. Both start by applying the mMDT to
  the anti-$k_t$ ($R=0.8$) jets with $\zcut=0.1$ and require the mMDT
  mass to be between 65 and 105~GeV.
  At the beginning of Run-II, CMS was then computing the
  $N$-subjettiness $\tau_{21}^{(\beta=1,\text{plain})}$ ratio, using
  exclusive $k_t$ axes to define the axes, on the plain
  jet, and imposing a $\tau_{21}^{(\beta=1,\text{plain})}$.
  More recently, they replaced the $N$-subjettiness cut by a cut on
  $N_2^{(\beta=1,\text{mMDT})}$ i.e. \ they impose instead a cut on an
  $N_2$ ratio computed of the groomed jet (see
  \eg~\cite{Sirunyan:2018ikr} for a recent analysis).
  In both cases, they used a decorrelated version of the shape (see
  below).
  
  \paragraph{Decorrelated taggers (DDT).} Let us consider the
  combination of the mMDT with a cut on $N$-subjettiness.
Because of the correlation between these two observables, a cut on the shape can significantly sculpt the jet mass distribution of the background, leading to a deterioration in performance. 
  The idea behind the DDT procedure~\cite{Dolen:2016kst} is to instead substitute the cut on $N$-subjettiness, with a cut on a suitable
  combination of $\tau_{21}$ and of a function of the
  $\rho_{\text{mMDT}}=m^2_{\text{mMDT}}/(p_{t,\text{mMDT}})^2$.
This function is chosen such that
  the final background mass spectrum, after imposing a fixed cut on the
  decorrelated shape, is flat. The flatness of the background makes it
  easier for searches where the mass of the signal is unknown (or when
  the $p_t$ of the jet can widely vary).
  In Ref.~\cite{Dolen:2016kst}, it was shown that
  $\tau_{21}-\text{cst.}\times\log(\rho_{\text{mMDT}})$, with the
  constant determined from the $\rho_{\text{mMDT}}$ dependence of the
  average $\tau_{21}$ value was giving good results.
  This can easily be extended to other combinations. For example, CMS
  has recently used a decorrelated $N_2$ variable defined as
  $N_2^{\text{DDT}}=N_2-N_2(\text{cut at }5\%)$ where $N_2(\text{cut at
  }5\%)$ corresponds to the value of a cut on $N_2$ that would give a
  5\% background rate.
  We also refer to~\cite{Moult:2017okx,Napoletano:2018ohv} for
  examples where decorrelated shapes are built analytically.
  
\paragraph{Dichroic ratios.} There is a conceptual difference between
  imposing the shape cut on the plain jet or on the groomed jet.
  Since shapes measure the soft radiation at large angles, one should
  expect a better performance when the cut is imposed on the plain
  jet, since any grooming algorithm would have, by definition,
  eliminated some of the soft-and-large-angle radiation.
  Conversely, this very same soft-and-large-angle part of the
  phase-space is the one which is most sensitive to the UE and pileup,
  so computing the shape on the groomed jet would be more resilient to
  these effects.
  Recently, it was proposed to adopt a hybrid, \emph{dichroic},
  approach.
  The starting point is the observation that the shapes are meant to
  constrain additional radiation, on top of the two hard prongs. For
  ratios the sensitivity to the extra radiation is usually captured
  by the numerator, \eg $\tau_2$, while the denominator (\eg $\tau_1$)
  is mostly sensitive to the two hard prongs.
  
  That said, the first step of a full two-prong tagger is usually to
  apply a groomer/prong-finder, say the mMDT, in order to resolve the
  two-prong structure of the jet and impose a cut on the mass.  One
  then imposes a radiation constrain.
  For the latter it is therefore natural to compute the denominator of
  the shape, here $\tau_1$, (sensitive to the two hard prongs) on the
  result of the groomer/prong-finder jet.
  In order to retain information about the soft-and-large-angle
  radiation in the jet (where one expects discriminating power), one
  then wishes to compute the numerator of the shape, here $\tau_2$, on
  a larger jet. The latter can be either the plain jet or, if we want
  a compromise between performance and soft resilience, a
  lightly-groomed jet like a \SD jet with a positive $\beta$
  (typically $\beta=2$) and a smallish $\zcut$.
  This defines the dichroic $N$-subjettiness ratio~\cite{Salam:2016yht}
  \begin{equation}\label{eq:def-dichroic}
    \tau_{21}^{(\beta=2,\text{dichroic})}
      = \frac{\tau_2^{(\beta=2,\text{loose grooming})}}{\tau_1^{(\beta=2,\text{tight grooming})}},
  \end{equation}
  which has been shown to give good results on Monte-Carlo simulation
  and analytic calculations.
  Although it was initially introduced for $\beta=2$ $N$-subjettiness,
  it can be applied to other shapes as well.

\paragraph{Additional remarks.} 
Besides the specific prescriptions discussed above, it is helpful to
keep a few generic ideas in mind when combining different substructure
tools:
\begin{itemize}
\item When the $M$, $N$ and $U$ series of generalised angularities
  have been introduced, their combination with a grooming procedure
  was also discussed. We therefore encourage the reader interested in
  additional details to refer to Ref.~\cite{Moult:2016cvt}.
\item In a similar spirit, combining a Y-splitter cut, computed on the
  plain jet, with a grooming technique, such as trimming or the mMDT,
  for the measurement of the jet mass has been shown
  \cite{Dasgupta:2015yua,Dasgupta:2016ktv} to provide nice
  improvements both over Y-splitter alone --- owing to a reduced
  sensitivity to soft non-perturbative effects --- and over grooming
  alone --- owing to a larger suppression of the QCD background.
\item When one uses tagging techniques based on radiation constraints,
  one may want to first run a \SD grooming procedure with positive
  $\beta$, i.e. as a groomer, so as to limit the sensitivity to pileup
  and the Underlying Event, while keeping some of the
  soft-and-large-angle radiation for the radiation
  constraint.\footnote{Overall, it appears natural to use in parallel
    negative, or zero, $\beta$ as a tool to identify the two-prong
    structure and positive $\beta$ with a jet shape, to impose a cut
    on radiation.}
\end{itemize}

Finally, we note that a systematic and extensive investigation of the
tagging performance and resilience to non-perturbative effects
obtained when combining one of many prong finders with one of many
radiation constraints has been investigated in the context of the Les
Houches {\em Physics at TeV colliders} workshop in 2017. We will
briefly cover that study in chapter~\ref{chap:calc-two-prongs}, but we
refer to Section~III.2 of~\cite{Bendavid:2018nar} for more details
(cf.\ also our discussion on performance assessment in
Section~\ref{sec:performance-intro}.

\section{Other important tools}

As all classifications, separation of substructure tools in prong
finders and radiations constraints has its limits and some methods do
not obviously fall in either category.
In this section we list the most important ones.

\subsection{Shower deconstruction}\label{shower_dec}

Given a set of four-momenta $p_N$ of the $N$ measured final state
objects, one can associate probabilities $P(p_N|S)$ and
$P(p_N|B)$ that it was initiated by a signal ($S$) or background
($B$) process respectively.
From these probabilities one can build an ideal classifier\footnote{The Neyman-Pearson Lemma proves formally that $\chi$, as defined in Eq.~(\ref{eq:chi}), is an ideal classifier.} 
\begin{equation}
\chi(p_N) =\frac{P(p_N|S)}{P(p_N|B)}.
\label{eq:chi}
\end{equation} 
This fundamental observation is also the foundation of the so-called
matrix-element method~\cite{Kondo:1988yd, Abazov:2004cs}, used in
various applications in particle phenomenology with fixed-order matrix
elements~\cite{Artoisenet:2013vfa, Cranmer:2006zs, Andersen:2012kn}.

Shower deconstruction also relies on Eq.~(\ref{eq:chi}) to separate
 signal jets from background ones.
As discussed in Chapter~\ref{chap:calculations-jets}, the probabilities
$P(p_N|S)$ and $P(p_N|B)$ cannot reliably be computed at fixed order due
to the disparate scales in the process.
Instead one makes use of all-order calculations in QCD to compute
$\chi(p_N)$.

In practice, shower deconstruction considers all possible splittings
of the set $\{p_N\}=\{p_I\}\cup \{p_F\}$ into initial an final-state
radiation. For each such splitting it then considers all possible
shower histories, taking into account all possible parton-flavor
assignments, that could lead to the final state $\{p_N\}$.
A weight can then be calculated in perturbative QCD (see below) for
each history and the probabilities $P(p_N|S,B)$ are taken as the sum of
all the weights associated with $\{p_N\}$ under the signal or
background hypothesis.
To compute the weight for a given history, one uses a
Feynman-diagrammatic approach~\cite{Soper:2011cr,Soper:2012pb} where
each vertex receives a factor of the form $H e^{-R}$ with $H$ a
partonic splitting probability at a given virtuality and $e^{-R}$ is a
Sudakov factor, built from the splitting probability $H$ which
accounts for the fact that the splitting did not happen at a larger
virtuality. The specific form of $H$ depends on the splitting at hand,
using \eg Eq~(\ref{eq:eikonalfactorisation}) and
Eqs.~(\ref{eq:quarksplitting})-(\ref{eq:gluonsplitting}) for QCD
branchings, however retaining full mass dependency for the partons involved, thereby reaching a modified leading-logarithmic accuracy and the
full LO matrix element for the decay of W/Z/H bosons or top
quarks.

At the moment, probabilities are available for massive or massless
quark, gluons, hadronically-decaying electroweak W/Z/H bosons and
hadronically-decaying top quarks.
This makes shower deconstruction readily available for quark-gluon
discrimination, W/Z/H boosted bosons tagging and top tagging.

Note also that including all the constituents of the jet can quickly
become prohibitive due to the large number of possible histories.
A workaround is to first recluster the jet into small subjets and use
those subjets as an input to shower deconstruction.

\begin{figure}
  \centerline{\includegraphics[width=0.79\textwidth]{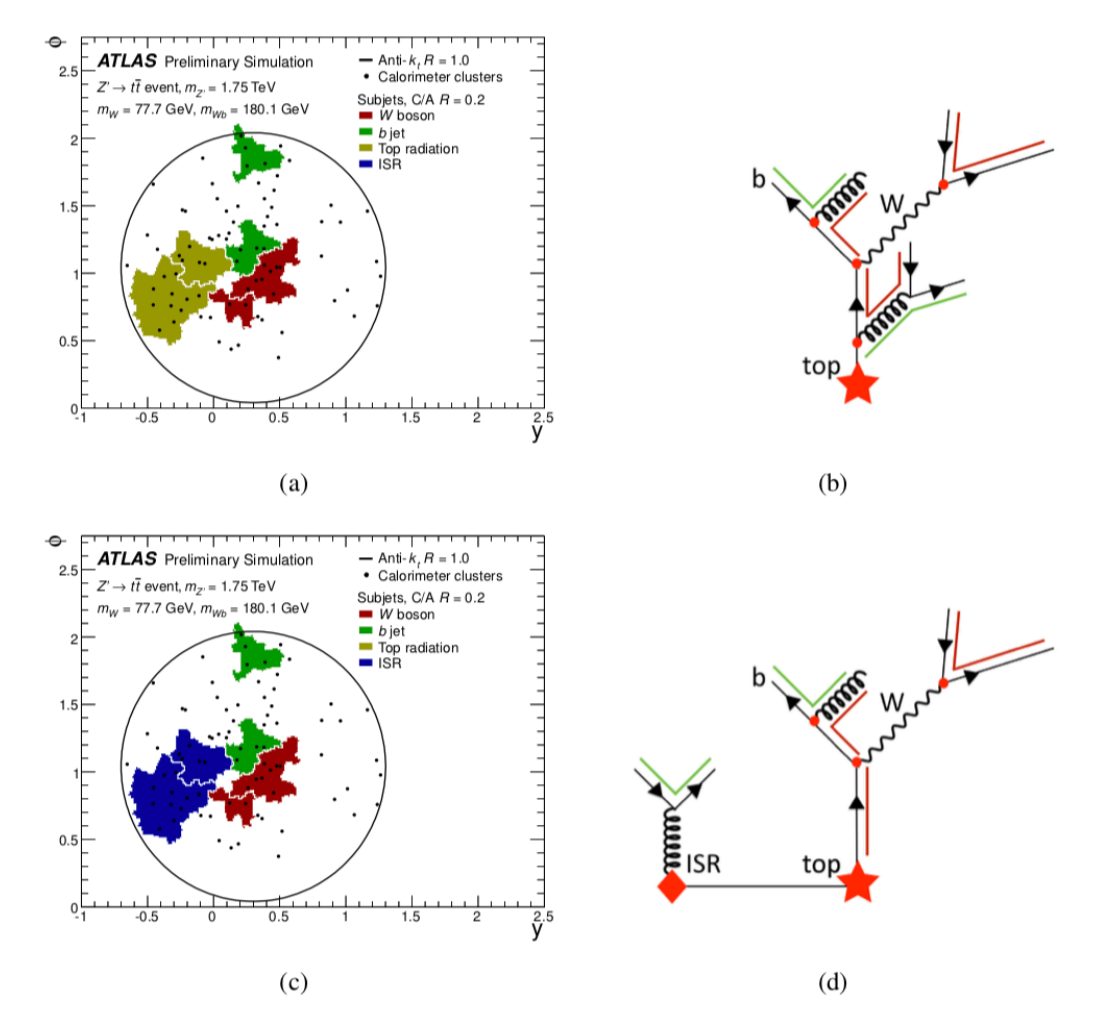}}
  \caption{The figure
    illustrates how shower deconstruction works as a top tagger. The
    left-hand panel shows the energy depositions in the
    rapidity-azimuth plane, while the left-hand panel shows the
    corresponding most-likely shower histories. The coloured lines in
    the right panels indicate which partons are colour-connected in
    the respective shower histories. Figure taken from Ref.~\cite{ATLAS:2014twa}, licensed under CC BY 4.0.}\label{fig:sd_histories}
\end{figure}

To illustrate the process, Fig.~\ref{fig:sd_histories} (taken from
Ref.~\cite{ATLAS:2014twa}) shows the two histories out of more than
1500 with the largest probabilities for a particular simulated
$\text{Z}' \to t \bar{t}$ event, where the leading large-radius jet
(anti-$k_t$, with $R = 1$) in this event was reclustered into six
subjets (using the Cambridge/Aachen algorithm $R = 0.2$). The left
plots show energy deposits, while the right panels show the actual
histories.

\subsection{HEP top tagger}

The HEP top tagger was first designed to reconstruct mildly boosted
top quarks in a busy event environment, i.e.\ for the reconstruction
of top quarks in the process $pp \to \bar{t} t h$ with semi-leptonic
top quark decays and $\text{H} \to \bar{b}b$~\cite{Plehn:2009rk}. The
hadronically top was expected to be boosted in the $p_t$ range around
250-500~GeV. This first incarnation of the tagger was augmented by
cuts on observables that were manifestly Lorentz-invariant, and thus
boosting between reference frames were no longer necessary. It
proceeds as follows (see Appendix~A of~\cite{Plehn:2010st}): 
\begin{enumerate}
\item one first defines the fat jets with the Cambridge/Aachen
  algorithm with $R=1.5$,
\item for a given fat jet jet, one recursively undoes the last step of
  the clustering, i.e. decluster the jet $j$ into subjets $j_1$ and
  $j_2$ with $m_{j_1}>m_{j_2}$, until we observe a mass-drop
  $m_{j_1}<0.8 m_{j}$. When the mass-drop condition is not met, one
  carries on with the declustering procedure with $j_1$.
\item For subjets which have passed the mass-drop condition and which
  satisfy $m_j>30$~GeV, one further decomposes the subjet recursively
  into smaller subjets.
\item The next step is to apply a filter similarly to what is done by
  the Mass-Drop Tagger. One considers all pairs of hard subjets,
  defining a filtering radius
  $R_{\text{filt}}=\text{min}(0.3,\Delta R_{ij})$. We then add a third
  hard subjet --- considering again all possible combinations --- and apply
  the filter on the three hard subjets keeping (at most) the 5 hardest
  pieces and use that to compute the jet mass.
  Amongst all possible triplets of the original hard subjets, we keep
  the combination for which the jet mass --- calculated after
  filtering --- gives the mass closest to the top mass and is within a mass window around the true top mass, e.g. in the range $150-200$ GeV.
\item Out of the 5 filtered pieces, one extracts a subset of 3
  pieces, $j_1$, $j_2$, $j_3$, ordered in $p_t$ and accept it as a
  top candidate if the masses satisfy at least one of the following
  3 criteria:
\begin{align}\label{eq:heptoptagger-cut}
&0.2<\text{arctan}\Big(\frac{m_{13}}{m_{12}}\Big)<1.3
\qquad\text{and}\qquad
R_{\text{min}}<\frac{m_{23}}{m_{123}}<R_{\text{max}}\\
&R_{\text{min}}^2\bigg(1+\frac{m_{13}^2}{m_{123}^2}\bigg)
  < 1-\frac{m_{23}^2}{m_{123}^2}
  < R_{\text{max}}^2\bigg(1+\frac{m_{13}^2}{m_{123}^2}\bigg)
\qquad\text{and}\qquad
\frac{m_{23}}{m_{123}}>0.35\nonumber\\
&R_{\text{min}}^2\bigg(1+\frac{m_{12}^2}{m_{123}^2}\bigg)
  < 1-\frac{m_{23}^2}{m_{123}^2}
  < R_{\text{max}}^2\bigg(1+\frac{m_{12}^2}{m_{123}^2}\bigg)
\qquad\text{and}\qquad
\frac{m_{23}}{m_{123}}>0.35,\nonumber
\end{align}
with $R_{\text{min}}=0.85\, m_W/m_t$ and $R_{\text{max}}=1.15\,
m_W/m_t$.
\item the combined $p_t$ of the 3 subjets constructed in the previous
  step is imposed to be at least 200~GeV.
\end{enumerate}

\begin{figure}
  \centerline{\includegraphics[width=0.65\textwidth]{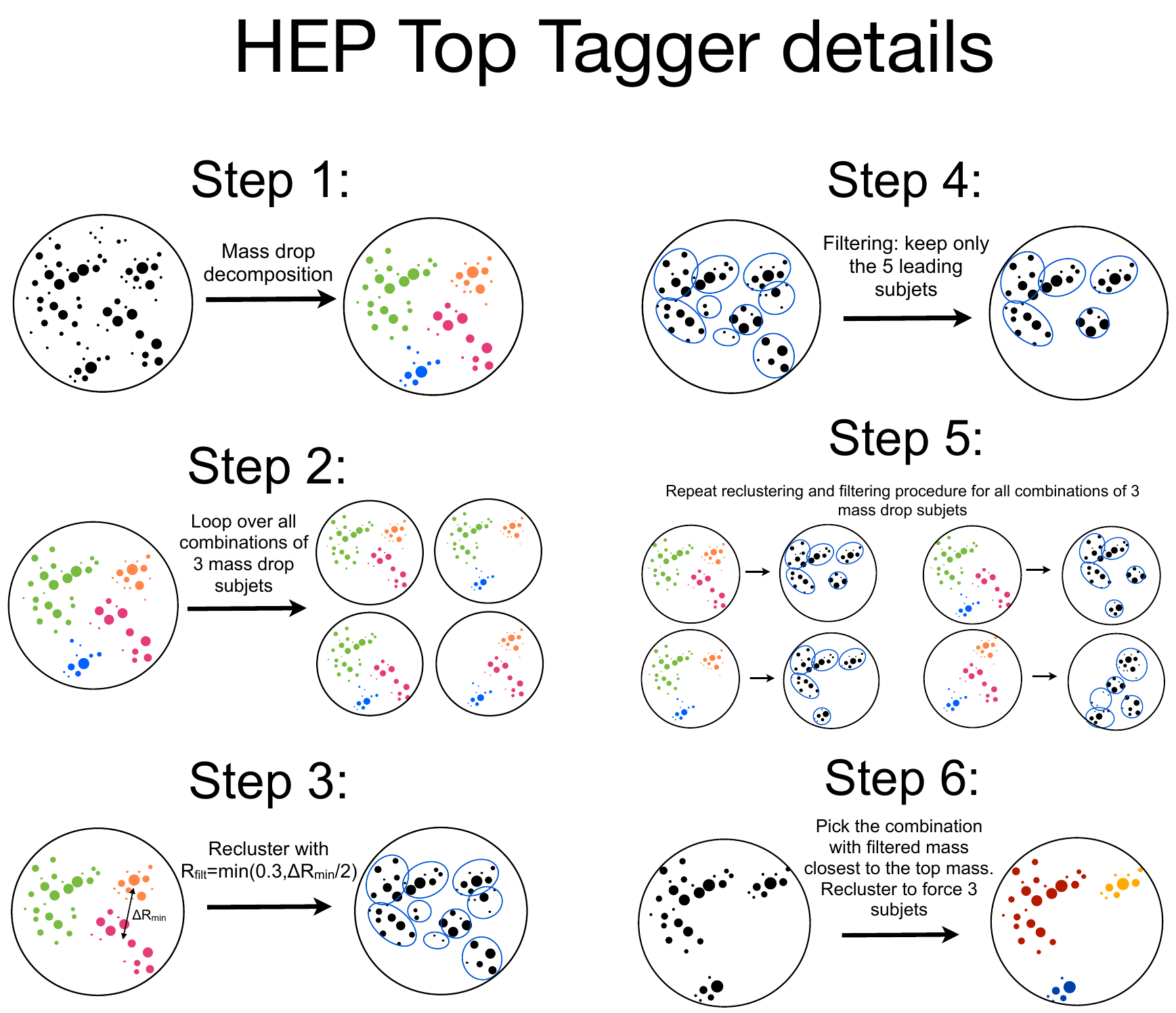}}
  \caption{Visualisation of the HEP top tagger algorithm.}\label{fig:heptoptagger}
\end{figure}

Physically, the first three steps above try to decompose a massive
object into its hard partons, in a spirit similar to what the
mass-drop condition used in the MassDrop tagger does.
The filtering step also plays the same role of further cleaning the
contamination from the Underlying Event as in the MassDrop tagger.
Finally, the set of constraints in (\ref{eq:heptoptagger-cut}) is
meant as a cut on the 3-subjets, mimicking a 3-parton system, to match
the kinematics of a top decay and further suppress the QCD
background. The whole procedure can be visualised as shown in Fig.~\ref{fig:heptoptagger}.

%picture taken from:
%https://twiki.cern.ch/twiki/pub/CMSPublic/PhysicsResultsJME13007/HEP_one_slide_description.pdf

Version 2 of the HEPTopTagger~\cite{Kasieczka:2015jma} brings several
improvements by using an extended set of variables and cuts. We just
list those modifications without entering into the details.
First, it introduces a variable radius by repeatedly reducing the jet
radius, starting from $R=1.5$, until we see a drop in the
reconstructed top mass. This is meant to reduce possible combinatorial
effects where the softest of the W decays is mistaken with a hardish
QCD subjet in the fat top candidate jet.
Then, the tagger includes additional shape variables:
\begin{itemize}
\item $N$-subjettiness values for $\beta=1$ computed both on the
  plain, ungroomed, jet and on the filtered jet
\item $Q$-jet information: the reconstructed top mass obtained from
  100 $Q$-jet histories based on the Cambridge/Aachen algorithm with $\alpha=1$,
  as well as the fraction of positive top tags one would obtain with
  version 1 of the HEPTopTagger.
\end{itemize}
In the end, the tagger uses a multivariate (Boosted Decision Tree)
analysis based on the series of kinematic variables --- subjet
transverse momenta and masses --- the optimal jet radius, and the
shape values.

\subsection{Energy Correlators}\label{energy_correlators}
\begin{figure}
\includegraphics[width=0.49\textwidth]{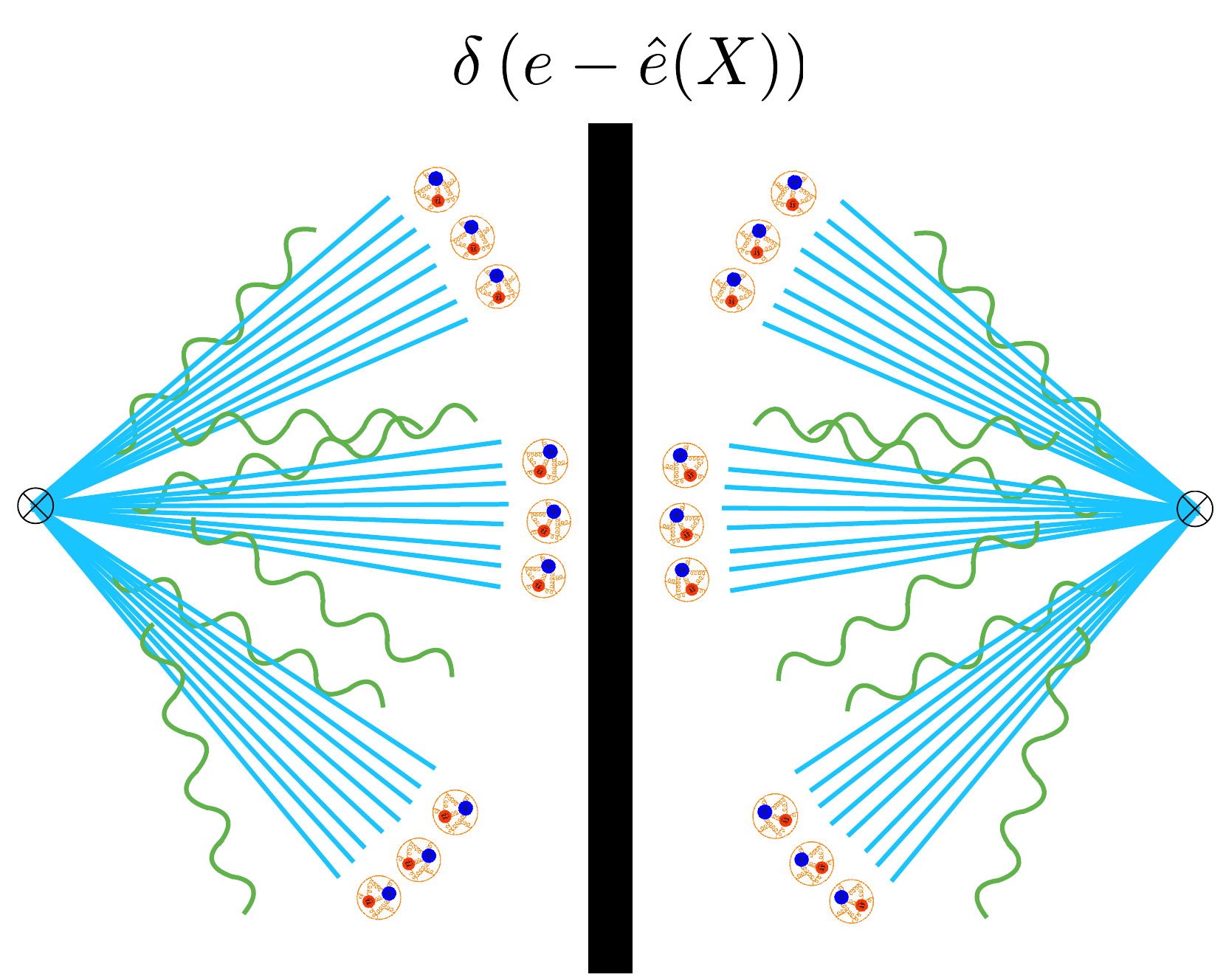}
\includegraphics[width=0.49\textwidth]{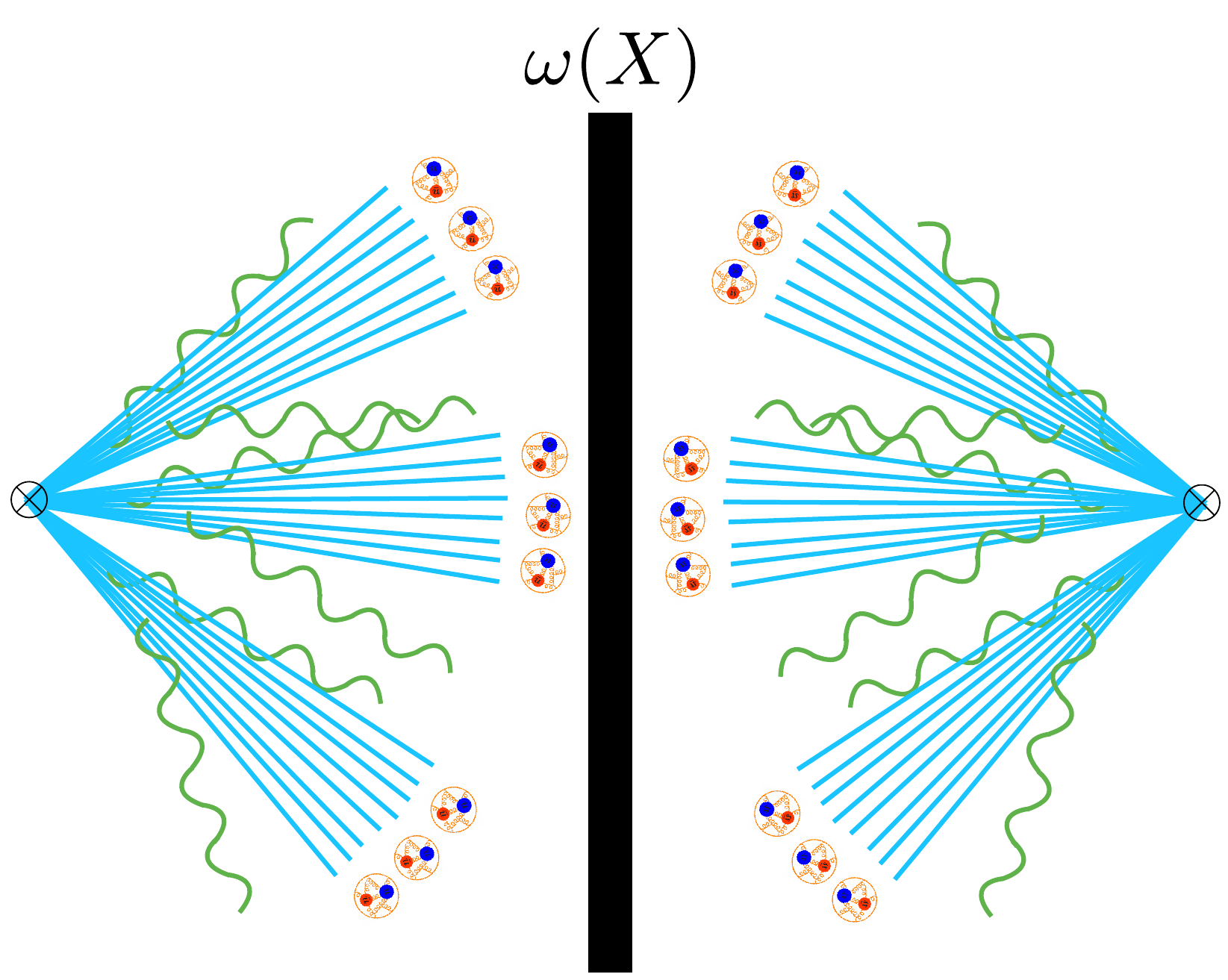}
\caption{Schematic illustration of the difference between a weighted cross section, and a more standard jet observable. 
For a standard jet observable, shown on the left, the final state is constrained by an operator $\hat e(X)$, and the cross section is calculated as a function of this constraint.
For a weighted cross section, shown on the right, a weighting function $\omega(X)$ is applied to the final state. In both cases, the cut  is illustrated by the black bar.  
Figure taken from Ref.~\cite{Chen:2020vvp}, licensed under CC BY 4.0.}\label{fig:weighted_v_delta}
\end{figure}
%%%%%%%%%%%%
The basic question that jet substructure is trying to address is the characterisation of the energy flow resulting from a hard interaction. To be more precise, 
for an event with $M$ particles with energies $E_i$ and directions $\hat{n}_i$ the energy flow is defined as
\begin{equation}\label{eq:energyflow_def}
\mathcal{E}(\hat{n}) = \sum_{i=1}^M E_i \delta(\hat{n}-\hat{n}_i).
\end{equation}
The most natural objects that we can construct (and compute) in field theory are correlation functions of energy-flow operators
$\langle \mathcal{O} \mathcal{E}(\hat{n}_1) \mathcal{E}(\hat{n}_2) \dots \mathcal{E}(\hat{n}_N)\rangle$, with $N\le M$ and $\mathcal{O}$ is a source operator, such as the electromagnetic current for $e^+e^-$ collisions. Because energy correlators are directly built from energy flows, they enjoy many theoretical properties that arise directly from the underling field theory. For this reason, they have been the subject of many formal studies that exploit, among other things, techniques from conformal field theories and the AdS/CFT correspondence~\cite{Maldacena:1997re}, see e.g.\ \cite{Hofman:2008ar}. 
Energy correlators can be used to study the QCD structure of an entire event but also to probe the internal structure of jets, by simply considering correlations between directions $\hat n_i$ that become closer in angles.

Observables built using directly energy correlators are somewhat different with respect to the ones discussed so far.  
Following Ref.~\cite{Chen:2020vvp}, let us consider a process that produces an hadronic final state $X$. The observables that we have been studying so far, for instance the jet mass, constrain the four-momenta of the final-state particles to produce a given value. Then, the cross section is studied as a function of this constraint. From a theoretical viewpoint, this means that the differential distribution is built by integrating the squared matrix element for the final-state phase-space, with an additional constraint given by the observable, as schematically shown in Fig.~\ref{fig:weighted_v_delta}, on the left. Experimentally, this means that a given event is an entry to the histogram of the observable we are considering. 
In order to make the connection between these ``delta-function" observables and energy correlators more clear, it is useful to introduce the concept of weighted cross-section observables. In this case, the final-state phase-space integral is supplemented with a weight function $\omega(X)$, rather than $\delta(e- \hat e(X))$, as shown in Fig.~\ref{fig:weighted_v_delta}, on the right. We have
\begin{align}
\sigma_\omega&= \int d^4x \,e^{iq\cdot x} \sum_{X}  \langle 0 | \mathcal{O} (x) |X\rangle \omega(X) \langle X | \mathcal{O}^\dagger(0) |0 \rangle = \int d^4x \, e^{iq\cdot x}  \langle 0 | \mathcal{O} (x)  \hat \omega \mathcal{O}^\dagger(0) |0 \rangle,
\end{align}
where we weighting operator $\hat \omega |X\rangle=\omega(X) |X\rangle$ is just the product of energy flow operators:
\begin{align}
\hat \omega=  \mathcal{E}(\hat n_1) \cdots \mathcal{E}(\hat n_N),
\end{align}
so that
\begin{align}
\sigma_\omega&= \int d^4x \, e^{i q\cdot x}  \langle 0 | \mathcal{O} (x)  \mathcal{E}(\hat n_1) \cdots \mathcal{E}(\hat n_N) \mathcal{O}^\dagger(0) |0 \rangle\,.
\end{align}
Thus, weighted observables can be expressed directly as matrix elements of energy flow operators. The simple field-theoretic definition of these objects has allowed significant recent progress in their understanding and a very active area of research has been developing in the past few years. 
On the other hand, standard ``delta-function" observables do not enjoy directly this property and in order to write them as products of energy flows, we have to consider their moments:
\begin{align}
\int de \, e^n \, \frac{d\sigma}{de}&= 
\int de \, e^n\,  \int d^4x \, e^{iq\cdot x}  \langle 0 | \mathcal{O} (x)  \delta(e-\hat e) \mathcal{O}^\dagger (0) |0 \rangle \nonumber\\
&=\int d^4x\,  e^{iq\cdot x}  \langle 0 | \mathcal{O} (x)  \hat e^n \mathcal{O}^\dagger(0) |0 \rangle.
\end{align}
Thus, standard observables are sensitive to an infinite number of energy flows.
This observation has very interesting practical consequences. Because of angular resolution and pileup mitigation, jet substructure measurements are often performed using track information only. This greatly improves experimental results at the expense of calculability because considering final-states with only charged particles renders most observables IRC unsafe. For this reasons, track-functions must be introduced~\cite{Chang:2013rca,Chang:2013iba,Jaarsma:2022kdd,Chen:2022pdu,Jaarsma:2023ell,Lee:2023xzv} in order to absorb collinear singularities. While for standard observables, one needs to determine (from data) the full functional form of the track functions, only the first moment of the track function appears in the case of energy correlators~\cite{Chen:2020vvp}.

The simplest energy correlator is the two-point energy-energy correlator (EEC)~\cite{Basham:1978bw} that has been widely studied in $e^+e^-$ collisions. This observable measures the correlation between any two directions $\hat n_i, \hat n_j$ separated by an angle $\chi$:
\begin{align}\label{eq:EEC}
\frac{1}{\sigma}\frac{d \sigma}{d \chi}&= \sum_{i \neq j} \int d^3 p_i \, d^3 p_j \left(\frac{2 E_i E_j}{s} \right)\frac{d^6 \sigma}{d^3 p_i d^3 p_j}\delta(\hat n_i \cdot \hat n_j-\cos \chi)
\nonumber \\&+\sum_{i} \int d^3 p_i \left(\frac{E_i^2}{s} \right)\frac{d^3 \sigma}{d^3 p_i }\delta(1-\cos \chi),
 \end{align}
 where the first sum runs over all distinct pairs of hadrons, while the second one ensures normalisation: $\int_{-1}^1 d \cos \chi \, \frac{1}{\sigma}\frac{d \sigma}{d \chi}=1$. We note that Eq.~(\ref{eq:EEC}) is indeed a weighted cross-section.
Recently, higher-points correlators have been studied theoretically and many new interesting observables have been derived from them. We refer the interested reader to more specialised review on this topic, e.g.~\cite{Moult:2025nhu}.
 %\cite{Chen:2019bpb,Moult:2018jzp,Dixon:2019uzg,Kologlu:2019mfz,Korchemsky:2019nzm,Chen:2019bpb,Belitsky:2013ofa,Dixon:2018qgp,Luo:2019nig,Henn:2019gkr}.  

\subsection{Event Geometry}
\begin{table}
\centering
\begin{tabular}{|c|c|c|}
\hline\hline
\bf Concept &  \bf Geometry & \bf Picture  \\
\hline\hline
&  &  \multirow{5}{*}{\raisebox{-2em}{\includegraphics[scale=0.21]{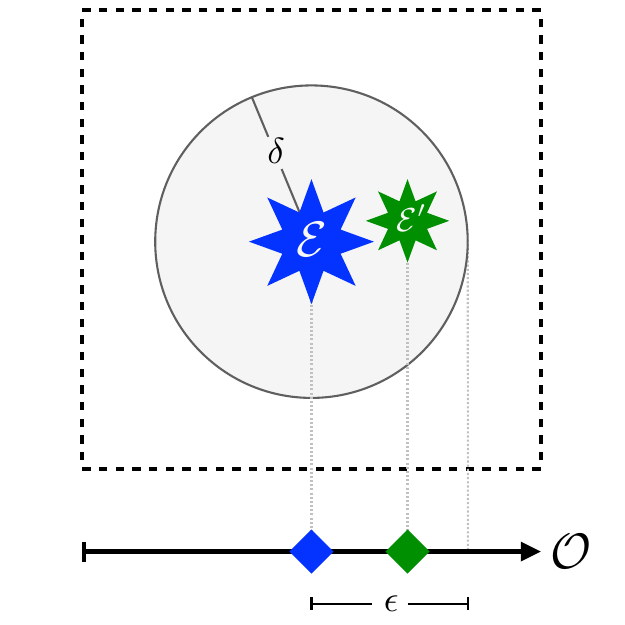}}} \\
 {\bf Infrared and} & An observable $\mathcal{O}$ is IRC safe if it is   & \\ 
  {\bf Collinear Safety} &EMD continuous for all energy flows, except &  \\
 &   potentially on a negligible set of events. & \\
 &  $ \displaystyle\text{EMD}(\mathcal{E},\mathcal{E}')<\delta  \Rightarrow |\mathcal{O}(\mathcal{E}) - \mathcal{O}(\mathcal{E}')| < \epsilon$ &  \\ \hline
 &  & \multirow{5}{*}{\raisebox{-2em}{\includegraphics[scale=0.21]{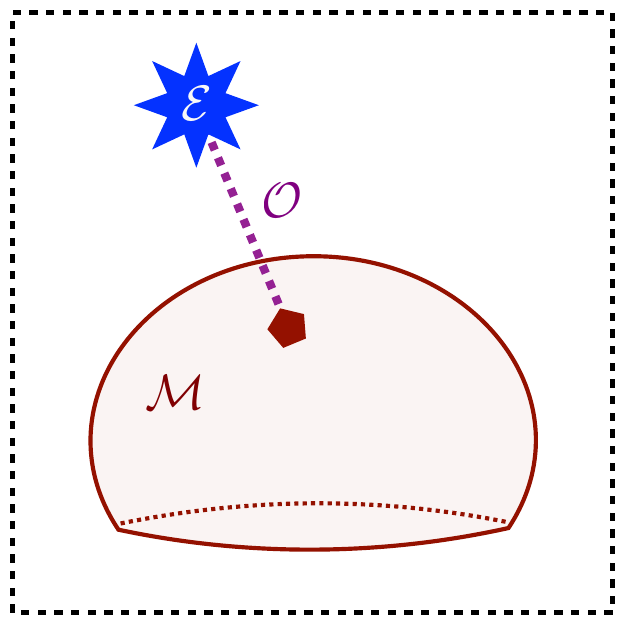}}} \\ 
{\bf Observables} & Closest distance between event (jet)  &   \\
 Event Shapes
  &  $\mathcal{E}$ and manifold $\mathcal{M}$:&  
 \\
 Jet Shapes
 &  $\displaystyle\mathcal{O}(\mathcal{E}) = \min_{\mathcal{E}'\in\mathcal M}\text{EMD}(\mathcal{E},\mathcal{E}') $&    \\
 &  Thrust: $\mathcal{M}=\mathcal{P}_2^\text{BB}$; $N$-jettiness:  $\mathcal{M}=\mathcal{P}_N$ &  \\ \hline
 &  & \multirow{5}{*}{\raisebox{-2em}{\includegraphics[scale=0.21]{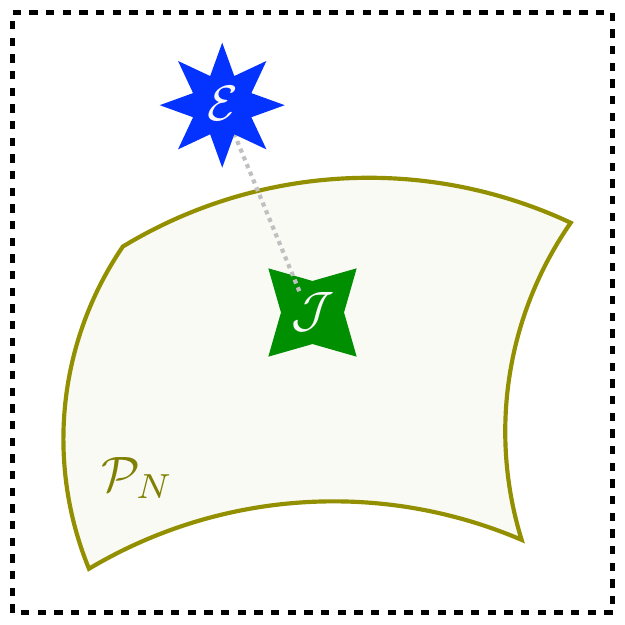}}} \\ 
  {\bf Jets} & &   \\
  XCone~\cite{Stewart:2015waa,Thaler:2015xaa}
  &  $\displaystyle\mathcal J_{N,\beta,R}(\mathcal{E})=\argmin_{\mathcal J\in\mathcal P_N}\,\text{EMD}_{\beta,R}(\mathcal{E},\mathcal J)$& \\
Gen.~$k_t$ with WTA
 &   $\displaystyle\mathcal J_{\beta,R}(\mathcal{E})=\mathcal{E}-\argmin_{ \mathcal{E}'\in\mathcal P_{M-1}}\,\text{EMD}_{\beta,R}(\mathcal{E},\mathcal{E}')$ & \\
 &  & \\ \hline
 &  & \multirow{5}{*}{\raisebox{-2em}{\includegraphics[scale=0.21]{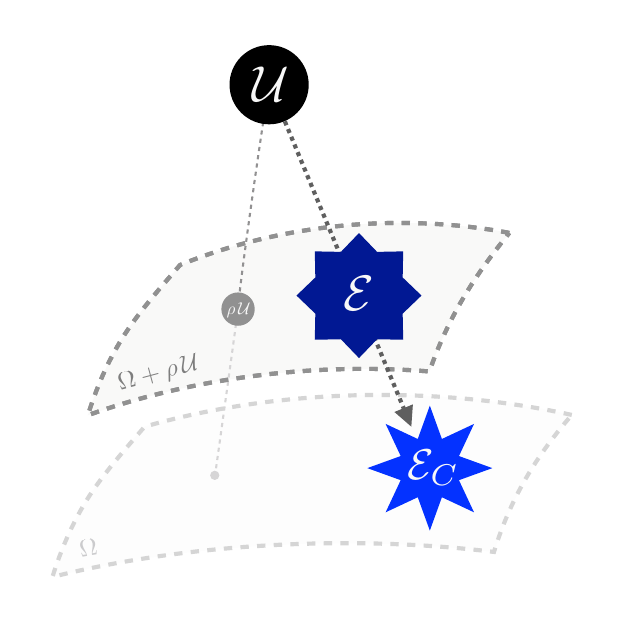}}} \\ 
 {\bf Pileup } &  Best approximation to an event with & \\
 {\bf Substraction}& added uniform energy distribution & \\
 &   $\mathcal E_C(\mathcal E, \rho)  = \displaystyle\argmin_{\mathcal{E}' \in \Omega}\,\text{EMD}(\mathcal E,\mathcal E' + \rho\,\mathcal U)$ & \\
 %&  & \\ 
 \hline\hline
\end{tabular}
\caption{\label{tab:outlinefigs}
Summary of the most significant concepts discussed in this book that can be rephrased in a geometrical language exploiting the distance between events. Adapted from Ref.~\cite{Komiske:2020qhg}, licensed under CC BY 4.0.
}
\end{table}

In the previous section, we have seen that from a theoretical viewpoint, what characterises events in particle collisions is the energy flow $\mathcal{E}$, defined in Eq.~(\ref{eq:energyflow_def}). It is then natural to think about two events as being similar if their energy flows are not too different. This idea can be made rigorous by equipping the space of energy flows, i.e.\ of events, with a metric. The Authors of Refs~\cite{Komiske:2019fks,Komiske:2020qhg}, introduced such a metric by taking inspiration from theory of optimal transport and, in particular, from the concept of 
 the earth mover's distance.
 %~\cite{DBLP:journals/pami/PelegWR89,Rubner:1998:MDA:938978.939133,Rubner:2000:EMD:365875.365881,DBLP:conf/eccv/PeleW08,DBLP:conf/gsi/PeleT13}.
 %
The Energy Mover's Distance (EMD) between two events measures the amount of ``work'' required to rearrange one event to the other.
Its value can be obtained by solving the following optimal transport problem between energy flows $\mathcal{E}$ and $\mathcal{E}'$:
\begin{equation}
\label{eq:emd}
\text{EMD}_{\beta,R} (\mathcal E, \mathcal E') = \min_{\{f_{ij}\ge0\}} \sum_{i=1}^M\sum_{j=1}^{M'} f_{ij} \left( \frac{\theta_{ij}}{R} \right)^\beta + \left|\sum_{i=1}^M E_i - \sum_{j=1}^{M'}E_j'\right|,
\end{equation}
with
\begin{equation}
\label{eq:emdconstraints}
\sum_{i=1}^M f_{ij} \le E_j', \quad\quad \sum_{j=1}^{M'} f_{ij} \le E_i, \quad\quad \sum_{i=1}^M\sum_{j=1}^{M'} f_{ij} = \min\left(\sum_{i=1}^M E_i,\sum_{j=1}^{M'}E_j'\right),
\end{equation}
where $\theta_{ij}$ is a pairwise distance between particles known as the ground metric. This is naturally taken as an angular distance
$
\theta_{ij} = \sqrt{2 (1 - \hat{n}_i \cdot \hat{n}_j)}$, which reduces, in the collinear limit, to the opening angle.
The parameter $R>0$ controls the tradeoff between transporting energy and destroying it, and $\beta > 0$ is an angular weighting exponent.
Let us briefly discuss the physical interpretation of the two terms that appear in Eq.~(\ref{eq:emd}). The first one quantifies the difference in radiation patterns between the two energy flows under consideration, while the second one accounts for differences in the total energies of the two events.  Thus, if we wish to compare events with the same total energy, the latter vanishes. 
Furthermore, the constraints in Eq.~(\ref{eq:emdconstraints}) specify that the amount of energy moved to or from a particle cannot exceed its initial energy.

The $\text{EMD}$ is positive, symmetric and, it is easy to see that $\text{EMD}_{\beta,R} (\mathcal E, \mathcal E)=0$.
Furthermore,  it can be shown that $\text{EMD}^{1/\beta}$, with the additional constraint that $\theta_{ij}<2R$ indeed satisfies the triangle inequality.
It is therefore a metric. 
In such a case, $\text{EMD}^{1/\beta}$, coincides with the $p$-Wasserstein metric~\cite{kantorovich1942translocation,wasserstein1969markov}, with $p=\beta$.
Equipping collider events with a metric allows us to explore interesting geometric ideas in the space of events.
Thanks to this observation,  we are able to rephrase in this language many of the topics in jet physics that we have explored throughout this book. The most important examples are reported in Table~\ref{tab:outlinefigs}. Let us briefly discuss them.

We start by considering the $\text{EMD}$ distance between two events. Because observables are functionals of the energy-flows $\mathcal{O}=\mathcal{O}(\mathcal{E})$, we can study their analytic properties. In particular, intuitively, IRC safe observables as those that change little, when we consider events that do not differ too much. Consequently, we can characterise IRC safe observables as being continuous according to the EDM metric.
We can proceed further and introduce manifolds of events. These are sets of events with some defined properties. For instance, $\mathcal{P}_2^\text{BB}$ is the manifold of idealised events with 2 back-to-back particles, while $\mathcal{P}_N$ is the manifold of events with $N$ particles. Many of the observables discussed in this book such as event shapes and jet shapes can be then introduced as the distance between the event $\mathcal{E}$ under consideration and a manifold $\mathcal{M}$. The event shape thrust~\cite{Farhi:1977sg} measures how much an event looks pencil-like. Thus, it does not come as a surprise that thrust can be geometrically defined as the EMD distance between $\mathcal{E}$ and the manifold $\mathcal{P}_2^\text{BB}$. Similarly, $N$-jettiness~\cite{Stewart:2010tn} can be cast as the distance between an event and $\mathcal{P}_N$.
As a consequence, the jet algorithm  XCone~\cite{Stewart:2015waa,Thaler:2015xaa}, which is based on $N$-jettiness, has a straightforward geometric interpretation. This jet-finding procedure looks for the collection of jets, which are elements of the $N$-particle manifold ($\mathcal{J} \in \mathcal{P}_N$), that best approximates the event under scrutiny.  It is less obvious that also sequential-recombination algorithms (see Section~\ref{sec:jetalgs-recombination}), e.g. the members of the generalised $k_t$ family, can be interpreted geometrically. Indeed these algorithms compare events with $M$ particles to the manifold $\mathcal{P}_{M-1}$, essentially because at each step  two particles are either merged together or one particle is removed.
Finally, let us mention that the EMD offers a natural way to describe pileup contamination and offers insights to mitigation strategy. To first approximation, pileup can be modelled as an event with uniform distribution $\mathcal{U}$ of energy with density $\rho$. Thus, pileup subtraction aims to find the corrected event $\mathcal{E}_C$ that best approximates a given event $\mathcal{E}$, when pileup is included. These ideas can be also applied to jet grooming~\cite{Alipour-Fard:2023prj}, overcoming some of the non-smooth behaviour in transition points  that will be discussed in the next chapter.

\section{Code Availability}

An essential component of a successful jet substructure algorithm, is its availability.
Therefore, for completeness, we list below where one can find the implementation
of the tools presented above.

\vspace{0.5cm}

\noindent
 \begin{tabular}{lp{9cm}}
  \hline
  {\bf Tool} & {\bf Code} \\
  \hline
  Mass-Drop Tagger          & \ttt{MassDropTagger} class in \fastjet \\
  modified Mass-Drop Tagger & \ttt{ModifiedMassDropTagger} class in the \\
                            & \ttt{RecursiveTools} \fastjet contrib\\
  SoftDrop                  & \ttt{SoftDrop} class in
                              the \ttt{RecursiveTools} \fastjet
                              contrib\\
  Recursive SoftDrop        & \ttt{RecursiveSoftDrop} class in
                              the \ttt{RecursiveTools} \fastjet
                              contrib\\
  Filtering                 & \ttt{Filter} class in \fastjet (use \ttt{SelectorNHardest})\\
  Trimming                  & \ttt{Filter} class in \fastjet\\ & (use \ttt{SelectorPtFractionMin})\\
  Pruning                   & \ttt{Pruner} class in \fastjet\\
  I and Y-Pruning           & Not available per se but can be
                              implemented as a derived class of
                              \ttt{Pruner}\\
  Johns Hopkins top tagger  & \ttt{JHTopTagger} class in \fastjet \\
  CMS top tagger            & as part of CMS-SW (see Ref.~\cite{CodeCMSTopTagger})\\
  Generalised angularities  & no know public standard implementation\\
  $N$-subjettiness          & \ttt{Nsubjettiness} \fastjet contrib\\
  Energy Correlation Functions & \ttt{EnergyCorrelator} \fastjet contrib\\
  HEPTopTagger  & code available from Ref.~\cite{CodeHEPTopTagger} \\
  Shower Deconstruction & code available from Ref.~\cite{CodeShowerDeconstruction} \\
  \hline
\end{tabular}

\vspace{0.5cm}

Let us conclude this chapter with a more general remark.
Grooming techniques
might at first sight be similar to pileup mitigation techniques. They
however target a different goal: while pileup mitigation techniques
aim at correcting for the average effect of pileup, grooming
techniques reduce the overall sensitivity to pileup.
In practice, this means that, unless one first applies an event-wide
pileup mitigation technique such as
SoftKiller~\cite{Cacciari:2014gra} or PUPPI~\cite{Bertolini:2014bba},
grooming techniques should in principle be supplemented by pileup
subtraction, like the
area--median~\cite{Cacciari:2007fd,Cacciari:2008gn,AlcarazMaestre:2012vp,Soyez:2012hv}. 
Many tools provide hooks to combine them with pileup subtraction.

%% GS helper for auctex
%%% Local Variables:
%%% mode: latex
%%% TeX-master: "notes"
%%% End:

%  LocalWords:  pronginess performant MassDrop WTA Eq Nsubjettiness
%  LocalWords:  ECFs ij Houches Neyman Eqs HEPTopTagger hardish PUPPI
%  LocalWords:  RecursiveTools contrib JHTopTagger EnergyCorrelator

% $Id: calculations-grooming.tex 531 2022-01-31 11:32:19Z smarzani $
%
% This chapter contains the pQCD calculations for the grommed jet mass
%------------------------------------------------------------------------

\chapter{Calculations for the jet mass with grooming}\label{calculations-substructure-mass}

In this chapter we will revisit the calculations performed in Chapter~\ref{chap:calculations-jets} and extend them 
in order to describe jet mass distributions with grooming algorithms. 
In what follows, we are not going to present state-of-the art theoretical calculations, but instead we aim to keep the 
our discussion as simple as possible. Therefore, the theoretical accuracy of the calculations that we will present will be the minimum one which is required to capture the essential feature of the distributions. 
We will mostly concentrate of QCD jets, which present the most interesting and intricate features, while a discussion about jets originated to a boosted heavy particles will be presented in Sec.~\ref{sec:calc-groomed-mass-signal}.

%%========================================================================
\section{mMDT/ \SD mass}\label{sec:calc-groomed-mass}

The first calculation we perform is that of the invariant mass distribution of a jet after the mMDT / \SD algorithm
has been applied. As we have already mentioned, the \SD algorithm reduces to mMDT when the angular exponent
$\beta$ is set to zero. Therefore, in order to keep our notation light we are going to generically refer to the algorithm as \SD (SD) and it is understood that the $\beta=0$ case corresponds to mMDT.

In the next subsection, we do the calculation at leading order in the
strong coupling constant. This simple example will allow us to see the
large logarithms that appear and we will turn to their resummation in
the next subsection.

\subsection{LO calculation}\label{sec:calc-groomed-mass-LO}

\begin{figure}[t!]
  \centerline{\includegraphics[width=0.45\textwidth]{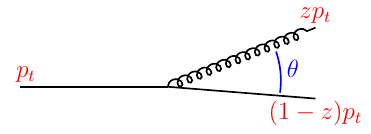}}
  \caption{Diagram contributing to the leading-order mass
    distribution.}\label{fig:mmdt-mass-one-gluon}  
\end{figure}

At zeroth order in $\alpha_s$, the jet mass is always zero. To obtain
a non-trivial mass, we therefore need to consider a high-energy
parton, say a quark for definiteness, radiating an extra gluon, as
depicted in Fig.~\ref{fig:mmdt-mass-one-gluon}.
We want to focus on the boosted jet limit and highlight large
logarithms of $m/p_t$, with $p_t$ the transverse momentum of the
initial quark, which arise in the perturbative series expansion.
At the leading-logarithmic accuracy we are interested in, we can work
in the collinear approximation where the gluon emission angle $\theta$
is small.\footnote{Alternatively, we can assume a small jet radius $R$
  so that corrections beyond the collinear approximation are
  suppressed by powers of $R$. Note also that in the case of the mMDT
  jet mass, the SD condition actually gets rid of this
  contribution so that the collinear approximation remains valid at
  higher logarithmic accuracy.}
The gluon is set to carry a fraction $z$ of the quark momentum,
leaving a fraction $1-z$ for the recoiling quark after the emission.

When applying SD, the jet is split into two subjets, one
with the quark and one with the gluon which is tested for the
SD condition. Two situations can occur: (i) either the splitting
passes the SD condition, i.e. \ $z>\zcut (\theta/R)^\beta$ in
which case the quark-gluon system is retained by the SD procedure
and the (squared) jet mass is given by
\begin{equation}\label{eq:mmdt-qg-mass}
m^2 = z(1-z)\theta^2 p_t^2,
\end{equation}
or (ii) the condition is failed in which case only the harder of the quark
and the gluon is kept and the jet mass vanishes.
The mass distribution at LO is therefore given by
\begin{equation}\label{eq:mmdt-lo-full}
  \frac{m^2}{\sigma}\frac{d\sigma^{\text{(LO)}}}{dm^2}
  = \frac{\alpha_s}{2\pi} \int_0^{R^2}\frac{d\theta^2}{\theta^2}\int_0^1dz\,P_q(z)
  m^2\delta(m^2-z(1-z)\theta^2 p_t^2)\Theta\big(z>\zcut(\theta/R)^\beta\big),
\end{equation}
where $P_q(z)$ is the quark splitting function.

The mass constraint can be used to perform the integration over
$\theta$, and the constraint $\theta<R$ means we have to impose
$z(1-z)>\rho$ where we have introduced the dimensionless variable
\begin{equation}\label{eq:rho-definition}
  \rho = \frac{m^2}{p_t^2R^2}.
\end{equation}
Up to power corrections in $\rho$, i.e. in the groomed jet mass, we
can neglect the factor $1-z$ in this constraint.
We are therefore left with
\begin{equation}\label{eq:LOmmdtSD-zinteg}
  \frac{\rho}{\sigma}\frac{d\sigma^{\text{(LO)}}}{d\rho}
  = \frac{\alpha_s}{2\pi} \int_\rho^1dz\,P_q(z)
  \Theta\big(z>\zcut^{2/(2+\beta)}\rho^{\beta/(2+\beta)}\big).
\end{equation}
In the remaining integration over $z$, the SD constraint is only
relevant for $\rho<\zcut$ and we get 
\begin{equation} \label{eq:sigma-SD-LO}
  \frac{\rho}{\sigma}\frac{d\sigma^{\text{(LO)}}_\text{SD}}{d\rho} =
  \begin{cases}
   \frac{\alpha_sC_F}{\pi}
    \bigg[\log\Big(\frac{1}{\rho}\Big)-\frac{3}{4}\bigg],
  \quad \text{if} \quad \rho>\zcut,  \\
    \frac{\alpha_sC_F}{\pi}
    \bigg[\frac{\beta}{2+\beta}\log\Big(\frac{1}{\rho}\Big)+\frac{2}{2+\beta}\log\Big(\frac{1}{\zcut}\Big)-\frac{3}{4}\bigg],
  \quad \text{if} \quad \rho<\zcut,
  \end{cases}
\end{equation}
again up to power corrections in $\rho$.\footnote{Technically, for
  mMDT, this result is valid up to power corrections in $\zcut$. These
  corrections can be included and
  resummed~\cite{Dasgupta:2013ihk,Marzani:2017mva} but we will assume
  small $\zcut$ here and neglect them.}
The above result exhibits two different regimes: when the jet mass is
not very small $\rho>\zcut$, SD is inactive and one recovers the
plain, i.e.\ ungroomed jet-mass distribution discussed in Chapter~\ref{chap:calculations-jets}. However, when the mass becomes
smaller, $\rho<\zcut$, SD becomes active, as manifested here
under the form of a larger cut on the $z$ integration in Eq.~(\ref{eq:LOmmdtSD-zinteg}).

As mentioned in Chapter~\ref{chap:calculations-jets}, it is usual to
work with the cumulative distribution. At ${\cal O}(\alpha_s)$, we
find
\begin{align}
  \Sigma_{\text{SD}}^{\text{(LO)}}(\rho)
  & = \frac{1}{\sigma_0}\int_0^\rho d\rho' \frac{d\sigma}{d\rho'}
  = 1-\frac{1}{\sigma_0}\int_\rho^1 d\rho' \frac{d\sigma}{d\rho'}  \label{eq:mmdtsd-lo-cumul} \\
  & =
  \begin{cases}
     1- \frac{\alpha_sC_F}{\pi}
    \Big[\frac{1}{2}\log^2\big(\frac{1}{\rho}\big)-\frac{3}{4}\log\big(\frac{1}{\rho}\big)\Big], &
  \text{if} \quad \rho>\zcut, \\
    1- \frac{\alpha_sC_F}{\pi}
    \Big[\frac{1}{2}\log^2\big(\frac{1}{\rho}\big)-\frac{1}{2+\beta}\log^2\big(\frac{\zcut}{\rho}\big)-\frac{3}{4}\log\big(\frac{1}{\rho}\big)\Big], &
  \text{if} \quad \rho<\zcut.
  \end{cases}\nonumber
\end{align}

In going from the first to the second equality, one could either argue that
the probability is conserved (\ie the mass is either larger or
smaller than $\rho$), or realise that $\frac{d\sigma}{d\rho'}$ also
has a virtual contribution at $\rho'=0$ which, up to subleading power
corrections, can be written as
\[
  \left. \frac{d\sigma^{\text{(LO)}}}{d\rho'}\right|_{\text{virt.}}
   = -\bigg(\int_0^1 d\rho \frac{d\sigma}{d\rho}\bigg)\,\delta(\rho').
\]
More importantly, the results above clearly show that a gluon emission
comes with large logarithms of the jet mass on top the expected power
of $\alpha_s$. When the jet mass becomes sufficiently small, this is
no longer a small quantity and one needs to resum gluon emissions to
all orders. We do that in the next section.
There are however a few interesting points we can already highlight now.
For example, we see that the dominant logarithms in $\Sigma(\rho)$ are
double logarithms of the jet mass. These are associated with the
emission of a gluon which is both soft and collinear. The subleading
single-logarithmic contribution comes here from a hard and collinear
gluon emission.
Then, one expects the SD condition to be less effective as
$\beta$ increases. This is indeed what one sees here since one tends
to the plain jet mass distribution in the limit
$\beta\to\infty$. Conversely, for $\beta= 0$, the double logarithm of
the jet mass disappears --- going back to
Eq.~\eqref{eq:LOmmdtSD-zinteg} the $z$ integration is cut at $\zcut$
for $\beta=0$, meaning that the soft emissions only produce a
logarithm of $\zcut$ instead of a combination of $\log(\zcut)$ and
$\log(\rho)$ for the generic case --- leaving a single-logarithmic
dominant term, which is purely collinear.  

We conclude this section with a discussion about soft emissions at
large angles. These have not been included in the calculation above
where we have worked in the collinear, small $R$,
approximation. However, as seen in
Chapter~\ref{chap:calculations-jets} (see \eg
Eq.~(\ref{eq:sigma-1-loop-cntd})), soft emissions at finite angles can
also give single-logarithmic contributions.
This will no longer be the case in the region where SD is
active. To see this, imagine that we have a soft emission passing the
SD condition and dominating the jet mass. This implies
$\rho=z(\theta/R)^2$ and $z>\zcut (\theta/R)^\beta$, from which one
easily deduces $\theta<R (\rho/\zcut)^{1/(2+\beta)}$. A contribution
at a finite angle (\ie not enhanced by a collinear $d\theta/\theta$)
would therefore be suppressed by a power of $\rho$.
Similarly, one can show that non-global logarithms are also suppressed
by SD.
This is a fundamental analytic property of SD, namely that it
suppresses soft-and-large-angle gluon emissions so that observables can
(usually) be computed in the collinear limit. We will come back to
that point in the next section.

%%========================================================================
\subsection{Resummation of the mMDT/\SD mass distribution}\label{sec:calc-groomed-mass-LL}

We now move to the all-order resummation of the logarithms of the SD
jet mass distribution. We target a modified leading-logarithmic
accuracy, \ie include the leading double-logarithmic terms as well as
the hard-collinear single-logarithmic contributions.

In an all-order calculation, one has two types of contributions to
consider. First, real emissions which fail the SD condition will be
groomed away by the SD procedure~\footnote{Strictly speaking, since SD
  stops the first time the condition is passed, this is only true for
  gluons at angles larger than the first emission passing the SD
  condition. However, such gluons cannot dominate the jet mass and so
  can be neglected. It is worth noting that for more complicated
  quantities, like jet shapes computed on a SD jet, this effect would
  have to be taken into account.}  and will therefore not contribute
to the jet mass. They will therefore cancel explicitly against the
corresponding virtual corrections.
We are therefore left with the case of the real gluons which pass the
SD condition and the associated virtual emissions. These gluons
will contribute to the jet mass.
The situation here is therefore exactly as the one discussed in
Sec.~\ref{sec:jet-mass-res} for the case of the plain jet mass but
now restricted to the gluons passing the SD condition.

At the end of the day, this means that, if we want to compute the
cumulative distribution $\Sigma_{\text{SD}}(\rho)$, we have to
veto all real emissions that, while passing the SD condition, would give a ``mass'' larger than $\rho$.
Real emissions outside the SD
region and emissions at smaller mass do not contribute to the jet
mass~\footnote{At full single-logarithmic accuracy, one would also get
  a contribution with multiple emissions contributing to the jet mass,
  These emission would again have to pass the SD condition and
  their resummation goes exactly as for the plain jet, yielding a
  factor $\exp(-R')/\Gamma(1+R')$ with $R'$ the derivative of the
  SD radiator given below.} and cancel against virtual
corrections. We are therefore left with a ``standard'' Sudakov-type
factor
\begin{equation}\label{eq:mMDTSD-ll}
 \Sigma_{\text{SD}}(\rho) = \exp\big[-R_{\text{SD}}(\rho)\big],
\end{equation}
with (measuring the angles in units of the jet radius $R$ for
convenience and $i=q,g$)
\begin{equation}\label{eq:mMDTSD-radiator}
 R_{\text{SD}}(\rho) = \int_0^1
 \frac{d\theta^2}{\theta^2}\,dz\,P_i(z) \frac{\alpha_s(z\theta p_tR)}{2\pi}
 \Theta(z\theta^2>\rho) \Theta(z>\zcut \theta^\beta).
\end{equation}
In a fixed-coupling approximation, $R_{\text{SD}}$ is the same as
the one-gluon emission result, Eq.~\eqref{eq:mmdtsd-lo-cumul}. Including
running-coupling corrections is straightforward. We choose the hard scale to be $p_t R$ and we write
\begin{equation}
\alpha_s(z\theta p_tR) = \frac{\alpha_s (p_tR)}{1+2\alpha_s\beta_0\log(z\theta)},
\end{equation}
and we perform the integration keeping only the leading double-logarithmic contributions from
soft-and-collinear emissions as well as hard-collinear branchings. For $\rho<\zcut$, we obtain
\begin{align}\label{eq:mMDTSD-radiator-modll}
  R_{\text{SD}}^{\text{(LL)}}(\rho)
    & = \frac{C_i}{2\pi\alpha_s\beta_0^2}
      \bigg[\frac{2+\beta}{1+\beta}W\Big(1-\frac{\lambda_c+(1+\beta)\lambda_\rho}{2+\beta}\Big)
      -\frac{W(1-\lambda_c)}{1+\beta}-2W\Big(1-\frac{\lambda_\rho}{2}\Big)\nonumber\\
    & \phantom{=\frac{C_i}{2\pi\alpha_s\beta_0^2} \quad}
      -2\alpha_s\beta_0B_i\log\Big(1-\frac{\lambda_\rho}{2}\Big)\bigg],
\end{align}
with
\[
  \lambda_\rho = 2\alpha_s\beta_0\log(1/\rho),\qquad
  \lambda_c = 2\alpha_s\beta_0\log(1/\zcut),\qquad
  \text{and}\quad
  W(x)=x\log(x).
\]
The first line in Eq.~(\ref{eq:mMDTSD-radiator-modll}) corresponds to
the double logarithms, while the second line comes from hard-collinear
splittings.
This expression covers both the case of quark- and gluon-initiated
jets, with the only difference between the two are the overall colour
factor ($C_i=C_F$ for quarks and $C_i=C_A$ for gluons) and the
contribution from hard-collinear splittings ($B_i=B_q$ or $B_i=B_g$,
see Appendix~\ref{chap:app-analytic-details}).
As before, we recover the plain-jet case in the limit
$\beta\to\infty$, while the distribution becomes single-logarithmic
for the mMDT case, i.e.\ $\beta=0$.
Note that it might be convenient to reabsorb the contribution
from hard-collinear splittings, the last term
of Eq.~(\ref{eq:mMDTSD-radiator-modll}), directly into the
double-logarithmic contribution. This gives an expression equivalent
to Eq.~(\ref{eq:mMDTSD-radiator-modll}) up to NNLL corrections:
\begin{align}\label{eq:mMDTSD-radiator-modll-altB}
  R_{\text{SD}}^{\text{(LL)}}(\rho)
    & = \frac{C_i}{2\pi\alpha_s\beta_0^2}
      \bigg[\frac{2+\beta}{1+\beta}W\Big(1-\frac{\lambda_c+(1+\beta)\lambda_\rho}{2+\beta}\Big)
      -\frac{W(1-\lambda_c)}{1+\beta}-2W\Big(1-\frac{\lambda_\rho+\lambda_B}{2}\Big)\nonumber\\
    & \phantom{=\frac{C_i}{2\pi\alpha_s\beta_0^2} \quad}
      +W(1-\lambda_B)\bigg],
\end{align}
with $\lambda_B=-2\alpha_s\beta_0B_i$. 
The pros and cons of this alternative treatment of the $B$ term are
further discussed in Appendix~\ref{chap:app-analytic-details}.
More generally, the $B$ terms can systematically be inserted in the LL
contributions by replacing the $z<1$ kinematic boundary by
$z<\exp(B_i)$.
This is the approach we have adopted for all the plots obtained from
analytic calculations in this chapter.

\begin{figure}[t!]
    \centering
  \subfloat[]{\includegraphics[width=0.45\textwidth]{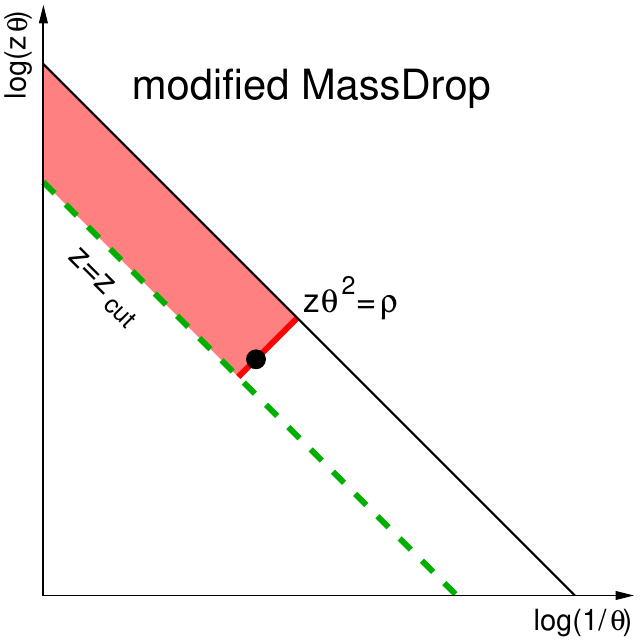}%
    \label{fig:lund-mmdt}}%
  \qquad
  \subfloat[]{\includegraphics[width=0.45\textwidth]{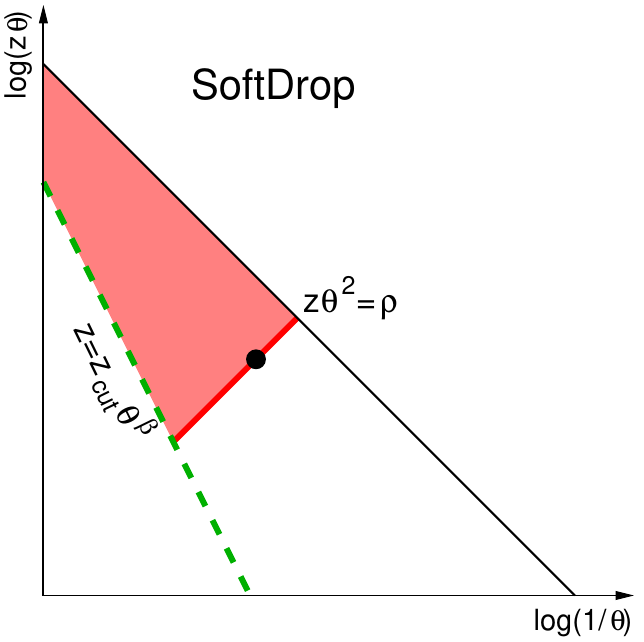}%
    \label{fig:lund-sd}}%
  \caption{Lund diagrams for the groomed jet mass distribution at LL
    for mMDT (left) and generic SD (right). The solid green line
    represents the edge of the SD region, corresponding to the
    condition $z=\zcut\theta^\beta$. The solid red line corresponds to
    emissions yielding the requested jet mass, \ie satisfying
    $z\theta^2=\rho$. The shaded red area is the vetoed area
    associated with the Sudakov suppression.}\label{fig:lund-sd-mmdt}
\end{figure}

The above results can easily be represented using Lund diagrams
(cf.~\ref{sec:jet-mass-res}). This is done in
Fig.~\ref{fig:lund-sd-mmdt}. compared to the plain jet mass, only the
emissions above the SD condition have to be vetoed.
This corresponds to the shaded red region on the plot, therefore
corresponding to the radiator $R_{\text{SD}}$.
Similarly, its derivative with respect to $\log(1/\rho)$, $R'_{\text{SD}}$,
is the weight associated with having an emission passing the SD
condition and satisfying $z\theta^2=\rho$, and is represented by the
solid red line in Fig.~\ref{fig:lund-sd-mmdt}.
From both the analytic results and the simple Lund diagrams, one
clearly sees that the smaller $\beta$, the more aggressively one
grooms soft-and-large-angle emissions. Furthermore, when $\beta$ decreases,
both $R_{\text{SD}}$ and $R'_{\text{SD}}$ decrease.

%%========================================================================
\section{Other examples: trimming and pruning}\label{sec:calc-groomed-mass-others}

Amongst the taggers and groomers introduced in Chapter~\ref{tools},
the modified Mass-Drop Tagger and Soft~Drop are the ones with the
simpler analytic structure.
It is however possible to obtain results for other groomers/taggers as
well. In this section we give a brief overview of the mass
distribution one would obtain after applying trimming or pruning, as
initially calculated in Ref.~\cite{Dasgupta:2013ihk}.
We refer to Sec.~\ref{sec:tools-prong-finders-groomers} for a
description of the substructure tools.

\subsection{Trimming}

\paragraph{Leading-order result.} As above, we start with
${\cal O}(\alpha_s)$ calculation. Therefore, we consider a single soft
and collinear gluon emission in the jet, emitted from a high-energy
quark at an angle $\theta$ and carrying a fraction $z$ of the leading
parton's momentum.
For the jet mass to be non-zero, the emission needs to be kept in the
trimmed jet. 
If the emission is clustered in the same subjet as the leading parton,
it will automatically be kept; otherwise, if it is in its own subjet,
it will only be kept if it carries a fraction of the total jet $p_t$
larger than $f_\text{trim}$. After adding together real and virtual
contribution. the LO contribution to the cumulative distribution
is:\footnote{For brevity, the notation $\Theta(a\text{ or }b)$ is one
  if either $a$ or $b$ is satisfied and 0 is none of $a$ and $b$ are
  satisfied. It can be rewritten as
  $\Theta(a\text{ or
  }b)=\Theta(a)+(1-\Theta(a))\Theta(b)=\Theta(b)+(1-\Theta(b))\Theta(a)$.}
\begin{equation}\label{eq:trim-lo-full}
  \Sigma^{\text{(LO)}}_{\text{trim}}(\rho)
  = 1-\frac{\alpha_s}{2\pi} \int_0^1\frac{d\theta^2}{\theta^2}\int_0^1dz\,P_q(z)
  \,\Theta(z\theta^2>\rho)\,\Theta\big(z>\ftrim\text{ or }\theta<\rtrim\big),
\end{equation}
where we have introduced $\rtrim=\Rtrim/R$.
We note that the above expression  differs from the mMDT/SD case only by the tagger/groomer condition.
Therefore, if we are only interested in terms enhanced by logarithms of $\rho$,
$\ftrim$ or $\rtrim$, we can easily follow the same approach as in
Sec.~\ref{sec:calc-groomed-mass-LO} and get
\begin{align} \label{eq:trim-LO-result-cumulative}
 \Sigma_{\text{trim}}^{\text{(LO)}}(\rho)
  = 1- \frac{\alpha_sC_F}{\pi}
  &\bigg[\frac{1}{2}\log^2\Big(\frac{1}{\rho}\Big)-\frac{1}{2}\log^2\Big(\frac{\ftrim}{\rho}\Big)\Theta(\rho<\ftrim)\\
  &+\frac{1}{2}\log^2\Big(\frac{\ftrim\rtrim^2}{\rho}\Big)\Theta(\rho<\ftrim\rtrim^2)-\frac{3}{4}\log\Big(\frac{1}{\rho}\Big)\bigg],\nonumber
\end{align}
This results is very similar to what was obtained for the mMDT, i.e.\
Eq.~(\ref{eq:mmdtsd-lo-cumul}) with $\beta=0$, with one striking
difference: there is an additional transition point at
$\rho=\ftrim\rtrim^2$.
For $\ftrim\rtrim^2<\rho<\ftrim$, the distribution is
single-logarithmic and is the same as what one gets for $\rho<\zcut$
in the mMDT case with the replacement $\zcut\to \ftrim$. However, at lower
$\rho$, one has an extra contribution,
$\frac{1}{2}\log^2(\ftrim\rtrim^2/\rho)$, corresponding to a typical
plain-jet double-logarithmic contribution (albeit for a jet of
smaller radius).

For completeness, we also give the results for the differential mass
distribution at leading order, which reads
\begin{equation}  \label{eq:sigma-trimming-lo}
\frac{\rho}{\sigma} \frac{d\sigma^{\text{(LO)}}_\text{trim}}{d\rho}=
\begin{cases}
   \frac{\alpha_sC_F}{\pi}
    \Big[\log\big(\frac{1}{\rho}\big)-\frac{3}{4}\Big] \;
  &\text{if } \rho\ge\ftrim,\\
    \frac{\alpha_sC_F}{\pi}
    \Big[\log\big(\frac{1}{\ftrim}\big)-\frac{3}{4}\Big] \;
  &\text{if } \ftrim\rtrim^2\le\rho<\ftrim \\
   \frac{\alpha_sC_F}{\pi}
    \Big[\log\big(\frac{\rtrim^2}{\rho}\big)-\frac{3}{4}\Big] \;
  &\text{if } \rho<\ftrim\rtrim^2.
  \end{cases}
\end{equation}

\begin{figure}[t!]
  \centering
  \includegraphics[width=0.45\textwidth]{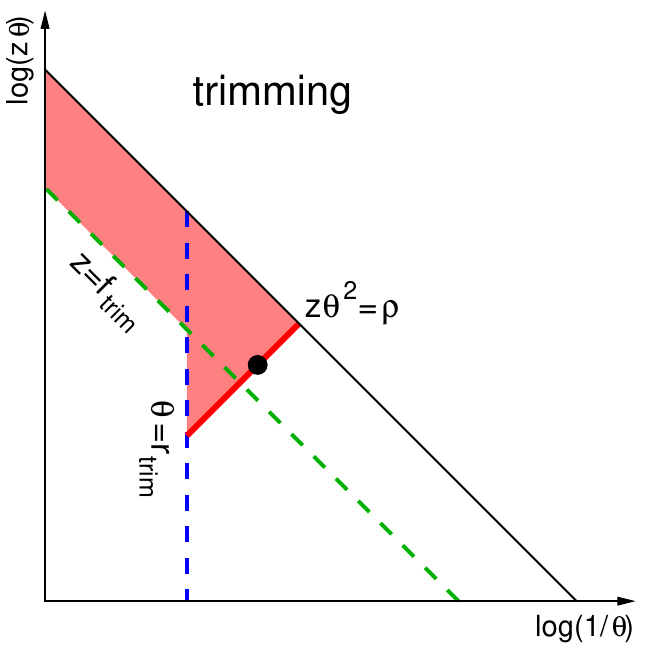}%
  \caption{Lund diagrams for the trimmed jet mass distribution at LL
    for mMDT (left) and generic SD (right). The solid green and blue
    lines represents the edge of the trimming region, respectively
    representing the $z=\ftrim$ and $\theta=\Rtrim$ conditions.
    The solid red line corresponds to emissions yielding the requested
    jet mass, \ie satisfying $z\theta^2=\rho$. The shaded red area is
    the vetoed area associated with the Sudakov
    suppression.}\label{fig:lund-trim}
\end{figure}

\paragraph{All-order resummation.} As for the \SD case, it is
relatively easy to show that the all-order resummed result is simply
the exponential of the one-gluon emission result (including
running-coupling corrections which we shall not explicitly calculate
here). We therefore get
\begin{equation}\label{eq:trim-ll}
 \Sigma_{\text{trim}}^{\text{(LL)}}(\rho) = \exp\big[-R_{\text{trim}}(\rho)\big],
\end{equation}
with, up to running-coupling corrections, 
\[
 R_{\text{trim}}(\rho) = 1-\Sigma_{\text{trim}}^{\text{(LO)}}(\rho).
\]
It is also informative to look at the corresponding Lund diagram,
plotted in Fig.~\ref{fig:lund-trim}. Compared to
Fig.~\ref{fig:lund-mmdt}, we explicitly see the emergence of a
transition point at $\rho=\ftrim\Rtrim^2$ and a double-logarithmic
behaviour in $\rho$ at smaller masses. This is associated with the
trimming radius $\Rtrim$ and the fact that emissions at angles smaller
than $\Rtrim$ will be kept in the groomed jet regardless of their
momentum fraction. This was different in the mMDT case where these emissions would still be
subject to the mMDT $\zcut$ constraint.

Finally, we can argue that this extra transition point is pathological
and a strong motivation to prefer the mMDT and \SD over trimming. Indeed, this
transition point produces a kink in the mass spectrum (see also
Sec.~\ref{sec:calc-groomed-mass-mc} below), smeared by subleading
contributions. Finding a possible signal in this region, or using this
mass domain as a side-band for a signal in an adjacent mass window,
would then become much more complex, if not impossible.
Additionally, this region would also receive single-logarithmic
contributions from soft-and-large angle emissions and non-global
logarithms (albeit suppressed by $\Rtrim^2$) which were absent in the
\SD case. 

Thus, all these factors render the calculation of the trimmed mass spectra of the same degree of complexity as the plain jet mass, if not worse because of the presence of the transition points. On the other hand, the analytic structure we have found for \SD was remarkable simpler and therefore amenable for precision calculations.

\subsection{Pruning}

In this section, we show explicitly that the case of pruning is more
complex but can be simplified by introducing instead the
Y-pruning variant. Since the main issue of pruning does not appear
in a LO calculation, we will briefly discuss its origin at NLO,
without providing an explicit calculation.
For simplicity, we take $f_{\text{prune}}=\frac{1}{2}$, so that the
pruning radius is given by $\Rprune=m_{\text{jet}}/p_{t,\text{jet}}$
and we introduce $\rprune=\Rprune/R$.

\paragraph{Leading-order result.} For a single soft-and-collinear
emission of momentum fraction $z$ and emission angle $\theta$, the jet
mass is given by $z\theta^2$, meaning that the pruning radius will be
set to $R_{\text{prune}}=\sqrt{z}\theta$ which is always smaller than
$\theta$. The emission will therefore be kept in the pruned jet only
if $z>\zprune$. This give exactly the same result as for mMDT, with
$\zcut$ replaced by $\zprune$:
\begin{equation}\label{eq:prune-lo-full}
  \Sigma^{\text{(LO)}}_{\text{prune}}(\rho)
   = \left.\Sigma^{\text{(LO)}}_{\text{mMDT}}(\rho)\right|_{\zcut\to\zprune},
\end{equation}
where we recall that mMDT corresponds to \SD with angular exponent $\beta=0$.

\paragraph{Behaviour at higher orders.} The pruning behaviour becomes significantly more complicated beyond LO. Let us give an explicit example. At NLO, we should consider
situations where we have two real emissions, 1 and 2, with respective
momentum fractions $z_1$ and $z_2$ and emission angles $\theta_1$ and
$\theta_2$ with respect to the leading parton (one should as well include the
cases with one or two virtual emissions).
Without loss of generality, we can assume that $z_1\theta_1^2\gg
z_2\theta_2^2$, with the strong ordering sufficient to capture the
leading logarithms of the jet mass we are interested in.
Emission 1 therefore dominates the (plain) jet mass and sets the
pruning radius to $R_{\text{prune}}=\sqrt{z_1}\theta_1$.
The complication comes from the fact that emission 1 itself may be
groomed away by pruning, \ie have $z_1<\zprune$, in which case, the
jet mass will only be non-zero if emission 2 is kept by pruning and
this is ensured by the condition
\[
  \Theta(z_2>\zprune\text{ or }\theta_2^2<z_1\theta_1^2),
\]
which depends on $z_1$. As we will see below, this is not a
show-stopper to resum the pruned jet mass distribution to all orders
but we definitely depart from the simple Sudakov exponentiation seen
for \SD, Eq.~(\ref{eq:mMDTSD-ll}), and trimming,
Eq.~(\ref{eq:trim-ll}).

\begin{figure}[t!]
    \centering
  \subfloat[]{\includegraphics[width=0.32\textwidth]{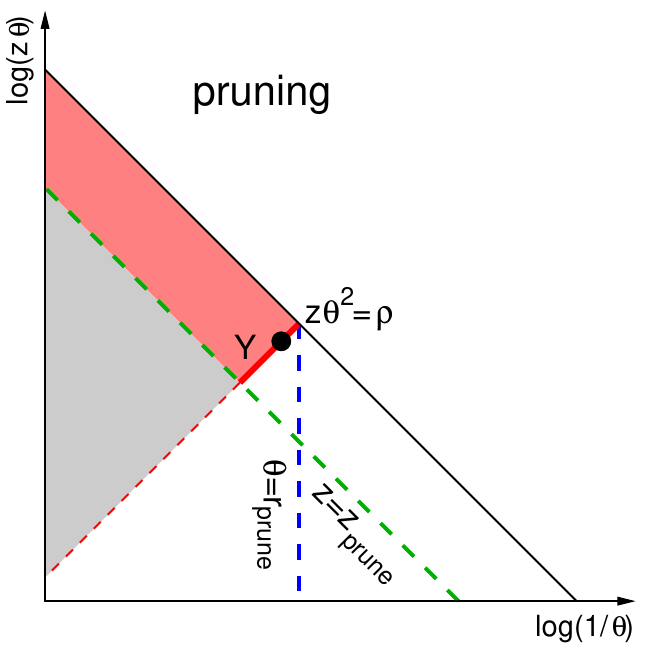}%
    \label{fig:lund-pruning-1}}\hfill%
  \subfloat[]{\includegraphics[width=0.32\textwidth]{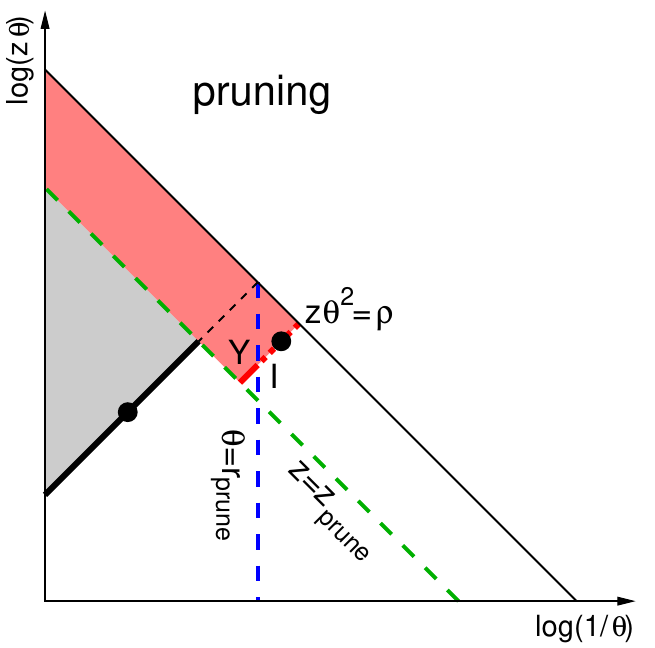}%
    \label{fig:lund-pruning-2a}}\hfill%
  \subfloat[]{\includegraphics[width=0.32\textwidth]{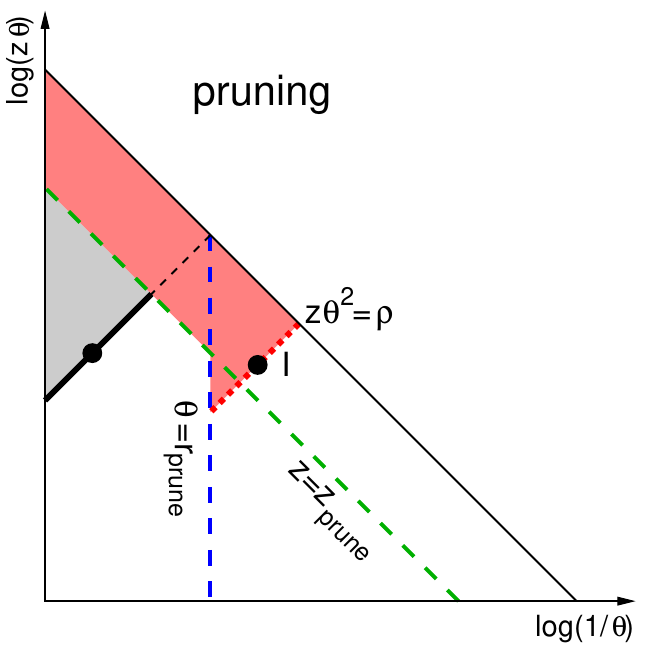}%
    \label{fig:lund-pruning-2b}}%
  \caption{Lund diagrams for the groomed jet mass distribution at LL
    with pruning in three different kinematic configurations. In
    case~(a), the emission that dominates the plain jet mass (and
    hence sets the pruning radius) also has $z>\zprune$.
    In cases~(b) and~(c), the emission that dominates the plain mass
    and sets the pruning radius has $z<\zprune$ and does not pass the
    pruning condition.
    Another emission at lower mass dominates the pruned jet mass. This
    emission can either be constrained by the condition $z>\zprune$,
    case~(b), or by the condition $\theta>r_\text{prune}$, case~(c).
    For each of the three cases, we indicate the contributions to Y-
    and I-pruning.}\label{fig:lund-pruning}
\end{figure}

\paragraph{All-order resummation.} 
To construct the all-order result, it is easier to consider the
differential jet mass distribution. Let us then denote by ``in'' the
emission that dominates the pruned jet, carrying a fraction $z_\text{in}$ of
the jet $p_t$ and emitted at an angle $\theta_\text{in}$, such that
$\rho=z_\text{in}\theta_\text{in}^2$.

The pruning radius in units of the original jet radius is given by
$\rprune^2=R_{\text{prune}}^2/R^2=m_{\text{jet}}^2/(p_{t,\text{jet}}R)^2$ which is set by the
emission dominating the plain jet mass.
We thus need to consider two cases: (i) there are no emissions in the
plain jet with $z\theta^2>z_\text{in}\theta_\text{in}^2$, (ii) there
is at least an emission in the plain jet with
$z\theta^2>z_\text{in}\theta_\text{in}^2$, and we call emission
``out'' the one with the largest $z\theta^2$, introducing
$\rho_\text{out}=z_\text{out}\theta_\text{out}^2$.
The corresponding Lund diagram is shown in
Fig.~\ref{fig:lund-pruning-1}.
In the first case, the pruning radius is set by emission one,
$r_{\text{prune}}^2=\rho<\theta_\text{in}^2$. To be in the pruned jet,
the ``in'' emission should therefore satisfy $z_\text{in}>\zprune$. We
get an associated Sudakov suppression $\exp(-R_\text{plain}(\rho))$
since we must veto emissions at larger mass than $\rho$ both in the
pruned jet and in the plain jet.
In the second case, the pruning radius is set by the ``out'' emission,
\ie $r_{\text{prune}}^2=\rho_\text{out}>\rho$. For
$\rho>\zprune\rho_{\text{out}}$, the pruning condition is then
$z_\text{in}>\zprune$ (shown in Fig.~\ref{fig:lund-pruning-2a}), while
for $\rho<\zprune\rho_{\text{out}}$ it becomes
$z_\text{in}>\rprune=\rho_\text{out}$
(Fig.~\ref{fig:lund-pruning-2b}).
The Sudakov receives two different contributions: one from inside the
pruning region, down to the scale $\rho$, represented by the red
shaded are in Fig.~\ref{fig:lund-pruning}, and one from outside the
pruning region, the grey area in Fig.~\ref{fig:lund-pruning}.
Note that since $\rho_\text{out}<\zprune$, the situation
$\rho<\zprune\rho_{\text{out}}$ only happens for $\rho<\zprune^2$,
yielding a transition point at $\rho=\zprune^2$.

For $\zprune^2<\rho<\zprune$, the sum over the two regions can be
written as
\begin{align}\label{eq:pruning-largerho-int}
  \frac{\rho}{\sigma}\frac{d\sigma_{\text{prune}}}{d\rho}
  &= \int_{\zprune}^1 dz_\text{in} P_i(z_\text{in})
    \frac{\alpha_s}{2\pi} e^{-R_\text{in}(\rho)}
    \bigg[
    e^{-R_\text{out}(\rho)}
    +\int_\rho^{\zprune}\frac{d\rho_\text{out}}{\rho_\text{out}}R'_\text{out}(\rho_\text{out})
    e^{-R_\text{out}(\rho_\text{out})}
    \bigg]\nonumber \\
  & = R'_\text{in}(\rho) e^{-R_\text{in}(\rho)}, \quad \text{with }\rho> \zprune^2,
\end{align}
where we have introduced the radiators
\begin{align}
  R_\text{in}(\rho) & = R_{\text{mMDT}}(\rho),\\
  R_\text{out}(\rho) & = R_{\text{plain}}(\rho)-R_{\text{mMDT}}(\rho),
\end{align}
where $R_{\text{mMDT}}$ is obtained by setting $\beta=0$ in Eq.~(\ref{eq:mMDTSD-radiator-modll}).
The radiators correspond respectively to the region kept (the shaded red area of
Fig.~\ref{fig:lund-pruning}) and rejected (the grey area of
Fig.~\ref{fig:lund-pruning}) by pruning. As long as the pruning
condition is only $z>\zprune$, as in the above case, the $R_\text{in}$
Sudakov is the same as the mMDT Sudakov,
Eqs.~(\ref{eq:mMDTSD-radiator}) and~(\ref{eq:mMDTSD-radiator-modll}).
The $R'_\text{out}$ factor in the first line of Eq.~(\ref{eq:pruning-largerho-int}) corresponds to the integral over the
momentum fraction of the emission outside the pruning region,
represented by the solid black line in Fig.~\ref{fig:lund-pruning}.
After integration, we find that the pruned jet mass distribution is
identical to the mMDT mass distribution for $\rho>\zprune^2$.

The situation for $\rho<\zprune^2$ is more involved as one now has to
include the situation from Fig.~\ref{fig:lund-pruning-2b} as well. In
that case, the $R_\text{in}$ Sudakov gets an additional contribution
and the lower bound of the $z_\text{in}$ integration extends down to
$\rprune$. We then write
\begin{align}\label{eq:pruning-smallrho-int}
  \frac{\rho}{\sigma}\frac{d\sigma_{\text{prune}}}{d\rho}
   = &\int_{\zprune}^1 dz_\text{in} P_i(z_\text{in})
    \frac{\alpha_s}{2\pi} e^{-R_\text{in}(\rho)}
    \bigg[
    e^{-R_\text{out}(\rho)}
    +\int_\rho^{\rho/\zprune}\frac{d\rho_\text{out}}{\rho_\text{out}}R'_\text{out}(\rho_\text{out})
    e^{-R_\text{out}(\rho_\text{out})}
    \bigg]\nonumber\\
  & + \int_{\rho/\zprune}^{\zprune}\frac{d\rho_\text{out}}{\rho_\text{out}}R'_\text{out}
    e^{-R_\text{out}(\rho)-R_\text{in}(\rho;\rho_\text{out})}
    \int_{\rho_\text{out}}^1 dz_\text{in} P_i(z_\text{in})
    \frac{\alpha_s}{2\pi} \nonumber \\
 =   & \, R'_\text{in}(\rho) e^{-R_\text{in}(\rho)-R_\text{out}(\frac{\rho}{\zprune})} 
  \nonumber \\
 &+ \int_{\rho/\zprune}^{\zprune}\frac{d\rho_\text{out}}{\rho_\text{out}}R'_\text{out}
    e^{-R_\text{out}(\rho)-R_\text{in}(\rho;\rho_\text{out})}
    \int_{\rho_\text{out}}^1 dz_\text{in} P_i(z_\text{in})
    \frac{\alpha_s}{2\pi}
\end{align}
with the new radiator
\begin{equation}
  R_\text{in}(\rho;\rho_\text{out})
  = \int_0^1 \frac{d\theta^2}{\theta^2}dz  P_i(z)\frac{\alpha_s}{2\pi}
  \Theta(z\theta^2>\rho) \Theta(z>{\text{min}}(\zprune,\rho_\text{out})),
\end{equation}
corresponding to the shaded red region of
Fig.~\ref{fig:lund-pruning-2b}.
Some simplifications and approximations can be done at fixed coupling
but the main message here is that at small $\rho$, $\rho<\zprune^2$,
the pruned mass distribution no longer involves a Sudakov which is the
simple exponentiation of the one-gluon-emission result. In that
region, one is left with an additional integration over the plain jet
mass $\rho_\text{out}$ which gives a Sudakov with double logarithms of
the pruned jet mass $\rho$.

\paragraph{Y-pruning and I-pruning.} The main complication of
pruning originates from the situation depicted in Fig.~\ref{fig:lund-pruning-2b},
where the pruning radius is set by an emission which is groomed away. 
and the pruned mass is dominated by an emission at an angle smaller
than the pruning radius. 
In this situation the prune radius is anomalously large because it is not set by hard splitting, as one would physically expect from pruning, especially when it is used as a two-prong tagger and the pruned jet is characterised by just one hard prong, hence the name I-pruning. 
Conversely, Y-pruning configurations are characterised by a hard $1\to 2$ splitting. 
It is therefore interesting to compute the jet mass for Y-pruning.

The situation of Fig.~\ref{fig:lund-pruning-2b} where the emission that
dominates the pruned jet mass always has $\theta<\rprune$ is of the
I-pruning type and does not contribute at all to Y-pruning.
This is already a great simplification since, for example, all the
expressions will now involve the simple $R_\text{in}(\rho)$ Sudakov
and no longer $R_\text{in}(\rho';\rho_\text{out})$.
Furthermore, for the cases where the emission setting the pruned mass
also sets the plain jet mass, Fig.~\ref{fig:lund-pruning-1}, we always
have $z_\text{in}>\zprune$ and
$\theta_\text{in}>\rprune=\sqrt{z_\text{in}}\theta_\text{in}$, meaning that
this situation is always of the Y-pruning type.

Unfortunately, there is a price to pay for the remaining contribution,
Fig.~\ref{fig:lund-pruning-2a} for which, as indicated on the figure,
one only gets a jet contributing to Y-pruning for smaller values of
$z_\text{in}$, namely for
$\zprune<z_\text{in}<\rho/\rho_\text{out}$.\footnote{This argument is
  not entirely true since even for $z_\text{in}>\rho/\rho_\text{out}$
  we could still have another emission with $z>\zprune$,
  $\theta>\rprune$ and $z\theta^2<\rho$. Such a contribution would
  only give terms proportional to $\alpha_s\log^2(\zprune)$ i.e.\ not
  enhanced by any logarithm of the jet mass. We therefore neglect
  these contributions here.}
Taking this into account, the Y-pruned jet mass distribution can
be written as  (assuming $\rho<\zprune$)
\begin{align}
  \frac{\rho}{\sigma}\frac{d\sigma}{d\rho}
   & = \int_{\zprune}^1 dz_\text{in} P_i(z_\text{in})
   \frac{\alpha_s}{2\pi} e^{-R_\text{plain}(\rho)}\\
   & + \int_\rho^{\text{min}(\zprune,\rho/\zprune)}\frac{d\rho_\text{out}}{\rho_\text{out}}R'_\text{out}(\rho_\text{out})
    e^{-R_\text{out}(\rho_\text{out})-R_\text{in}(\rho)}
     \int_{\zprune}^{\rho/\rho_\text{out}} dz_\text{in} P_i(z_\text{in}).\nonumber
\end{align}
Inverting the two integrations on the second line, one can perform
explicitly the integration over $\rho_\text{out}$ and keep only an
integration over $z_\text{in}$:
\begin{equation}\label{eq:Ypruning}
  \frac{\rho}{\sigma}\frac{d\sigma}{d\rho}
    =  \int_{\zprune}^1 dz_\text{in} P_i(z_\text{in})
   \frac{\alpha_s}{2\pi} e^{-R_\text{in}(\rho)-R_\text{out}({\text{min}}(\zprune,\rho/z_\text{in}))}.
 \end{equation}
 The Sudakov in the $z_\text{in}$ integrand has a few interesting
 properties. First, for $\rho/z_\text{in}>\zprune$, it involves
 $R_\text{out}(\zprune)=0$ and we recover a behaviour similar to what
 was seen for the mMDT. For $\rho/z_\text{in}<\zprune$, which is
 always the case for $\rho<\zprune^2$, $R_\text{out}$ then becomes
 double-logarithmic in $\rho$.

\subsection{Non-perturbative corrections in groomed distributions}\label{sec:groomed-mass-hadronisation}

In Sec.~\ref{sec:plain-mass-hadronisation} we have provided a rough
estimate of the value of the jet mass at which the distribution
becomes sensitive to non-perturbative physics.
It is instructive to study revisit that calculation and see how this non-perturbative transition point changes if grooming techniques are applied. 

We start by considering trimming. Assuming that we are in the
$\rho < \ftrim \rtrim^2$ region, the situation is analogous to the
plain jet mass and the mass $m$ at which one becomes sensitive to
non-perturbative effects is the same as
Eq.~(\ref{eq:mass-NP-estimate}) but with the jet radius substituted by
the trimming radius
\begin{equation}\label{eq:mass-NP-estimate-trimming}
m^2\simeq \frac{\muNP}{p_t \Rtrim} p_t^2 \Rtrim^2= \muNP p_t \Rtrim,
\end{equation}
where, compared to Eq.~(\ref{eq:mass-NP-estimate}), we have switched
to hadron-collider variables and used $p_t$ rather than $E_J$.
For pruning (both Y- and I-configuration), the non-perturbative
transition point is formally the same as the plain jet mass,
essentially because it is the latter that sets the pruning
radius. Note however, that the size of non-perturbative corrections
can differ with respect to the plain mass and one does expect pruning
to achieve a significant reduction.

For the \SD case we need a new calculation. We have to work out when
an emission of constant $\rho = z \theta^2$, and passing the \SD
condition, first crosses into the non-perturbative region
$ z \theta< \tilde{\mu}=\frac{\muNP}{p_tR}$.
This happens at the maximum allowed (rescaled) angle ($\theta=1$ for
the plain mass) which is determined by the \SD condition
$z=\zcut \theta^\beta$. We obtain
$\rho\simeq \tilde{\mu} \left(\tfrac{\tilde{\mu}}{\zcut}
\right)^\frac{1}{1+\beta}$, which implies
\begin{equation}\label{eq:non-perturative-mass-SD}
m^2\simeq \muNP^\frac{2 +\beta}{1+\beta} \zcut^\frac{-1}{1+\beta}(p_t R)^\frac{\beta}{1+\beta}.
\end{equation}
Compared to the plain jet mass case,
Eq.~(\ref{eq:mass-NP-estimate}), the (squared) mass at which
one becomes sensitive to non-perturbative effects is therefore smaller
by a factor $\big(\tfrac{\muNP}{\zcut p_tR}\big)^{\frac{1}{1+\beta}}$.
Once again, we note that the mMDT limit $\beta=0$ is particularly
intriguing as the $p_t$ dependence disappears from
Eq.~\eqref{eq:non-perturative-mass-SD}.

\subsection{Summary and generic overview}

\begin{table}
  \centering
  \begin{tabular}{r|cccccc}
    Groomer/  & transition & exponen- & largest & soft & non-global & non-pert\\
    tagger    &  points & -tiates &  logs & logs &  logs  & $m^2$ scale\\
    \hline
    Plain     & -                         & yes & $\alpha_s^nL^{2n}$ &  yes  & yes & $\muNP p_tR$\\
    mMDT      & $\zcut$                   & yes & $\alpha_s^nL^{n}$ & no  & no & $\muNP^2/\zcut$\\
    SoftDrop  & $\zcut$                   & yes & $\alpha_s^nL^{2n}$ & no  & no & $(\tfrac{\muNP^{2+\beta}(p_tR)^\beta}{\zcut}\big)^{\frac{1}{1+\beta}}$\\
    trimming  & $\ftrim$,$\ftrim\rtrim^2$ & yes & $\alpha_s^nL^{2n}$ & yes & yes & $\muNP p_t\Rtrim$\\
    pruning   & $\zprune$,$\zprune^2$     & no  & $\alpha_s^nL^{2n}$ & yes & yes & $\muNP p_tR$\\
    Y-pruning & $\zprune$                 & no  & $\alpha_s^nL^{2n-1}$
                                                & yes & yes & $\muNP p_tR$\\
    \hline
  \end{tabular}
  \caption{Summary of the basic analytic properties of taggers. Here,
    $L=\log(\rho)$. By soft logs we mean logarithmic contributions
    originating from soft emissions at finite angle. We note that \SD
    does retain soft/collinear contributions (hence the double
    logarithmic behaviour), while mMDT only keeps hard-collinear
    radiation. 
  }\label{tab:groomers-analytic-props}
\end{table}

To conclude this section on analytic calculations, we summarise the
basic analytic properties of the groomers/taggers in
Table~\ref{tab:groomers-analytic-props}.

A few key observations can be made.
\begin{itemize}
\item The modified MassDropTagger and SoftDrop groom soft radiations
  at all angular scales, \ie without stopping at a given subjet
  radius. This has the consequence that they are insensitive to soft
  gluon emissions at finite angles and have no non-global logarithms.
\item Another consequence of the absence of a subjet radius for mMDT
  and SD is that they are free of transition points beyond the one at
  $\rho=\zcut$. This is also the case of Y-pruning.
  Transition points can have subtle consequences in phenomenological
  applications and are therefore best avoided if possible.
  Furthermore, as we shall see explicitly in comparisons to Monte
  Carlo simulation in Sec.~\ref{sec:calc-groomed-mass-ps} below,
  for heavily boosted bosons these transition points can be around the
  electroweak scale and therefore have delicate side-effects when used
  in tagging boosted electroweak bosons.
\item the simple symmetry cut of the mMDT, independent on the emission
  angles, translates into a perturbative logarithmic series where
  there are no double logarithms and the leading contributions are
  single logarithms of the jet mass.
  Although this translates into a smaller Sudakov suppression of the
  QCD background, this has the advantage of being theoretically
  simple. This, together with the fact that the mMDT strongly reduces
  the sensitivity to non-perturbative effects (see
  Sec.~\ref{sec:calc-groomed-mass-mc} below), is why it is a tool
  with a great potential for precision physics at the LHC.
\item For many of the groomers we have studied, the resummed result
  has a simple structure where the one-gluon-emission expression
  simply exponentiates. The main exception to that is pruning which
  does not exponentiate. The situation is partially alleviated in the
  Y-pruning case.
\end{itemize}

%%========================================================================
\section{Comparison to Monte Carlo}\label{sec:calc-groomed-mass-mc}

Now that we have obtained resummed results it is instructive to
compare our findings to Monte Carlo simulations, which are ubiquitously used in phenomenology.
 We first do that at
leading order to explicitly test the appearance of logarithms of the
jet mass and check our control over the associated coefficients. We
then move to a comparison to parton-shower simulations. In
this case we will also discuss the impact of non-perturbative effects.

\subsection{Comparisons at leading order}\label{sec:calc-groomed-mass-mc-lo}

An simple test of the above substructure calculations is to verify
that they do reproduce the logarithmic behaviour of a
fixed-order calculation.
To this purpose, we can use the
Event2~\cite{Catani:1996jh,Catani:1996vz} generator.
 Although the program generates $e^+e^-$ collisions,
one can simulate quark jets of a given $p_t$ (at $y=\pi=0$) by
rotating the whole event so that the thrust axis (or, alternatively
the axis of the reference $q\bar q$ event generated by Event2) aligns
with the $x$ axis.
We then cluster the jets with the anti-$k_t$
algorithm~\cite{Cacciari:2008gp}\footnote{At leading order,
  ${\cal {O}}(\alpha_s)$, one could equivalently use any algorithm in
  the generalised-$k_t$ family.} with $R=1$ (cf.
Chapter~\ref{chap:jets-and-algs}).
We then apply any groomer to the resulting jets and measure the
groomed jet mass.
In practice, we have used mMDT with $\zcut=0.1$, \SD with
$\beta=2$ and $\zcut=0.1$ and trimming with $\ftrim=0.1$ and
$\Rtrim=0.2$.
In this section, we focus on the lowest non-trivial order of
perturbation theory, ${\cal{O}}(\alpha_s)$. since we need at least 2
partons in a jet if we want a non-zero mass, it is sufficient to
consider the real gluon emissions, \ie $e^+e^-\to q\bar
q g$ events.

\begin{figure}[t!]
  \includegraphics[width=0.5\textwidth]{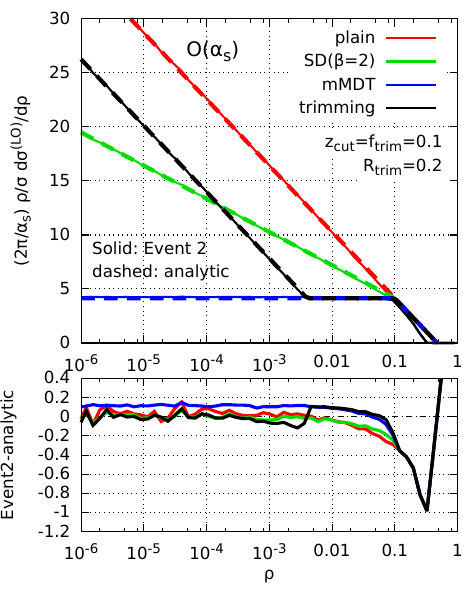}\hfill%
  \begin{minipage}[b]{0.42\textwidth}
    \caption{Comparison of the (normalised) mass distribution obtained
    at leading order, ${\cal{O}}(\alpha_s)$, between Event2 (solid
    lines) and our analytic expectations (dashed lines). The
    distribution is shown for both the plain jet (red) and a series of
    groomers: SoftDrop with $\beta=2$ (green), mMDT (blue) and
    trimming with $\Rtrim=0.2$ (black).
    The lower panel shows the difference between Event2 and the
    associated analytic expectation.
  }\label{fig:groomers-event2}
    \vspace*{1.4cm}
  \end{minipage}\hspace*{0.3cm}
\end{figure}

Fig.~\ref{fig:groomers-event2} shows the mass distribution for a few
selected groomers, together with the analytic calculations from above,
expanded at order $\alpha_s$. For \SD, this is given by
Eq.~\eqref{eq:sigma-SD-LO}, while for trimming by
Eq.~\eqref{eq:sigma-trimming-lo}. At ${\cal {O}}(\alpha_s)$, pruning
and Y-pruning coincide with the mMDT and are therefore not showed.
The bottom panel of the plot shows the difference between the Event2
simulations and the analytic results.

All the features discussed in this chapter are clearly visible on this
plot: the transition points, at $\rho=\zcut$ for
SD and at $\rho=\ftrim$ and $\rho=\ftrim\rtrim^2$ for trimming,
are present in the exact Event2 simulation; the effect of grooming
is clearly visible at small $\rho$, with a reduction of the
cross-section; the reduced $\log(\rho)$ contribution with \SD and
the absence of the $\log(\rho)$ enhancement for mMDT; the equivalence of
trimming and mMDT in the intermediate $\rho$ region; and the
reappearance of the plain-mass-like $\log(\rho)$ contribution at small
$\rho$ for trimming.

Comparing the asymptotic behaviour at small mass to our
analytic calculation, we first see that the leading logarithmic
behaviour, \ie the $\log(\rho)$ contribution, is correctly
reproduced. This is visible on the bottom panel of
Fig.~\ref{fig:groomers-event2} where all curves tend to a constant at
small $\rho$.
Furthermore, for trimming and \SD, the analytic calculation also
captures the constant term --- $B_q=-\tfrac{3}{4}$ coming from
hard-collinear branchings --- and the difference between Event2 and the
analytic results vanishes at small $\rho$.
Although it is a bit delicate to see it on the figure, in the case of the
plain, ungroomed, jet, this difference is  only going to a non-zero
constant at small $\rho$, because our calculation is missing a finite
$R^2$ contribution coming from the emission of a soft gluon at a large
angle.
Finally, in the case of the mMDT, this difference is clearly different from 0 at small
$\rho$. This originates from the fact that our analytic calculation in
Sec.~\ref{sec:calc-groomed-mass} has assumed $\zcut\ll 1$. For a
finite value of $\zcut$, one has to keep the full $z$ dependence in
the splitting function which, at ${\cal{O}}(\alpha_s)$ means
\begin{equation}
  \frac{\rho}{\sigma}\frac{d\sigma_{\text{mMDT}}}{d\rho}
   =
   \frac{\alpha_sC_F}{2\pi}\int_{\zcut}^{1-\zcut}dz\,\frac{1+(1-z)^2}{z}
   = \frac{\alpha_sC_F}{\pi}\bigg[\log\Big(\frac{1-\zcut}{\zcut}\Big)-\frac{3}{4}(1-2\zcut)\bigg].
\end{equation}
Finite $\zcut$ effects are of then given by
$\tfrac{\alpha_sC_F}{\pi}\left[ \tfrac{3}{2}\zcut-\log(1-\zcut)\right]$. Pulling out
an $\tfrac{\alpha_s}{2\pi}$ factor as done in Event2 and in
Fig.~\ref{fig:groomers-event2}, this gives a difference around 0.12
for our choice of $\zcut=0.1$, which corresponds to what is observed
on the plot.
  
\subsection{Comparisons with parton shower}\label{sec:calc-groomed-mass-ps}

\begin{figure}[t!]
  \subfloat[]{\includegraphics[width=0.48\textwidth,page=1]{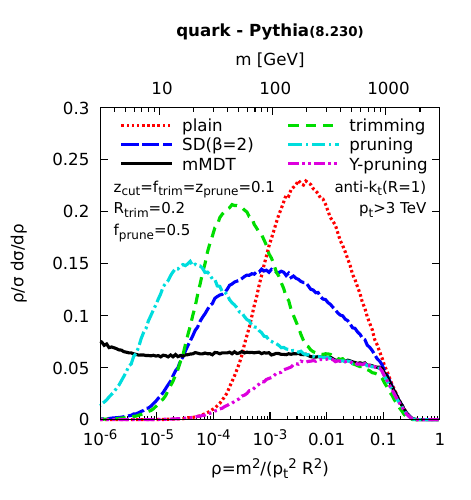}\label{fig:groomers-pythia}}%
  \hfill%
  \subfloat[]{\includegraphics[width=0.48\textwidth,page=1]{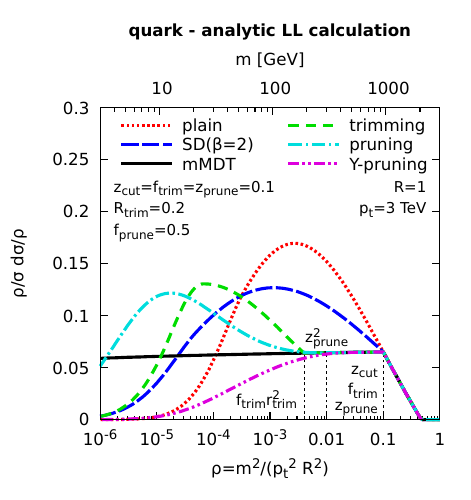}\label{fig:groomers-analytic}}
  \caption{Mass distribution obtained for the ungroomed jet (dotted,
    red) as well as with different groomers: SoftDrop($\beta=2$)
    (long-dashed, blue), mMDT (solid, black), trimming (short-dashed,
    green), pruning (dot-dashed, cyan) and Y-pruning (dot-dot-dashed,
    magenta). The left plot is the result of a Pythia parton-level
    simulation and the right plot is the analytic results discussed in
    this chapter.}\label{fig:groomers-pythia-v-analytic}
\end{figure}

\begin{figure}[t!]
  \subfloat[]{\includegraphics[width=0.48\textwidth,page=1]{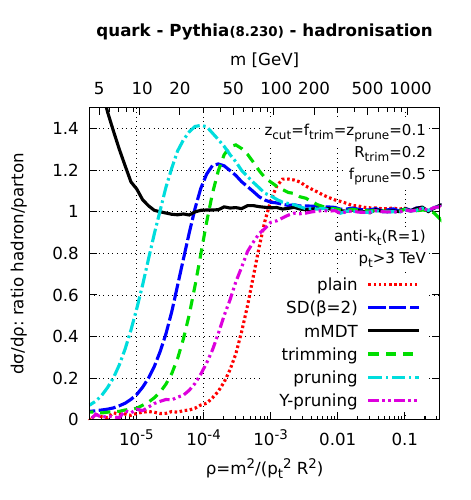}\label{fig:groomer-pythia-hadr}}%
  \hfill%
  \subfloat[]{\includegraphics[width=0.48\textwidth,page=2]{figures/groomed-rho-pythia-np.pdf}\label{fig:groomer-pythia-ue}}
  \caption{Non-perturbative effects on the groomed jet mass
    distribution. The lines are as in
    Fig.~\ref{fig:groomers-pythia-v-analytic}. All results are
    obtained from Pythia8 simulations. The left plot corresponds to
    hadronisation effects, \ie the ratio of hadron-level to
    parton-level distributions. The right plot shows the effects of the
    UE, \ie the ratio of the mass distribution with UE
    effects on and off.}\label{fig:groomers-pythia-npeffects}
\end{figure}

\paragraph{Setup.}
We now compare our all-order results, including running coupling, to a
full parton-shower simulation.
For this, we use the Pythia8~\cite{Sjostrand:2014zea} generator, in
its Monash13 tune~\cite{Skands:2014pea} at parton level. We generate
dijet events at $\sqrt{s}=13$~TeV, restricting the hard matrix element
to $qq\to qq$ processes.
Jets are reconstructed with the anti-$k_t$ algorithm, as implemented
in FastJet~\cite{Cacciari:2005hq,Cacciari:2011ma}, with $R=1$,
keeping only jets with $p_t>3$~TeV and $|y|<4$,
We study the same groomers as for the Event2 study, as well as pruning
and Y-pruning with $\zprune=0.1$ and $f_\text{prune}=0.5$.

\paragraph{Parton-level study.}
The distributions obtained from Pythia and the analytic results from
above are presented in Fig.~\ref{fig:groomers-pythia-v-analytic}. As
for the case of the fixed-order studies in the previous section, the
features observed in the parton-level simulation are very well
reproduced by the analytic results, including the various transition
points.
The Pythia distributions tend to be more peaked than what is predicted
from the analytic calculation, in particular in the regions where the
distributions have a large double-logarithmic contribution. This
effect would be (at least partially) captured by subleading, NLL,
contributions, and in particular by contributions from multiple
emissions which tend to increase the Sudakov and produce more peaked
distributions. The latter should be present in the Pythia simulation
but are absent from the above calculation.\footnote{They can easily be
  added to the ungroomed, \SD and trimming calculations. We have
  not done it here because it clearly goes beyond the scope of these
  lecture notes.}

Finally, we see in Fig.~\ref{fig:groomers-pythia-v-analytic} that for
heavily-boosted jets, the transition points of trimming and pruning
can be close to the electroweak scale. This is to keep in mind when
using substructure techniques to tag boosted electroweak bosons. 

\paragraph{Non-perturbative corrections.}
While the analytic calculations do a good job at reproducing the
features observed in a parton-level Pythia simulation, the jet mass
will also be affected by non-perturbative effects such as
hadronisation and the UE.
Ideally, we want these effects to be as small as possible to reduce
the dependence on model-dependent, tuned, aspects of soft physics,
which are not usually under good control and they can therefore obscure the partonic picture.

We therefore switch on non-perturbative effects in Pythia8 and
study how the reconstructed mass distributions are affected.
Fig.~\ref{fig:groomer-pythia-hadr} shows the effects of
hadronisation and it is obtained by taking
the ratio of the mass distribution with and without hadronisation
effects. 
Fig.~\ref{fig:groomer-pythia-ue} instead aims to study the impact of UE and it obtained by taking the ratio of the
distribution with and without multiple-parton interactions (but with
hadronisation).

Focusing first on UE effects, we clearly see the main idea behind
grooming at play: by removing  soft radiation at large angles, one
significantly reduces the sensitivity to the UE, whereas the plain jet
mass distribution shows a large distortion when this contribution is switched on.
Furthermore, while all the groomers show almost no sensitivity to the
UE at large mass ($\rho\gtrsim 0.002$ in
Fig.~\ref{fig:groomer-pythia-ue}), differences start to appear at
smaller masses.  Y-pruning shows a relatively large sensitivity to the UE
for $\rho\lesssim 0.002$, followed by pruning. This is likely due to
UE effects on the plain jet mass affecting the determination of the
pruning radius. Since the pruning radius will tend to be increased by
UE effects, jets that would perturbatively be deemed as Y-pruning will
fall in the I-pruning category once the UE is switched on. This is
expectably the main source behind the drop observed in the Y-pruning
curve in Fig.~\ref{fig:groomer-pythia-ue}.
For the other groomers, trimming shows a smaller sensitivity, \SD 
an even smaller one and the mMDT which is the most efficient at
grooming away soft radiation shows almost no
sensitivity to the UE.

This trend is similar when it comes to hadronisation corrections,
Fig.~\ref{fig:groomer-pythia-hadr}.
While all the groomed jet mass distributions show a significantly
smaller sensitivity to hadronisation than the plain jet mass distribution,
one sees potentially sizeable effects at small values of $\rho$.
As for the UE, Y-pruning shows the largest sensitivity
amongst the groomers and mMDT clearly exhibits
the smallest non-perturbative corrections.

Finally, by inspecting the mass scale on the upper horizontal axis, we note that 
for heavily boosted jets ($p_t=3$~TeV in this case) it is worth keeping in mind that the
non-perturbative effects can still be non-negligible around the
electroweak scale.

Note finally that some degree of analytic control over the
non-perturbative corrections to groomed jets can be achieved. This can
be done either qualitatively by inspecting the expected
non-perturbative scales to which each groomer is sensitive (see
\eg~\cite{Dasgupta:2013ihk}), or more quantitatively using analytic
models of hadronisation (see
\eg~\cite{Dasgupta:2007wa,Dasgupta:2013ihk,Marzani:2017kqd}).

%%========================================================================
\section{Calculations for signal jets}\label{sec:calc-groomed-mass-signal}

Thus far, we have only discussed the case of QCD jets, which are initiated by
high-energy quarks and gluons. Since the substructure tools discussed
above are used extensively in the context of tagging boosted bosons ---
either as prong finders or as groomers ---, it is also interesting to
discuss their behaviour for signal jets. Here, we will focus on
electroweak bosons decaying to a quark-antiquark pair, leaving the more complicated case of
the top quark aside.
Our goal here is to give a very brief overview of how the tools
discussed so far behave on signal jets. We will therefore only give
analytic results at leading-order and rely mostly on Monte Carlo
simulations to highlight the desired features associated with
parton-shower and non-perturbative effects. Some degree of analytic
calculation can be achieved for these effects as well but we will only
highlight their main features here.
More extensive analytic calculations, of both the perturbative and
non-perturbative contributions, can be found
in~\cite{Dasgupta:2015yua}.

\paragraph{Zeroth-order behaviour.}
At the lowest order in perturbation theory, we just have an
electroweak boson decaying to a $q\bar q$ pair. When this two-parton system
is passed to the groomer, the latter can either keep both partons in
the groomed jet, in which case the jet is kept/tagged as a signal jet,
or groom away one or both prongs in which case the jet is not tagger
as a signal jet. In this simple  situation, the signal efficiency ---
\ie the fraction of signal jets kept after applying the jet substructure algorithm --- is simply given
by the rate of jets for which the two partons are kept by the
groomer.
This can be written as
\begin{equation}\label{eq:grm-signal-base}
\epsilon_S^{\text{(tagger)}} = \int_0^1 dz\,P_X(z) \Theta^{\text{(tagger)}}(z),
\end{equation}
where $P_X(z)$ is the probability that the electroweak boson $X$
decays into a quark carrying a momentum fraction $z$ of the boson and
an anti-quark carrying a momentum fraction $1-z$ of the boson.
Crucially, the splitting function $P_X(z)$ does not exhibit the $1/z$
singularity at small $z$ which we have encountered in the QCD case. This is nothing but our
original argument that signal jets have a hard quark and a hard
anti-quark, while QCD jets are dominated by a hard parton emitting
soft gluons. Here, we will assume for simplicity a flat splitting
probability $P_X(z)=1$. This is correct for a heavily-boosted Higgs
boson but only approximate for W and Z. For the latter,
$P_\text{W/Z}(z)$ also depends on the polarisation of the boson. We refer the reader, for example, to
the discussion in Section~III.2.7 of~\cite{Bendavid:2018nar} for a
study of W polarisation in the context of jet substructure.

In Eq.~(\ref{eq:grm-signal-base}), $\Theta^{\text{(tagger)}}(z)$
denotes the action of the tagger on the $q\bar q$ pair.
For a massive object X of mass $m_\text{X}$, the decay angle is given by
$\theta^2=\tfrac{m^2}{p_t^2z(1-z)}$, or, again assuming that the
angles are normalised to the jet radius $R$,
$\theta^2=\tfrac{\rho}{z(1-z)}$.
The action of each tagger is then easy to write:
\begin{align}
\Theta^{\text{(plain)}}(z) & = \Theta(\theta<1),\nonumber\\
\Theta^{\text{(mMDT)}}(z) & = \Theta(\theta<1)\,\Theta(\text{min}(z,1-z)>\zcut),\nonumber\\
\Theta^{\text{(SD)}}(z) & = \Theta(\theta<1)\,\Theta(\text{min}(z,1-z)>\zcut\theta^\beta),\nonumber\\
\Theta^{\text{(trim)}}(z) & = \Theta(\theta<1)\,\Theta(\text{min}(z,1-z)>\zcut\text{ or }\theta<\rtrim),
\end{align}
with pruning and Y-pruning showing the same behaviour as the mMDT at
this order of the perturbation theory. Expressing $\theta$ as a
function of $z$, we can rewrite all the above constraints as a cut on
$z$ and find (up to subleading power corrections in $\rho$)
\begin{align}
\epsilon_S^{\text{(plain)}}(z) & = 1-2\rho,\nonumber\\
\epsilon_S^{\text{(mMDT)}}(z) & = 1-2\,\text{max}(\rho,\zcut),\nonumber\\
\epsilon_S^{\text{(SD)}}(z) & = 1-2\,\text{max}(\rho,\zcut (\rho/\zcut)^{\beta/(2+\beta)}),\nonumber\\
\epsilon_S^{\text{(trim)}}(z) & = 1-2\,\text{max}(\rho, \text{min}(\ftrim,\rho/\rtrim^2)).\label{eq:lo-grm-sig-eff}
\end{align}
These results show the same transition point as for the signal jet (at
least at the lowest order of perturbation theory). Except at low $p_t$
(or large mass), the mMDT (and (Y-)pruning) have a $\rho$-independent
behaviour, with $\epsilon_S=1-2\zcut$; the other taggers/groomers have
an efficiency going asymptotically to 1 like a power of $\rho$,
although in the case of trimming, this only happens at very small
$\rho$, $\rho\ll \zcut\rtrim^2$.

\begin{figure}[t!]
  \subfloat[]{\includegraphics[width=0.48\textwidth,page=1]{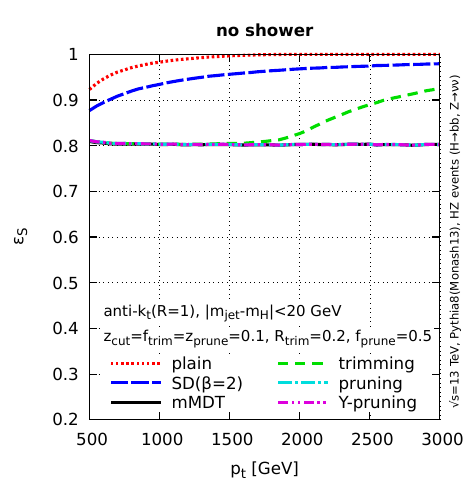}}%
  \hfill%
  \subfloat[]{\includegraphics[width=0.48\textwidth,page=1]{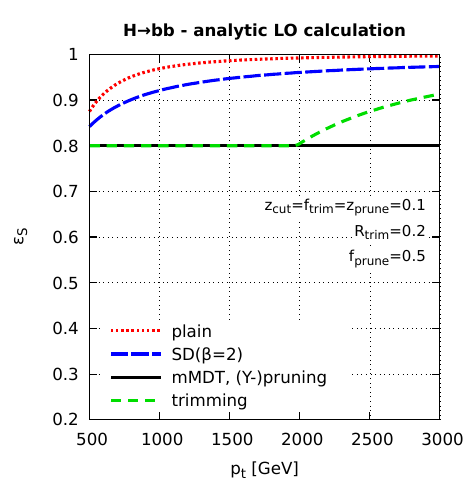}}
  \caption{Higgs reconstruction efficiency as obtained from Pythia8
    (left) and a LO analytic calculation (right). The Pythia8
    simulation is done at parton level with both the initial-state and
    final-state shower switched off. Different curves correspond to
    different taggers (see \eg
    Fig.~\ref{fig:groomers-pythia-v-analytic} for
    details).}\label{fig:groomers-pythia-sig-hard}
\end{figure}

We can compare these results to Monte Carlo simulations. 
For simplicity, we use the Pythia8 generator, simulating the
associated production of a Higgs and a Z boson, where the latter decays into (invisible)
neutrinos and the Higgs boson decays to a $b\bar b$ pair. We
reconstruct the jets using the anti-$k_t$ algorithm with $R=1$ and
select the hardest jet in the event, imposing a cut on the jet
$p_t$. The jet is then tagged/groomed and we deem the
jet as tagged if the jet mass after grooming is within
$\delta M=20$~GeV of the Higgs mass, \ie between 105 and 145~GeV,
with $m_H=125$~GeV.
We study the Higgs tagging efficiency as a function of the $p_t$ cut
applied to the initial jet.

To compare to the analytic results, Eq.~\eqref{eq:lo-grm-sig-eff}, we
simulate parton-level results switching off both the initial and
final-state showers.
Results are presented in Fig.~\ref{fig:groomers-pythia-sig-hard}
(left) together with our simple analytic results (right).
The analytic results capture very well the behaviour
observed in the Monte Carlo simulations. In particular, all the
features discussed above can be observed: the mMDT and (Y-)pruning
remain constant as a function of the jet $p_t$ and the efficiency of
the other taggers/groomers increases with $p_t$, with the plain jet
efficiency increasing more rapidly than the \SD one.
With our choices of parameters, the transition $\rho=\zcut$ (or
$\ftrim$) corresponds to a $p_t\approx 400$~GeV and is thus not
visible on the plot.
For trimming one sees the transition between the region dominated by
the $z>\ftrim$ condition at lower $p_t$ and the region dominated by
the $\theta<\rtrim$ condition at larger $p_t$. The transition between
the two regions happens at
$p_t=m_H/(R_\text{trim}\sqrt{\ftrim})\approx 2$~TeV, in agreement with
what is observed on the plot.

\begin{figure}[t!]
  \subfloat[]{\includegraphics[width=0.48\textwidth,page=2]{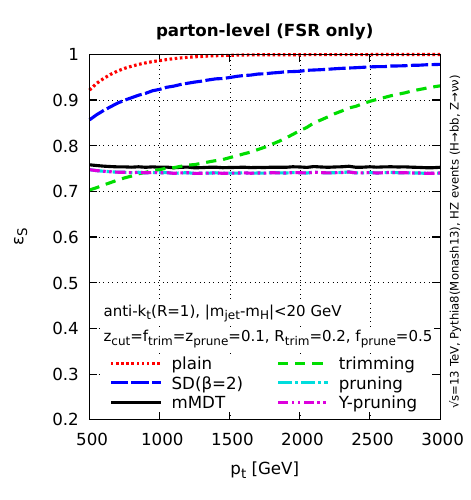}}%
  \hfill%
  \subfloat[]{\includegraphics[width=0.48\textwidth,page=6]{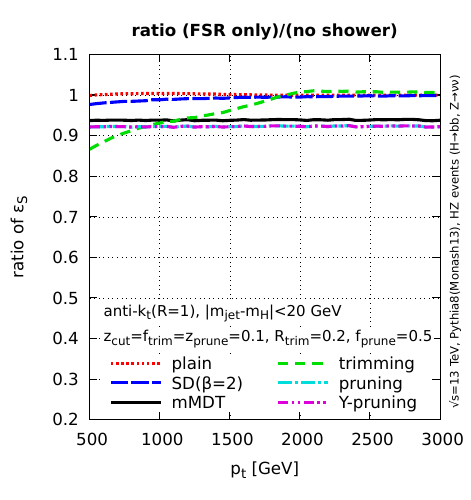}}
  \caption{Higgs reconstruction efficiency as obtained from
    Pythia8. The simulation is done at parton level, including only
    final-state radiation. The right plot shows the effects of
    final-state shower, \ie the ratio to the efficiencies obtained
    with no final-state shower. See
    Fig.~\ref{fig:groomers-pythia-sig-hard} for other
    details.}\label{fig:groomers-pythia-sig-fsr}
\end{figure}

\paragraph{Final-state radiation.}
We now move to consider the effects of final-state radiation (FSR) on
signal efficiency. The final-state gluons radiated by the $q\bar q$
pair can be groomed away, resulting in a
decrease of the reconstructed jet mass. The jet mass can therefore fall
below our lower cut $m_H-\delta M$ on the mass meaning that FSR is
expected to reduce the signal efficiency.
We know from our discussion of QCD jets in the previous sections that
the emissions in a final-state shower can have
logarithmically-enhanced effects on jet substructure observable. From
an analytic viewpoint, these emissions would then have to be resummed
to all orders.

While in practice it would be insightful to first consider the
${\cal {O}}(\alpha_s)$ case where a single gluon is emitted by the
$q\bar q$ pair --- similarly to what was done for the one-gluon
emission case for QCD jets at LO ---, we directly turn to the
situation where we include the full parton shower. We first discuss
the case of final-state radiation --- by the $q\bar q$ pair --- and
discuss initial-state radiation below.
We therefore run Pythia8 simulations, still at parton level, but this
time including final-state shower (and with the initial-state shower
still disabled).
The resulting efficiencies are plotted in
Fig.~\ref{fig:groomers-pythia-sig-fsr}.
If one focuses on the right-hand plot, showing the ratio of the
efficiencies obtained with final-state radiation to the efficiencies
obtained without, we see a relatively small effect of FSR for all the
substructure algorithms, even very small for the plain jet and \SD. This is not true for
trimming, for which the effect of FSR is to a large extend constant in $p_t$.
In the case of trimming, we see that at small $p_t$, more precisely
for $p_t<m_H/(R_\text{trim}\sqrt{\zcut})\approx 2$~TeV, \ie
$\rho>\zcut\rtrim^2$, the effect of FSR increases when
decreasing $p_t$.

From an analytic perspective, the emission of FSR gluons can come with
an enhancement proportional to $\log(\delta M^2/M_H^2)$ for a
small-width mass window, or a logarithm of $\zcut$, $\ftrim$ or
$\zprune$, all associated with soft gluon emissions. This is what
drives the $p_t$ independent loss of signal efficiency in the case of
mMDT and (Y-)pruning in Fig.~\ref{fig:groomers-pythia-sig-fsr}. For
the plain jet and \SD, this effect becomes suppressed by a power of
$M_H/p_t$.
Furthermore, in the case of trimming, due to the fixed trimming
radius, the effect of final-state radiation is also enhanced by
collinear logarithms of $\rho/\rtrim^2$ for $\rtrim^2\ll\rho\ll 1$,
\ie in the intermediate $p_t$ region. This logarithmically-enhanced
effect is the main reason for the slow rise of the trimming signal
efficiency between 500~GeV and 2~TeV.

\begin{figure}[t!]
  \subfloat[]{\includegraphics[width=0.48\textwidth,page=3]{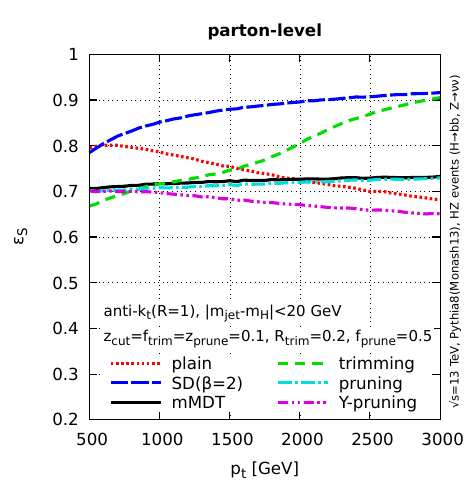}}%
  \hfill%
  \subfloat[]{\includegraphics[width=0.48\textwidth,page=7]{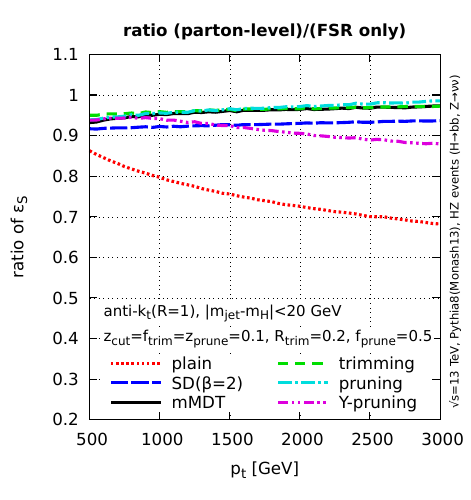}}
  \caption{Left: Higgs reconstruction efficiency obtained with Pythia8
    at parton level. Right: effects of initial-state radiation, \ie
    ratio to the efficiencies obtained with only final-state
    shower. See Fig.~\ref{fig:groomers-pythia-sig-hard} for other
    details.}\label{fig:groomers-pythia-sig-isr}
\end{figure}

\paragraph{Initial-state radiation.}
Next, we discuss the effect of initial-state radiation
(ISR). Compared to the case of FSR, capturing an ISR gluon in the
(groomed) jet shifts its mass up, meaning that it can go above
$M_H+\delta M$, again lowering the efficiency.
This effect is again potentially enhanced by a logarithm of
$\delta M^2$.
The results of our Monte Carlo study of ISR effects are presented in
Fig.~\ref{fig:groomers-pythia-sig-isr}, where we see a small effect for
mMDT, SoftDrop, trimming and pruning, a slightly larger effect for
Y-pruning and a sizeable loss of efficiency in the case of the plain
jet.

In the case of the plain jet mass, one does get an enhancement of ISR
effects by a logarithm of $M_H\,\delta M/p_t^2$, responsible for the
loss of signal efficiency when increasing $p_t$.
For groomed jets, one can show (see~\cite{Dasgupta:2015yua}) that this
logarithm is typically suppressed by a power of $M_H/p_t$ (related to
the fact that the groomed jet radius decreases with $p_t$) and is
replaced by a less harmful logarithm of $\zcut$, $\ftrim$ or $\zprune$
coming from situations where a large-angle ISR gluon passes the
grooming condition.
The case of Y-pruning is a bit more complex as even when an ISR
emission fails the pruning condition, it could have still affected
(increased) the pruning radius and cause the Y-pruning condition to
fail.  This is the main source of the decrease of the signal
efficiency observed for Y-pruning at large $p_t$ in
Fig.~\ref{fig:groomers-pythia-sig-isr}.

\begin{figure}
  \subfloat[]{\includegraphics[width=0.48\textwidth,page=4]{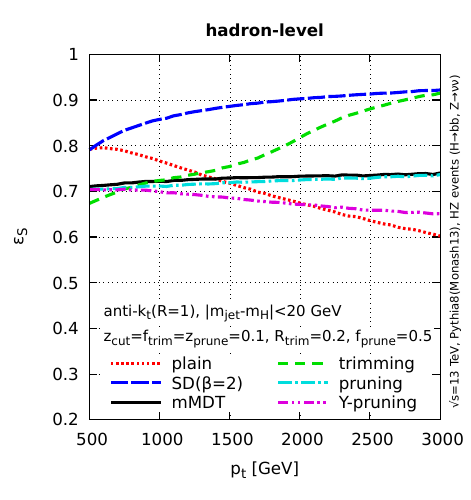}}%
  \hfill%
  \subfloat[]{\includegraphics[width=0.48\textwidth,page=8]{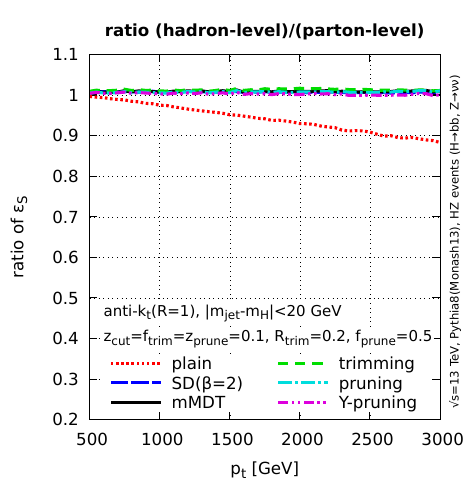}}
  \caption{Left: Higgs reconstruction efficiency obtained from
    Pythia8 at hadron level. Right: hadronisation effects, \ie ratio
    to parton-level efficiencies. See
    Fig.~\ref{fig:groomers-pythia-sig-hard} for
    details.}\label{fig:groomers-pythia-sig-hadr}
\end{figure}

\begin{figure}
  \subfloat[]{\includegraphics[width=0.48\textwidth,page=5]{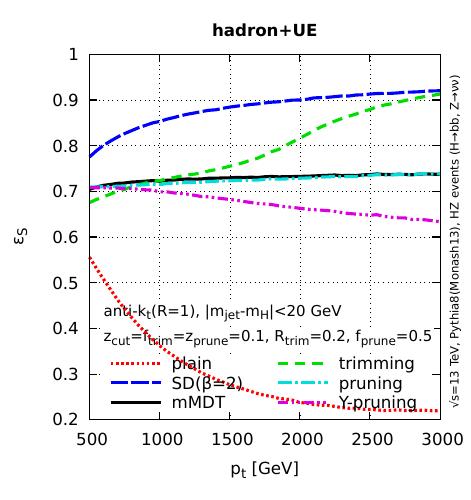}}%
  \hfill%
  \subfloat[]{\includegraphics[width=0.48\textwidth,page=9]{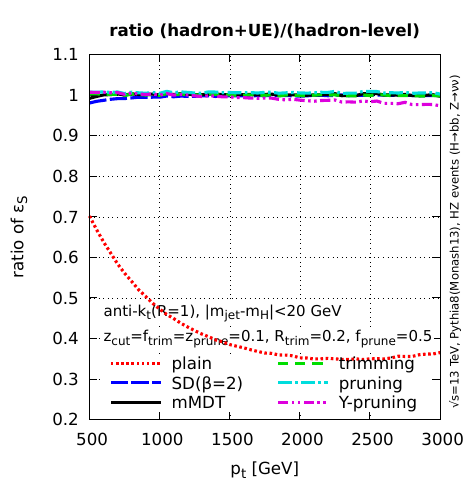}}
  \caption{Left: Higgs tagging efficiency obtained from a full
    Pythia8 simulation. Right: UE effects, \ie ratio to
    efficiencies with UE switched off. See
    Fig.~\ref{fig:groomers-pythia-sig-hard} for
    details.}\label{fig:groomers-pythia-sig-ue}
\end{figure}

\paragraph{Non-perturbative effects.} The effects of hadronisation and
of the UE are presented in
Figs.~\ref{fig:groomers-pythia-sig-hadr}
and~\ref{fig:groomers-pythia-sig-ue}, respectively.
Hadronisation corrections are generally small, especially for groomed
jets where they are almost negligible. In the case of the plain jet,
hadronisation effects tend to increase at large $p_t$ but the
correction remains within 10\%.
The case of UE corrections is more striking: the signal
efficiency in the case of the plain jet is severely affected by
UE contamination.
After grooming, the UE correction becomes very small
across the whole range of $p_t$ studied. This is directly related to
the initial idea behind grooming, namely to reduce soft contamination
--- and hence UE effects --- by removing soft and
large-angle emissions in the jet.

Once all effects are taken into account, the efficiency for groomed
jets is found to be close to the initial prediction at leading order,
with small corrections from ISR, FSR and non-perturbative
effects. Trimming has a small extra $p_t$ dependence at intermediate
$p_t$ coming from final-state radiation, and Y-pruning has a small
loss of signal efficiency at large $p_t$ due to initial-state
radiation.
This picture is contrasted by what happens in the case of the plain
jet where ISR and, in particular, the UE have a sizeable
effect, and hadronisation corrections are larger than for groomed
jets.
A consequence of this resilience of groomed jets is that, despite the
smaller signal efficiency at leading-order,
cf.~Fig.~\ref{fig:groomers-pythia-sig-hard}, the groomed jet signal
efficiency is clearly larger than the ungroomed signal efficiency once
all effects beyond LO are included.

%% GS helper for auctex
%%% Local Variables:
%%% mode: latex
%%% TeX-master: "notes"
%%% End:

%  LocalWords:  Eq eq NNLL parton's Eqs UE Monash NLL expectably FSR
%  LocalWords:  ISR

% $Id: calculations-quark-gluon.tex 531 2022-01-31 11:32:19Z smarzani $
%
% This contains calculations for quark-gluon discrimination
%------------------------------------------------------------------------

%%========================================================================
\chapter{Quark/gluon discrimination}\label{sec:calc-shapes-qg}

\begin{figure}
  \centerline{\includegraphics[width=0.75\textwidth]{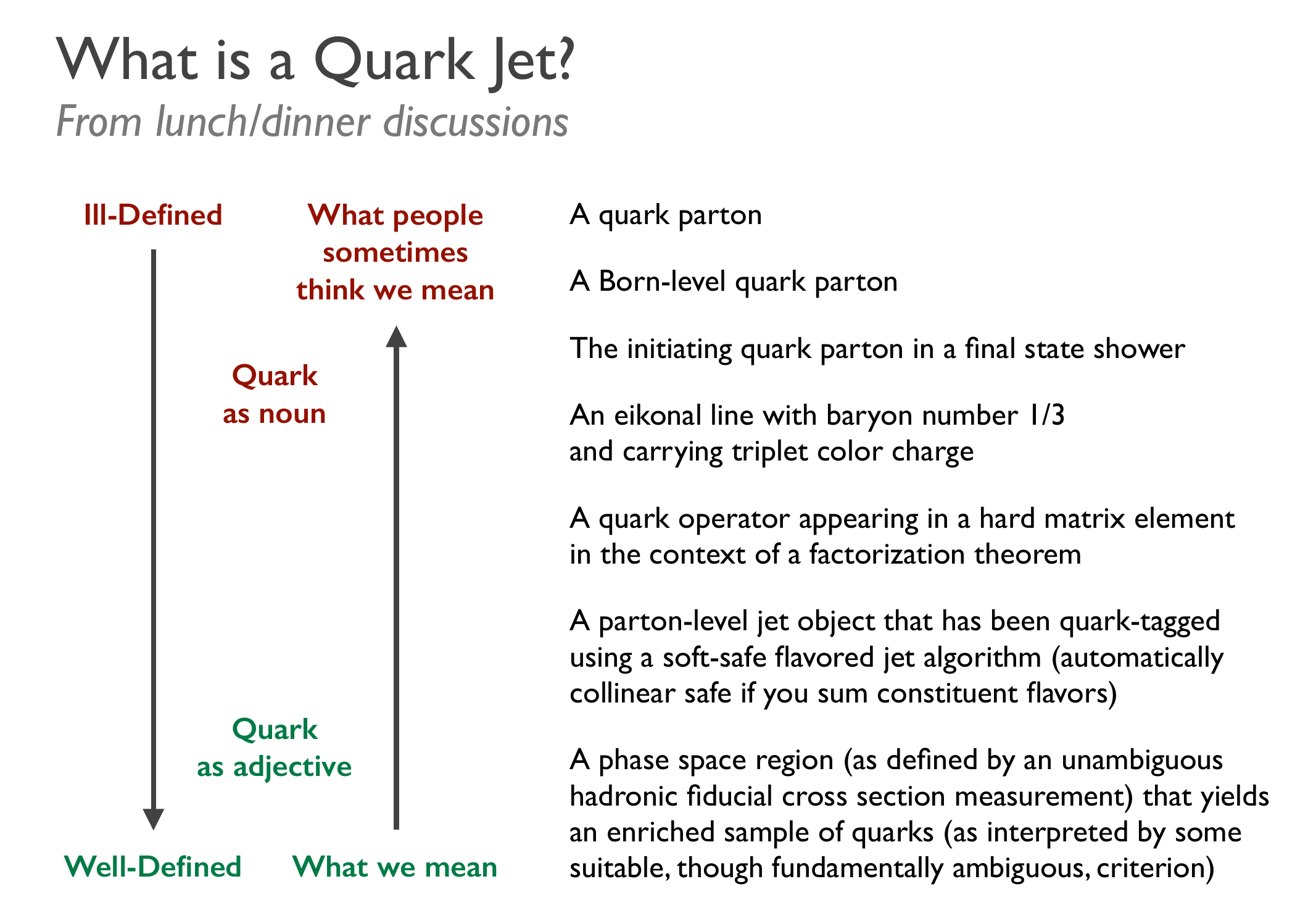}}
  \caption{Possible definitions of a ``quark jet'' or a
    ``gluon jet'' (from Ref.~\cite{Badger:2016bpw}, see also~\cite{Gras:2017jty}).}\label{fig:qgjet-defs}
\end{figure}

In this chapter we discuss the application of jet substructure tools
for discriminating between quark- and gluon-initiated jets.
Before digging into the substructure aspects of the matter, let us
briefly mention that there are many ways to define what a ``quark
jet'' or a ``gluon jet'' is. Several possibilities are listed in
Fig.~\ref{fig:qgjet-defs}. Amongst these possible definitions, many
are clearly pathological, simply because a parton is not a physically
well-defined object (cf. also our discussion about jets in
Chapter~\ref{chap:jets-and-algs}).
What is well-defined is a measurable quantity, that one can associate
(in an inevitably ambiguous way) to an enriched sample of quarks or
gluons.
For simplicity, we often rely on event samples involving hard quarks
or gluons in the Born-level process, but one has to be aware that this
is not unambiguously defined approach and keep this in mind when interpreting
the results.
This is what we have already done in the previous chapter when
generating $qq\to qq$ Pythia8 events as a proxy for quark jets and
this is again what we will do here.
Note that the better-defined definition in Fig.~\ref{fig:qgjet-defs}
depends on which sample is used. An investigation of this dependence
can be found in~\cite{Bright-Thonney:2018mxq}.

That said, several processes one wants to measure at the LHC, like
Higgs production through vector-boson-fusion, or new-physics events, such as cascades of supersymmetric particles, tend to
produce quark jets while QCD backgrounds are gluon-dominated. This
motivates the use of substructure tools to try and discriminate between
the two.
Some years ago, a wide range of discriminants
has been systematically studied and
compared~\cite{Gallicchio:2012ez}.
It is not our goal to go through all the details of this
study. Instead, we have selected a few representative discriminators
and discussed their performance and their basic analytic properties.
We focus on two main categories of tools: jet shapes, namely
angularities and energy-correlation functions, and multiplicity-based
observables, namely the iterated \SD multiplicity.
We conclude this chapter with a comparison of their performance (in
the sense of Sec.~\ref{sec:performance-intro}) using Monte Carlo
simulations.

Our Monte Carlo studies use Pythia8 (with the Monash13 tune). We
generate ``quark-initiated jets'' using $qg\to \text{Z}q$ hard matrix
elements and ``gluon-initiated jets'' using $q\bar q\to \text{Z}g$
events. In both cases, the Z boson is made to decay into invisible
neutrinos and we focus on the hardest anti-$k_t$($R=0.5$) jet in the
event requiring $p_t>500$~GeV.

\section{Angularities, ECFs and Casimir scaling}

The motivation behind using jet shapes for quark-gluon discrimination
is the observation that gluons tend to radiate more than quarks and jet shapes are
precisely a measure of this radiation.
Typical examples of shapes that can be used in this context are the
angularities $\lambda_\alpha$ and the energy-correlation functions (ECFs)
$e_2^{(\alpha)}$, introduced in
Sec.~\ref{sec:tools-radiation-constraints}.
In both cases, one would expect a larger value of
$v=\lambda_\alpha,e_2^{(\alpha)}$ for gluon jets than for quark jets
and one can build an enhanced quark sample by simply imposing a cut
$v<v_\text{cut}$.

We will first perform some analytic calculations for angularities and ECFs,
before discussing their performance as quark-gluon separators. We will
come back to this in Sec.~\ref{sec:qg-perf-robustness}, where we
also discuss their robustness against non-perturbative effects.

\paragraph{Analytic behaviour.}
For the purpose of the physics discussion we want to have, we will
need a resummed calculation at NLL accuracy.
At this accuracy, angularities and ECFs have the same structure,
provided one uses a recoil-insensitive jet axis definition
for angularities with $\alpha\le 1$.
This is easy to explain from a simple one-gluon emission argument (cf.\ \eg
Fig.~\ref{fig:mmdt-mass-one-gluon}). If $\theta$ denotes the angle
between the emitted soft gluon and the recoiling hard parton, a
standard four-vector recombination scheme, e.g.\ the E-scheme, would give an angle
$(1-z)\theta$ between the soft gluon and the jet axis and an angle
$z\theta$ between the recoiling hard parton and the jet axis.  This
gives
\begin{equation}
\lambda_\alpha^\text{(E-scheme)}=z[(1-z)\theta]^\alpha + (1-z)[z\theta]^\alpha=[z(1-z)^\alpha+(1-z)z^\alpha]\theta^\alpha,
\end{equation}
where the first contribution comes from the soft gluon and the second
from the recoiling parton.
For $\alpha=1$, both partons contribute equally to give
$\lambda_1^\text{(E-scheme)}=2z(1-z)\theta\approx 2z\theta$. This
leaves the LL behaviour unaffected but introduces recoil effects at
NLL (with a resummation structure more complex than the simple
exponentiation in~(\ref{eq:Sigma-plain-jet-mass}).
For $\alpha<1$, $\lambda_\alpha^\text{(E-scheme)}
\approx z^\alpha\theta^\alpha$ dominated by the recoil of the hard
parton, so recoil effects are already present at LL.
If we use the winner-takes-all (WTA) axis --- what we did in
practice in our Monte Carlo simulations --- angularities become
recoil-free and we have
\begin{equation}
  \lambda_\alpha^\text{(WTA)}=z\theta^\alpha.
\end{equation}
This effect is not present for ECFs for which we have
$e_2^\alpha = z(1-z)\theta^\alpha\overset{z\ll
  1}{\approx}z\theta^\alpha$, independently of the recombination
scheme.

For $\alpha=2$, angularities and ECFs are essentially equivalent to
the mass --- more precisely $m^2/(p_tR)^2$ --- and we can reuse the
same results as in Chapter~\ref{chap:calculations-jets}. 
These results can almost trivially be extended to a generic value of
the angular exponent $\alpha$. First, we need expressions for the
radiators valid at NLL. This requires including the two-loop
running-coupling corrections in the CMW scheme (see the discussion
before Eq.~\eqref{eq:radiator-nll-expansion}).
For the plain jet, one finds a generalisation of Eqs.~(\ref{eq:quark})
and~\eqref{eq:radiator-nll-contribution}:
\begin{align}\label{eq:plain-angularity-radiator-nll}
  R_\text{plain}^{\text{(NLL)}}(v)
   = \frac{C_i}{2\pi\alpha_s\beta_0^2}&\Bigg\{
    \bigg[\frac{1}{\alpha-1}W(1-\lambda)-\frac{\alpha}{\alpha-1}W(1-\lambda_1)+W(1-\lambda_B)\bigg]\\
  & +\frac{\alpha_s\beta_1}{\beta_0}\bigg[
\frac{1}{\alpha-1}V(1-\lambda)-\frac{\alpha}{\alpha-1}V(1-\lambda_1)+V(1-\lambda_B)
    \bigg]\nonumber\\
    & -\frac{\alpha_sK}{2\pi}\bigg[
\frac{1}{\alpha-1}\log(1-\lambda)-\frac{\alpha}{\alpha-1}\log(1-\lambda_1)+\log(1-\lambda_B)
    \bigg]\Bigg\},\nonumber
\end{align}
where $W(x)=x\log(x)$, $V(x)=\tfrac{1}{2}\log^2(x)+\log(x)$ and we have introduced 
\begin{equation}
  \lambda = 2\alpha_s\beta_0\log(1/v),\qquad
  \lambda_B = -2\alpha_s\beta_0 B_i,\qquad\text{and}\quad
  \lambda_1  = \frac{\lambda+(\alpha-1)\lambda_B}{\alpha}.
\end{equation}
Before discussing these results, let us point out that one can also
apply grooming to the jet, using mMDT or \SD, and compute the shape on
the groomed jet. In this case, we get the same as 
Eq.~(\ref{eq:plain-angularity-radiator-nll}) for $v>\zcut$ and a generalisation of
Eq.~(\ref{eq:mMDTSD-radiator-modll}) for $v< \zcut$:
\begin{align}\label{eq:sd-angularity-radiator-nll}
  \MoveEqLeft[1] R_\text{mMDT/SD}^{\text{(NLL)}}(v) = \\
  & = \frac{C_i}{2\pi\alpha_s\beta_0^2}\Bigg\{
    \bigg[\frac{(\alpha+\beta)W(1-\lambda_2)}{(\beta+1)(\alpha-1)}-\frac{\alpha
    W(1-\lambda_1)}{\alpha-1}-\frac{W(1-\lambda_c)}{\beta+1}+W(1-\lambda_B)\bigg]\nonumber\\
  & \phantom{=\frac{C_i}{2\pi\alpha_s\beta_0^2}}
    +\frac{\alpha_s\beta_1}{\beta_0}\bigg[
\frac{(\alpha+\beta)V(1-\lambda_2)}{(\beta+1)(\alpha-1)}-\frac{\alpha
    V(1-\lambda_1)}{\alpha-1}-\frac{V(1-\lambda_c)}{\beta+1}+V(1-\lambda_B)    \bigg]\nonumber\\
  & \phantom{=\frac{C_i}{2\pi\alpha_s\beta_0^2}}
    -\frac{\alpha_sK}{2\pi}\bigg[
\frac{(\alpha+\beta)\log(1-\lambda_2)}{(\beta+1)(\alpha-1)}-\frac{\alpha
    \log(1-\lambda_1)}{\alpha-1}-\frac{\log(1-\lambda_c)}{\beta+1}+\log(1-\lambda_B) \bigg]\Bigg\},\nonumber
\end{align}
with 
\begin{equation}
 \lambda_c = 2\alpha_s\beta_0\log(1/\zcut),
 \qquad\text{and}\quad
 \lambda_2 = \frac{(\beta+1)\lambda+(\alpha-1)\lambda_c}{\alpha+\beta}.
\end{equation}

These expressions require a few comments. First of all,
Eq.~(\ref{eq:plain-angularity-radiator-nll}), for $\alpha=2$, slightly
differs from Eqs.~(\ref{eq:quark})
and~\eqref{eq:radiator-nll-contribution}.
The difference is in the treatment of the $B$ term which corresponds
to hard collinear splittings where, as in
Chapter~\ref{calculations-substructure-mass}
(cf.~(\ref{eq:mMDTSD-radiator-modll-altB})), we have inserted the
contribution from hard-collinear splittings in the double-logarithmic
terms (see also Appendix~\ref{chap:app-analytic-details} for a
discussion on how to do this in practice).
One can also notice that the limit $\beta\to\infty$
of~Eq.~(\ref{eq:sd-angularity-radiator-nll}) gives back
Eq.~(\ref{eq:plain-angularity-radiator-nll}) as expected.
Furthermore, taking $\alpha=2$ in the mMDT/\SD case and neglecting the
two-loop corrections, one recovers
Eq.~(\ref{eq:mMDTSD-radiator-modll-altB}).
Finally, we note that, although the above results have factors of $\alpha-1$ in the
denominator, they are finite for $\alpha\to 1$ (corresponding to the
specific case of broadening or girth for angularities). 

Given the above radiators, we can compute the probability that the
angularity (or ECF) has a value smaller than $v$, i.e.\ the cumulative distribution, at NLL:
\begin{equation}\label{eq:angularities-nll-cumul}
  \Sigma^\text{(NLL)}(v) = \frac{e^{-R(v)-\gamma_E R'(v)}}{\Gamma(1+R'(v))},
\end{equation}
where the factor $e^{-\gamma_E R'(v)}/[\Gamma(1+R'(v))]$ accounts for
multiple emissions (cf.~(\ref{eq:mult-emission-nll})), and $R'(v)$ is
the derivative of $R(v)$ with respect to $\log(1/v)$. Since the multiple-emission
correction is already subleading, $R'$
in~\eqref{eq:angularities-nll-cumul} can be computed from the LL terms
in $R$ and we get (again, keeping the $B$ term only to guarantee an
endpoint at $\log(v)=B_i$)\footnote{In practice, this definition of
  $R'_\text{mMDT/SD}(v)$ introduces a discontinuity in the
  differential distribution at $v=\zcut$. This discontinuity is
  strictly-speaking subleading and can be avoided by defining $R'$
  using a finite-difference derivative:
  $R'(v)=[R(v e^{-\Delta})-R(v)]/\Delta$, with $\Delta$ a constant
  number, which respects NLL accuracy
  (see~\cite{Larkoski:2014wba}). This is what we have done for the
  results presented below, using $\Delta=0.5$.}
\begin{align}\label{eq:angularities-Rp}
R'_\text{plain}(v) & = \frac{C_i}{\pi\beta_0}\frac{1}{\alpha-1}\log\bigg(\frac{1-\lambda_1}{1-\lambda}\bigg),\\
R'_\text{mMDT/SD}(v) & = \frac{C_i}{\pi\beta_0}\frac{1}{\alpha-1}\log\bigg(\frac{1-\lambda_1}{1-\lambda_2}\bigg).
\end{align}
Note finally that while Eq.~\eqref{eq:angularities-nll-cumul} is only
correct in the small jet radius limit and should include soft
wide-angle emissions and non-global logs to reach full NLL accuracy,
Eq.~\eqref{eq:angularities-nll-cumul} includes all the NLL
contributions for Soft-Dropped angularities which are insensitive to soft
wide-angle emissions.

\begin{figure}
  \includegraphics[width=0.48\textwidth,page=1]{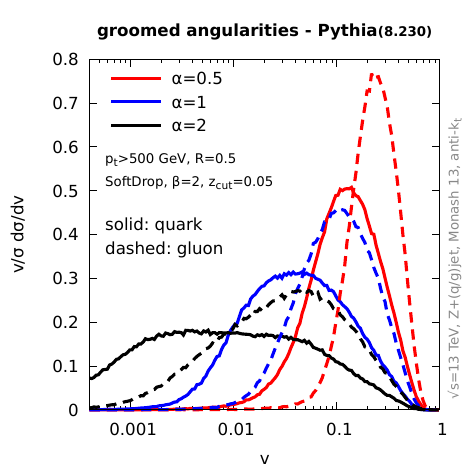}%
  \hfill%
  \includegraphics[width=0.48\textwidth,page=1]{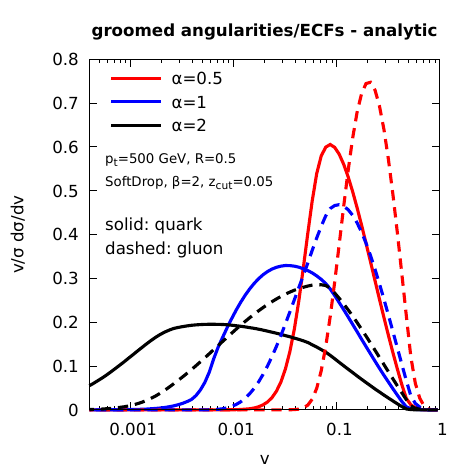}%
  \caption{Distribution of a sample of groomed angularities for quark
    (solid lines) and gluon (dashed lines) jets. The left plot
    corresponds to parton-level Pythia simulations and the right plot
    to the analytic results obtained in these lecture
    notes.}\label{fig:grm-ang-v-alpha}
\end{figure}  

\begin{figure}
  \includegraphics[width=0.48\textwidth,page=2]{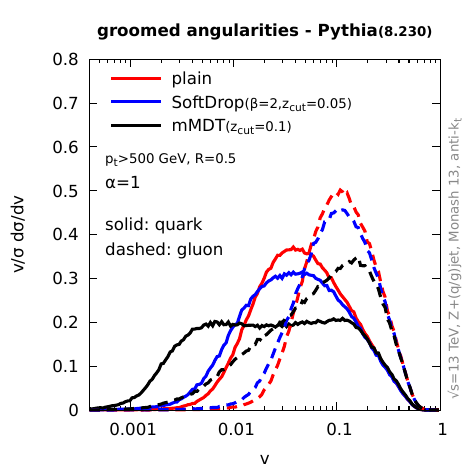}%
  \hfill%
  \includegraphics[width=0.48\textwidth,page=2]{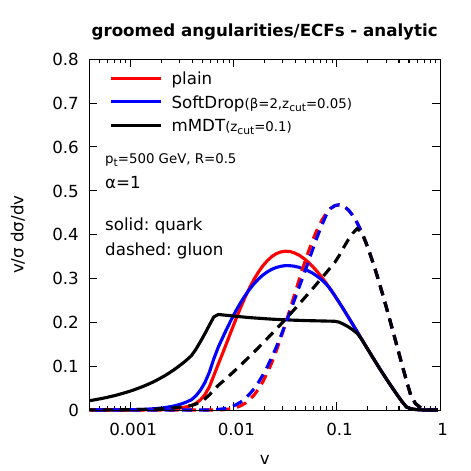}%
  \caption{Same as Fig.~\ref{fig:grm-ang-v-alpha}, this time for a
    fixed angularity $\lambda_1$, varying the
    groomer.}\label{fig:grm-ang-v-beta}
\end{figure}  

\paragraph{Comparison to Monte Carlo.}
A comparison between the above
analytic predictions and parton-level Monte Carlo simulations are
shown in Fig.~\ref{fig:grm-ang-v-alpha}, for different values of the angularity
exponent for \SD jets, and in Fig.~\ref{fig:grm-ang-v-beta}, for different
levels of grooming for $\lambda_{\alpha=1}$.
Overall, we see that there is a good agreement between the analytic
calculation and the Monte Carlo simulations.
We recall that our resummed calculation should not be trusted in the
region of large $v$ where an exact fixed-order calculation would be
needed. This could be obtained from NLO Monte Carlo generators like
NLOJet++~\cite{Nagy:2003tz} for dijet hard processes (here one would
need a 3-jet NLO calculation for the angularity distribution) and
MCFM~\cite{Campbell:1999ah,Campbell:2011bn,Campbell:2015qma} for
W/Z+jet events (here we would need W/Z+2 jets at NLO for the
angularity distribution).
The NLO distributions could then be matched to the resummed calculation
to obtain a final prediction which is valid at the same time in the
resummation-dominated region (small angularity) and in the
fixed-order-dominated region (large angularity).
More importantly, Figs.~\ref{fig:grm-ang-v-alpha}
and~\ref{fig:grm-ang-v-beta} show the expected clear separation
between the quark and gluon samples, with smaller values of the
angularity for the quark jets.

\begin{figure}
  \centering
  \subfloat[]{\includegraphics[width=0.48\textwidth,page=1]{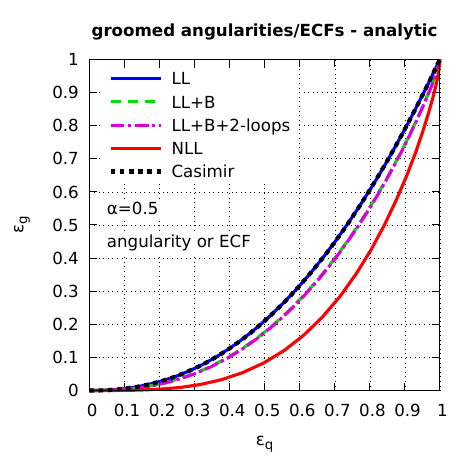}\label{fig:grm-ang-casimir}}%
  \hfill%
  \subfloat[]{\includegraphics[width=0.48\textwidth,page=3]{figures/groomed-angularities-roc-pythia.pdf}\label{fig:grm-ang-ptdep}}%
  \caption{Left: analytic predictions for the quark-gluon separation
    ROC curve using different approximations. Right: ROC curve for
    different values of $p_t$, shown for both Pythia8 simulations
    (solid) and our analytic calculation (dashed).}\label{fig:grm-ang-basicprops}
\end{figure}

\paragraph{Quark-gluon discrimination and Casimir scaling.}
With the above results at hand, we can finally discuss the performance
of angularities and energy-correlation functions to separate quark
jets from gluon jets. This is simply done by imposing a cut
$v<v_\text{cut}$ on angularities or ECFs.
On the analytic side, the quark and gluon efficiencies are therefore
directly given by $\Sigma_{q,g}$  computed above.
An interesting behaviour emerges from these analytic results. If we
look at Eqs.~(\ref{eq:plain-angularity-radiator-nll})
and~(\ref{eq:sd-angularity-radiator-nll}) at leading-logarithmic
accuracy, the only difference between quark and gluon jets is the
colour factor --- $C_F$ for quarks and $C_A$ for gluons, in front of
the radiators. This means that we have
\begin{equation}\label{eq:casimir-scaling}
\epsilon_g \overset{\text{LL}}{=} (\epsilon_q)^{C_A/C_F}.
\end{equation}
This relation is often referred to as {\em Casimir scaling}~(see
\eg~\cite{Larkoski:2013eya}). This means that the leading behaviour of
quark-gluon tagging will follow Eq.~(\ref{eq:casimir-scaling}) regardless
of the angularity (of ECF) exponent and of the level of grooming.

Departures from Casimir scaling will start at NLL accuracy. In our
collinear/small-$R$ limit, these means that there can be three sources of
Casimir-scaling violations: hard-collinear corrections (the $B$ term),
two-loop running-coupling corrections, and multiple-emissions
(cf. Eq.~(\ref{eq:angularities-nll-cumul})). 
Of these three effects, only the first and the last give scaling
violations since two-loop running coupling corrections are also simply
proportional to $C_i$.
This is illustrated in Fig.~\ref{fig:grm-ang-casimir}, where we see
that the LL result gives perfect Casimir scaling, and the inclusion of
the hard collinear splitting and the multiple-emission corrections
both slightly increase the quark-gluon discrimination performance.
The correction due to the $B$-term is proportional to $B_g-B_q$ which
is small and positive. The effect of multiple emissions starts at
${\cal {O}}(\alpha_s^2)$ in the perturbative expansion and is
proportional to $(C_A-C_F)$.
In practice, this last effect appears to have the largest impact.
A direct consequence of Casimir scaling is that the quark-gluon
discriminative power remains relatively independent of the jet $p_t$
as shown on Fig.~\ref{fig:grm-ang-ptdep} for both our analytic
calculation (dashed lines) and Pythia8 parton-level simulations (solid
lines).

\begin{figure}
  \includegraphics[width=0.48\textwidth,page=1]{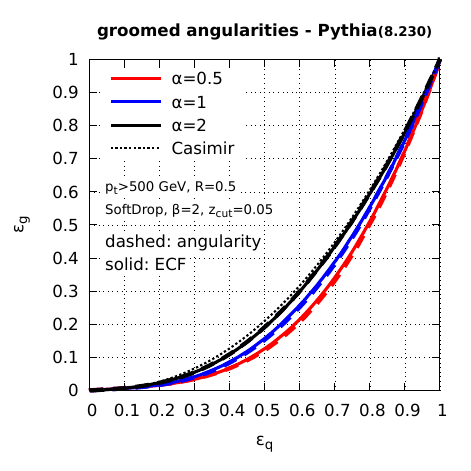}%
  \hfill%
  \includegraphics[width=0.48\textwidth,page=1]{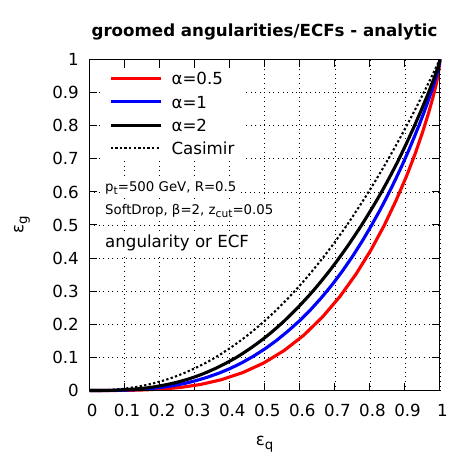}%
  \caption{ROC curves for quark/gluon separation using ECFs (solid lines) and angularities (dashed lines). The left plot
    corresponds to parton-level Pythia simulations and the right plot
    to the analytic results obtained in these lecture
    notes.}\label{fig:grm-ang-roc-v-alpha}
\end{figure}

All these effects are discussed at length in
Ref.~\cite{Larkoski:2013eya} and we refer the reader to this
discussion for further details.
ROC curves for quark-gluon discrimination
are shown in Fig.~\ref{fig:grm-ang-roc-v-alpha} for both Pythia (at
parton level) and our analytic calculation, for jets groomed with \SD.
We see a good level of agreement between the two although the
analytic results tend to produce a slightly larger quark-gluon
discrimination that Pythia.
It is however notorious that different Monte Carlo generators tend to predict
relatively different deviations from Casimir scaling, both at parton and hadron level. We refer to Ref.~\cite{Gras:2017jty} for more
details about this. 
Above all, we conclude from Fig.~\ref{fig:grm-ang-roc-v-alpha} that smaller
values of $\alpha$ give better discrimination, with very similar
results obtained for angularities and energy-correlation functions.
We will come back to this in Sec.~\ref{sec:2prongs-perf-robustness}
when discussing the performance and robustness of quark-gluon
discriminators.

\section{Beyond Casimir scaling with Iterated SoftDrop}\label{sec:isd}

Given the observation made in the previous section that angularities
and energy-cor\-relation functions produce quark-gluon discriminators
which depart from Casimir scaling only due to subleading corrections,
it is natural to wonder if it is possible to find substructure tools
which have a different behaviour already at leading-logarithmic
accuracy.

The behaviour one would want to obtain is a Poisson-like
behaviour like what the particle multiplicity in a jet, or the
charged-track multiplicity, typically achieve.
In this section, we discuss a tool, namely the {\em Iterated SoftDrop (ISD)
  multiplicity} introduced in Sec.~\ref{sec:def-other-shapes} and
show that it achieves a Poisson-like behaviour already at LL while
remaining infrared-and-collinear safe (contrary to particle or
charged-track multiplicity).
As above, we will first briefly discuss the analytic structure of ISD
multiplicity and compare the resulting performance with Monte-Carlo
simulations.

\paragraph{ISD Multiplicity at LL.}
The main interesting features of ISD multiplicity already arise at
leading logarithmic accuracy, so we will focus on this in what
follows.
The key observation is that at LL, all the emissions from the hard
(leading) parton are soft and collinear, strongly ordered in angle and
independent from one another. The fact that the emissions are
independent automatically guarantees that, if $\nu$ is the probability
that one emission is counted by the ISD de-clustering procedure, \ie
passes the \SD condition, then the probability to have $n$ emission
passing the \SD condition follows a Poisson distribution
\begin{equation}\label{eq:isd-poisson}
\frac{1}{\sigma} \frac{d\sigma}{dn_\text{ISD}} = e^{-\nu} \frac{\nu^n_\text{ISD}}{n_\text{ISD}!}.
\end{equation}

We now need to compute $\nu$ explicitly.
This is straightforward since, at LL, the probability to have an emission that passes the \SD condition is simply
given by (measuring angles in units of the jet radius as usual)
\begin{equation}
\nu = \int_0^1
\frac{d\theta^2}{\theta^2}dz\,P_i(z)\frac{\alpha_s(z\theta p_tR)}{2\pi} \Theta(z>\zcut\theta^\beta).
\end{equation}
For the ISD multiplicity to be IRC-safe, $\nu$ has to remain
finite. This can easily be achieved by using a negative value for
$\beta$, guaranteeing a finite phase-space for the emissions (cf. \eg
the Lund diagram of Fig.~\ref{fig:lund-sd}).
Alternatively, we can manually impose a minimum $k_t$ cut,
$z\theta>\kappa_\text{cut}$, on the emissions which pass the \SD
condition, or stop the iterative de-clustering procedure at a minimum
angle $\theta_\text{cut}$.

\begin{figure}
    \centering
    \includegraphics[width=0.32\textwidth]{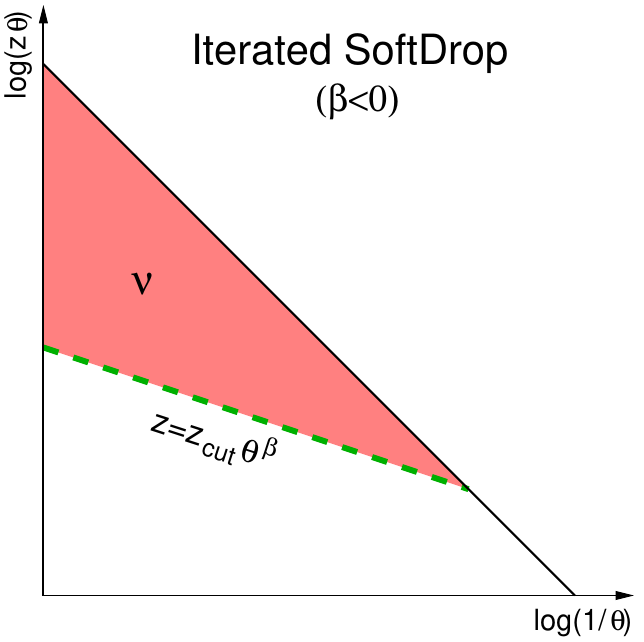}%
    \hfill%
    \includegraphics[width=0.32\textwidth]{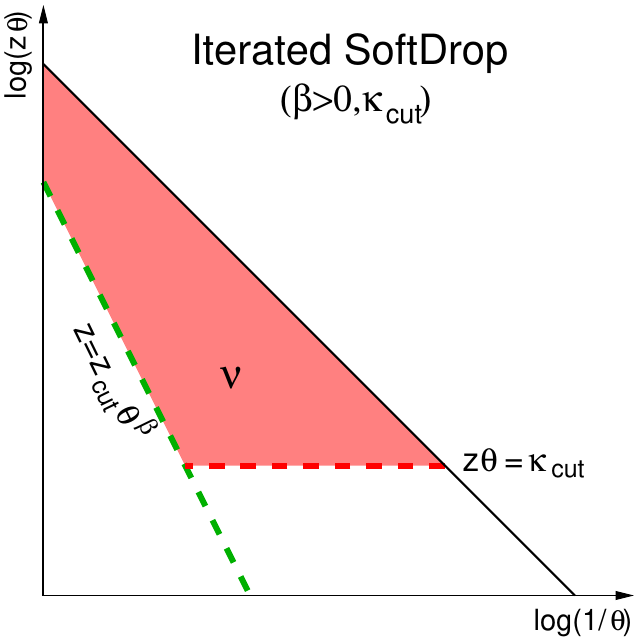}%
    \hfill%
    \includegraphics[width=0.32\textwidth]{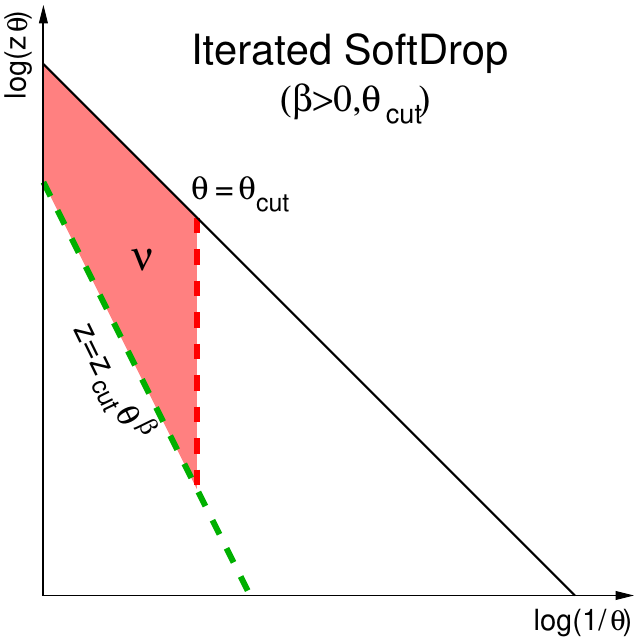}
    \caption{Lund diagrams representing the regions in which Iterated
      \SD counts the emissions. From left to right we have $\beta<0$,
      $\beta>0$ with a cut on $k_t$ and $\beta>0$ with an angular cut.
    }\label{fig:lund-isd-lund}
\end{figure}

These three options correspond to the three regions of the Lund
diagram shown in Fig.~\ref{fig:lund-isd-lund}.
The corresponding analytic expressions for $\nu$ can be obtained
exactly as for the radiators computed for angularities in the previous
Section (this time keeping only LL term, and hard collinear
splittings). One finds (assuming $\kappa<\zcut$ for the second
case):\footnote{These first two results can be directly derived
  from Eq.~(\ref{eq:plain-angularity-radiator-nll})
  and Eq.~(\ref{eq:sd-angularity-radiator-nll}). The third corresponds to
  the radiator for the \SD grooming radius originally computed in
  Ref.~\cite{Larkoski:2014wba} and discussed in Sec.~\ref{sec:thetag}
  below.}
\begin{align}
  \nu_{\beta<0} & 
   = \frac{C_i}{2\pi\alpha_s\beta_0^2}
    \bigg[\frac{-1}{1+\beta}W(1-\lambda_c)-\frac{\beta}{1+\beta}W\Big(1+\frac{\lambda_c}{\beta}\Big)\bigg]-\frac{C_i}{\pi\beta_0}
                   \log\Big(1+\frac{\lambda_c}{\beta}\Big)B_i,\label{eq:isd-nu-betaneg}\\
  \nu_{\beta>0,\kappa} & 
   = \frac{C_i}{2\pi\alpha_s\beta_0^2(1+\beta)}
                         \bigg[-W(1-\lambda_c)-(\lambda_c+\beta)\log(1-\lambda_\kappa)-\lambda_c-\beta\lambda_\kappa\bigg]\nonumber\\
  & -\frac{C_i}{\pi\beta_0} \log(1-\lambda_\kappa)B_i,\\
  \nu_{\beta>0,\theta} & 
   = \frac{C_i}{2\pi\alpha_s\beta_0^2(1+\beta)}
    \bigg[-W(1-\lambda_\theta)-\frac{W(1-\lambda_c)}{1+\beta}+\frac{W(1-\lambda_c-(1+\beta)\lambda_\theta)}{1+\beta}
     \bigg]\nonumber\\
  & -\frac{C_i}{\pi\beta_0} \log(1-\lambda_\theta)B_i,
\end{align}
with
\[
  \lambda_c = 2\alpha_s\beta_0\log\Big(\frac{1}{\zcut}\Big),\quad
  \lambda_\kappa = 2\alpha_s\beta_0\log\Big(\frac{1}{\kappa_\text{cut}}\Big),\quad\text{and}\quad
  \lambda_\theta =
  2\alpha_s\beta_0\log\Big(\frac{1}{\theta_\text{cut}}\Big),
\]
Counting logarithms of $\zcut$, $\kappa_\text{cut}$ and
$\theta_\text{cut}$, all the above expressions show a
double-logarithmic behaviour. An easy way to see this is to compute
$\nu$ using a fixed-coupling approximation (equivalent to taking the
limit $\beta_0\to 0$ in the above results). For example, for the
representative $\beta<0$ case we will use in what follows, one has
\begin{equation}
  \nu_{\beta<0} \overset{\text{f.c.}}{=}
   \frac{\alpha_sC_i}{\pi}\frac{-1}{\beta}\bigg[
   \log^2\Big(\frac{1}{\zcut}\Big)+2B_i\log\Big(\frac{1}{\zcut}\Big)\bigg].
\end{equation}

\begin{figure}
  \includegraphics[width=0.48\textwidth]{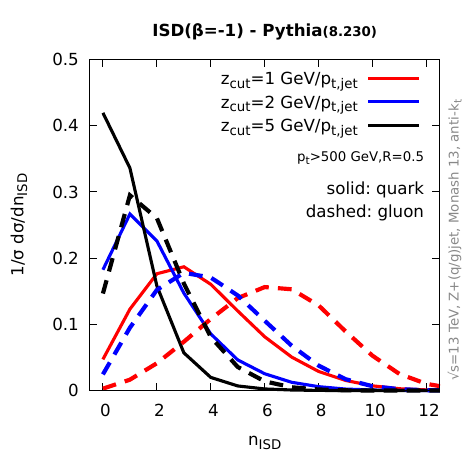}%
  \hfill%
  \includegraphics[width=0.48\textwidth]{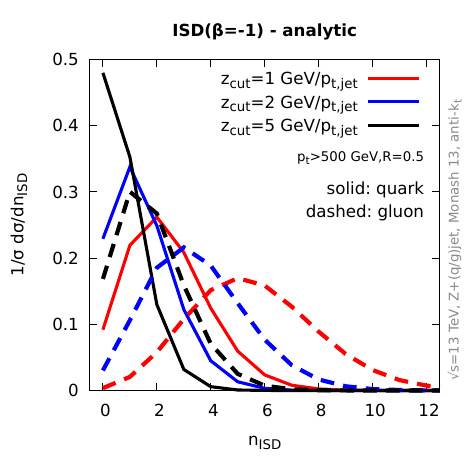}%
  \caption{Distribution of ISD multiplicity for $\beta=-1$, varying
    $\zcut$. The value of $\zcut$ is given as a dimensionful $k_t$
    scale, normalised to $p_tR$. The left plot
    corresponds to parton-level Pythia simulations (for which $\zcut$
    is re-calculated for each jet) and the right plot
    to the analytic calculation, Eqs.~(\ref{eq:isd-poisson})
    and~(\ref{eq:isd-nu-betaneg}). Solid lines correspond to quark
    jets, while dashed lines correspond to gluon jets.}\label{fig:isd-distrib-v-ktcut}
\end{figure}  

Fig.~\ref{fig:isd-distrib-v-ktcut} shows the ISD multiplicity
distributions for quark and gluon jets, obtained from (parton-level)
Pythia8 simulations (left) and using the analytic expressions above
(right). Each plot shows different values of $\zcut$.
For these plots, we have used $\beta=-1$, corresponding to a cut on
the relative $k_t$ of the emissions.
To make this more concrete, the value of $\zcut$ is given as a
function of the corresponding $k_t$ cut. In the case of the Pythia8
simulations, the cut has been adapted using the $p_t$ of each
individual jets.
Overall, we see that the analytic calculation captures the main
features of the Monte Carlo simulation, albeit with distributions
which tend to be peaked towards lower multiplicities than in Pythia8.
We note that NLL corrections, computed in the initial ISD study,
Ref.~\cite{Frye:2017yrw}, improves the agreement between the
two.
One particular effect that becomes relevant at NLL is that the flavour
of the leading branch followed through the ISD declustering can
change. This is included in Pythia8 via the DGLAP splitting functions
and can be tracked analytically as well.

\begin{figure}
  \includegraphics[width=0.48\textwidth,page=1]{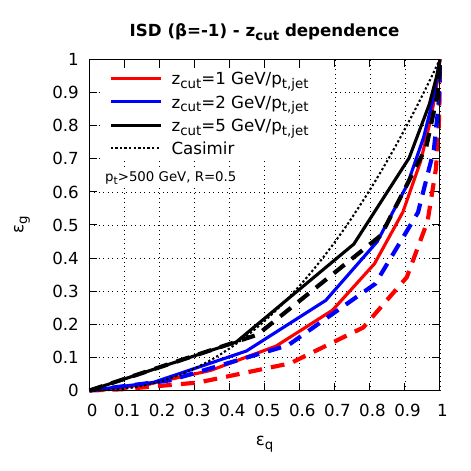}%
  \hfill%
  \includegraphics[width=0.48\textwidth,page=2]{figures/isd-roc.pdf}%
  \caption{Quark-gluon discrimination (ROC curve) using Iterated
    SoftDrop. The left plot uses a fixed jet $p_t$ cut and varies the
    Iterated \SD cut (defined as in
    Fig.~\ref{fig:isd-distrib-v-ktcut}. For the right plot, $\zcut$ is
    fixed to 2~GeV$/(p_tR)$ and the cut on the jet $p_t$ is
    varied.}\label{fig:isd-roc}
\end{figure}  

\paragraph{Quark-gluon discrimination.}
The ROC curves obtained for quark-gluon discrimination are presented
in Fig.~\ref{fig:isd-roc}, for Pythia (solid) and the LL analytic
calculations (dashed).
The left plot corresponds to the distributions shown in
Fig.~\ref{fig:isd-distrib-v-ktcut}.
First, we see that the discriminating power improves with lower
$\zcut$. This is expected since the phase-space for emissions
increases and so does $\nu$.
Then, although the analytic calculation tends to over-estimate the
discriminating power, the generic trend remains decently reproduced
and we see, in particular, that the agreement is better at larger
$\zcut$ where the distribution is expected to have smaller
non-perturbative corrections.
It is  worth pointing  out that  the flavour-changing  effects briefly
mentioned above  and appearing at  NLL accuracy would have  the effect
that  quark and  gluon jets  would  become more  similar as  we go  to
smaller angles, hence reducing the discriminating power.

Finally, the right plot of Fig.~\ref{fig:isd-roc} shows that the
discriminating power of ISD multiplicity improves at larger
$p_t$ (for a fixed $k_t$ cut).
This is again a consequence of the fact that the phase-space
available for emissions, and hence $\nu$, increases.
This contrasts with the angularities discussed
previously: while the latter remain close to Casimir scaling at any
energy, the performance of ISD multiplicity improves for
larger jet $p_t$.

\section{Performance and robustness}\label{sec:qg-perf-robustness}

To conclude this study of quark-gluon tagging, we compare several
quark-gluon discriminators in terms of both their performance and
their robustness.
This is based on Pythia8 Monte Carlo simulations and we reiterate the
caveat that the quark-gluon separation varies between Monte Carlo
(cf.~\cite{Badger:2016bpw}), so this should be taken as a highlight of
the main features rather than a full study.
The main goal of this discussion is to stress more explicitly that, as
introduced in Sec.~\ref{sec:performance-intro}, a high-quality
substructure tool needs obviously to have a strong discriminating
power, but at the same time it small sensitivity to non-perturbative effects is
also desirable.

We first specify our quality measures for performance and
robustness.
For this, let us consider a given quark-gluon discriminator at a fixed
working point (\ie a given cut value). To treat quarks and gluons
symmetrically, we define performance as the geometric mean of the quark
significance and the gluon significance:
\begin{equation}
\Gamma_\text{sym} = \sqrt{\frac{\epsilon_q}{\sqrt{\epsilon_g}}\frac{1-\epsilon_g}{\sqrt{1-\epsilon_q}}},
\end{equation}
where one has used the fact that to tag gluon jets, one would impose a
cut $v>v_\text{cut}$ and $\epsilon_{v>v_\text{cut}}=1-\epsilon_{v<v_\text{cut}}$.
Robustness is then quantified through resilience, as introduced in
Sec.~\ref{sec:performance-intro}, Eq.~(\ref{eq:resilience}). For
simplicity, we will focus here on the resilience against
non-perturbative effects including both hadronisation and the
Underlying Event (UE). These effects could be studied separately but this
goes beyond the scope of this book. We note however that in our case,
resilience is dominated by hadronisation effects, with UE having a much smaller impact.
Finally, note that both the performance $\Gamma_\text{sym}$ and the
resilience $\zeta$ can be computed for any fixed cut on a shape or
multiplicity.

\begin{figure}[t]
  \includegraphics[width=0.48\textwidth,page=3]{figures/qg-rocs.pdf}%
  \hfill%
  \includegraphics[width=0.48\textwidth,page=1]{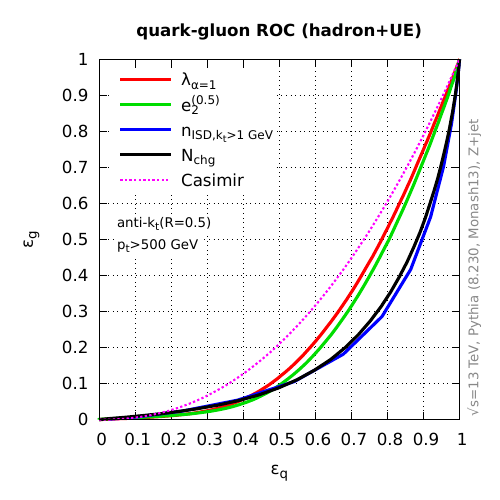}
  \caption{ROC curves for a representative series of quark-gluon
    taggers: broadening, $\lambda_{\alpha=1}$ (red),
    energy-correlation function $e_2^{(\alpha=0.5)}$ (green), Iterated
    \SD with $\beta=-1$ and $\zcut=1~\text{GeV}/p_{t,\text{jet}}$, and the
    charged track multiplicity. All the results are shown for Pythia8
    simulations with a jet $p_t$ cut of 500~GeV. The left plot
    corresponds to parton-level events while the right plot
    corresponds to full simulations including hadronisation and the
    Underlying Event. The charged-track multiplicity is not shown at 
    parton level.}\label{fig:qg-roc-summary}
\end{figure}

First, we compare the performance of a few representative tools
discussed earlier in this section: girth or broadening, equivalent to
the angularity $\lambda_{\alpha=1}$, energy-correlation function
$e_2^{(\alpha=0.5)}$, the ISD multiplicity with
$\zcut=1~\text{GeV}/p_{t,jet}$ (corresponding to a $k_t$ cut of
1~GeV), and the charged track multiplicity.
The ROC curves are shown on Fig.~\ref{fig:qg-roc-summary} for Pythia8
simulations at parton level (left) and at hadron level including the
Underlying Event (right).
At small quark efficiency ($\epsilon_q\lesssim 0.5$) angularities and
energy correlation functions tend to give a better discriminating
power. At larger quark efficiency multiplicity-based discriminators show a better performance, with the
ISD and charged-track multiplicities behaving similarly.

\begin{figure}[t!]
  \centerline{\includegraphics[width=0.48\textwidth,page=1]{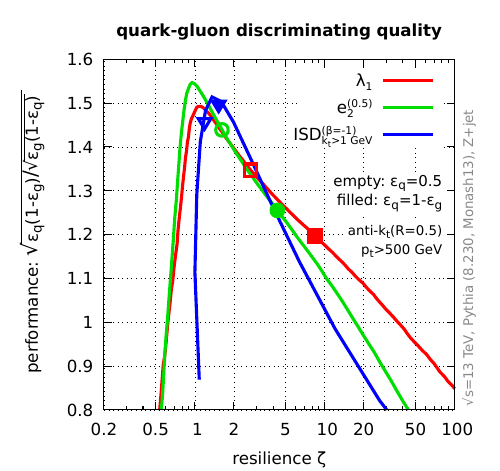}}
  \caption{Quark-gluon tagging quality: performance v.\ resilience for
    the taggers used in Fig.~\ref{fig:qg-roc-summary}. The curves
    correspond to varying the cut on the jet shape or
    multiplicity. Solid (empty) points correspond to the specific
    working point for which $\epsilon_q=1-\epsilon_g$
    ($\epsilon_q=0.5$). Performance is computed at hadron+UE level and
    resilience includes both hadronisation and UE
    effects.}\label{fig:qg-quality}
\end{figure}

We now discuss both the performance and resilience of our
representative sample of quark-gluon taggers. This is first shown on
Fig.~\ref{fig:qg-quality} for the full ROC curves corresponding to
Fig.~\ref{fig:qg-roc-summary}, \ie where the lines are obtained by
varying the cut on the shape or multiplicity. The empty symbols
correspond to a fixed quark efficiency of 0.5 at hadron+UE level,
while the solid symbols correspond to a symmetric working point where
$\epsilon_q=1-\epsilon_g$ (at hadron+UE level).\footnote{For
  multiplicity-based observables, we have interpolated linearly
  between the discrete multiplicities.} The charged-track
multiplicity is not plotted simply because it is not well-defined at
parton level.

We see that angularities and ECFs give their best performance at
relatively low quark efficiency, corresponding to a fairly low
resilience. As the quark efficiency decreases (going to
$\epsilon_q=0.5$, then $\epsilon_q=1-\epsilon_g$) performance
decreases but one gains resilience. A similar behaviour is seen for
ISD although the highest performance is observed for larger quark
efficiencies and large resilience at yet larger quark efficiencies.
For our 500-GeV sample, the best performance is achieved by
ECF($\alpha=0.5$) closely followed by ISD, with the latter showing a
slightly better resilience against non-perturbative effects.
At lower $\Gamma_\text{sym}$ this is inverted, with shape-based
variables becoming more resilient than ISD. 

The crucial observation one draws from Fig.~\ref{fig:qg-quality} is
that, generally speaking, there is a trade-off between performance and
resilience.
This pattern is seen repeatedly in substructure studies (we will see
another example in our two-prong-tagger study in the next chapter) and can be
understood in the following way: tagging constrains patterns of
radiation inside a jet; usually, increasing the phase-space over which
we include the radiation, and in particular the region of soft
emissions, means increasing the information one includes in the tagger
and hence increasing the performance; at the same time, the region of soft
emissions being the one which is most sensitive to hadronisation and
the Underlying Event, one also reduces resilience.

\begin{figure}[t!]
  \includegraphics[width=0.48\textwidth,page=3]{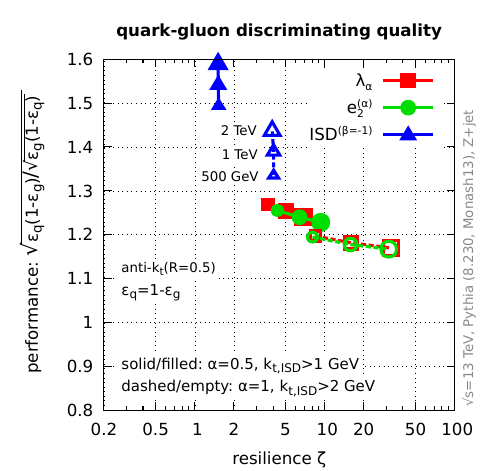}%
  \hfill%
  \includegraphics[width=0.48\textwidth,page=2]{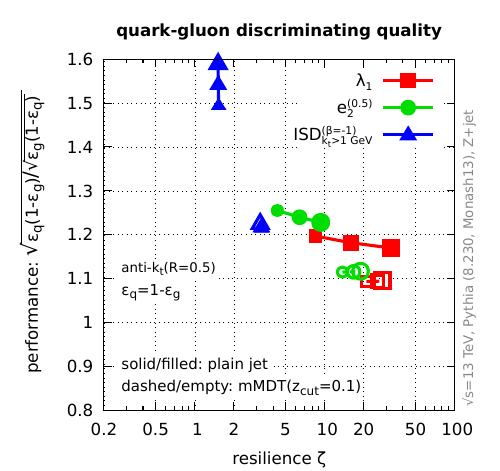}
  \caption{Plot of performance v.\ resilience for quark-gluon taggers,
    as in Fig.~\ref{fig:qg-quality}, varying the cut on the jet $p_t$,
    using the working point $\epsilon_q=1-\epsilon_g$. The left plot
    shows two different choices of parameters for the taggers. The
    right plot shows two different levels of
    grooming. }\label{fig:qg-quality-ptdep}
\end{figure}

\begin{figure}[t!]
  \includegraphics[width=0.48\textwidth,page=7]{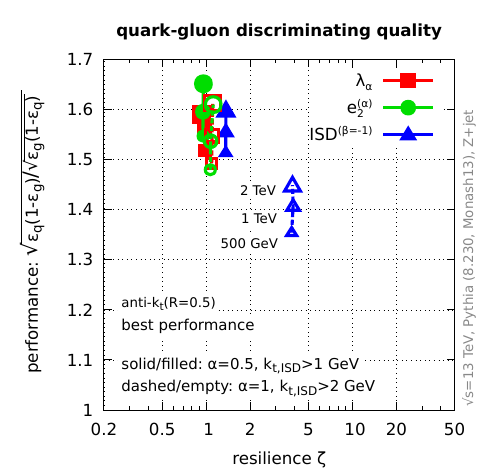}%
  \hfill%
  \includegraphics[width=0.48\textwidth,page=6]{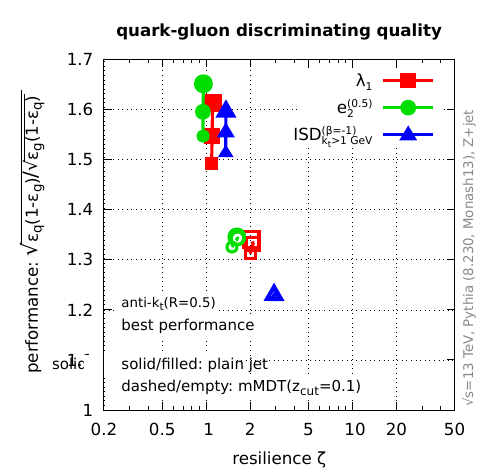}%
  \caption{Plot of performance v.\ resilience for quark-gluon taggers,
    as in Fig.~\ref{fig:qg-quality-ptdep} but now using the working
    point that maximises performance for each
    setup.}\label{fig:qg-quality-ptdep-best-perf}
\end{figure}

To finish this study of quark-gluon taggers, we show in
Fig.~\ref{fig:qg-quality-ptdep} how the quark-gluon tagging quality
varies with the jet $p_t$. From small to big symbols, we have used
$p_t>500$~GeV, $p_t>1$~TeV and $p_t>2$~TeV, and we have focused on the
point for which $\epsilon_q=1-\epsilon_g$.
The left plot shows this for two different choices of parameters (two
exponents for angularities and ECFs and two $\zcut$ for ISD).
We see clearly that, as expected from our earlier studies, the
performance of ISD increases with the jet $p_t$ while that of
shape-based taggers remains roughly constant. Conversely, shape-based
taggers become more resilient at larger $p_t$, highlighting again a
trade-off between performance and resilience.

The right plot of Fig.~\ref{fig:qg-quality-ptdep} shows two different
levels of grooming: the plain jet and a jet groomed with
mMDT\footnote{In the case of ISD, we have applied mMDT recursively,
  giving a behaviour equivalent to using $\beta=0$ and a $k_t$ cut as
  shown in the middle plot of Fig.~\ref{fig:lund-isd-lund}.}
The dependence on the jet $p_t$ is the same as what was already
observed for the left plot (although, for mMDT jets, the performance
of ISD only increases marginally). What is more interesting is that
one clearly sees that grooming has the effect of reducing the
performance and increasing the resilience.
Since grooming is (almost by definition) removing soft emissions at
large angles, this is another textbook example of a trade-off between
performance and resilience.
We note however that these conclusions are relatively sensitive to the
choice of working point. For example,
Fig.~\ref{fig:qg-quality-ptdep-best-perf} shows the same result as
Fig.~\ref{fig:qg-quality-ptdep} but now selecting for each method the
working point which maximises performance. In this case, we see that
all methods give similar results both in terms of performance and in
terms of resilience, with even a small preference for ECFs (with
$\alpha=0.5$) if one is looking for sheer performance. It is worth
pointing out that in this case the quark and gluon efficiencies are
relatively low, meaning that (i) one might be affected by issues
related to lower statistics and (ii) we are in a region where the
discreteness of ISD can have large effects which one would need to
address in a more complete study.\footnote{For the results of
  Fig.~\ref{fig:qg-quality-ptdep-best-perf} we have simply
  interpolated between different points in the distribution.}

As a final comment, we point out that, given the different
behaviours seen between shape-based taggers and multiplicity-based
taggers, it would be interesting to study their combination in a
multivariate analysis. It would also be interesting to see how recent
quark-gluon taggers based on deep learning techniques use the
information relevant for ECFs and ISD.

%% GS helper for auctex
%%% Local Variables:
%%% mode: latex
%%% TeX-master: "notes"
%%% End:

%  LocalWords:  supersymmetric Monash ECFs Casimir NLL WTA CMW Eq eq
%  LocalWords:  Eqs NLOJet MCFM ECF ISD dimensionful UE

% $Id: calculations-two-prongs.tex 516 2019-04-15 15:01:34Z gsoyez $
\chapter{Two-prong tagging with jet shapes}\label{chap:calc-two-prongs}

 Two-prong taggers aim at
discriminating massive objects that decay into two hard QCD partons
(usually quarks), from the background of QCD jets. This signal is
often an electroweak boson (H/W/Z) but it can also be a new
particle (see Chapter~\ref{searches-measurements} for examples).

Our goal in this chapter is two-folded and it closely follows what was
done in the previous chapter for quark-gluon tagging. First, we want to
give a brief insight into analytic properties of two-prong taggers,
mainly selecting a few representative substructure tools and comparing their
behaviour in Monte Carlo simulations with analytic results. Then, we
will perform a comparative Monte Carlo study of the taggers
discriminating properties, assessing both their performance and their
resilience against non-perturbative effects. 

%%========================================================================
\section{A dive into analytic properties}\label{sec:two-prongs-analytic}

Two-prong taggers used for Run-II of the LHC tend to combine two major
ingredients: a two-prong finder also acting as a groomer, and a cut on
a jet shape for radiation constraint. 
Since groomers have already been extensively discussed in
Chapter~\ref{calculations-substructure-mass}, in this chapter we are going to focus on
the understanding of jet shapes and of their
interplay with grooming.
Note that a variety of jet shapes can be used in the context of tagging
two-pronged boosted objects: Y-splitter, $N$-subjettiness, ECFs,
pull, and so on.  We will only select a few for our discussion.
 
While computations for groomers and prong-finders, such as the
modified MassDrop Tagger or \SD, have seen a lot of
development towards precision calculations in the last few years and
one can say that they are under good analytic control, the
situation for jet shapes is more complex.
This can be understood as follows: imagine one wants to tag a boosted
object around a mass $M_\text{X}$; one would typically first require that the jet
mass (groomed or ungroomed) is in a window close to $M_\text{X}$ and then
that the cut on the jet shape is satisfied; for QCD jets, which constitute the background, this means that we need to consider at least two emissions inside the jet --- one
setting the jet mass, the second setting the value of the shape --- so
calculations for the QCD background will start at order $\alpha_s^2$ in
the perturbative expansion, compared to $\alpha_s$ for groomers or
quark/gluon taggers.
That said, calculations now exist for a range of jet shapes~(see
\eg~\cite{Larkoski:2013eya,Larkoski:2014tva,Larkoski:2015kga,Dasgupta:2015lxh,Larkoski:2017iuy,Napoletano:2018ohv}),
noticeably ECFs and $N$-subjettiness, in both the direct QCD approach
used in this book and in SCET.

To keep the discussion simple, we will assume that, on top of working
in the boosted limit $m\ll p_{t,\text{jet}}$, the cut on the jet shape,
$v<v_\text{cut}$, is also small so we can study the effect of the
shape in the leading-logarithmic approximation.
Technically, since we expect signal jets to mostly exhibit  small values of $v$
--- \ie there is less radiation in a signal jet than in QCD background
jets --- this approximation seems reasonable. For practical
phenomenological applications however cuts on jet shapes are not much
smaller than one and so finite $v$ corrections are potentially sizeable.
The leading-logarithmic approximation we will adopt in what follows,
treating logarithms of $m/p_{t.\text{jet}}$ and $v_\text{cut}$ (and,
optionally of the grooming $\zcut$ parameter) on an equal footing, is
nevertheless sufficient to capture the main properties of two-prong
taggers and differences between them.

For the purpose of this book, we will focus on three different shapes: the
 $N$-subjettiness ratio $\tau_{21}$, with $\beta=2$, which has a fairly
simple structure and has been used at the LHC (albeit with $\beta=1$). We will then move to the dichroic version of the $\tau_{21}$ ratio (see Eq.~\eqref{eq:def-dichroic}) in order to illustrate how separating the
grooming and prong-finding parts of the tagger could be helpful.
Finally, we will discuss the ECFs $C_2^{(\beta=2)}$ and $D_2^{(\beta=2)}$. The latter in particular shows a very good discriminating power and  it is used at the LHC (albeit with $\beta=1$).

A typical LL calculation involves two steps: (i) compute an
expression for the shape valid at LL and (ii) use it to derive an
expression for the mass distribution with a cut on the jet shape, or
the distribution of the shape itself.
The calculations for QCD jets will be followed by a calculation for
signal (W/Z/H) jets and a comparison to Monte Carlo simulations done
using the Pythia8 generator.
Note that the analytic calculations below focus on computing the jet
mass distribution imposing a cut on the jet shape:
$(\rho/\sigma\,d\sigma/d\rho)_{v<v_\text{cut}}$. We can deduce the
cumulative and differential distribution for the shape itself:
\begin{equation}
  \Sigma(v)=\frac{(d\sigma/d\rho)_{v<v_\text{cut}}}
                 {(d\sigma/d\rho)_{\text{no cut}}}
  \qquad\text{ and }\qquad
  \frac{v}{\sigma}\frac{d\sigma}{dv} = v\frac{d\Sigma}{dv}.
\end{equation}
The background efficiency in a given mass window can also be obtained
from the mass distribution with a cut on the shape via
\begin{equation}
  \epsilon_B(\rho_\text{min},\rho_\text{max};v_\text{cut})
  =\int_{\rho_\text{min}}^{\rho_\text{max}} d\rho \left.\frac{d\sigma}{d\rho}\right|_{v<v_\text{cut}}.
\end{equation}

\subsection{$N$-subjettiness $\tau_{21}^{(\beta=2)}$ ratio}

\paragraph{Approximate $\tau_{21}$ value at LL.}
To fully specify the definition of the $\tau_{21}$ ratio we are working with, it is not
sufficient to specify the value of the $\beta$ parameter, one also
needs to specify the choice of axes. For our choice of $\beta=2$, it
is appropriate to work either with minimal axes, \ie the axes that
minimise the value of $\tau_N$, or exclusive generalised-$k_t$ axes
with $p=1/\beta=1/2$. 
Let us consider a set of $n$ emissions. For the purpose of our LL
calculation, we can assume that they are strongly ordered in ``mass'' (or to be more precise in their contribution to the mass)
\ie $\rho_1\gg\rho_2\gg\dots\gg\rho_n$, with $\rho_i=z_i\theta_i^2$,
and strongly ordered in energy and angle (\ie, for example,
$\theta_i\gg\theta_j$ or $\theta_i\ll\theta_j$ for any two emissions
$i$ and $j$).
For the sake of definiteness, let us work with axes defined using the
generalised-$k_t$~($p=1/2$) exclusive subjets.
We should thus first go through how our set of emissions is clustered.
The generalised-$k_t$ clustering will proceed by identifying the
smallest $d_{ij}=\text{min}(z_i,z_j)\theta_{ij}^2$ distance.
Using $i=0$ to denote the leading parton and assuming $z_i\ll z_j$, we have
\begin{align}
  d_{i0} & = z_i\theta_i^2 = \rho_i,\\
  d_{ij} & = z_i\theta_{ij}^2
           \approx z_i\,\text{max}(\theta_i^2,\theta_j^2)
           \ge  z_i\theta_i^2 \equiv \rho_i.\label{eq:tau-genkt-ij-distance}
\end{align}
The overall minimal distance will therefore be the smallest of the
$\rho_i$'s, \ie $\rho_n$. This can be realised in two ways: either the
distance between emission $n$ and the leading parton ($d_{n0}=\rho_n$)
of the distance between emission $n$ and any emission $k$ with
$\theta_k\ll\theta_n$ (for which Eq.~\eqref{eq:tau-genkt-ij-distance}
gives $d_{nk}\approx\rho_n$). In the second case, we also have
$z_k\gg z_n$. Due to the energy ordering --- and the fact that for
$\beta=2$ recoil effects can be neglected --- after clustering
particle $n$ with either the leading parton or emission $k$, one gets
a situation with the leading parton and emissions $1,\dots,n-1$. The
above argument can then be repeated, clustering particles
$n-1,n-2,\dots,2,1$ successively.
This means that the $\tau_1$ axis will be the jet axis --- equivalent
to the leading parton in this case --- and the two exclusive
generalise-$k_t$ axes used for $\tau_2$ will be aligned with the
leading parton and with the largest $\rho_i$ emission, \ie with
emission 1.\footnote{The argument can be extended to the $N$ exclusive
axes used for $\tau_N$ which would be aligned with the leading parton
and with emissions $1,\dots,N-1$.}

With these axes, it is easy to deduce the value of
$\tau_1$ and $\tau_2$ for our set of emissions:
\begin{align}
\tau_1 & = \sum_{i=1}^n z_i\theta_i^2 = \rho \approx \rho_1,\\
\tau_2 & = \sum_{i=1}^n z_i\text{min}(\theta_i^2,\theta_{i1}^2) \approx \rho_2,\label{eq:tau2-LL-value}
\end{align}
where, in the second line, the contribution from emission $1$ vanishes.

Note that the above derivation is slightly incomplete: on top of the
$n$ emissions from the leading parton, we can also have secondary
emissions from the leading emissions $1,\dots,n$, \ie, in our
angular-ordered limit, emissions ``$j$'' from the leading parton $i$
with $z_j\ll z_i$ and $\theta_{ij}\ll\theta_i$.
These will not affect the finding of the two axes needed to compute
$\tau_2$ but secondary emissions from emission $1$ can dominate
$\tau_2$. Specifically, an emission with a momentum fraction $z_2$
relative to $z_1$ emitted at an angle $\theta_{21}$ from emission 1
would give
\begin{equation}\label{eq:tau21-value-secondary}
  \tau_{2,\text{secondary}} \approx z_1 z_2 \theta_{12}^2\qquad\text{\ie}\quad
  \tau_{21,\text{secondary}} \approx z_2\frac{\theta_{12}^2}{\theta_1^2}.
\end{equation}
Another way to view this is to consider that the two axes used to
compute $\tau_2$ define a partition of the jet in two subjets (one
around the leading parton, the second around emission $1$).
The total $\tau_2$ is therefore the sum of the individual
contributions from these two subjets, \ie from the sum of
$z_i\theta_{i,\text{axis}}^2$ in these two subjets and the dominant
contribution can come from either subjet.
This is in contrast with all the calculations done previously in this
book, which were only sensitive to primary emissions.
It should however not come as a surprise since we are discussing tools
which measure the radiation pattern around a two-prong structure so
one should expect a contribution from both prongs.

Note finally that the same result is obtained with the one-pass
generalised-$k_t$ axes or with the minimal axes. However, if we were
to use exclusive $k_t$ axes, which contrary to the above arguments
orders emission's in $z_i\theta_i$, we could have situations where the
emission with the largest $z_i\theta_i$ is different from the emission
with the largest $\rho_i$. This inevitably leads to additional complexity.

\begin{figure}[t]
  \centering
  \subfloat[]{\includegraphics[width=0.48\textwidth]{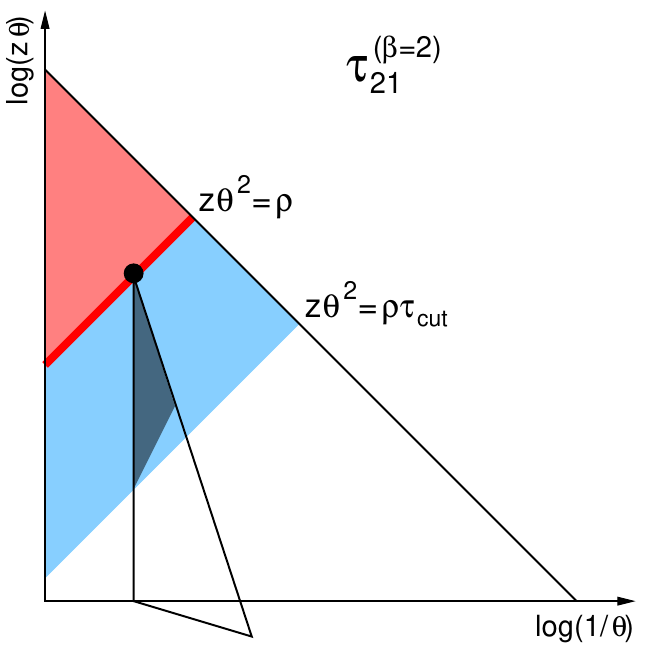}\label{fig:lund-tau21-plain}}%
  \hfill%
  \subfloat[]{\includegraphics[width=0.48\textwidth]{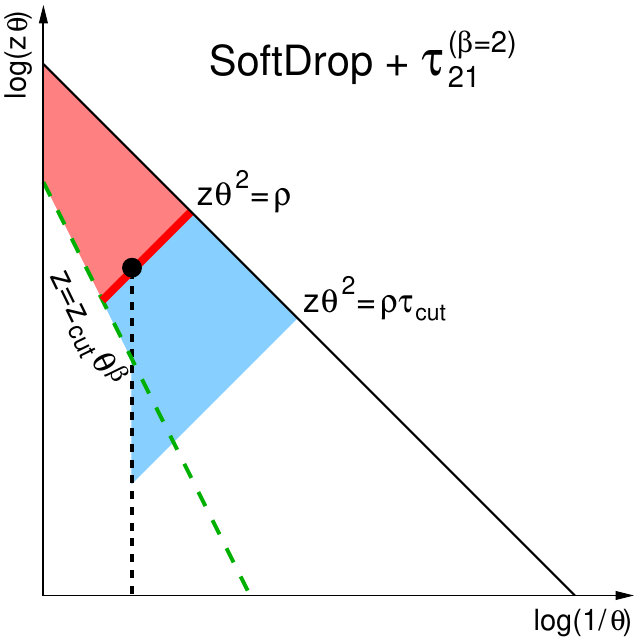}\label{fig:lund-tau21-SD}}%
  \caption{Lund diagram for the LL mass distribution with a cut on the
    $\tau_{21}$ $N$-subjettiness ratio. The solid red line corresponds
    to the desired jet mass. Real emissions are vetoed in the shaded
    light red region because they would yield a larger mass and in the
    light blue region because they would not pass the cut on
    $\tau_{21}$. The left plot (a) corresponds to the plain jet and the
    right plot (b) to a jet previously groomed with \SD. The left
    plot shows also the plane for secondary (gluon) emissions. An
    identical secondary plane should also be present on the right plot
    but has been omitted for clarity.}\label{fig:lund-tau21}
\end{figure}

\paragraph{LL mass distribution with a cut $\tau_{21}<\tau_\text{cut}$.}
Once an expression has been found it is straightforward to understand
the structure of the jet mass distribution with a cut
$\tau_{21}<\tau_\text{cut}$.
Since $\tau_{21}$ is given by the second ``most massive'' emission
(either from the leading parton or from the emission which dominates
the jet mass), imposing a cut on $\tau_{21}$ vetoes such emissions,
leaving a Sudakov factor corresponding to virtual emissions in that region of
phase-space.
This is represented on the Lund plane in
Fig.~\ref{fig:lund-tau21-plain} and one gets
\begin{align}
  \frac{\rho}{\sigma} \frac{d\sigma}{d\rho}\Big|_{\tau_{21}<\tau_\text{cut}}
  & = \int_0^1 \frac{d\theta_1^2}{\theta_1^2}\frac{dz_1}{z_1}
      \frac{\alpha_s(z_1\theta_1)C_i}{\pi}\rho\delta(\rho-\rho_1)
      \exp[-R_\tau^{\text{(primary)}}-R_\tau^\text{(secondary)}]\label{eq:tau21-plain-mass}\\
  R_\tau^\text{(primary)}
    & =  \int_0^1 \frac{d\theta_2^2}{\theta_2^2}\frac{dz_2}{z_2}
      \frac{\alpha_s(z_2\theta_2)C_i}{\pi}\Theta\Big(\frac{\rho_2}{\rho}>\tau_\text{cut}\Big)\label{eq:tau21-plain-Rprimary},\\
  R_\tau^\text{(secondary)}
    & = \int_0^{\theta_1^2} \frac{d\theta_{12}^2}{\theta_{12}^2}\int_0^1\frac{dz_2}{z_2} \frac{\alpha_s(z_1z_2\theta_{12})C_A}{\pi}\Theta\Big(\frac{z_2\theta_{12}^2}{\theta_1^2}>\tau_\text{cut}\Big),\label{eq:tau21-plain-Rsecondary}
\end{align}
where angles are measured in units of the jet radius $R$ and  the
arguments of the strong couplings are in units of $p_tR$.

The integration in Eq.~\eqref{eq:tau21-plain-mass} corresponds to the
particle which dominates the jet mass, \ie constrained so that
$\rho=\rho_1$. Eq.~\eqref{eq:tau21-plain-Rprimary} is the Sudakov veto
on primary emissions. It includes a standard jet-mass Sudakov,
$\rho_2>\rho$, from the fact that emission 1 dominates the mass (the
light red region in Fig.~\ref{fig:lund-tau21-plain}), as well as an
additional Sudakov veto $\rho>\rho_2>\rho\tau_\text{cut}$ coming from
the extra constraint on $\tau_{21}$, the light blue region in
Fig.~\ref{fig:lund-tau21-plain}.
Finally, Eq.~(\ref{eq:tau21-plain-Rsecondary}) corresponds to the extra
Sudakov veto imposing that secondary emissions with
$\tau_{21}>\tau_\text{cut}$ (cf.~Eq.~(\ref{eq:tau21-value-secondary}))
also have to be vetoed.
As before, one can obtain the ``modified'' LL results, including hard
collinear splittings, by setting the upper limits of the $z$ integrations
to $\exp(B_i)$, which is what we do in practical applications below.

In the fixed-coupling approximation, the integrations can be
done analytically, and one obtains
\begin{equation}\label{eq:mass-distrib-tau21-fc}
  \frac{\rho}{\sigma} \frac{d\sigma}{d\rho}\Big|_{\tau_{21}<\tau_\text{cut}}
  \overset{\text{f.c.}}{=}
  \frac{\alpha_s C_i}{\pi} (L_\rho+B_i)
  \exp\Big[-\frac{\alpha_sC_i}{\pi}(L_\rho+L_\tau+B_i)^2
  - \frac{\alpha_sC_A}{\pi}(L_\tau+B_g)^2\Big] ,
\end{equation}
where we have defined
\begin{equation}
  L_\rho = \log(1/\rho)\qquad\text{ and }\qquad
  L_\tau=\log(1/\tau_\text{cut}).
\end{equation}
This has to be compared to the jet mass distribution without the cut
on $\tau_{21}$ which has the same prefactor but only
$\tfrac{\alpha_sC_i}{\pi}(L_\rho+B_i)^2$ in the Sudakov exponent.
The cut on $\tau_{21}$ brings an additional Sudakov suppression,
double-logarithmic in $\tau_\text{cut}$ with contributions from both
primary and secondary emissions and, more interestingly, a
contribution proportional to $\log(1/\rho)\log(1/\tau_\text{cut})$,
meaning that with a fixed cut on $\tau_{21}$, the QCD background will
be more suppressed when increasing the jet boost, i.e.\ decreasing $\rho$.
We provide more physical discussions below, once we also have
results for the signal and ROC curves.

The calculation of the jet mass with a cut on $\tau_{21}$ can also be
performed for groomed jets, \ie one grooms the jet before measuring
its mass and $\tau_{21}$ on the groomed jet.
Here we consider the case of \SD. 
As discussed in Sec.~\ref{sec:calc-groomed-mass}, emission 1, which
dominates the \SD mass, has to satisfy the \SD condition and the
associated Sudakov is given by Eq.~(\ref{eq:mMDTSD-radiator-modll}).
One small extra complication compared to the case of the \SD jet mass
is that one should
remember that the \SD de-clustering procedure stops once some hard
structure has been found, \ie once the \SD condition is met.
Since the de-clustering procedure uses the Cambridge/Aachen jet algorithm, this
means that once the procedure stops, all emissions at smaller angles
are kept, whether or not they pass the \SD condition.

In our LL calculation for $\tau_{21}$, it is sufficient to realise
that one can consider that the \SD procedure keeps all emissions at
angles smaller than $\theta_1$. Thus, the resulting phase-space is depicted in
Fig.~\ref{fig:lund-tau21-SD} and one gets:
\begin{align}
  \frac{\rho}{\sigma} \frac{d\sigma}{d\rho}\Big|_{\tau_{21}<\tau_\text{cut}}^\text{SD}
  & = \int_0^1 \frac{d\theta_1^2}{\theta_1^2}\frac{dz_1}{z_1}
      \frac{\alpha_s(z_1\theta_1)C_i}{\pi}\rho\delta(\rho-\rho_1)
      \,\Theta(z_1>\zcut\theta_1^\beta)
      e^{-R_{\tau,\text{SD}}^\text{(primary)}-R_\tau^\text{(secondary)}}\label{eq:tau21-SD-mass}\\
  R^\text{(primary)}_{\tau,\text{SD}}
    & =  \int_0^1 \frac{d\theta_2^2}{\theta_2^2}\frac{dz_2}{z_2}
      \frac{\alpha_s(z_2\theta_2)C_i}{\pi}\Theta\Big(\frac{\rho_2}{\rho}>\tau_\text{cut}\Big)
      \,\Theta(z_2>\zcut\theta_2^\beta\text{ or }\theta_2<\theta_1).\label{eq:tau21-SD-Rprimary}
\end{align}
The Sudakov corresponding to secondary emissions is the same as for
the plain jet, since all emissions at angles smaller than $\theta_1$
are kept in the groomed jet.
Keeping the running-coupling contributions, one finds the following
expressions for the radiators:
\begin{align}
  R^\text{(primary)}_{\tau,\text{SD}}&(\rho,\tau_\text{cut},\theta_1)
  =R_{\text{SD}}^{\text{(LL)}}(\rho\tau_\text{cut}) + \delta
    R_{\tau,\text{SD}}(\rho,\tau_\text{cut},\theta_1)  \\
  \delta R_{\tau,\text{SD}}&(\rho,\tau_\text{cut},\theta_1)
  = \frac{C_i}{2\pi\alpha_s\beta_0^2}\bigg[
    W(1-\lambda_\rho-\lambda_\tau+\lambda_1)
    +\frac{W(1-\lambda_c-(1+\beta)\lambda_1)}{1+\beta}\\
 & -\frac{2+\beta}{1+\beta}W\Big(1-\frac{\lambda_c+(1+\beta)(\lambda_\rho+\lambda_\tau)}{2+\beta}\Big)    
  \bigg]\Theta(\lambda_c+(2+\beta)\lambda_1 > \lambda_\rho+\lambda_\tau)\nonumber\\
  R_\tau^\text{(secondary)}&(\rho,\tau_\text{cut},\theta_1)
  = \frac{C_i}{2\pi\alpha_s\beta_0^2}\bigg[
    W(1-\lambda_\rho-\lambda_{B_g}+\lambda_1)
    +W(1-\lambda_\rho-\lambda_\tau+\lambda_1)\\
  & -2W(1-\lambda_c-\frac{\lambda_\tau+\lambda_{B_g}}{2}+\lambda_1)
    \bigg]\Theta(\lambda_\tau>\lambda_{B_g}),\nonumber
\end{align}
with $\lambda_\rho$ and $\lambda_c$ defined as in
Eq.~(\ref{eq:mMDTSD-radiator-modll}),
$\lambda_\tau=2\alpha_s\beta_0\log(1/\tau_\text{cut})$ and, 
$\lambda_1=2\alpha_s\beta_0\log(1/\theta_1)$ and
$\lambda_{B_g}=-2\alpha_s\beta_0B_g$.
$\delta R_{\tau,\text{SD}}$ is the additional contribution from
$\theta_2<\theta_1$ and $z_2<\zcut\theta_2^\eta$.
$R_\tau^\text{(primary)}$ can be easily obtained from
$R^\text{(primary)}_{\tau,\text{SD}}$ by taking the limit
$\beta\to\infty$ and it is nothing else than the plain (ungroomed) jet
mass Sudakov  evaluated at the scale $\rho\tau_\text{cut}$.
Contrary to the fixed-coupling limit, $\delta R_{\tau,\text{SD}}$ and
$R_\tau^\text{(secondary)}$ explicitly depend on $\theta_1$ and the
integration in Eq.~\eqref{eq:tau21-SD-mass} cannot be performed analytically.

\subsection{$N$-subjettiness dichroic $\tau_{21}^{(\beta=2)}$ ratio.}\label{sec:2prongs-analytic-dichroic}

The idea behind dichroic observables arises when combining a prong
finder and a shape constraint.
The identification of two hard prongs in a jet, is usually achieved by applying tools
like the mMDT, trimming or pruning to the jet. These algorithms are also active,
(and tight) groomers, meaning that they groom away a large fraction of soft and large-angle radiation in the jet. However, the region of phase-space which is groomed away does carry a lot of information about the radiation pattern, which would be potentially exploited by the shape constraint.
The idea is therefore to to compute the shape constraint on a larger,
less tightly groomed jet, that we call the {\em large} jet below. For
shapes which are expressed as a ratio, like $\tau_{21}$, and for
$\beta=2$, the denominator of the shape is a measure of the jet mass
--- recall $\tau_1=\rho$ in the previous section --- which is
naturally computed on the tight jet found by the prong finder,
referred to as the {\em small} jet in what follows.
This hints at the following combination
\begin{align}
  \text{mass constraint: }&\text{ use }\rho_{\text{small}},\\
  \text{shape constraint: }&\text{ use }\tau_{21}^{\text{(dichroic)}}=\frac{\tau_{2,\text{large}}}{\tau_{1,\text{small}}}.
\end{align}
We will assume that the small jet is obtained using mMDT with the
condition $z>x_\text{cut}$, and the large jet is either the plain jet
or a \SD jet with positive $\beta$ and a given $\zcut$.
We first derive LL analytic results similar to the ones obtained
in the previous section for $\tau_{21}$ and then come back to the
benefits of the dichroic variant.

\paragraph{Approximate $\tau_{21}^\text{(dichroic)}$ value at LL}
The value of $\tau_{21}^\text{(dichroic)}$ for a given set of
emissions in a jet can be readily obtained from the results in the
previous section. First, $\tau_{1,\text{small}}$ is
equivalent to the small-jet (dimensionless squared) mass:
$\tau_{1,\text{small}}=\rho_\text{small}$. We will denote by $a$ the
emission that sets the mass of the small jet.

For $\tau_{2,\text{large}}$, we need to use
Eq.~\eqref{eq:tau2-LL-value}, \ie $\tau_{2,\text{large}}$ is dominated
by the emission with the second-largest $\rho_i=z_i\theta_i^2$ in the
large jet. We will therefore denote by $b$ and $c$, the emissions with the
largest and second-largest $\rho_i$ in the large jet,
respectively. With these notations, we get
\begin{equation}
  \tau_{21}^{\text{(dichroic)}} \approx \frac{\rho_c}{\rho_a}\qquad\quad
  \text{($\rho_a$ largest in small, $\rho_c$ $2^\text{nd}$ largest in large)}.
\end{equation}
Note that, contrary to the standard $\tau_{21}$ ratio, the dichroic
ratio can be larger than one.
More specifically, three situations can arise: (i) the emission which
dominates the mass of the small jet also dominates the one of the
large jet, \ie $\rho_a=\rho_b>\rho_c$, yielding
$\tau_{21}^{\text{(dichroic)}}<1$; (ii) the emission which dominates
the mass of the small jet is the $2^\text{nd}$ largest in the large
jet, \ie $\rho_b>\rho_a=\rho_c$ yielding
$\tau_{21}^{\text{(dichroic)}}=1$; and (iii) there are at least two
emissions with a larger $\rho_i$ in the large jet than in the small
jet, \ie $\rho_b>\rho_c>\rho_a$ yielding
$\tau_{21}^{\text{(dichroic)}}>1$.
It is easy to check that the value of $\tau_{21}^\text{(dichroic)}$ is
always equal or larger than the value of the $\tau_{21}$ ratio obtained 
with approaches frequently used in experimental contexts.
 This is a desired feature since increasing the
value of $\tau_{21}$ means rejecting more QCD jets when imposing a
cut.\footnote{As we will see below, this increase of $\tau_{21}$ for
  QCD jets in the dichroic case comes with no modifications for signal
  jets.}

\begin{figure}[t]
  \centering
  \subfloat[]{\includegraphics[width=0.48\textwidth]{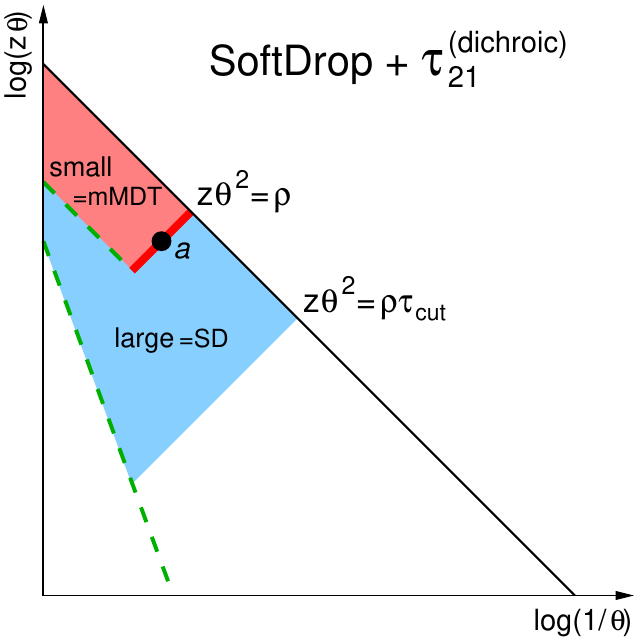}\label{fig:lund-tau21-dichroic-SD}}%
  \hfill%
  \subfloat[]{\includegraphics[width=0.48\textwidth]{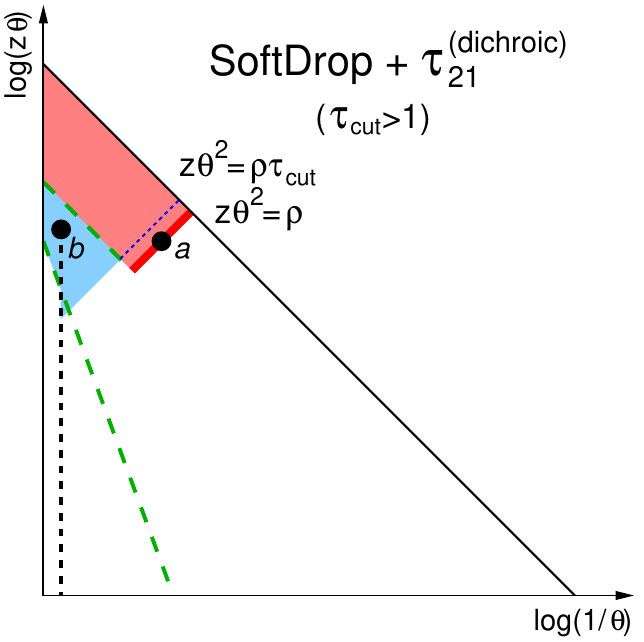}\label{fig:lund-tau21-dichroic-SD-gt1}}%
  \caption{Lund diagrams for a cut
    $\tau_{21}^{\text{(dichroic)}}<\tau_\text{cut}$, assuming that mMDT
  is used for the small jet and \SD for the large jet. Emissions $a$
  and $b$ are the emissions with the largest $z_i\theta_i^2$ in the
  mMDT and \SD jet respectively. The shaded red region corresponds to
  the vetoed region from the requirement on the (small) jet mass, and
  the shaded blue region is the extra Sudakov veto from the constraint
  on $\tau_{21}^{\text{(dichroic)}}$.
  Figure (a) corresponds to a cut
  $\tau_\text{cut}<1$ for which emissions $a$ and $b$ are identical.
  Figure (b) corresponds to $\tau_\text{cut}>1$, where one has an
  emission $\rho_b$ in the large jet such that $\rho_b>\rho$ and one
  has to veto real emissions with $z\theta^2>\rho\tau$.
  In both cases, we omitted a contribution from secondary emissions
  for readability. It corresponds to a secondary plane originating
  from emission $a$ (resp. $b$) in case (a) (resp. (b)), with a
  Sudakov veto extending down to $z\theta^2=\rho\tau_\text{cut}$ with
  $z$ measured with respect to the initial
  jet.}\label{fig:lund-dichroic-tau21}
\end{figure}  

\paragraph{LL mass distribution with a cut $\tau_{21}^\text{(dichroic)}<\tau_\text{cut}$.}
The calculation of the jet mass distribution with a cut on
$\tau_{21}^\text{(dichroic)}$ has to be separated in the same three
possible of mass orderings as before, corresponding to
$\tau_{21}^\text{(dichroic)}$smaller, equal or larger than $1$.
The three situations are represented in
Fig,~\ref{fig:lund-dichroic-tau21} for the case where the large jet
has been groomed with \SD using a positive $\beta$.

The case of a cut $\tau_\text{cut}<1$ is the most interesting as it is
the situation relevant for phenomenology --- the other cases would, as we
show below, also kill the signal --- and where the effect of adopting
a dichroic ratio can be explicitly seen. As for the case of the
standard $\tau_{21}$, one as to integrate over the emission $a$ which
dominates the small jet mass and veto any additional real emission
which would give a value of $\tau_{21}^\text{(dichroic)}$ larger than
$\tau_\text{cut}$, \ie any emission in the large jet with
$z\theta^2>\rho_a\tau_\text{cut}$. This gives
\begin{equation}\label{eq:distrib-tau21dichroic-less1}
  \frac{\rho}{\sigma} \frac{d\sigma}{d\rho}\Big|_{\tau_{21}<\tau_\text{cut}}^\text{dichroic}
  \overset{\tau_\text{cut}<1}{=}
  \int_0^1 \frac{d\theta_a^2}{\theta_a^2}\frac{dz_a}{z_a}
  \frac{\alpha_s(z_a\theta_a)C_i}{\pi}\rho\delta(\rho-\rho_a)
  \,\Theta(z_a>x_\text{cut})
  e^{-R_{\tau,\text{SD}}^\text{(primary)}-R_\tau^\text{(secondary)}},
\end{equation}
with $R_{\tau,\text{SD}}^\text{(primary)}$ and
$R_\tau^\text{(secondary)}$ again given
by~(\ref{eq:tau21-SD-Rprimary}) and~(\ref{eq:tau21-plain-Rsecondary}).
Compared to Eq.~(\ref{eq:tau21-SD-mass}), one clearly sees that the
lower bound of the $z_a$ ($z_1$ in~(\ref{eq:tau21-SD-mass}))
integration has been increased, corresponding to a reduction of the
QCD cross-section in the dichroic case.

For completeness, we briefly discuss the case $\tau_\text{cut}\ge 1$.
Situations with zero or one emissions in the large jet with
$\rho_b>\rho$ give $\tau_{21}^\text{(dichroic)}\le 1$ and are
therefore accepted. For situations with (at least) two emissions
$\rho_b>\rho_c>\rho_a$, one only accepts the cases with
$\rho_c/\rho<\tau_\text{cut}$.
Thus, the only situation which has to be vetoed is
$\rho_b>\rho_c>\rho\tau_\text{cut}$.

This can be reorganised in a slightly more convenient way.
First, if there is no emission $\rho_b$ with $\rho_b>\rho\tau_\text{cut}$, the
veto condition cannot be satisfied, meaning the case always
contributes to the cross-section.
For cases with at least one emission such that $\rho_b>\rho\tau_\text{cut}$, one
needs an additional veto on emissions $c$ such that
$\rho_b>\rho_c>\rho\tau_\text{cut}$.
This situation corresponds to
Fig.~\ref{fig:lund-tau21-dichroic-SD-gt1}.
If one assumes that the small jet is obtained using mMDT and the large
jet using \SD, and if we denote by $R_\text{out}$ the radiator
corresponding to the region in the large jet but outside the small one
(\ie the shaded blue region in Fig.~\ref{fig:lund-dichroic-tau21}),
this yields
\begin{align}\label{eq:tau21dichroic-largetau}
  \frac{\rho}{\sigma} \frac{d\sigma}{d\rho}\Big|_{\tau_{21}<\tau_\text{cut}}^\text{dichroic}&
    \overset{\tau_\text{cut}>1}{=}
  \int_0^1 \frac{d\theta_a^2}{\theta_a^2}\frac{dz_a}{z_a}
  \frac{\alpha_s(z_a\theta_a)C_i}{\pi}\rho\delta(\rho-\rho_a)
  \,\Theta(z_a>x_\text{cut})\\
  &\bigg[ e^{-R_\text{out}(\rho\tau_\text{cut})}+\int_0^1 \frac{d\theta_b^2}{\theta_b^2}\frac{dz_b}{z_b}
  \frac{\alpha_s(z_b\theta_b)C_i}{\pi}\Theta(\rho_b>\rho\tau_\text{cut})
  \,\Theta(x_\text{cut}>z_b>\zcut\theta_b^\beta)\nonumber\\
  & \phantom{e^{-R_\text{out}(\rho\tau_\text{cut})}+\int_0^1 \frac{d\theta_b^2}{\theta_b^2}\frac{dz_b}{z_b}\frac{\alpha_s(z_b\theta_b)C_i}{\pi}}
  e^{-R_{\text{out}}(\rho\tau_\text{cut},\rho_b,\theta_b)-R_\tau^\text{(secondary)}(\rho_b,\rho\tau_\text{cut}/\rho_b,\theta_b)}\bigg]\nonumber
\end{align}
In this expression, $R_\text{out}(\rho\tau_\text{cut})$ is trivially
given by
$R_\text{SD}(\rho\tau_\text{cut})-R_\text{mMDT}(\rho\tau_\text{cut})$.
In the presence of an emission $b$, one has to be careful that \SD
will keep emissions at angles smaller than $\theta_b$, and therefore
$R_{\text{out}}(\rho\tau_\text{cut},\rho_b,\theta_b)=R_{\tau,\text{SD}}^{\text{(primary)}}(\rho_b,\rho\tau_\text{cut}/\rho_b,\theta_b)-R_\text{mMDT}(\rho\tau_\text{cut})$.
We note that in \eqref{eq:tau21dichroic-largetau}, the integration
over $z_a$ can be done explicitly and gives an overall factor
$R_\text{mMDT}'(\rho)$.
Finally, \eqref{eq:tau21dichroic-largetau} does not coincides with
\eqref{eq:distrib-tau21dichroic-less1} when $\tau_\text{cut}\to
1$. This is simply because situations with a single emission
$\rho_b>\rho$ give $\tau_{21}^\text{(dichroic)}=1$, yielding a
discontinuity at $\tau_\text{cut}=1$, or, equivalently, a contribution
to the $\tau_{21}$ distribution proportional to $\delta(\tau_{21}-1)$.

\subsection{Energy-Correlation functions $C_2^{(\beta=2)}$ or $D_2^{(\beta=2)}$}

The last shape we want to discuss is the energy-correlation-function
ratio $D_2$, or, almost equivalently, $C_2$ (which differs from $D_2$
by a factor $\rho$).
As before, we first give an analytic expression, valid in the
leading-logarithmic approximation, for the value of $D_2$ for a given
jet. We then compute the mass distribution with a cut
$D_2<D_\text{cut}$.

\paragraph{Approximate $D_2$ value at LL}
Consider once again a set of $n$ emissions with momentum fractions
$z_i$ and emitted at angles $\theta_i$ from the parent parton, and
define $\rho_i=z_i\theta_i^2$.
We can assume, as before, that the jet mass is dominated by emission
1, \ie the jet mass is $\rho\approx\rho_1$.
From Eq.~\eqref{eq:ecf-e3} we then have
\begin{align}
  e_3^{(\beta=2)}
  & = \sum_{i<j<k\in\text{jet}} z_i z_j z_k
      \theta_{ij}^2 \theta_{ik}^2 \theta_{jk}^2
    \approx \sum_{i<j} z_i z_j
      \theta_{ij}^2 \theta_i^2 \theta_j^2\\
  & \approx \sum_{i<j} z_i z_j
      \text{max}(\theta_i^2,\theta_j^2) \theta_i^2 \theta_j^2
    \approx \sum_{i<j} \rho_i \rho_j
      \text{max}(\theta_i^2,\theta_j^2),
\end{align}
where, for the second equality we have used the fact that all
emissions are soft so we can neglect triplets which do not involve the
leading parton, and the third equality comes from the strong angular
ordering between emissions, valid at LL.

For pairs $i,j$ which do not include emission 1, we have, assuming
$\theta_i\ll \theta_j$ , $\rho_i \rho_j \theta_j^2\ll \rho_1 \rho_j
\theta_j^2< \rho_1 \rho_j
\max(\theta_1^2,\theta_j^2)$. These contributions can therefore be
neglected and we have
\begin{equation}
  e_3^{(\beta=2)}
  \approx \rho \sum_{i,\theta_i<\theta_1} \rho_i \theta_1^2 + \rho \sum_{i,\theta_i>\theta_1} \rho_i \theta_i^2.
\end{equation}
At LL accuracy, only one emission, that we will denote by ``2'' will
dominate the sum and we have
\begin{equation}
  e_3 \approx \rho \rho_2 \text{max}(\theta_1^2,\theta_2^2)
  \quad \Rightarrow \quad
  D_2 = \frac{e_3}{e_2^3} \approx \frac{\rho_2}{\rho^2} \text{max}(\theta_1^2,\theta_2^2),
\end{equation}
which has an extra factor $\text{max}(\theta_1^2,\theta_2^2)/\rho$
compared to the $\tau_{21}$ ratio. Alternatively, we can work with
$C_2=\rho D_2$.
Note that $D_2$ can be larger than 1 and, in this case, the LL
approximation refers to $C_2\ll 1$ which is dominated by
$\rho_2\ll \rho$ and $\theta_{1,2}^2\ll 1$.

Finally, when imposing a constraint on $D_2$, we are also sensitive to
secondary emissions from 1. A gluon ``2'' emitted with a momentum
fraction $z_2$ (measured with respect to to $z_1$) at an angle $\theta_{12}$ from
1, will give a (dominant) contribution $z_1^2z_2\theta_{12}^2\theta_1^4$
to $e_3$ (taking the leading parton, and emissions 1 and 2 as $i$, $j$
and $k$). We therefore have
\begin{equation}
  D_{2,\text{secondary}} \approx \frac{z_2\theta_{12}^2}{\rho}.
\end{equation}

\paragraph{LL mass distribution with a cut $D_2<D_\text{cut}$}
The LL mass distribution with a cut on $D_2$ proceeds as for
$\tau_{21}$ above except that now the constraint on the shape will
impose a Sudakov vetoing emissions for which
$\rho_2 \text{max}(\theta_1^2,\theta_2^2)>\rho^2D_\text{cut}$, with
$\rho_2<\rho$.

\begin{figure}[t]
  \centering
  \subfloat[]{\includegraphics[width=0.48\textwidth]{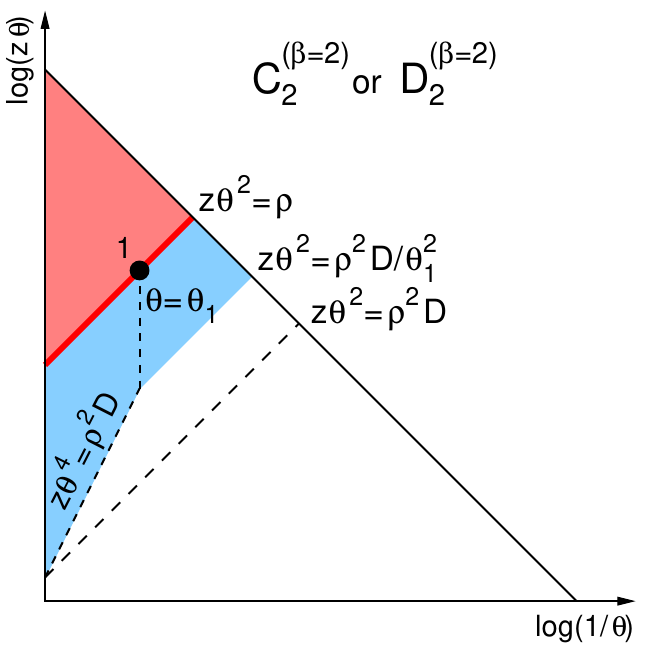}\label{fig:Lund-D2}}%
  \hfill%
  \subfloat[]{\includegraphics[width=0.48\textwidth]{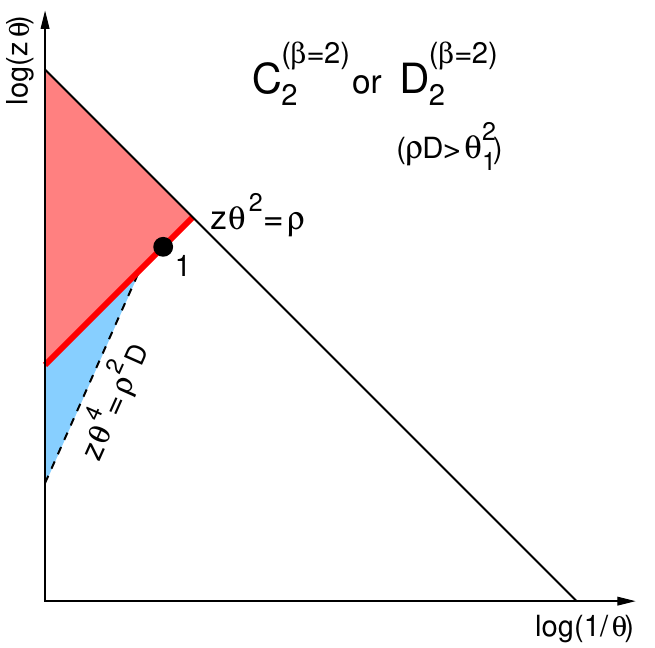}\label{fig:Lund-D2-large}}%  
  \caption{Lund diagrams for a constraint $D_2<D_\text{cut}$. For
    $D_\text{cut}<1$, we are always in the situation depicted on
    Fig.~(a), while for $1<D_\text{cut}<1/\rho$, we have either the case
    of Fig.~(a) for $\rho D_\text{cut}<\theta_1^2$ or the case of
    Fig.~(b) for $\rho D_\text{cut}>\theta_1^2$. As above, an extra
    veto for secondary emissions from emission 1 (only in case (a)) is
    not shown for clarity.}\label{fig:Lund-D2-all}
\end{figure}

The corresponding phase-space is represented in
Fig.~\ref{fig:Lund-D2-all}.
We have to consider two regimes. First, for $D_\text{cut}<1$, we have
$\rho^2D/\theta_1^2<\rho$ for any $\rho<\theta_1^2<1$, resulting in
the phase-space represented in Fig.~\ref{fig:Lund-D2}. Then. for
$1<D_\text{cut}<1/\rho$, one can either have $\rho
D_\text{cut}<\theta_1^2$ or $\rho
D_\text{cut}>\theta_1^2$. For the former corresponds one again recovers
Fig.~\ref{fig:Lund-D2}, but for the latter, only the region
$\rho^2D_\text{cut}<\rho_2\theta_2^2<\rho$ (\ie $\theta_2^2>\rho D$),
shown in Fig.~\ref{fig:Lund-D2-large}.

The mass distribution with a cut on $D_2$ can be written as 
\begin{align}
  \frac{\rho}{\sigma} \frac{d\sigma}{d\rho}\Big|_{D_2<D_\text{cut}}
  & = \int_0^1 \frac{d\theta_1^2}{\theta_1^2}\frac{dz_1}{z_1}
      \frac{\alpha_s(z_1\theta_1)C_i}{\pi}\rho\delta(\rho-\rho_1)
      \exp[-R_D^\text{(primary)}-R_D^\text{(secondary)}]\label{eq:D2-plain-mass}\\
  R_D^\text{(primary)}
    & =  \int_0^1 \frac{d\theta_2^2}{\theta_2^2}\frac{dz_2}{z_2}
      \frac{\alpha_s(z_2\theta_2)C_i}{\pi}\Theta\Big(\frac{\rho_2}{\rho} \frac{\text{max}(\theta_1^2,\theta_2^2)}{\rho}>D_\text{cut}\Big)\label{eq:D2-plain-Rprimary},\\
  R_D^\text{(secondary)}
    & = \int_0^{\theta_1^2} \frac{d\theta_{12}^2}{\theta_{12}^2}\int_0^1\frac{dz_2}{z_2} \frac{\alpha_s(z_1z_2\theta_{12})C_A}{\pi}\Theta\Big(\frac{z_2\theta_{12}^2}{\rho}>D_\text{cut}\Big).\label{eq:D2-plain-Rsecondary}
\end{align}
For the two cases above, one finds, at LL (including both the
mass and shape vetoes)
\begin{align}
  R_D^\text{(primary)} = \frac{C_i}{2\pi\alpha_s\beta_0^2}
    & \begin{cases}
    \frac{1}{3} W(1-2\lambda_\rho-\lambda_D)
    +\frac{2}{3} W(1-2\lambda_\rho-\lambda_D+\frac{3}{2}\lambda_1)\\
    \qquad -2 W(1-\frac{2\lambda_\rho+\lambda_D-\lambda_1+\lambda_B}{2})
    + W(1-\lambda_B) & \text{if }\rho D<\theta_1^2\\
    \frac{1}{3} W(1-2\lambda_\rho-\lambda_D)
    +\frac{2}{3} W(1-\frac{\lambda_\rho-\lambda_D}{2})\\
    \qquad -2 W(1-\frac{\lambda_\rho+\lambda_B}{2})
    + W(1-\lambda_B) & \text{if }\rho D>\theta_1^2\\    
  \end{cases}\\
  R_D^\text{(secondary)} = \frac{C_A}{2\pi\alpha_s\beta_0^2} &
  \bigg[                                                              
     W\Big(1-2\lambda_\rho-\lambda_D+\frac{3}{2}\lambda_1\Big)
    -2 W\Big(1-\frac{3\lambda_\rho+\lambda_D-2\lambda_1+\lambda_{B_g}}{2}\Big)\nonumber\\
    & + W\Big(1-\lambda_\rho-\frac{\lambda_1}{2}+\lambda_{B_g}\Big)\bigg]
      \Theta(2\lambda_\rho+\lambda_D-\lambda_1>\lambda_{B_g})
\end{align}
where $\lambda_\rho$ and $\lambda_B$ are defined as before and we have
introduced $\lambda_D=2\alpha_s\beta_0\log(1/D_\text{cut})$ and
$\lambda_1=2\alpha_s\beta_0\log(1/\theta_1^2)$.
$R_D^\text{(primary)}$ is manifestly continuous at $\rho D=\theta_1^2$.

As for the case of $\tau_{21}$, similar expressions can be obtained
with \SD. In this case, the integration over emission 1 in
Eq.~(\ref{eq:D2-plain-mass}) has to be restricted to the region where
emission 1 passes the \SD condition. Focusing on the case
$\rho<\zcut$, one has ,for a given $z_1\theta_1^2=\rho$,
$z_1>(\zcut^2\rho^\beta)^{\frac{1}{2+\beta}}$ or
$\theta_1<(\rho/\zcut)^{\frac{1}{2+\beta}}$.
The Sudakov for primary emissions also gets modified by \SD as one
only needs to veto emissions for which either
$z_2>\zcut\theta_2^\beta$ or $\theta_2<\theta_1$. The veto on
secondary emissions is unchanged compared to the plain-jet case.
After some relatively painful manipulations, one gets
\begin{align}
  R_{D,\text{SD}}^\text{(primary)} = R_D^\text{(primary)}
  - \frac{C_i}{2\pi\alpha_s\beta_0^2}
  \bigg[\frac{1}{3}W(1-2\lambda_\rho-\lambda_D)
  +\frac{W(1-\lambda_c)}{1+\beta}\nonumber\\
  -\frac{1}{3}W\Big(1-2\lambda_\rho-\lambda_D+\frac{3}{2}\lambda_1\Big)
  -\frac{1}{1+\beta}W\Big(1-\lambda_c-\frac{1+\beta}{2}\lambda_1\Big)
  \bigg]  
\end{align}
if $\rho D<\theta_1^2$ and  $\rho^2 D< \zcut \theta_1^{4+\beta}$,
\begin{align}
  R_{D,\text{SD}}^\text{(primary)} = R_D^\text{primary)}
  - \frac{C_i}{2\pi\alpha_s\beta_0^2}
  \bigg[\frac{1}{3}W(1-2\lambda_\rho-\lambda_D)
  +\frac{W(1-\lambda_c)}{1+\beta}\nonumber \\
  -\frac{4+\beta}{3(1+\beta)}W\Big(1-\frac{(1+\beta)(2\lambda_\rho+\lambda_D)+3\lambda_c}{4+\beta}\Big)
  \bigg],
\end{align}
if either $\rho D<\theta_1^2$ and $\rho^2 D> \zcut
\theta_1^{4+\beta}$, or  $\rho D>\theta_1^2$ and $\zcut^2 \rho^\beta
D^{2+\beta}<1$, and
%
% If we want to write the 1st case above as the 2nd - a correction
% corresponding to the extra triangle:
%
%   -\bigg[
%   \frac{1}{3}W\Big(1-2\lambda_\rho-\lambda_D+\frac{3}{2}\lambda_1\Big)+\frac{1}{1+\beta}W\Big(1-\lambda_c-\frac{1+\beta}{2}\lambda_1\Big)\\
%   -\frac{4+\beta}{3(1+\beta)}W\Big(1-\frac{(1+\beta)(2\lambda_\rho+\lambda_D)+3\lambda_c}{4+\beta}\Big)
%   \bigg]\\
%
\begin{align}
  R_{D,\text{SD}}^\text{(primary)} = R_{\text{SD}}^{\text{(LL)}},
\end{align}
if $\rho D>\theta_1^2$ and $\zcut^2 \rho^\beta D^{2+\beta}>1$.

The first result corresponds to the situation where one has a
contribution similar to $\delta R_{\tau}^\text{(SD)}$ in the
$\tau_{21}$ case, coming from the extra small triangle
$z<\zcut\theta^\beta$, $\theta<\theta_1$. The existence of this
extra region requires $\rho^2 D> \zcut \theta_1^{2+\beta}$.
The second result with ``normal'' \SD grooming, covering both
kinematic configurations in Fig.~\ref{fig:Lund-D2-all}.
The third result corresponds to the case of
Fig.~\ref{fig:Lund-D2-large} where the shaded blue region is fully
outside the region allowed by the \SD condition, in which case, the
shape cut has no effects and one recovers a \SD mass Sudakov.

This finishes our calculations for our sample of shapes in the case of
QCD jets. Before comparing our results with Monte Carlo simulations,
we briefly discuss the case of signal jets so as to be able to discuss
the performance when tagging $2$-prong boosted objects.

\subsection{Calculations for signal jets}\label{sec:two-prongs-analytic-signal}

In order to be able to study the performance of two-prong taggers
analytically, we also need expressions for signal jets. Generally
speaking, signal jets are dominated by the decay of a colourless heavy
object of mass $m_X$ into two hard partons. Here, we will assume the
decay is in a $q\bar q$ pair, which is valid for electroweak bosons
W/Z/H and for a series of BSM candidates.
If the decay happens at an angle $\theta_1$ (measured in units of the
jet radius $R$) and the quark carries a fraction $1-z_1$ of the
boson's transverse momentum, we have
\begin{equation}\label{eq:signal-mass-constraint}
  m_X^2=z_1 (1-z_1) \theta_1^2 (p_tR)^2
  \qquad\text{\ie}\quad
  \rho_X = \frac{m_X^2}{p_t^2R^2}=z_1 (1-z_1) \theta_1^2.
\end{equation}
Furthermore, we will use index 0 (resp. 1) to refer to the quark
(resp. antiquark).

\begin{figure}
  \centerline{\includegraphics[width=0.48\textwidth]{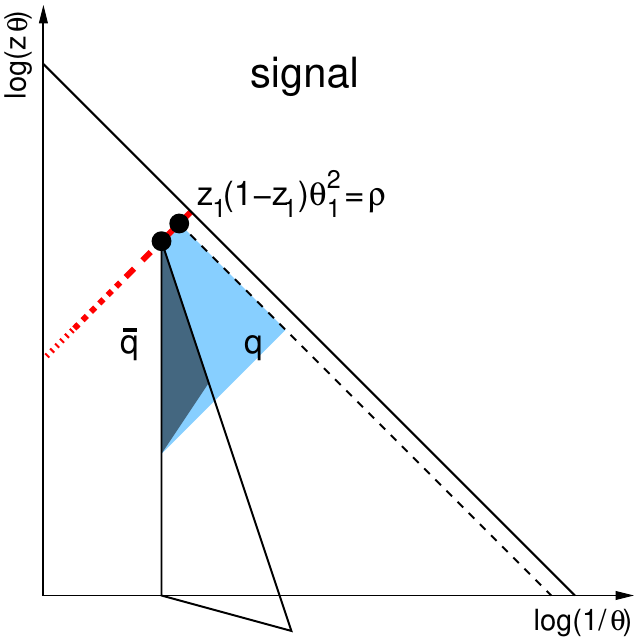}}
  \caption{Lund plane for signal jets. The two solid dots correspond
    to the initial $a\bar{a}$ splitting which satisfies
    $z_1(1-z_1)\theta_1^2=\rho$. A
    % secondary-like
    Lund plane originates from each of the two quarks and the shape
    constraints impose Sudakov vetos (represented as the shaded areas)
    in each of them.}\label{fig:lund-shape-signal}
\end{figure}  

The effect of a cut on a jet shape is similar to what we have just
discussed for QCD jets: it constrains additional radiation in the
jet.
The key difference with QCD jets is that now the radiation, is only
coming from the $q\bar q$ dipole. In the collinear limit sufficient
for our purpose here this is equivalent to having two secondary-like
Lund planes associated with the quark and antiquark respectively, as
depicted on Fig.~\ref{fig:lund-shape-signal}.

\paragraph{Calculation of the shape value.}
The calculation for a given shape proceeds as before by first
computing an expression for the shape value. Say that emission 2,
emitted at an angle $\theta_{02}$ from the quark (or $\theta_{12}$ from
the antiquark) and carrying a fraction $x_2$ of the jet's transverse
momentum, dominates the shape value (at LL). For the $N$-subjettiness
$\tau_{21}$ ratio, the two axes will align with the quark and
antiquark and we find
\begin{equation}
  \tau_2 = x_2 \text{min}(\theta_{02}^2,\theta_{12}^2)
    \quad \Rightarrow\quad
  \tau_{21}\approx\frac{x_2  \text{min}(\theta_{02}^2,\theta_{12}^2)}{\rho}.
\end{equation}
This expression is also valid for the dichroic ratio (just like the
contribution from secondary emission for QCD jets).
For ECFs, we get
\begin{equation}
  e_3 = z_1 (1-z_1) x_2 \theta_{01}^2\theta_{02}^2\theta_{12}^2
  \approx \rho\theta_{01}^2 x_2
  \text{min}(\theta_{02}^2,\theta_{12}^2)
  \quad \Rightarrow\quad
  D_2\approx \frac{\theta_{01}^2}{\rho}
  \frac{x_2  \text{min}(\theta_{02}^2,\theta_{12}^2)}{\rho}.
\end{equation}

\paragraph{Signal efficiency.}
In the case of signal jets with a fixed jet mass, one should compute
directly the signal efficiency, \ie the fraction of signal jets that
are accepted by the tagger and the cut on the shape.
Assuming again that the two hard prongs are identified using \SD as a
prong finder, one can write
\begin{equation}
  \epsilon_S(v<v_\text{cut}) = \int_0^1 dz_1 P_X(z_1)
  \Theta(\text{min}(z_1,1-z_1)>\zcut\theta_{01}^\beta)
  %\exp[-R_v^\text{($q$)}(v_\text{cut};z_1)-R_v^\text{($\bar q$)}(v_\text{cut};z_1)]\label{eq:signal-eff-basic}\\
  \,e^{-R_v^\text{($q$)}(v_\text{cut};z_1)-R_v^\text{($\bar q$)}(v_\text{cut};z_1)},\label{eq:signal-eff-basic}\\
\end{equation}
where $P_X(z_1)$ is the probability density for the quark to carry a
fraction $1-z_1$ of the boson's transverse momentum (for simplicity,
we will assume $P_X(z)=1$ in what follows), $\theta_{01}$ is
constrained by Eq.~\eqref{eq:signal-mass-constraint}, and the veto on
radiations in the quark and antiquark prongs takes the form of a
Sudakov suppression, with the two related by a
$z_1\leftrightarrow 1-z_1$ symmetry
$R_v^\text{($\bar q$)}(v;z_1)=R_v^\text{($q$)}(v;1-z_1)$.
As already discussed in Sec.~\ref{sec:calc-groomed-mass-signal}, an
important aspect of signal jets is that $P_X$ is finite when $z_1$ or
$1-z_1$ goes to 0.

Note that from the above signal efficiency, one can recover the
differential distribution of the shape value using
\begin{equation}
  \left. \frac{v}{\sigma}\frac{d\sigma}{dv}\right|_\text{signal}
  = \frac{1}{\epsilon_S(\text{no $v$ cut})}
  \left.\frac{d\epsilon_S(v<v_\text{cut})}{d\log(v_{\text{cut}})}\right|_{v_\text{cut}=v}.
\end{equation}

The Sudakov exponents can be computed explicitly for the $\tau_{21}$
ratio and $D_2$.
For $\tau_{21}$ (``standard'' or dichroic), we find, using
$x_2=(1-z_1)z_2$
\begin{equation}
  R_v^\text{($\bar q$)}(v;z_1)
  = \int_0^{\theta_{01}^2} \frac{d\theta_{02}^2}{\theta_{02}^2}
  \int_0^1 \frac{dz_2}{z_2}
  \frac{\alpha_s(x_2\theta_{02})C_F}{\pi}
  \Theta((1-z_1)z_2\theta_{02}^2>\rho\tau_\text{cut})
  \Theta((1-z_1)^2z_2\theta_{02}^2<\rho),
\end{equation}
where the last condition of the first line imposes that emission 2
does not dominate the mass.\footnote{This is mostly an artefact of our
  approximations. In the case of signal jets with $z_1\ll 1$, this is
  equivalent to saying that the effect of the shape corresponds to the
  shaded blue region in Fig.~\ref{fig:lund-tau21} which extends up to
  $z\theta^2=\rho$, with the region above corresponding to the
  structure which gives the mass. In practice, this condition is valid
  up to finite squared logarithms of $1-z_1$ when $1-z_1>z_1$, \ie
  well beyond our current accuracy.}
At leading logarithmic accuracy, including as well
hard-collinear splittings by imposing $z_2<\exp(B_q)$ as before, one
gets (with $\log(1/\tau)+B_q>0$)
\begin{align}
  R_\tau^\text{($\bar q$)}(v;z_1) \overset{\text{LL}}=
  & \frac{C_F}{2\pi\alpha_s\beta_0^2}\bigg\{
    \bigg[ W\Big(1-\frac{\lambda_\rho-\lambda_z+\lambda_-}{2}-\lambda_B\Big)
        -2 W\Big(1-\frac{\lambda_\rho+\lambda_-+\lambda_\tau+\lambda_B}{2}\Big)
    \nonumber\\
  & + W\Big(1-\frac{\lambda_\rho+\lambda_z+\lambda_-}{2}-\lambda_\tau\Big)\bigg]
    -\bigg[W\Big(1-\frac{\lambda_\rho-\lambda_z+\lambda_-}{2}-\lambda_B\Big)\\
  & -2 W\Big(1-\frac{\lambda_\rho+\lambda_B}{2}\Big)
    + W\Big(1-\frac{\lambda_\rho+\lambda_z-\lambda_-}{2}\Big)\bigg]
    \Theta(\lambda_z-\lambda_->\lambda_B)\bigg\},\nonumber
\end{align}
where $\lambda_z=2\alpha_s\beta_0\log(1/z_1)$ and $\lambda_-=2\alpha_s\beta_0\log(1/(1-z_1))$.

For $D_2$ we find similarly (with $\log(1/\tau)+B_q>\log(z_1^2(1-z_1))$
\begin{align}
  R_D^\text{($\bar q$)}(v;z_1) 
  & = \int_0^{\theta_{01}^2} \frac{d\theta_{02}^2}{\theta_{02}^2}
  \int_0^1 \frac{dz_2}{z_2}
  \frac{\alpha_s(x_2\theta_{02})C_F}{\pi}
  \Theta\Big(\frac{z_2\theta_{02}^2}{z_1}>\rho D\Big)\Theta((1-z_1)^2z_2\theta_{02}^2<\rho)
  \nonumber\\
  & \overset{\text{LL}}= \frac{C_F}{2\pi\alpha_s\beta_0^2}\bigg\{
    \bigg[ W\Big(1-\frac{\lambda_\rho-\lambda_z+\lambda_-}{2}-\lambda_B\Big)
        -2 W\Big(1-\frac{\lambda_\rho+\lambda_z+2\lambda_-+\lambda_D+\lambda_B}{2}\Big)
    \nonumber\\
  & + W\Big(1-\frac{\lambda_\rho+3\lambda_z+3\lambda_-}{2}-\lambda_D\Big)\bigg]
    -\bigg[W\Big(1-\frac{\lambda_\rho-\lambda_z+\lambda_-}{2}-\lambda_B\Big)\\
  & -2 W\Big(1-\frac{\lambda_\rho+\lambda_B}{2}\Big)
    + W\Big(1-\frac{\lambda_\rho+\lambda_z-\lambda_-}{2}\Big)\bigg]
    \Theta(\lambda_z-\lambda_->\lambda_B)\bigg\},\nonumber
\end{align}

These expressions will be compared to Monte Carlo simulations in the
next section where we also discuss key phenomenological observations.

\section{Comparison to Monte Carlo simulations}

In this section, we compare our analytic results to Monte Carlo
simulations obtained with Pythia. For all the Monte Carlo simulations in
this chapter, we have relied on the samples used in the ``two-prong
tagger study'' performed in the context of the Les Houches
Physics at TeV Colliders workshop in 2017 (Section~III.2 of
Ref.~\cite{Bendavid:2018nar}).
Background jets are obtained from a dijet sample while signal jets
are obtained from a WW event sample.

In order to streamline our discussion, we focus on a selection of five working
points:
\begin{itemize}
  \setlength\itemsep{0cm}
  \item $\tau_{21}^\text{(SD)}$: \SD jet mass with a cut on
    $\tau_{21}^{(\beta=2)}$ computed on the \SD jet;
  \item $\tau_{21}^\text{(mMDT)}$: mMDT jet mass with a cut on
    $\tau_{21}^{(\beta=2)}$ computed on the mMDT jet;
  \item $\tau_{21}^\text{(dichroic)}$: mMDT jet mass with a cut on
    $\tau_{21}^{(\text{dichroic})}=\tau_2^\text{(SD)}/\tau_1^\text{(mMDT)}$;
  \item $D_2^\text{(SD)}$: \SD jet mass with a cut on
    $D_2^{(\beta=2)}$ computed on the \SD jet;
  \item $D_2^\text{(mMDT)}$: mMDT jet mass with a cut on
    $D_2^{(\beta=2)}$ computed on the mMDT jet.
\end{itemize}
Note that above selection of working points never includes the plain jet. Although using ungroomed jets can show good tagging performances (as expected from the discussion below),
they usually have poor resilience against non-perturbative effects (see next section) and are therefore
less relevant for a comparison to analytic calculations.

\begin{figure}
  \centering
  \subfloat[]{\includegraphics[width=0.48\textwidth]{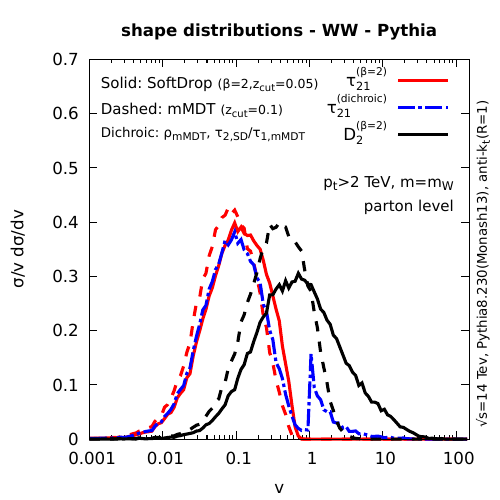}\label{fig:shape-distribs-sig-pythia}}%
  \hfill%
  \subfloat[]{\includegraphics[width=0.48\textwidth]{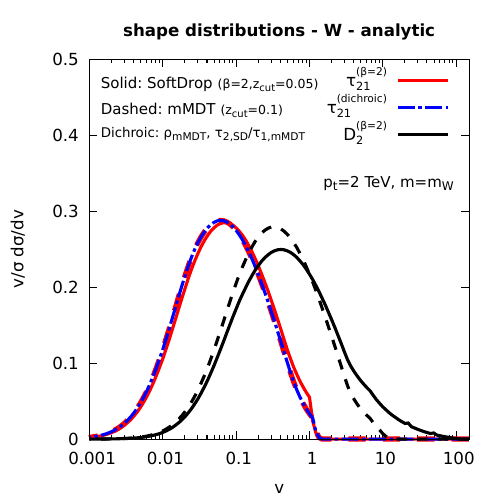}\label{fig:shape-distribs-sig-analytic}}\\
  \subfloat[]{\includegraphics[width=0.48\textwidth]{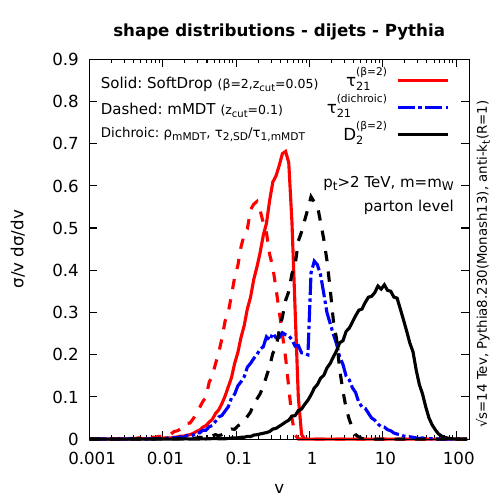}\label{fig:shape-distribs-qcd-pythia}}%
  \hfill%
  \subfloat[]{\includegraphics[width=0.48\textwidth]{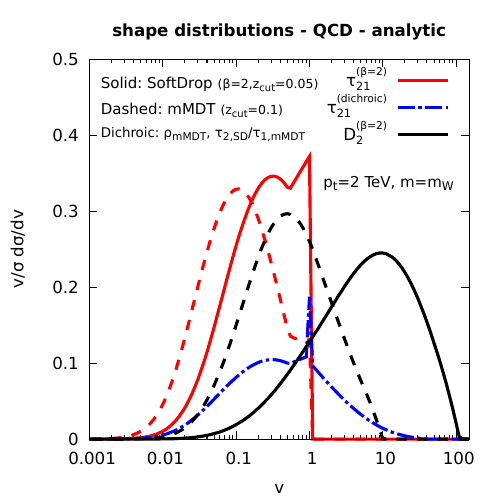}\label{fig:shape-distribs-qcd-analytic}}
  \caption{Distributions for our representative set of shapes as
    obtained from Pythia (left) and from the analytic calculations of
    Sec.~\ref{sec:two-prongs-analytic} (right). The top row
    corresponds to signal (WW) jets while the bottom row shows results for
    background (QCD) jets.}\label{fig:shape-distribs}
\end{figure}

We first focus on the shape distributions, shown for QCD and signal (W) jets in
Fig.~\ref{fig:shape-distribs}.
Globally speaking, we see that the main features observed in the Monte
Carlos simulations are well reproduced by our simple analytic
calculations, although the former exhibit distributions that are generally more peaked than the ones obtained with the analytics. 
We observe that the signal distribution is, to a large
extend, independent of the level of grooming (\SD or
mMDT). Analytically, this comes from the fact that the grooming
procedure stops at the angle $\theta_{01}$ of the $\text{W}\to q\bar q$
decay, keeping the full radiation inside the two prongs unaffected by
the groomer.
The small differences seen in the Pythia simulations are likely due to
radiation outside the $q\bar q$ prongs and to initial-state radiation
which is less efficiently groomed by \SD (with $\beta=2$) than by
mMDT, shifting the former to slightly larger values than the latter.
In the case of $D_2$, the differences between the \SD and mMDT results
also involve the fact that the $D_2$ Sudakov has a stronger dependence
on the $p_t$ sharing between the quark and antiquark than $\tau_{21}$.
A specific case of this independence of signal distributions to grooming
is that the distribution for the dichroic $\tau_{21}$ ratio is very
close to the ``standard'' ones, again with little differences seen \eg
by the presence of a small peak at $\tau_{21}>1$.

Turning to QCD jets, the situation is clearly different: distributions
shift to smaller values when applying a tighter grooming \ie when
going from \SD to mMDT. This shift is reasonably well reproduced in
the analytic calculation and it is due to the fact that jet shapes are
sensitive to radiations at large angles --- larger than the angle of
the two-prong decay dominating the jet mass --- which is present in
QCD jets, but largely absent in W jets.
{\em This has a very important consequence: one expects the tagging
  efficiency to increase for lighter grooming on the jet shape} as the
signal is largely unmodified and the background peak is kept at large
values of the shape.
In this context, the case of the dichroic $N$-subjettiness ratio is
also interesting: the dichroic distribution (mixing mMDT and \SD
information) has larger values than both the corresponding \SD and mMDT
distributions.
In other words, at small $\tau_{21}$, relevant for tagging purposes,
the dichroic distribution is lower than the \SD and mMDT ones.
From an analytic viewpoint, one expects the dichroic distribution to
be smaller than the \SD distribution because, for the same Sudakov
suppression, it imposes a tighter condition on the emission that gives
the mass, and smaller than the mMDT distribution because keeping more
radiation at larger angles increases the Sudakov suppression
(cf.~Figs.~\ref{fig:lund-tau21-SD}
and~\ref{fig:lund-tau21-dichroic-SD}).
{\em This is our second important observation: one expects dichroic
  ratios to give a performance improvement.}

One last comment about Fig.~\ref{fig:shape-distribs} is the presence
of peaks for $\tau_{21}^\text{(dichroic)}\gtrsim 1$ in the Pythia
simulation and spikes at $\tau_{21}^\text{(dichroic)}=1$ in our
analytic calculation.
As discussed in the analytic calculation of
Sec.~\ref{sec:2prongs-analytic-dichroic}, the cumulative
$\tau_{21}^\text{(dichroic)}$ distribution is discontinuous at
$\tau_{21}^\text{(dichroic)}=1$ and this directly gives a
$\delta(\tau_{21}^\text{(dichroic)}-1)$ contribution to
Fig.~\ref{fig:shape-distribs-qcd-analytic}.\footnote{For readability,
  the  peak has been scaled down on the plot.}
Once we go beyond leading logarithmic approximation --- for example,
following the technique introduced in~\cite{Napoletano:2018ohv} ---
this is replaced by a Sudakov peak corresponding to what is seen in the
Pythia simulations from Fig.~\ref{fig:shape-distribs-qcd-pythia}.
We also note a kink in the $\tau_{21}$ and
$\tau_{21}^\text{(dichroic)}$ distributions around 0.5. This
corresponds to the point below which secondary emissions start to
contribute, namely $\log(\tau_{21})=B_g$.
Since this is in a region where our approximation $\tau_{21}\ll 1$ is
not clearly satisfied, subleading corrections play an important role.

\begin{figure}
  \centering
  \subfloat[]{\includegraphics[width=0.48\textwidth]{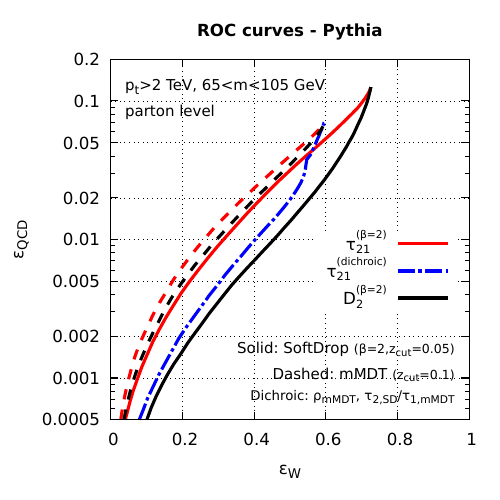}\label{fig:shape-rocs-pythia}}%
  \hfill%
  \subfloat[]{\includegraphics[width=0.48\textwidth]{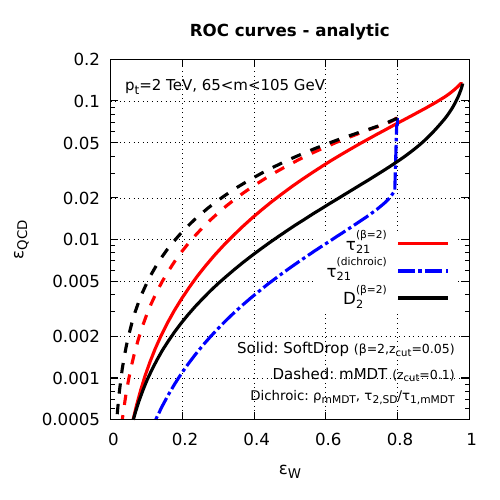}\label{fig:shape-rocs-analytic}}%  
  \caption{ROC curves corresponding to our representative set of
    shapes as obtained from Pythia (left) and from the analytic
    calculations (right).}\label{fig:shape-rocs}
\end{figure}

We now turn to a direct analysis of the tagging performance of our
tools with the ROC curves shown on Fig.~\ref{fig:shape-rocs}.
Note that the tagging efficiencies include both the effect of the
requirement on the jet mass and of the cut on the jet shape. For signal
jets, we have assumed that the jet mass is exactly the W mass if
the jet passes the $\zcut$ (or \SD) condition on the $\text{W} \to q\bar q$
decay.
The two important features highlighted above are indeed seen here:
decreasing the level of grooming results in an increased tagging
performance, as does using dichroic ratios. In the first case, note
that the situation is more delicate at large signal efficiency (close
to the endpoint of the ROC curves corresponding to no constraint on
the jet shape) since one also has to include the effect of the groomer
on constraining the jet mass.
Note also that our analytic calculation generally overestimates the
signal efficiency, which is likely due to various oversimplifications
mentioned earlier.

{\em The other important observation (our third) is that a constraint
  on $D_2$ outperforms a constraint on $\tau_{21}$.}\footnote{We refer
  here to the standard definition of $\tau_{21}$. A proper assessment
  of the dichroic ratio would also require using a dichroic version of
  $D_2$ which is done in the next section.}
Although there is only a small gain (that the simple analytic
calculation fails to capture) with tight (mMDT) grooming, there is a
clear gain in using $D_2$ when using a looser grooming (\SD) \ie when
opening to larger angles. This is seen in both the analytic
calculation and Monte Carlo simulations.
This feature can be explained from our analytic approach.
Based on Fig.~\ref{fig:lund-shape-signal}, fixing the signal
efficiency (say for a given $z_1$ or, equivalently, $\theta_1$) is
equivalent to selecting how much of the radiation is vetoed, \ie fixing
the lower end of the shaded blue region. This, in turns, determines 
the behaviour at small angles ($\theta<\theta_1$) in the case of
background jets. The remaining differences between $\tau_{21}$ and
$D_2$ therefore comes from radiation at angles larger than
$\theta_1$. For the latter, $D_2$ clearly imposes a stronger
constraint (related to its $z\theta^4$ behaviour) than $\tau_{21}$
(with a lighter $z\theta^2$) behaviour,
cf.~Figs~\ref{fig:lund-tau21-plain} and~\ref{fig:Lund-D2}.

To conclude this section, we want to make a final comment on two other
observations emerging from the analytic results.
First, in the case of groomers (used here to find the two prongs
dominating the jet mass) we had a strong Sudakov suppression of QCD
jets for a relatively mild (typically $1-2\zcut$) suppression for
signal jets (cf.~Chapter~\ref{calculations-substructure-mass}). In
contrast, imposing a cut on a jet shape yields a Sudakov suppression
for both the background and the signal. This means that if we want to
work at a reasonable signal efficiency, the cut on the shape should
not be taken too small. Our analytic calculations, strictly valid in
the limit $v_\text{cut}\ll 1$ are therefore only valid for
qualitative discussions and a more precise treatment is required for
phenomenological predictions. We refer to
Refs.~\cite{Larkoski:2015kga,Napoletano:2018ohv} for practical
examples.

Our last remark is also {\em our last important point: for a fixed
  mass and cut on the jet shape, the signal efficiency will remain
  mostly independent of the jet $p_t$ but background jets will be
  increasingly suppressed for larger $p_t$.} Analytically, the
associated Sudakov suppression in the signal is independent of
$\rho$. For background jets, the Sudakov exponent increases with the
boost like $\log(1/\rho)$ (cf.~Eq.~\eqref{eq:mass-distrib-tau21-fc},
with a similar result for $D_2$).
Note that this dependence on $\rho$ of the background efficiency is
not always desirable. In particular, it might complicate the
experimental estimation of the background, thus negatively impacting searches for
bumps on top of it.
An alternative strategy consists of designing a ``decorrelated
tagger''~\cite{Dolen:2016kst} (see description in Sec.~\ref{sec:tools-combinations}), \eg built from $\rho$ and $\tau_{21}$,
yielding a flat background, hence facilitating searches.

%%========================================================================
\section{Performance and robustness}\label{sec:2prongs-perf-robustness}

The last set of studies we want to perform in this chapter is along
the lines of our quality criteria introduced in
Sec.~\ref{sec:performance-intro}, namely looking at two-prong
taggers both in terms of their performance and in terms of their
resilience against non-perturbative effects.
An extensive study has been pursued in the context of the Les Houches
Physics at TeV Colliders workshop in 2017 (LH-2017), where a
comparison of a wide range of modern two-prong taggers was
performed. Here, we want to focus on a subset of these results,
highlighting the main features and arguments one should keep in mind
when designing a two-prong tagger and assessing its performance. We
refer to Section~III.2 of Ref.~\cite{Bendavid:2018nar} for additional
details and results.

The study is done at three different values of $p_t$ (500~GeV, 1 and
2~TeV) and here we focus on jets reconstructed with the anti-$k_t$
algorithm with $R=1$ (the LH-2017 study also includes $R=0.8$).
Crucially, we are going to discuss in detail the resilience with respect to
non-perturbative effects, including both hadronisation and the
Underlying Event. We refer to the extensive study for a separate
analysis of hadronisation and the UE, as well as for a study of
resilience against detector effects and pileup.

To make things concrete, we consider a wide set of two-prong taggers
which can be put under the form
\begin{equation}
  m_\text{min}<m<m_{max}\qquad
  \text{shape }v=\frac{\text{3-particle observable}}{\text{2-particle observable}}<v_\text{cut},
\end{equation}
where the mass, the two-particle observable and the three-particle observable can
potentially be computed with different levels of grooming.
We will focus on four levels of grooming
\begin{itemize}
  \setlength\itemsep{0cm}
\item {\em plain ($p$)}: no grooming,
\item {\em loose ($\ell$)}: \SD with $\beta=2$ and $\zcut=0.05$,
\item {\em tight ($t$)}: mMDT with $\zcut=0.1$,
\item {\em trim}: trimming with $k_t$ subjets using
  $R_\text{trim}=0.2$ and $f_\text{trim}=0.05$,
\end{itemize}
and four different shapes: the $\tau_{21}$ $N$-subjettiness ratio and the
$D_2$, $N_2$ and $M_2$ ECF ratios either with $\beta=1$ or $\beta=2$.
\begin{table}[t!]
\begin{center}
\begin{tabular}{| c | c | c |c |c|c|c |c|c | }
  \hline                       
  Notation: $m \otimes n/d$ & $m$ (mass) & $n$ (numerator) & $d$ (denominator)\\
  \hline
  $p    \otimes    p/p$    & plain  &  plain & plain \\
  $\ell \otimes    p/p$    & loose  &  plain & plain \\
  $\ell \otimes    p/\ell$ & loose  &  plain & loose \\
  $\ell \otimes \ell/\ell$ & loose  &  loose & loose \\
  $t    \otimes    p/p$    & tight  &  plain & plain \\
  $t    \otimes \ell/\ell$ & tight  &  loose & loose \\
  $t    \otimes    p/t$    & tight  &  plain & tight \\
  $t    \otimes \ell/t$    & tight  &  loose & tight \\
  $t    \otimes    t/t$    & tight  &  tight & tight \\
  $\text{trim}$            & trim   &  trim  & trim \\
  \hline  
\end{tabular}
\end{center}
\caption{List of the different tagging strategies considered with the
  corresponding level of grooming for the mass, and
  numerator and denominator of the shape variable.}\label{table:lh-grooming-levels}
\end{table}
A generic tagger can then be put under the form
\begin{equation}
v\Big[m\otimes\frac{n}{d}\Big],
\end{equation}
where $v$ is one of our three shapes, $m$ is the level of grooming used
to compute the jet mass and $n$ and $d$ are the levels of grooming
used respectively for the numerator and denominator of the shape. We
consider the combinations listed in
Table~\ref{table:lh-grooming-levels}.

In order to study the tagging quality, we impose the reconstructed mass to
be between 65 and 105~GeV and we vary the cut on the jet shape.
We select a working point so that the signal efficiency (at truth, \ie
hadron+UE level) is 0.4, which fixes the cut on the jet shape. For
that cut, we can compute both the signal and background efficiencies
at parton level and at hadron+UE level, which allows us to compute the
tagging performance and robustness using the significance
$\epsilon_S/\sqrt{\epsilon_B}$ and resilience $\zeta$ introduced in
Sec.~\ref{sec:performance-intro}.
\begin{figure}
  \centering
  \subfloat[]{\includegraphics[width=0.48\textwidth,page=1]{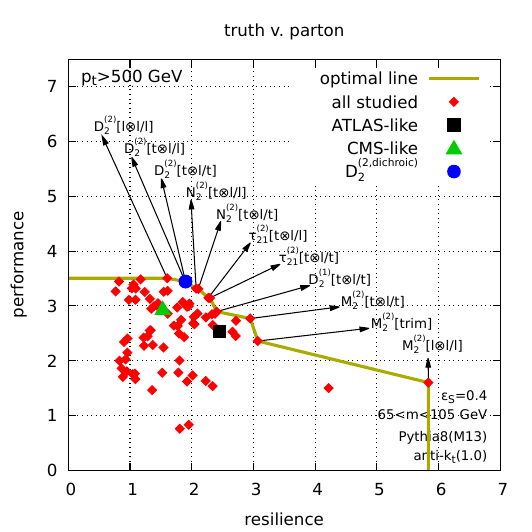}\label{fig:lh2017-optimal-500}}%
  \hfill%
  \subfloat[]{\includegraphics[width=0.48\textwidth,page=3]{figures/lh2017-optimal.pdf}\label{fig:lh2017-optimal-2000}}%  
  \caption{Summary of the performance (significance) v. robustness
    (resilience) of a set of two-prong taggers based on the
    combination of a prong finder and a shape
    cut. }\label{fig:lh2017-optimal}
\end{figure}
The resulting tagging qualities are summarised in
Fig.~\ref{fig:lh2017-optimal} for the two extreme $p_t$ values.
Each point on the plot represents a different tagger.
The ``ATLAS-like'' tagger, i.e. \ trimmed mass with $D_2^{(1)}$ computed on the
trimmed jet), and ``CMS-like'' tagger, i.e.\ mMDT mass with $N_2^{(1)}$ computed on
the mMDT jet, correspond to the working points defined in
Sec.~\ref{sec:tools-combinations}.  The $D_2^\text{(2,dichroic)}$ tagger
corresponds to a working point which appears to show a large
performance without sacrificing too much resilience. This tagger features
$t\otimes \ell/t$ dichroic $D_2$ variable (with angular exponent
$\beta=2$) with the mass computed on the tight jet, the shape
numerator $e_3/(e_e^2)$ computed on the loose jet, and the shape
denominator $e_2$ computed on the tight jet.
The plot also shows the line corresponding to the envelope which
maximises resilience for a given performance (and vice versa).

There are already a few interesting observations we can draw from
Fig.~\ref{fig:lh2017-optimal}.
\begin{itemize}
\item As $p_t$ increases, the discriminating power increases as
  well. This can be explained by the fact that when $p_t$ increases,
  the phase-space for radiation becomes larger, providing more
  information that can be exploited by the taggers;
\item The main observations from the previous section still largely
  hold: dichroic variants and variants based on $D_2$ give the best
  performance.
  One possible exception is the case of
  $D_2^{(2)}[\ell\otimes\ell/\ell]$ (\ie both the mass and $D_2$
  computed on the loose (\SD) jet), which shows a slightly larger
  performance than our $D_2^\text{(2,dichroic)}$ working point, albeit
  with a smaller resilience.\footnote{If we were seeking absolute
    performance without any care for resilience, this suggests that
    even looser groomers, possibly combined with a dichroic approach,
    could yield an even greater performance.}
  One aspect which is to keep in mind here is that using a looser
  grooming to measure the jet mass could have the benefit of avoiding
  the $1-2\zcut$ signal efficiency factor before any shape cut is
  applied, of course probably at the expense of more distortion of the
  W peak.
\item Generically speaking, {\em there is a trade-off between
    resilience and performance}.
  This is particularly striking if one looks along the optimal
  line.
  This is an essential feature to keep in mind when designing
  boosted-object taggers: keeping more radiation in the jet (by using
  a looser groomer) or putting tighter constraints on soft radiation
  at larger angles typically leads to more efficient taggers but at
  the same time yields more sensitivity to the regions where
  hadronisation and the Underlying Event have a larger impact, hence
  reducing resilience.
  This is seen both in terms of the shape, when going from $M_2$ to
  $\tau_{21}$ and $N_2$ and then to $D_2$, and in terms of grooming,
  when going from tight to loose jets. 
\item Apart from a few exceptions at relatively lower significance and
  high resilience, the taggers on the optimal are dominated by shapes
  with angular exponent $\beta=2$ rather than the current default at
  the LHC which is $\beta=1$.
\end{itemize}

\begin{figure}
  \centering
  \subfloat[]{\includegraphics[width=0.48\textwidth,page=1]{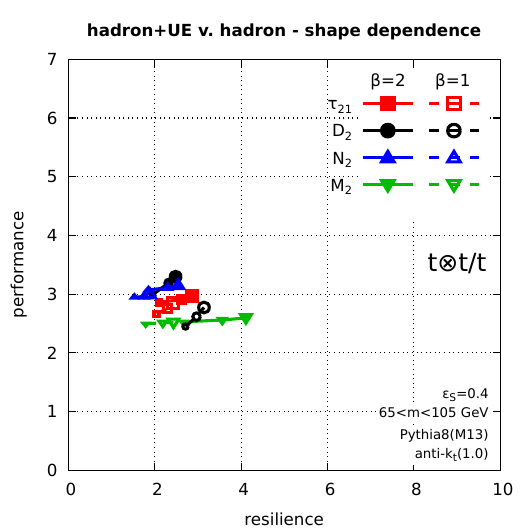}\label{fig:lh2017-shape-ttt}}%
  \hfill%
  \subfloat[]{\includegraphics[width=0.48\textwidth,page=5]{figures/lh2017-shape-dependence.pdf}\label{fig:lh2017-shape-trim}}\\
  \subfloat[]{\includegraphics[width=0.48\textwidth,page=3]{figures/lh2017-shape-dependence.pdf}\label{fig:lh2017-shape-tlt}}%
  \hfill%
  \subfloat[]{\includegraphics[width=0.48\textwidth,page=2]{figures/lh2017-shape-dependence.pdf}\label{fig:lh2017-shape-lll}}%  
  \caption{Dependence of the tagging quality (performance
    versus robustness) on the choice of jet shape.
    Results are shown for different grooming strategies indicated on
    each plot.
    Each curve has three points with increasing symbol size corresponding
    to $p_t=500$~GeV, $1$ and $2$~TeV.
    Each panel corresponds to a different grooming level as indicated on
    the plots.}\label{fig:lh2017-shape}
\end{figure}

\begin{figure}
  \centering
  \subfloat[]{\includegraphics[width=0.48\textwidth,page=2]{figures/lh2017-grooming-dependence.pdf}\label{fig:lh2017-grooming-ttt}}%
  \hfill%
  \subfloat[]{\includegraphics[width=0.48\textwidth,page=6]{figures/lh2017-grooming-dependence.pdf}\label{fig:lh2017-grooming-trim}}\\
  \subfloat[]{\includegraphics[width=0.48\textwidth,page=3]{figures/lh2017-grooming-dependence.pdf}\label{fig:lh2017-grooming-tlt}}%
  \hfill%
  \subfloat[]{\includegraphics[width=0.48\textwidth,page=1]{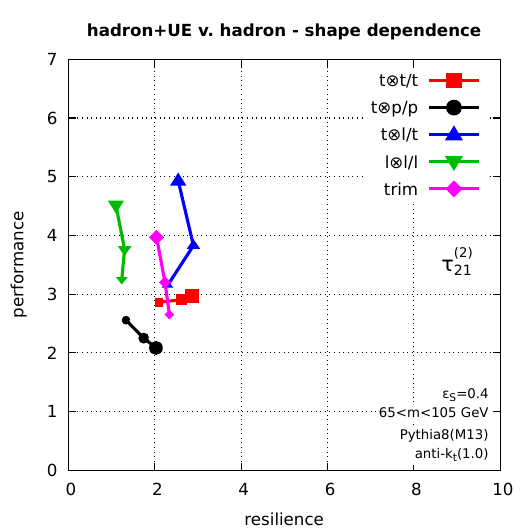}\label{fig:lh2017-grooming-lll}}%  
  \caption{Dependence of the tagging quality (performance
    versus robustness) on the choice of grooming strategy.
    Results are shown for a representative set of jet shapes.
    Each curve has three points with increasing symbol size corresponding
    to $p_t=500$~GeV, $1$ and $2$~TeV.
    Each panel corresponds to a different choice of shape
    as indicated on the plots.}\label{fig:lh2017-grooming}
\end{figure}

In order to gain a little more insight than what is presented in the
summary plot from Fig.~\ref{fig:lh2017-optimal}, we have extracted a
few representative cases in Fig.~\ref{fig:lh2017-shape}, where each
plot shows different shapes for a fixed grooming strategy and
Fig.~\ref{fig:lh2017-grooming} where each plot shows different
grooming strategies for a fixed shape.
All of the key points made above are visible on these plots. We
highlight here a few additional specific examples.

On Fig.~\ref{fig:lh2017-shape}, one sees that the performance of
the taggers increases with $p_t$, with $D_2$ having the best performance,
followed by $\tau_{21}$ and $N_2$ which show a similar pattern, and
$M_2$ which shows a (much) lower performance.
With tight grooming, Fig.~\ref{fig:lh2017-shape-ttt}, the phase-space
available for radiation constraint is limited and the differences
between the shapes are not large. Conversely, when opening more
phase-space, \eg Figs.~\ref{fig:lh2017-shape-tlt}
and~\ref{fig:lh2017-shape-lll}, the differences between shapes becomes
more visible.
The trade-off between performance and resilience is visible in each
plot, with the exception of $D_2^{(2)}[\ell\otimes\ell/\ell]$ in
Fig.~\ref{fig:lh2017-shape-lll}.
We also see that shapes with angular exponent $\beta=2$ show a
better performance than their $\beta=1$ counterparts. In terms of
resilience which can be either smaller (\eg
$D_2^{(2)}[\ell\otimes\ell/\ell]$), similar, or larger (\eg
$N_2^{(2)}[t\otimes\ell]/t$).
We note that for plain jets, we would expect $\beta=1$ shapes,
typically behaving like a $k_t$ scale, to be
more resilient than shapes with $\beta=2$, behaving like a mass scale
instead, since they can maximise the available perturbative
phase-space before hitting the hadronisation scale (which corresponds
to a soft $k_t$ scale).
And a similar argument hold for the Underlying Event.
Conversely, from a perturbative QCD point of view, $\beta=2$ has often
be shown (see \eg \cite{Larkoski:2013eya,Salam:2016yht}) to have a
larger discriminating power.
A natural expectation is therefore that once jets are groomed,
non-perturbative these effects are expected to be reduced, giving more
prominence to the perturbative QCD tendency to favour $\beta=2$.
Turning finally to Fig.~\ref{fig:lh2017-grooming}, we clearly see for
all four shapes, that using a looser groomer for the shape (either via
the ``all-loose'' $\ell\otimes \ell/\ell$ or the ``dichroic'' $t\otimes\ell/t$
combination) comes with large gains in terms of performance.
However, using the plain jet typically shows bad performance, an
effect which can be attributed to an enhanced sensitivity to the
Underlying Event.
Comparing the ``all-loose'' and the ``dichroic'' variants, we see that
they show a similar performance, with the dichroic variant having
a larger resilience.

To conclude, we stress once again that, in order to get a complete
picture, the above discussion about performance versus resilience should
be supplemented by a study of the resilience against detector effects
and pileup. Even though we will not do this study here, one can at
least make the educated guess that pileup effects would be reduced by
using a tighter grooming.

%% GS helper for auctex
%%% Local Variables:
%%% mode: latex
%%% TeX-master: "notes"
%%% End:

%  LocalWords:  ECFs SCET Eq eq BSM Houches ECF

% $Id: calculations-substructure.tex 569 2025-06-19 13:07:43Z smarzani $

%%========================================================================
\chapter{Curiosities: Sudakov Safety}\label{sec:curiosities}
In Chapter~\ref{tools} we have introduced the modified Mass Drop
Tagger/\SD and in Chapter~\ref{calculations-substructure-mass} we have
discussed at length the analytic properties of the jet mass
distribution after mMDT or \SD. Furthermore, we have just analysed
some aspects of applying this grooming technique to jet shapes used
for quark/gluon and W-boson discrimination.
However, if we go back to its original definition, we notice that the
\SD condition Eq.~(\ref{eq:soft-drop-condition}) does not involve
directly the jet mass or any jet shape, but rather the distance
between two prongs in the azimuth-rapidity plane $R_{ij}$ and the
momentum fraction
$ z=\tfrac{{\rm min}(p_{t,i},p_{t,j})}{p_{t,i}+p_{t,j}} $. It is quite
natural to ask ourselves if we can apply the calculation techniques
described for jet masses and jet shapes to better characterise the
distributions of these two quantities. To be precise, let us define the
two observables $\theta_g$ and $z_g$ as follows. We start with a jet
which has been re-clustered with Cambridge/Aachen and we apply \SD. When we find the
first declustering with subjets $j_1$ and $j_2$ that passes the \SD
condition Eq.~(\ref{eq:soft-drop-condition}), we define the groomed radius
and the groomed momentum fraction as
\begin{align}\label{eq:thetag-zg-def}
\theta_g& = \frac{R_{12}}{R}, \\
z_g &= \frac{{\rm min}(p_{t,1},p_{t,2})}{p_{t,1}+p_{t,2}},
\end{align}
where $R$ is the original jet radius.
We note that these variables are interesting for a number of
reasons. 
The groomed jet radius is of interest because the groomed jet area is
of the order of $\pi \theta_g^2$.  Thus, $\theta_g$ serves as a proxy for the sensitivity of the groomed jet to possible contamination from pileup~\cite{Cacciari:2008gn,Sapeta:2010uk}.
Furthermore, as we shall shortly see, $z_g$ provides us with an almost unique perturbative access to one of the most fundamental building blocks of QCD, namely the Altarelli-Parisi splitting function~\cite{Larkoski:2015lea,Larkoski:2017bvj}.

This last observation has drawn the interest of the scientific
community in particular with the study of heavy-ion collisions. In
particular, an observable such as $z_g$ provides information about how
perturbative QCD evolution is modified by the interaction between the
high-energy jet and the quark-gluon plasma, thus providing a new probe
of the latter.
Different experiments have now measured $z_g$ distribution. For
instance the STAR collaboration at the Relativistic Heavy Ion Collider
of the U.S. Brookhaven National Laboratory performed this measurement
using gold-gold collisions~\cite{Kauder:2017mhg}. Furthermore, the CMS
experiment and the ALICE experiments studied this variable, at the
Large Hadron Collider, in lead-lead heavy-ion
collisions~\cite{Chen:2017rhw,Caffarri:2017bmh}. We will describe some
of these measurements in more detail in
Chapter~\ref{searches-measurements}.
In parallel, this line of research triggered noticeable interest in
the theoretical nuclear physics and heavy-ion communities,
e.g.~\cite{Lapidus:2017dek,Zapp:2017ria,Tywoniuk:2017dzi,Casalderrey-Solana:2017mjg,Mangano:2017plv,Qin:2017roz,Milhano:2017nzm,Chang:2017gkt,KunnawalkamElayavalli:2017hxo}.

In this chapter, we focus on a baseline description of the $\theta_g$
and $z_g$ observables in proton-proton collision, leaving aside the
extra complications due to interactions of jets with the quark-gluon
plasma in the heavy-ion case.
In this context, we anticipate that while we will be able to apply the
standard techniques presented so far in this book in order to obtain a
perturbative prediction for the $\theta_g$ distribution for, the
situation will be very different for $z_g$, where interesting features
emerge.

\section{The groomed jet radius distribution $\theta_g$} \label{sec:thetag}
\begin{figure}[t]
    \centering
    \includegraphics[width=0.32\textwidth]{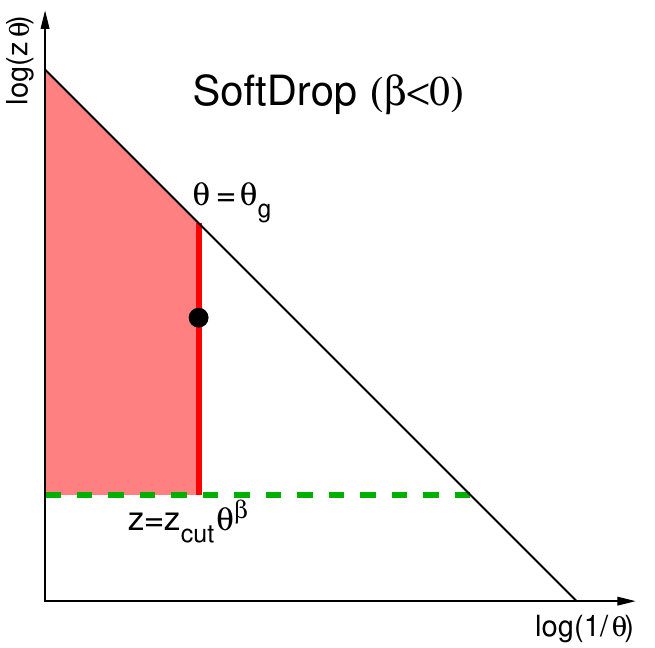}
    \includegraphics[width=0.32\textwidth]{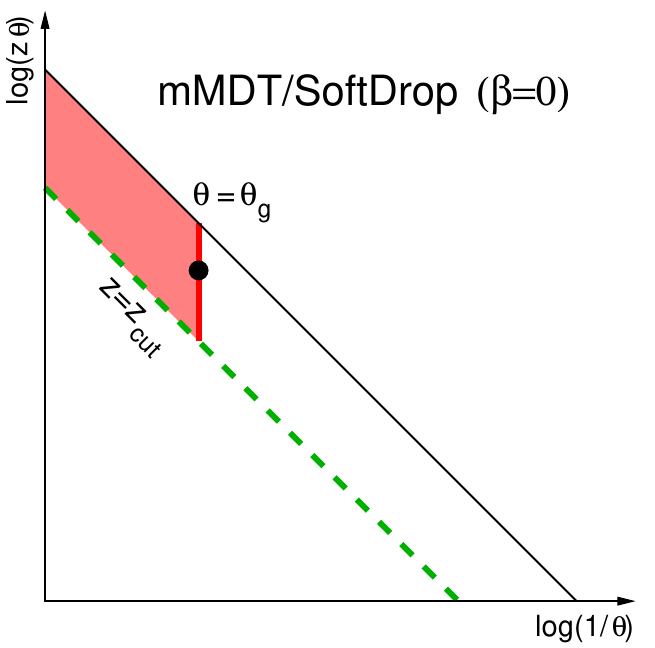}
    \includegraphics[width=0.32\textwidth]{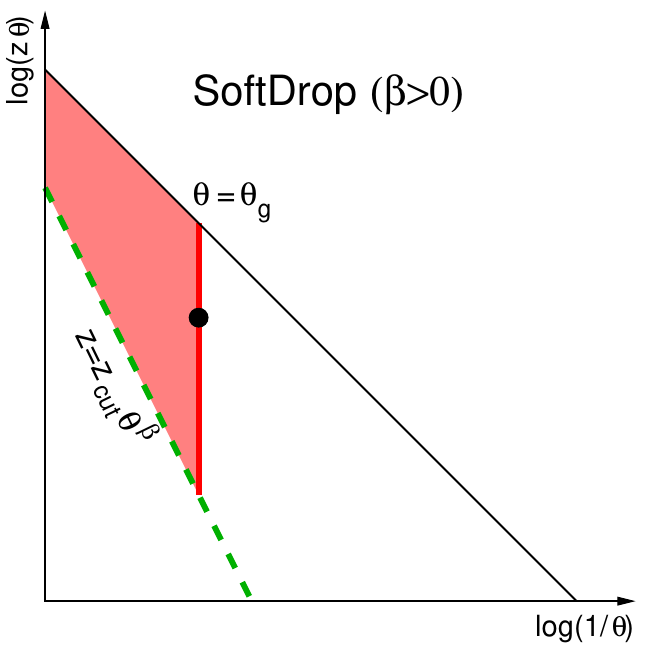}
  \caption{Lund diagrams for the $\theta_g$ distribution for three representative values of the SD angular exponent $\beta$. From left to right we have $\beta<0$, $\beta=0$ (mMDT) and $\beta>0$. 
  The dashed green line
    represents the edge of SD region, the solid red line corresponds to
    emissions yielding the requested groomed jet radius and the shaded red area is the vetoed area
    associated with the Sudakov suppression. We note that the latter
    is finite in all three cases, as it should be for an IRC
    observable. 
   }\label{fig:lund-sd-thetag}
\end{figure}

We start by calculating the cumulative distribution for the groomed jet radius. In doing so, we are going to exploit the techniques developed in the previous chapters. In particular, we begin by drawing the Lund plane for the observables at hand. We do this in Fig.~\ref{fig:lund-sd-thetag}, where we distinguish three cases according to the sign of the \SD angular exponent $\beta$. From left to right we have $\beta<0$, $\beta=0$ and $\beta>0$. We remind the reader that \SD with $\beta=0$ corresponds to mMDT. 

The dashed green line
    represents the edge of phase-space region where emissions pass the \SD condition, while the solid red line corresponds to
    emissions yielding the requested groomed jet radius. Finally, the
    shaded red area is the region we have to veto in order not to
    exceed the requested groomed radius. With these considerations and
    the expertise gained from the previous chapters, we can almost immediately arrive at an all-order cumulative distribution, which resums leading logarithms and next-to-leading ones but limited to the collinear sector. We have
\begin{equation}\label{eq:grad_exp}
\Sigma_\text{SD}(\theta_g) = \exp\left[
   - \int_{\theta_g}^{1} \frac{d\theta}{\theta}\int_0^{1} dz\, P_{i}(z)\,
   \frac{\alpha_s(z\theta p_tR)}{\pi}
   \Theta\left( z> z_\text{cut}\theta^\beta \right)
\right]\equiv \exp \left[ -R(\theta_g) \right],
\end{equation}
where the integral in the exponent again corresponds to vetoed emissions and $i=q,g$ depending on the jet flavour.
We note that the integrals in Eq.~(\ref{eq:grad_exp}) are finite
(modulo the question of the Landau pole) for all values of $\beta$.
This is the case because the integral in the exponent arises after adding together real and virtual contributions and therefore its finiteness is guaranteed by the IRC safety of the observable.
The integrals in Eq.~(\ref{eq:grad_exp}) can be easily evaluated to
leading-logarithmic accuracy, leading to the following
radiator\footnote{Note that we have used the same approach as for the
  rest of this book and included it in the double-logarithmic
  terms. In this specific case, this is less relevant as the endpoint
  of the distribution does not depend on it, so we could have left it
  explicitly as a separate correction.}
\begin{equation} \label{radiator-rad}
R(\theta_g) 
= \frac{C_i}{2\pi \alpha_s \beta_0^2} \bigg[
     W(1-\lambda_B)
     -W(1-\lambda_g-\lambda_B)
     -\frac{W(1-\lambda_c)}{1+\beta}
     +\frac{W(1-\lambda_c-(1+\beta)\lambda_g)}{1+\beta}\bigg],
   \end{equation}
where $\lambda_g= 2 \as \beta_0 \log\big(\frac{1}{\theta_g}\big)$,
$\lambda_c= 2 \as \beta_0 \log\big(\frac{1}{z_\text{cut}}\big)$ and
$\lambda_B=-2\alpha_s\beta_0B_i$ as before.

For $\beta<0$, this distribution has an
endpoint at $\theta_g^\text{(min)}=\zcut^{-1/\beta}$ (modulo corrections from
hard-collinear splittings). Correspondingly, there is a finite probability,
$\exp[-R(\theta_g^\text{(min)})]$, that the \SD de-clustering
procedure does not find a two-prong structure passing the \SD
condition, in which case we set $\theta_g=0$.

\begin{figure}[t!]
  \centering
  \includegraphics[width=0.48\textwidth,page=1]{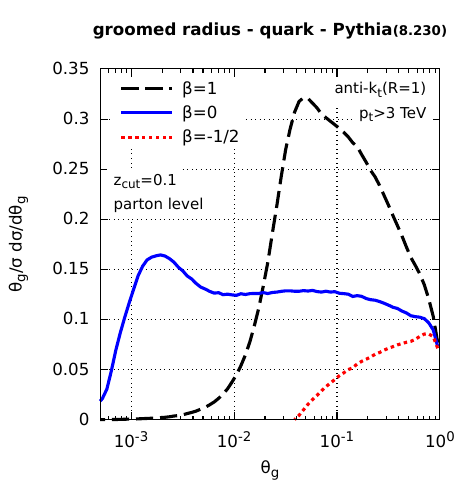}%
  \hfill%
  \includegraphics[width=0.48\textwidth,page=1]{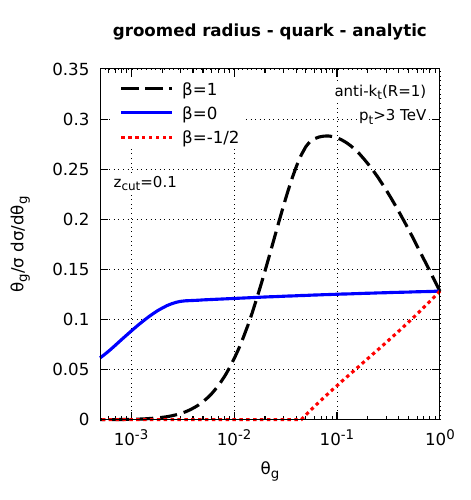}%
  \caption{The groomed radius distribution The left plot is the result of a Pythia parton-level
    simulation and the right plot is the analytic results discussed in
    this chapter.}\label{fig:radius-pythia-v-analytic}
\end{figure}

The theoretical calculation is compared to the Monte Carlo prediction,
at parton level, in Fig.~\ref{fig:radius-pythia-v-analytic}, showing
that it captures the main features of the distribution.
In particular, we notice that the $\theta_g$ distribution has an
endpoint for negative values of $\beta$, related to the finiteness of
the available phase-space.
Furthermore, as $\beta$ decreases, the distribution is shifted towards
smaller angles.\footnote{Here, the case of negative $\beta$ can be seen
  as shifting a whole part of the distribution to
  $\theta_g=0$.} Since the groomed jet area is proportional to
$\pi\theta_g^2$, this agrees with the expectation that smaller $\beta$
corresponds to more aggressive grooming, meaning a smaller jet area or
a smaller sensitivity to pileup and the Underlying Event.

It is
also worth noting that a few complications would arise if we wanted to
extend Eq.~\eqref{radiator-rad} to full NLL accuracy.
Since $\theta_g$ is only sensitive to the first emission being
de-clustered that passes the \SD condition, one could expect that it
does not get any correction from multiple emissions at NLL.
However, if one has multiple emissions at a similar angle and
strongly-ordered in energy, which emission emission is de-clustered
first will depend on the details of the Cambridge/Aachen
clustering. This situation, which occurs only for $\beta>0$, is
reminiscent of the non-global and clustering logarithms discussed in
Secs.~\ref{sec:non-global} and \ref{sec:clustering}.
This type of effect has been discussed, for instance, in
Refs.~\cite{Neill:2018yet} and~\cite{Dreyer:2018nbf}.

\section{The $z_g$ distribution} \label{sec:zg}
\begin{figure}
    \centering
    \includegraphics[width=0.32\textwidth]{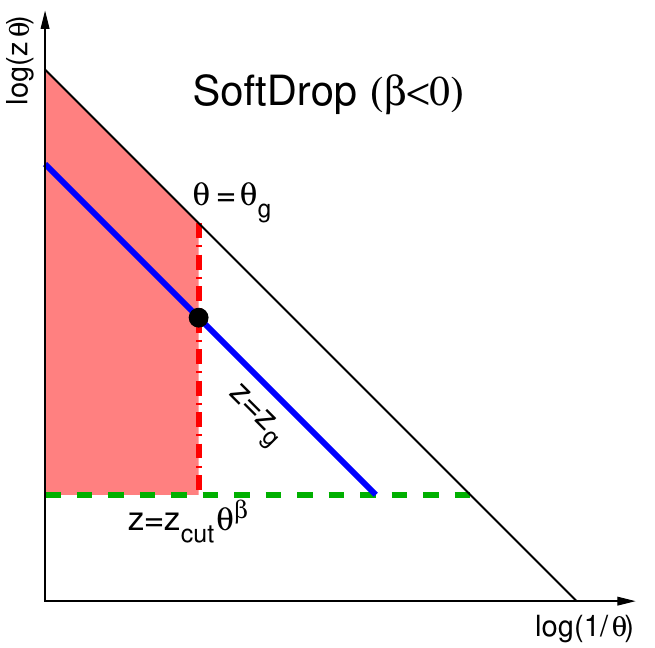}
    \includegraphics[width=0.32\textwidth]{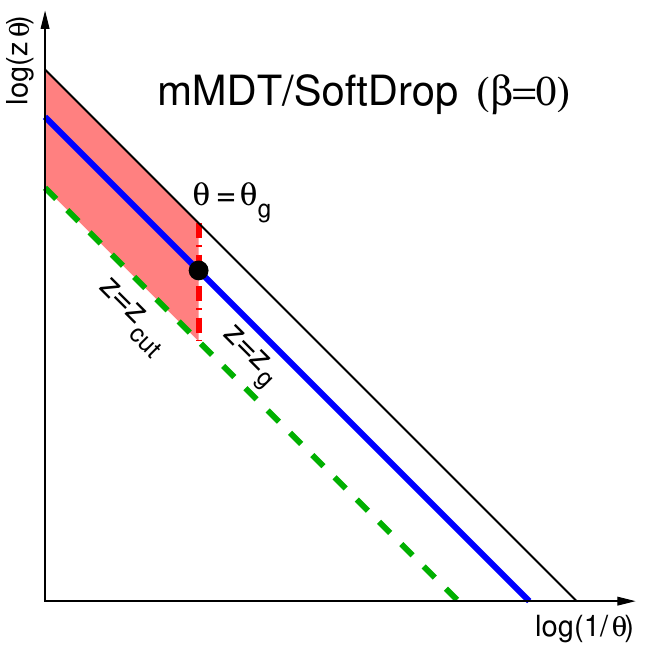}
    \includegraphics[width=0.32\textwidth]{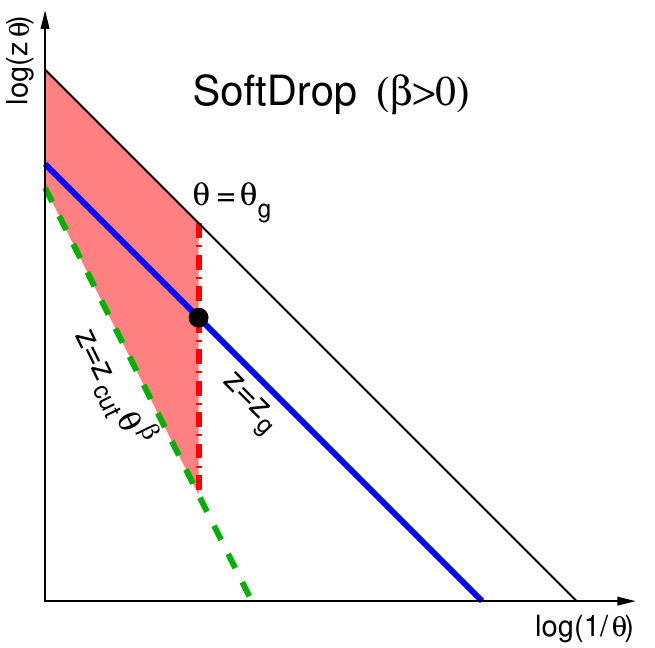}
  \caption{Lund diagrams for the $z_g$ distribution for three representative values of the SD angular exponent $\beta$. From left to right we have $\beta<0$, $\beta=0$ (mMDT) and $\beta>0$. 
  The dashed green line
    represents the edge of SD region
     The dot-dashed red line corresponds to
    emissions yielding a given groomed jet radius and the shaded red area is the vetoed area
    associated with the Sudakov suppression. Finally, the solid blue line corresponds to the requested value of $z_g$. Because we have to integrate over all possible values of $\theta_g$, only the $\beta<0$ case showed on the left exhibits IRC safety.
    %    We note that the latter is finite only when $\beta<0$, while it indefinitely extends in the soft/collinear region when $\beta\ge 0$, signalling the fact that the observable is not IRC safe.  
   }\label{fig:lund-sd-zg}
\end{figure}

We would like now to study the observable $z_g$. This immediately
faces a difficulty: $z_g$ is fixed by the first de-clustering of the
jet that passes the \SD condition. Because we are completely inclusive
over the splitting angle $\theta_g$ we must take into account all
possible values of $\theta_g$ including configurations where the two
prongs become collinear. Indeed collinear splittings always pass the
\SD condition, if $\beta \ge 0$ (strictly speaking, soft-collinear
emissions fail \SD $\beta=0$/mMDT). These configurations are not
cancelled by the corresponding virtual corrections, for which $z_g$ is
undefined, and herald the fact that the observable is not IRC safe.
At this point a possible approach would be to just stop this analysis because the observable we are dealing with does not respect the very basic set of properties set out in Chapter~\ref{chap:qcd-colliders}. However, we have just argued that $z_g$ is a very interesting observable for jet substructure and therefore, we decide to be stubborn and push forward with our study. In order to do that, we must generalise the concept of IRC safety and introduce \emph{Sudakov safety}~\cite{Larkoski:2013paa}.
 
 Following~\cite{Larkoski:2015lea}, we introduce a general definition of Sudakov safety which exploits conditional probabilities.
Let us consider an IRC unsafe observable $u$ and a companion IRC safe observable $s$.  The observable $s$ is chosen such that its measured value regulates all singularities of $u$.  That is, even though the probability of measuring $u$,
\begin{equation}
p(u) = \frac{1}{\sigma_0} \frac{d \sigma}{d u},
\end{equation}
is ill-defined at any fixed perturbative order, the probability of measuring $u$ given $s$, $p(u|s)$, is finite at all perturbative orders, except possibly at isolated values of $s$ e.g., $s=0$ for definiteness.  Given this companion observable $s$, we want to know whether $p(u)$ can be calculated in perturbation theory.
Because $s$ is IRC safe, $p(s)$ is well-defined at all perturbative
orders and one can define the joint probability distribution
\begin{equation}\label{eq:cond_prob}
p(s,u) = p(s)\, p(u|s) ,
\end{equation}
which is also finite at all perturbative orders, except possibly at isolated values of $s$.  To calculate $p(u)$, we can simply marginalise over $s$:
\begin{equation}
\label{eq:sudsafeone}
p(u) = \int d s\, p(s) \, p(u|s) \,.
\end{equation}
If $p(s)$ regulates all (isolated) singularities of $p(u|s)$, thus ensuring that the above integral is finite, then we deem $u$ to be Sudakov safe.

Clearly, we cannot just evaluate $p(s)$ at fixed-order in the strong coupling, but we need some information about its all-order behaviour.  If we consider the resummed distribution for the observable $s$, its distribution will exhibit a Sudakov form factor (hence the name Sudakov safety) that can make the integral in Eq.~(\ref{eq:sudsafeone}) convergent. 
In the case that one IRC safe observable is insufficient to regulate all singularities in $u$, we can measure a vector of IRC safe observables $\vec{s}=\{s_1,\dotsc,s_n\}$.  This gives a more general definition of Sudakov safety:
\begin{equation}\label{eq:sudsafedef}
p(u)= \int d^n\vec{s} \, p({\bf s})\, p(u| \vec{s}) \,.
\end{equation}
Only if the vector of safe observables has a finite number of elements, then $u$ is Sudakov safe. For example, particle multiplicity does not fall in this category as it would require an infinite number of safe observables to regulate the arbitrary number of soft/collinear splittings. Thus, particle multiplicity is neither IRC safe, nor Sudakov safe.

We can now go back to the observable $z_g$ and check whether it is Sudakov safe. To this purpose, we need to introduce a safe companion observable. The \SD procedure itself suggests to use the groomed angle $\theta_g$, which we have calculated in Eq.~(\ref{eq:grad_exp}). 
Therefore, we imagine to measure a value of $z_g$, given a finite angular separation between the two prongs $\theta_g$. This situation is illustrated by the Lund diagrams in Fig.~\ref{fig:lund-sd-zg}. As usual, the dashed green line represents the edge of \SD region. The black dot is the emission passing \SD that provides $z_g$ (solid blue line) and $\theta_g$ (dot-dashed red line). The shaded red area is the vetoed area  associated with the Sudakov suppression for the groomed radius $\theta_g$, i.e.\ it is the same as in Fig.~\ref{fig:lund-sd-thetag}.
In order to obtain the $z_g$ distribution, we have to integrate over all possible values of $\theta_g$, which corresponds to all allowed positions for the dot-dashed red line. For $\beta<0$, the area we swipe as we move the red dot-dashed line is bounded by the \SD line in dashed green and it is therefore finite. In this case we expect IRC safety to hold. On the other hand, the $\beta=0$ and $\beta>0$ cases are remarkably different as the resulting area is unbounded. This situation is not IRC safe, but the Sudakov form factor for the groomed radius is enough to regulate (suppress) the resulting divergence. 
We have
\begin{equation}\label{eq:zg-cond-prob}
\frac{1}{\sigma_0}\frac{d \sigma}{d z_g} = \int_0^1 d \theta_g \, p(\theta_g) \, p(z_g|\theta_g),
\end{equation}
where $p(\theta_g)$ is the resummed distribution computed in the
previous section, i.e.\ the derivative of Eq.~(\ref{eq:grad_exp}),
while the conditional probability is calculated at fixed-order in the
strong coupling. In the collinear limit, it reads, for a jet of
flavour $i=q,g$,
\begin{equation}\label{eq:cond-explicit}
p(z_g|\theta_g ) =  \frac{P_{\text{sym},i}(z_g)\as(z_g \theta_g p_t R)}{\int_{z_\text{cut}\theta_g^\beta}^{1/2} d z \, P_{\text{sym},i}(z) \as(z \, \theta_g p_t R)} \Theta(z_g>\zcut \theta_g^\beta)\,,
\end{equation}
where $0 < z_g < 1/2$ and following the approach of Refs.~\cite{Larkoski:2015lea,Larkoski:2017bvj,Tripathee:2017ybi}, we have introduced a symmetrised notation of the splitting function
\begin{equation}
P_{\text{sym},i}(z)=P_i(z)+P_i(1-z).
\end{equation}
Crucially, the integral in Eq.~(\ref{eq:zg-cond-prob}) is finite for
all values of $\beta$, provided we introduce a prescription for the
Landau pole, and it can be evaluated numerically.\footnote{In
  practice, we have frozen the coupling at a scale
  $\mu_\text{NP}=1$~GeV, cf.\
  Appendix~\ref{chap:app-analytic-details}.}
We also note that the $z_g$ distribution in~(\ref{eq:zg-cond-prob}) is
normalised to the ungroomed jet rate. This means that jets for which
the \SD procedure fails to find a two-prong structure, and so do not
have a well-defined $z_g$, are still included in the normalisation of
Eq.~(\ref{eq:zg-cond-prob}). This is obviously relevant for $\beta<0$
where, even perturbatively, there is a finite probability for this to
happen. For $\beta\ge 0$, this can also happen \eg due to
non-perturbative effects, or finite cut-offs in Monte Carlo
simulations.

\begin{figure}[t!]
  \centering
  \includegraphics[width=0.48\textwidth,page=2]{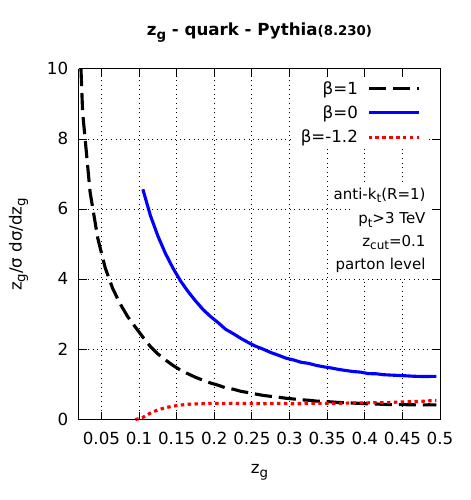}%
  \hfill%
  \includegraphics[width=0.48\textwidth,page=2]{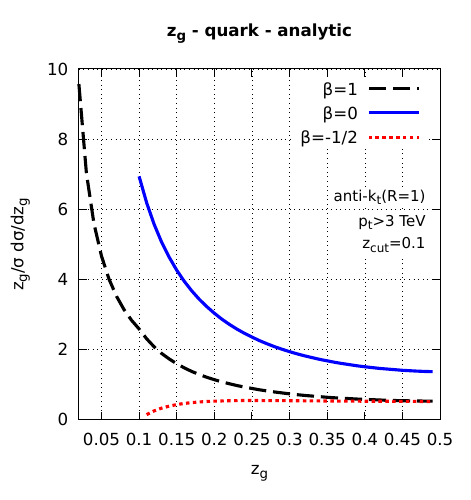}%
  \caption{The $z_g$ distribution The left plot is the result of a Pythia parton-level
    simulation and the right plot is the analytic results discussed in
    this chapter. We note that for $\beta<0$ the observable is IRC safe, while for $\beta\ge 0$ it is only Sudakov safe.}\label{fig:zg-pythia-v-analytic}
\end{figure}

A comparison to parton-level Monte Carlo simulations is shown in
Fig.~\ref{fig:zg-pythia-v-analytic}, showing a remarkably good
agreement given the collinear unsafety of the observable (for
$\beta\ge 0$).
What is perhaps more interesting is to try and understand explicitly
the dominant behaviour of a Sudakov-safe observable.
For this, we first work in the fixed-coupling limit. This means that,
when evaluating Eq.~(\ref{eq:zg-cond-prob}), we can factor out
$P_{\text{sym},i}(z_g)$ from Eq.~(\ref{eq:cond-explicit}) and $z_g$ only
enters as a phase-space constraint in the remaining integration.
Next, we consider the soft limit. In this limit we can neglect
hard-collinear splittings (\ie the $B$ terms) in the $\theta_g$
probability, and in Eq.~(\ref{eq:cond-explicit}) we can simplify the
denominator by setting the upper bound of integration to $1$ and set
$P_{\text{sym},i}(z)\approx 2C_i/z$.
The derivative of Eq.~(\ref{eq:grad_exp}) needed for $p(\theta_g)$ brings
a factor $R'(\theta_g)$ which, with our assumptions, cancels the
denominator of Eq.~(\ref{eq:cond-explicit}) up to a factor
$C_i/(2\pi)$.\footnote{This is easy to understand from a physical
  viewpoint: both $R'(\theta_g)$ and the denominator
  of Eq.~(\ref{eq:cond-explicit}) correspond to the probability for having
  a real emission passing the \SD condition at a given $\theta_g$.}
Writing $R(\theta_g)$ at fixed coupling, we are thus left with the
following integration
\begin{equation}\label{eq:res_zf}
  \frac{1}{\sigma_0}\frac{d\sigma}{dz_g}
  = P_{\text{sym},i}(z_g)\frac{\alpha_sC_i}{\pi}
  \int_0^1\frac{d\theta_g}{\theta_g}
  \exp\bigg[-\frac{\alpha_sC_i}{\pi\beta}\Big(\log^2(\zcut\theta_g^\beta)-\log^2(\zcut)\Big)\bigg]
  \Theta(\zcut\theta_g^\beta<z_g).
\end{equation}
Let us first consider the case $\beta<0$ for which $z_g>\zcut$ and we
get\footnote{Note that the assumptions used in this book slightly
  differ from the ones originally used in~\cite{Larkoski:2015lea}.}
\begin{align} \label{eq:res_zf_betaneg}
  \frac{1}{\sigma_0}\frac{d \sigma}{d z_g}
  \approx&\sqrt{\frac{\as}{4|\beta| C_i}}
           e^{-{\frac{\as C_i}{\pi |\beta|}\log^2(z_\text{cut})}}\\
  &
    \bigg[ \text{erfi} \bigg( \sqrt{\frac{\as C_i}{\pi|\beta|} }
                              \log\bigg(  \frac{1}{\zcut}\bigg) \bigg)
         - \text{erfi} \bigg( \sqrt{\frac{\as C_i}{\pi|\beta|} }
                              \log\bigg(  \frac{1}{z_g}\bigg)\bigg) \bigg]
    P_{\text{sym},i}(z_g),\nonumber
\end{align}
where $\text{erfi}(x)=-i\, \text{erf}(ix)$ is the imaginary error
function.
For $\beta<0$, $z_g$ is an IRC-safe observable and, accordingly, the
above result admits an expansion in powers of the strong coupling:
\begin{equation}
\beta < 0: \quad \frac{1}{\sigma_0}\frac{d \sigma}{d z_g}  = \frac{\alpha_s}{\pi |\beta| } \, P_{\text{sym},i}(z_g) \log\Big(\frac{z_g}{z_\text{cut}}\Big) \Theta(z_g -\zcut)+{\cal O}(\alpha_s^2).
\end{equation}
Moving now to $\beta>0$, the evaluation of Eq.~(\ref{eq:res_zf}) gives
\begin{equation} \label{eq:res_zf_betapos}
  \frac{1}{\sigma_0}\frac{d \sigma}{d z_g}
  \approx\sqrt{\frac{\as}{4\beta C_i}}
    e^{{\frac{\as C_i}{\pi \beta}\log^2 (z_\text{cut})}}
    \bigg[ 1 - \text{erf} \bigg( \sqrt{\frac{\as C_i}{\pi\beta} }
               \log\bigg(\frac{1}{\text{min}(z_g,\zcut)}\bigg)\bigg) \bigg]
    P_{\text{sym},i}(z_g).
\end{equation}
Although at first sight this looks similar to what was previously obtained, 
Eq.~\eqref{eq:res_zf_betapos} (for positive $\beta$)
shows a significantly different behaviour compared to
Eq.~\eqref{eq:res_zf_betaneg} for negative $\beta$, as a direct consequence
of the fact that $z_g$ is only Sudakov safe for $\beta>0$.
Indeed, for $\beta > 0$, the distribution has the expansion
\begin{equation}
\beta > 0: \quad \frac{1}{\sigma_0}\frac{d \sigma}{d z_g}  =
\sqrt{\frac{\alpha_s}{4 \beta C_i}}\, P_{\text{sym},i}(z_g)+{\cal O}\left(\alpha_s\right) , 
\end{equation}
and the presence of $\sqrt{\alpha_s}$ implies non-analytic dependence
on $\alpha_s$.
This behaviour is associated with the ``1'' in the square bracket
of Eq.~\eqref{eq:res_zf_betapos}, which can be traced back to the
contribution from $\theta_g\to 0$ in Eq.~(\ref{eq:res_zf}), \ie to the
region where the observable is collinear unsafe (though Sudakov safe).

Finally, it is interesting to consider the specific case $\beta=0$. At
fixed coupling, $p(z_g|\theta_g)$ (Eq.~(\ref{eq:cond-explicit})) is
independent of $\theta_g$ and factors out of the $\theta_g$
integration in Eq.~(\ref{eq:zg-cond-prob}) to give
\begin{equation}
\label{eq:beta_zero_pzg}
\beta = 0: \quad \frac{1}{\sigma_0}\frac{d \sigma}{d z_g} = \frac{P_{\text{sym},i}(z_g)}{\int_{z_\text{cut}}^{1/2} dz \, P_{\text{sym},i}(z)}\Theta(z_g>\zcut)\,.
\end{equation}
It is not difficult to see that the $\beta = 0$ case does have a valid
perturbative expansion in $\alpha_s$, despite being
$\alpha_s$-independent at lowest order.
This case is also only Sudakov safe, as the integration in
Eq.~(\ref{eq:res_zf}) includes the collinear-unsafe region
$\theta_g\to 0$.
More generally, $\beta=0$ marks the boundary between Sudakov-safe and
IRC-safe situations.  Eq.~(\ref{eq:beta_zero_pzg}) has remarkable
properties. Despite having being calculated in perturbative QCD, it
exhibits a lowest-order behaviour which does not depend on the strong
coupling, nor on any colour charge (in the small $z_g$ limit). 
Instead, as anticipated, the distribution is essentially driven by the
QCD splitting function, thus offering a unique probe of the dynamics
of QCD evolution.

There exist now several examples of other Sudakov safe observables. These include ratios of angularities~\cite{Larkoski:2013paa}, the transverse momentum spectrum of a \SD $\beta=0$ (mMDT) jet~\cite{Marzani:2017mva}, or equivalently the amount of energy which has been groomed away~\cite{Larkoski:2014wba}, and the pull angle~\cite{Gallicchio:2010sw,Larkoski:2019urm,Larkoski:2019fsm}, which is an observable that aims to measure colour-flow in a multi-jet event. 
Despite the very interesting results obtained thus far, the study of Sudakov safety is still in its infancy. 
Questions regarding the formal perturbative accuracy of the results,
with related estimate of perturbative uncertainties,  its dependence
upon the choice of the safe companion, the inclusion of running
coupling corrections, as well as the role of non-perturbative
uncertainties are interesting theoretical aspects of perturbative QCD
which are still actively investigated.

%% GS helper for auctex
%%% Local Variables:
%%% mode: latex
%%% TeX-master: "notes"
%%% End:

%  LocalWords:  Eq Altarelli Parisi Brookhaven NLL

% This describes the Lund jet plane
%------------------------------------------------------------------------
\chapter{The Lund jet plane: an overarching tool}\label{lundplane}
In Sec.~\ref{sec:jet-mass-res}, we have seen that Lund
diagrams are very convenient
graphical representations to understand how to organise  all-order
calculations. It has actually been realised recently that, in the
context of jet substructure, it was possible to promote this idea to a
genuine observable~\cite{Dreyer:2018nbf}.
In this chapter, we will introduce this observable, henceforth the Lund jet plane, and discuss its main properties and applications. As usual, we will apply perturbative QCD methods to gain an analytic understanding of this observable in order to be able to judge its performance as a tagger. 

\section{Definition and main underlying ideas}\label{sec:lund-def}

\subsection{Constructing the Lund jet plane}\label{sec:lund-construction}

The main conceptual idea behind the Lund jet plane construction is that clustering a
jet with the C/A algorithm generates a tree of
recombinations that mimics angular ordering and, therefore, it provides us with a
practical picture similar to the conceptual one underpinning Lund
diagrams.
The procedure given in~\cite{Dreyer:2018nbf} focused on the primary Lund plane, but for the discussion in the next few
pages, it is helpful to actually first construct a structure that
captures the full Lund diagram.

Let us therefore start with a given jet that one has clustered
with the C/A algorithm. If the jet was initially
reconstructed with a different algorithm, e.g.\ the anti-$k_t$
algorithm, we can simply recluster its constituents with the
C/A algorithm.
We build the Lund tree $\mathcal{L}$ by applying the following
iterative procedure, starting with the full jet $j$:
\begin{enumerate}
\item undo the last step of the clustering to get two subjets
  $j_\text{hard}\equiv j_\text{hard}(j)$ and
  $j_\text{soft}\equiv j_\text{soft}(j)$ from $j$, such that
  $p_{t,\text{hard}}>p_{t,\text{soft}}$. The idea behind this
  separation is that it corresponds to a physical branching
  $j\to j_\text{hard},j_\text{soft}$ where, in the soft limit, one has
  an emission $j_\text{soft}$ from an emitter $j$;
\item define the following kinematic
  variables:
  \begin{subequations}\label{eq:lund-kinematic-variables}
  \begin{align}
    \Delta &\equiv \Delta R_{j_\text{hard},j_\text{soft}},& k_t&\equiv p_{t,\text{soft}}\Delta,&
    m^2&\equiv (p_\text{hard}+p_\text{soft})^2\\
    z&\equiv\frac{p_{t,\text{soft}}}{p_{t,\text{hard}}+p_{t,\text{soft}}}, &\kappa&\equiv z\Delta,&
    \psi&\equiv\tan^{-1}\frac{y_\text{soft}-y_\text{hard}}{\phi_\text{soft}-\phi_\text{hard}},
  \end{align}
  \end{subequations}
  which one can then group into a tuple
  \begin{equation}\label{eq:lund-coordinates}
    \mathcal{T}(j) \equiv\big\{\Delta,k_t,\dots\big\};
  \end{equation}
\item iterate the procedure with $j_\text{hard}$ and $j_\text{soft}$
  (separately). The Lund tree associated with jet $j$ is then defined
  as
  \begin{equation}
    \mathcal{L}(j)\equiv \Big[\mathcal{T}(j), \left(\mathcal{L}(j_\text{hard}), \mathcal{L}(j_\text{soft})\right)\Big].
  \end{equation}
\end{enumerate}
From this full tree of nested kinematic properties associated with each
node of C/A clustering history, we can extract the {\it
  primary} Lund plane by keeping only the ones where the iterative
procedure follows the hard subjet. This gives the following
(angular-)ordered set of tuples:
\begin{equation}
  \mathcal{L}_{\text{primary}} \equiv \big[\mathcal{T}(j_1),\dots,\mathcal
  {T}(j_n)\big]
  \quad\text{ with }
  \quad
  \mathcal{T}(j_1)\equiv\mathcal{T}(j)
  \text{ and }
  \mathcal{T}(j_{i+1})\equiv\mathcal{T}(j_\text{hard}(j_i)).
\end{equation}
In particular, the set of pairs
$(\log(1/\Delta^{(i)}),\log(k_t^{(i)}))$ corresponds to a
representation of all the primary emissions of a given jet in the
Lund-plane representation of Sec.~\ref{sec:jet-mass-res}
(cf.~Fig.~\ref{fig:lund-plain}).
Similarly, for each jet $j_i$ in the above primary list, one can
define a {\it secondary} Lund plane.
Subsidiary planes can be built similarly at will.

\begin{figure}
  \centering
  \subfloat[][Initial clustered jet and primary Lund plane.]{\includegraphics[width=0.32\textwidth,page=5]{figures/lund_sketch.pdf}\label{fig:lund-clustering-example-init}}\hfill
  \subfloat[][First declustering and new secondary plane.]{\includegraphics[width=0.32\textwidth,page=4]{figures/lund_sketch.pdf}\label{fig:lund-clustering-example-declust1}}
  \subfloat[][Second declustering and new secondary plane.]{\includegraphics[width=0.32\textwidth,page=3]{figures/lund_sketch.pdf}\label{fig:lund-clustering-example-declust2}}\\\hfill
  \subfloat[][Third declustering and new secondary plane.]{\includegraphics[width=0.32\textwidth,page=2]{figures/lund_sketch.pdf}\label{fig:lund-clustering-example-declust3}}
  \subfloat[][Fourth declustering and new ternary plane.]{\includegraphics[width=0.32\textwidth,page=1]{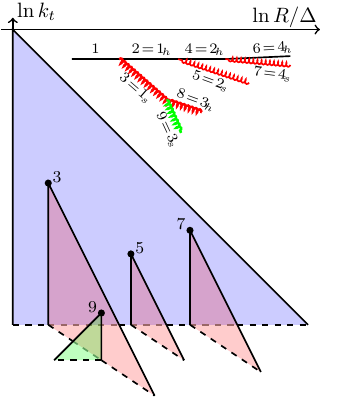}\label{fig:lund-clustering-example-final}}\hfill\\
  \caption{Illustration of how the Lund plane(s) and tree are built
    from a simple jet.}\label{fig:lund-clustering-example}
\end{figure}

The full procedure is illustrated in
Fig.~\ref{fig:lund-clustering-example}. Our starting point is the jet
clustered with the C/A algorithm of
Fig.~\ref{fig:lund-clustering-example-init}.
The first step, represented in
Fig.~\ref{fig:lund-clustering-example-declust1} is then to consider
the declustering where the initial jet ``$1$'' is separated into two
subjets ``$2\equiv 1_h$'' (harder) and ``$3\equiv 1_s$'' (softer). The
Lund coordinates of this declustering (here $\log \Delta$ and $\log k_t$) give a point on the primary Lund plane. This new point starts a
secondary plane associated with ``$3$''.
If one carries on with the declusterings following the hard branch,
one then successively gets two new points in the primary Lund plane,
``$5$'' and ``$7$'', as shown in
Figs~\ref{fig:lund-clustering-example-declust2}
and~\ref{fig:lund-clustering-example-declust3}. Each of these two
points start their own secondary Lund plane.
Finally, ``$3$'' can be further declustered into ``$8\equiv 3_h$''
(harder) and ``$9\equiv 3_s$'' (softer). This corresponds to a
secondary declustering on the plane initiated by ``$3$'', with Lund
coordinates indicated by ``$9$'' and starting a tertiary Lund plane.

Finally, the construction provided above for a jet produced in hadron
collisions can trivially be extended to $e^+e^-$
collisions.\footnote{Extensions to DIS are also possible but will not
  be discussed here.}
In this case, the full event is clustered with the $e^+e^-$ version of
the C/A algorithm and the procedure above is adapted to use kinematic
variables appropriate for $e^+e^-$ collisions. This means that the
labelling of the hard and soft branches is done based on the energy of
the two subjets and that the kinematic properties of
Eq.~\eqref{eq:lund-kinematic-variables} are adapted to use energy and
angles (or equivalently,
pseudorapidities):\footnote{Ref.~\cite{Karlberg:2021kwr} also
  introduced a new definition of the azimuthal angle $\psi$, based on
  the difference in angle between the planes defined by consecutive
  Lund declusterings.}
\begin{align}\label{eq:lund-kinematic-variables-ee}
  \eta &\equiv -\log\tan\frac{\theta_{ij}}{2},&
  k_t&\equiv E_{\text{soft}}\sin\theta_{ij},&
  z&\equiv\frac{E_{\text{soft}}}{E_{\text{hard}}+E_{\text{soft}}}.
\end{align}

\subsection{Physics with the Lund jet plane: generic overview}\label{sec:lund-generic-physics}

\begin{figure}
  \centerline{\includegraphics[width=0.5\textwidth,page=1]{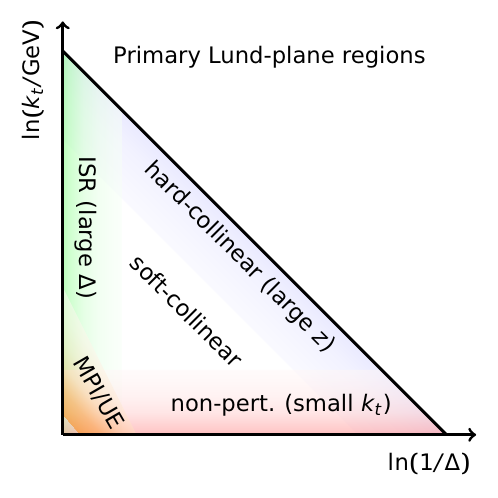}}
  \caption{Illustration of how different regions in the (primary) Lund
    jet plane are dominated by different types of kinematic contributions. Figure taken from Ref.~\cite{Dreyer:2018nbf}, licensed under CC BY 4.0.}\label{fig:lund-regions}
\end{figure}

The most appealing feature of the Lund plane construction is
that it provides one with a separation of the phase space available for
radiation within a jet into regions where different physical
contributions dominate, as illustrated in
Fig.~\ref{fig:lund-regions}.
This is a feature that we have already relied upon, see
Fig.~\ref{fig:lund-plain}, at least conceptually, across this book and which can now be extended to a practical construction.
More specifically:
\begin{itemize}
\item emissions in the bulk of the plane are at the same time soft and
  collinear, they correspond to the region of double-logarithmic
  enhancement in perturbative QCD;
\item emissions close to the upper kinematic edge ($z\lesssim 1$)
  correspond to hard-collinear perturbative branchings;
\item similarly, emissions at the edge of the jet, $\Delta\sim R$ are
  soft but finite-angle (e.g.\ associated with non-global logarithms
  in perturbative QCD);
\item emissions close to the top of the Lund plane (not indicated on
  the figure), are at the same time hard and with a finite angle, they
  would therefore correspond to fixed-order calculations with no
  logarithmic enhancement;
\item for $k_t\lesssim \Lambda_\text{QCD}$, emissions become
  non-perturbative --- an easy way to see this is that $k_t$ is the physical
  scale of $\alpha_s$ for a given emission, in such a way that at
  small $k_t$ we are eventually reaching the Landau pole --- and hence
  the bottom part of the Lund plane is dominated by hadronisation
  effects;
\item finally, the region which is at the same time very soft and at
  large angle receives dominant contributions from soft
  non-perturbative radiation in the event, i.e.\ from multi-parton
  interactions (MPI) and the Underlying Event (UE).
\end{itemize}
This feature of the Lund plane can be exploited in various ways in jet
substructure applications. For example, applying a minimal $k_t$ cut
on Lund declusterings allows one to reduce the sensitivity of a
Lund-plane-based observable to non-perturbative effects.

Another hallmark of the Lund plane construction is that it provides a
very physically-intuitive picture of the substructure of a jet --- one
with direct ties to resummations --- which can be applied to almost
all aspects where jet substructure is relevant.
This includes jet tagging and machine-learning
applications~\cite{Dreyer:2018nbf,Lifson:2020gua,Dreyer:2021hhr,Mehtar-Tani:2020oux,Cavallini:2021vot}, the design of new
jet substructure observables~\cite{Dreyer:2021hhr,Medves:2022ccw,Medves:2022uii,Hamilton:2021dyz,Dasgupta:2020fwr,vanBeekveld:2025zjh,Karlberg:2021kwr,Mehtar-Tani:2019rrk},
experimental
measurements~\cite{ATLAS:2020bbn,Ehlers:2020piz,Havener:2021yhb,CMS:2023lpp,ATLAS:2024wrd} (see also the dedicated discussion in
section~\ref{sec:exp-lund-plane} of the next
chapter), precision
analytic calculations~\cite{Lifson:2020gua,Caucal:2021bae,Hamilton:2021dyz,Dreyer:2021hhr,Medves:2022ccw,Medves:2022uii,vanBeekveld:2025zjh,Ghira:2025nym}, Monte
Carlo studies and accuracy assessment~\cite{Dasgupta:2020fwr,Karlberg:2021kwr,Hamilton:2021dyz,vanBeekveld:2022ukn,vanBeekveld:2023chs,vanBeekveld:2024wws}, studies of hard probes in heavy-ion
collisions~\cite{Andrews:2018jcm,Caucal:2021bae}, etc.
The rest of this chapter is devoted to illustrating this statement,
putting a special emphasis on analytic approaches as done in the rest
of these lectures.

Before diving head-first into this matter, another generic comment is
due. Over the past few years, another jet substructure tool with a broad range of applications has emerged, namely energy
correlators, briefly discussed in Sec.~\ref{energy_correlators}.
It is not our aim here to provide an in-depth discussion of the pros
and cons of each of these two approaches to characterising jet
substructure.\footnote{One can, for example, quote interesting
  properties of energy correlators in quantum field theory, or the
  flexibility of building tailored observables using Lund-plane
  coordinates.}
Instead, we want to point out that they are each rooted in different
ways of viewing a jet: energy correlators tend to view a jet as a flow
of energy, while Lund-plane constructions tend to view it as
successive branchings.
It is interesting to notice that these two pictures were already
around in the early days of jet physics, when jet algorithms could also be divided into two broad families based on the same views. Indeed,
cone and pairwise recombination algorithms have been conceptually
based on these two views.
It is also worth noting that the two approaches are not incompatible
with one another. We could, for example, imagine defining
energy-correlator-based observables using Lund declusterings as
effective particles.

\section{Lund-plane-based substructure observables}\label{sec:lund-observables}

\begin{figure}
  \centerline{\includegraphics[width=0.5\textwidth,page=1]{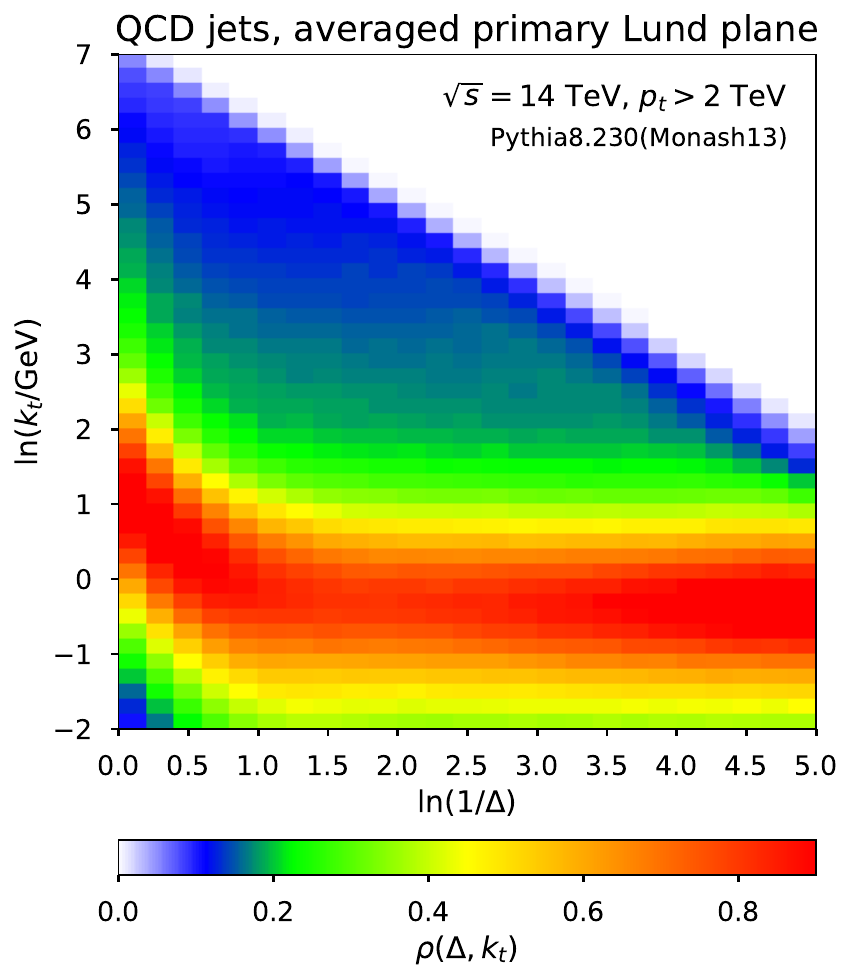}}
  \caption{The average primary Lund plane density, $\rho$, for jets
    clustered with the C/A algorithm with $R=1$. We selected jets having
    $p_t>2$~TeV and $|y|<2.5$. Figure taken from Ref.~\cite{Dreyer:2018nbf}, licensed under CC BY 4.0.}\label{fig:lund-primary-plane}
\end{figure}

The construction introduced earlier in this chapter provides us with a full
tree of kinematic properties given an initial jet.
They can be used to define a whole zoo of new jet substructure
observables.
We only give a small subset here, in order to illustrate typical
possibilities.

Probably the most natural observable one can consider it the
{\it (primary) Lund plane density}
\begin{equation}\label{eq:lund-density-def}
  \rho(\Delta,k_t) \equiv \frac{1}{N_\text{jet}}\frac{dn_\text{emissions}}{d\log(1/\Delta)d\log(k_t)},
\end{equation}
which corresponds to the density of emissions in the Lund plane, we have
already considered several times through these lectures.

Fig.~\ref{fig:lund-primary-plane} shows the average Lund plane density
obtained from Pythia simulations using a dijet event sample. This
shows in a very intuitive way the distribution of radiation in various
regions of phase-space. In particular, the main regions labelled in
Fig.~\ref{fig:lund-regions} (see also Fig.~\ref{fig:lund-plain}) can
be clearly identified in Fig.~\ref{fig:lund-primary-plane}.
At first sight, when lowering $k_t$, one sees a density of emissions
slowly rising in the perturbative region ($k_t$ of at least a few GeV)
and showing a large peak with $k_t$ slightly below 1~GeV.
An increase of density attributed to multi-parton interactions is also
observed in the soft-large-angle region.
We will further discuss several properties of this observable in the
forthcoming sections.

A particularly simple observable one can then consider is the {\it Lund
  multiplicity}, $N^\text{(Lund)}$, defined as the number of Lund
declusterings in a jet~\cite{Medves:2022ccw,Medves:2022uii}. This is usually computed using the full
(de)clustering tree and including only the declusterings above a
minimal $k_{t,\text{cut}}$. Multiplicity observables have always been
considered in collider phenomenology, mostly for their simplicity, and
the Lund plane construction allows one to define an
infrared-and-collinear-safe version also valid for hadronic
collisions.

Finally, many of the jet substructure observables and tools introduced
throughout this book, can be reformulated, exactly or as a new
similar observable, using Lund variables.
As a typical example, consider \SD.
The \SD jet is given by the first jet that satisfies the condition $z\ge z_ \text{cut}(\Delta/R)^\beta$ in the ordered
primary Lund sequence, $\mathcal{L}_\text{primary}$.
Recursive versions can also be obtained, for example, by applying the
\SD condition on all the declustering in the primary Lund
sequence or in the full Lund tree.

Furthermore, a set of Lund declusterings can be used as
pseudo-particles in order to define jet substructure observables. As a
specific example, we can consider the following Lund observables~\cite{Dasgupta:2020fwr,vanBeekveld:2022ukn,vanBeekveld:2025zjh}
\begin{align}
  S_\beta
  & = \sum_{\mathcal{T}\in\mathcal{L}_\text{primary}} k_t\Delta^\beta,\\
  M_\beta
  & = \max_{\mathcal{T}\in\mathcal{L}_\text{primary}} k_t\Delta^\beta,
\end{align}
where the sum (max) is taken over all the primary Lund declusterings.
The first of these observables, $S_\beta$ is very similar to the
angularities $\lambda_\alpha$ and energy-correlation functions
$e_2^{(\alpha)}$ (with $\alpha=\beta+1$) introduced in
Sec.~\ref{sec:tools-radiation-constraints}. The fact that $S_\beta$
is defined on clustering variables also means that it is a recoil-free
observable, at least at NLL accuracy, for $\beta\le 0$, as
energy-correlation functions.
The $M_\beta$ observable is also relatively simple and remains
infrared-and-collinear safe thanks to the use of declusterings
instead of jet constituents.

These two observables admit several easy generalisations. Here is a
non-exhaustive list of interesting examples:
\begin{itemize}
\item one can define groomed versions of the $S_\beta$ and $M_\beta$
  observables by computing them only on declusterings that pass a grooming
  condition, like the \SD one,
\item one can select the primary declustering with the largest value
  of $v_\beta=k_t\Delta^\beta$ and use $v_{\beta'}$ with
  $\beta'\neq\beta$ as an observable. This is referred to as {\it
    dynamical grooming}~\cite{Mehtar-Tani:2019rrk,Mehtar-Tani:2020oux,Caucal:2021bae} and reduces to $M_\beta$ for
  $\beta'=\beta$.
\item if one orders all the declusterings according to the ordering
  variable $v_\beta$, we can consider their sum or max excluding the
  $n$ larger ones. This is similar to what is achieved by
  $N$-subjettiness $\tau_{n+1}$.
\end{itemize}

\section{Tagging and machine-learning applications}\label{sec:lund-tagging}

As already seen multiple times throughout these lecture notes, a core
application of jet-substructure tools is to help discriminate
between boosted jets of different origins.
This is not different for the Lund plane.
A specific aspect that we will try to highlight here is that Lund
plane variables have the flexibility to interpolate between basic
substructure variables (cf.\ the previous section) and a much more complete set of variables that captures the full clustering
kinematics.
Simpler variables have the potential to be understood from first principles in QCD, using techniques similar to what we covered in
chapters~\ref{calculations-substructure-mass}-\ref{chap:calc-two-prongs},
while more extensive sets are suited for taggers using an approach
based on machine learning.

To illustrate various approaches, we will discuss three specific
applications: (i) the tagging of massive two-prong colourless bosons
using information from the primary Lund plane,
(ii) ($W$ and top) jet tagging using information from the full Lund tree,
and (iii) the simpler case of quark/gluon tagging, where, on top of
deep-learning-based taggers, one can use analytic approaches using
information from the full Lund tree.

\subsection{Using the primary Lund plane: two-prong tagging}\label{sec:lund-W-tagging}

\begin{figure}[t]
  \centering
  \subfloat[][Equivalent of Fig.~\ref{fig:lund-primary-plane} for jets
  originated from boosted $W$ bosons.]{\includegraphics[width=0.48\textwidth,page=2]{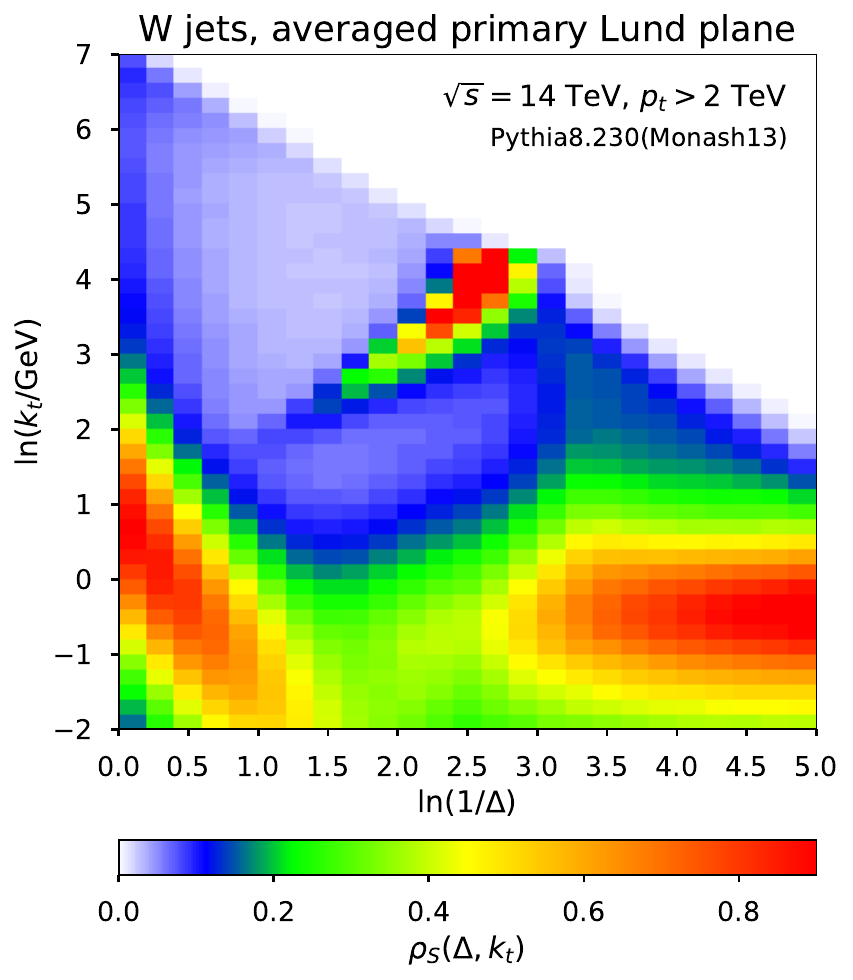}\label{fig:primary-lund-W}}\hfill
  \subfloat[][Ratio between Lund density for $W$ and QCD jets after
  removal of the leading emission.]{\includegraphics[width=0.48\textwidth,page=3]{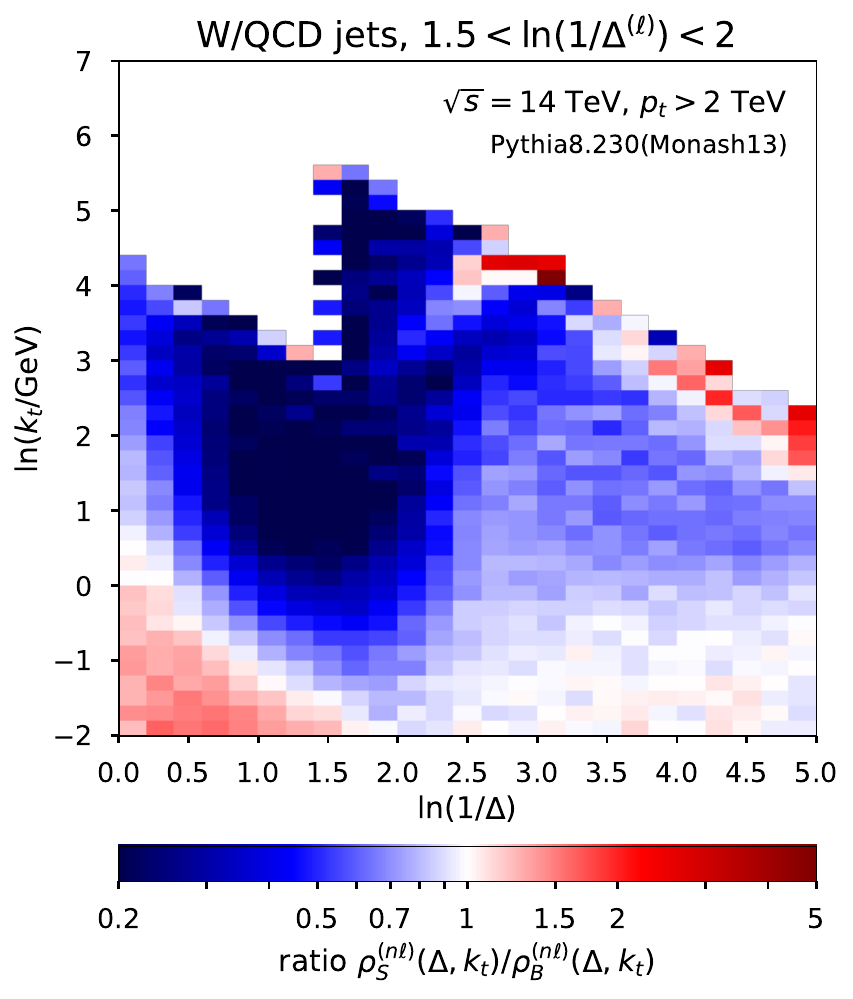}\label{fig:primary-lund-Wratio}}  
  \caption{The average primary Lund plane density, $\rho$, for jets
    clustered with the C/A algorithm with $R=1$. We selected jets having
    $p_t>2$~TeV and $|y|<2.5$. Figures taken from Ref.~\cite{Dreyer:2018nbf}, licensed under CC BY 4.0.}\label{fig:lund-primary-plane-W-tagging}
\end{figure}

We start by illustrating how the Lund plane density $\rho(\Delta,k_t)$
can be used to discriminate boosted hadronic $W$ jets from the QCD
background.
The primary Lund plane density for $W$ jets is shown in
Fig.~\ref{fig:primary-lund-W} and should be compared to the one of QCD
jets in Fig.~\ref{fig:lund-primary-plane}.
The massive 2-prong structure of boosted $W$ jets is clearly visible
from the overdensity on a line around the $W$ mass in the Lund plane,
as well as from an underdensity at larger masses.
We note also that large-angle and MPI effects populate the region of
the $W$ Lund plane that would otherwise be vetoed. This partially
justifies the use of grooming techniques to eliminate this region in
basic two-prong-tagging approaches.

To dive deeper into $W$-jet tagging, we proceed by identifying the
leading emission ``$\ell$'' in the jet, defined as the first
declustering with $z>z_\text{cut}=0.0025$ (equivalent to running mMDT
or SoftDrop).
We then consider the Lund plane density of all other primary emissions
(in bins of $\Delta^{(\ell)}$) and plot the ratio between $W$ and QCD
jets in Fig.~\ref{fig:primary-lund-Wratio}.
Two main features are visible on this plot.
The light blue region in the small-angle (right) region of the Lund
plane shows a depletion in $W$-initiated jets compared to QCD
jets. This can be attributed to the fact that, in the $W$ case, one is
strongly dominated by radiation off quark legs, while in the QCD case,
one has a mixture of quarks and gluons (mainly originating from the
leading parton initiating the jet).
More importantly, one sees a strong depletion in a region of angles
$\Delta\gtrsim\Delta^{(\ell)}$, except, as before, in the
large-angle/MPI regions.
This is a clear trace of the fact that QCD radiation at angles larger
than the hard $W\to q\bar q$ opening angle is strongly suppressed
because the $W$ is a colourless object, again a feature extensively
exploited and discussed in previous chapters.

Based on these observations, one can construct a very simple
Lund-plane-based tagger by taking the Likelihood ratio of the expected
average density for $W$ and QCD jets:\footnote{In practice, one would
  actually separate the density for the leading emission from the one
  of all the other emissions. We have omitted this aspect in this
  equation for simplicity.}
\begin{equation}
  \mathbb{L} = \sum_{i\in\mathcal{L}_\text{primary}} \log\frac{\rho_W(\Delta_i,k_{t,i})}{\rho_\text{QCD}(\Delta_i,k_{t,i})}.
\end{equation}
In practice, the expected averages can be computed from a Monte Carlo
simulation and we use Pythia(v8.230). 
If all the emissions in the primary Lund plane sequence were
independent, this would be the optimal discriminant.

Alternatively, we can train a neural network with the list of all
primary Lund declusterings. In practice, one feeds the ordered list
to a Long Short-Term Memory (LSTM) neural network~\cite{lstm}
(see~\cite{Dreyer:2018nbf} for details).
We have also included a more standard approach using the \SD
($\beta=2$, $z_\text{cut}=0.05$) jet mass and $D_2^{(\beta=2)}$, which
essentially corresponds to
$D_2^\text{[loose]}\equiv D_2^{(2)}[l\otimes l/l]$ in
Sec.~\ref{sec:2prongs-perf-robustness}, as inputs to a boosted
decision tree (BDT).

\begin{figure}
  \centering
  \includegraphics[width=0.48\textwidth]{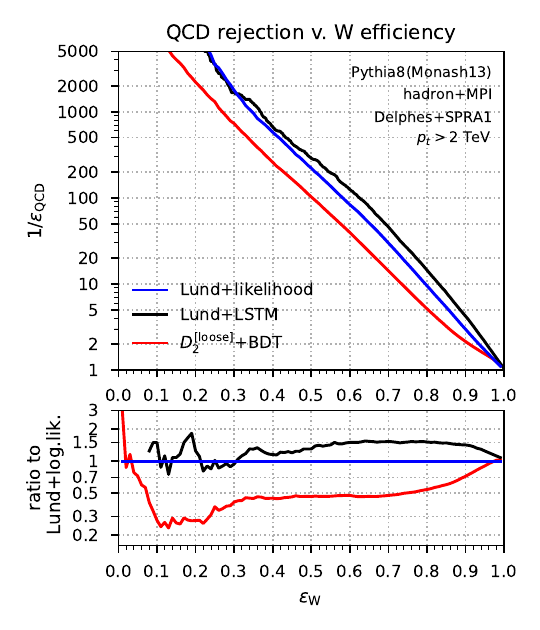}
  \caption{Background rejection as a function of the $W$ tagging
    efficiency for several tagging methods. Figure adapted from Ref.~\cite{Dreyer:2018nbf}, licensed under CC BY 4.0.
    }\label{fig:lund-Wtag}
\end{figure}

The resulting tagging performance is plotted in Fig.~\ref{fig:lund-Wtag}
(together with other machine-learning-based approaches,
see~\cite{Dreyer:2018nbf} for details), which shows the QCD background
rejection rate as a function of the $W$ tagging efficiency.
Focusing on the methods described above, we see that using the full
primary Lund plane information yields a gain in performance compared
to the $D_2^\text{[loose]}$+BDT approach.
Furthermore, the LSTM-based approach, which is in principle able to
exploit correlations between Lund declusterings outperforms the
likelihood-based approach.

\begin{figure}
  \centering
  \includegraphics[scale=0.75]{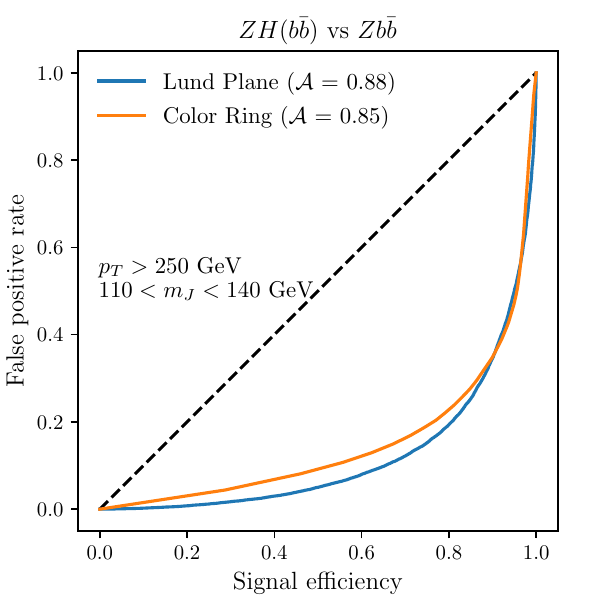}
  \includegraphics[scale=0.75]{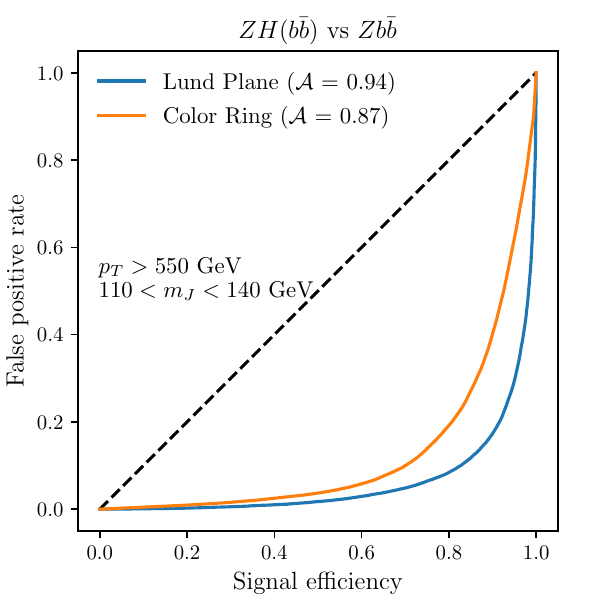}
  \caption{False positive rate as a function of the signal efficiency for $H \to b \bar b$, as obtained with the primary Lund plane density and the colour ring. Two different transverse-momentum regions are considered: moderate boost (on the left) and high boost (on the right). Plots taken from~\cite{Khosa:2021cyk}, licensed under CC BY 4.0.
    }\label{fig:lund-Hbb}
\end{figure}
  
Another interesting application of the Lund plane in the context of
two-prong jet tagging is in the case of boosted Higgs boson identification.
In particular, the decay of the Higgs boson into a pair of a pair bottom ($b$) quarks that subsequently fragment into two $b$-(sub)jets  ($H\to b\bar b$) allows us to study the mechanism of electroweak symmetry breaking in the fermionic sector. 

The main background to $H\to b\bar b$ is given by $g\to b\bar
b$. However, unlike the case of vector boson tagging discussed at
length in this book, once an invariant mass window is selected, the kinematics of the $H\to b\bar b$ and $g\to b\bar
  b$ branchings look rather similar because the splitting of a
gluon into a quark pair has no soft singularity. Thus, one has to rely
on different properties to distinguish signal and background, such as
the colour representation of the decaying particle, i.e.\ colour
singlet versus colour octet.
To this purpose, the jet colour ring ($\mathcal{O}$) is a
colour-singlet tagger for boosted two-prong decays that was introduced
in~\cite{Buckley:2020kdp}. It is defined as an approximation of the optimal classifier, which is in turn given by the likelihood ratio:
\begin{equation}\label{eq:higgs-likelihood-ratio}
  \mathbb{L}(\mathcal L) = \frac{p_{g}(\mathcal{L})}{p_{H}(\mathcal{L})}.
\end{equation}
In particular, the colour ring is defined by evaluating the above ratio with leading-order matrix elements for the emission of a soft and collinear gluon, arriving at
\begin{equation}
 \mathcal{O} = \frac{\Delta_{ka}^2+\Delta_{kb}^2}{\Delta_{ab}^2},
 \end{equation}
where $a$ and $b$ are primary subjets, i.e.\ the subjets that have been $b$-tagged, while $k$ is leading remaining subjet. This third subjet is taken as a proxy for soft-gluon emission in the jet. As usual, $\Delta$ measures the separation, in the rapidity-azimuth plane, between the subjet pairs. 
Colour conservation dictates that $a$ and $b$ are colour-connected if the decaying state is a colour-singlet. In such a case, $k$ will be predominantly emitted in between the legs of the $ab$ dipole and, as a result, the distribution of the colour ring will be peaked at small $\mathcal{O}$.

The Lund plane offers an alternative way of contrasting the radiation patterns due to the different colour-connections in signal and background jets.
In particular, by considering the primary Lund plane as a proxy for the two-dimensional phase space of the leading emission in the jet, one can build Lund jet plane images for signal and background, which are used as inputs to a classifier, such as, for instance, a CNN~\cite{Khosa:2021cyk}.
While both the Lund jet plane and the jet colour ring are inspired by a first-principle analysis of the physical process we are interested in, the former provides us with more information than the jet colour ring. 
A comparison of the tagging performance of the Lund jet plane, used as the input to a CNN, is compared to the one obtained with the colour ring in Fig.~\ref{fig:lund-Hbb} in two different transverse momentum regions: moderate boost on the left and high boost, on the right. 
A standard metric used to assess the classification performance is the Area Under the ROC Curve (AUC). With our definition of ROC curves, optimal performance corresponds to AUC=0. In the plots we show the metric $\mathcal{A}=1-\text{AUC}$, where now $\mathcal{A}=1$ corresponds to optimal performance. 
We see that, in these cases, the Lund-plane-based tagger brings some moderate improvement in performance. Actually, this improvement becomes much more prominent when detector effects are included~\cite{Cavallini:2021vot}.

\subsection{Using the full Lund tree: $W$ and top tagging}\label{sec:lund-Wt-tagging}

The full information from the Lund tree can be used in a
machine-learning approach thanks to graph neural networks, see e.g.~\cite{hamilton2020graph}.
The jet tagging method is called LundNet~\cite{Dreyer:2020brq}. It works by
creating a graph network where nodes carry kinematic Lund variables
and the connections between nodes follow the Lund clustering
tree.
We note that this approach is similar to ParticleNet~\cite{Qu:2019gqs}, which also relies on a graph network where nodes
carry the 4-vectors and PDGids of the jet constituents, and are
connected according to their nearest neighbours.
In practice, for each node, two different sets of kinematic variables
have been considered'' 
\begin{align}
  &(\log k_t, \log \Delta, \log z) & \text{[LundNet-3]},\\
  &(\log k_t, \log \Delta, \log z, \log m, \Psi) & \text{[LundNet-5]}.
\end{align}
\begin{figure}
  \subfloat[][Background rejection for $W$ tagging]{%
    \includegraphics[width=0.48\textwidth,page=2]{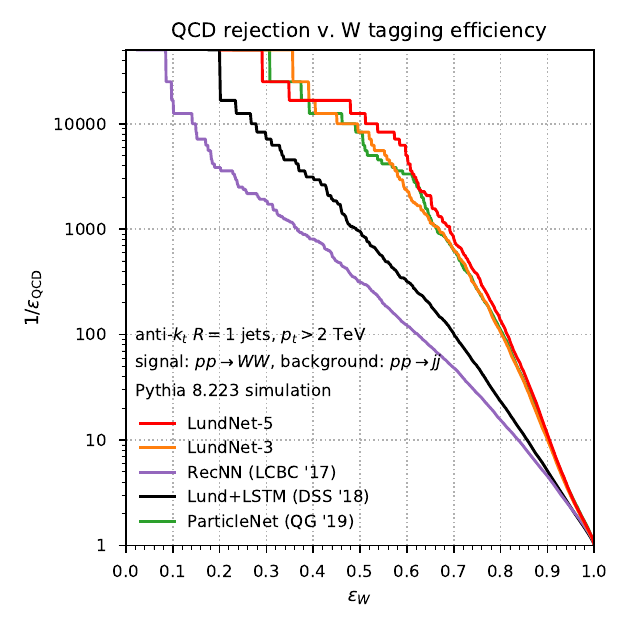}
    \label{fig:lundnet-roc-W}}\hfill
  \subfloat[][Background rejection for top tagging]{%
    \includegraphics[width=0.48\textwidth,page=3]{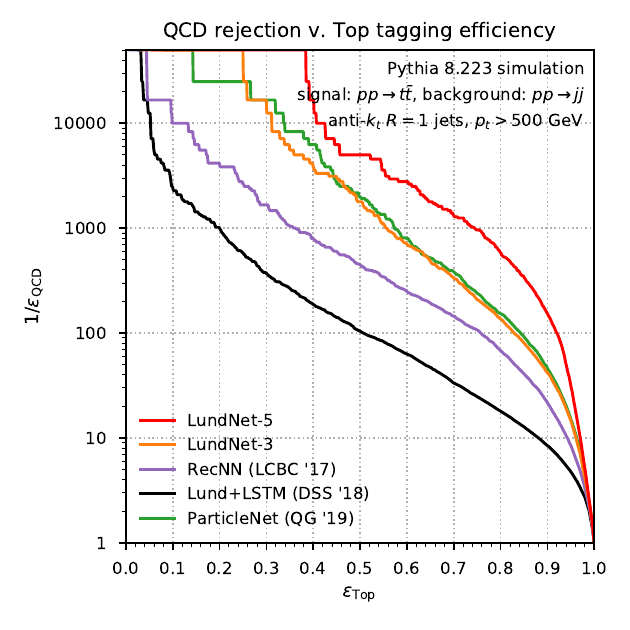}
    \label{fig:lundnet-roc-t}}
  \caption{ROC curves for boosted jet tagging with the LundNet graph
    network using the full Lund tree information. Plots taken from
    Ref.~\cite{Dreyer:2020brq}, licensed under CC BY 4.0.}\label{fig:lundnet-rocs}
\end{figure}
The neural networks are trained on Pythia(v.8.223) samples of either
$pp\to WW$ event to study boosted $W$ jets with $p_t>2$~TeV, or
$pp\to t\bar t$ events to study boosted top jets of $p_t>500$~GeV,
each time with the corresponding QCD dijet sample as
background.
The resulting tagging efficiencies are plotted in
Fig.~\ref{fig:lundnet-rocs}, together with the ParticleNet results, as well as the recursive neural network,
RecNN, from Ref.~\cite{Louppe:2017ipp}.
For the case of $W$ tagging, we also include the results using primary
Lund plane with an LSTM network presented in the previous section.
In both the $W$ and top tagging cases, we see that including
information from the full Lund tree gives a substantial performance
gain compared to using only the primary Lund plane.
Furthermore, the performance of LundNet-3 is on par with using the
full 4-momentum+PDGid information in ParticleNet.
Furthermore, LundNet-5 gives a small but visible additional
enhancement, likely related to the additional information brought by
the inclusion of the azimuthal angle $\psi$.
Finally, we note that one can also add PDGid information to LundNet.

\subsection{Analytic and deep-learning approaches to quark/gluon tagging}\label{sec:lund-qg-tagging}

While the approaches introduced above also apply to the case of
quark/gluon tagging, this latter case also exhibits several
interesting analytic possibilities.

We start from the fact that the best discriminant between quark and
gluon jets is the likelihood ratio
\begin{equation}\label{eq:qg-likelihood-ratio}
  \mathbb{L}(\mathcal L) = \frac{p_{g}(\mathcal{L})}{p_{q}(\mathcal{L})},
\end{equation}
where $p_{q,g}(\mathcal{L})$ is the probability to have a quark or
gluon jet given a set of Lund declusterings $\mathcal{L}$, with
$\mathcal{L}$ being either the set of primary declusterings or the
full Lund tree. 

In principle, a neural network is supposed to learn this ratio during the training phase.
The new approach we want to address here is to compute
$\mathbb{L}(\mathcal L)$ analytically in perturbative QCD.
For this to be infrared-and-collinear safe, we only consider the Lund
declusterings above a minimal $k_t\equiv k_{t,\text{cut}}$ scale. We
can then consider the likelihood ratio in a perturbative series
expansion.

\subsubsection{Double-logarithmic limit: Lund multiplicity}

In the double logarithmic approximation, all emissions are soft gluons
strongly ordered in both angle and energy.
This considerably simplifies the computations of
$p_{q,g}(\mathcal{L})$. In particular, in this limit, primary
emissions are independent, and all the non-primary emissions are the
same in quark and gluon jets. Furthermore, the only difference in
primary emissions between quark and gluon jets is their colour factor:
$C_F$ for quarks and $C_A$ for gluon.
We therefore have
\begin{equation}\label{eq:qg-likelihood-doublelog}
  \mathbb{L}
  = \frac{p_{g}(\mathcal{L})}{p_{q}(\mathcal{L})}
  = \prod_{i\in\mathcal{L}_\text{prim}} \frac{C_A}{C_F}
  = \left(\frac{C_A}{C_F}\right)^{n_\text{prim}(\mathcal{L})},
\end{equation}
where $n_\text{prim}(\mathcal{L})$ denotes the multiplicity of
primary gluon emissions.
This means that, in the soft-collinear limit, the best discriminant
between quark and gluon jets is the multiplicity of primary Lund
emissions.
This recovers a result initially derived in~\cite{Frye:2017yrw}.

\subsubsection{Beyond the double-logarithmic limit}

We now extend the previous result beyond the soft-collinear limit.
Working, as we did throughout most of these lecture notes, in the
boosted jet limit, it is natural to consider the perturbative
corrections arising from subleading logarithmic contributions.
In this limit, one would typically have to consider three main sources
of effects: (i) subleading soft-collinear corrections, (ii) collinear
emissions, and (iii) soft emissions at an angle commensurate with either
the jet radius $R$, or to an earlier emission.
We note that in all these cases, and at NLL accuracy, the flavour of the jet is well-defined and corresponds
to the flavour of the initial hard parton, see our discussion in chapter~\ref{sec:calc-shapes-qg}.

The inclusion of contributions that originate from the radiation of soft gluons at commensurate
angles is not trivial. Since it is not the dominant phenomenological
effect, namely an increase of $\mathcal{O}(10\%)$ in tagging
efficiency, we will neglect it here and refer to Ref.~\cite{Dreyer:2021hhr}.
We will briefly touch upon the underlying physics aspects later in this
chapter, when discussing analytic aspects of the primary Lund plane
density.

From now on, we therefore work in the collinear limit.
For further simplicity, we consider the case where we only use the
primary Lund declusterings. The extension to the full clustering tree
only brings additional technicalities.
The idea is therefore that we have a list of primary Lund
declusterings, strongly ordered from large angle to small angle.
The initial hard parton, at step ``0'', has a given flavour $f_0$.
Through the succession of collinear branchings, the flavour of the
hard branch can vary, and we denote by $f_i$ the flavour of the leading
parton after branching $i$, $i=1,\dots,N$.

In the leading (collinear) logarithmic approximation, the
probabilities associated with the successive branchings are given by
the Altarelli-Parisi splitting kernels. There are, however, two subtle
points to take into account: (a) the primary Lund declustering
procedure recurses into the harder branch, meaning that if a parton of
momentum fraction $x$ branches into two partons of momentum fractions
$zx$ and $(1-z)x$, the leading parton will be the one with momentum
fraction $\max(z,1-z)x$; (b) between two branchings, one should
include a Sudakov factor accounting for the probability of no
emissions.

To put this in a practical equation, we introduce the matrix
\begin{equation}
  p_{ab}^{(i)} =
  \begin{pmatrix}
    p(q_i|q_0) & p(q_i|g_0) \\
    p(g_i|q_0) & p(g_i|g_0) 
  \end{pmatrix},
\end{equation}
where $p(f'_i|f_0)$ denotes the probability to have a leading parton
of flavour $f'_i$ after the $i$th branching if the original flavour
was $f_0$.
If we take into account a DGLAP-like probability at each branching and
a Sudakov factor in between each branching, we get
\begin{equation}
  p^\text{(final)}
  = S^{(N+1,N)}
  \tilde P^{(N)}S^{(N,N-1)}
  \dots
  \tilde P^{(i)}S^{(i,i-1)}
  \dots
  \tilde P^{(1)}S^{(1,0)}
  p^{(0)},
\end{equation}
where both the splitting kernel and the Sudakov factor are matrices in
flavour space:
\begin{align}
  P^{(N)}
  & = \frac{\alpha_s(k_{ti})}{\pi\Delta_i}\begin{pmatrix}
    \tilde P_{qq}(z_i) & \tilde P_{qg}(z_i) \\
    \tilde P_{gq}(z_i) & \tilde P_{gg}(z_i) 
  \end{pmatrix}
  & S^{(i,i-1)}
  & = \begin{pmatrix}
    S_q^{(i,i-1)} & . \\
    . & S_g^{(i,i-1)}
  \end{pmatrix}.
\end{align}
In these expressions, we have explicitly factored the running coupling
and collinear enhancement out of $\tilde P$, and made the Sudakov
matrix diagonal as the flavour of the leading parton is unchanged
between successive emissions.
More explicitly, we have introduced the non-standard splitting functions $\tilde{P}_{ij}$, which in terms of the usual leading-order splitting function matrix (see appendix~\ref{app:splitting_functions}) read
\begin{subequations}\label{eq:splitting_matrix}
\begin{align}
  \tilde P_{qq}(z)
  & = P_{gq}(z)\Theta\big(z<\tfrac{1}{2}\big)
    = C_F\frac{1+(1-z)^2}{z}\Theta\big(z<\tfrac{1}{2}\big),\\
 \tilde P_{gq}(z)
  & = P_{qq}(z)\Theta\big(z<\tfrac{1}{2}\big)
    = C_F\frac{1+z^2}{1-z}\Theta\big(z<\tfrac{1}{2}\big),\\
  \tilde P_{qg}(z)
  & = [P_{gq}(z)+P_{gq}(1-z)]\Theta\big(z<\tfrac{1}{2}\big) 
    = 2 n_f T_R [z^2+(1-z)^2]\Theta\big(z<\tfrac{1}{2}\big),\\
  \tilde P_{gg}(z)
  & = [P_{gg}(z)+P_{gg}(1-z)]\Theta\big(z<\tfrac{1}{2}\big)
    = 2C_A\left[\frac{1-z}{z}+\frac{z}{1-z}+z(1-z)\right]\Theta\big(z<\tfrac{1}{2}\big),
\end{align}
\end{subequations}
and
\begin{equation}
  \log S^{(i,i-1)}_f
  = -\int_{\Delta_i}^{\Delta_{i-1}}\frac{d\Delta}{\Delta}
  \int dz\,\frac{\alpha_s(p_{t,i-1}z\Delta)}{\pi}
  P_f(z)\Theta(p_{t,i-1}z\Delta>k_{t,\text{cut}}),
\end{equation}
where $f=q,g$, see Eqs.~(\ref{eq:quarksplitting}) and~(\ref{eq:gluonsplitting})
The expression we have used for the Sudakov explicitly uses the fact
that the transverse momentum of the emission relative to its emitter
can be written as $k_t=p_{t,i-1}z\Delta$, with
$p_{t,i-1}\equiv xp_{t,\text{jet}}$ the momentum of the leading parton
between branchings $i-1$ and $i$.
If we use a 2-loop expression for $\alpha_s$ in the CMW scheme, we can
evaluate $S$ at the NLL accuracy as done repeatedly in the previous
chapters. We get
\begin{align}
  \log S^{(i,i-1)}_f
  = -\frac{C_f}{2\pi\alpha_s\beta_0^2}
  & \bigg[(1-\lambda_{i-1}\log\frac{1-\lambda_{i-1}}{1-\lambda_\text{cut}}
    -(1-\lambda_{i}\log\frac{1-\lambda_i}{1-\lambda_\text{cut}}
    +(\lambda_{i-1}-\lambda_i)\nonumber\\
  & -\frac{\alpha_s\beta_1}{\beta_0}\left(
    \frac{1}{2}\log^2(1-\lambda_i)-\frac{1}{2}\log^2(1-\lambda_{i-1})
    +\frac{\lambda_i-\lambda_{i-1}}{1-\lambda_\text{cut}}\log(1-\lambda_\text{cut})
    \right)\nonumber\\
  & +\left(\frac{\alpha_sK}{2\pi}-\frac{\alpha_s\beta_1}{\beta_0}\right)
    \left(\frac{\lambda_i-\lambda_{i-1}}{1-\lambda_\text{cut}}-\log\frac{1-\lambda_{i-1}}{1-\lambda_i}\right)\bigg],
\end{align}
with $\alpha_s\equiv\alpha_s(p_{t,\text{jet}}R)$, and
\begin{align}
  \lambda_j & =
    2\alpha_s\beta_0\left(\log\frac{R}{x\Delta_j}-B_f\right),&
  \lambda_\text{cut} & =
  2\alpha_s\beta_0\log\frac{p_{t,\text{jet}}R}{k_{t,\text{cut}}}.
\end{align}
In practice, these expressions can be evaluated iteratively, assuming
that we start either from an initial quark or gluon, i.e.\ with
$p^{(0)}_{ab}=\delta_{ab}$, and summing over all possible flavours for
the final parton, i.e.
\begin{equation}
  p_f(\mathcal{L_\text{prim}})
  = p^\text{(final)}(q|f_0) + p^\text{(final)}(g|f_0).
\end{equation}
These probabilities can, in turn, be inserted into the likelihood ratio
\eqref{eq:qg-likelihood-ratio} and used as a quark-gluon discriminant,
extending the simple double-logarithmic expressions to a more precise
version accounting for the full hard-collinear structure at leading
log.
A similar approach can be derived for the Lund clustering tree,
essentially by keeping track of the flavour of both branches and
performing the iteration procedure on both branches as well.

\subsubsection{Validation}

\begin{figure}[t]
  \subfloat[][AUC varying the size of the training sample.]{\includegraphics[width=0.472\textwidth]{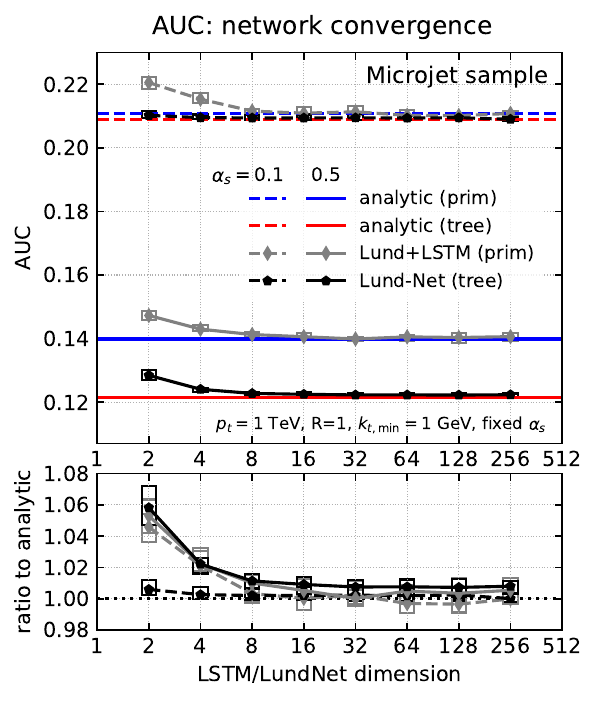}\label{fig:qg-lund-coll-auc}}\hfill
  \subfloat[][ROC curve for a series of taggers.]{\includegraphics[width=0.48\textwidth]{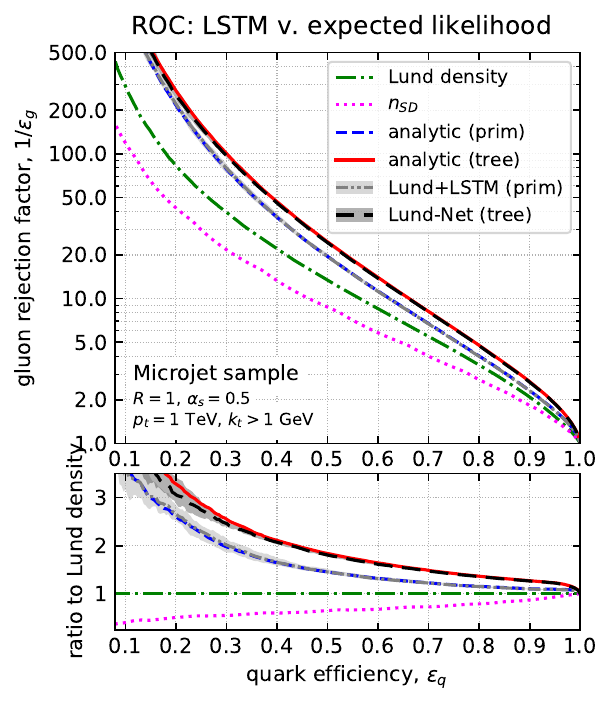}\label{fig:qg-lund-coll-roc}}
  \caption{Performance plots for quark/gluon taggers on a sample with
    exact leading-order collinear evolution (as obtained with
    MicroJet). Plots taken from Ref.~\cite{Dreyer:2021hhr}, licensed under CC BY 4.0.}\label{fig:qg-lund-coll}
\end{figure}

The above calculation derives an optimal quark/gluon tagger at a given
order of perturbation theory.
A nice side aspect of this construction is that it provides a way to
check that a neural-network-based tagger correctly learns the optimal
discriminant.

For this, we generate quark and gluon samples in the exact
leading-order collinear limit using the MicroJet
code~\cite{Dasgupta:2014yra}.
For these events, we can reconstruct the primary Lund declusterings or
the full Lund tree, and keep only the ones above a minimum
$k_t\equiv k_{t,\text{cut}}$.
We can then use this information to train a neural network, an LSTM for the
primary tree, or a LundNet for the full tree.
We can then compare the quark/gluon tagging efficiency of this network
against the performance of the analytic tagger derived above, knowing
that the latter is the optimal one.

The results are shown in Fig.~\ref{fig:qg-lund-coll}.
First, Fig.~\ref{fig:qg-lund-coll-auc} shows the area under the ROC
curve as a function of the dimension of the neural network. It
includes results for analytic (blue and red curves) and deep-learning
taggers (grey and black curves), using either information from the
primary Lund plane (blue and grey curves) or from the full Lund tree
(red and black curves), each time with two values of $\alpha_s$.
In absolute terms, we see an increase in performance from using
information from the full Lund tree.
In the bottom panel, we plot the ratio between the ML and analytic
taggers. We see that, when the network is large enough, it correctly
converges to the optimal (analytic) tagger.

The right plot, Fig.~\ref{fig:qg-lund-coll-roc}, shows the ROC curves
obtained with the $\alpha_s=0.5$ MicroJet sample.
The same methods as for the left panel are included, together with the
ratio of the average primary Lund density (as what was presented for
$W$ tagging in section~\ref{sec:lund-W-tagging}), and with the primary
Lund plane multiplicity, which is optimal in the soft-collinear limit.
We first see that the performance increases as one includes more
information for tagging. We also see that, as for the AUC, the methods
based on training a neural network successfully capture the best
discriminant (given by the corresponding analytic tagger).

\subsubsection{Performance}

\begin{figure}[t]
  \subfloat[][Including all Lund declusterings.]{\includegraphics[width=0.48\textwidth,page=1]{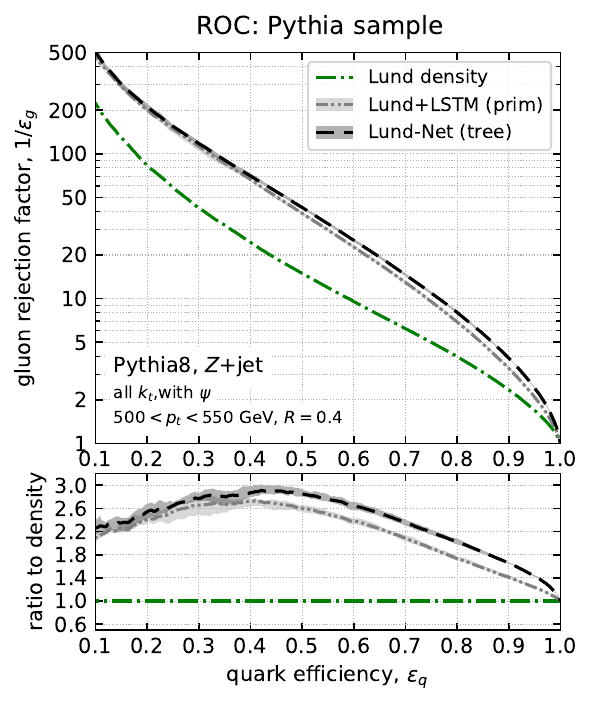}\label{fig:qg-lund-py-allkt}}\hfill
  \subfloat[][Including only declusterings with $k_t\ge 1$~GeV.]{\includegraphics[width=0.48\textwidth,page=2]{figures/lund-qg-rocs.pdf}\label{fig:qg-lund-py-kt1}}
  \caption{Performance plots for quark/gluon taggers on a $Z$+jet
    sample obtained with the Pythia8 Monte Carlo generator. Plots taken from Ref.~\cite{Dreyer:2021hhr}, licensed under CC BY 4.0.}\label{fig:qg-lund-py}
\end{figure}

\begin{figure}[t]
  \subfloat[][ML-based approaches.]{\includegraphics[width=0.48\textwidth,page=1]{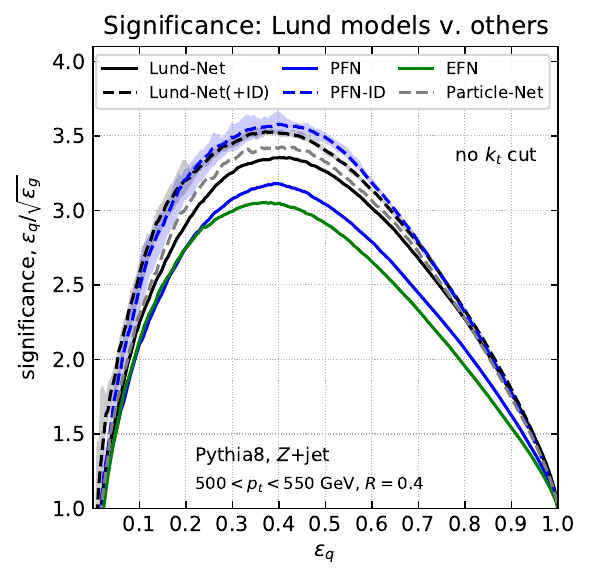}\label{fig:qg-lund-py-allkt}}\hfill
  \subfloat[][Analytic approaches.]{\includegraphics[width=0.48\textwidth,page=2]{figures/lund-qg-rocs-v-others.pdf}\label{fig:qg-lund-py-kt1}}
  \caption{Performance comparison with other jet substructure methods. Plots taken from Ref.~\cite{Dreyer:2021hhr}, licensed under CC BY 4.0.}\label{fig:qg-lund-py-v-others}
\end{figure}

We conclude this section with an overview of the performance of the
above taggers on full Monte-Carlo simulations.
For this, we consider $Z$+jet events simulated with Pythia8, where,
for simplicity, the flavour of the jet is taken as the one from the
Born-level process (see Ref.~\cite{Dreyer:2021hhr} for details).

First, ROC curves for the discrimination of quark- and
gluon-initiated jets are shown in Fig.~\ref{fig:qg-lund-py}. The left
plot shows the results obtained including all the Lund declustering, where
the perturbative-QCD-based analytic taggers cannot be applied, while
the right plot only includes all declusterings with $k_t\ge 1$~GeV. 
Both plots show a significant gain in performance in using detailed
Lund information compared to using either the Lund multiplicity or the
Lund density from section~\ref{sec:lund-W-tagging}.

Deep-learning-based methods also show a large performance gain
compared to analytic approaches, which may, for example, indicate
significant subleading effects compared to the analytic approach
described above, which is valid in the collinear limit and at the
single-logarithmic limit.
Furthermore, the addition of the full tree information, compared to
using only primary information also comes with a substantial
performance gain, although this gain is only visible at large quark
efficiency in the analytic case.

To put this in a broader perspective and connect with other methods
introduced in these lecture notes, we show on
Fig.~\ref{fig:qg-lund-py-v-others} a comparison between the Lund-based
taggers discussed in this section and a selection of other
jet-substructure-based taggers. The vertical axis of these plots is
the significance $\varepsilon_q/\sqrt{\varepsilon_g}$ with larger
numbers indicating a better performance.

Let us start with the left plot, which focuses on approaches relying on
machine-learning techniques.
Besides Lund-Net, we have included the Particle-Net
method~\cite{Qu:2019gqs} based on point clouds, as well as the
Particle-Flow Network (PFN) and Energy-Flow Network
(EFN)~\cite{Komiske:2018cqr} based on jet constituents.\footnote{Both
  map each particle onto a latent space and the main difference
  between the two is that the latter imposes an IRC-safety constraint
  on this latent space.}
The methods marked with ``ID'' also include information based on the
PDG-id of the constituents on top of the kinematic properties.
The main message is that, while some differences are visible, the
taggers which include the most information available in the jet
perform similarly, albeit with a small favour for ``PFN-ID'' and
``Lund-Net+ID'' over Particle-Net.
Not including PDG-id of the jet constituents (Lund-Net and PFN), or
limiting the network to use IRS-safe information (EFN) reduces the
performance.

We now move to the right plot, which instead focuses on analytic methods,
comparing the analytic approach described earlier in this section to
the \SD multiplicity, $n_{\rm SD}$, and two representative
jet shapes, namely the energy-energy correlation function (EEC) here
taken with $\alpha=1/2$, and the jet angularity, here taken with
$\alpha=1$.
Two incantations of the latter have been considered.
The ones represented by dashed lines are the standard versions based
on all the jet constituents. The ones shown as solid lines instead
compute the jet shape using only the Lund-declusterings passing a
$k_t$ cut of 1~GeV. This is meant to produce a fair comparison based
on the information used to tag the jets.
It is interesting to see that the standard jet shapes tend to be more
efficient at low quark efficiency, while the \SD multiplicity
performs well at larger quark efficiency.
The Lund-based approach (red solid curve) shows a further performance
gain over the whole range of quark efficiencies. This is a trace of
the fact that this observable has been constructed to be optimal in
this case, at least from a perturbative QCD point of view and at a
given order.
Looking instead at the dashed curves, which include all the
constituents in the jet, a clear performance gain is observed, also
signalling that a significant part of the information in quark-gluon
tagging comes from the non-perturbative region.

We conclude by saying that the usual trade-off between tagging
performance and resilience to modelling effects (non-perturbative
effects, Monte-Carlo dependence, etc.), discussed several times
earlier in these notes, also applies in this context.
The interested reader will find more details about this point in
Ref.~\cite{Dreyer:2021hhr}.

\section{A brief analytic overview}\label{sec:lund-analytic}

As with many jet substructure methods introduced in this book, a series of observables based on Lund-plane variables are
amenable to an analytic treatment in perturbative QCD.
Here, we discuss two examples: the Lund multiplicity and the primary
Lund plane density.
In both cases, we will not provide the full derivation of the
state-of-the-art results available in the literature, but instead
illustrate some aspects of the resummed calculation that work
different from what has been done previously in these lecture
notes.\footnote{Furthermore, the Lund jet plane representation of the
  QCD radiation pattern can be used to define new sets of observables,
  e.g. the Lund-Tree Shapes~\cite{vanBeekveld:2025zjh}.
  Their definition applies to scattering processes with any number of
  resolved jets in the final state, as well as to groomed jets.
  These observables share many desirable properties of the jet shapes
  defined in this book and, from a theoretical viewpoint, feature a
  simple all-order structure. For instance, they are free of
  non-global logarithmic corrections.}
We note that this analytic resummation structure is not specific to
multiplicity-related observables and can also be applied to jet
shapes, as done for dynamical-grooming in Ref.~\cite{Caucal:2021bae}.

Before diving into the actual calculations, let us mention that both
of these observables have been measured by the LHC experiments.
The primary Lund plane density was measured by the
ATLAS~\cite{ATLAS:2020bbn} and the CMS~\cite{CMS:2023lpp}
collaborations, both in good agreement with the analytic calculation
of Ref.~\cite{Lifson:2020gua}, as well as by the ALICE
collaboration~\cite{Havener:2021yhb}.
The Lund multiplicity was measured by the ATLAS
collaboration~\cite{ATLAS:2024wrd} and successfully compared to the
calculation in Refs.~\cite{Medves:2022ccw,Medves:2022uii}.
We will discuss this a bit more in chapter~\ref{searches-measurements}.

\subsection{(Average) Lund multiplicity}

Multiplicity is one of the most basic observables one can consider for
final-state studies. 
However, in general, it is  infrared and collinear
unsafe.
Jets, and jet substructure in particular, open possibilities to
define multiplicity observables which are infrared and collinear safe
and can therefore be computed in perturbative QCD.
Some typical examples date back to the LEP era, where one would study
the multiplicity of exclusive jets obtained from running the Cambridge
or Durham $k_t$) jet algorithm with a given cutoff scale
$y_\text{cut}$.\footnote{As we shall briefly discuss below, the former
  is very similar to the Lund multiplicity discussed here.}
Furthermore, the Iterated \SD multiplicity introduced in
Sec.~\ref{sec:isd} is equivalent to the primary Lund plane multiplicity, provided that we replace the condition
$z>z_\text{cut}\theta^\beta$ with a cut on the $k_t$ of the emissions
(which roughly corresponds to $\beta=1$ and the use of an absolute
energy scale).

Here, we will consider the Lund plane multiplicity, which, for a given
jet $j$, is defined as one (for the ``leading'' particle) plus the
number of Lund declustering, in the full Lund tree, with $k_t$ above a
certain $k_{t,\text{cut}}$. Formally, we have
\begin{equation}
  N^\text{(Lund)}(j) = 1 + \sum_{\mathcal{T}_i \in \mathcal{L}(j)} \Theta(k_{t,i} > k_{t,\text{cut}}).
\end{equation}
This definition of the Lund multiplicity can trivially be extended to
the case of $e^+e^-$ collisions, where the sum is carried over all the
Lund declusterings in the event.
\begin{equation}
  N^\text{(Lund)}(Q) = 2 + \sum_{\mathcal{T}_i \in \mathcal{L}} \Theta(k_{t,i} > k_{t,\text{cut}}).
\end{equation}

In the rest of this section, we want to provide a brief description of
how to compute the average Lund multiplicity analytically.
For the simplicity of the discussion, we will focus on the event
multiplicity in $e^+e^-$ collisions and work in the limit
$k_{t,\text{cut}}\ll Q$, with $Q$ the collision centre-of-mass energy.
In this limit, we expect factors of $L=\log(Q/k_{t,\text{cut}})$ and all-order resummation is needed.
In contrast to all the cases we have met so far, where the resummation was
mostly driven by a Sudakov form factor that exponentiates, this is not
the case for multiplicity.
For this reason, we will adopt a different strategy to perform the
resummation. More specifically, we will work in the following limit:
\begin{equation}\label{eq:nkdl-parametric-limit}
  \alpha_s\ll 1,\qquad
  L\gg 1,\qquad
  \xi=\alpha_sL^2\sim 1.
\end{equation}
In this limit, the dominant contribution resums all terms with
double-logarithmic (DL)  enhancement, $\alpha_s^nL^{2n}$. The first
subleading corrections, next-to-double-logs (NDL), would be down
by one power of $L$, i.e.\ resum terms proportional to
$\alpha_s^nL^{2n}$, etc.
In practice, one can therefore write\footnote{In contrast, for
  exponentiating observables like jet shapes, one would typically work
  in the limit $\alpha_s\ll 1$, $\lambda=\alpha_s L\sim 1$ and write
  $\log\Sigma(L) = g_1(\lambda) L +g_2(\lambda) + \alpha_s g_3(\lambda)
  + \dots$. The successive terms correspond to the leading logs (LL),
  next-to-leading logs (NLL), next-to-next-to-leading logs (NNLL), etc.}
\begin{equation}\label{eq:multiplicity-expansion}
  \avg{N^\text{(Lund)}}
  = h_1(\xi)
  + \sqrt{\alpha_s} h_2(\xi)
  + \alpha_s h_3(\xi)
  + \dots
\end{equation}
The current state-of-the-art for the average Lund multiplicity in
$e^+e^-$ events or within a boosted jet is NNDL, i.e.\ including the
$h_1$, $h_2$ and $h_3$ contributions.
Here, we will give a full derivation in the double-logarithmic limit
and sketch a strategy that can be used systematically to compute
subleading contributions.

\subsubsection{Double-logarithmic limit}

In the double-logarithmic limit, all the emissions in an event are
soft and collinear. Each of them is emitted either by the original
hard $q\bar q$ pair, or by a previous gluon emission.
One can also have additional virtual corrections, also computed in the
soft-and-collinear limit.
In this limit of perturbative QCD, all the emissions (real or associated virtual
corrections) from a given emitter are independent from one another and
therefore factorise. Each emission has a weight $\frac{2\alpha_s
  C_i}{\pi}$ per unit of $\log k_t$ and rapidity $\eta$, with $C_i=C_F$
($C_i=C_A$) for emissions from a quark (gluon). Virtual corrections
have the same weight with an opposite sign.

The Lund multiplicity is a (linear) sum over all the emissions with
$k_t>k_{t,\text{cut}}$ and any kinematically-allowed rapidity.
Let us therefore identify all the kinematic configurations that
contribute to emitting a gluon at transverse momentum
$k_t>k_{t,\text{cut}}$ and Lund rapidity $\eta$.

A specific contribution comes from what we will call {\it irreducible
  chains of nested emissions}. These correspond to successive series of
$n\ge 1$ gluon emissions $g_1,\dots,g_n$ of Lund coordinates
$(k_{t1}, \eta_1)$, $(k_{t2}, \eta_2)$, $\dots$, $(k_{tn}, \eta_n)$,
with $k_{t,i}\ll k_{t,i-1}$, $\eta_i>\eta_{-1}$, such that $g_1$ is
emitted from the original hard quark (or anti-quark), all the
following gluons are emitted from their predecessor, $g_i$ is radiated
collinearly to $g_{i+1}$, and $g_n$ is the measured gluon, i.e.\
$k_{tn}=k_t$ and $\eta_n=\eta$.
In a Lund-plane picture, each emission in the chain is in the bulk of
the Lund plane of the previous emission.

The key observation for our computation is that all the configurations
of particles which are not irreducible chains of nested emissions do
not contribute to the average Lund multiplicity.
The main idea behind this result is that in the presence of additional
real emissions, one can always find virtual corrections which cancel
exactly the contribution from real emissions.

Since the argument is not totally trivial, let us provide a more
rigorous derivation.
Every configuration of real emissions that contributes to the multiplicity
at Lund coordinates $(k_t,\eta)$ can be seen as the irreducible chain
that ultimately radiates the measured gluon $g_n$, accompanied by an
arbitrary number of additional soft-collinear emissions.
Within these additional emissions, one can always identify a subset of
``final'' emissions, i.e. all the gluons that have not radiated
another gluon with $k_t>k_{t,\text{cut}}$. These gluons are all
radiated from the original $q$ (or $\bar q$) or from one of the gluons in the
irreducible chain, either directly or via another irreducible chain
of nested emissions.
Assuming our configuration has $p$ such gluons, there will be $2^p-1$
other configurations where an arbitrary subset of the $p$ gluons are
replaced by the corresponding virtual correction. In this ensemble of
$2^p$ configurations, all have the same absolute weight, $2^{p-1}$
with a positive sign and $2^{p-1}$ with a negative sign. Their sum
therefore cancels.\footnote{For a more in-depth understanding, what we
  are really exploiting here is the fact that each virtual correction
  can be associate with a corresponding real emission that does no
  further branch, at the same kinematic point. This way of pairing
  real and virtual contributions may not be the most intuitive one but
  can be helpful beyond the current calculation.}
The only way one can get a non-zero net contribution is if there are no
additional gluons which, in practice, corresponds to the case $p=0$,
that is the case of a single irreducible nested chain of emissions.

To compute the average Lund multiplicity at DL accuracy, one therefore
have to sum over all irreducible chains of emissions strongly ordered
both in $k_t$ and in rapidity, or, equivalently, in energy and
rapidity. Focusing on a single hemisphere with a leading parton of
flavour $i$, this gives
\begin{align}
  \avg{N^\text{(Lund)}_i}_\text{DL}
  & = 1 + \sum_{n=1}^\infty \frac{C_i}{C_A}\left(\frac{2\alpha_sC_A}{\pi}\right)^n
    \int_0^\infty d\eta_n\dots\int_0^{\eta_{i+1}}d\eta_i\dots\int_0^{\eta_2}d\eta_1\nonumber\\
  & \phantom{= 1 + \sum_{n=1}^\infty}
    \int_0^Q\frac{dE_n}{E_n}\dots\int_{E_{i+1}}^Q\frac{dE_i}{E_i}\dots\int_{E_2}^Q\frac{dE_1}{E_1}
    \Theta(E_ne^{-\eta_n}>Q e^{-L})\nonumber\\
  & = 1 + \frac{C_i}{C_A}\sum_{n=1}^\infty \int_0^\infty d\eta_n
    \frac{\eta_n^{n-1}}{(n-1)!}\frac{(L-\eta_n)^n}{n!}\left(\frac{2\alpha_sC_A}{\pi}\right)^n
    \nonumber\\
  & = 1 + \frac{C_i}{C_A}\sum_{n=1}^\infty
    \left(\frac{2\alpha_sC_A}{\pi}\right)^n \frac{(L)^{2n}}{(2n)!}
  \nonumber\\
  & = 1 + \frac{C_i}{C_A}\left[\cosh\left(\sqrt{\frac{2\xi C_A}{\pi}}\right)-1\right].
    \label{eq:lund-multiplicity-dl}
\end{align}
where we have used
\begin{equation}
  L = \log\frac{Q}{k_{t,\text{cut}}}
  \quad\text{ and }\quad
  \xi = \alpha_s L^2.
\end{equation}
We note that in all the discussions so far, the argument of the strong
coupling can be taken at the hard scale $Q$.
Indeed, for each emission, the difference between $\alpha_s(k_t)$ and
$\alpha_s(Q)$ is of the order of $\alpha_s L$ which is a
single-logarithmic term is negligible in our double-log approximation.
Furthermore, in our kinematic limit, $\xi$ is not necessarily large,
meaning that we can not approximate the hyperbolic cosine by an
exponential.

Finally, the average multiplicity can trivially be obtained
differentially in $k_t$ by taking the derivative of the above
result. We will denote this using a lower case $n$:
\begin{equation}
  \avg{n^\text{(Lund)}_i}_\text{DL}
  = \frac{C_i}{C_A}\sqrt{\frac{2\alpha_sC_A}{\pi}}\sinh\left(\sqrt{\frac{2\xi C_A}{\pi}}\right).
    \label{eq:lund-multiplicity-diff-dl}
\end{equation}

\subsubsection{Next-to-double-logarithmic contributions}

Instead of providing a full derivation of the Lund multiplicity at the
NDL accuracy, we would rather want to highlight a strategy to
systematically compute subleading corrections.
This strategy has so far been used to rederive NDL corrections and
compute the average multiplicity at NNDL accuracy, but given enough
technical effort could be extended beyond.

The main logic is that any emission that is only enhanced by a single logarithm costs one power of $L$, and any emission that only occurs in an $\mathcal{O}(1)$ region of the Lund plane --- that is an
emission proportional to $\alpha_s$ without any logarithmic enhancement
would cost two powers of $L$.

At NDL, one should therefore consider configurations where only one
emission has a single-logarithmic enhancement, with all the other
emissions being soft and collinear.
At NNDL, we would then either have a single emission with no
logarithmic enhancement, or exactly two emissions with a
single-logarithmic enhancement, accompanied by an arbitrary number of
soft-collinear emissions.
As we progress to an increasingly refined accuracy, a growing number
of subleading configurations will have to be taken into account, but
since there are only a finite number of sources for constant and
single-logarithmic emissions, only a finite number of contributions has
to be computed order by order.

Once one has identified all the relevant kinematic configurations at a
given order, the calculation of each contribution would involve an
integration over a handful of emissions associated with the subleading
terms, together with an arbitrary number of soft-collinear emissions.
Since we know that configurations with soft-collinear
emissions can always be described through irreducible chains, we can
heavily recycle the results of the previous section. This considerably simplifies the calculation.

At NDL, only two contributions have to be taken into account:
running-coupling corrections (as already alluded to above) and
situations where one of the emissions is a hard-collinear
branching.\footnote{In the case of the Lund multiplicity in a jet, one
  would have a third contribution coming from soft emissions at the
  edge of the jet.}
To illustrate the method, we will compute the hard-collinear
correction.
It can happen in two places: either a hard-collinear branching of the
leading parton, or a hard-collinear branching of any subsidiary
emission.
In the second case, the hard-collinear branching has to be one of
a gluon radiated through an irreducible chain of nested gluons, and
the number of such gluons at a given $k_t$ scale is given by
Eq.~(\ref{eq:lund-multiplicity-diff-dl}).
An extra subtlety is that we further have to consider real and virtual
collinear splittings.
For all the emissions following the two branches of the hard-collinear
splitting one can then reuse Eq.~(\ref{eq:lund-multiplicity-dl}) to
obtain their contribution to the multiplicity.
For definiteness, we consider all possible $a\to bc$ collinear
branchings, we denote by $\eta_\text{hc}$ and $z$ the rapidity and
momentum fraction of the hard-collinear branching and, if any, by
$k_t$ the relative transverse momentum of the parent gluon that
undergoes a hard-collinear branching.
For simplicity, we consider the case of an initial $q\bar q$ event.
We also use the shorthand notation
$N_\text{DL}^{(i)}\equiv \avg{N^\text{(Lund)}_i}_\text{DL}$ and
$n_\text{DL}^{(i)}\equiv \avg{n^\text{(Lund)}_i}_\text{DL}$.
At NDL accuracy, we can therefore write
\begin{align}
  \avg{N^\text{(Lund)}_q}_\text{NDL}^\text{(hc)}
  & = \int_0^L d\eta_\text{hc} \int_0^1 dz\, P^\text{hc}_{q\to qg}(z)
    \left[N_\text{DL}^{(q)}(L-\eta_\text{hc}) + N_\text{DL}^{(g)}(L-\eta_\text{hc}) -
    N_\text{DL}^{(q)}(L-\eta_\text{hc})\right]\nonumber \\
  & + \int_0^L d\ell\, n_\text{DL}^{(q)}(\ell)
    \int_0^{L-\ell} d\eta_\text{hc} \int_0^1 dz\, P^\text{hc}_{g\to
    gg}(z)\nonumber \\
  & \phantom{= \int_0^L d\eta_\text{hc} \int_0^1}
    \left[N_\text{DL}^{(g)}(L-\ell-\eta_\text{hc}) + N_\text{DL}^{(g)}(L-\ell-\eta_\text{hc}) -
    N_\text{DL}^{(g)}(L-\ell-\eta_\text{hc})\right]\nonumber \\
  & + \int_0^L d\ell\, n_\text{DL}^{(q)}(\ell)
    \int_0^{L-\ell} d\eta_\text{hc}
    \int_0^1 dz\, P^\text{hc}_{g\to q\bar q}(z)\nonumber \\
  & \phantom{= \int_0^L d\eta_\text{hc} \int_0^1}
    \left[N_\text{DL}^{(q)}(L-\ell-\eta_\text{hc}) + N_\text{DL}^{(q)}(L-\ell-\eta_\text{hc}) -
    N_\text{DL}^{(g)}(L-\ell-\eta_\text{hc})\right],\nonumber
\end{align}
where $P^\text{hc}_{a\to bc}$ corresponds to the purely hard-collinear
part of the DGLAP splitting (i.e.\ where we have removed the soft
divergence).
The first term in this equation corresponds to a hard-collinear
branching of the hard quark at a rapidity $\eta_\text{hc}$. The first
two terms in the square bracket correspond to further branchings of
the resulting quark and gluon branches for a real $q\to qg$
hard-collinear splitting, while the third term corresponds to further
branchings for the virtual correction. In each case, we can use the
double-logarithmic multiplicity result which, at NDL
accuracy\footnote{Explicit factors of the momentum fraction,
  $\log(1-z)$ and $\log(z)$ for the two terms associated with real
  branchings, describing the fact that the slightly lower energy of
  each branch would result in a smaller phase-space for further
  emissions, is an example of corrections which would have to be
  included at NNDL.} can be evaluated at the scale
$L-\eta_\text{hc}$. In this case, the virtual correction cancels
exactly the first term in the square bracket.

The next two terms in the above expression correspond to a
hard-collinear branching, still at a rapidity $\eta_\text{hc}$, of a
gluon emitted, through an irreducible chain of nested emissions, at a
scale $k_t = Q e^{-\ell}$. The factor $n_\text{DL}^{(q)}(\ell)$ gives
the average number of such gluons at DL accuracy. The rest of the
terms are similar to what we have already seen for the first line,
except that now the gluon can either branch into gluons or into
a $q\bar q$ pair. In this latter case, we no longer have an explicit
cancellation between the virtual correction and one of the real terms.

The key observation here is that, plugging in the DL expressions for
$N_\text{DL}^{(i)}$ and $n_\text{DL}^{(i)}$, only a handful of
integrations have to be performed to get the NDL contributions. In
practice, the above equation evaluates to
\begin{multline}
  \avg{N^\text{(Lund)}_q}_\text{NDL}^\text{(hc)}
  = \sqrt{\frac{\alpha_sC_A}{2\pi}}\bigg\{ 2 B_q \sinh \nu
  + B_{gg} (\nu\cosh \nu-\sinh \nu)\\
  + B_{gq}\left[ \left(\frac{2C_F}{C_A}-1\right)\nu\cosh\nu
  +\left(\frac{6C_F}{C_A}-5\right)\sinh\nu+4\left(\frac{C_F}{C_A}-1\right)\nu\right]
  \bigg\},
\end{multline}
with the short-hand notation $\nu = \sqrt{2\xi C_A/\pi}$, and where the
$B_i$ coefficients are the integrals over the finite part of the DGLAP
splitting functions that we have already encountered several times
throughout these lectures.

The NDL correction coming from the running of the coupling can be
computed using a similar approach.

\subsection{Primary Lund plane density}

As discussed previously, the density of the primary Lund plane, defined in Eq.~(\ref{eq:lund-density-def}), is among the most informative observables for characterising the radiation pattern inside a jet. In this section, we present a perturbative evaluation of this quantity, beginning at leading order (LO), i.e.\ considering the contribution from a single gluon emission, and then moving to its all-order resummation. 

By construction, the Lund-plane density measures the effective intensity of radiation per unit logarithmic interval in transverse momentum $k_t$ and angular scale $\Delta$. In the bulk of the Lund plane, corresponding to the soft and collinear limit, the result takes a particularly simple form. In this regime, QCD radiation is uniform in these logarithmic variables, leading to a constant LO density, which is given by
\begin{equation}\label{eq:lund-density-lo-SC}
 \rho^{(\text{soft-coll},\as)}_{i}(k_t, \Delta) = \frac{2 C_i \as}{\pi},
\end{equation}
where $C_i$ denotes the Casimir operator associated with the flavour $i$ of the hard parton initiating the jet, with $C_i = C_A$ for gluon-initiated jets and $C_i = C_F$ for quark-initiated jets.

This result can be refined by relaxing the soft approximation and considering a generic collinear splitting process \(i \to k(y) + j(1-y)\), where the emitted parton carries an energy fraction $y$ of the parent. Remember that in the collinear (though not necessarily soft) limit, the squared matrix element factorises according to
\begin{align}\label{eq:QC-fact-begin}
    |\mathcal{M}|^2\,  d \Phi^{(2)} \simeq
    |\mathcal{M}_0|^2\frac{\as}{2 \pi} P_{ki}(y) \, d y \frac{d q_t^2}{q_t^2}  \, ,
\end{align}
the transverse momentum of the splitting is given by
\[
q_t = y(1 - y)\, \theta\, Q,
\]
with $\theta$ denoting the splitting angle and $y$ the energy fraction with respect to the emitter.
The hard scale $Q$ may be identified with the energy of a jet produced in $e^+e^-$ annihilation, or with the jet transverse momentum relative to the beam axis in hadronic ($pp$) collisions.
At small angles, the Lund plane density is obtained from Eq.~(\ref{eq:QC-fact-begin}) performing the change of variables:
\begin{align}
      k_t=\frac{q_t}{\max(y,1-y)}, \quad
    \Delta=\frac{k_t}{\min(y,1-y)Q}=\frac{q_t}{y(1-y)Q}.
\end{align}
In particular, the density of the primary plane is constructed by recording the momentum
fraction of the softer branch. Thus,  assuming $j$ to be the softest parton, i.e.\ $y>1/2$,
we have
\begin{equation}
\label{eq: rho_coll_alphas}
    \rho_i^{(\text{coll},\as)}(\Delta,\kt)= \frac{\as}{\pi}\sum_{k=q,g}(1-y)
   P_{ki}(y)\Big |_{1-y= \frac{k_t }{Q \Delta}}=  \frac{\as}{\pi} z \,\mathcal{P}_i(1-z)\Big |_{z= \frac{k_t }{Q \Delta}},
\end{equation}
where we have introduced the Lund plane variable $z=1-y$ and $\mathcal{P}_i(y)= \sum_{k=q,g}P_{ki}(y)$, with the sum running over all the possible flavours of the
leading parton after the emission.

\begin{figure}
\begin{minipage}{0.50\linewidth}
\includegraphics[width=\linewidth,page=2]{figures/cartoon.pdf}    \hfill
\end{minipage}
\begin{minipage}{0.50\linewidth}
  \includegraphics[width=\linewidth,page=1]{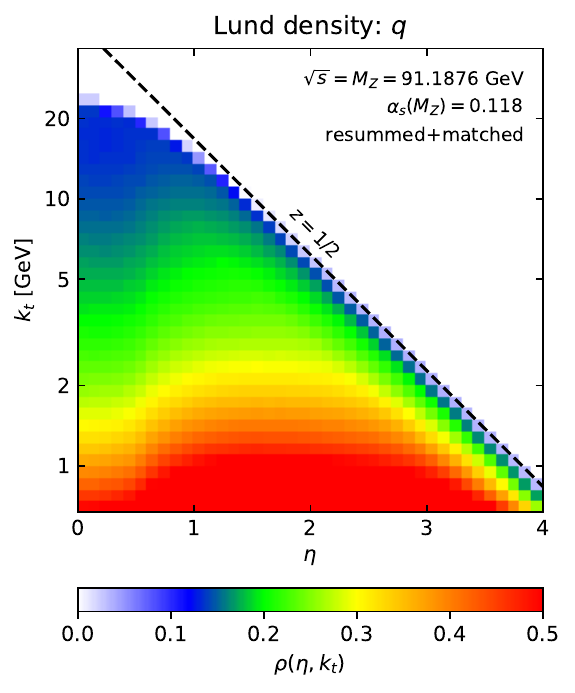}  
    \end{minipage}
    \caption{
    On the left, a schematic illustration of the various contributions entering the resummation of the primary Lund plane density is shown. The collinear evolution of the initiating parton $i_0$, from the initial angular scale $\Delta_0$ down to a parton $i$ at an angle $\Delta$, is depicted by a green arrow. Along this evolution, the transverse momentum decreases from $Q\,\Delta_0$ to $x\,Q\,\Delta$. The subsequent soft evolution, occurring at fixed angular scale and represented by a red arrow, further reduces the transverse momentum from $x\,Q\,\Delta$ to $k_t$. The black blob denotes the splitting that is ultimately recorded in the Lund plane.
On the right, we display the resummed and matched predictions for the primary Lund plane density in $e^+e^-$ collisions at centre-of-mass energy $\sqrt{s}=M_Z$.
Figures taken from Ref.~\cite{Ghira:2025nym}, licensed under CC BY 4.0.}
    \label{fig:cartoon}
\end{figure}

Beyond leading order, the above behaviour receives logarithmic corrections. They have several physical origins:
\begin{itemize}
    \item[(i)] running-coupling logarithms of the transverse momentum \(k_t\);
    \item[(ii)] hard-collinear logarithms of the angular resolution variable $\Delta$, which induce flavour-changing effects;
    \item[(iii)] soft emissions at large and commensurate angles, enhanced by logarithms of $z=\frac{k_t}{Q\Delta}$, including the intricate structure of clustering logarithms.
\end{itemize}
These effects were resummed to all orders, at single-logarithmic accuracy, in Ref.~\cite{Lifson:2020gua} for jets initiated by massless quarks or gluons, while the extension to the case of heavy-flavour jets was done in~\cite{Ghira:2025nym}.

Running coupling corrections are universal, i.e.\ they do not depend on the type of collision we are considering, nor on the details of the final state we are selecting. They can be straightforwardly implemented in Eq.~(\ref{eq: rho_coll_alphas}), by employing the CMW scheme for the two-loop running coupling, cf. Eq.~(\ref{eq:CMW}),
\begin{equation}
     \rho_i^{(\text{r.c.})}(\Delta,\kt)=  \frac{\as^{\text{CMW}}(k_t^2)}{\pi}  z \, \mathcal{P}_i(1-z)\Big |_{z= \frac{k_t }{Q \Delta}}.
\end{equation}

The inclusion of hard-collinear logarithms as well as soft ones instead pose more challenges. In order to describe the main feature of this combined resummation, it is useful to exploit the cartoon in left-hand panel of Fig.~\ref{fig:cartoon}.
We start from a hard parton of flavour $i_0$ at an initial angular resolution $\Delta_0$ and Lund transverse momentum $Q\Delta_0$.
We first perform collinear evolution, indicated by the green arrow, to obtain parton $i$ at the scale $(\Delta, xQ \Delta)$, with $x<1$.
At this point, we use soft resummation to evolve from $xQ \Delta \to k_t$, at fixed $\Delta$, as shown by the red arrow. Finally, the splitting $\mathcal{P}_i$ (indicated by the black blob) is recorded.
In formulae, this reads
\begin{align}\label{eq:final-resummation}
  \rho_{i_0}^{(\text{res})}(\Delta,\kt)= \sum_{i=q,\mathcal{Q},g}\frac{\as^{\text{CMW}}(x_K^2 k_t^2)}{\pi}\int^1_0& d x \, p_{i|i_0}(x, t(\Delta,\Delta_0))\, z \, \mathcal{P}_i(1-z)\Theta(1-2z)\Big|_{z= \frac{k_t }{x Q \Delta}} \nonumber\\ 
  &\cdot {\bar \rho}^{(\text{soft})}_{i} \left(\Delta, \tilde{z}_\text{eff}(x)\, Q \Delta\right),
\end{align}
where we have introduced, for each flavour
$i$, the function $p_i(x,t)$ as the probability for the leading parton at an angular scale $\Delta$
to have flavour $i$ and energy fraction $x$. Note that the angular dependence is encoded in the  evolution time $t$:
\begin{align}
\label{eq:evolution_time}
t\left(\Delta, \Delta_0\right)= \int^{Q^2 \Delta_0^2}_{Q^2 \Delta^2} \frac{d k_t^2}{k_t^2} \frac{\as^{\text{CMW}}(k_t^2)}{2\pi}.
\end{align}
Thus, as $t$ increases, we probe smaller angular scales. 
In Eq.~(\ref{eq:final-resummation}), we have also introduced the all-order soft density  ${\bar \rho}^{(\text{soft})}_{i}$,
where the bar indicates that we have divided out the small-angle limit at $\order{\as}$ in order to avoid double counting with $\mathcal{P}_i$.

Let us now briefly discuss the main features of the collinear and soft resummations, starting with the former. 
While on the soft and collinear limit, only soft-gluon emissions contribute and, therefore, the flavour of the emitting parton cannot be altered, at single-logarithmic accuracy, we also have to consider hard-collinear splittings, which can change the flavour of the harder branch. For instance, this happens through \(q \to qg\) splittings in which the daughter gluon carries more than half of the momentum of the parent quark, or via \(g \to q\bar{q}\) splittings.
These effects give rise to collinear logarithms corresponding to a
sequence of emissions strongly ordered in angle and not enhanced by
soft logarithms. 
Additionally, through collinear evolution, the energy of the leading
parton will decrease, translating into a smearing of the Lund plane
density close to its kinematic boundary. 

The resummation of collinear logarithms is based on the observation that the probability functions $p_i(x,t)$ fulfil a DGLAP-type evolution equation:
\begin{align}
\label{eq: DGLAP type}
    \frac{d}{d t}p_i(x,t)= \int^1_0 \frac{d y}{y} \left(P_{ij}^{(\text{R})}(y) \,p_j\left(\frac{x}{y},t\right)-y P_{ij}
    ^{(\text{V})}(y) \,p_j\left(x,t\right)\right).
    \end{align}
The splitting kernels $P^{(R)}$ and $P^{(V)}$ in Eq.~(\ref{eq: DGLAP
  type}) account for real and virtual emissions, respectively, and
they can be expressed in terms of the standard massless splitting
functions times $\Theta$-functions, which enforce the condition that
at each (real) splitting, we always follow the harder branch, see appendix~\ref{app:splitting_functions}.
Eq.~(\ref{eq: DGLAP type}) can be solved using the numerical method developed in~\cite{Lifson:2020gua} and originally presented in~\cite{Dasgupta:2014yra}. Alternatively, one can solve the equation exactly in Mellin space and subsequently perform a numerical inverse transform to return to \( x \)-space, as discussed in~\cite{Ghira:2025nym}.

The resummation of soft emission at wide or commensurate angles is notoriously a difficult problem, even at single-logarithmic accuracy and analytic approaches exist only in limited cases, see e.g.~\cite{Banfi:2002hw}.
The method followed in~\cite{Lifson:2020gua} takes inspiration from the resummation of non-global~\cite{Dasgupta:2001sh} and clustering logarithms~\cite{Appleby:2003sj,Banfi:2005gj}.
Namely, we work in the large-$\nc$ limit and we construct the all-order result through a dipole shower ordered in the transverse momentum of the emissions. 

To achieve this, we start with a Born event and we decompose it, in the large-$\nc$ limit into a series of dipoles ($i,j$), each with appropriate weight $w_{ij}$. For a given initial dipole ($i,j$), we generate the emission of a soft gluon $k$ with probability given by the eikonal factor. The procedure is then repeated for each of the daughter dipoles $(i,k)$ and $(j,k)$, and so on, until one reaches a cutoff scale. 
We perform this operation for each initial dipole in the event. All resulting particles in a given event are then
clustered to produce the Lund plane density.
In general, a given process can receive contributions from different
initial dipole configurations, i.e.\ colour flows $\mathcal{C}_i$, each with a weight
$w_i$, which needs to be summed over:
\begin{align}\label{eq:rho_soft_dipole_sum}
    \rho^{(\text{soft})}(\Delta,\kt)&= \sum_{\substack{\text{partonic}
  \\ \text{channel}}} \sum_{\mathcal{C}_i} w_{i}\,
    \left[\rho^{(\text{soft})}(\Delta,\kt)\right]_{\mathcal{C}_i}.
\end{align}
Note that because the dipole evolution only allows for the emission of soft gluons, flavour can never change in the soft sector, as opposed to the collinear one. 

From a technical point of view, the evolution of the dipoles is performed at fixed-coupling.
That is, from the dipole shower we obtain the Lund plane coordinates $(\Delta,z)$ and hence the fixed-coupling density in the soft limit as $\rho^{(\text{soft})}(\Delta,z Q \Delta)$.
To account for running-coupling effects, we determine an effective energy fraction $z_\text{eff}$ as
\begin{equation}
   z_\text{eff}= \exp \left[ -\frac{\pi}{\as}t\left(\kt,  Q \Delta\right)\right],
\end{equation}
where the evolution time is given by the integral over the running coupling, see Eq.~(\ref{eq:evolution_time}). We then evaluate the soft-resummed density with the effective energy fraction, $z_\text{eff}$, i.e.\
\begin{equation}
\rho^{(\text{soft})}(\Delta,z Q \Delta)
\to \rho^{(\text{soft})}(\Delta,z_\text{eff} \, Q \Delta).
\end{equation}
This result is then used in the resummation master formula Eq.~(\ref{eq:final-resummation}).
Finally, we note that in Eq.~(\ref{eq:final-resummation}), the flavour label $i$ is attached to the soft density. While this is well-defined in the small-angle approximation, this assignment is ambiguous for the full $\rho^{(\text{soft})}$. 
However, we can still combine all the dipole configurations that contribute to a given jet and thus separate $\rho^{(\text{soft})}$ according to the flavour label, $i =q, g$.
There is an additional subtlety. In Eq.~(\ref{eq:final-resummation}), soft evolution starts at the scale $x Q \Delta$ for a parton of flavour $i$ that does not have to be present at Born level. Thus, we also need to generate the evolution for all possible QCD dipoles, not just the one present in the Born process.
In particular, one sees that at large angles,
where the details of the dipole configuration matter, collinear flavour changing effects can be neglected and only the dipole configurations present at Born level contribute. In the opposite regime, i.e.\ at small
angles, where collinear evolution matters, the flavour assignment to $\rho_i^{(\text{soft})}$ is unambiguous. 

We conclude our discussion by showing in the right-hand panel of Fig.~\ref{fig:cartoon}, the resummed Lund plane density matched to tree-level matrix element, computed for hemisphere jets in $e^+e^-$
collisions at $\sqrt{s}=M_Z$.

%% GS helper for auctex
%%% Local Variables:
%%% mode: latex
%%% TeX-master: "notes"
%%% End:

% LocalWords:  Lund recombinations recluster declustering correlators
% LocalWords:  resummations branchings correlator TeV Pythia dijet Eq
% LocalWords:  GeV NLL subjettiness bosons fermionic ROC ParticleNet
% LocalWords:  LSTM LundNet Altarelli Parisi recurses DGLAP CMW AUC
% LocalWords:  iteratively MicroJet exponentiates resums NNLL NNDL
% LocalWords:  exponentiating kinematically collinearly NDL rederive
% LocalWords:  differentially Casimir massless Mellin

% $Id: searches-measurements.tex 608 2026-03-15 14:25:57Z smarzani $
%
% This contains applications and more experimental aspects of jet
% substructure
%------------------------------------------------------------------------
\chapter[Searches and Measurements]{Searches and Measurements with jet substructure}\label{searches-measurements}

The previous chapters have focused on the theoretical description of jet substructure variables, e.g.\ the jet mass, jet shapes and the classification of the jet-sourcing particles, together with some phenomenological studies performed with simulated data. In this chapter, we will give a brief overview of existing experimental performance studies, measurements and searches using jet substructure performed by ATLAS and CMS. As alluded to in Chapters~\ref{chap:jets-and-algs} and~\ref{chap:calculations-jets} all theoretical predictions of jet substructure observables can potentially deviate from experimental measurements for various reasons. For instance, theoretical calculations may fall short in capturing all relevant contributions or experimental effects, e.g.\ imperfect reconstruction of particle momenta, become important. Thus, it is of interest to see how well the theoretical predictions discussed in this book agree with experimental measurements.
Furthermore,  jet substructure tools and ideas are used in tens of different searches for new physics, 
Here, we are not going to attempt to provide a comprehensive discussion of all searches and measurements performed by LHC experiments, but we will select and showcase results with a close connection to the topics discussed before~\footnote{In this chapter, we show many experimental results. All the plots we include are properly referenced, and they are licensed under CC BY 4.0 (or 3.0). }. 

\section{Tagging performance studies}

Many taggers have been proposed have been proposed in the literature and we have reviewed a selection of them in Chapter~\ref{tools}. Often jet shapes or prong-finders are combined with other jet observables to perform a classification of the jet's initiating particle. Such a procedure can be augmented using machine-learning techniques to find the region of highest significance in the multi-dimensional parameter space of jet substructure observables. Different observables are used by ATLAS and CMS and their individual approaches have significantly evolved over the years. It is highly likely that the development of increasingly powerful classifiers, i.e. taggers, for jets will continue. Thus, in this brief review we will predominantly focus on ATLAS' and CMS' latest public performance comparisons.

ATLAS bases its W and top taggers on a set of techniques, rooted in jet shape observables, to determine a set of optimal cut-based taggers for use in physics analyses~\cite{Aaboud:2018psm, Aad:2016pux, Aad:2015rpa}. The first broad class of observables studied for classification rely on constituents of the trimmed jet to combine the topoclusters and tracks to a so-called combined jet mass $m^\mathrm{comb}$. In addition to the jet mass, a set of jet shape observables are constructed: $N$-subjettiness ratios ($\tau_{21}$ and $\tau_{32}$), splitting measures ($\sqrt{d_{12}}$ and $\sqrt{d_{23}}$), planar flow and energy correlation functions ($C_i$ or $D_i$). Various subsets of these and similar observables are then combined in a boosted decision tree (BDT) or a deep neural network (DNN), see Table~\ref{tab:obs} for more details.

\begin{figure}[t]
 \includegraphics[width=0.485\textwidth]{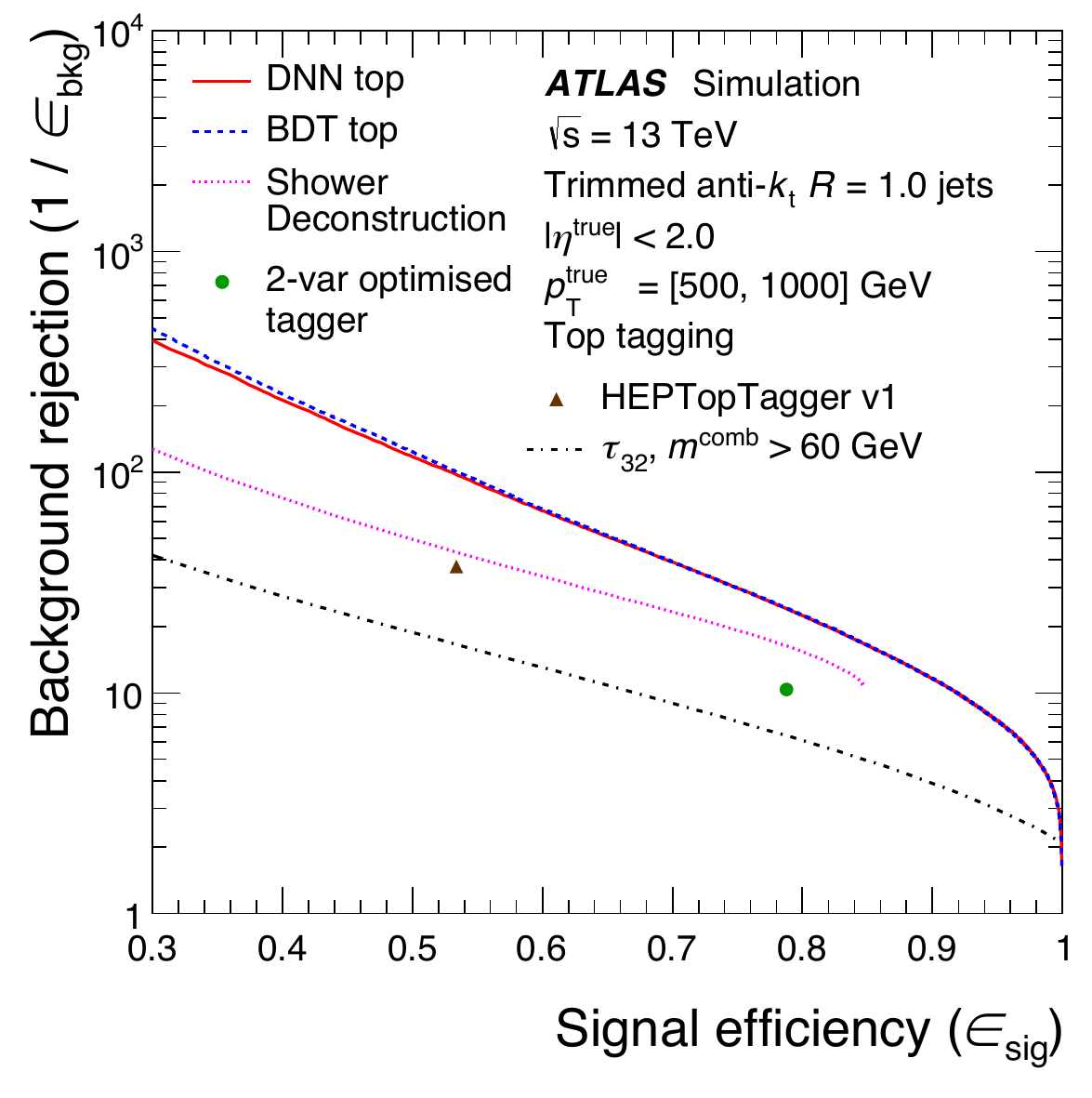}
  \includegraphics[width=0.515\textwidth]{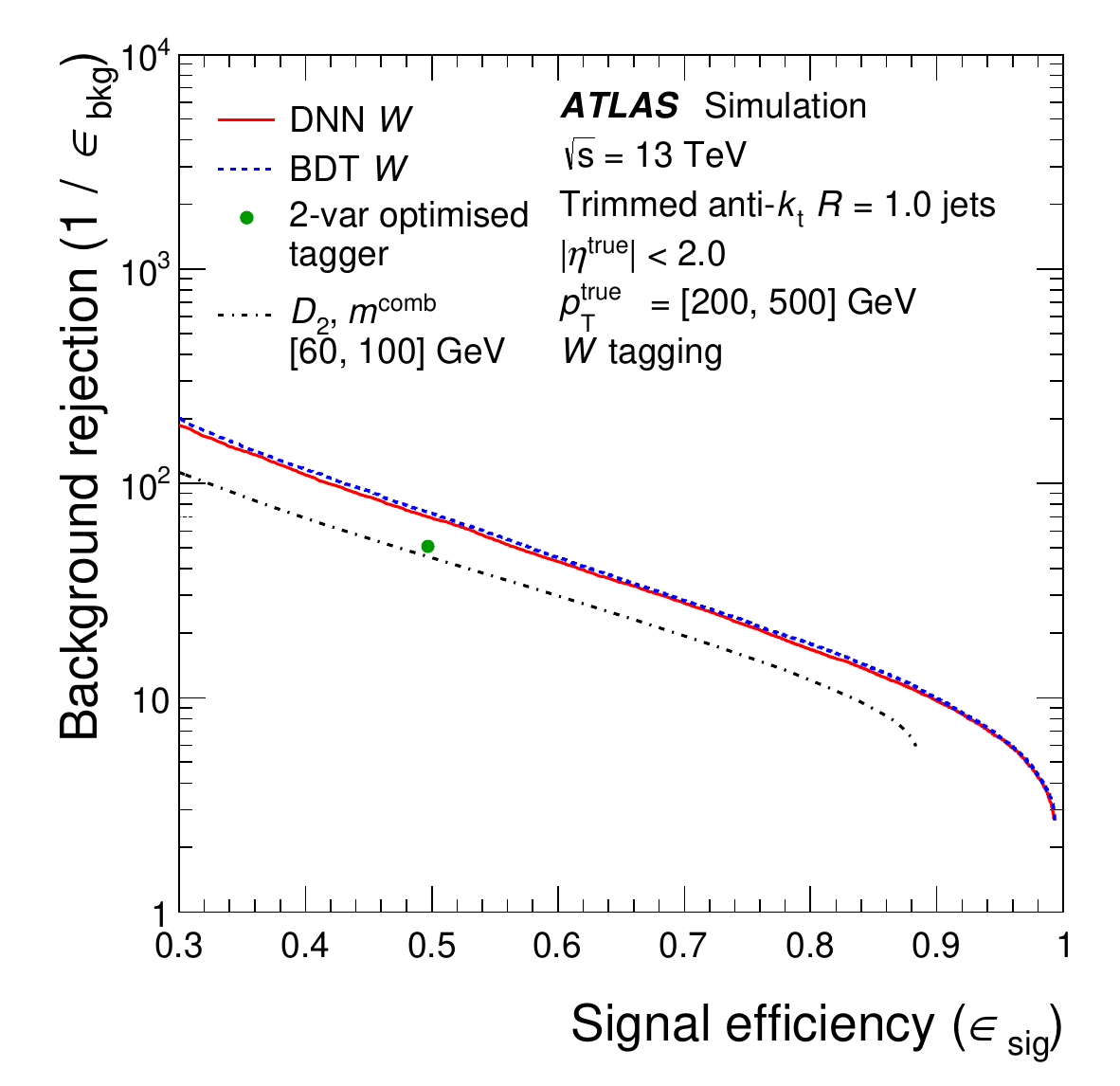}
  \caption{Top quark (left) and W boson (right) tagging efficiencies for various tagging approaches used by ATLAS \cite{Aaboud:2018psm}.}\label{fig:roc_tag_atlas}
\end{figure}

\begin{table}[tp]
\centering
%\scriptsize
\tiny
\begin{tabular}{ c| c c c c c c c c c |c c|| c c c c c c c c c |c c| }    % 23 in total
\hhline{~|-|-|-|-|-|-|-|-|-|-|-||-|-|-|-|-|-|-|-|-|-|-}
                                     & \multicolumn{11}{c||}{W Boson Tagging}                                                            & \multicolumn{11}{c|}{Top Quark Tagging}            \\ 
\hhline{~|-|-|-|-|-|-|-|-|-|-|-||-|-|-|-|-|-|-|-|-|-|-}
                                     & \multicolumn{9}{c|}{DNN Test Groups}                                                            & \multicolumn{2}{c||}{Inputs} & \multicolumn{9}{c|}{DNN Test Groups}                                                            & \multicolumn{2}{c|}{Inputs} \\
\hhline{|-||-|-|-|-|-|-|-|-|-|-|-||-|-|-|-|-|-|-|-|-|-|-}
\hhline{=::=:=:=:=:=:=:=:=:=:=:=::=:=:=:=:=:=:=:=:=:=:=}
\multicolumn{1}{|c||}{}    & 1   & 2   & 3   & 4   & 5   & 6   & 7   & 8   & 9   & BDT & DNN & 1   & 2   & 3   & 4   & 5   & 6   & 7   & 8   & 9   & BDT & DNN \\
\multicolumn{1}{|c||}{$m^{\mathrm{comb}}$}        & $\circ$ & $\circ$ &     & $\circ$ & $\circ$ & $\circ$ & $\circ$ & $\circ$ & $\circ$ & $\circ$ & $\circ$ &     & $\circ$ & $\circ$ & $\circ$ &     & $\circ$ & $\circ$ & $\circ$ & $\circ$ & $\circ$ & $\circ$   \\
\multicolumn{1}{|c||}{$p_t$}           & $\circ$ & $\circ$ &     &     & $\circ$ & $\circ$ &     & $\circ$ & $\circ$ & $\circ$ & $\circ$ &     &     & $\circ$ & $\circ$ &     &     & $\circ$ & $\circ$ & $\circ$ & $\circ$ & $\circ$   \\
\multicolumn{1}{|c||}{$e_3$}   & $\circ$ & $\circ$ &     &     &     & $\circ$ &     &     & $\circ$ &     &     &     &     &     & $\circ$ &     &     & $\circ$ &     & $\circ$ & $\circ$ & $\circ$   \\
\multicolumn{1}{|c||}{$C_2$}         &     &     & $\circ$ & $\circ$ & $\circ$ &     & $\circ$ & $\circ$ & $\circ$ &     & $\circ$ & $\circ$ & $\circ$ & $\circ$ &     & $\circ$ & $\circ$ &     & $\circ$ & $\circ$ &     & $\circ$   \\
\multicolumn{1}{|c||}{$D_2$}         &     &     & $\circ$ & $\circ$ & $\circ$ &     & $\circ$ & $\circ$ & $\circ$ & $\circ$ & $\circ$ & $\circ$ & $\circ$ & $\circ$ &     & $\circ$ & $\circ$ &     & $\circ$ & $\circ$ &     & $\circ$   \\
\multicolumn{1}{|c||}{$\tau_1$}       & $\circ$ & $\circ$ &     &     &     & $\circ$ &     &     & $\circ$ & $\circ$ &     &     &     &     & $\circ$ &     &     & $\circ$ &     & $\circ$ &     & $\circ$   \\
\multicolumn{1}{|c||}{$\tau_2$}       & $\circ$ & $\circ$ &     &     &     & $\circ$ &     &     & $\circ$ &     &     &     &     &     & $\circ$ &     &     & $\circ$ &     & $\circ$ & $\circ$ & $\circ$   \\
\multicolumn{1}{|c||}{$\tau_3$}       &     &     &     &     &     &     &     &     &     &     &     &     &     &     & $\circ$ &     &     & $\circ$ &     & $\circ$ &     & $\circ$   \\
\multicolumn{1}{|c||}{$\tau_{21}$}    &     &     & $\circ$ & $\circ$ & $\circ$ &     & $\circ$ & $\circ$ & $\circ$ & $\circ$ & $\circ$ & $\circ$ & $\circ$ & $\circ$ &     & $\circ$ & $\circ$ &     & $\circ$ & $\circ$ & $\circ$ & $\circ$   \\
\multicolumn{1}{|c||}{$\tau_{32}$}    &     &     &     &     &     &     &     &     &     &     &     & $\circ$ & $\circ$ & $\circ$ &     & $\circ$ & $\circ$ &     & $\circ$ & $\circ$ & $\circ$ & $\circ$   \\
\multicolumn{1}{|c||}{$R_2^{\mathrm{FW}}$} &     &     & $\circ$ & $\circ$ & $\circ$ & $\circ$ & $\circ$ & $\circ$ & $\circ$ & $\circ$ & $\circ$ &     &     &     &     &     &     &     &     &     &     & \\
\multicolumn{1}{|c||}{$\mathcal{P}$}   &     &     & $\circ$ & $\circ$ & $\circ$ & $\circ$ & $\circ$ & $\circ$ & $\circ$ & $\circ$ & $\circ$ &     &     &     &     &     &     &     &     &     &     & \\
\multicolumn{1}{|c||}{$a_3$}   &     &     & $\circ$ & $\circ$ & $\circ$ & $\circ$ & $\circ$ & $\circ$ & $\circ$ & $\circ$ & $\circ$ &     &     &     &     &     &     &     &     &     &     & \\
\multicolumn{1}{|c||}{$A$}   &     &     & $\circ$ & $\circ$ & $\circ$ & $\circ$ & $\circ$ & $\circ$ & $\circ$ & $\circ$ & $\circ$ &     &     &     &     &     &     &     &     &     &     & \\
\multicolumn{1}{|c||}{$z_\mathrm{cut}$}         &     &     & $\circ$ & $\circ$ & $\circ$ &     & $\circ$ & $\circ$ & $\circ$ &     & $\circ$ &     &     &     &     &     &     &     &     &     &     & \\
\multicolumn{1}{|c||}{$\sqrt{d_{12}}$}      &     & $\circ$ &     &     &     & $\circ$ & $\circ$ & $\circ$ & $\circ$ & $\circ$ & $\circ$ &     &     &     &     & $\circ$ & $\circ$ & $\circ$ & $\circ$ & $\circ$ & $\circ$ & $\circ$   \\
\multicolumn{1}{|c||}{$\sqrt{d_{23}}$}      &     &     &     &     &     &     &     &     &     &     &     &     &     &     &     & $\circ$ & $\circ$ & $\circ$ & $\circ$ & $\circ$ & $\circ$ & $\circ$   \\
\multicolumn{1}{|c||}{$KtDR$}         &     & $\circ$ &     &     &     & $\circ$ & $\circ$ & $\circ$ & $\circ$ & $\circ$ & $\circ$ &     &     &     &     &     &     &     &     &     &     & \\
\multicolumn{1}{|c||}{$Q_w$}           &     &     &     &     &     &     &     &     &     &     &     &     &     &     &     & $\circ$ & $\circ$ & $\circ$ & $\circ$ & $\circ$ & $\circ$ & $\circ$   \\
\hhline{|-||-|-|-|-|-|-|-|-|-|-|-||-|-|-|-|-|-|-|-|-|-|-}
\end{tabular}
\caption{%
A summary of the set of observables that were 
tested for W-boson and top-quark tagging for the final set of DNN and BDT input observables \cite{Aaboud:2018psm}. $p_t$ and $m^{\mathrm{comb}}$ are the transverse momentum of the jet and the combined jet mass \cite{ATLAS:2016vmy}. $e_3$, $C_2$ and $D_2$ are energy correlation ratios \cite{Larkoski:2013eya, Larkoski:2014gra}. $\tau_i$ and $\tau_{ij}$ are $N$-subjettiness variables and ratios respectively. $R_2^{\mathrm{FW}}$ is a Fox-Wolfram moment \cite{Fox:1978vu}. Splitting measures are denoted $z_\mathrm{cut}$, $\sqrt{d_{12}}$ and $\sqrt{d_{23}}$ \cite{Thaler:2008ju, Aad:2013ueu}. The planar flow variable $\mathcal{P}$ is defined in \cite{Almeida:2008tp} and the angularity $a_3$ in \cite{Aad:2012meb}. Definitions can be found for aplanarity $A$ \cite{Chen:2011ah}, $KtDR$ \cite{Catani:1993hr} and $Q_w$ \cite{Thaler:2008ju}.}\label{tab:obs}
\end{table}
The performance of such multivariate BDT and DNN taggers is then
compared to perturbative-QCD inspired taggers, i.e.\ the HEPTopTagger
and the Shower Deconstruction tagger, using trimmed anti-$k_t$ $R=1.0$
fat jets. While the inputs to construct the observables of
Table~\ref{tab:obs} consist of all jet constituents, the HEPTopTagger
and Shower Deconstruction tagger are restricted to be used on
calibrated Cambridge/Aachen subjets of finite size, i.e.\
$R_\mathrm{subjet} \geq 0.2$ . Thus, ROC curves, as shown in
Fig.~\ref{fig:roc_tag_atlas}, have to be taken with a grain of salt,
as systematic uncertainties of the input objects have not been
propagated consistently into the performance
curves.\footnote{Systematic~\cite{Louppe:2016ylz, Shimmin:2017mfk} and
  theoretical~\cite{Englert:2018cfo} uncertainties can be taken into
  account in the performance evaluation of a neural net classifier by
  adding an adversarial neural network.} However, in particular for
highly boosted top quarks, see Fig.~\ref{fig:roc_tag_atlas}, the combination of multiple jet shape observables shows a very strong tagging performance over the entire signal efficiency range.

\begin{figure}
 \includegraphics[width=0.5\textwidth]{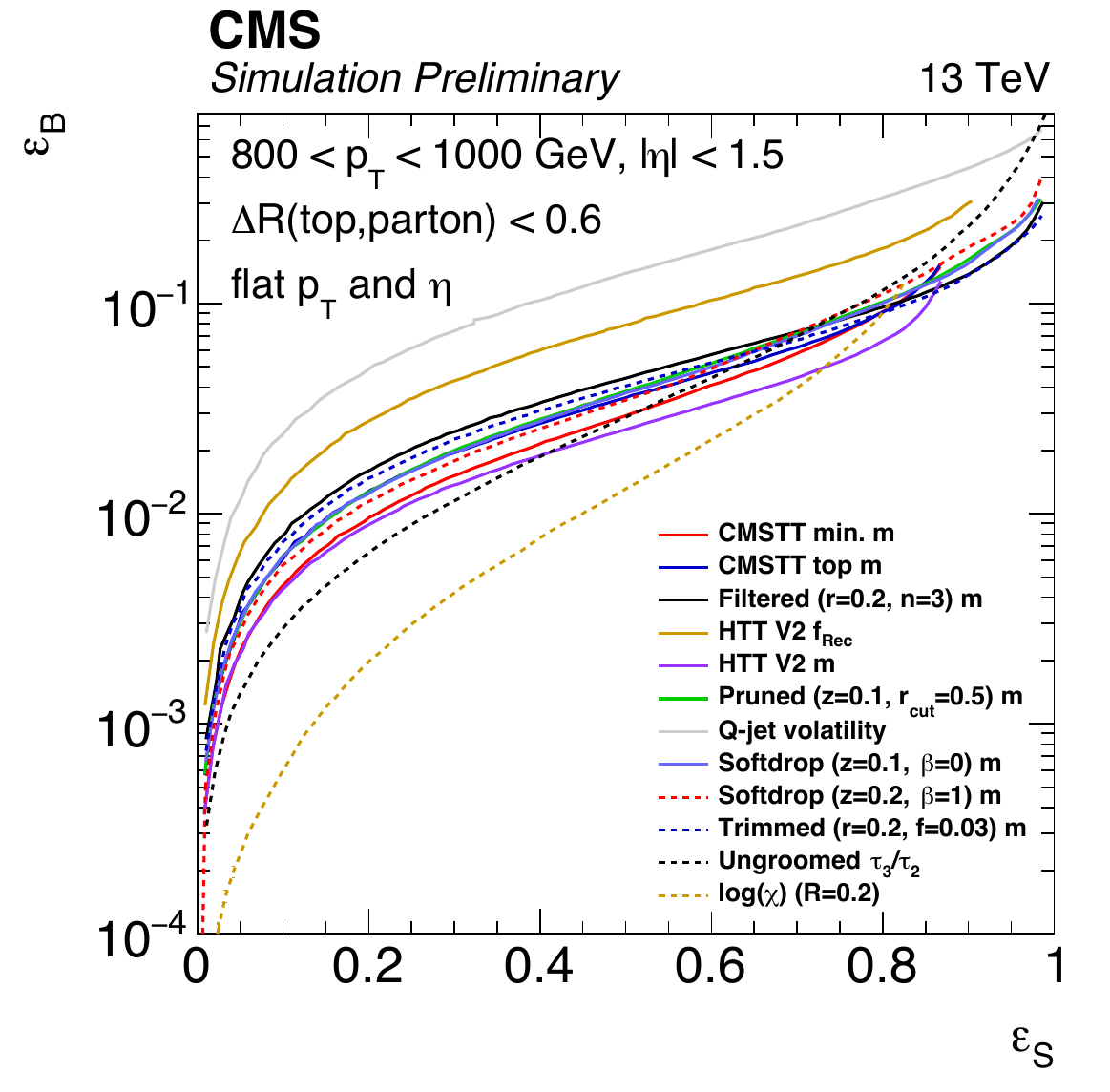}
  \includegraphics[width=0.5\textwidth]{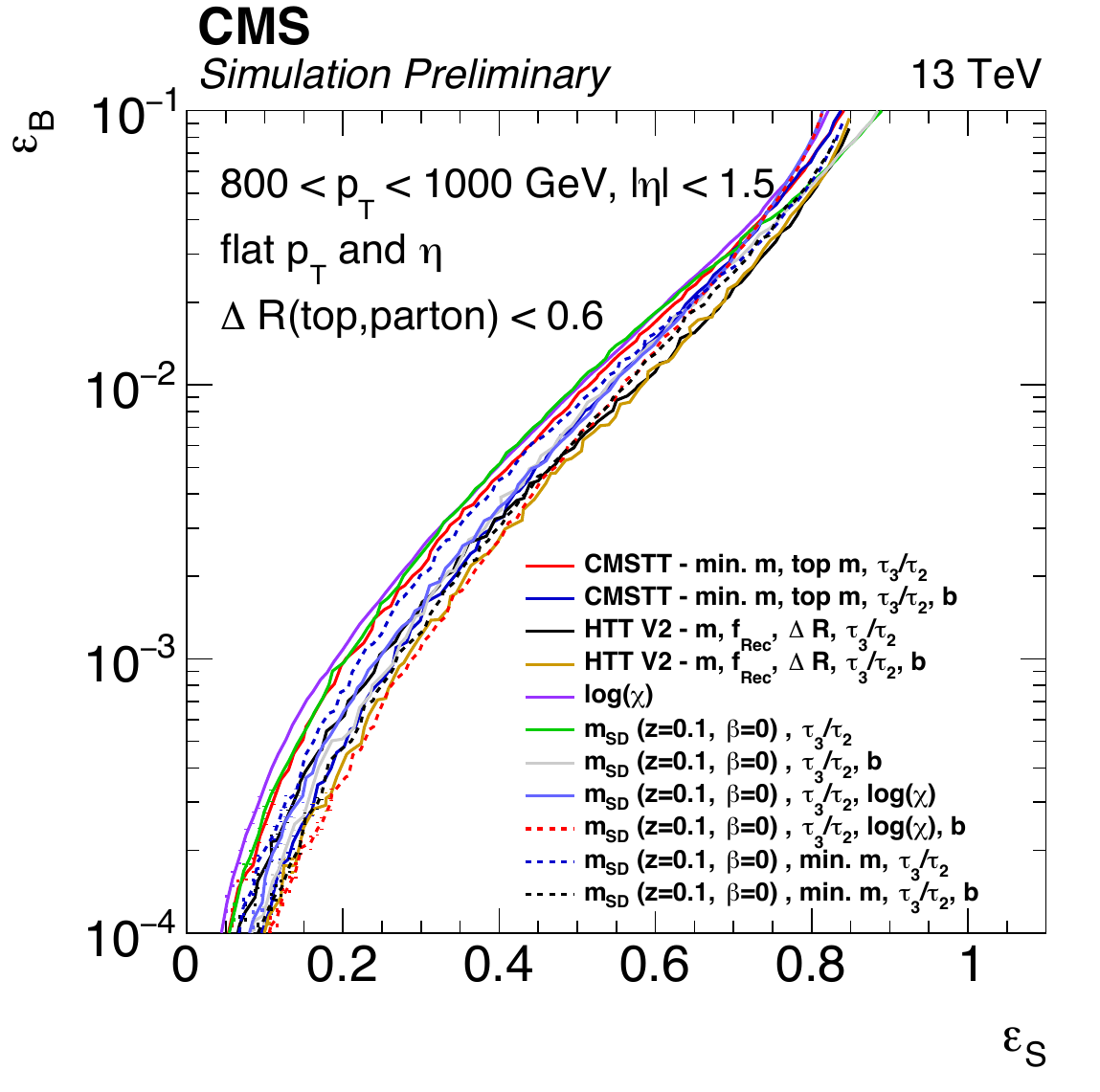}
  \caption{Top quark tagging performance comparison from CMS \cite{CMS:2016tvk}.}\label{fig:roc_tag_cms}
\end{figure}

CMS \cite{CMS:2016tvk, Khachatryan:2014vla} takes a similar approach
to W boson and top quark tagging as ATLAS. CMS uses a subset of the
observables of Table~\ref{tab:obs}, and extends it by including Qjet
volatility \cite{Ellis:2012sn} and $b$-tagging\footnote{For jets,
  $b$-tagging is meant to separate jets originating from a $b$ quark
  from light-quark and gluon jets. $b$-tagging algorithms are using
  the fact that $B$ hadrons decay with a displaced vertex together
  with a list of variables included in a BDT or neural network (with
  details depending on the experiment).} in their performance
analysis. In addition to the Shower Deconstruction tagger, an updated
version of the HEPTopTagger (V2) and the CMS top tagger are included
in the comparison. The results of Fig.~\ref{fig:roc_tag_cms} (left)
show that the performance of individual observables and taggers can
vary a lot, with Shower Deconstruction performing best in the signal
efficiency region of $\varepsilon_\mathrm{S} \leq 0.7$. However, when
various tagging methods are combined in a multivariate approach,
Fig.~\ref{fig:roc_tag_cms} (right), their performance become very
similar and the potential for further improvements seems to saturate
for the scenario at hand.
\begin{figure}
  \includegraphics[width=0.5\textwidth]{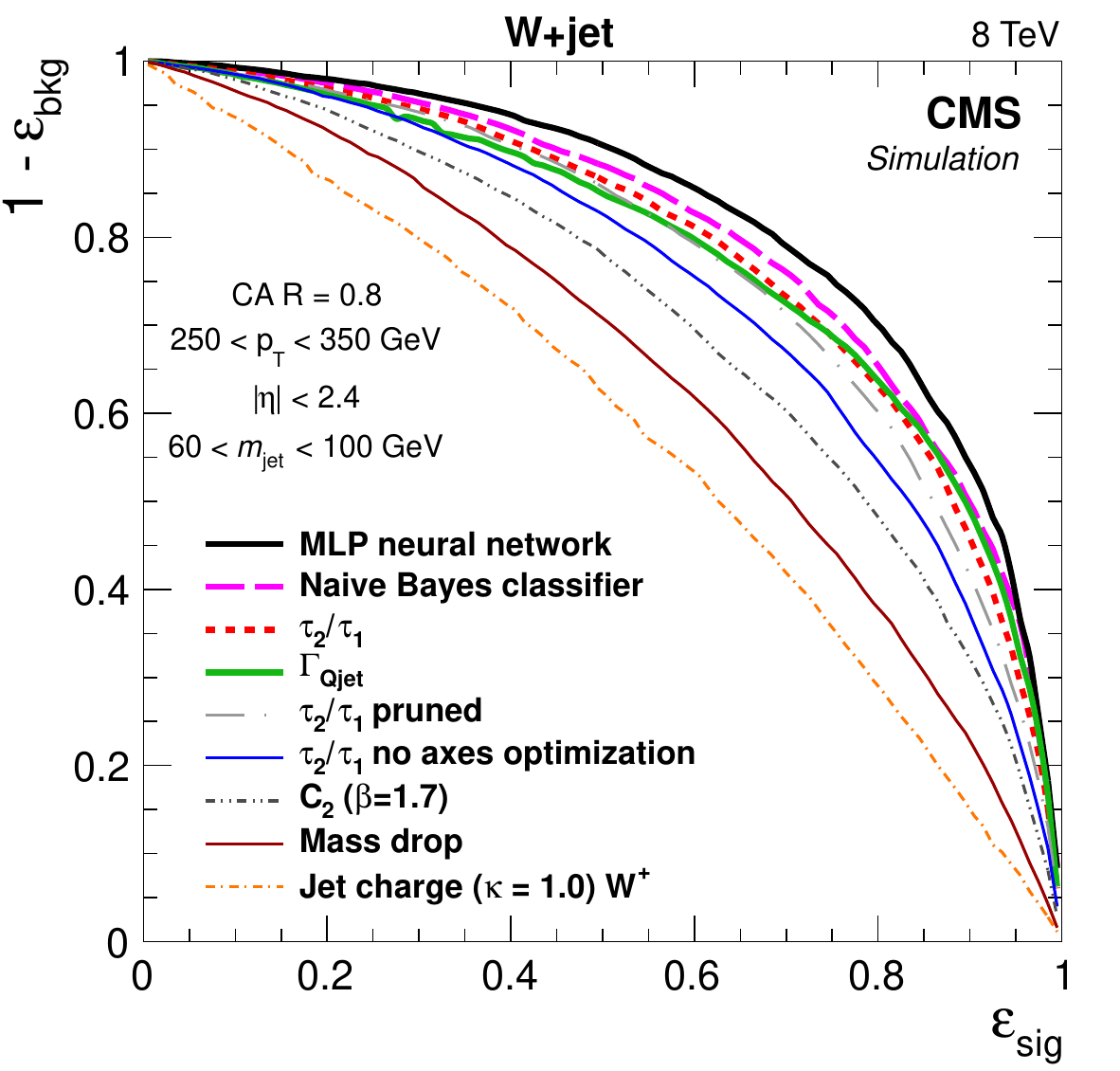}\hfill%
  \begin{minipage}[b]{0.42\textwidth}
    \caption{W boson tagging performance comparison from CMS~\cite{Khachatryan:2014vla}. 
  }\label{fig:cms_w_tag}
    \vspace*{2.4cm}
  \end{minipage}\hspace*{0.3cm}
\end{figure}
For W tagging, see Fig.~\ref{fig:cms_w_tag}, CMS combines several jet shape observables using either a naive Bayes classifier or a Multilayer Perceptron (MLP) neural network discriminant. When comparing to individual jet shape observables, such as $N$-subjettiness ratios or Qjet volatility, mild improvements can be achieved.

The discrimination between quark and gluon-initiated jets can have profound phenomenological implications. A large class of processes associated with the production of new particles have a strong preference to result in quarks, e.g.\ the production and subsequent decay of squarks in the Minimal Supersymmetric Standard Model, while Standard Model QCD backgrounds are more likely to result in gluon-initiated jets. Thus, the ability to separate these two classes of jets reliably could boost our sensitivity in finding new physics. However, as discussed at length in Chapter~\ref{sec:calc-shapes-qg}, the discrimination between a jet that was initiated by a gluon from a jet that was initiated by a quark is subtle. 
Consequently, sophisticated observables which attempt to exploit small features between quarks and gluon jets can potentially be sensitive to limited experimental resolution and experimental uncertainties in the construction of the jet constituents.
In their performance studies, ATLAS~\cite{Aad:2014gea} and CMS~\cite{CMS:2017wyc} aim to exploit the differences in the radiation profiles between quarks and gluons using observables such as the number of charged tracks $n_{\mathrm{trk}}$, calorimeter $w_\mathrm{cal}$ or track width $w_\mathrm{trk}$ with
\begin{equation}
w = \frac{\sum_i p_{T,i} \times \Delta R\mathrm{(i,jet)}}{\sum_i p_{T,i}}\;,
\end{equation}
where $i$ runs either over the calorimeter energy clusters to form $w_\mathrm{cal}$ or over the charged tracks for $w_\mathrm{trk}$. Further observables are the track-based energy-energy-correlation (EEC) angularities
\begin{equation}
\mathrm{ang}_{\mathrm{EEC}} = \frac{\sum_i \sum_j p_{T,i} ~ p_{T,j} ~ (\Delta R(i,j))^\beta}{(\sum_i p_{T,i})^2}\;,
\end{equation}
where the index $i$ and $j$ run over the tracks associated with the jet, with $j>i$, and $\beta$ is a tunable parameter, the jet minor angular opening $\sigma_2$ of the $p_t^2$-weighted constituents distribution in the lego plane and the jet fragmentation distribution $p_{T}D$, defined as
\begin{equation}
p_{T}D = \frac{\sqrt{\sum_i p_{T,i}^2}}{\sum_i p_{T,i}},
\end{equation}
where $i$ runs over all jet constituents.
\begin{figure}[t]
  \includegraphics[width=0.47\textwidth]{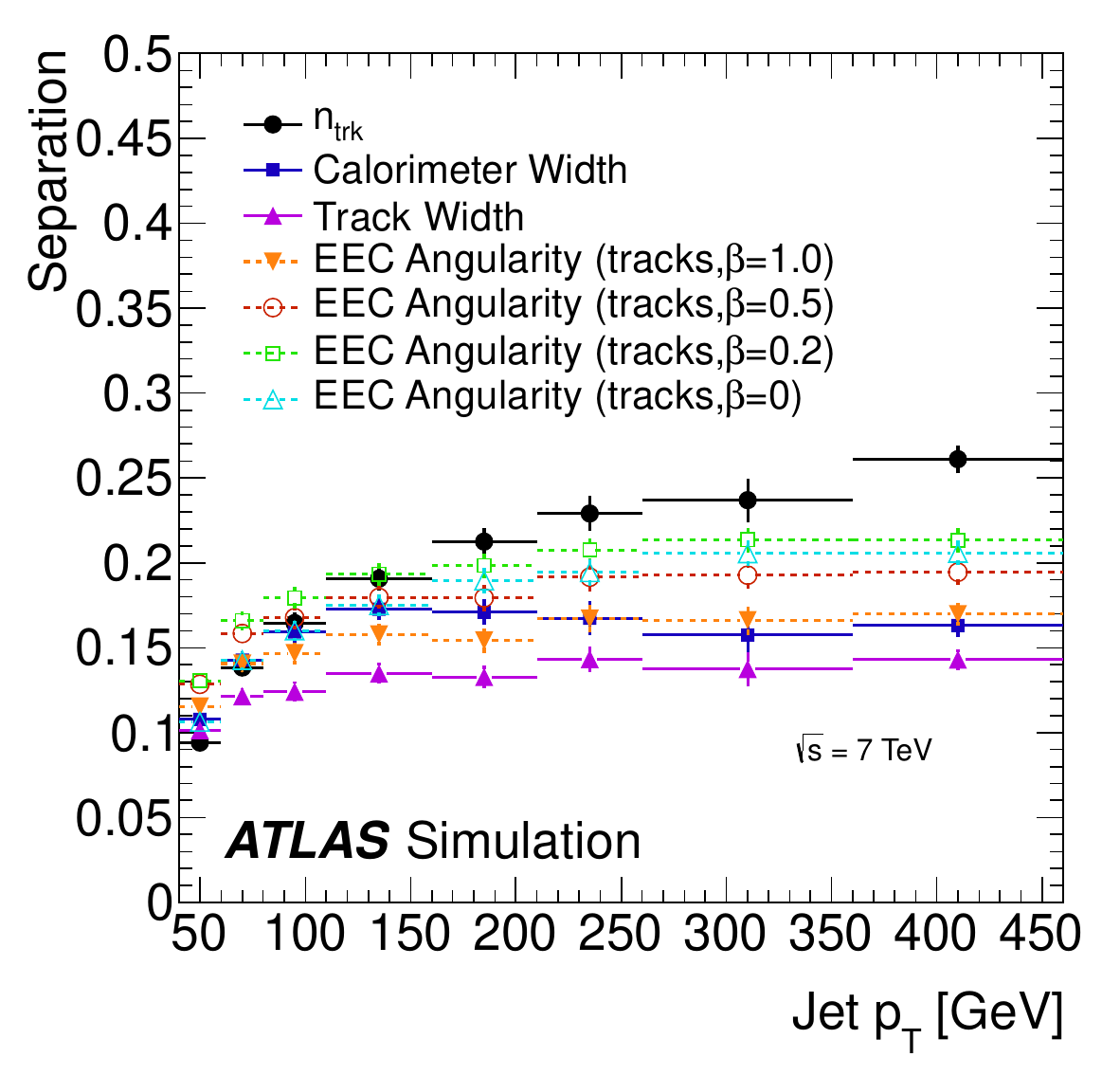}
 \includegraphics[width=0.51\textwidth]{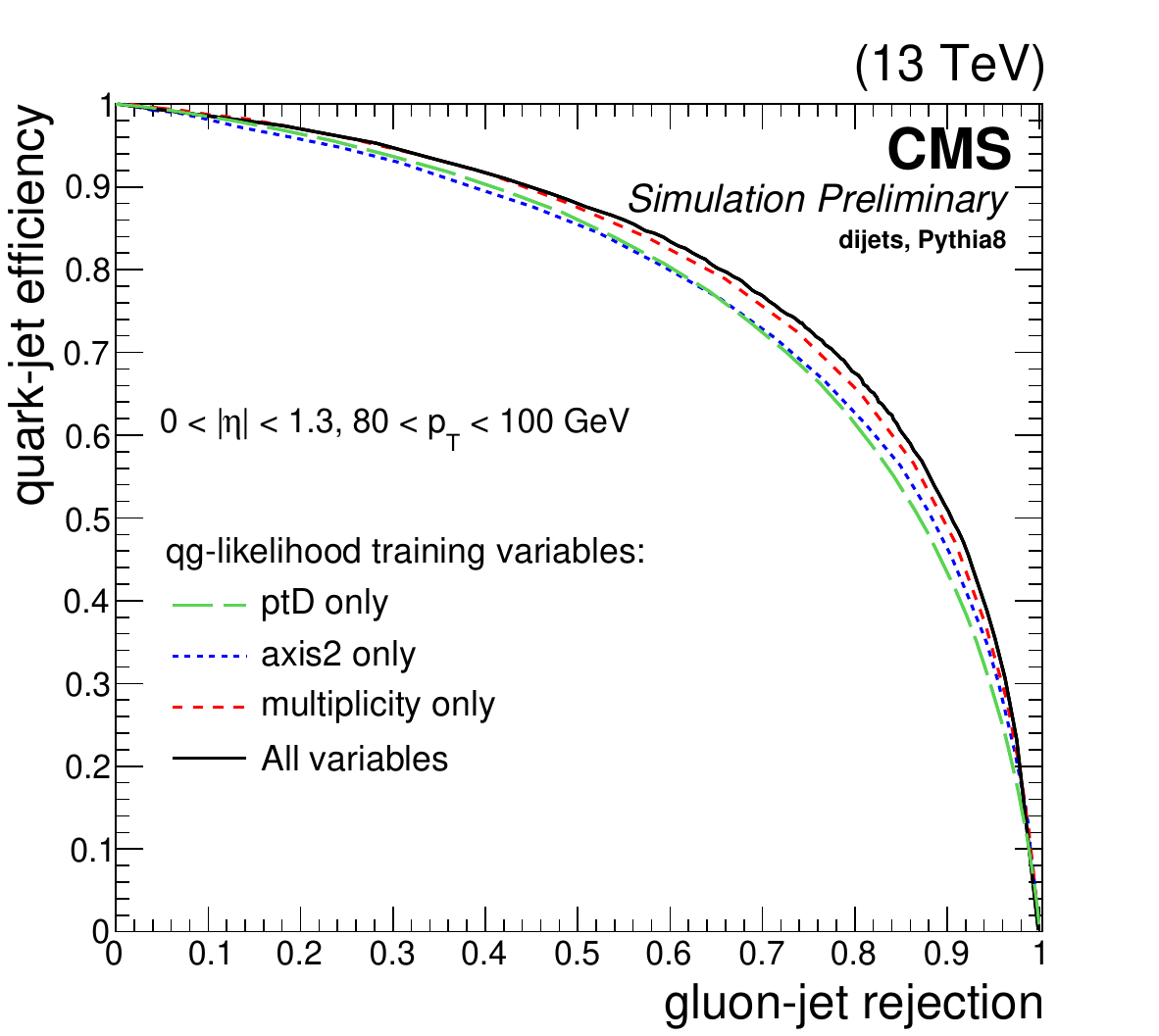}
  \caption{ATLAS (left) and CMS (right) studies of quark/gluon discrimination.
  The plots are taken from, respectively, Ref.~\cite{Aad:2014gea} and~\cite{CMS:2017wyc}. }\label{fig:exp_roc_qg}
\end{figure}

ATLAS results for quark-gluon tagging~\cite{Aad:2014gea, ATL-PHYS-PUB-2017-009} are reported in Fig.~\ref{fig:exp_roc_qg} on the left, in terms of the variable ``separation", which is defined as:
\begin{equation}
\mathrm{Separation} = \frac{1}{2} \int \frac{(p_q(x) - p_g(x))^2}{p_q(x) + p_g(x)} dx
\end{equation}
where $p_q(x)$ and $p_g(x)$ are normalised distributions of the variables used for discrimination between quark and gluon jets. 
Both experiments achieve a good separation between quark and gluon jets for the observables used and the $p_t$-windows studied. For example, CMS achieves for a $50\%$ quark jet acceptance a rejection of roughly $90\%$ of gluon jets.

\section{Measurements of jet observables}
\begin{figure}[t!]
 \includegraphics[width=0.5\textwidth]{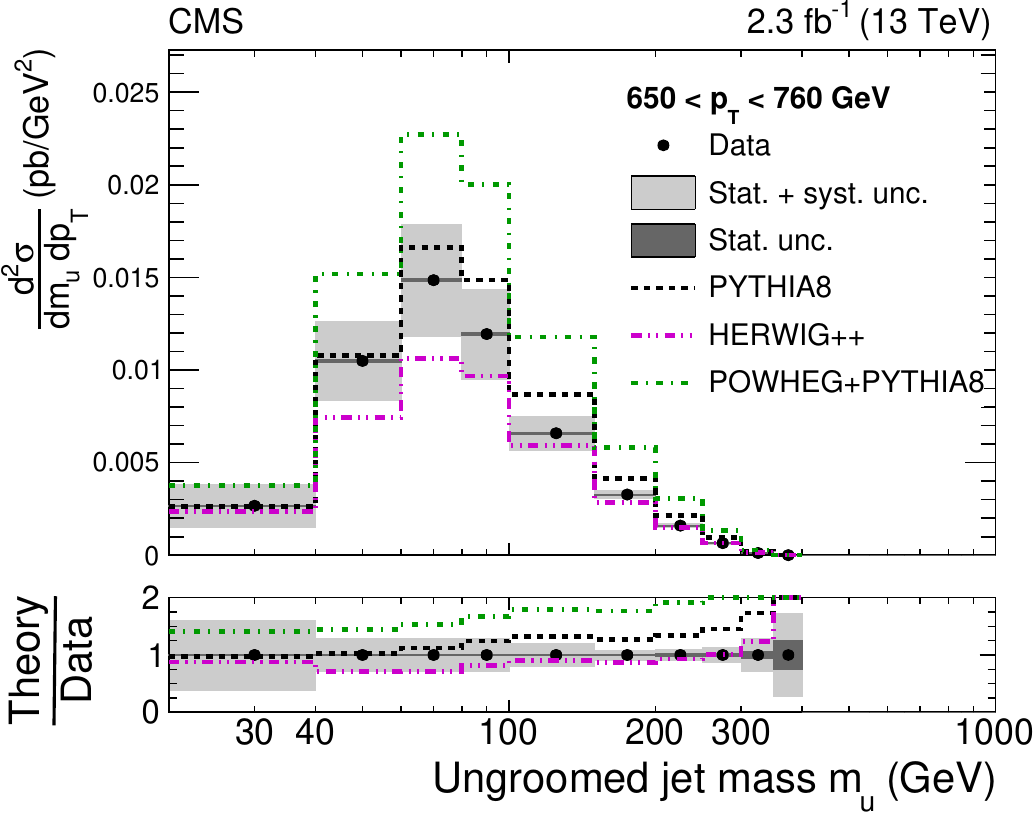} 
 \includegraphics[width=0.5\textwidth]{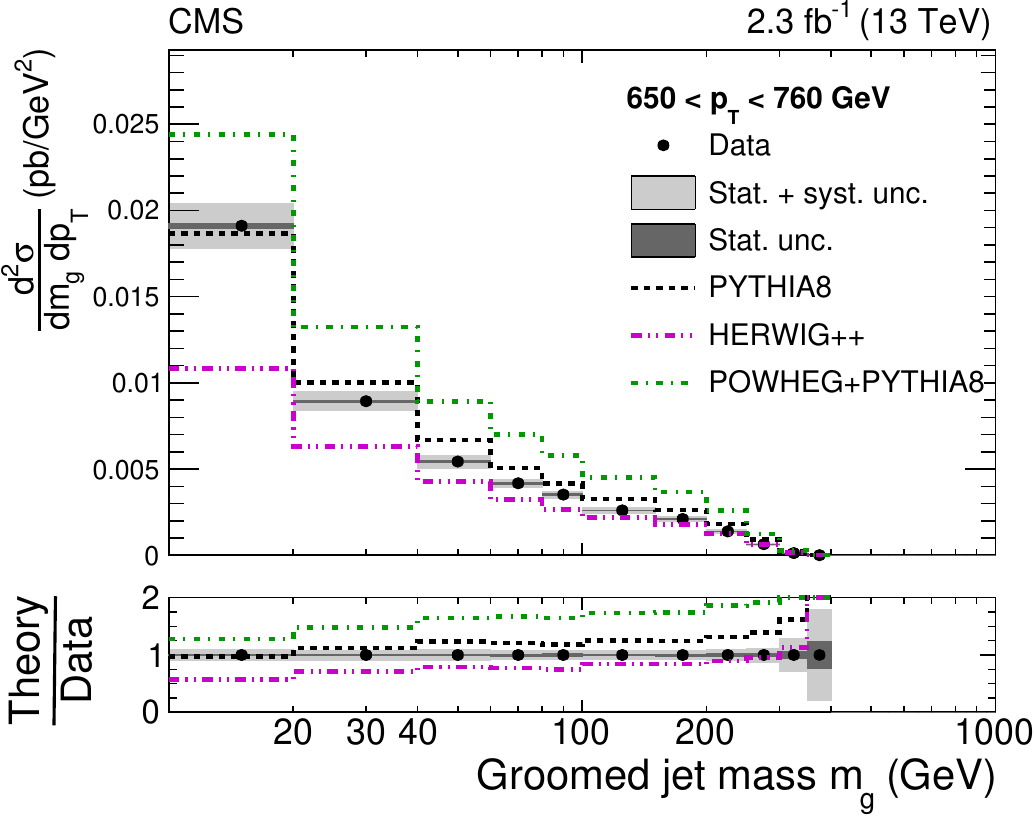} \\
 \includegraphics[width=0.45\textwidth]{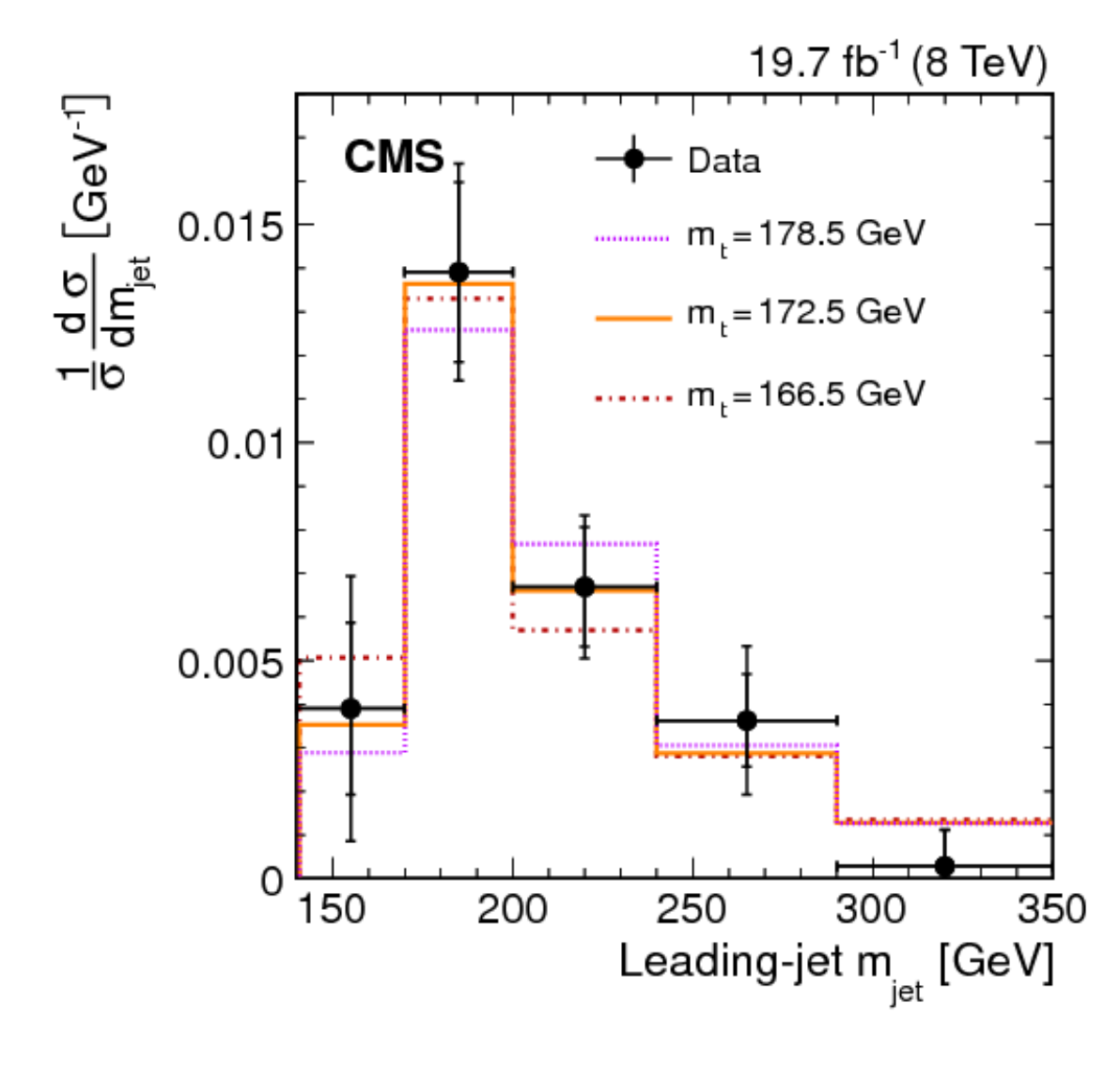} 
 \includegraphics[width=0.55\textwidth]{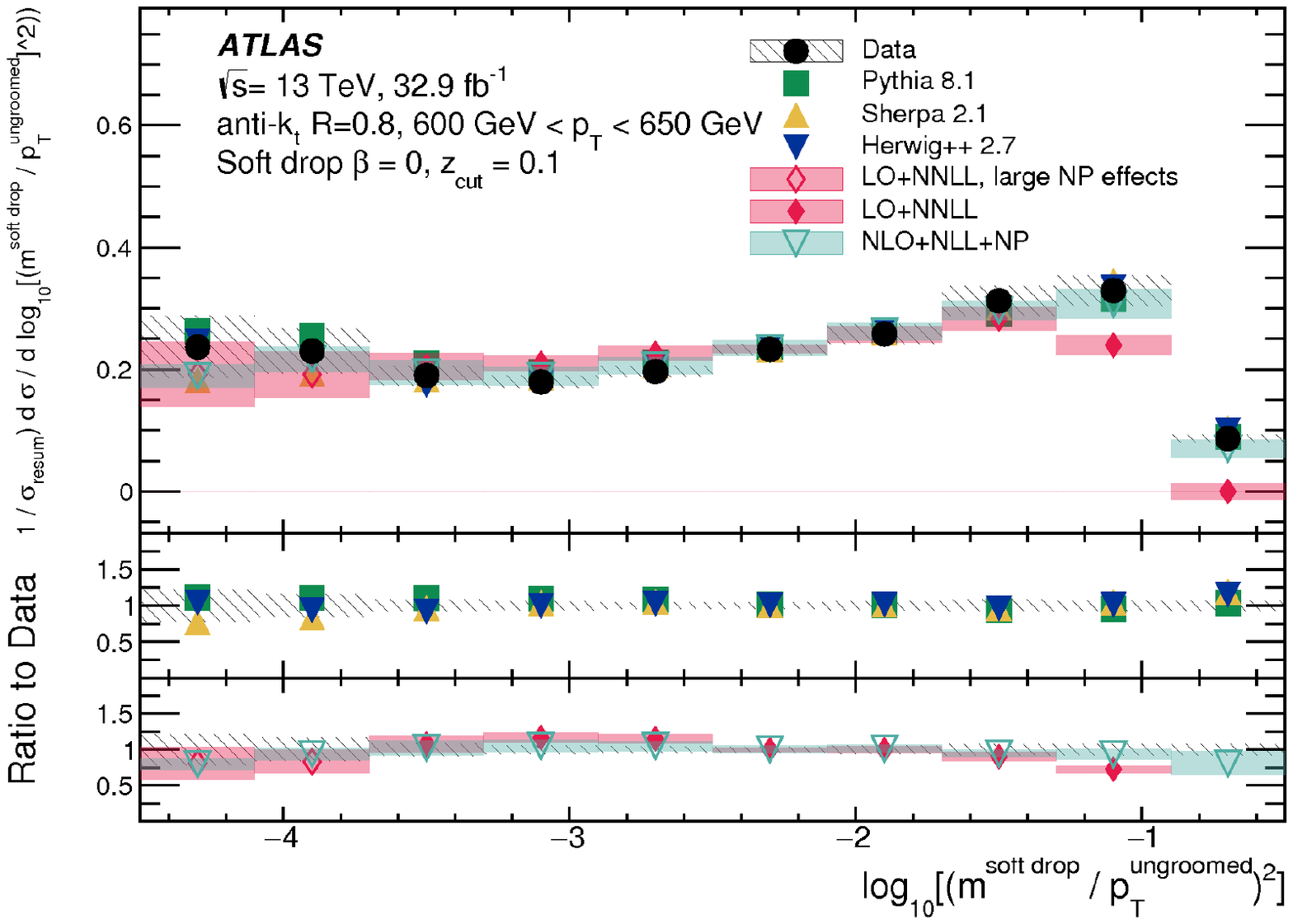}
  \caption{Jet mass measurements at the LHC, starting from the top
    left and going clockwise, we have: plain jet mass by CMS
    \cite{Sirunyan:2018xdh}, \SD (mMDT) jet mass by CMS \cite{Sirunyan:2018xdh}, \SD mass measurement by
    ATLAS~\cite{Aaboud:2017qwh}, and top jet mass by CMS \cite{Sirunyan:2017yar}.}\label{fig:jet_mass}
\end{figure}

Various jet observables discussed in Chapters~\ref{tools}-\ref{sec:curiosities} have been measured by the LHC experiments. In the following we will discuss a selection of the measurements performed for these observables and we will focus on measurements for large-$R$ jets. Where possible, we will favour measurements that have been compared to first-principle calculations that have been described in the previous chapter. The up-to-date list of jet substructure measurement is maintained by the \href{https://twiki.cern.ch/twiki/bin/view/LHCPhysics/LHCJetSubstructureMeasurements}{LHC Electro-Weak Working Group}.

\subsection{Jet mass}\label{sec:exp-jet-mass}

The mass of a jet is one of the most basic observables associated with a jet. As such, it was discussed in great detail in Chapters~\ref{chap:calculations-jets} and \ref{calculations-substructure-mass}, with and without the application of various grooming methods to the jet. As the mass is sensitive to the energy distribution in the jet, it can also be thought of as a jet-shape observable.

ATLAS \cite{ATLAS:2012am, Aaboud:2017qwh} and CMS~\cite{Sirunyan:2017yar, Sirunyan:2018xdh} have both measured the mass of jets under various conditions. In \cite{ATLAS:2012am} ATLAS has measured the jet mass, amongst other jet shape and jet substructure observables, in $pp$ collisions at a centre-of-mass energy of 7~TeV. 

The \SD mass has been measured in \cite{Aaboud:2017qwh} and \cite{
  Sirunyan:2018xdh} by ATLAS and CMS, respectively. After requiring a
jet with $p_t > 600$ GeV and imposing the dijet topology cut
$p_{T,1}/p_{T,2} < 1.5$, ATLAS runs the soft-drop algorithm on the two
leading jets in the events. Three different values of $\beta \in
{0,1,2}$ are considered, while the value on the $z_\mathrm{cut}$ is
fixed at $0.1$. Then the dimensionless ratio $m^{\mathrm{soft~drop}}/p_t^{\mathrm{ungroomed}}$ is constructed and
shown in Fig~\ref{fig:jet_mass} in the lower right panel. The measured data is in good agreement with various theoretical predictions, including resummed analytic calculations and full event generators.
CMS selects similar event kinematics for this measurement, but fixes $\beta=0$. In Fig.~\ref{fig:jet_mass} in the upper right and upper left panel the groomed and ungroomed jet mass, measured by CMS at 13 TeV centre-of-mass energy, is shown respectively. Data is compared with theory predictions from Pythia8, Herwig++ and Powheg+Pythia8, showing significant differences between the three event generators. While the overall normalisation of the cross sections predicted by the event generators is quite different, with Pythia8 being closest to data, the shape of the theoretically predicted distributions agree well with data. Thus, when the distributions are normalised to the total cross section, the difference between data and all three theory predictions is small.

The precision with which a boosted top quark's mass can be measured by analysing a fat jet is a crucial parameter for many tagging algorithms. In \cite{Sirunyan:2017yar} CMS purifies the final state with respect to semi-leptonic $t\bar{t}$ events and reconstructs Cambridge/Aachen $R=1.2$ fat jets with $p_t>400$ GeV. The mass of the leading fat jet is sown in the lower left panel of Fig.~\ref{fig:jet_mass}. No special grooming procedure has been used, yet the measured jet mass agrees well with the physical top mass. 

\subsection{Jet angularities}\label{sec:exp-jet-ang}
\begin{figure}[t]
  \includegraphics[width=0.45\textwidth]{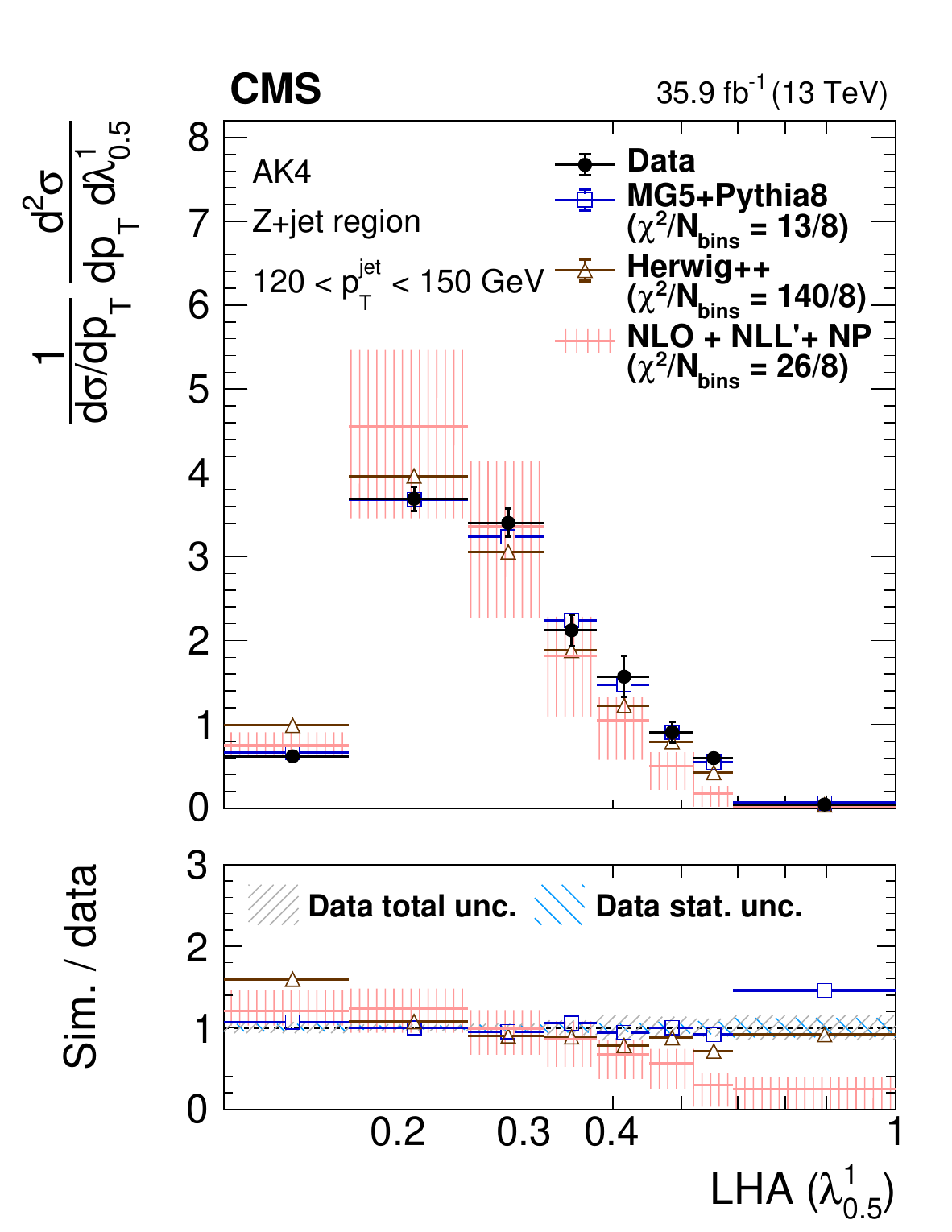}
  \includegraphics[width=0.45\textwidth]{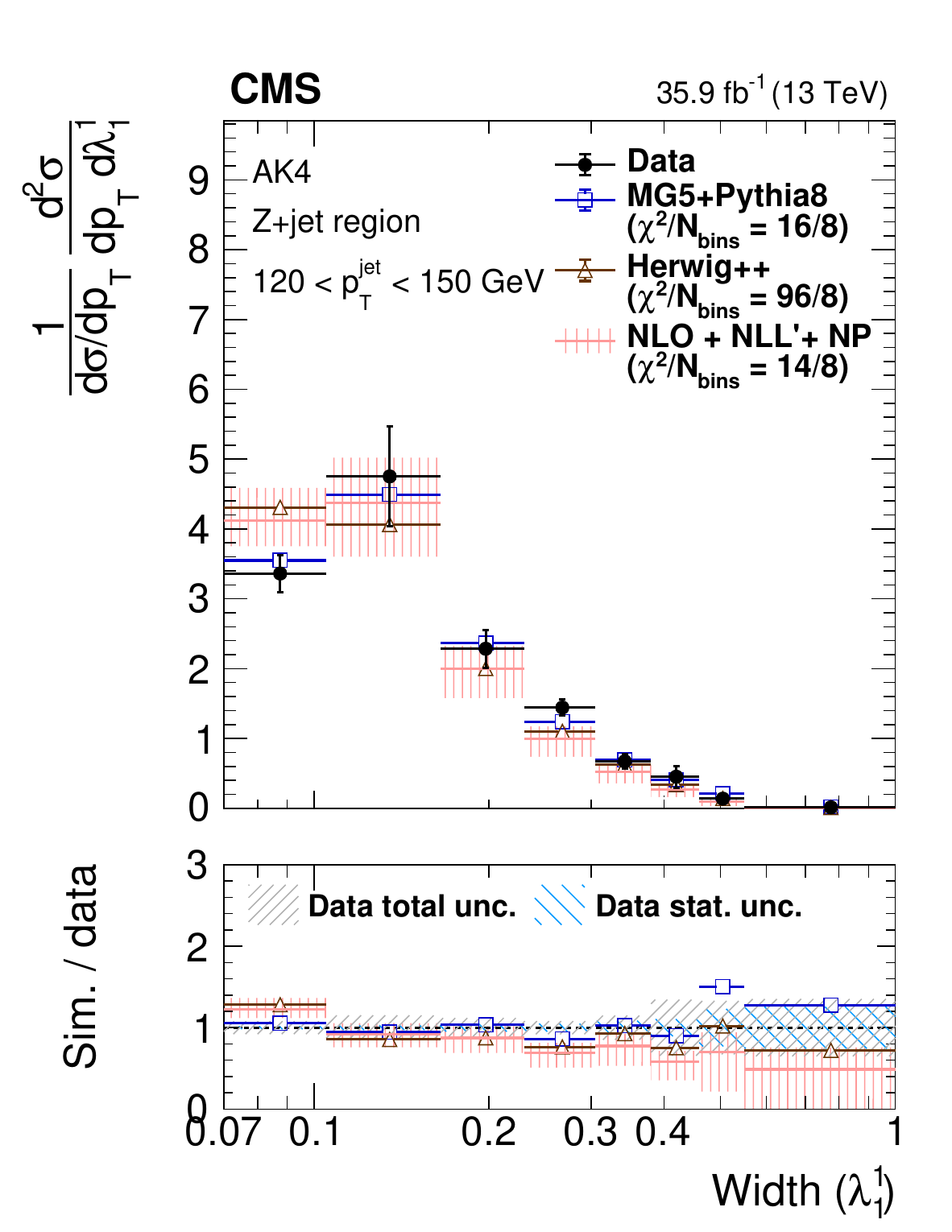}
  \caption{Measurements of the Les Houches Angularity (LHA), on the left, and tje jet width, on the right, by CMS~\cite{CMS:2021iwu}.
  The data are compared to state-of-the-art theoretical calculations, as well as from the resummed and matched calculation of Ref.~\cite{Caletti:2021oor,Reichelt:2021svh}.
  }\label{fig:exp_ang}
\end{figure}

Jet angularities, discussed in~\ref{sec:def-angularities}, represent another important class of jet shapes.
Jet angularities have been measured by CMS at 13~TeV centre-of-mass energy~\cite{CMS:2021iwu} and by ALICE at 5.02~TeV~\cite{ALICE:2021njq}. 
The experimental results have been compared to state-of-the art Monte Carlo simulations as well as dedicated resummed and matched  calculations~\cite{Caletti:2021oor,Reichelt:2021svh,Kang:2018vgn}.
The CMS measurements is performed for two distinct samples, Z+jet, which is  enriched in quark-initiated jets, and dijets, which is enriched in gluon jets, allowing to test the theoretical modelling of different jet flavours. 
Measurements were performed in different transverse momentum and rapidity bins, for standard and for \SD jets. In Fig.~\ref{fig:exp_ang}, we show results from CMS for IRC safe angularities ($\kappa=1)$ with angular exponents 0.5 and 1, i.e.\ the Les Houches Angularity (LHA) and the jet width, respectively.  We note that the NLO+NLL calculation, once supplemented with non-perturbative corrections (extracted from Monte Carlo) gives a good description of the data. On the other hand, the theoretical uncertainties are rather large, motiving the quest for higher theoretical precision.

\subsection{Jet charge}
\begin{figure}[t]
  \includegraphics[width=0.45\textwidth]{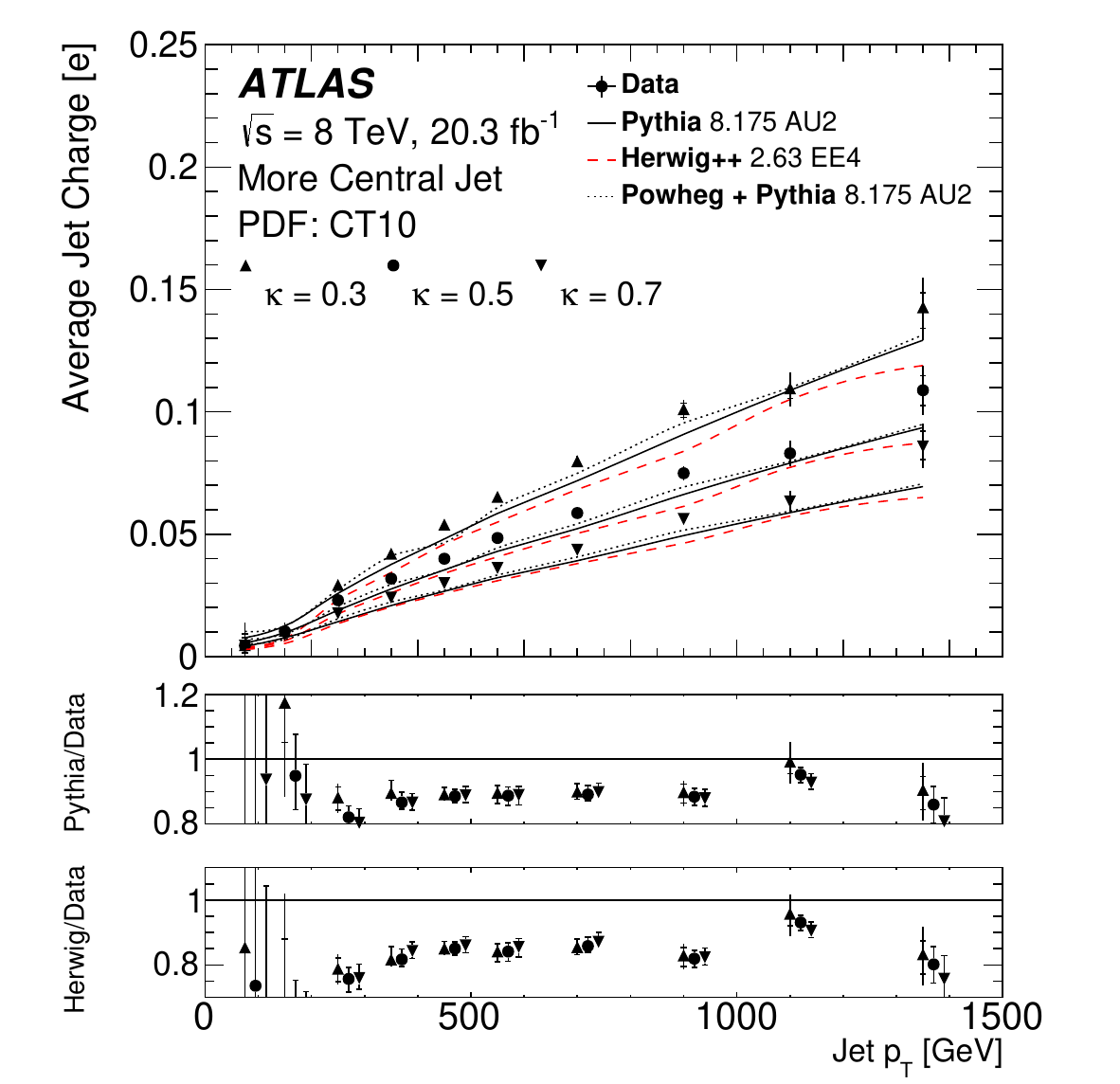}
 \includegraphics[width=0.65\textwidth]{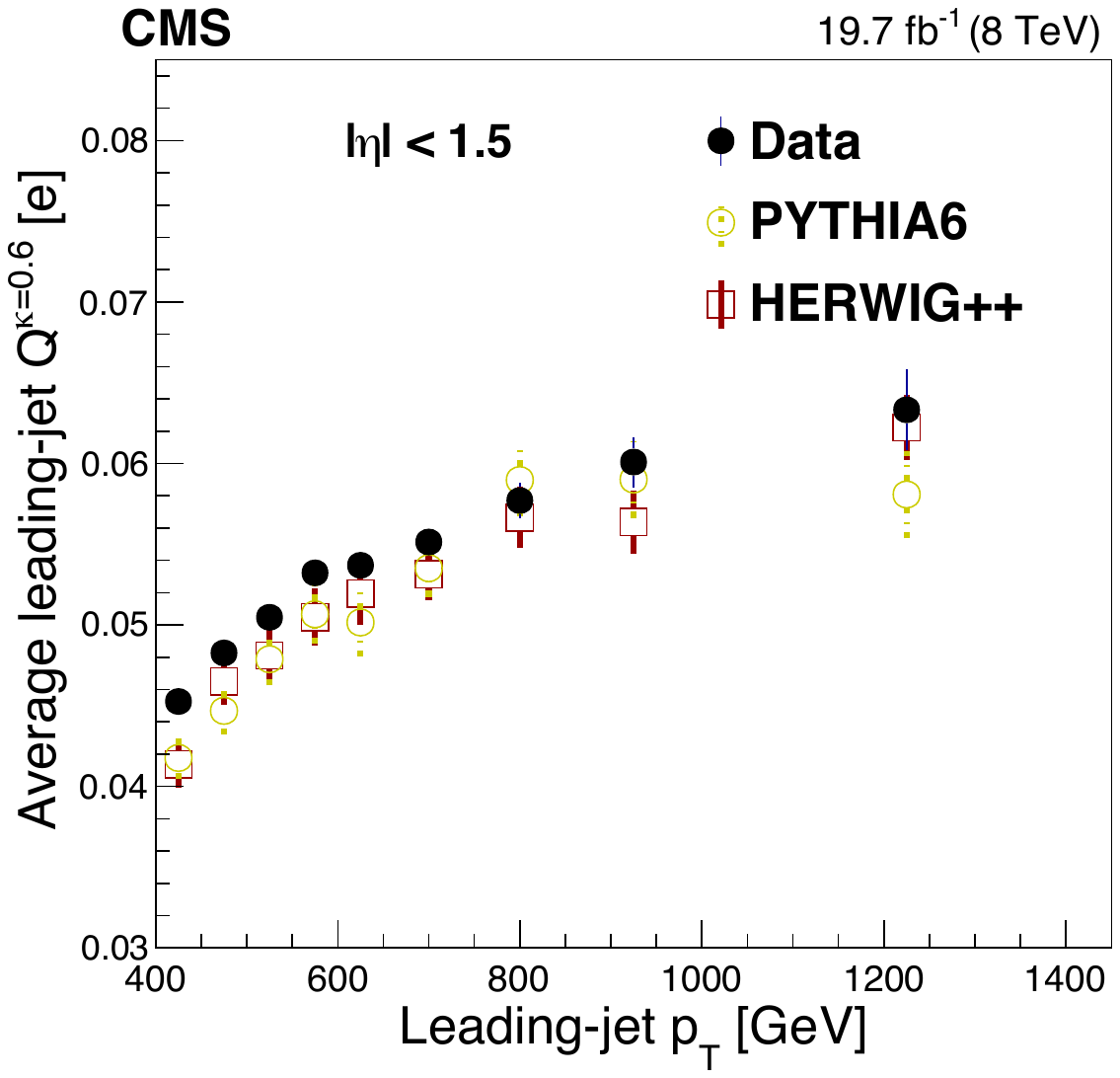}
  \caption{Measurements of the jet charge by ATLAS~\cite{Aad:2015cua} (left) and CMS~\cite{Sirunyan:2017tyr} (right).}\label{fig:exp_charge}
\end{figure}

The energy deposits and tracks associated with a jet can originate from dozens of charged particles, depending on the size and on transverse momentum of the jet. If charged particles become too soft, e.g. $p_t \ll 1$ GeV, they can curl up in the magnetic field of the detector and might not even be measurable in the calorimetry or the tracker. Thus, it is useful to define the jet charge as a $p_t$-weighted sum of the charge of the jet constituents.
As the number of charged particles amongst the jet constituents is neither an infrared-safe nor a perturbatively calculable quantity, experimental measurements of these observables have to be compared to fitted hadronisation models included in full event generators.
A natural and common definition for the jet charge is  \cite{Aad:2015cua,Sirunyan:2017tyr} 
\begin{equation}
Q_J = \frac{1}{(p_{T,J})^\kappa} \sum_{i \in \mathrm{tracks}} q_i ~(p_{T,i})^\kappa \;,
\end{equation}
where $i$ runs over all tracks associated with jet $J$. $q_i$ is the
measured charge of track $i$ with associated transverse momentum
$p_{T,i}$, and $\kappa$ is a free regularisation
parameter.\footnote{There are alternative definitions of jet
  charge. For examples and how their theoretical prediction compares
  to experimental measurements, see~\cite{Sirunyan:2017tyr}.} In this
definition the charge associated with individual tracks, \ie
individual charged particles, is weighted by their transverse
momentum.
That way $Q_J$ is less sensitive to experimental and theoretical uncertainties. 
ATLAS and CMS both find good agreement between theoretical predictions and data over a large range of transverse momenta of the jets, when calculating their average charge, see Fig.~\ref{fig:exp_charge}.

\subsection{Splitting functions}

\begin{figure}[t]
  \includegraphics[width=0.49\textwidth]{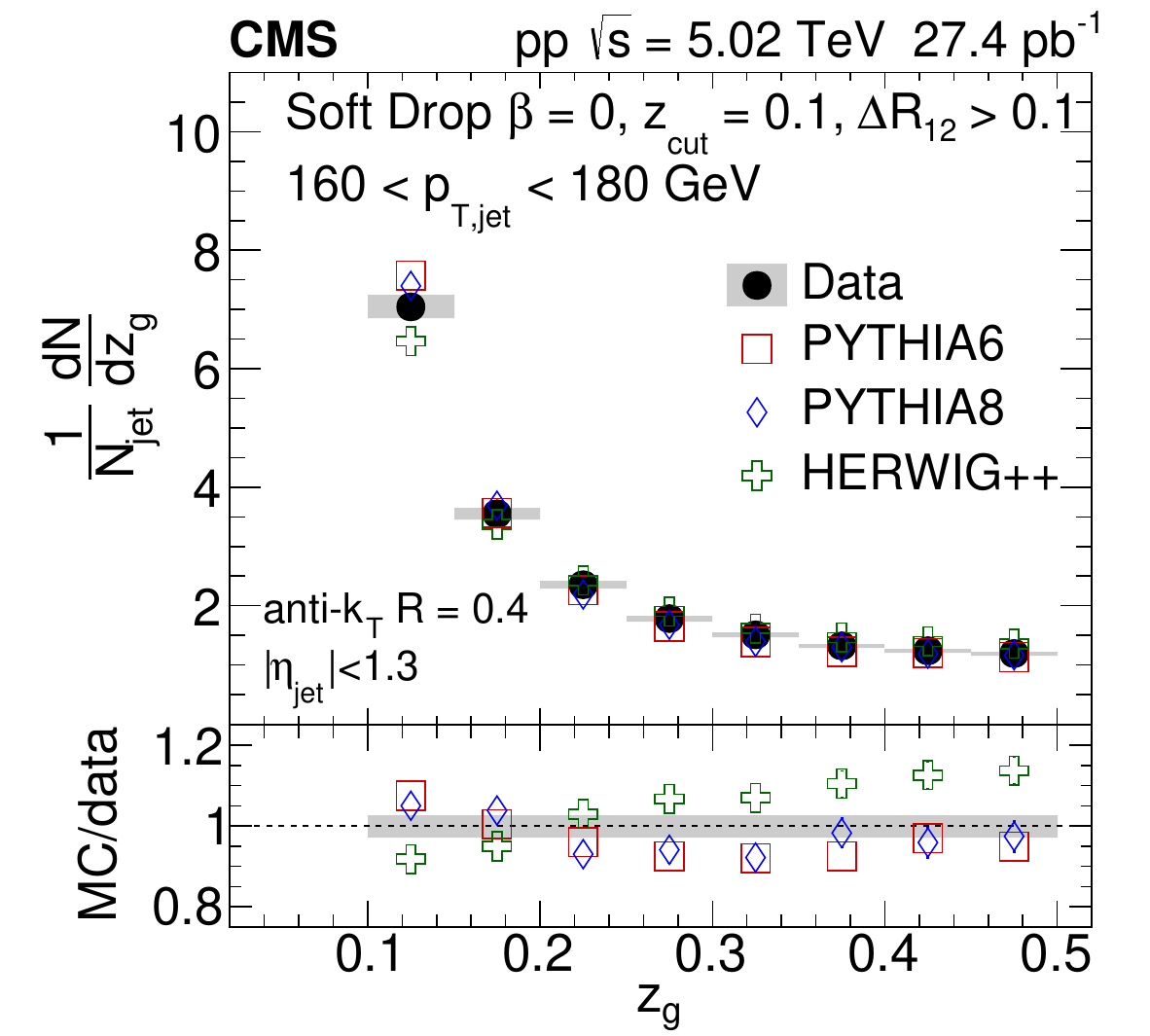}
  \includegraphics[width=0.49\textwidth]{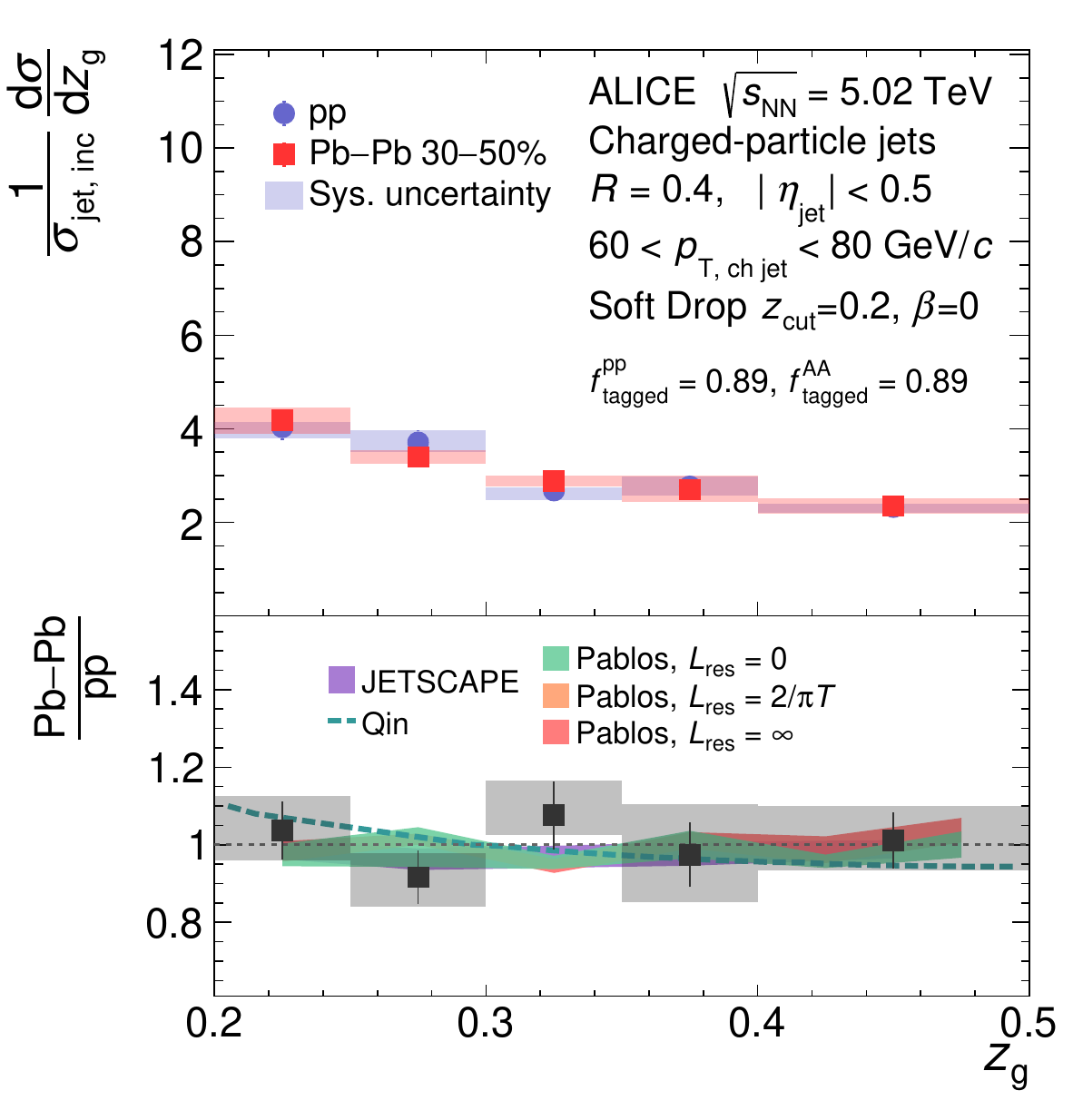}
     \caption{The groomed momentum sharing $z_g$ measured by CMS~ \cite{Sirunyan:2017bsd} (on the left) and by ALICE~\cite{ALargeIonColliderExperiment:2021mqf} (on the right).
     Both measurements are performed at 5.02~TeV on \SD jets with $\beta=0$. However, note the different jet radii and $\zcut$ values employed. Fiducial cuts in transverse momentum and pseudorapidities also differ.     
     }
    \label{fig:exp_splitting}
    \vspace*{1.4cm}
\end{figure}

The momentum sharing $z_g$ between the two subjets that pass the \SD condition was introduced in Sec.~\ref{sec:zg}. The variable $z_g$ can be taken as a proxy of the ``most important" partonic splitting in the jet evolution and thus its distribution is governed by the QCD splitting functions.
A measurement\footnote{Note that we refer here to observables that have not been unfolded. Thus, a comparison of data to theoretical predictions requires the knowledge of detector effects on the reconstructed observable.} of the $z_g$ distribution in $pp$ collisions, using CMS open data, was reported in \cite{Larkoski:2017bvj, Tripathee:2017ybi}. 
Using data obtained during LHC's heavy-ion runs, CMS has studied $z_g$ in PbPb and $pp$ collisions \cite{Sirunyan:2017bsd}. A measurement in of $z_g$ in PbPb collisions reflects how the two colour-charged partons produced in the first splitting propagate through the quark-gluon plasma, thereby probing the role of colour coherence of the jet in the medium. In the $pp$ case all particle-flow anti-$k_t$ jets with $R=0.4$ and $p_{t,j} > 80$ GeV were recorded. To identify the hard prongs of a jet and to remove soft wide-angle radiation, \SD grooming is applied to the jets with $\beta=0$ and $z_\mathrm{cut}=0.1$. 
Fig.~\ref{fig:exp_splitting} shows the comparison of $z_g$ between the measured CMS data and the theoretical predictions from Pythia6, Pythia8 and Herwig++, including a full simulation of detector effects. While in general good agreement is observed, both Pythia simulations have a slightly steeper $z_g$ distribution than the data, whereas Herwig++ shows an opposite trend.
The $z_g$ distribution in $pp$ and PbPb collisions has was also measured by ALICE~\cite{ALargeIonColliderExperiment:2021mqf}. Interestingly, ALICE has also performed this measurements in $pp$ at 13~TeV on jets identified as \emph{charm} jets, thus opening a window on the so-called massive splitting function~\cite{Catani:2002hc}.

\subsection{Primary Lund jet plane}\label{sec:exp-lund-plane}

\begin{figure}
  \includegraphics[width=0.48\textwidth]{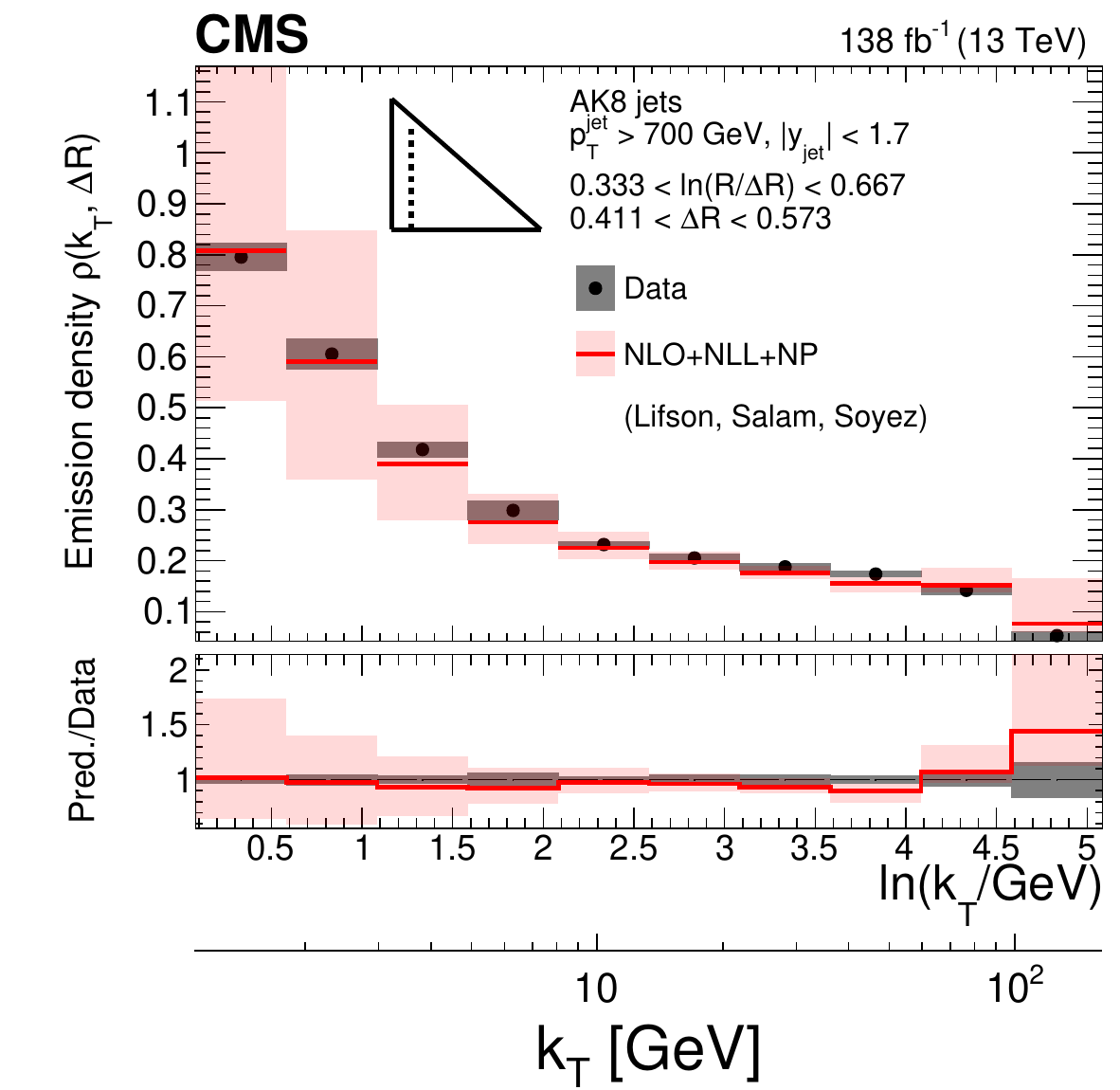}
    \includegraphics[width=0.48\textwidth]{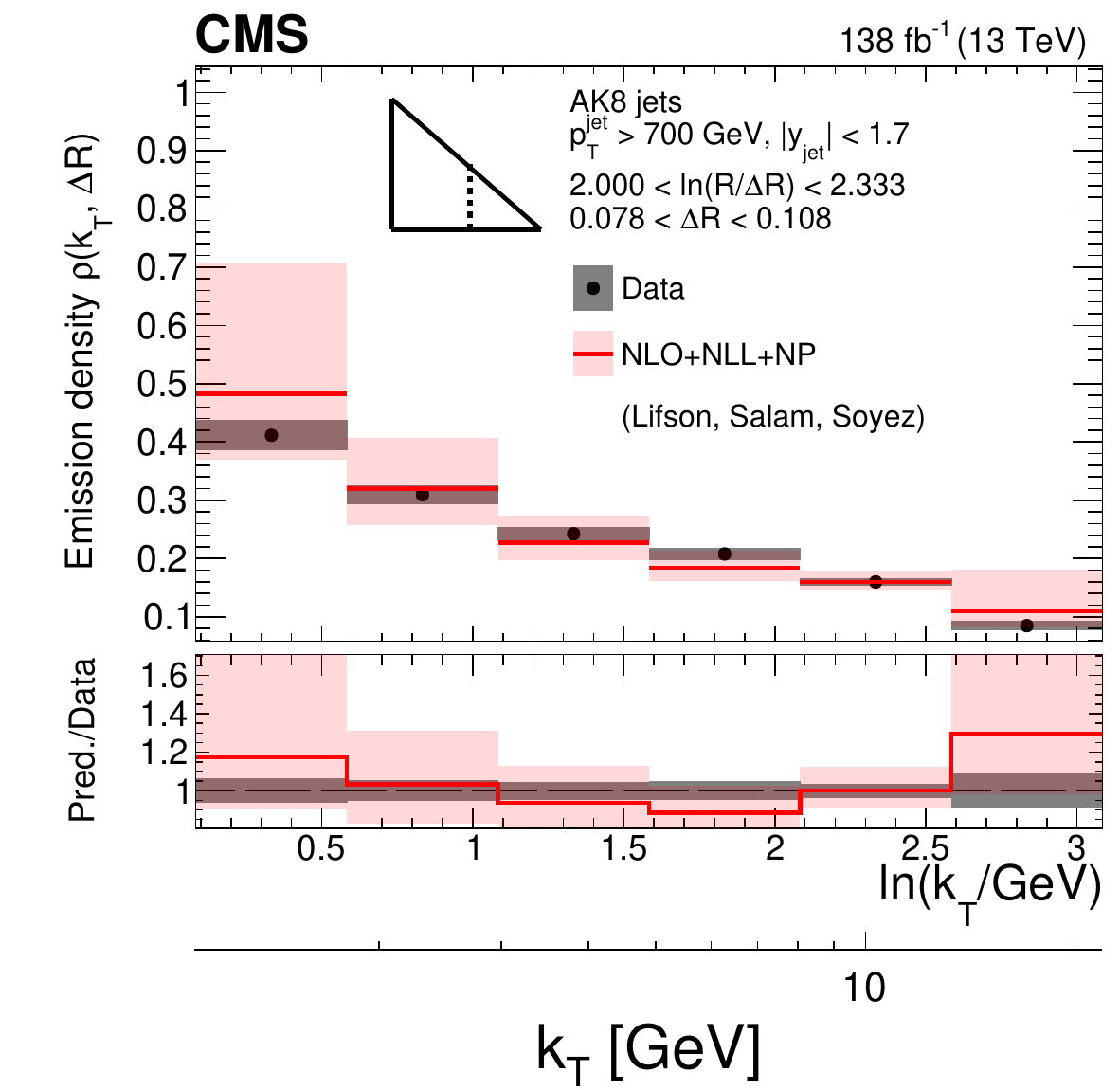} \\
  \includegraphics[width=0.48\textwidth]{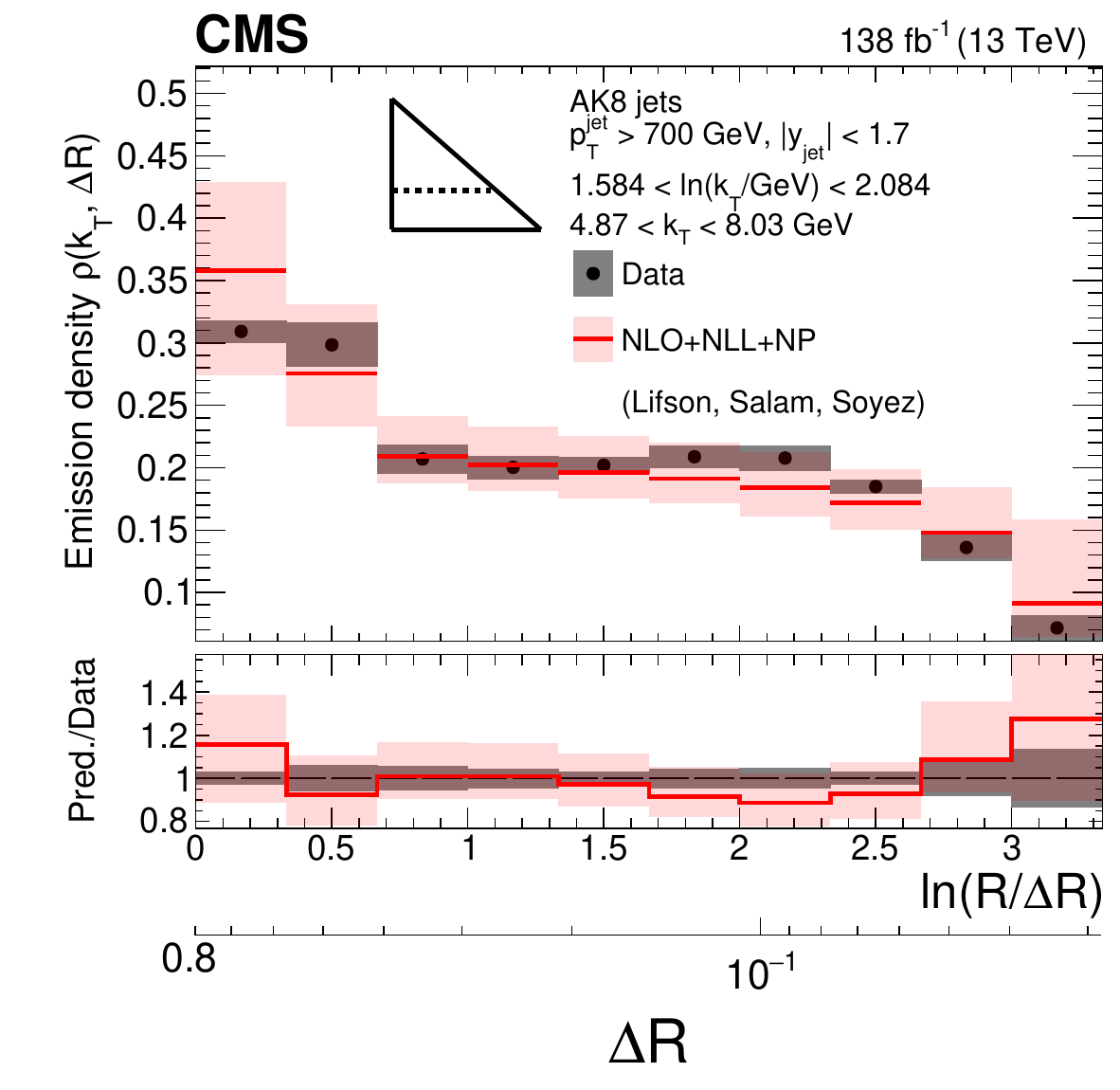}
    \includegraphics[width=0.48\textwidth]{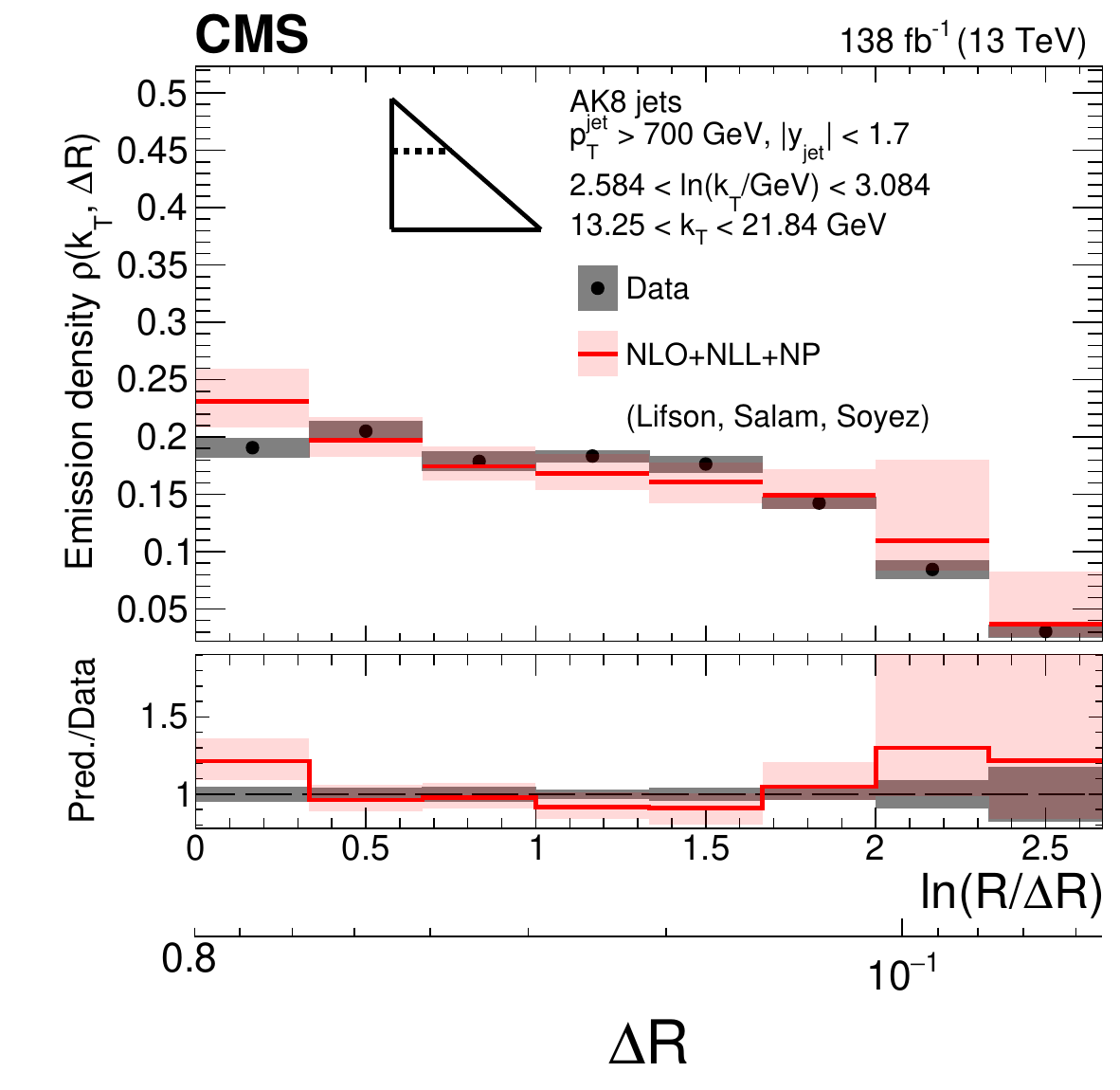}       
  \caption{Different $\Delta R$ (upper plots) and $k_t$ (lower plots) slices of the primary Lund jet plane, as measured by the CMS collaboration~\cite{CMS:2023lpp}. Data are compared to the resummed theoretical prediction of Ref.~\cite{Lifson:2020gua}.}\label{fig:cms_lund}
\end{figure}
As discussed in detail in Chapter~\ref{lundplane}, the Lund jet plane has proved itself a powerful tools in different contexts of jet substructures. Thus, it does not come as a surprise that all four LHC collaborations have measured (or are measuring) the Lund jet plane, focussing on the primary Lund plane density. The first measurement was performed by ATLAS in $pp$ collisions at 13~TeV~\cite{ATLAS:2020bbn}, closely followed by ALICE~\cite{Havener:2021yhb} and CMS~\cite{CMS:2023lpp}., while the measurement by LHCb is work in progress.
The ATLAS and CMS measurements focussed on the high-$p_t$ region, above 675 (700)~GeV, where one can explore a wide portion of the jet phase space exploiting perturbative methods, while ALICE concentrated on a lower transverse momentum region, $p_t \in [20,120]$~GeV, where non-perturbative contributions are expected to give a sizeable contribution. Thus, these measurements can be used to test and tune different parts of Monte Carlo event generators, i.e.\ the parton shower and hadronisation models. 
Data can also be compared to first-principle resummed calculations, such as the ones described in  Chapter~\ref{lundplane}, as illustrated in Fig.~\ref{fig:cms_lund}.

Finally, we find it relevant to mention that the ALICE collaboration, using de-clustering techniques similar to ones employed for the Lund jet plane but applied to charm-tagged jets, have performed a measurement of the so-called dead cone effect~\cite{Dokshitzer:1991fd,Dokshitzer:1995ev} the suppression of radiation collinear to a massive quark~\cite{ALICE:2021aqk}.

\section{Search for boosted Higgs boson in the SM}

The possibility to search for the Standard Model Higgs boson in the decay to $b\bar{b}$ at the LHC using jet substructure techniques gave the field of jet physics a tremendous boost \cite{Butterworth:2008iy}. The projected sensitivity for a discovery of the Higgs boson with only $\sim 30~\mathrm{fb}^{-1}$, however, requires a centre-of-mass energy of 14 TeV for LHC proton-proton collisions. Due to technical issues of the LHC to reach its design energy of 14 TeV during Runs I and~II, the decay of a Higgs boson into a $b\bar{b}$-pair was never a contender to contribute to its discovery. Still, the measurement of the Higgs boson coupling to bottom quarks, while notoriously difficult, is of crucial importance as it is a dominant contributor to the total width of the Higgs boson, which in turn affects the branching ratios of all available decay modes. 

\begin{figure}
  \includegraphics[width=0.45\textwidth]{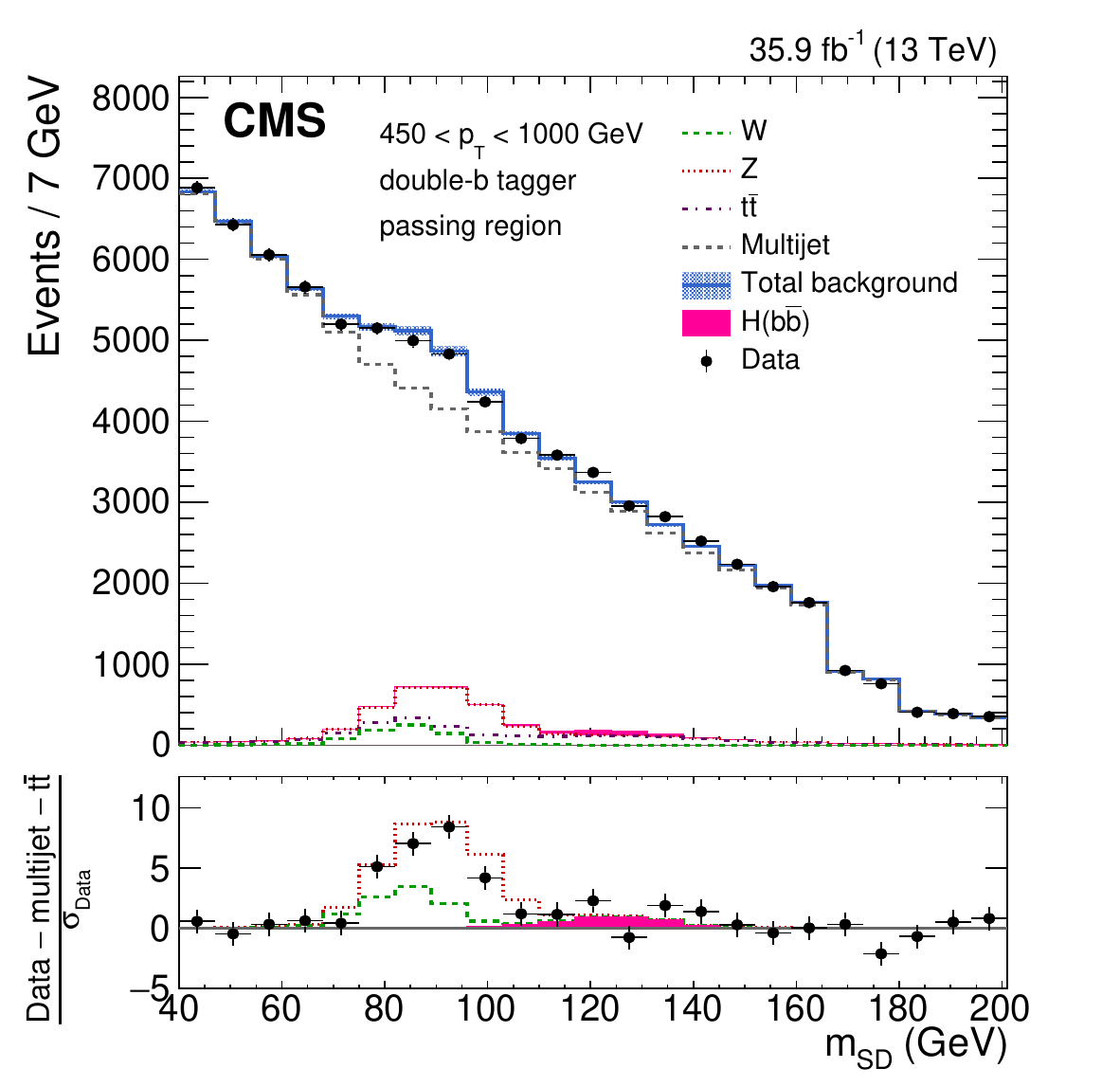}
    \includegraphics[width=0.55\textwidth]{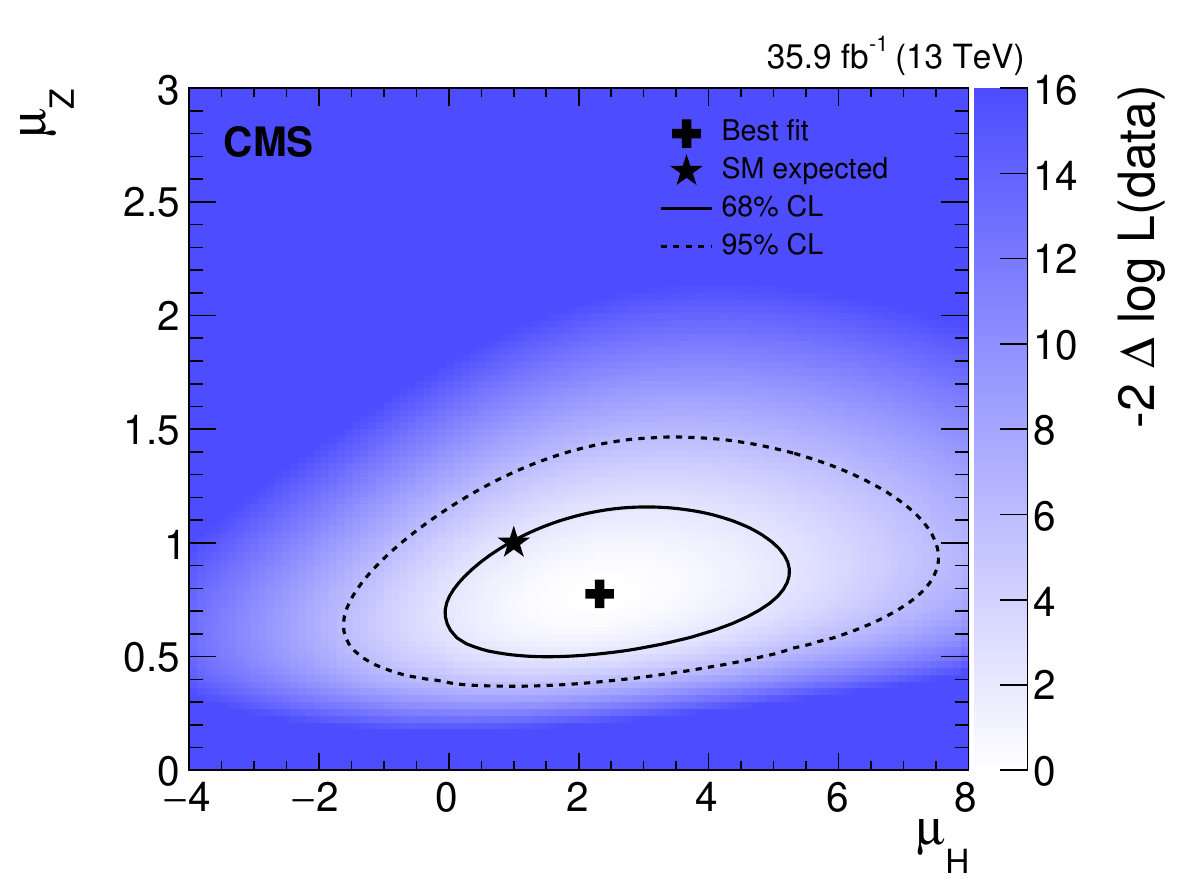} \\
  \includegraphics[width=0.45\textwidth]{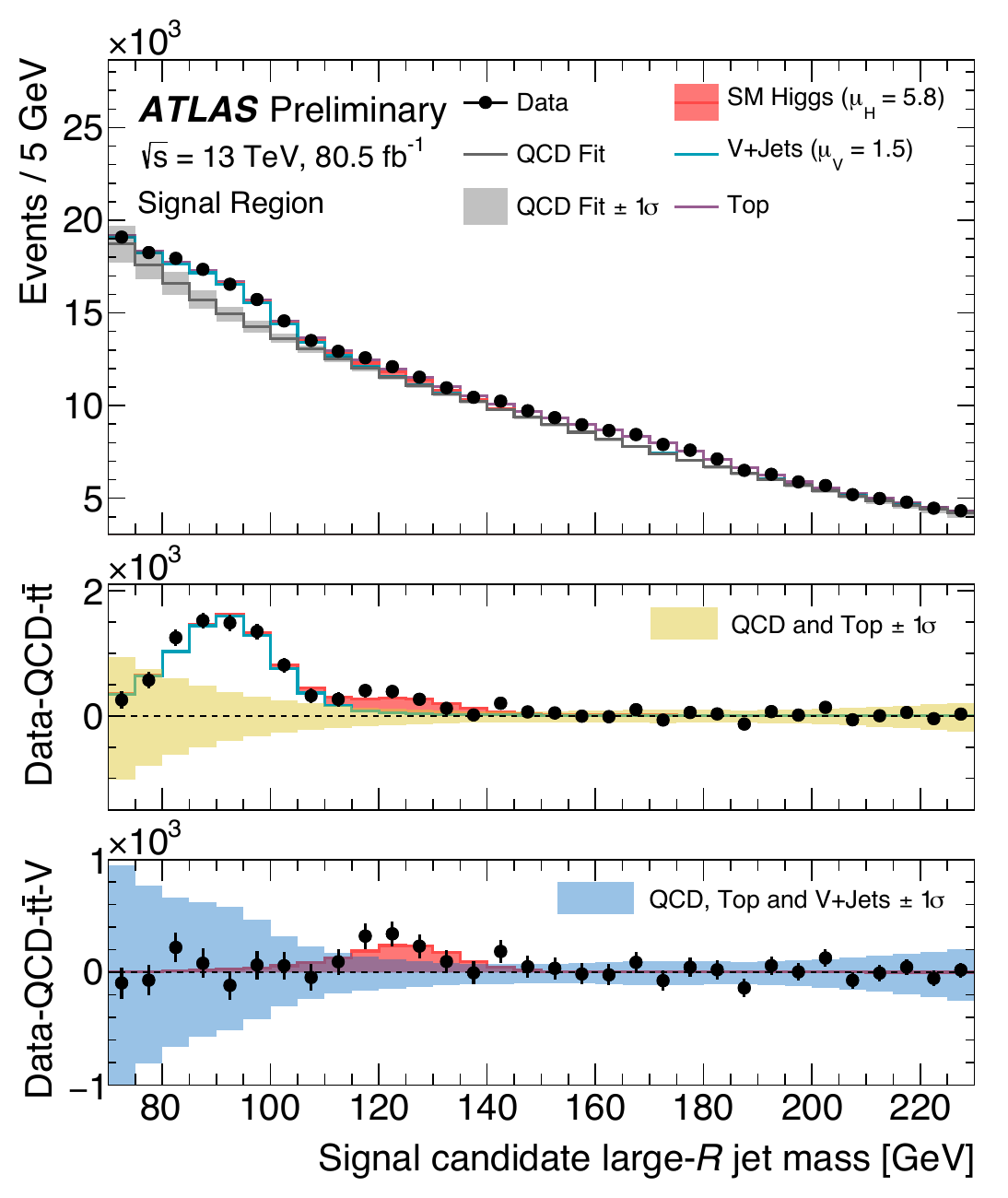}
    \includegraphics[width=0.55\textwidth]{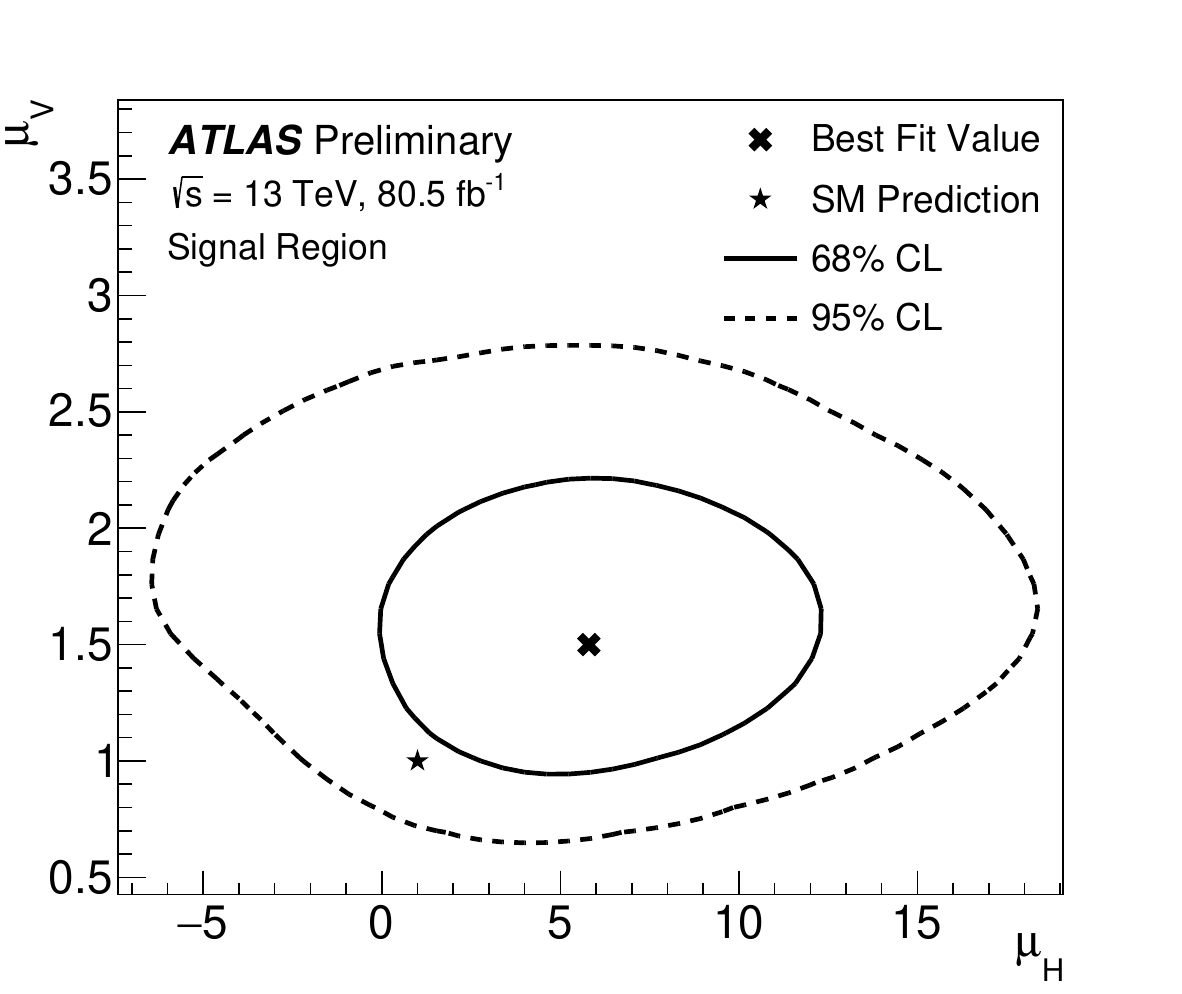}       
  \caption{Searches for boosted Higgs boson decaying into $b \bar b$. The plots show the invariant mass distribution and the signal-strength modification, top for CMS~\cite{Sirunyan:2017dgc}, bottom for ATLAS~\cite{ATLAS:2018hzj}.  }\label{fig:cms_boostedH}
\end{figure}

CMS performed an inclusive search for a Higgs boson decaying to $b\bar{b}$ pair, which is expected to result in an anti-$k_t$ $R=0.8$ jet, with $p_t \geq 450$ GeV \cite{Sirunyan:2017dgc}. The main experimental challenge originates in the large cross section for background multijet events at low jet mass. To increase the sensitivity for the reconstruction of the Z and Higgs boson \SD grooming is applied to the jet before two- and three-point generalised energy correlation functions are exploited to determine how consistent a jet is with having a two-prong structure. While a peak is clearly visible for the reconstruction of the Z boson, Fig.~\ref{fig:cms_boostedH} (top left) shows that the sensitivity to the Higgs boson still remains weak. However, it is already possible to set a limit on large signal-strength modifications to the production of either resonance, see Fig.~\ref{fig:cms_boostedH} (top right).

ATLAS has provided a similar measurement with an increased data set of $\mathcal{L} = 80.5~\mathrm{fb}^{-1}$ \cite{ATLAS:2018hzj}. To select the event, an anti-$k_t$ $R=1$ fat jet with $p_t \geq 480$ GeV is required. ATLAS is not showing the soft-drop groomed mass of the fat jet, but the invariant mass of trimmed jets. After subtracting the rather large QCD background a clear excess around the Higgs mass of $m_\text{H}=125$ GeV is observed, see Fig.~\ref{fig:cms_boostedH} (lower left panel). Small excesses around the Higgs and Z masses are indicative of an enhanced signal strength compared to the Standard Model predicted cross sections. Thus, ATLAS central value for the fit, allowing the signal strength for the V+jets and H+jets independently to float, is above the Standard Model value for either process, see Fig.~\ref{fig:cms_boostedH} (lower right panel). Yet, ATLAS and CMS 95\% exclusion contours both still contain the Standard Model value.

\section{Searches for new physics}

The kinematic situation outlined at the beginning of this book, cf Fig.~\ref{fig:resonance}, is common to many scenarios where the Standard Model is extended by heavy degrees of freedom. If such degrees of freedom descend from a model that addresses the hierarchy problem of the Higgs boson, they are likely to couple to the top quark and the bosons of the electroweak sector of the Standard Model, which in turn have a large branching ratio into jets. The conversion of energy from the heavy particle's rest mass into kinetic energy of the much lighter electroweak resonances causes them to be boosted in the lab frame. Thus, searches for new physics using jet substructure methods applied to fat jets can be amongst the most sensitive ways to probe new physics.

\subsection{Resonance decays into top quarks}
Top-tagging is the most active playground for the development of jet substructure classification techniques. A top jet has a rich substructure, providing several handles to discriminate it from the large QCD backgrounds, and due to the top quark's short lifetime its dynamics are to a large degree governed by perturbative physics. Thus, ATLAS and CMS have performed searches using a large variety of top-reconstruction techniques. 

New physics scenarios that are the focus of ATLAS and CMS searches contain models with extra dimensions or extended gauge groups, which give rise to heavy $\text{Z}'$ bosons, Kaluza-Klein gluons $g_{KK}$ and spin-2 Kaluza-Klein gravitons $G_{KK}$. The hadronic activity, and hence the tagging efficiency, depends on the quantum numbers of the heavy decaying resonance, in particular its colour charge~\cite{Joshi:2012pu}. These three resonances provide interesting benchmark points which can arise in many classes of new physics models. 

While ATLAS \cite{Aaboud:2018mjh} separates between a resolved and a boosted analysis in the semi-leptonic top-decay channel, i.e.\ with one top decaying leptonically ($t\to b \nu l^+$) and the other one hadronically ($t \to b j j$), CMS \cite{Sirunyan:2018ryr} focuses on the boosted regime but also considers the dileptonic and purely hadronic top decay modes. For the purpose of these notes, we are mostly interested in the boosted semi-leptonic top-decay mode, which suffers less from large dijet backgrounds, yet providing a larger signal cross section than the dileptonic channel.
ATLAS varies the resonance masses for the colour-singlet and colour-octet bosons with spin 1 or spin 2 between $0.4$ to 5 TeV and respectively their width between $1\%$ and $30\%$. To reconstruct the hadronic top, a large-$R$ jet is formed using the anti-$k_t$ algorithm with radius parameter $R=1.0$. This jet is trimmed to mitigate the effects of pileup and underlying event, using $R_\mathrm{sub} = 0.2$ and $f_\mathrm{cut}=0.05$. The resulting jets are required to have $p_t> 300$ GeV and $|\eta| < 2.0$. Such jets are then identified as top-tagged using the $N$-subjettiness ratio $\tau_{32}$ and an algorithm based on the invariant mass of the jet. The signal efficiency for this algorithm is found to be $80\%$. 

\begin{figure}[t]
  \includegraphics[width=0.45\textwidth]{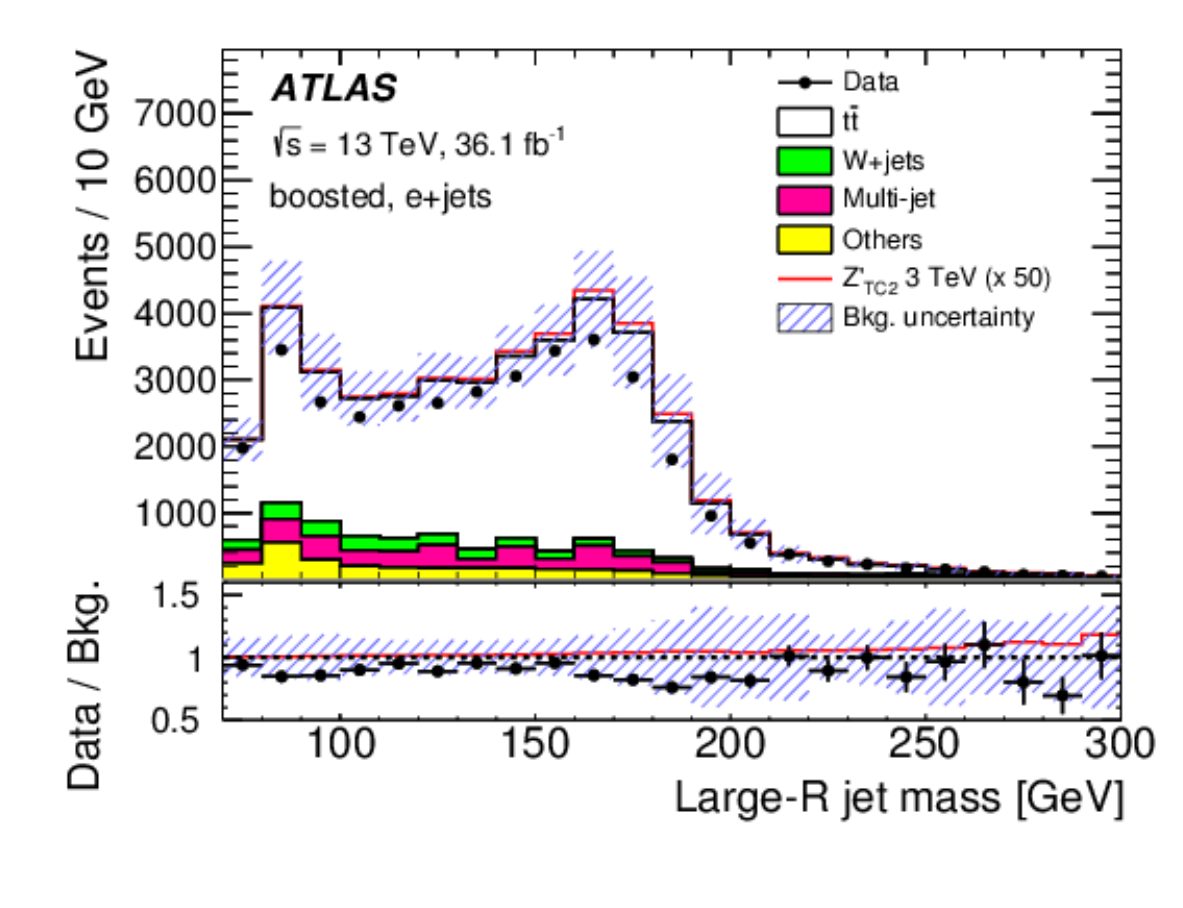}
  \includegraphics[width=0.5\textwidth]{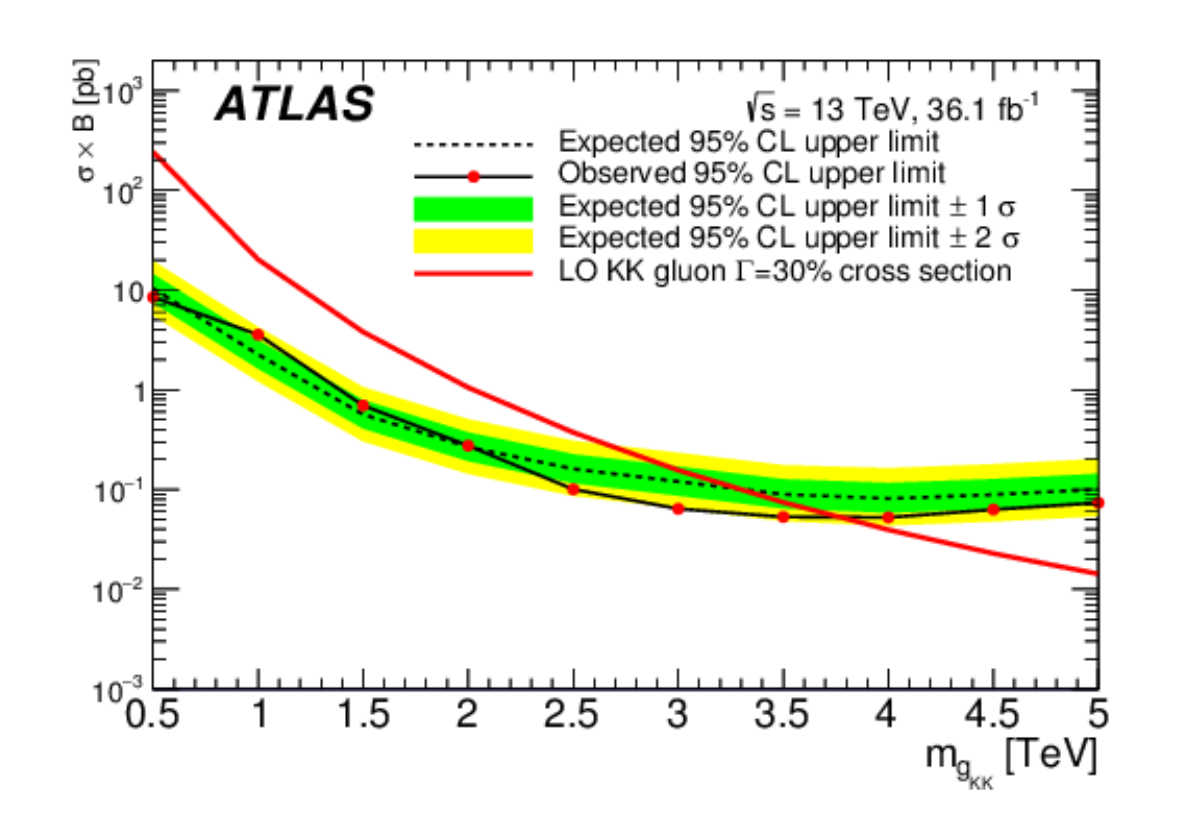} \\
  \includegraphics[width=0.45\textwidth]{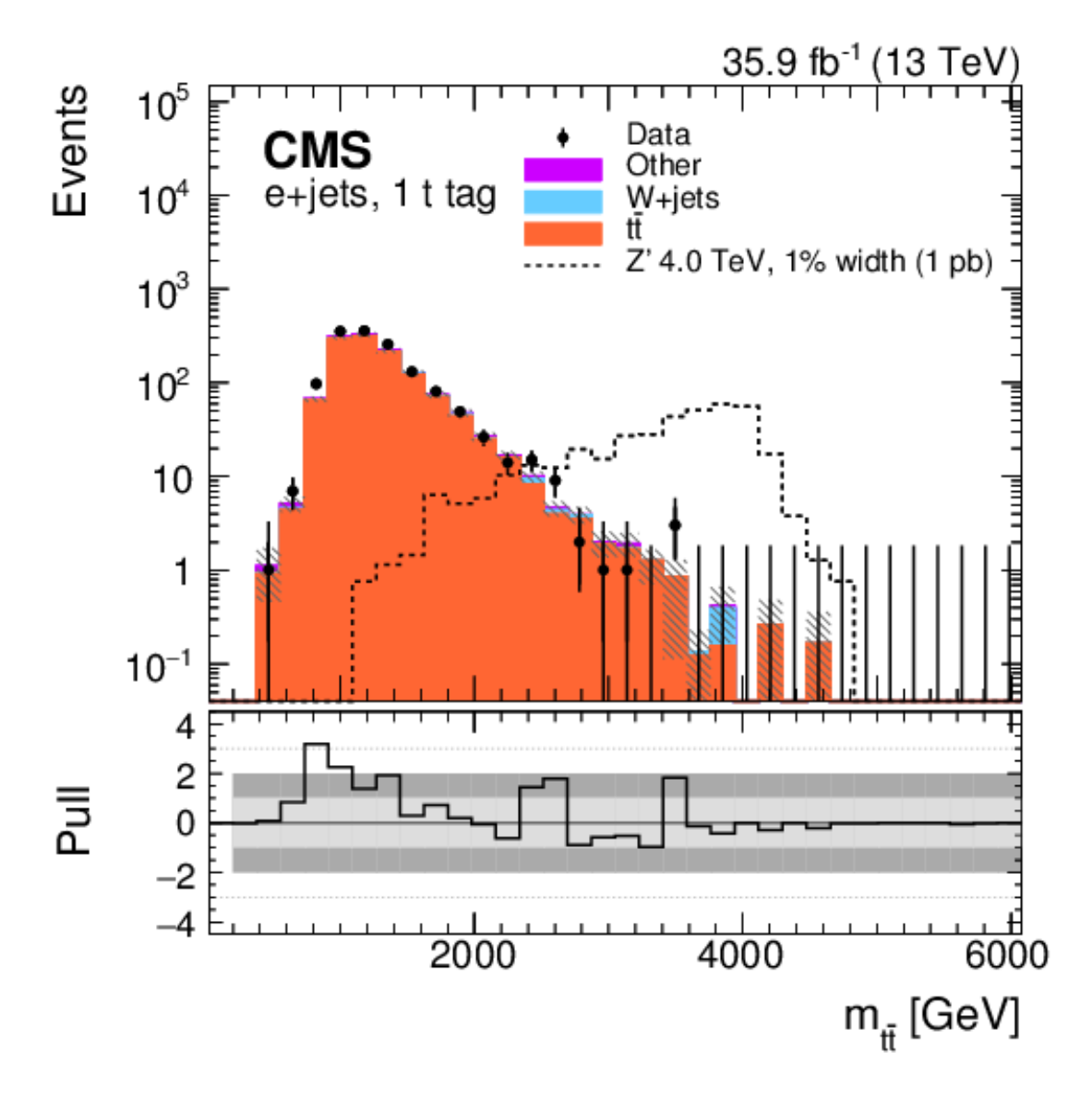}
  \includegraphics[width=0.5\textwidth]{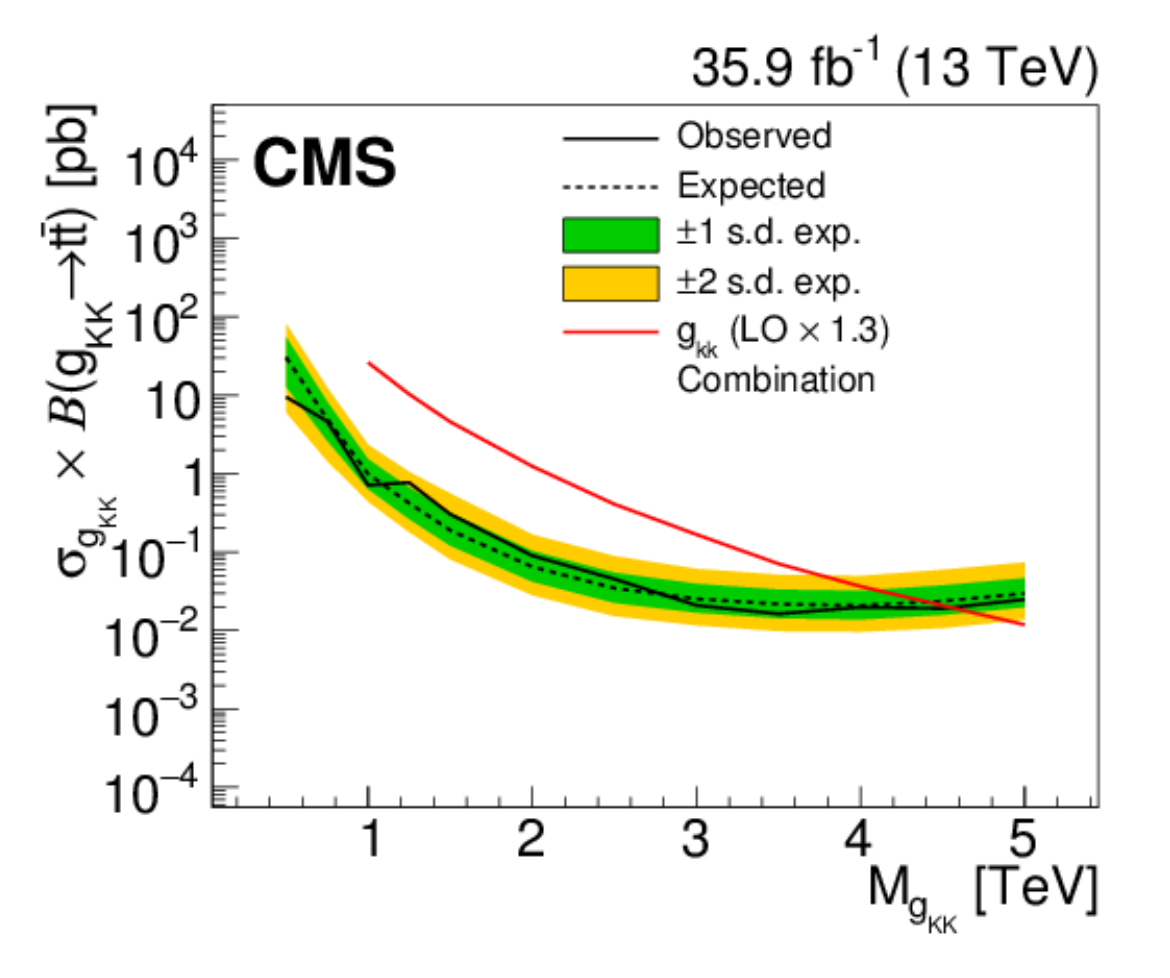} 
  \caption{Searches for heavy resonances that involve top tagging from ATLAS~\cite{Aaboud:2018mjh}, on the top, and CMS~\cite{Sirunyan:2018ryr} on the bottom. }\label{fig:res_ttdecays}
\end{figure}

For the same task CMS uses somewhat smaller anti-$k_t$ $R=0.8$ jets. These jets receive pileup per particle identification (PUPPI) \cite{Bertolini:2014bba} corrections. The top tagging algorithm then only considers jets with $p_t > 400$ GeV to ensure a collimated decay of the top quark. 
The top tagging algorithm then includes a grooming step, performed with \SD $\beta=0$, i.e.\ mMDT, algorithm, with $z_\mathrm{cut} = 0.1$ and $R_0 = 0.8$ and a cut on the $N$-subjettiness ratio $\tau_{32}$. The \SD mass is then required to be close enough to the true top mass, i.e. $105 < m_{\mathrm{SD}} < 210$ GeV and $\tau_{32}$ must be less than 0.65. 

The reconstruction techniques applied show a very good agreement between the measured data and the Monte-Carlo predicted pseudo-data, Fig.~\ref{fig:res_ttdecays} (left panels). With only $36~\mathrm{fb}^{-1}$, depending on the resonance's couplings and width, heavy resonances decaying into top quarks can be excluded up to mass of 3.5 TeV, Fig.~\ref{fig:res_ttdecays} (right panels). For such large masses jet-substructure methods are not optional. Without using the internal structure of jets the QCD-induced dijet backgrounds would overwhelm the signal.

In models where the $\text{Z}'$ arises from to a SU(N) gauge group, it will be accompanied by a $\text{W}'$. ATLAS \cite{Aaboud:2018juj} has performed searches for heavy $\text{W}'$ decaying into a hadronic top and a bottom quark, i.e. $\text{W}' \to t  \bar{b} \to q \bar{q} b \bar{b}$. This search is somewhat more intricate than the searches for decays into two top quarks as there are fewer handles to suppress the backgrounds. Thus, ATLAS uses the shower deconstruction top tagging algorithm, discussed in Sec.~\ref{shower_dec}, which has a strong rejection power of QCD jets while maintaining a large signal efficiency. ATLAS finds a working point of the tagger with $50\%$ signal efficiency and a background rejection factor of 80, thus improving the signal-to-background ratio by a factor 40, for anti-$k_t$ $R=1.0$ jets with $p_t>450$ GeV. 
\begin{figure}
  \includegraphics[width=0.4\textwidth]{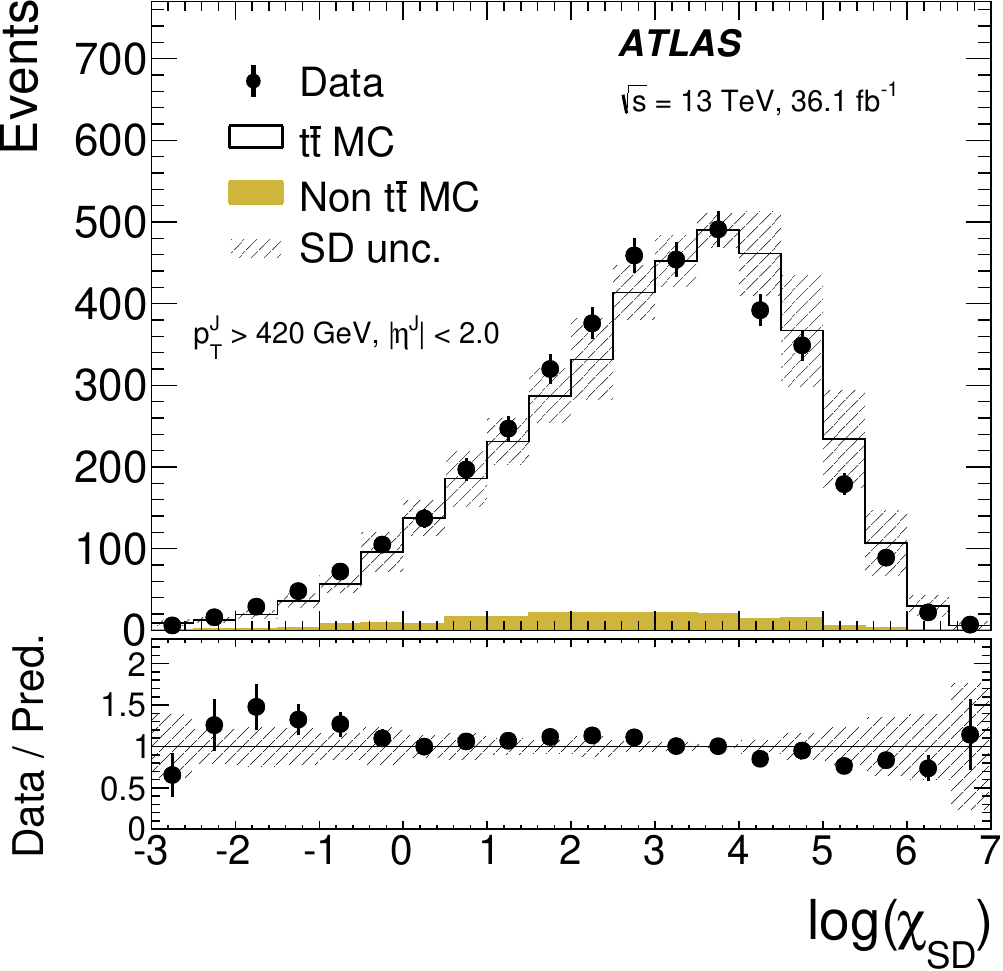} 
  \includegraphics[width=0.6\textwidth]{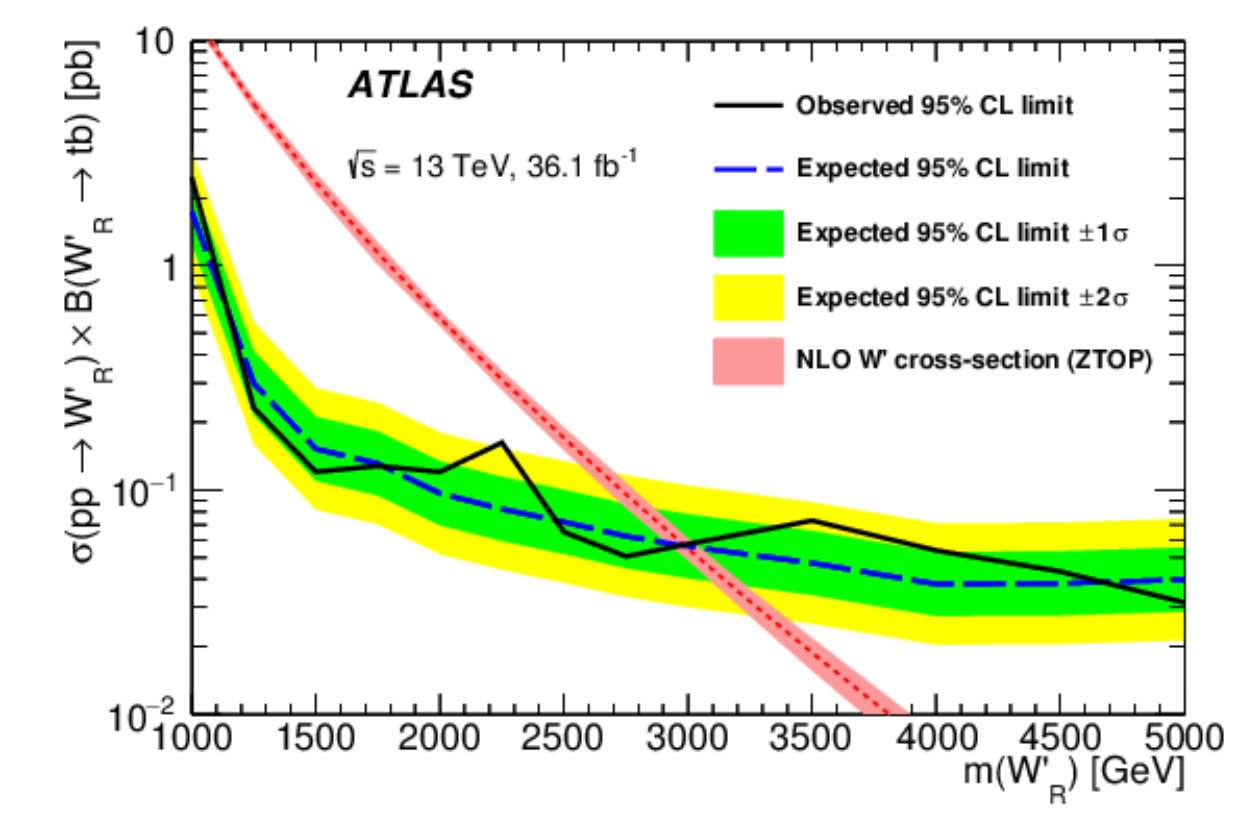}
  \caption{Search for a $\text{W}'$ performed by ATLAS~\cite{Aaboud:2018juj} using shower deconstruction.}\label{fig:Wprime_excl}
\end{figure}
Applied to $t\bar{t}$ events the shower deconstruction algorithm shows very good agreement between data and Monte-Carlo simulated pseudo-data, Fig.~\ref{fig:Wprime_excl} (left). After applying the top tagger and reconstructing the final state in the search for a $\text{W}'$, ATLAS can exclude masses up to 3 TeV, Fig.~\ref{fig:Wprime_excl} (right). The sensitivity does not depend on whether the $\text{W}'$ couples to left or right-handed quarks.

\subsection{Resonance decays into Higgs and gauge bosons}
Currently, most of the resonance searches into electroweak bosons that are using jet substructure techniques are focusing either on heavy resonances decaying into Higgs bosons, with subsequent decay into bottom quarks \cite{Aad:2015uka, Sirunyan:2017isc, Sirunyan:2018qca}, or decays into W and Z bosons, with subsequent decay into quarks \cite{Aad:2015owa, Khachatryan:2014hpa}.
The new physics scenarios studied range from the decay of a Kaluza-Klein excitation of the graviton in the bulk Randall-Sundrum model with a warped extra dimension \cite{Agashe:2007zd}, over the decay of a $CP$-even heavy Higgs boson, as present in two-Higgs double models (2HDM) \cite{Branco:2011iw}, to a heavy scalar from a triplet-Higgs model, e.g.\ the so-called Georgi-Machacek model \cite{Georgi:1985nv}. In general one assumes a heavy resonance in the range $m_X \gtrsim 500$ GeV is produced, with a short lifetime. While the spin of the resonance could be in principle studied by reconstructing and analysing the decay planes of the quark pairs \cite{Hackstein:2010wk, Englert:2010ud}, at this point such attempts are not being made and the separation between signal and background, after reconstructing the electroweak bosons, relies entirely on the presence of a bump in their invariant mass distribution. Thus, the width in combination with the mass of the decaying resonance is of great importance for its discovery or exclusion.

\begin{figure}
  \includegraphics[width=0.48\textwidth,trim=0 20 0 40]{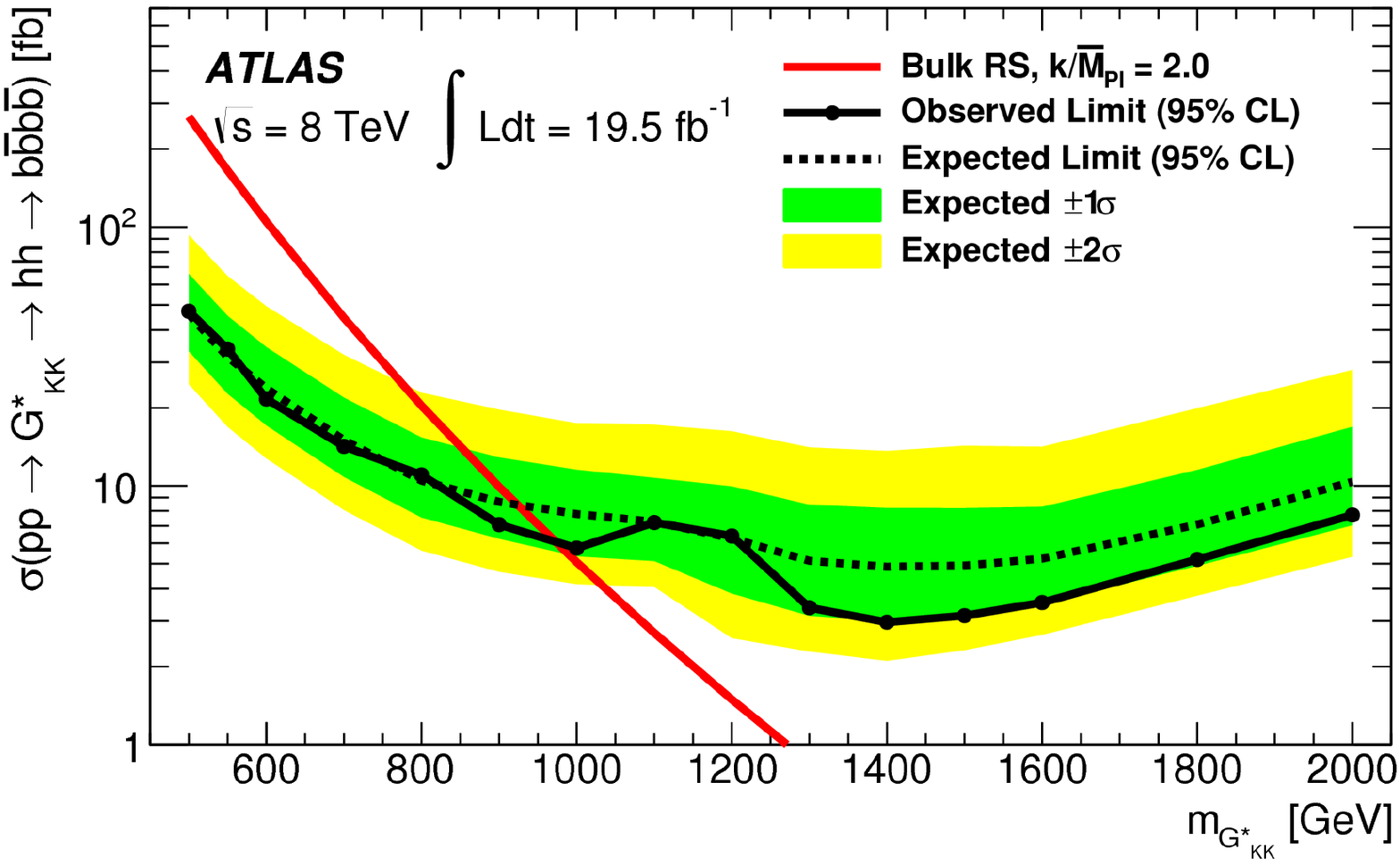}\hfill
  \includegraphics[width=0.45\textwidth,trim=0 37 0 20]{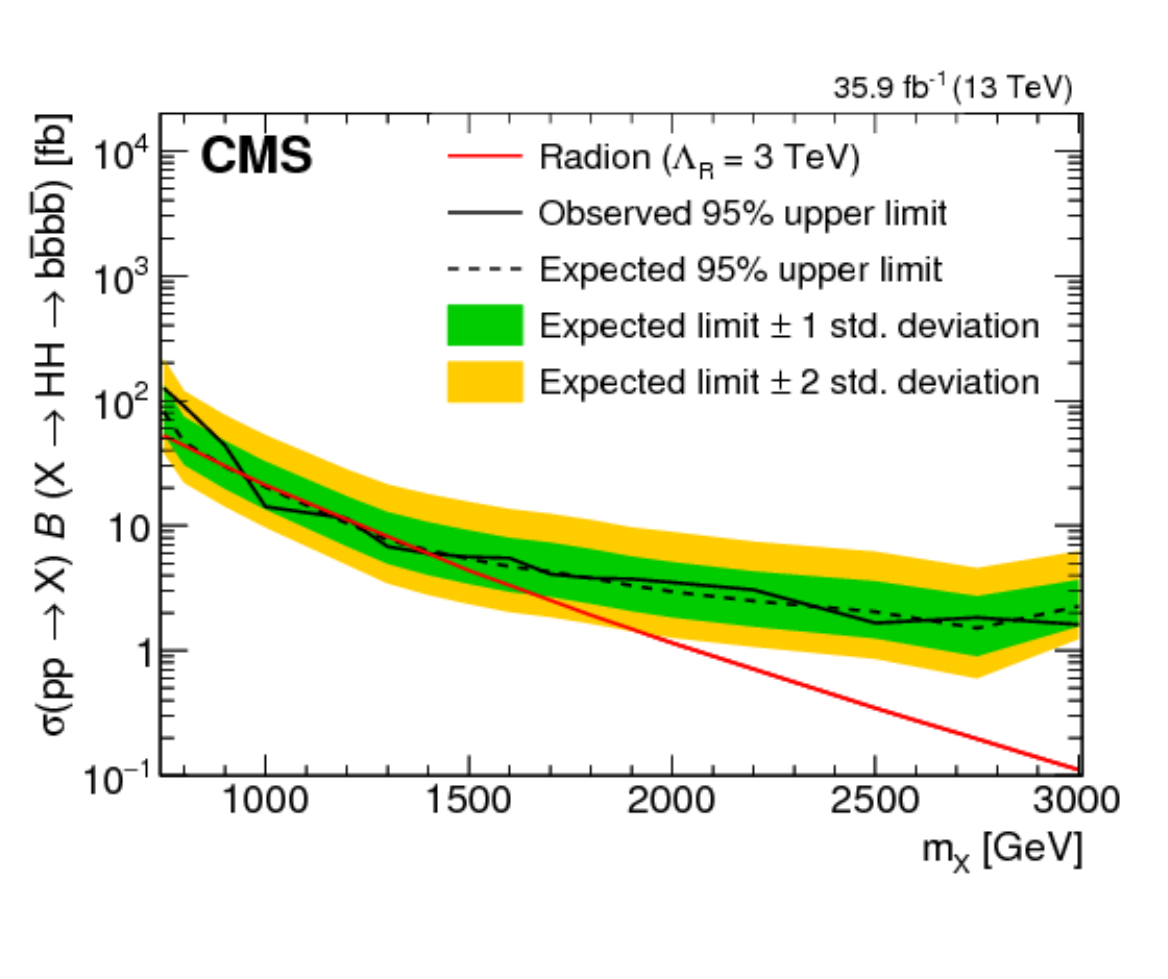} 
  \caption{Example of exclusions limits from searches of resonances decaying into two Higgs bosons, the ATLAS search of Ref.~\cite{Aad:2015uka} is shown on the left, while the CMS search of Ref.~\cite{Sirunyan:2018qca} on the right. }\label{fig:X_HH}
\end{figure}

To reconstruct the Higgs boson pairs from a heavy resonance decay ATLAS \cite{Aad:2015uka, ATLAS:2016ixk} considers a resolved and a boosted analysis. In the boosted case, Higgs bosons are selected by requiring that two large-$R$, i.e. anti-$k_t$ $R=1.0$ with $p_t \geq 250$ GeV, have each two $b$-tags, and that the leading fat jet has in addition $p_t \geq 350$ GeV. The backgrounds are derived from Monte-Carlo simulations. For a resonance width of $\Gamma = 1$ GeV, in combination with the resolved analysis, this results in an 95\% C.L. exclusion for a bulk Randall-Sundrum graviton with coupling value $k/\tilde{M}_{PL}=2$, where $\tilde{M}_{\mathrm{PL}}$ is the reduced Planck mass, of $500 \leq m_{G^*_{KK}} \leq 990$ GeV, see Fig.~\ref{fig:X_HH} (left). With the integrated luminosity used in this analysis, the resolved analysis is more sensitive than the boosted analysis up to $m_{G^*_{KK}} \leq 1100$ GeV.

CMS \cite{Sirunyan:2018qca} in a search at $\sqrt{s} = 13~\mathrm{TeV}$ with 35.9 $\mathrm{fb}^{-1}$ aims for the exclusion of heavier masses and instead uses anti-$k_t$ $R=0.8$ jets with $p_t \geq 350$ GeV. The resulting fat jet is groomed using the \SD algorithm with $z=0.1$ and $\beta=0$ (mMDT). The groomed jet mass is required to have $105 \leq m_{\mathrm{sd}} \leq 135$ GeV. To suppress QCD backgrounds further, the N-subjettiness algorithm is used, requiring $\tau_{21} < 0.55$. Eventually, each of the jets is double-$b$ tagged, which has the largest impact on the backgrounds. After searching for a bump in the invariant mass spectrum of the two fat jets, CMS obtains a 95\% C.L. exclusion for a bulk radion with mass $970 < m_R < 1400$ GeV, see Fig.~\ref{fig:X_HH} (right).

\begin{figure}
  \includegraphics[width=0.45\textwidth]{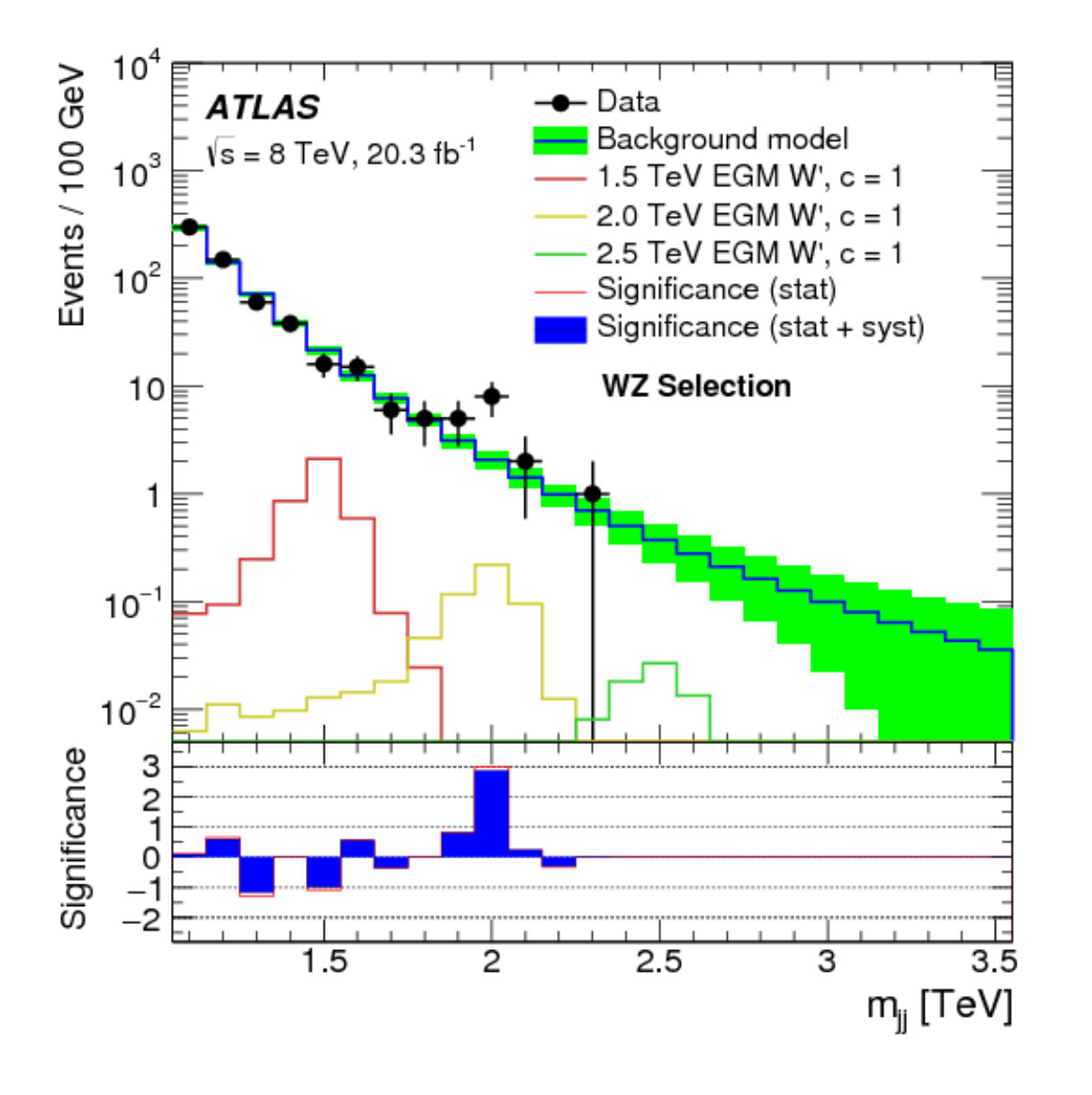} 
  \includegraphics[width=0.45\textwidth]{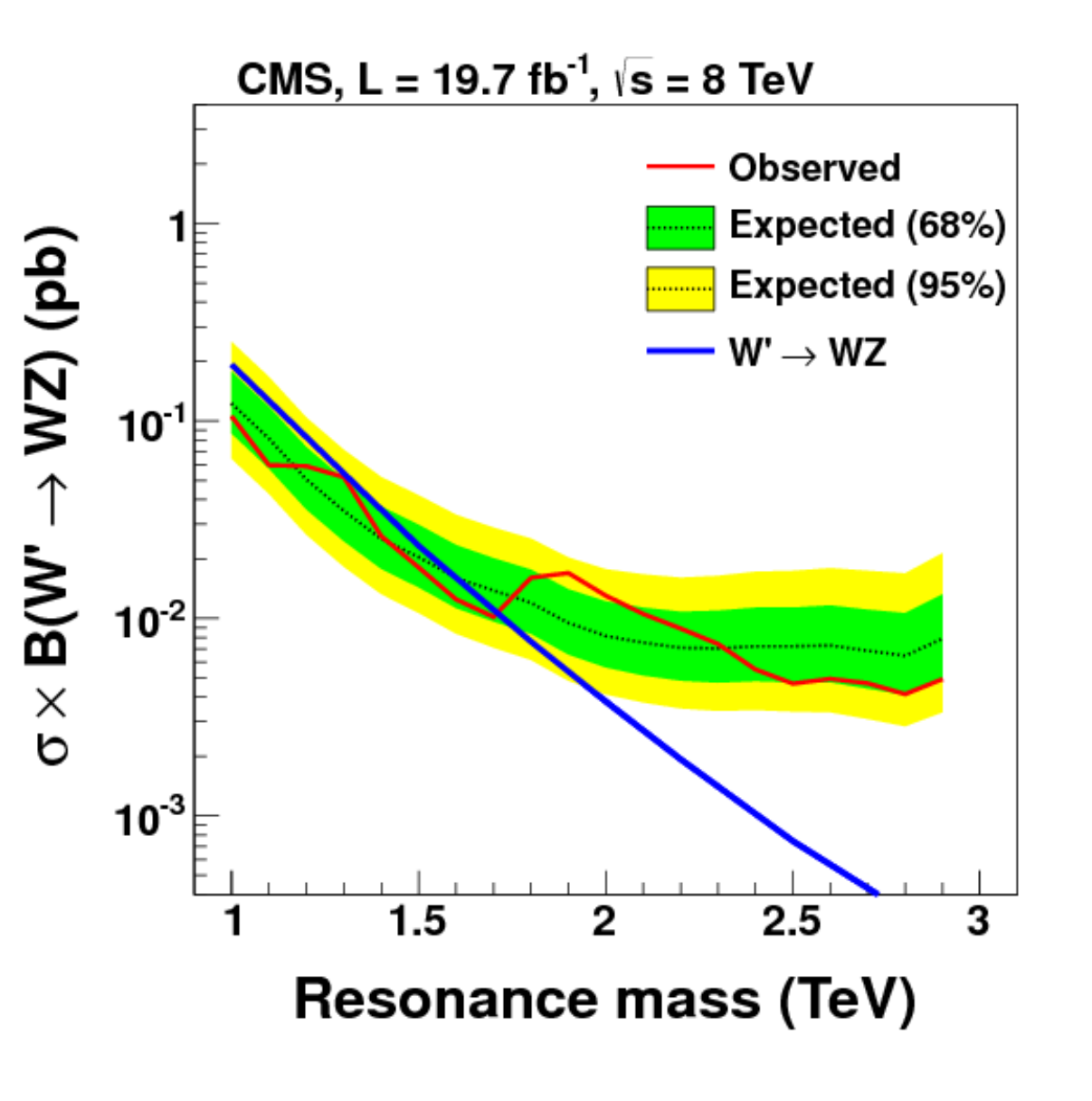} 
  \caption{Invariant mass distribution of the reconstructed gauge bosons as measured by ATLAS \cite{Aad:2015owa}, on the left, and the exclusion limit for a $\text{W}'$ decaying into WZ set by CMS \cite{Khachatryan:2014hpa} on the right.}\label{fig:X_VV}
\end{figure}

The decay of a heavy resonance into gauge bosons is a frequent feature of many extensions of the Standard Model. For example, the aforementioned bulk graviton could decay into a pair of W or Z bosons, or a heavy gauge boson of an additional or extended gauge group, a so-called W$'$ or Z$'$, can decay into the pairs of Standard Model gauge bosons. In \cite{Aad:2015owa, Khachatryan:2014hpa} ATLAS and CMS have both observed a small excess in dijet final states, where each jet was W/Z tagged. The excess resided at a similar invariant mass range of $m_{jj} \sim 2$ TeV, but was slightly more significant in the ATLAS analysis.
In this search, at $\sqrt{s}=8$ TeV, ATLAS selected two fat jets with Cambridge/Aachen algorithm $R = 1.2$, with a minimal transverse momentum of $p_t  \geq 540$ GeV. The reconstruction of the gauge bosons relied on a combination of a jet-mass cut around the masses of the weak gauge bosons and grooming techniques, where a modified version of the BDRS reconstruction technique \cite{Butterworth:2008iy} was employed\footnote{As discussed in \cite{Krauss:2014yaa}, the way the BDRS approach was modified in the search by ATLAS could result in shaping the $m_{VV}$ distribution in the region of 2 TeV, where the excess was observed.}, a method initially designed for the reconstruction of a Higgs boson with $p_{t,\text{H}} \geq 200$ GeV. Using this approach, the mass resolution of the reconstructed gauge boson is not good enough to discriminate between W and Z bosons. Eventually, to improve on the separation of signal and background, cuts were applied on the momentum ratios of subjets, the number of charged particles within a subjet and the mass of the reconstructed gauge bosons. After recombining the four-momenta of the two reconstructed gauge bosons an excess was observed in the mass range $1.9 \leq m_\text{VV} \leq 2.1$ TeV over the data-driven (fitted) background estimate, mainly driven by the QCD background, see Fig.~\ref{fig:X_VV}.

CMS reconstructed the W and Z bosons by applying the pruning algorithm as a groomer and tagger for the fat jet. To further improve the separation between $W/Z$ bosons and QCD jets $\tau_{21}$ was used. Both experiments find an excess at $\sim 1.9$ TeV, whereas the excess in ATLAS with $2.8 \sigma$ is more pronounced than in CMS with $1.8 \sigma$. Both experiments have updated this search using different reconstruction strategies and with more statistics, which eventually dampened the excess strongly \cite{Aaboud:2017eta, Sirunyan:2017acf}.

\subsection{Resonance decays into new particles}

If a heavy resonance decays via electroweak resonances of the Standard Model into quarks or gluons the masses of the Standard Model particles themselves provide an important handle to separate signal from background. This task is however complicated if the intermediate resonances are not known, e.g.\ when a squark decays into four quarks through an intermediate Higgsino with a hadronic R-parity-violating coupling \cite{Butterworth:2009qa}. CMS has performed searches for squarks via such a decay mode \cite{CMS:2018sek} and ATLAS a similar study in \cite{Aad:2016kww}. 

\begin{figure}
  \includegraphics[width=0.48\textwidth]{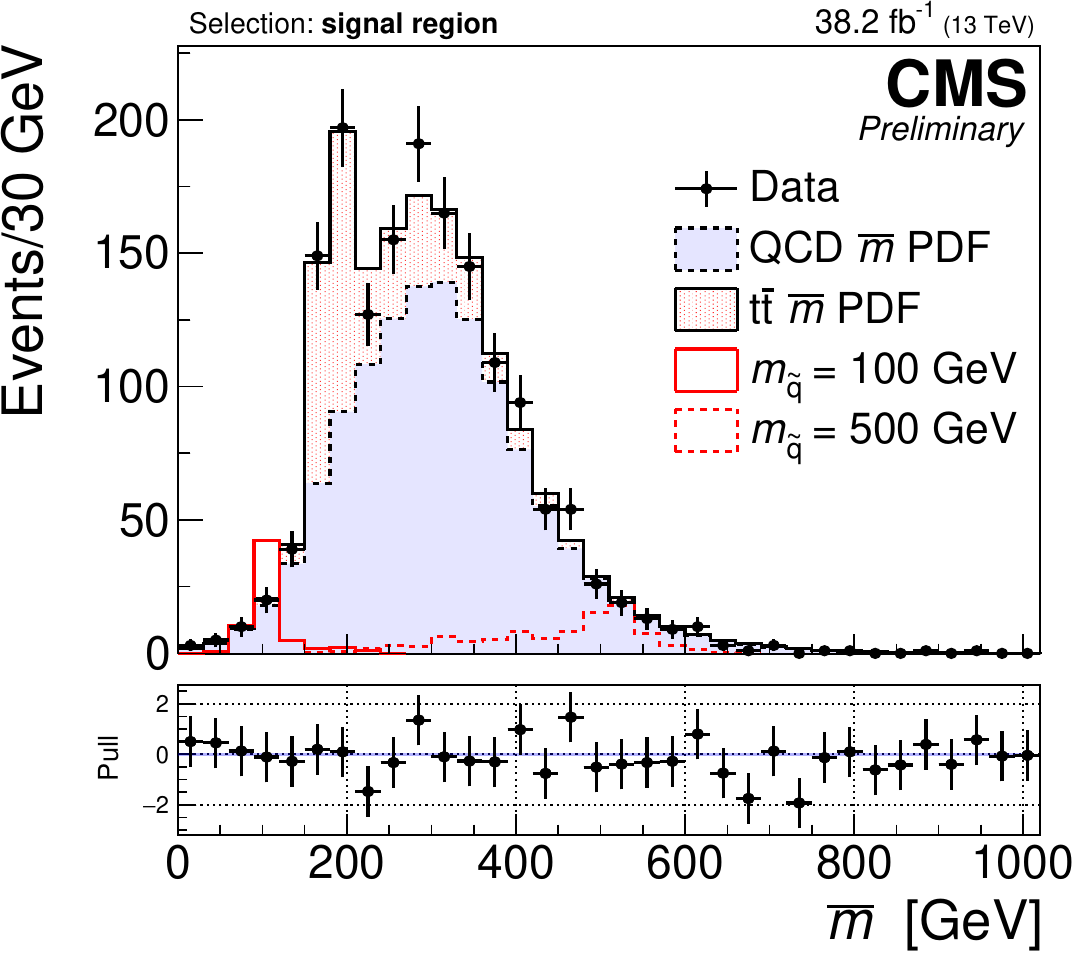} \hfill
  \includegraphics[width=0.47\textwidth]{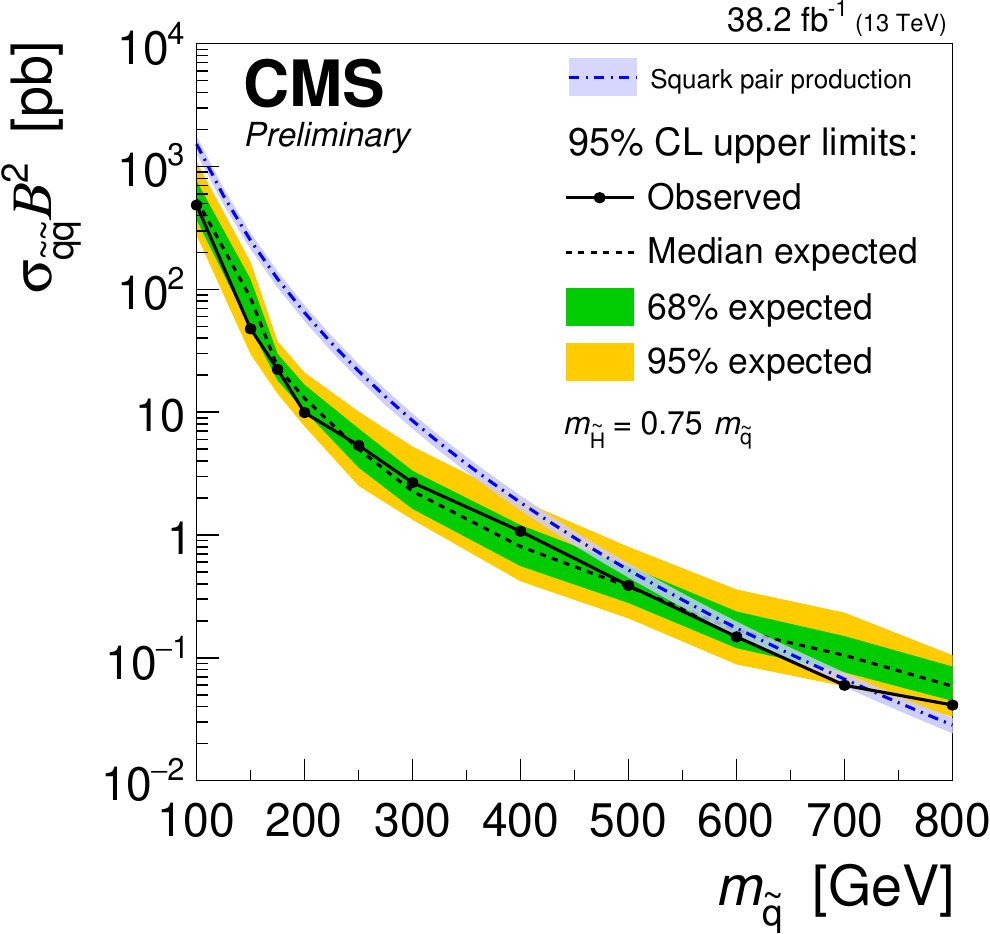} 
  \caption{Search for a squark decaying into four quarks (via a Higgsino) performed by CMS \cite{CMS:2018sek}. The plot on the right show the average invariant mass of the fat jets and the exclusion limits.}\label{fig:X_search}
\end{figure}

Without knowing the mass of the Higgsino, CMS in \cite{CMS:2018sek} uses the fact that a squark decays into a four-pronged object to discriminate signal from background. In order to capture as many of the final-state constituents of the squark decay products as possible, large Cambridge/Aachen ($R=1.2$) jets are formed. These jets are analysed using the $N$-subjettiness ratios, requiring $\tau_{43}<0.8$ and $\tau_{42}<0.5$ for each jet. As pair production of the squarks is assumed, the masses of the reconstructed fat jets should not be too asymmetric, i.e. $|m_1 - m_2| / (m_1+m_2) < 0.1$. After defining the average jet mass as $\bar{m} = (m_1+m_2)/2$ CMS finds very good agreement between the theoretically predicted background cross sections and the measured data, see Fig.~\ref{fig:X_search} (left). This allows to set an exclusion limit in this channel requiring squark masses to be $m_{\tilde{q}}  > 720$ GeV, when assuming that the Higgsino mass is $m_{\tilde{H}} = 0.75 m_{\tilde{q}}$, Fig.~\ref{fig:X_search} (right).

In \cite{CMS:2024nsz}, CMS searches model-agnostically for narrow dijet resonances with anomalous substructure, using data at $\sqrt{s} = 13$ TeV with an integrated luminosity of $138~\text{fb}^{-1}$. The analysis targets mass ranges between $1.8$ and $6$ TeV and employs machine learning-based anomaly detection methods, including unsupervised, weakly supervised, and semi-supervised approaches. These methods aim to detect resonances whose jet substructures deviate from quantum chromodynamics (QCD) expectations.
The benchmark signal models explored in this work feature the decay topology $A \to BC$, where $A$ is a heavy resonance decaying into boosted resonances $B$ and $C$, each producing multiple subjets. Example scenarios include:
\begin{itemize}
    \item $W' \to B't \to bZt$ (3+3 prongs),
    \item $G_\text{KK} \to HH \to 4t$ (6+6 prongs).
\end{itemize}
The five anomaly detection methods enhance sensitivity by utilising substructure features such as N-subjettiness and soft-drop mass. The analysis fits the dijet mass spectra using a smoothly falling background model and a double Crystal Ball function for potential signal peaks. No significant excesses above the SM background were observed, with the largest local significance being $2.9\sigma$.
Compared to traditional methods, anomaly detection improves sensitivity by factors up to $7.1$ for some benchmark signals. For example, the VAE-QR method achieved a five-sigma discovery potential at significantly lower cross sections for signals with complex jet substructure. These methods also broaden the search coverage for previously unexplored resonances.

Semi-visible jets are known to pose a challenge for reconstruction methods. \cite{ATLAS:2023swa} presents the first search for semi-visible jets in the $t$-channel production mode, utilising $139~\text{fb}^{-1}$ of data at $\sqrt{s} = 13~\text{TeV}$ collected by ATLAS. SVJs arise in strongly interacting dark sector models, where dark quarks undergo hadronisation, producing jets composed of stable and unstable dark hadrons alongside SM hadrons. This analysis explores mediator masses between $1$ and $5~\text{TeV}$ and various fractions ($R_{\text{inv}}$) of stable dark hadrons.
The search focuses on events with significant missing transverse momentum ($E_T^{\text{miss}} > 600~\text{GeV}$) and a scalar sum of jet transverse momenta ($H_T > 600~\text{GeV}$). Events are required to have at least two jets, with the leading jet having $p_t > 250~\text{GeV}$. Background contributions include $W/Z$+jets, top-quark, diboson, and multijet processes. Control regions are defined to constrain background uncertainties using events with leptons and $b$-jets. Based on observables such as the azimuthal separation between jets and the transverse momentum balance, a simultaneous fit to both signal and control regions is performed. No significant excess over SM expectations is observed. The search sets 95\% confidence-level exclusion limits on the mediator mass, varying between $2.4$ and $2.7~\text{TeV}$ depending on $R_{\text{inv}}$. Limits are also placed on the coupling strength $\lambda$ between the mediator, SM quarks, and dark quarks. This search exemplifies the potential of collider searches for probing non-resonant dark sector phenomena and complements previous searches focused on $s$-channel production modes.

%% GS helper for auctex
%%% Local Variables:
%%% mode: latex
%%% TeX-master: "notes"
%%% End:

%  LocalWords:  topoclusters BDT DNN aplanarity HEPTopTagger squarks
%  LocalWords:  Supersymmetric trk tunable lego Powheg Kaluza
%  LocalWords:  calorimetry gravitons leptonically dileptonic PUPPI
%  LocalWords:  Sundrum Georgi Machacek Higgsino

% $Id: summary.tex 590 2026-02-27 08:19:59Z smarzani $
%
% Summary and discussions of our results
%------------------------------------------------------------------------
\chapter{Take-home messages and perspectives}\label{summary}

Since many facets and applications of jet substructure have been
covered in this book, it is useless to try and summarise them
all individually.
Instead, in this concluding chapter, we will briefly summarise the
main lessons we have learned from about a decade of jet
substructure studies and from the aspects covered in this book.

The first observation is that jet substructure has been a great 
success, both from a theoretical viewpoint and from an
experimental viewpoint. 
It took only a few years for the initial idea of
looking at the internal dynamics of jets to grow and develop a myriad
of new tools, opening doors to explore all sorts of new physics
domains.
Furthermore, as searches and measurements probe larger and larger
energy scales, boosted-object and jet substructure algorithms are
increasingly relied upon. In particular, at a possible future circular
hadron collider with $\sqrt{s}$ as large as 100~TeV, boosted jets
would be almost omnipresent.

In practice, jet substructure tools are rooted in the theory of strong
interactions. The first generation of substructure techniques were
designed based on core concepts and features of QCD: a QCD jet is
usually made of a single hard core accompanied with soft particles
corresponding to soft-gluon radiation, while boosted massive objects
decay into several hard prongs accompanied by further soft radiation
(at smaller angles if the initial particle is colourless).
Key techniques, many of which still in use today, have been developed
starting from these fundamental observations, allowing to establish
jet substructure as a powerful and promising field.
A few years later, the introduction of a new generation of
substructure tools was made possible by a better understanding of the
QCD dynamics inside jets using analytic techniques.
This first-principle approach has allowed for a more fine-grained
description of the underlying physics, which seeded either simpler
and cleaner tools (e.g.\ the modified MassDropTagger and \SD) or
tools with improved performance (e.g.\ the $D_2$ energy-correlation
functions or dichroic ratios), all under good  theoretical control. 

One of the key features repeatedly appearing when studying jet
substructure from first principles in QCD is the necessity of a
trade-off between performance and robustness. Here, by performance, we
mean the discriminating power of a tool when extracting a given signal
from the QCD background, and by robustness we mean the ability to
describe the tool from perturbative QCD, i.e.\ being as little
sensitive as possible to model-dependent effects such as
hadronisation, the Underlying Event, pileup or detector effects, all
of which likely translate into systematic uncertainties in an
experimental analysis.
This trade-off has been seen on multiple occasions throughout this
book.
When designing new substructure techniques, we therefore think that it
is helpful to keep in mind both these aspects.

In this context, it was realised that some tools like
\SD or the modified MassDropTagger are amenable to precise
calculations in perturbative QCD, while maintaining small
hadronisation and Underlying Event corrections.
This is particularly interesting since jet substructure tools are
often sensitive to a wide range of scales --- between the TeV scale
down to non-perturbative scales --- offering an almost unique
laboratory for QCD studies.
It has opened new avenues for future jet
substructure studies. 
 A typical example is a potential for
an extraction of the strong coupling constant from substructure
measurements (see e.g.\ Ref.~\cite{Bendavid:2018nar}), but other
options include the improvement of Monte Carlo parton showers,
measurements of the top mass, or simply a better control over QCD
background for new physics searches.

Another interesting research direction is the study of jets featuring heavy flavours, namely charm ($c$) and beauty ($b$), which are of interest for a variety of studies at the LHC.
On the one hand, they play a crucial role in studies of the Higgs boson. 
On the other hand, they provide us with many possibilities to study Standard Model and QCD effects, such as the heavy-quark component of the proton wave function.
Studies that focus on heavy-quark fragmentation and the formation of heavy-flavour jets are equally interesting.
One of the most fascinating QCD effects affecting the substructure of heavy-flavour jets is the so-called dead-cone effect~\cite{Dokshitzer:1991fd,Dokshitzer:1995ev}, i.e.\ the suppression of collinear radiation around heavy quarks,   
the first direct observation of which was recently reported by the ALICE collaboration~\cite{ALICE:2021aqk}, while indirect observations were also reported in the past~\cite{DELPHI:1992pnf, OPAL:1994cct,  OPAL:1995rqo, SLD:1999cuj, DELPHI:2000edu, ALEPH:2001pfo, ATLAS:2013uet}.

In this context, the recent development of IRC safe flavour-jet algorithms~\cite{Banfi:2006hf,Caletti:2022hnc,Czakon:2022wam,Gauld:2022lem,Caola:2023wpj} (to NNLO~\cite{Caletti:2022hnc} or to all orders~\cite{Czakon:2022wam,Gauld:2022lem,Caola:2023wpj}) provides us with the possibility of setting up a yet-unexplored flavour-jet substructure program at the LHC, see for a comparative study~\cite{Behring:2025ilo}.
From a theoretical point of view, calculations for identified heavy flavours can be performed essentially because the quark mass sets a perturbative scale for the running coupling and simultaneously removes collinear singularities. 
However, as opposed to the top quark, $b$ and $c$ quark form heavy-flavour hadrons. Therefore, the heavy-flavour production process can serve as a bridge between the perturbative and non-perturbative regimes of QCD.
From an experimental point of view, the lifetime of $B$ (or $D$) hadrons is long enough for their decay to occur away from the interaction point. Dedicated $b$- and $c$-tagging techniques that exploit this property to identify $B$ and $D$ hadrons or $b$ and $c$ jets are widely used in collider experiments, see e.g.~\cite{ATLAS:2017bcq, ATLAS:2018nnq}.

Resummed calculations for jets initiated by heavy quarks were first performed in the context of studies focusing on $B$-hadron decays~\cite{Aglietti:2006wh,Aglietti:2007bp,Aglietti:2008xn,Aglietti:2022rcm}
and top jets~\cite{Fleming:2007qr,Fleming:2007xt,Bachu:2020nqn,Jain:2008gb,Hoang:2019fze,Bris:2020uyb}, using the formalism of effective field theory.
More recently, a growing number of studies addressing the substructure of heavy-flavour jets have appeared~\cite{Maltoni:2016ays, Lee:2019lge,Llorente:2014bha,Li:2017wwc,Li:2021gjw, Craft:2022kdo,Cunqueiro:2022svx,Fedkevych:2022mid,Caletti:2023spr,Blok:2023ugf, Zhang:2023jpe,Dhani:2024gtx,Dhani:2025fbk,Ghira:2025nym}. 
In particular,  many of the observables discussed at length in this paper, such as jet angularities, ECFs, EEC, \SD $\theta_g$ and $z_g$, and the Lund plane density, when measured on heavy-flavour jets provides us with a powerful tool to explore the dynamics of heavy-flavour and dead-cone effects. Consequently, heavy-flavour jet substructure has become an active area of theoretical and experimental research. 

Because of its potential for interesting Standard Model measurements
across a wide range of scales, jet substructure has also recently
found applications in heavy-ion collisions.
One of the approaches to study the quark-gluon plasma is by analysing
how high-energy objects are affected by their propagation through it.
The LHC is the first collider where jets are routinely used for this
type of studies and an increasing interest for jet substructure
observables has been seen very recently in the heavy-ion community.
This will for sure be an important avenue in the future of jet
substructure, including the development of specific observables to
constraint the properties of the quark-gluon plasma and their study in
QCD.

The analysis of cosmic ray interactions is a further area of research
where jet substructure techniques were introduced to study the
detailed structure of complicated objects \cite{Brooijmans:2016lfv,
  Aab:2018jpg}. Ultra-high-energy cosmic rays, e.g. protons, can
produce interactions with very high momentum transfer between when
they scatter of atoms of Earth's atmosphere. Such interactions produce
a collimated high-multiplicity shower of electrons, photons and
muons. Their spacial distribution and penetration depth can be
analysed to inform the nature of the incident particle and interaction
in the collision. It is likely that in the near future, with the
increased interest in so-called beam-dump experiments, more ideas are
going to be introduced where jet substructure techniques can become of
importance.

Finally, one should also expect the future to deliver its fair share
of new tools for searches and measurements.
We believe that there are two emblematic directions worth exploring.
An obvious direction is the one of artificial intelligence.
Novel machine learning techniques are having a profound impact on all aspects of particle physics, ranging from analysis techniques to numerical simulations. Particularly relevant for jet physics is the use of algorithms to perform classification.
Indeed, this is an increasingly hot topic in the jet substructure community as testified by the popularity of the workshop ML4Jets and one
should expect it to continue growing in importance. 
In the context of the first-principle understanding used throughout
this book, one should highlight that it is important to keep in mind
that applying machine-learning techniques to jet substructure is not
just a problem for computer scientists. These algorithms are to a large
extent dealing with QCD and therefore a good control of the QCD
aspects of jet substructure is crucial. Several examples of this have
appeared very recently --- like QCD-aware
networks~\cite{Louppe:2017ipp}, energy-flow polynomials and
networks~\cite{Komiske:2017aww,Komiske:2018cqr} or the Lund jet
plane~\cite{Dreyer:2018nbf} --- and we should definitely expect more
in the future.
One can even imagine to extend concepts developed for jets to be
applied to the full event, i.e. a full-information approach to study
the whole radiation profile of an event. This could maximise the
sensitivity of collider experiments in searches for new physics.
The second direction we want to advocate for is the development of
additional tools which are theory-friendly, i.e.\ that are under
analytical control and are amenable for precision calculations.
As shown in this book, basic substructure tools have now been
understood from first-principles, including the main physics aspects
responsible for the trade-off between performance and
robustness. However, modern boosted jet taggers involve several of
these tools in order to maximise performance (cf.\ our discussion in
chapter~\ref{tools}). 

We think that new tools offering a combination of grooming,
prong-finding and radiation constraints will always be of great value.
Compared to a deep-learning-based tool, this
might show a small loss in performance, but it would offer the
advantage of a better control of its behaviour across a wide range of
processes and studies. One of the key ingredients here is that these
new tools should remain as simple as possible to facilitate their
calibration in an experimental context, hopefully resulting in small
systematic uncertainties. This would make them usable for the
precision programme at the LHC, including both measurements and
searches.
From an analytic perspective, achieving precision for such
substructure algorithms will also require further developments in
resummation techniques and fixed-order (amplitude) calculations, where
many promising results have already been obtained recently. 

% extensions to other fields

All this being said, we hope that we have conveyed the idea that jet
substructure has been a fascinating field for more than a decade, with an
ever-growing range of applications. Over this time-span, the field has
managed to stay open to new ideas and new approaches. One should
therefore expect more exciting progress in the years to come.
We therefore hope that this book will constitute a good introduction
for newcomers to the field.

\vspace{0.5cm}

\begin{center}
\emph{If you ain't boostin' you ain't living}\\
\emph{\textexclamdown Boostamos!}~\cite{boostamos}
\end{center}

%% GS helper for auctex
%%% Local Variables:
%%% mode: latex
%%% TeX-master: "notes"
%%% End:

\appendix

% $Id: appendix-analytic-details.tex 610 2026-08-04 15:41:32Z smarzani $
\chapter{Details of analytic calculations}\label{chap:app-analytic-details}

In this appendix we detail the analytic calculations that we have to
perform in order to obtain the resummed exponents discussed in the
main text. As an example we consider the plain jet mass distribution
discussed in Chapter~\ref{chap:calculations-jets}. The generalisation
to other jet substructure observables merely adds additional
phase-space constraints, yielding longer expressions without changing
the steps of the calculation. It is left as an exercise for the
interested reader.

We therefore consider the resummed expression
Eq.~(\ref{eq:res-mass-cont}) and we focus on the resummed exponent
(focusing here on a quark-initiated jet, although similar results can
trivially be obtained for gluon-initiated jets)
\begin{align}\label{app:analytic-start}
R(\rho)=\int_\rho^1 \frac{d \rho'}{\rho'} \int_{\rho'}^1 dz  P_{q}(z)
  \frac{\as( \sqrt{z \rho'} R \mu)}{2 \pi} ,
\end{align}
where $\mu$ is the hard scale of the process, i.e.\ $\mu=\frac{Q}{2}$
for electron-positron collisions or $\mu=p_t$ for proton-proton
collisions, while as usual $R$ is the jet radius, and
$\rho=\tfrac{m^2}{\mu^2R^2}$. For the above expression to capture the
resummed exponent to NLL accuracy in the small-$R$ limit, we need to
make sure that 
\begin{itemize}
\item the running of the coupling is considered at two loops, i.e.\ with $\beta_0$ and $\beta_1$:
\begin{align}\label{app:running-coupling}
\as(k_t)=\frac{\as(R\mu)}{1+\tlambda}\left[ 1- \as(\mu R)
  \frac{\beta_1}{\beta_0} \frac{\log(1+\tlambda)}{1+\tlambda}\right],
  \quad \tlambda=2\as(R\mu) \beta_0 \log\left(\frac{k_t}{R\mu}\right),
\end{align}
where the $\beta$ function coefficients
$\beta_0$ and $\beta_1$ are 
\begin{equation}
\beta_0 = \frac{11 C_A - 2 n_f }{12 \pi}, \quad \beta_1 = \frac{17 C_A^2 - 5 C_A n_f -3 C_F n_f}{24 \pi^2}.
\end{equation}
\item the splitting function is considered at one loop;
\item the coupling is considered in the CMW scheme (or equivalently
  the soft contribution to the two-loop splitting function is
  included), cf.~Eq.~(\ref{eq:CMW}).
\end{itemize}

As a warm up, let us first evaluate the above integral to LL where, we
can limit ourselves to the soft limit of the splitting function and to
the one-loop approximation for the running coupling. We have (with
$\lambda'=\as \beta_0\log(\rho')$ and $\lambda"=\as \beta_0\log(z)$)
\begin{align}\label{app:analytic-LL}
R^{\text{(LL)}} & = \int_\rho^1 \frac{d \rho'}{\rho'} \int_{\rho'}^1 \frac{d
                  z}{z}\frac{\as( \sqrt{z \rho'} R\mu)C_F}{\pi}  \\
  &=
\frac{ \as C_F}{\pi} \ \int_\rho^1 \frac{d \rho'}{\rho'}
    \int_{\rho'}^1 \frac{d z}{z} \frac{1}{1+ \as \beta_0 \log ( z
    \rho')}  \nonumber \\
& =\frac{C_F}{\as \pi \beta_0^2}  \int_\frac{-\lambda}{2}^0 d \lambda'
    \int_{\lambda'}^0 d \lambda'' \frac{1}{1+\lambda'+\lambda'' } \nonumber\\
&= \frac{C_F}{2 \pi \beta_0^2 \as} \left[(1- \lambda) \log(1-\lambda)-2\left (1-\frac{\lambda}{2}\right) \log \left(1-\frac{\lambda}{2}  \right)\right],\nonumber\\
&= \frac{C_F}{2 \pi \beta_0^2 \as} \left[W(1- \lambda)-2W\left (1-\frac{\lambda}{2}\right)\right],\nonumber
\end{align}
where $\as\equiv\as(R\mu)$ is the $\MSb$ coupling,
$\lambda=2 \as \beta_0 \log \big(\frac{1}{\rho}\big)$, and $W(x)=x\log(x)$. The
above result can be then easily recast in the form of the $f_1$
function Eq.~(\ref{eq:quark}), which appears in the expression for the
resummed exponent Eq.~(\ref{eq:radiator-nll-expansion}).

Next, we consider the inclusion of the hard-collinear contribution. For this we have to include regular part of the splitting function. Thus, we have to evaluate the following integral:
\begin{align}\label{app:analytic-B-term}
  \delta R^\text{(hard-collinear)}
  &= \int_\rho^1 \frac{d \rho'}{\rho'} \int_{\rho'}^1 \frac{d z}{z}\left[P_q(z)- \frac{2}{z} \right] \frac{\as( \sqrt{z \rho'} R\mu)}{\pi}
\nonumber \\
&= \frac{2C_F \as }{\pi}  \int_\rho^1 \frac{d \rho'}{\rho'} \int_{\rho'}^1 d z\left[-1+\frac{z}{2} \right] \frac{1}{1+ \as \beta_0 \log ( z \rho')}.
\end{align}
When evaluating the expression above to NLL we can make the further
simplifications that, since we are working in the hard-collinear
limit, we can set $z=1$ in the running-coupling contribution.
We are left with an integral over $z$ with no logarithmic
enhancement so, up to power corrections in $\rho$, we can safely set
the lower limit of integration to $z=0$. The two integrals decouple
and we find
\begin{align}\label{app:analytic-B-term-ctd}
\delta R^\text{(hard-collinear)}
  &= \frac{C_F \as }{\pi}  \int_{0}^1 dz
    \left[-1+\frac{z}{2} \right]  \int_\rho^1 \frac{d
    \rho'}{\rho'}\frac{1}{1+ \as \beta_0 \log(\rho')}
  =-\frac{C_F }{\pi \beta_0}B_q \log\left( 1-\frac{\lambda}{2}\right),
\end{align}
with
\begin{equation}\label{eq:definition-B-quark}
  B_q = \int_0^1 dz \left[\frac{P_q(z)}{2C_F}-\frac{1}{z} \right] =  \int_{0}^1 dz \left[-1+\frac{z}{2} \right] = -\frac{3}{4},
\end{equation}
already defined in Eq.~(\ref{eq:B1}).
Note that for a gluon-initiated jet one should instead use the gluon
splitting function, Eq.~(\ref{eq:gluonsplitting}), which includes a
contribution from $g\to gg$ splitting and one from $g\to q\bar q$
splitting:
\begin{equation}\label{eq:definition-B-gluon}
  B_g =  \int_0^1 dz \left[\frac{P_g(z)}{2C_A}-\frac{1}{z} \right] =
  -\frac{11 C_A-2n_f}{12 C_A}.
\end{equation}
Since hard-collinear splittings often have a large numerical impact
and are relatively easy to include, one often works in the {\em
  modified LL approximation} where one includes the LL contribution
$R^\text{(LL)}$ as well as hard-collinear splittings,
$\delta R^\text{(hard-collinear)}$.

Before moving on to the other NLL contributions to the Sudakov
exponent, we would like to comment on an alternative way to achieve
modified leading logarithmic accuracy and include the ``$B$-term'' in the LL
expressions. We note that if we replace the actual splitting function
by any other expressions which behaves like $\tfrac{2C_i}{z}$ at small
$z$ and reproduces the correct $B_i$ term in
Eqs.~\eqref{eq:definition-B-quark} and~\eqref{eq:definition-B-gluon},
we would then recover the same modified-LL behaviour.
In particular, we can use
\begin{equation}\label{eq:splitting-B-term}
  P_i^\text{(modified-LL)}(z) =
  \frac{2C_i}{z}\Theta\big(z<e^{B_i}\big).
\end{equation}
This is equivalent to imposing a cut on $z$ in the LL integrals.
For example, Eq.~(\ref{app:analytic-LL}) would become
\begin{align}\label{app:analytic-LL-modB}
  R^{\text{(modified-LL)}}
  & = \int_\rho^1 \frac{d \rho'}{\rho'}
    \int_{\rho'}^{e^{B_i}} \frac{d z}{z}
    \frac{\as( \sqrt{z \rho'} R\mu)C_F}{\pi}  \\
  &= \frac{C_i}{2 \pi \beta_0^2 \as} \left[
    W(1- \lambda)
    -2W\left (1-\frac{\lambda+\lambda_B}{2}\right)
    +W(1-\lambda_B)\right],\nonumber
\end{align}
with $\lambda_B = -2\as\beta_0B_i$.
It is straightforward to show that if we expand this to the first
non-trivial order in $\lambda_B$, one indeed recovers
$R^\text{(LL)}+\delta R^\text{(hard-collinear)}$.
This approach is what we have adopted for most of the results and
plots presented in this book.

Coming back to the full NLL accuracy for the resummed exponent, we also have to consider the
contribution of the two-loop running coupling:
\begin{align}\label{app:analytic-beta1}
  \delta R^{\text{(2-loop)}}
  & = -\frac{\as^2 C_i}{\pi}\frac{\beta_1}{\beta_0}
    \int_\rho^1 \frac{d \rho'}{\rho'} \int_{\rho'}^1 \frac{d z}{z} 
    \frac{\log(1+\as\beta_0\log(z\rho'))}{(1+\as\beta_0\log(z\rho'))^2}\\
  & = -\frac{C_i \beta_1 }{ \pi \beta_0^3}  \int_\frac{-\lambda}{2}^0 d \lambda' \int_{\lambda'}^0 d \lambda'' \; \frac{\log(1+\lambda'+\lambda'' )}{(1+\lambda'+\lambda'' )^2}\nonumber \\
&=\frac{C_i \beta_1}{2 \pi \beta_0^3} \left [ \log \left (1-\lambda \right )-2 \log 
\left (1-\frac{\lambda}{2} \right ) + \frac{1}{2} \log^2 \left (1-  \lambda \right ) 
- \log^2 \left (1-\frac{\lambda}{2} \right ) \right ],\nonumber
\end{align}
which provides the $\beta_1$ contribution to the NLL function $f_2$ defined in Eq.~(\ref{eq:radiator-nll-contribution}).

Finally, to NLL accuracy we also have to include the two-loop contribution to the splitting function in the soft limit. Because this contribution is universal it can be also expressed as a redefinition of the strong coupling, which give rise to the so-called CMW scheme Eq.~(\ref{eq:CMW}).
Thus, we have to evaluate the following integral
\begin{align}\label{app:analytic-CMW}
  \delta R^{\text{(CMW)}}
  &=\frac{2 C_i K}{4 \pi^2} \int_\rho^1 \frac{d \rho'}{\rho'} \int_{\rho'}^1 \frac{d z}{z} \as^2( \sqrt{z \rho'} R\mu) \\
&=\frac{C_i K }{ 2\pi^2 \beta_0^2}  \int_\frac{-\lambda}{2}^0 d \lambda' \int_{\lambda'}^0 d \lambda'' \frac{1}{(1+\lambda'+\lambda'' )^2}\nonumber \\
&=\frac{C_i K }{ 4\pi^2 \beta_0^2} \left[2 \log\left(1-\frac{\lambda}{2} \right) -\log(1-\lambda)\right],\nonumber
\end{align}
where the coupling in the first line can be evaluated at the one-loop
accuracy since higher-order corrections would be beyond NLL.
This contribution is the $K$ term in the NLL function $f_2$ defined in
Eq.~(\ref{eq:radiator-nll-contribution}).

The expressions in this appendix allow us to capture the global part
of resummed exponent to NLL, in the small-$R$ limit. Had we decided to
include finite $R$ correction, we would have considered also soft
emissions at finite angles, not just from the hard parton in the jet
but from all dipoles of the hard scattering process (see for instance
Sec.~\ref{sec:pp-collisions} and
Ref.~\cite{Dasgupta:2012hg}). Furthermore, we remind the reader that,
as discussed in Chapter~\ref{chap:calculations-jets}, in order to
achieve full NLL accuracy, one needs to consider non-global logarithms
as well as potential logarithmic contributions originating from the
clustering algorithm that is used to define the jet.

Finally, we note that the above expressions exhibit a singular behaviour at $\lambda=1$ and $\lambda=2$.
These singularities originate from the Landau pole of the perturbative QCD coupling and they signal the breakdown of perturbation theory. In phenomenological applications of analytic calculations, this infrared region is dealt by introducing a particular prescription. For instance, one could imagine freezing the coupling below  a non-perturbative scale $\mu_\text{NP}\simeq 1$~GeV
\begin{equation}\label{eq:coupling-freezing}
\bar{\alpha}_s (\mu)= \as(\mu)\Theta\left (\mu-\mu_\text{NP}\right)+\as(\mu_\text{NP})\Theta\left (\mu_\text{NP}-\mu\right).
\end{equation}
Other prescriptions are also possible. For example, in Monte Carlo
simulations, the parton showers is typically switched off at a cutoff
scale and the hadronisation model then fills the remaining
phase-space.

With the prescription Eq.~(\ref{eq:coupling-freezing}), the above
expressions for the Sudakov exponent are modified at large
$\lambda$. For completeness, we give the full expressions resulting
from the more tedious but still straightforward integrations.
To this purpose, it is helpful to introduce $W(x)=x\log(x)$,
$V(x)=\tfrac{1}{2}\log^2(x)+\log(x)$, and
$\lambda_\text{fr}=2\alpha_s\beta_0\log(\tfrac{\mu
  R}{\mu_\text{NP}})$.
For $\lambda<\lambda_\text{fr}$, \ie
$\rho>\tfrac{\mu_\text{NP}}{R\mu}$, we find
\begin{align}
  R^\text{(NLL)}(\lambda)
  & = R^\text{(modified-LL)}+\delta R^\text{(2-loop)}+\delta R^\text{(CMW)}\\
  & = \frac{C_i}{2\pi\alpha_s\beta_0^2}\Big[
    W(1-\lambda)-2W\big(1-\frac{\lambda+\lambda_B}{2}\big)+W(1-\lambda_B)
    \Big]\nonumber\\
  & +\frac{C_i\beta_1}{2\pi\beta_0^3}\Big[
    V(1-\lambda)-2V\big(1-\frac{\lambda+\lambda_B}{2}\big)+V(1-\lambda_B)
    \Big]\nonumber\\
  & -\frac{C_iK}{4\pi^2\beta_0^2}\Big[
    \log(1-\lambda)-2\log\big(1-\frac{\lambda+\lambda_B}{2}\big)+\log(1-\lambda_B)
    \Big],\nonumber
\end{align}
in agreement with Eqs.~(\ref{app:analytic-LL-modB}),
(\ref{app:analytic-beta1}) and~(\ref{app:analytic-CMW}) above.
We note that the above expressions have included the $B$ term using
the trick of Eq.~\eqref{eq:splitting-B-term} for all terms including
the two-loop and CMW corrections. In these terms, one can set
$\lambda_B=0$ at NLL accuracy. Although keeping these contribution has
the drawback of introducing uncontrolled subleading corrections, it
comes with the benefit of providing a uniform treatment of
hard-collinear splitting which places the endpoint of all the terms in
the resummed distribution at $\lambda=\lambda_B$.

For $\lambda_\text{fr}<\lambda<2\lambda_\text{fr}$, \ie
$\big(\tfrac{\mu_\text{NP}}{R\mu}\big)^2<\rho<\tfrac{\mu_\text{NP}}{R\mu}$,
we start being sensitive to the freezing of the coupling at
$\mu_\text{NP}$. In this case, we find
\begin{align}
  &R^\text{(NLL)}(\lambda)\\
  &\quad = \frac{C_i}{2\pi\alpha_s\beta_0^2}\Big[
    (1-\lambda)\log(1-\lambda_\text{fr})-2W\big(1-\frac{\lambda+\lambda_B}{2}\big)+W(1-\lambda_B)+\frac{1}{2}\frac{(\lambda-\lambda_\text{fr})^2}{1-\lambda_\text{fr}}
    \Big]\nonumber\\
  &\quad +\frac{C_i\beta_1}{2\pi\beta_0^3}\Big[
    \frac{1}{2}\log^2(1-\lambda_\text{fr})+\frac{1-\lambda}{1-\lambda_\text{fr}}\log(1-\lambda_\text{fr})-2V\big(1-\frac{\lambda+\lambda_B}{2}\big)+V(1-\lambda_B)\nonumber\\
  &\quad\phantom{+\frac{C_i\beta_1}{2\pi\beta_0^3}\Big[}
    -\frac{\lambda-\lambda_\text{fr}}{1-\lambda_\text{fr}}-\frac{1}{2}\frac{(\lambda-\lambda_\text{fr})^2}{(1-\lambda_\text{fr})^2}\log(1-\lambda_\text{fr})
    \Big]\nonumber\\
  &\quad -\frac{C_iK}{4\pi^2\beta_0^2}\Big[
    \log(1-\lambda_\text{fr})-2\log\big(1-\frac{\lambda+\lambda_B}{2}\big)+\log(1-\lambda_B)-\frac{\lambda-\lambda_\text{fr}}{1-\lambda_\text{fr}}-\frac{1}{2}\frac{(\lambda-\lambda_\text{fr})^2}{(1-\lambda_\text{fr})^2}
    \Big].\nonumber
\end{align}

Finally, for $\lambda>2\lambda_\text{fr}$, \ie
$\rho<\big(\tfrac{\mu_\text{NP}}{R\mu}\big)^2$, we have
\begin{align}
  R^\text{(NLL)}(\lambda)
  & = R^\text{(NLL)}(\lambda_\text{fr})\\
  & + \frac{C_i}{2\pi\alpha_s\beta_0^2}
    \frac{(\lambda-\lambda_B)^2-2(\lambda_\text{fr}-\lambda_B)^2}{4(1-\lambda_\text{fr})}
    \Big[
    1-\frac{\alpha_s\beta_1}{\beta_0}\frac{\log(1-\lambda_\text{fr})}{1-\lambda_\text{fr}}
    +\frac{\alpha_s K}{2\pi}\frac{1}{1-\lambda_\text{fr}}
    \Big].\nonumber
\end{align}

For all the analytic plots in this paper, we have used
$\alpha_s(M_Z)=0.1265$ (following the value used for the (one-loop)
running coupling in Pythia8 with the Monash 2013 tune), freezing
$\alpha_s$ at $\mu_\text{NP}=1$~GeV and used five active massless
flavours. 
Note finally that (modified)-LL results only include one-loop running
coupling effects.

\section{DGLAP splitting functions}\label{app:splitting_functions}
For most of the calculations described in this book, we only splitting functions that are inclusive over the final-state partons, see Eqs.~(\ref{eq:quarksplitting}) and~(\ref{eq:gluonsplitting}). However, when we discuss declustering observables such as the Lund plane in chapter~\ref{lundplane}, we have to keep track of the flavour evolution. To this purpose, we need the full matrix structure of DGLAP evolution in the singlet sector. 

We start by introducing the unregularised splitting functions:
\begin{align}
P_{qq}(z) &= C_F \left( \frac{1 + z^2}{1 - z} \right), 
&P_{gq}(z) = P_{qq}(1 - z), \nonumber\\
%P_{\mathcal{QQ}}(z) &= C_F \left( \frac{1 + z^2}{1 - z} - \frac{2 m^2 z(1 - z)}{q_t^2 + (1 - z)^2 m^2} \right), 
%&\quad P_{g\mathcal{Q}}(z) &= P_{\mathcal{QQ}}(1 - z), \nonumber\\
P_{gg}(z) &= 2C_A \left( \frac{z}{1 - z} + \frac{1 - z}{z} + z(1 - z) \right), 
%&\quad P_{\mathcal{Q}q}(z) &= P_{q\mathcal{Q}}(z) =0 , \nonumber\\
&P_{qg}(z) = T_R \left( z^2 + (1 - z)^2 \right), %\quad& P_{\mathcal{Q}g}(z) &= T_R \left( 1 - \frac{2 z(1 - z) q_t^2}{m^2 + q_t^2} \right),
\end{align}
where $C_F=\frac{4}{3}, \, C_A=3, \, T_R=\frac{1}{2}$ are the standard colour factors and Gell-Mann matrices normalisation, respectively.

The splitting kernels $P^{(R)}$ and  $P^{(V)}$ introduced in  Eq.~\eqref{eq: DGLAP type} read :
\begin{subequations}
\begin{align}
   P_{qq}^{(R)}(z) & = P_{qq}(z)\, \Theta(2z-1), &
   P_{qq}^{(V)}(z) & = P_{qq}(z), \\
   P_{gq}^{(R)}(z) & = P_{gq}(z)\,\Theta(2z-1),  &
   P_{gq}^{(V)}(z) & = 0, \\
   P_{gg}^{(R)}(z) & = P_{gg}(z)\,\Theta(2z-1), &
   P_{gg}^{(V)}(z) & = \frac{1}{2}P_{gg}(z) + n_f P_{qg}(z), \\
   P_{qg}^{(R)}(z) & = 2 n_fP_{qg}(z)\,\Theta(2z-1), &
   {P}_{qg}^{(V)}(z) & = 0,
   %{P}_{\mathcal{Q}\mathcal{Q}}^{(R)}(z) & = P_{qq}(z)\, \Theta(2z-1), &
   %{P}_{\mathcal{Q}\mathcal{Q}}^{(V)}(z) & = P_{qq}(z), \\
   % {P}_{\mathcal{Q}q}^{(R)}(z) & = 0, &
   %{P}_{\mathcal{Q}q}^{(V)}(z) & = 0, \\
   %{{P}}_{q\mathcal{Q}}^{(R)}(z) & = 0, &
   %{ {P}}_{q\mathcal{Q}}^{(V)}(z) & = 0, \\
   %{ {P}}_{g\mathcal{Q}}^{(R)}(z) & = P_{gq}(z)\, \Theta(2z-1),  &
   % P_{g\mathcal{Q}}^{(V)}(z) & = 0, \\
  %{{P}}_{\mathcal{Q}g}^{(R)}(z) & = 2 P_{q g}(z)\, \Theta(2z-1), & \quad  { {P}}_{\mathcal{Q}g}^{(V)}(z) & = 0.
\end{align}
\end{subequations}
Given the expressions of $P^{(R)}$ and $P^{(V)}$ we notice that Eq.~(\ref{eq: DGLAP type}) can also be expressed as:
\begin{align}
\label{eq: modified dglap}
    \frac{d}{dt} p_i(x,t)= \sum_{j=l,h,g} \int^1_x \frac{dz}{z} \left(\hat{P}_{ij}(z)-\delta \hat{P}_{ij}(z)\right) p_{j} \left(\frac{x}{z},t\right),
\end{align}
where $\hat{P}_{ij}(z)$ are the regularised splitting functions and $\delta\hat{P}_{ij}(z)= \hat{P}_{ij}\, \Theta(1-2z)$:
  \begin{align}
  \label{eq: reg splitting functions}
  \begin{array}{ll}
    \hat{P}_{qq}(z) = C_F\left(\frac{1+z^2}{1-z}\right)_+, &
    \hat{P}_{gq}(z) = P_{gq}(z), \\[1ex]
    \hat{P}_{gg}(z) = 2 C_A\left[\frac{z}{(1-z)_+}+\frac{1-z}{z}+z(1-z)\right]
    + 2\pi \beta_0 \delta(1-z), &
    \hat{P}_{qg}(z)= P_{qg}(z).
  \end{array}
\end{align}

We note that for $x>1/2$:
\begin{align}
    \int^1_x \frac{dz}{z} \delta\hat{P}_{ij}(z) p_{j} \left(\frac{x}{z},t\right)= \int^1_x \frac{dz}{z} \hat{P}_{ij}(z) p_{j} \left(\frac{x}{z},t\right) \Theta(1-2z)
=0.
\end{align}
Therefore, for $x>1/2$, Eq. (\ref{eq: modified dglap}) reduces to the standard DGLAP equation. 
Moreover, imposing the initial condition 
\begin{align}
\label{eq: initial cond}
    p_i(x,0) &= \delta_{ii_0} \delta(1-x),
\end{align}
 we can formally write the solution of the differential equation as:
\begin{align}
\label{eq: formal solution}
    p_{i|i_0}(x,t)= \delta_{i i_0}\delta(1-x)+\sum^\infty_{n=1} \frac{t^n}{n!} \bigotimes^n_{i=1}\hat{P}(z_i) \Theta(2z_i-1),
\end{align}
where:
\begin{align}
\label{eq: convolution def.}
    \bigotimes^n_{i=1}\hat{P}(z_i) \Theta(2z_i-1)= \int^1_{\frac{1}{2}} dz_1 \dots \int^1_{\frac{1}{2}} dz_n \hat{P}_{i j_1}(z_1) \dots \hat{P}_{j_{n-1} i_0}(z_n) \delta(z_1 \dots z_n-x).
\end{align}
From Eq. (\ref{eq: convolution def.}), we observe that the $n^{\text{th}}$ term of the series in eq. (\ref{eq: formal solution}) has support only for $x>2^{-n}$.
The actual solution can be obtained by either solving the evolution equation with numerical methods or analytically by considering Mellin moments and subsequently performing a numerical inverse transform to return to \( x \)-space.

%% GS helper for auctex
%%% Local Variables:
%%% mode: latex
%%% TeX-master: "notes"
%%% End:

%  LocalWords:  Eq NLL CMW Eqs eq Monash

% $Id: appendix-simulation-details.tex 435 2018-12-17 22:53:31Z mspannow $
\chapter{Details of Monte Carlo simulations}\label{chap:app-simulation-details}

In this appendix, we provide the details of the parton-shower Monte Carlo
simulations presented throughout this book.

For all the results shown in
Chapters~\ref{calculations-substructure-mass},
\ref{sec:calc-shapes-qg} and \ref{sec:curiosities}, we have used the
Pythia~8 generator~\cite{Sjostrand:2014zea,Sjostrand:2007gs} (version
8.230) with the Monash 2013 tune~\cite{Skands:2014pea}.
The analytic results are always compared to Monte Carlo results at
parton level, where both hadronisation and the Underlying Event have
been switched off.
The ``hadron level'' corresponds to switching on hadronisation but
keeping multi-parton interactions off, while the ``hadron+UE'' level
includes both hadronisation and the Underlying Event.
In the last two cases, $B$-hadrons have been kept stable for
simplicity.\footnote{Except for the groomed jet mass study in
  Sec.~\ref{sec:calc-groomed-mass-mc} where $B$-hadron decays are
  enabled.}
For the samples labelled as ``quark jets'', we have used Pythia's
dijet hard processes, keeping only the $qq\to qq$ matrix
elements. Similarly the ``gluon jet'' samples keep only the $gg\to gg$.

For all studies, jet reconstruction and manipulations are performed
using FastJet~\cite{Cacciari:2011ma,Cacciari:2005hq}
(version~3.3). Our studies include all the jets above the specified
$p_t$ cut and with $|y|<4$.
Substructure tools which are not natively included in FastJet are
available from {\tt {fastjet-contrib}}.

In the case of the discrimination between boosted W jets and QCD
jets in Chapter~\ref{chap:calc-two-prongs}, we have used the same
samples as those used in the initial Les-Houches 2017 Physics at TeV
colliders workshop.
These use essentially the same generator settings as above, but now
only up to the two hardest jets with $|y|<2.5$ are kept.

%% GS helper for auctex
%%% Local Variables:
%%% mode: latex
%%% TeX-master: "notes"
%%% End:

%  LocalWords:  Monash UE Pythia's qq gg natively Houches

\clearpage

\addcontentsline{toc}{chapter}{Bibliography}
\bibliographystyle{JHEP}
\bibliography{references}
\end{document}